\documentclass[epj]{svjour}
\pdfoutput=1 
\usepackage{preprintcover} 
\PreprintCoverPaperTitle{
Jet energy measurement and its systematic uncertainty in proton--proton collisions at $\sqrt{s}=7$~TeV 
with the ATLAS detector}

\PreprintIdNumber{CERN-PH-EP-2013-222}
\usepackage{atlasphysics} 
\usepackage{wrapfig}
\usepackage{graphicx}
\usepackage{mathptmx} 
\usepackage{amssymb}
\usepackage{amsmath}
\usepackage{helvet}
\usepackage{subfig}
\usepackage{epsfig}
\usepackage{multirow}
\usepackage{rotating}
\usepackage[switch]{lineno}
\usepackage{cite}

\usepackage{url}
\usepackage{hyperref}
\usepackage{stfloats} 
\usepackage{fixltx2e} 
\usepackage[mathcal]{eucal}

\newcommand{\antikt}{anti-$k_{t}$}
\newcommand{\Antikt}{Anti-$k_{t}$}
\newcommand{\asym}{\ensuremath{\mathcal{A}}}
\newcommand{\JER} {{\rm JER}}
\newcommand{\JES} {{\rm JES}}
\newcommand{\EMJES} {{\rm EM+JES}}

\newcommand{\LCWJES}{{\rm LCW+JES}}
\newcommand{\EM}    {{\rm EM}}

\newcommand{\LCW}   {{\rm LCW}}

\newcommand{\GS}    {{\rm GS}}

\newcommand{\MPF}   {{\rm MPF}}
\newcommand{\DB}   {{\rm DB}}
\newcommand{\MJB}   {{\rm MJB}}

\newcommand{\MPFbf}   {\textbf{MPF}}
\newcommand{\DBbf}   {\textbf{DB}}
\newcommand{\MJBbf}   {\textbf{MJB}}

\newcommand{\Etmiss}   {\ensuremath{{E}_{\mathrm{T}}^{\mathrm{miss}}}}
\newcommand{\vecEtmiss}{\ensuremath{\vec{E}_{\mathrm{T}}^{\mathrm{miss}}}}

\newcommand{\mw}    {\ensuremath{M_{\rm W}}}
\newcommand{\DRjj}  {\ensuremath{\Delta R_{\rm jj}}}

\newcommand{\BCHCORRCELL} {\ensuremath{\rm{BCH}_{\rm cor,cell}}}
\newcommand{\BCHCORRJET}  {\ensuremath{\rm{BCH}_{\rm cor,jet}}}

\newcommand{\Response}  {\ensuremath{\mathcal{R}}}

\newcommand{\RMPF}      {\ensuremath{\Response_\MPF}}

\newcommand{\rtrk}             {\ensuremath{r_{\rm trk}}}

\newcommand{\Rmin}     {\ensuremath{R_{\rm min}}}
\newcommand{\DeltaR}   {\ensuremath{\Delta R}}

\newcommand{\etaDet}{\ensuremath{\eta_{\rm det}}}

\newcommand{\etatrk}{\ensuremath{\eta^{\rm track}}}

\newcommand{\etaRange}[2]{\ensuremath{{#1}\leq|\eta|<{#2}}}
\newcommand{\AetaRange}[1]{\ensuremath{|\eta|<{#1}}}

\newcommand{\gammajet}{\ensuremath{\gamma\text{--jet}}} 
\newcommand{\Zjet}{\ensuremath{\Zboson}--jet}
\newcommand{\deltaphijetgamma}{\ensuremath{\Delta \phi_{{\rm jet}\mbox{-}\ensuremath{\gamma}}}}
\newcommand{\deltaphijetZ}{\ensuremath{\Delta \phi_{{\rm jet}\mbox{-}\ensuremath{Z}}}}

\newcommand{\insitu}{in situ}
\newcommand{\Insitu}{In situ}

\newcommand{\ptavg}  {\ensuremath{\pt^\mathrm{avg}}}
\newcommand{\ptjet}  {\ensuremath{\pt^\mathrm{jet}}}
\newcommand{\ptRecoil}{\ensuremath{\pt^\mathrm{Recoil}}}

\newcommand{\ptjetEM}{\ensuremath{p_{\rm T, \EM}^\mathrm{jet}}}
\newcommand{\ptjetLCW}{\ensuremath{p_{\rm T, \LCW}^\mathrm{jet}}}

\newcommand{\ptprobe}{\ensuremath{\pt^{\mathrm{probe}}}}
\newcommand{\ptref}  {\ensuremath{\pt^{\mathrm{ref}}}}
\newcommand{\ptRange}[2]{\ensuremath{{#1} \leq \ptjet<{#2} \GeV}}
\newcommand{\pttrue}  {\ensuremath{\pt^\mathrm{truth}}}
\newcommand{\pttruth} {\ensuremath{\pt^{\mathrm{truth}}}}
\newcommand{\pttrk}   {\ensuremath{\pt^{\mathrm{track}}}}

\newcommand{\pToff}{\ensuremath{\mathcal{O}}}

\newcommand{\pTsup}[1]{\ensuremath{p_{\mathrm{T}}^{\mathrm{#1}}}}
\newcommand{\pTind}[2]{\ensuremath{p_{\mathrm{T,#1}}^{\mathrm{#2}}}}

\newcommand{\pTrec}[1]{\pTind{#1}{jet}}
\newcommand{\pTgam}{\ensuremath{\pTsup{\gamma}}}
\newcommand{\ptgamma}{\ensuremath{\pTsup{\gamma}}}
\newcommand{\pTcor}[1]{\pTind{#1}{corr}}
\newcommand{\ptsecondjet}{\ensuremath{\pt^{\mathrm{jet2}}}}

\newcommand{\ptl}{\ensuremath{\pt^{\rm left}}}
\newcommand{\ptr}{\ensuremath{\pt^{\rm right}}}

\newcommand{\etajet}{\ensuremath{\eta}}

\newcommand{\TrkJet}  {\ensuremath{\mathrm{track\ jet}}}
\newcommand{\ptjetTrk}{\ensuremath{\pt^{\TrkJet}}}

\newcommand{\mvone}  {{\sc MV1}}

\newcommand{\alpgen}  {{\sc Alpgen}}
\newcommand{\pythia}  {{\sc Pyth\-ia}}
\newcommand{\Perugia} {{\sc Perugia}}
\newcommand{\geant}   {G{\sc eant}4}
\newcommand{\atlasfast}{{\sc ATLFAST}2}
\newcommand{\herwig}  {{\sc Herwig}}
\newcommand{\herwigpp}{{\sc Herwig++}}
\newcommand{\jimmy}   {{\sc Jimmy}}
\newcommand{\ACERMC}{{\sc ACERMC}}
\newcommand{\PowHeg}{{\sc POWHEG}}
\newcommand{\powheg}{{\sc POWHEG}}
\newcommand{\mcatnlo}  {{\sc MC@NLO}}
\newcommand{\MCAtNLO}  {{\sc MC@NLO}}
\newcommand{\madgraph}  {{\sc MadGraph}}

\newcommand{\topo} {topo-clust\-er}

\newcommand{\topos}{topo-clust\-ers}

\newcommand{\Npv}{\ensuremath{N_{{\rm PV}}}}
\newcommand{\Nref}{\ensuremath{\Npv^{\rm ref}}}

\newcommand{\axing}{\ensuremath{\mu}}
\newcommand{\axingRef}{\ensuremath{\axing^{\mathrm{ref}}}}

\newcommand{\NpvRef}{\Nref}

\newcommand{\ns}   {{\rm ns}}
\newcommand{\mm}   {{\rm mm}}

\newcommand{\ATLAS}   {ATLAS}
\newcommand{\Lone}    {\texttt{L1}}
\newcommand{\HLT}     {\texttt{HLT}}

\newcommand{\LHC}    {LHC}

\newcommand{\HEC}    {\texttt{HEC}}
\newcommand{\LAr}    {\texttt{LAr}}
\newcommand{\FCal}   {\texttt{FCal}}
\newcommand{\Tile}   {\texttt{Tile}}

\newcommand{\btagged}{{\ensuremath{b}\mbox{\rm-tagged}}}
\newcommand{\bquark }{{\ensuremath{b}\mbox{\rm-quark}}}
\newcommand{\bquarks}{{\ensuremath{b}\mbox{\rm-quarks}}}

\newcommand{\cquarks}{{\ensuremath{c}\mbox{\rm-quarks}}}
\newcommand{\bjet}{\text{\textit{b}-jet}}
\newcommand{\bjets}{\text{\textit{b}-jets}}

\newcommand{\cjets}{\text{\textit{c}-jets}}
\newcommand{\kooc}{\ensuremath{k_\mathrm{OOC}}}

\newcommand{\JVF}{\ensuremath{\mathrm{JVF}}}

\newcommand{\PAC}[1]{\ensuremath{\alpha^{#1}}}
\newcommand{\PBC}[1]{\ensuremath{\beta^{#1}}}

\newcommand{\alphaEM}{\PAC{\EM}}
\newcommand{\alphaLCW}{\PAC{\LCW}}
\newcommand{\betaEM}{\PBC{\EM}}
\newcommand{\betaLCW}{\PBC{\LCW}}
\newcommand{\alphaEMFct}{\ensuremath{\alphaEM(\etaDet)}}
\newcommand{\betaEMFct}{\ensuremath{\betaEM(\etaDet)}}
\newcommand{\alphaLCWFct}{\ensuremath{\alphaLCW(\etaDet)}}
\newcommand{\betaLCWFct}{\ensuremath{\betaLCW(\etaDet)}}
\newcommand{\alphaFct}{\ensuremath{\alpha(\etaDet)}}
\newcommand{\betaFct}{\ensuremath{\beta(\etaDet)}}

\newcommand{\ejets}{\ensuremath{\mbox{e+jets}}\xspace}
\newcommand{\mjets}{\ensuremath{\mu\mbox{+jets}}\xspace}
\newcommand{\ljets}{\ensuremath{\mbox{{\em l}+jets}}\xspace}
\newcommand{\Rjpesl}{\ensuremath{\mbox{$\alpha_l$}}}

\newcommand{\mylumi}{ $4.7$~\ifb}
\newcommand{\pp}{pro\-ton--pro\-ton}
\newcommand{\etaic}{\ensuremath{\eta}-in\-ter\-ca\-lib\-ra\-ti\-on}
\newcommand{\etacref}{central reference method}
\newcommand{\etamm}{matrix method}
\newcommand{\etaDetProbe}{\ensuremath{\eta_{\mathrm{det}}^{\mathrm{probe}}}}

\newcommand{\etaDetLR}[1]{\ensuremath{\eta_{\mathrm{det}}^{\mathrm{#1}}}}
\newcommand{\eqRef}[1]{Eq.~\eqref{#1}}

\newcommand{\figRef}[1]{Fig.~\ref{#1}}
\newcommand{\FigRef}[1]{Figure~\ref{#1}}
\newcommand{\myarraystretch}{1.25}
\newcounter{listctr}
\newcounter{subctr}
\newenvironment{mylist}{\begin{list}{\textbf{\arabic{listctr}}}{\usecounter{listctr}}}{\end{list}}
\newcommand{\myitem}[1]{\item{\textbf{#1}}\\}
\newenvironment{alphalist}{\begin{list}{(\alph{subctr})}{\usecounter{subctr}}}{\end{list}}

\newcommand{\jcloose}{\textsc{Loose}}
\newcommand{\jclooser}{\textsc{Looser}}
\newcommand{\jcmedium}{\textsc{Medium}}
\newcommand{\jctight}{\textsc{Tight}}
\newcommand{\cms}{centre-of-mass energy}
\newcommand{\rrextrap}{\ensuremath{\mathcal{R}_{\mathrm{extrap}}}}
\newcommand{\rrextrapfct}[2]{\ensuremath{\rrextrap(#1,#2)}}
\newcommand{\npdet}{{\sc Detector}}
\newcommand{\npmodel}{{\sc Model}}
\newcommand{\npstatmeth}{{\sc Stat/Meth}}
\newcommand{\npmixed}{{\sc Mixed}}
\newcommand{\trkjetid}{{\rm track\,jet}}
\renewcommand{\ptjetTrk}{\ensuremath{p_{\mathrm{T}}^{\trkjetid}}}
\newcommand{\pttrkvec}{\ensuremath{\vec{p}_{\mathrm{T}}^{\;\mathrm{track}}}}
\renewcommand{\DRjj}{\ensuremath{\Delta R_{\text{jj}}}}
\newcommand{\yphispace}{(\ensuremath{y,\phi}) space}
\newcommand{\etaphispace}{(\ensuremath{\eta,\phi}) space}
\newcommand{\ptrecoil}{\ensuremath{p_{\mathrm{T}}^{\mathrm{recoil}}}}

\newcommand{\ptrecoilvec}{\ensuremath{\vec{p}_{\mathrm{T}}^{\,\mathrm{recoil}}}}
\newcommand{\ptleadvec}{\ensuremath{\vec{p}_{\mathrm{T}}^{\,\mathrm{leading}}}}
\renewcommand{\ptRecoil}{\ptrecoil}
\newcommand{\deltaphi}[2]{\ensuremath{\Delta\phi(#1,#2)}}

\newcommand{\sumet}{\ensuremath{\Sigma E_{\mathrm{T}}}}
\newcommand{\MC}{MC}
\newcommand{\datatomc}{da\-ta-to-MC}
\newcommand{\Datatomc}{Da\-ta-to-MC}
\newcommand{\subleading}{sub-leading}
\newcommand{\nonleading}{non-leading}
\newcommand{\jetnum}[1]{\ensuremath{\text{jet}#1}}

\newcommand{\ds}{da\-ta\-set}
\newcommand{\Ds}{Da\-ta\-set}

\newcommand{\secRef}[1]{Sect. \ref{#1}}
\newcommand{\SecRef}[1]{Section \ref{#1}}

\newcommand{\ptordered}{\pt-ordered}
\newcommand{\sigpu}{\ensuremath{\sigma_{\text{noise}}^{\text{pile-up}}}}
\newcommand{\sigen}{\ensuremath{\sigma_{\text{noise}}^{\text{electronic}}}}
\newcommand{\mfakebf}[1]{\textit{\bfseries #1}}
\newcommand{\ZJET}{\mfakebf{Z}--jet}
\newcommand{\GAMMAJET}{\ensuremath{\boldsymbol{\gamma}}--jet}
\newcommand{\BJET}{\mfakebf{b}--jet}
\newcommand{\BTAGGING}{\mfakebf{b}-tagging}
\newcommand{\ETAIC}{\textbf{\ensuremath{\eta}}-in\-ter\-ca\-lib\-ra\-ti\-on}
\newcommand{\PTBF}{{\sffamily \textit{\bfseries p}$_{\textsf{T}}$}}
\newcommand{\ptbf}{\textit{\bfseries p}$_{\textrm{\bfseries T}}$}
\renewcommand{\mw}{\ensuremath{m_{\Wboson}}}
\newcommand{\mtw}{\ensuremath{m_{\text{T}}}(\Wboson)}
\newcommand{\mwrec}{\ensuremath{m_{\Wboson}^{\text{rec}}}}
\newcommand{\wmvone}{\ensuremath{w_{\text{\mvone}}}}
\newcommand{\ptjetn}[1]{\ensuremath{p_{\text{T}}^{\,\jetnum{#1}}}}
\PreprintCoverAbstract{
The jet energy scale (\JES) and its systematic uncertainty are determined for jets measured  
with the \ATLAS{} detector using \pp{} collision data with a centre-of-mass energy 
of $\sqrt{s}=7$~\TeV{} corresponding to an integrated luminosity of \mylumi. 
Jets are reconstructed 
from energy deposits forming
topological clusters of calorimeter cells using 
the \antikt{} algorithm with distance parameters $R=0.4$ or $R=0.6$, 
and are calibrated using \MC{} simulations.
A residual \JES{} correction is applied to account for differences between data and \MC{} simulations.
This correction and its systematic uncertainty 
are estimated using a combination of \insitu{} techniques exploiting the transverse momentum balance between a jet and 
a reference object such as a photon or a $Z$ boson, for \ptRange{20}{1000} and pseudorapidities \AetaRange{4.5}.  
The effect of multiple \pp{} interactions is corrected for, and an uncertainty is evaluated
using \insitu{} techniques. 
The smallest \JES{} uncertainty of less than $1\%$ is found in the central calorimeter region
(\AetaRange{1.2}) for jets with \ptRange{55}{500}. 
For central jets at lower \pt, the uncertainty is about $3 \%$.
A consistent \JES{} estimate is found using 
measurements of the calorimeter response of single hadrons in \pp{} collisions
and test-beam data,
which also provide the estimate for $\ptjet > 1$~\TeV. 
The calibration of forward jets is derived from dijet \pt{} balance measurements. 
The resulting uncertainty reaches its largest value of $6\%$ for low-\pt{} jets at $|\etajet|=4.5$. 
Additional \JES{} uncertainties due to specific event topologies, such as close-by jets
or selections of event samples with an enhanced content of jets originating from light quarks or gluons, are also discussed.
The magnitude of these uncertainties depends on the event sample used in a given physics analysis,
but typically amounts to 0.5\% to 3\%.

}
\PreprintJournalName{European Physics Journal C}  
\makeatletter
\g@addto@macro\bfseries{\boldmath}
\makeatother
\makeatletter
\renewcommand\dbltopfraction{0.75}
\renewcommand\dblfloatpagefraction{0.75}

\usepackage{xspace}
\usepackage{comment}
\usepackage{makeidx}
\makeindex
\hypersetup{
  colorlinks=true,
  linkcolor=blue,
  citecolor=red,
  urlcolor=red,
  pdftitle={
Jet energy measurement and its systematic uncertainty in proton--proton collisions at $\sqrt{s}=7$~TeV 
with the ATLAS detector}
  pdfauthor={ATLAS Collaboration},
  pdfpagemode={UseOutlines},
  bookmarksopen=true,
  bookmarksnumbered=true,
  pdfstartview={Fit}
}
\def\bfseries{\fontseries\bfdefault\selectfont\boldmath}
\def\itshape{\fontshape\itdefault\selectfont\let\mathrm=\mathit}
\begin{document}
\title{\vspace{-3.0cm}\flushleft{\normalsize\normalfont{CERN-PH-EP-2013-222}} 
\flushright{\vspace{-0.86cm}\normalsize\normalfont{Submitted Eur. Phys. J. C}}
\flushleft{
Jet energy measurement and its systematic uncertainty in proton--proton collisions at $\sqrt{s}=7$~TeV 
with the ATLAS detector
}}
\author{The \ATLAS{} Collaboration}
%
\newpage
%
%



%
\abstract{
The jet energy scale (\JES) and its systematic uncertainty are determined for jets measured  
with the \ATLAS{} detector using \pp{} collision data with a centre-of-mass energy 
of $\sqrt{s}=7$~\TeV{} corresponding to an integrated luminosity of \mylumi. 
Jets are reconstructed 
from energy deposits forming
topological clusters of calorimeter cells using 
the \antikt{} algorithm with distance parameters $R=0.4$ or $R=0.6$, 
and are calibrated using \MC{} simulations.
A residual \JES{} correction is applied to account for differences between data and \MC{} simulations.
This correction and its systematic uncertainty 
are estimated using a combination of \insitu{} techniques exploiting the transverse momentum balance between a jet and 
a reference object such as a photon or a $Z$ boson, for \ptRange{20}{1000} and pseudorapidities \AetaRange{4.5}.  
%
%
The effect of multiple \pp{} interactions is corrected for, and an uncertainty is evaluated
using \insitu{} techniques. 
%
%
The smallest \JES{} uncertainty of less than $1\%$ is found in the central calorimeter region
(\AetaRange{1.2}) for jets with \ptRange{55}{500}. 
For central jets at lower \pt, the uncertainty is about $3 \%$.
A consistent \JES{} estimate is found using 
measurements of the calorimeter response of single hadrons in \pp{} collisions
and test-beam data,
which also provide the estimate for $\ptjet > 1$~\TeV. 
The calibration of forward jets is derived from dijet \pt{} balance measurements. 
The resulting uncertainty reaches its largest value of $6\%$ for low-\pt{} jets at $|\etajet|=4.5$. 
Additional \JES{} uncertainties due to specific event topologies, such as close-by jets
or selections of event samples with an enhanced content of jets originating from light quarks or gluons, are also discussed.
The magnitude of these uncertainties depends on the event sample used in a given physics analysis,
but typically amounts to 0.5\% to 3\%.

\vspace{10.cm}
}
\authorrunning{\ATLAS{} Collaboration} 
\titlerunning{Measurements of jets with the \ATLAS{} detector}
%


%
%
\setcounter{tocdepth}{2}
\newpage
\tableofcontents
\newpage
\section{Introduction}
\label{sec:intro}
Jets are the dominant feature of high-energy, hard \pp{} interactions
at the Large Hadron Collider (\LHC) at CERN. 
They are key ingredients of many physics measurements and for searches 
for new phenomena.
In this paper, jets are observed as groups of topologically related energy deposits 
in the \ATLAS{} calorimeters, 
associated with tracks of charged particles as measured 
in the inner tacking detector.
They are reconstructed with the \antikt{} jet algorithm~\cite{Cacciari:2008gp}
and are calibrated using Monte Carlo (\MC) simulation.

A first estimate of the jet energy scale (\JES) uncertainty of 
about $5\%$ to $9\%$ depending on the jet transverse momentum ($\pt$),
 described in Ref.~\cite{atlasjet2010},
is based on information available before the first \pp{} collisions at the \LHC, 
and initial \pp{} collision data taken in $2010$.
A reduced uncertainty of about $2.5 \%$ in the central calorimeter region over a wide \pt{} range of  $60 \lesssim \pt < 800$~\GeV{} was achieved after applying the increased knowledge of the detector performance obtained during the analysis of this first year of \ATLAS{} data taking \cite{jespaper2010}. 
This estimation used single-hadron calorimeter response measurements, 
systematic variations of \MC{} simulation configurations,
and \insitu{} techniques, where the jet transverse momentum is compared to the \pt{} of a reference object.
These measurements were performed using the $2010$ \ds{}, corresponding
to an integrated luminosity of $38$~\ipb{} \cite{eppaper2010}.

During the year $2011$ the \ATLAS{} detector \cite{DetectorPaper} collected \pp{} collision data 
at a centre-of-mass energy of $\sqrt{s}=7$~\TeV, corresponding to an integrated luminosity of about \mylumi.
The larger \ds{} makes it possible to further improve the precision of the jet
energy measurement, and also to apply a correction derived from detailed comparisons
of data and \MC{} simulation using \insitu{} techniques.
This document presents the results of such an improved calibration of the jet energy measurement
and the determination of the uncertainties using the $2011$ \ds.

The energy measurement of jets produced in proton-proton and electron-proton collisions is also
discussed by other experiments \cite{cdf06,ref:D0_MPF,Abazov:2013hda,cmsJES,D0_jetcross,cdf_jetcross,ua183,h1_jes1,h1_jes2,zeus_jes1,zeus_jes2,Wing:2002fc}.

The outline of the paper is as follows.
\SecRef{sec:ATLAS} describes the ATLAS detector. The Monte Carlo simulation framework is presented in \secRef{sec:MC}, and the used \ds{} is described in \secRef{sec:dataset}. %
\SecRef{sec:jetrecocalib} summarises the jet reconstruction and calibration strategy.
The correction method for the effect of additional \pp{} interactions is discussed in 
\secRef{sec:pileupsection}. \SecRef{sec:InSitu} provides an overview of the techniques based on \pt{} balance that are described in
detail in 
Sects. \ref{sec:etaintercalibration} to \ref{sec:multijet}.
First the intercalibration between the central and the forward detector
using events with two high-\pt{} jets is presented in \secRef{sec:etaintercalibration}.
Then, \insitu{} techniques to assess differences of the jet energy measurement between data and Monte
Carlo simulation exploiting the \pt{} balance between a jet and a well-measured
reference object are detailed. The reference objects are \Zboson{} bosons in \secRef{sec:ZjetInSitu},
 photons in \secRef{sec:gammajetInSitu}, and a system of low-\pt{} jets in \secRef{sec:multijet}. 
The validation of the forward-jet energy measurements with \pt{} balance methods using \Zjet{} and \gammajet{} events follows in  \secRef{sec:closure}.
The strategy on how to extract a final jet calibration out of the combination of \insitu{} techniques, and the evaluation strategies for determining the corresponding systematic uncertainties, are discussed in \secRef{sec:strategy}. The same section also shows the final result of the jet calibration, including its systematic uncertainty,
from the combination of the \insitu{} techniques.

\SecRef{sec:SingleParticle} compares the \JES{} uncertainty as derived from the single-hadron 
calorimeter response measurements to that obtained from the \insitu{} method based on \pt{} balance discussed in the preceding sections.
Comparisons to \JES{} uncertainties using the \Wboson{} boson mass constraint in final states with hadronically decaying \Wboson{} bosons are presented in \secRef{sec:wmass}.  

Additional contributions to the systematic uncertainties of the jet measurement in \ATLAS{} are presented in Sects. \ref{sec:pileupsystematics} to \ref{sec:FlavorTopology}, where the correction for the 
effect of additional \pp{} interactions in the event, the presence of other close-by jets, and the response dependence on the jet fragmentation (jet flavour) are discussed.
The uncertainties for explicitly tagged jets with heavy-flavour content are outlined in \secRef{sec:bjets}.
A brief discussion of the correction of the calorimeter energy 
in regions with hardware failures and the associated uncertainty
on the jet energy measurement is presented in \secRef{sec:badjets}. 

A summary of the total jet energy scale uncertainty is given in \secRef{sec:Summary}. Conclusions follow in \secRef{sec:CONCLUSIONS}. A comparison of the systematic uncertainties of the \JES{} in \ATLAS{} with previous calibrations is presented in Appendix A.

\section{The ATLAS detector}
\label{sec:ATLAS}

\subsection{Detector description}
The \ATLAS{} detector consists of a tracking system (Inner Detector, or ID in the following),
sampling electromagnetic and hadronic calorimeters
and muon chambers. 
A detailed description of the \ATLAS{} experiment
can be found in Ref. \cite{DetectorPaper}. 

The Inner Detector has complete azimuthal coverage and spans the pseudorapidity\footnote{ATLAS uses a right-handed coordinate system with its origin at the nominal interaction point (IP) in the centre of the detector and the $z$-axis along the beam pipe. The $x$-axis points from the IP to the centre of the LHC ring, and the $y$-axis points upward. Cylindrical coordinates $(r,\phi)$ are used in the transverse plane, $\phi$ being the azimuthal angle around the beam pipe. The pseudorapidity is defined in terms of the polar angle $\theta$ as $\eta=-\ln\tan(\theta/2)$.} 
region $|\eta|<2.5$. It consists of layers of
silicon pixel detectors, silicon microstrip detectors and transition
radiation tracking detectors, all of which are immersed in a solenoid
magnet that provides a uniform magnetic field of 2~T. 

Jets are reconstructed using the ATLAS calorimeters, who\-se granularity and material 
varies as a function of $\eta$.
The electromagnetic calorimetry (\texttt{EM}) is provided by
high-granularity liquid-argon sampling calorimeters (\LAr), using lead as an absorber. It is 
divided into one barrel ($|\eta|<1.475$) and two end-cap ($1.375<|\eta|<3.2$) regions. The
hadronic calorimetry is divided into three distinct sections. 
The most central contains 
the central barrel region ($|\eta|<0.8$) and two extended barrel regions ($0.8<|\eta|<1.7$). 
These regions are instrumented with scintillator-tile/steel hadronic calorimeters (\Tile). 
Each barrel region consists of $64$ modules with individual $\phi$ coverages of $\sim 0.1$ rad. 
The two hadronic end-cap calorimeters (\HEC; $1.5<|\eta|<3.2$) feature liquid-argon/copper
calorimeter modules. The two forward calorimeters (\FCal; $3.1<|\eta|<4.9$) are instrumented with 
liquid-argon/copper and liquid-argon/tung\-sten modules to provide electromagnetic and hadronic energy measurements,
respectively.

The muon spectrometer surrounds the \ATLAS{} calorimeter. A system of three large 
air-core toroids, a barrel and two endcaps, generates a magnetic field in the
pseudorapidity range of $|\eta| < 2.7$. The muon spectrometer measures muon tracks 
with three layers of precision tracking chambers and is instrumented with separate trigger 
chambers.

The trigger system for the \ATLAS{} detector consists of a hardware-based Level~1~(\Lone) 
and a software-based High Level Trigger~(\HLT) \cite{triggerperformance}.  
At \Lone, jets are first built from coarse-granu\-la\-rity calorimeter towers using a sliding window algorithm, and then subjected
to early trigger decisions.
This is refined using jets reconstructed from
calorimeter cells in the \HLT, with algorithms similar to the ones applied offline.

\begin{figure*}[htp!]
\centering
\subfloat[$\sigma_{\mathrm{noise}}(\left|\eta\right|)$ in $2010$ ($\mu = 0$)]{\includegraphics[width=0.49\textwidth]{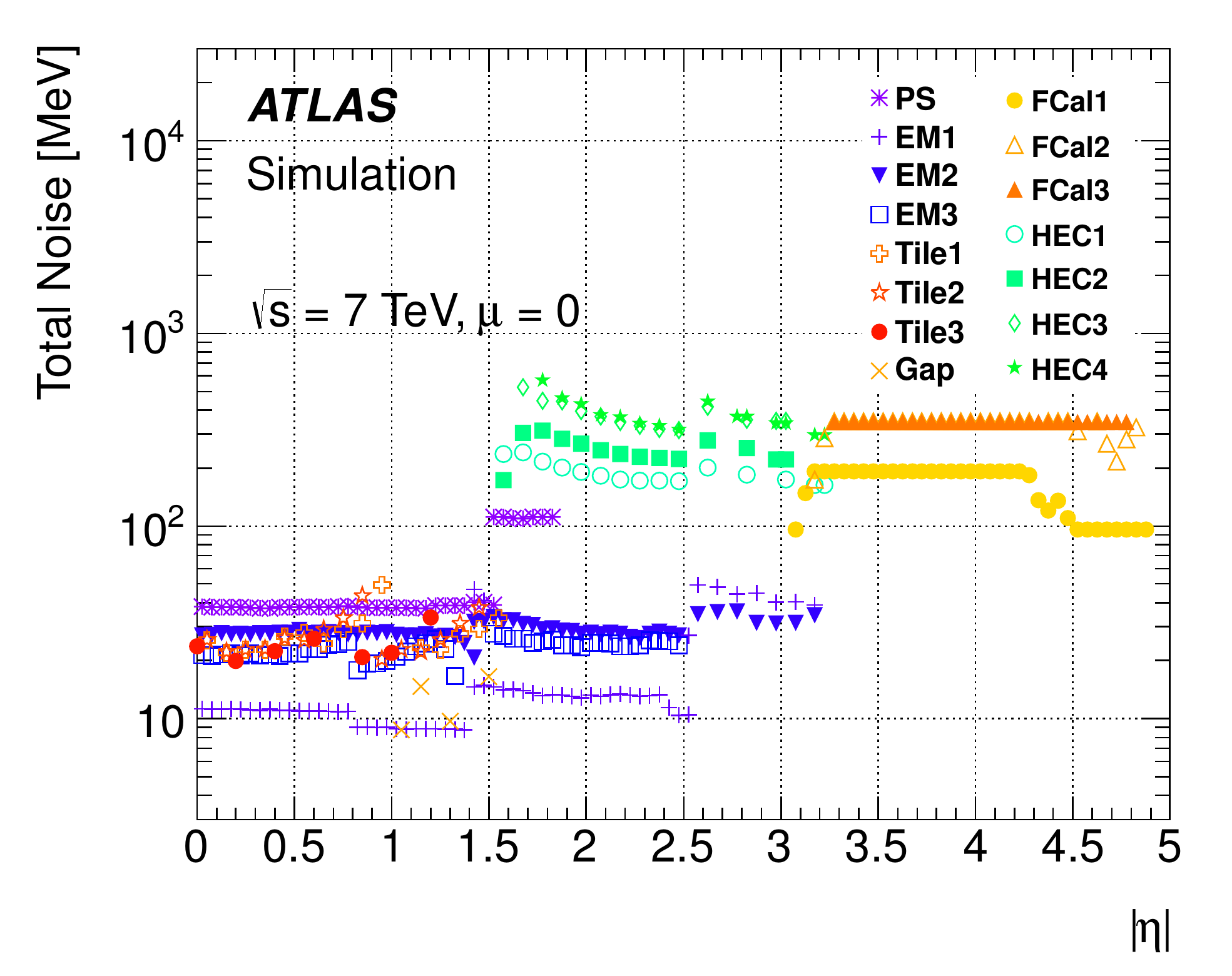} \label{fig:noise:2010}}
\subfloat[$\sigma_{\mathrm{noise}}(\left|\eta\right|)$ in $2011$ ($\mu = 8$)]{\includegraphics[width=0.49\textwidth]{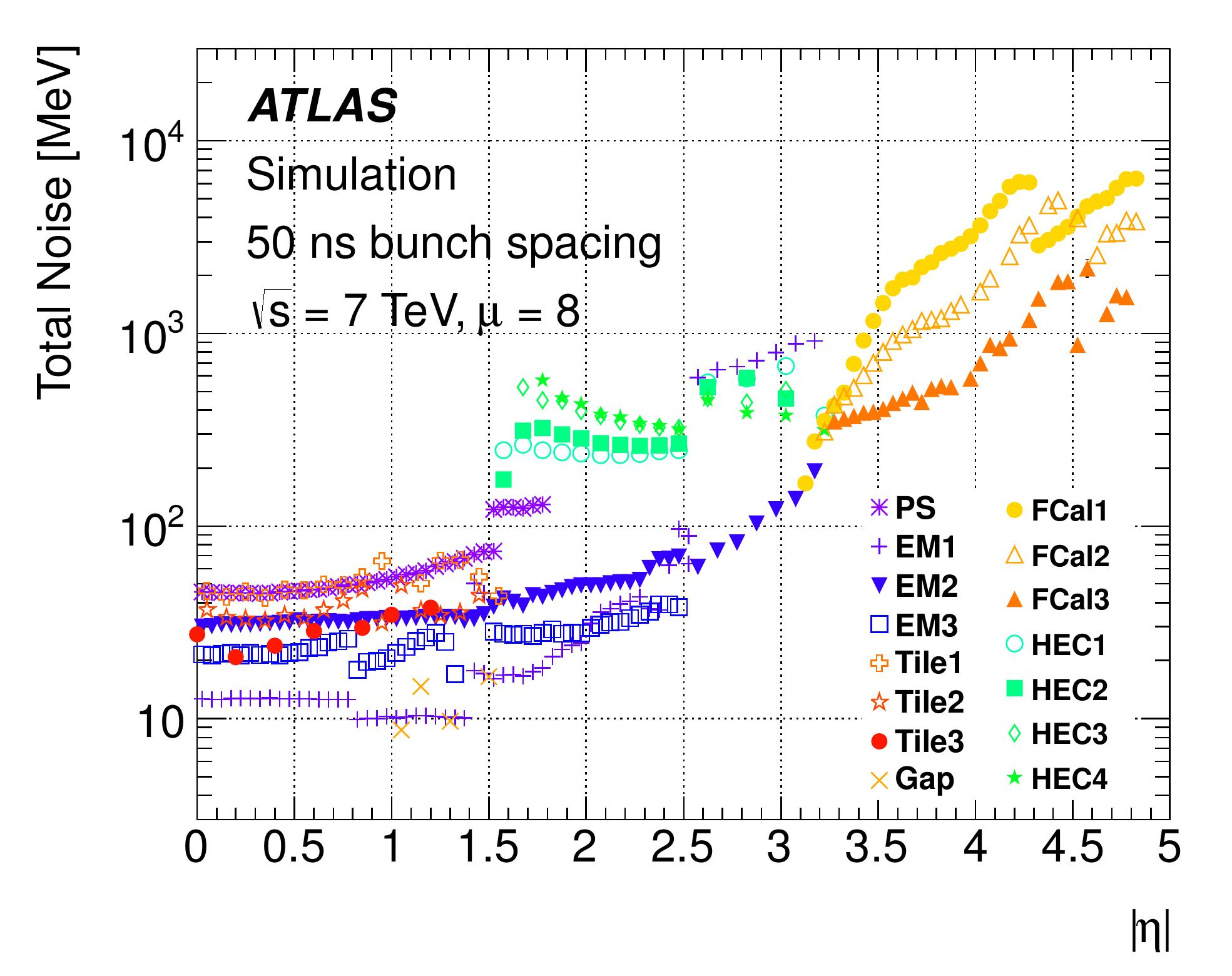} \label{fig:noise:2011}}
\caption[]{The energy-equivalent cell noise in the \ATLAS{} calorimeters on the electromagnetic (\EM) scale 
as a function of the direction $\left|\eta\right|$ in the detector, for the $2010$ configuration 
with \subref{fig:noise:2010} $\mu = 0$ and
the $2011$ configuration with   \subref{fig:noise:2011} $\mu = 8$. The various colours
indicate the noise in the pre-sampler (\texttt{PS}) and the up to three layers 
of the \LAr{} \texttt{EM} calorimeter, 
the up to three layers of the \Tile{} calorimeter, the four
layers for the hadronic end-cap (\texttt{HEC}) calorimeter, and the three 
modules of the forward (\texttt{FCal}) calorimeter. 
\label{fig:noise}}
\end{figure*}   

\subsection{Calorimeter pile-up sensitivity}
\label{sec:calopileup}
One important feature for the 
understanding of the contribution from additional \pp{} interactions (pile-up) to the signal in the $2011$ \ds{} is the sensitivity of
the \ATLAS{} liquid argon calorimeters to the bunch crossing history.
In any \LAr{} calorimeter cell, the reconstructed energy is sensitive to the \pp{} interactions occurring in approximately $12$ (2011 data, $24$ at \LHC{} design conditions) preceding and one immediately following bunch crossings (\emph{out-of-time pile-up}), in addition to pile-up interactions in the current bunch crossing (\emph{in-time pile-up}).
This is due to the relatively long charge collection time in these calorimeters
(typically $400- 600$~\ns), as compared to the bunch crossing intervals 
at the \LHC{} (design $25$~\ns{} and actually $50$~\ns{} in $2011$ data).
To reduce this sensitivity, a fast, bipolar shaped signal\footnote{The shaped pulse has a duration exceeding the charge collection time.} is used with net zero integral over time.

The signal shapes in the liquid argon calorimeters are optimised for this purpose,
leading to cancellation on average of in-time and out-of-time pile-up in any given calorimeter cell. By design of the shaping amplifier,
the most efficient suppression is achieved for $25$~ns bunch spacing in the LHC beams.
It is fully effective in the limit where, for each bunch crossing, about the same amount of energy is deposited
in each calorimeter cell.

The 2011 beam conditions, with $50$~ns bunch spacing and a relatively low cell occupancy from
the achieved instantaneous luminosities, do not allow for full pile-up
suppression by signal shaping, in particular in the central calorimeter region.
Pile-up suppression is further limited by large fluctuations in the number of additional
interactions from bunch crossing to bunch crossing, and in the energy flow patterns of the
individual collisions in the time window of sensitivity
of approximately $600$~ns. 
Consequently, the shaped signal extracted by digital filtering shows a principal sensitivity
to in-time and out-of-time pile-up,  in particular in terms of a residual non-zero cell-signal baseline. 
This baseline can lead to relevant signal offsets once the noise suppression, an important part of
the calorimeter signal extraction strategy presented in \secRef{sec:jetrecocalib}, is applied. 

Corrections mitigating the effect of these signal offsets on the reconstructed jet energy are discussed 
in the context of the pile-up suppression strategy in \secRef{sec:pileupmethod}.  
All details of the \ATLAS{} liquid argon
calorimeter readout and signal processing can be found in Ref.~\cite{LArReadiness_mod}.

The \Tile{} calorimeter shows very little sensitivity to pile-up since most of the associated (soft particle) energy flow is absorbed 
in the \LAr{} calorimeters in front of it.
Moreover, out-of-time pile-up is suppressed
by a short shaping time with sensitivity to only about $3$ bunch crossings \cite{TileReadiness}.

\section{Monte Carlo simulation of jets in the ATLAS detector}
\label{sec:MC}
The energy and direction of particles produced in \pp{} collisions
are simulated using various \MC{} event generators. An overview of these
generators for \LHC{} physics can be found in Ref.~\cite{mcforlhc}.
The samples using different event generators and theoretical models are described below. 
All samples are produced at $\sqrt{s}= 7$~\TeV.

\subsection{Inclusive jet Monte Carlo simulation samples}
\begin{enumerate}
\item \pythia{} (version 6.425) \cite{pythia} 
is used for the generation of the baseline simulation event samples. 
It models the hard sub-process in the final states of the generated \pp{}
collisions using a $2 \to 2$ matrix element
at leading order in the strong coupling \alphas. Additional radiation is modelled in the leading logarithmic (LL) approximation by \ptordered{} parton showers \cite{pythiapartonshower}.

Multiple parton interactions (MPI) \cite{Sjostrand:2004ef}, as well as fragmentation and hadronisation 
based on the Lund string mo\-del \cite{lundstring}, are also generated.
Relevant parameters for the modelling of the parton shower and multiple parton interactions in the underlying event (UE) are 
tuned to 
\LHC{} da\-ta (\ATLAS{} \pythia{} tune {\sc AUET2B} 
\cite{MC11c} 
with the MRST LO** 
parton density function (PDF)
\cite{mrstlostar}). Data from the LEP collider are included in this tune.

\item \herwigpp{} \cite{Herwigpp} is used to generate samples for evaluating 
systematic uncertainties.
This generator
uses a $2 \to 2$ matrix element and angular-ordered parton showers in the LL approximation \cite{herwig3,herwig2,herwig}. The cluster model  \cite{herwigclustermodel} is employed for the hadronisation.
The underlying event and soft inclusive interactions are described using a hard and soft MPI model \cite{HerwigppUI}.
The parton densities are provided by the MRST LO** PDF set.
 
\item \madgraph{} \cite{MadGraph} with the CTEQ6L1 PDF set \cite{PDF-CTEQ} is used to generate \pp{} collision samples
with up to three outgoing partons from the matrix element and with MLM matching \cite{MLM} applied in the parton shower, which is performed 
with \pythia{} using the {\sc AUET2B} tune. 
\end{enumerate}

\subsection[\Zjet{} and \gammajet{} Monte Carlo simulation samples]{\ZJET{} and \GAMMAJET{} Monte Carlo simulation samples}

\begin{enumerate}
\item \pythia{} (version 6.425) 
is used to produce \Zjet{} events with the modified leading-order PDF set 
MRST LO**.  The simulation uses a $2 \rightarrow 1$ matrix element to model the 
      hard sub-process,  and, as for the inclusive jet simulation, \ptordered{} parton showers to model additional parton radiation 
      in the LL approximation. In addition, weights are applied to the first branching of the shower, 
      so as to bring agreement with the matrix-element rate in the hard emission region. 
      The same tune and PDF is used as for the inclusive jet sample.

\item The \alpgen{} generator (version 2.13) \cite{alpgen} is used to produce \Zjet{} events, interfaced to 
         \herwig{} (version 6.510) \cite{herwig} for parton shower and fragmentation into particles, and to \jimmy{} (version 4.31) 
         \cite{jimmy} to model UE contributions using the \ATLAS{} {\sc AUET2} tune \cite{MC11}, here with the 
         CTEQ6L1 \cite{PDF-CTEQ} leading-order PDF set. \alpgen{} is a leading-order ma\-trix-element generator for hard multi-parton   
         processes ($2\rightarrow n$) in hadronic collisions. Parton showers are matched to the matrix element with the MLM matching 
         scheme. The CTEQ6L1 PDF set is employed. 

\item The baseline \gammajet{} sample is produced with \pythia{} (version 6.425). 
         It generates non-diffractive events using a $2 \to 2$ matrix element at leading order in \alphas{} to model the 
         hard sub-process. Again, additional parton radiation is modelled by \ptordered{} parton showers in the LL
         approximation. The modelling of non-perturbative physics effects arising in  MPI, fragmentation, and hadronisation is based 
         on the \ATLAS{} {\sc AUET2B} {\sc MRST} {\sc LO**} tune.
\item An alternative \gammajet{} event sample is generated with \herwig{} (version 6.510) and \jimmy{} 
using the \ATLAS{} {\sc AUET2} tune and the {\sc MRST} {\sc LO**} PDF.
It is used to evaluate the systematic uncertainty due to physics modelling.

\item The systematic uncertainty from jets which are misidentified as photons (fake photons) is studied with a dedicated \MC{} event sample. 
An inclusive jet sample is generated with \pythia{} (version 6.425)
with the same parameter tuning and PDF set 
as the \gammajet{} sample. 
         An additional filter is applied to the jets built from the stable generated particles to select events containing a narrow 
         particle jet, which is more likely to pass photon identification criteria. The surviving events are passed through the same 
         detector simulation software as the \MC{} \gammajet{} sample.
\end{enumerate}

\subsection{Top-quark pair Monte Carlo simulation samples}
Top pair (\ttbar) production samples are relevant for jet reconstruction performance studies, as they are a significant source of hadronically decaying \Wboson{} bosons and therefore important for light-quark jet response evaluations in a radiation environment very different from the inclusive jet and \Zjet/\gammajet{} samples discussed above. In addition, they provide jets from a heavy-flavour (\bquark) decay, the response to which can be studied in this final state as well. 
   
The nominal \ttbar\ event sample is generated using \mcatnlo{} (version 4.01) \cite{FRI-0201}, which implements a next-to-leading-order (NLO) matrix element for top-pair production. Correspondingly, the {\sc CT10} \cite{Lai:2010vv} NLO PDF set is used. This matrix-element generator is interfaced to parton showers 
from \herwig{} (ver\-sion 6.520) \cite{COR-0001} 
and the underlying event modelled by 
\jimmy{} (version 4.31), with the CT10 PDF and the \ATLAS{} {\sc AUET2} tune. 

A number of systematic variation samples use alternative \MC{} generators or different generator parameter sets.  Additional
\ttbar\ samples are simulated using the \powheg{} \cite{FRI-0701} generator interfaced with \pythia{}, as well as \herwig{} and
\jimmy{}. \powheg{} provides alternative implementations of the NLO matrix-element calculation and the interface to parton showers. The\-se samples allow comparison of two different parton shower, hadronisation and fragmentation models. In addition, the particular implementations of the NLO matrix-element calculations in \powheg{} and \mcatnlo{} can be compared. Differences in the $b$-hadron decay tables between \pythia{} and \herwig{} are also significant enough to provide a conservative uncertainty envelope on the effects of the decay model.

In addition, samples with more or less parton shower activity are generated with the leading-order generator \ACERMC{} \cite{KER-0401} interfaced to \pythia{} with the MRST LO** PDF set.  The\-se are used to estimate the model dependence of the 
event selection. In these samples the initial state radiation (ISR) and the final state radiation (FSR) parameters are varied in value ranges not excluded by the current experimental data, as detailed in Refs.~\cite{mtop2011paper, topvetopaper2011}.

\subsection{Minimum bias samples}
\label{sec:mc-mb}
Minimum bias events are generated using \pythia 8 \cite{pythia8} with the {\sc 4C} tune \cite{Corke:2010yf} and MRST LO** PDF set. These minimum bias events are used to form pile-up events, which are overlaid onto the hard-scatter events following a Poisson distribution around the average number $\langle\axing\rangle$ of additional \pp{} collisions per bunch crossing measured in the experiment. The \LHC{} bunch train structure with $36$ proton bunches per train and $50$~\ns{} spacing between the bunches,  
is also modelled by organising the simulated collisions into four such trains. This allows the inclusion of out-of-time pile-up effects driven by the distance of the hard-scatter events from the beginning of the bunch train. 
The first ten bunch crossings in each \LHC{} bunch train, 
approximately, are characterised by varying out-of-time pile-up contributions from the collision history, which is getting filled with an increasing number of bunch crossings with \pp{} interactions. For the remaining $\approx 26$ bunch crossings in a train, 
the effect of the out-of-time pile-up contribution is stable, i.e. it does not vary with 
the bunch position within the bunch train, if the bunch-to-bunch intensity is constant. 
Bunch-to-bunch fluctuations in proton intensity at the \LHC{} 
are not included in the simulation. 

\subsection{Detector simulation}
The \geant{} software toolkit \cite{Geant4} within the \ATLAS{} simulation framework \cite{simulation} propagates the stable 
particles\footnote{See the discussion of ``truth jets" in \secRef{sec:truthjets} for the definition of stable particles.}
produced by the event generators through the \ATLAS{} detector and simulates their interactions with the detector material. Hadronic showers are simulated with the QGSP\_BERT model 
\cite{Bertini,Bertini1,Bertini2,Bertini3,QGS,QGSP2,QGSP4,QGSP3,QGSP5}.
Compared to the simulation used in the context of the $2010$ data analysis, a newer version of \geant{} (version 9.4) is used and a more detailed description of the geometry of the \LAr{} calorimeter absorber structure is available. These geometry changes introduce an increase in the calorimeter response to pions below $10$~\GeV{} of about $2\%$.

For the estimation of the systematic uncertainties arising from detector simulation, several samples are also produced with the \ATLAS{} fast (parameterised) detector simulation \atlasfast{} \cite{ATLAS:2010bfa,Aad:2010ah}.

\section{\Ds}
\label{sec:dataset}

The data used in this study were recorded by \ATLAS{} between May and October $2011$, 
with all \ATLAS{} subdetectors operational. 
The corresponding total integrated luminosity is about \mylumi{} of \pp{} 
collisions at a centre-of-mass energy of $\sqrt{s}=7$~\TeV.

As already indicated in \secRef{sec:mc-mb}, the \LHC{} operated with bunch crossing intervals 
of $50$ \ns, 
and bunches organised in bunch trains. 
The average number of interactions per bunch crossing ($\axing$) 
as estimated from the luminosity measurement is
$3 \leq \axing \leq 8$ until Summer $2011$, with an average for this 
period of $\langle \axing \rangle \approx 6$. 
Between August $2011$ and the end of the proton run, \axing{} increased to about 
$5 \leq \axing \leq 17$, with an average $\langle \axing \rangle \approx 12$.   
The average number of interactions for the whole $2011$ \ds{} is $\langle \axing \rangle = 8$.
 
The specific trigger requirements and precision signal object selections applied to the data are analysis dependent. They are therefore discussed in the context of each analysis presented in this paper.

\begin{figure*}[hbp!!!]
  \centering
  \includegraphics[width=0.88\textwidth]{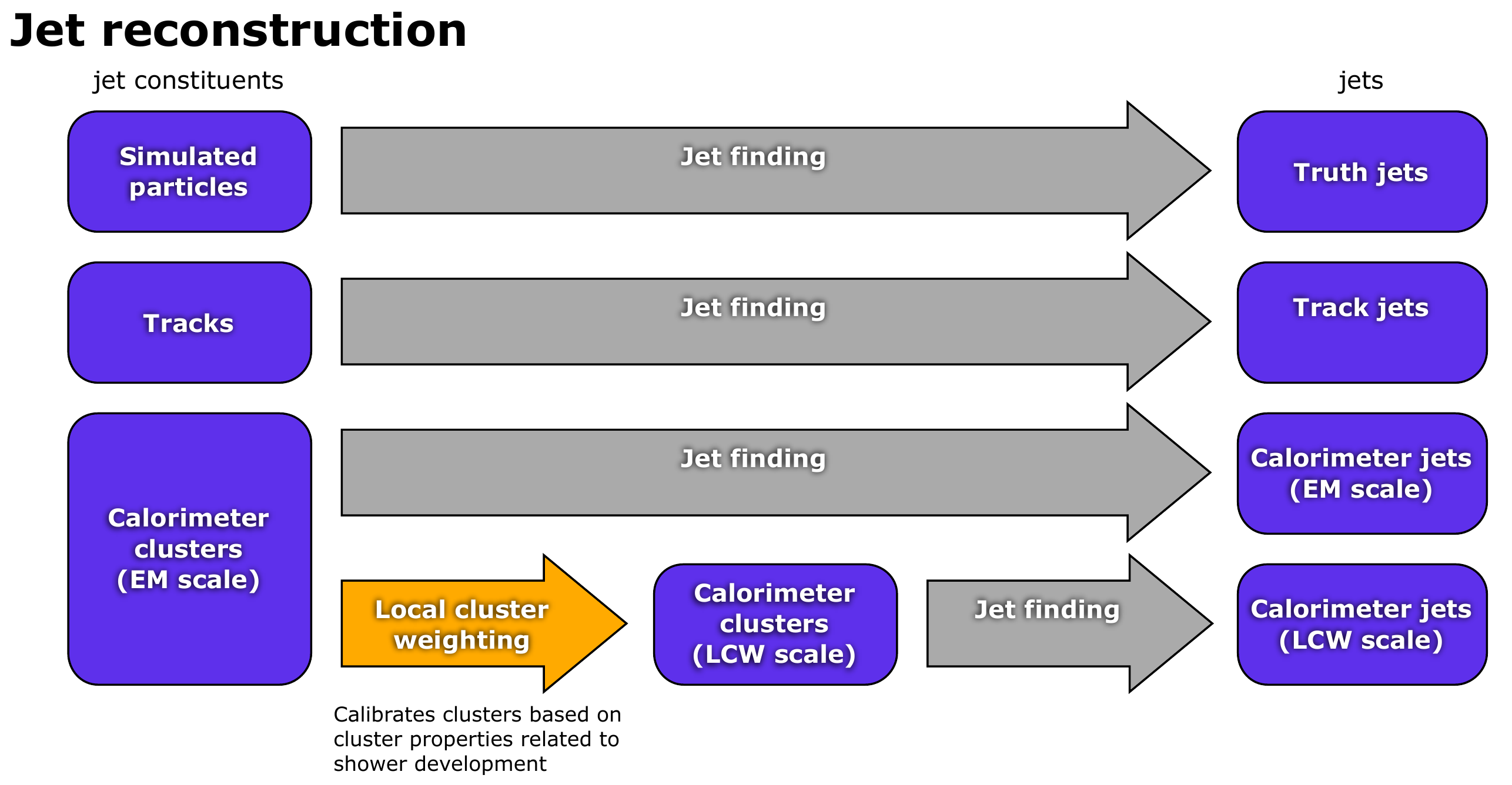}
  \caption{
    Overview of the ATLAS jet reconstruction.
    After the jet finding, the jet four momentum is defined as the four momentum sum of its constituents. 
    \label{fig:jet_reco}
  }
\end{figure*}

\begin{figure*}[hbp!]
  \centering
  \includegraphics[width=0.88\textwidth]{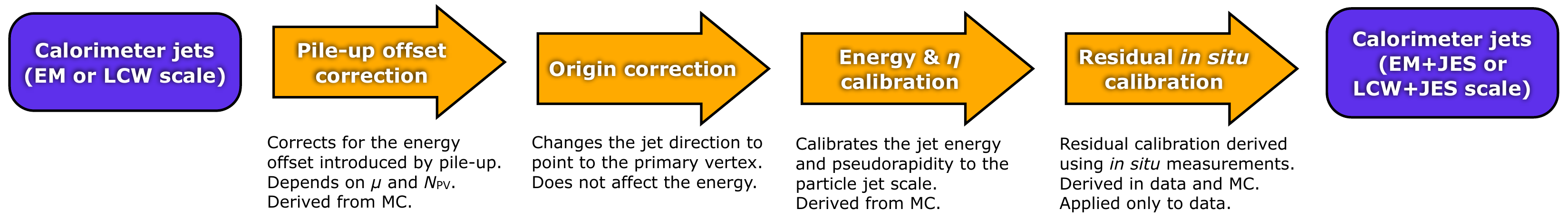}
  \caption{
    Overview of the ATLAS jet calibration scheme used for the $2011$ dataset. 
    The pile-up, absolute \JES{} and the residual \insitu{} corrections calibrate the scale of the jet, while the
    origin and the \eta{} corrections affect the direction of the jet.
    \label{fig:JetCalibrationScheme}
  }
\end{figure*}

  \begin{figure*}[htp!]
  \centering
  \subfloat[\EM scale ($\mathfrak{R}^{\EM}(\etaDet)$)] {\includegraphics[width=0.49\textwidth]{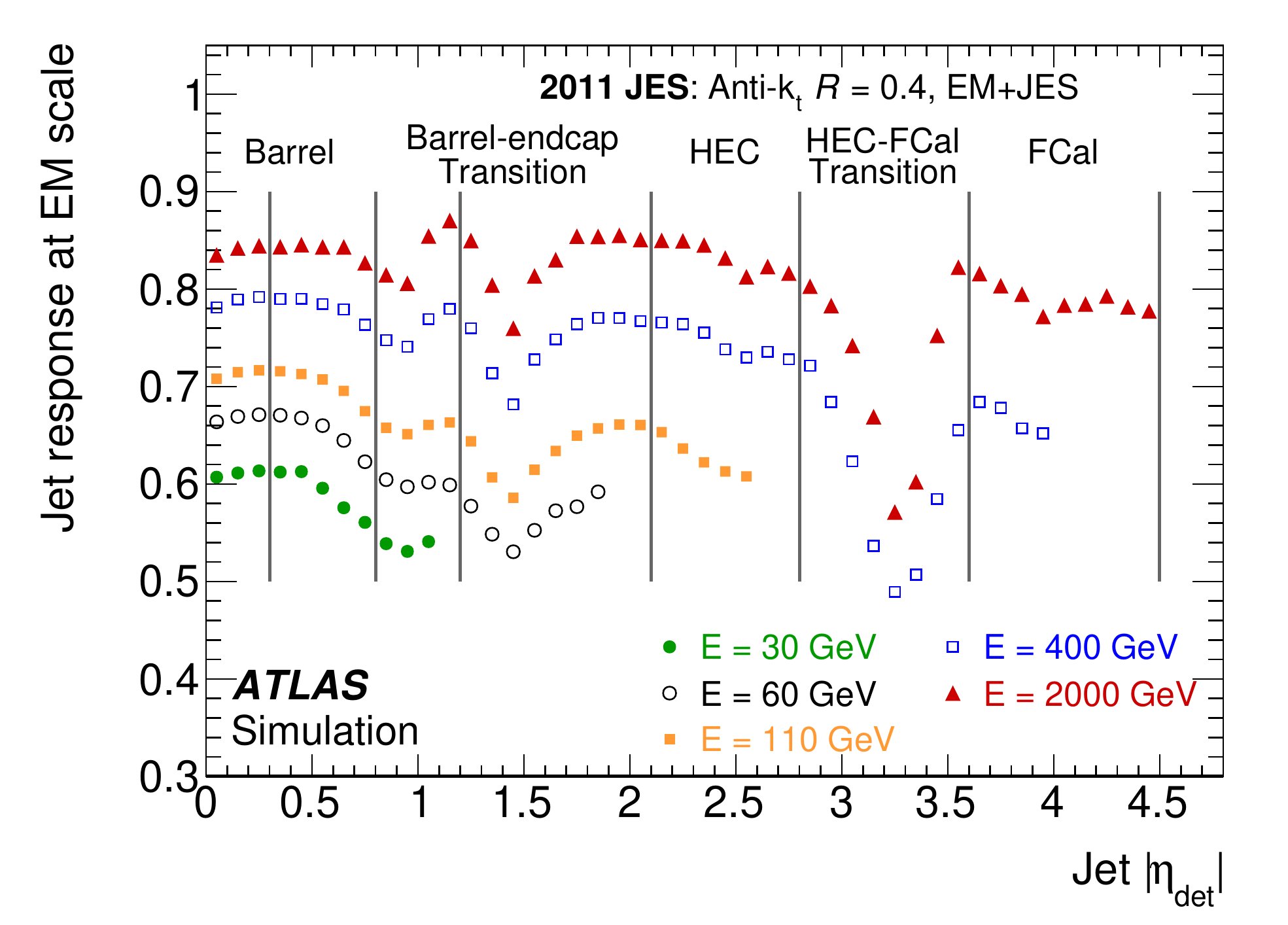}\label{fig:EtaJESEM}}
  \subfloat[\LCW scale ($\mathfrak{R}^{\LCW}(\etaDet)$)]{\includegraphics[width=0.49\textwidth]{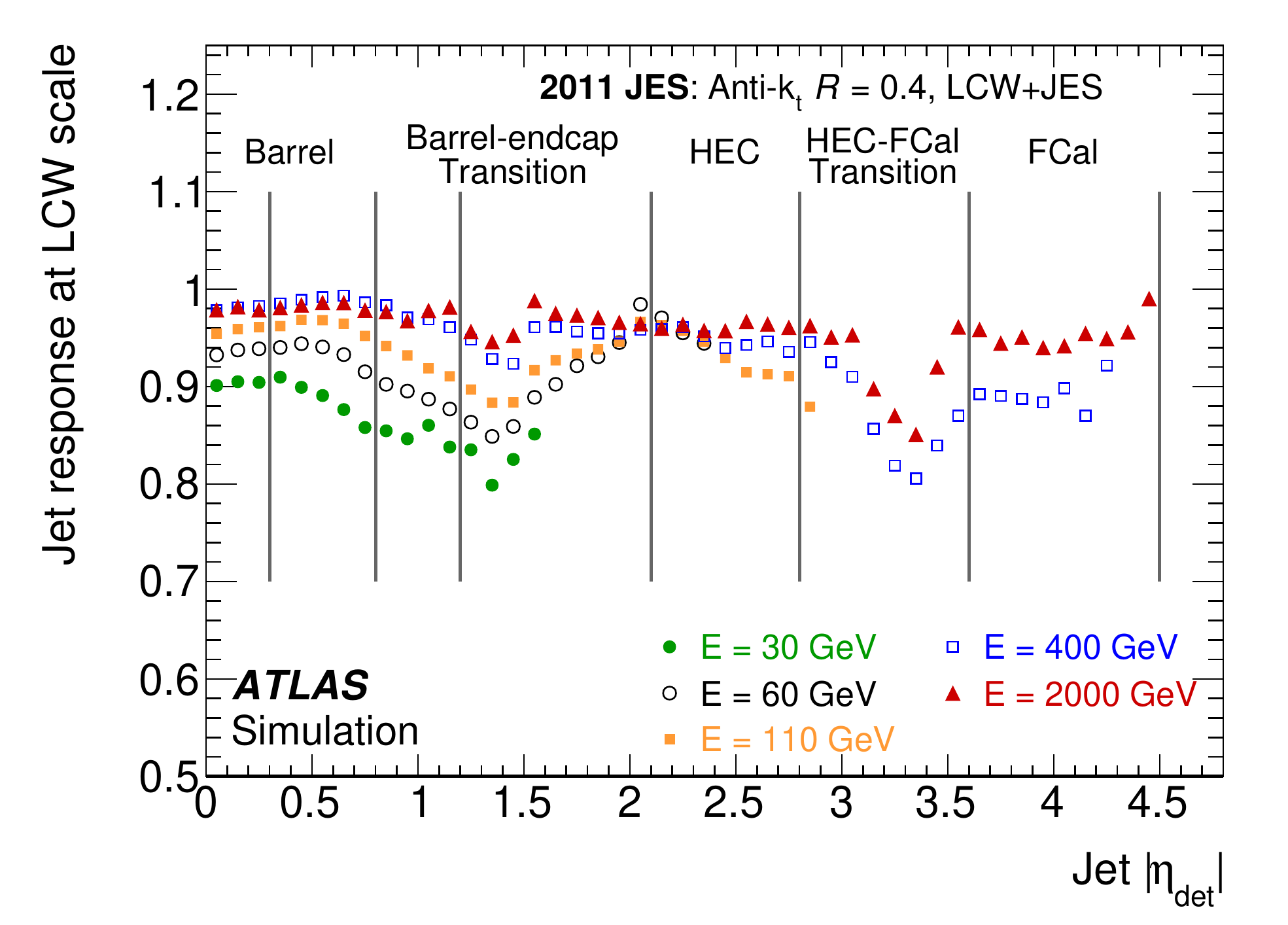}\label{fig:EtaJESLC}}
  \caption[]{
	Average response of simulated jets formed from \topos{}, calculated as defined in \eqRef{eq:mcjetresponse} and shown in 
       \subref{fig:EtaJESEM} for the 
        \EM{} scale ($\mathfrak{R}^{\EM}$)  and  in 
       \subref{fig:EtaJESLC} for the \LCW{} scale ($\mathfrak{R}^{\LCW}$).
       The response is shown separately for various truth-jet  
       energies as function of the uncorrected (detector) jet pseudorapidity \etaDet{}. Also indicated are the different calorimeter regions. 
       The inverse of $\mathfrak{R}^{\EM}$ ($\mathfrak{R}^{\LCW}$) corresponds to the average jet energy scale correction for 
\EM{} (\LCW) 
in each \etaDet{} bin. The results shown are based on the baseline \pythia{} inclusive jet sample.  
       \label{fig:EtaJESEMLC}
  }
\end{figure*}

\section{Jet reconstruction and calibration with the ATLAS detector}
\label{sec:jetrecocalib}

\subsection{Topological clusters in the calorimeter}
\label{sec:topos}
Clusters of energy deposits in the calorimeter
(\topos) are built from topologically connected calorimeter cells that contain a significant 
signal above noise, see Refs. \cite{jespaper2010,EndcapTBelectronPion2002,TopoClusters} for details.  
The \topo{} formation follows cell signal significance patterns in the \ATLAS{}
calorimeters. The signal significance is measured by the absolute ratio of the cell signal
to the energy-equivalent noise in the cell. 
The signal-to-noise thresholds for the cluster formation are not changed
with respect to the settings given in 
Ref. \cite{jespaper2010}.
However, the noise in the calorimeter increased due to the presence
of multiple proton-proton interactions, as discussed in \secRef{sec:calopileup}, and required the adjustments explained below.

While in \ATLAS{} operations prior to $2011$ the
cell noise was dominated by electronic noise, the short bunch crossing interval in $2011$
\LHC{} running added a noise component from bunch-to-bunch variations in the instantaneous
luminosity and in the energy deposited in a given cell from previous collisions inside the
window of sensitivity of the calorimeters. The cell noise thresholds steering the \topo{}
formation thus needed to be increased from those used in $2010$ to accommodate the
corresponding fluctuations, which is done by raising the nominal noise according to
\begin{displaymath}
        \renewcommand{\arraystretch}{2.0}
        \sigma_{\text{noise}}= \left\{\begin{array}{ll}
 	\sigen & \mathrm{(2010\ operations)} \\
	\sqrt{\left(\sigen\right)^{2} + \left(\sigpu\right)^{2}}   &
                                 \mathrm{(2011\ operations)}
                                 \end{array}\right. .
\end{displaymath} 
Here, \sigen{} is the electronic noise, and \sigpu{} the noise from pile-up, determined with \MC{} simulations and corresponding to 
an average of eight additional \pp{} interactions per bunch crossing ($\mu = 8$) in $2011$. 
The change of the total nominal noise $\sigma_{\text{noise}}$ and its dependence on the
calorimeter region in \ATLAS{} can be seen by comparing Figs. \ref{fig:noise}\subref{fig:noise:2010}
and \ref{fig:noise}\subref{fig:noise:2011}. In most calorimeter regions, the noise induced by pile-up 
is smaller than or of the same magnitude as the electronic noise, with the
exception of the forward calorimeters, where 
$\sigpu \gg \sigen$.

The implicit noise suppression implemented by the topological cluster
algorithm discussed above leads to significant improvements in the calorimeter performance for e.g.
the energy and spatial resolutions in the presence of pile-up. On the other
hand, contributions from larger negative and positive signal fluctuations introduced by
pile-up can survive in a given event. They thus contribute to the sensitivity to pile-up observed in
the jet response, in addition to the cell-level effects mentioned 
in \secRef{sec:calopileup}.

\subsection{Jet reconstruction and calibration}
\label{sec:jetrecocalibsequence}
Jets are reconstructed using the \antikt{} algorithm~\cite{Cacciari:2008gp} with distance parameters $R = 0.4$ or $R = 0.6$,
utilising the {\sc FastJet} software package~\cite{Cacciari200657,Fastjet}. 
The four-momentum scheme is used at each recombination step in the jet clustering.
The total jet four-momentum is therefore defined as the sum of the four-momenta sum of all its constituents.
The inputs to the jet algorithm are stable simulated particles ({\emph{truth jets}, see \secRef{sec:truthjets} for details), %
reconstructed tracks in the inner detector ({\emph{track jets}, see Ref. \cite{jespaper2010} and \secRef{sec:trackjets} for details)
or energy deposits in the calorimeter ({\emph{calorimeter jets}, see below for details).
A schematic overview of the ATLAS jet reconstruction is presented in \figRef{fig:jet_reco}.

The calorimeter jets are built from the \topos{} entering as massless particles 
in the jet algorithm as discussed in the previous section. Only clusters with positive energy are considered.
The \topos{} are initially re\-con\-struc\-ted at the \EM{} scale~\cite{EndcapTBelectronPion2002,ctb2004electronseoverp,ctb2004electrons,LArTB02uniformity,LArTB02linearity,Tile2002,Pinfold:2008zzb,LArTB02muons,Atlaselectronpaper}, 
which correctly measures the energy deposited in the calorimeter by particles produced in electromagnetic showers. 
A second \topo{} collection is built by calibrating the calorimeter cell such that the response
of the calorimeter to hadrons is correctly reconstructed.
This calibration uses the local cell signal weighting (\LCW) method that aims at an 
improved resolution compared to the \EM{} scale by correcting the signals from hadronic deposits, 
and thus reduces fluctuations due to the non-com\-pen\-sa\-ting nature of the \ATLAS{} calorimeter.
The \LCW{} method first classifies 
\topos{} as either electromagnetic or hadronic, primarily based on the measured energy density and the longitudinal shower depth. 
Energy corrections are derived according to this classification from single charged and neutral pion \MC{} simulations. 
Dedicated corrections address effects of calorimeter non-com\-pen\-sa\-ti\-on, 
signal losses due to noise threshold effects, and energy lost in non-instrumented regions close to the cluster~\cite{jespaper2010}.

\FigRef{fig:JetCalibrationScheme} shows an overview of the ATLAS calibration sche\-me for calorimeter jets used for the 2011 dataset, 
which restores the jet energy scale to that of jets reconstructed from stable simulated particles (truth particle level, see \secRef{sec:truthjets}). 
This procedure consists of four steps as described below.
\begin{mylist}
\item{\textbf{Pile-up correction}}\\
  Jets formed from \topos{} at the \EM{} or \LCW{} scale are first calibrated by applying
  a correction to account for the energy offset caused by pile-up interactions.
  The effects of pile-up on the jet energy scale are caused by both 
  additional proton collisions in a recorded event (in-time pile-up) and by past and future collisions influencing the
  energy deposited in the current bunch-crossing (out-of-time pile-up), and are outlined in \secRef{sec:pileupsection}. 
  This correction is derived from \MC{} simulations 
  as a function of the number of reconstructed primary vertices (\Npv, measuring the actual collisions in a given event) 
  and the expected average number of interactions (\axing, sensitive to out-of-time pile-up) in bins of 
  jet pseudorapidity and transverse momentum (see Section~\ref{sec:pileupsection}).
\item{\textbf{Origin correction}}\\
  A correction to the calorimeter jet direction is applied 
that makes the jet pointing back to the primary event vertex
instead of the nominal centre of the \ATLAS{} detector.
\item{\textbf{Jet calibration based on \MC{} simulations}}\\
Following the strategy presented in Ref. \cite{jespaper2010}, the calibration of the energy and pseudorapidity of a reconstructed jet is a simple correction derived from the relation of these quantities to the corresponding ones of the matching truth jet (see \secRef{sec:truthjets}) in \MC{} simulations. It can be applied to jets formed from \topos{} at \EM{} or at \LCW{} scale 
with the resulting jets being referred to as calibrated with the \EMJES{} or with the \LCWJES{} scheme.
This first \JES{} correction uses isolated jets 
from an inclusive 
jet \MC{} sample including pile-up events
  (the baseline sample described in \secRef{sec:MC}).
  \FigRef{fig:EtaJESEMLC} shows the average energy response 
 \begin{equation} \label{eq:mcjetresponse}\mathfrak{R}^{\EM(\LCW)}=E^{\EM(\LCW)}_{\rm jet}/E^{\rm truth}_{\rm jet}, \end{equation} 
  which is the inverse of the jet energy calibration function,
  for various jet energies as a function of the jet pseudorapidity \etaDet{} measured in the detector frame of reference (see  \secRef{sec:jetdirections}).
\item{\textbf{Residual \insitu{} corrections}}\\
  A residual correction derived \insitu{} is applied as a last step to jets reconstructed in data. 
  The derivation of this correction is described in 
in \secRef{sec:InSitu}.
\end{mylist}

%
\begin{figure*}[htp!]
\subfloat[\AetaRange{0.3}]{\includegraphics[width=0.33\textwidth]{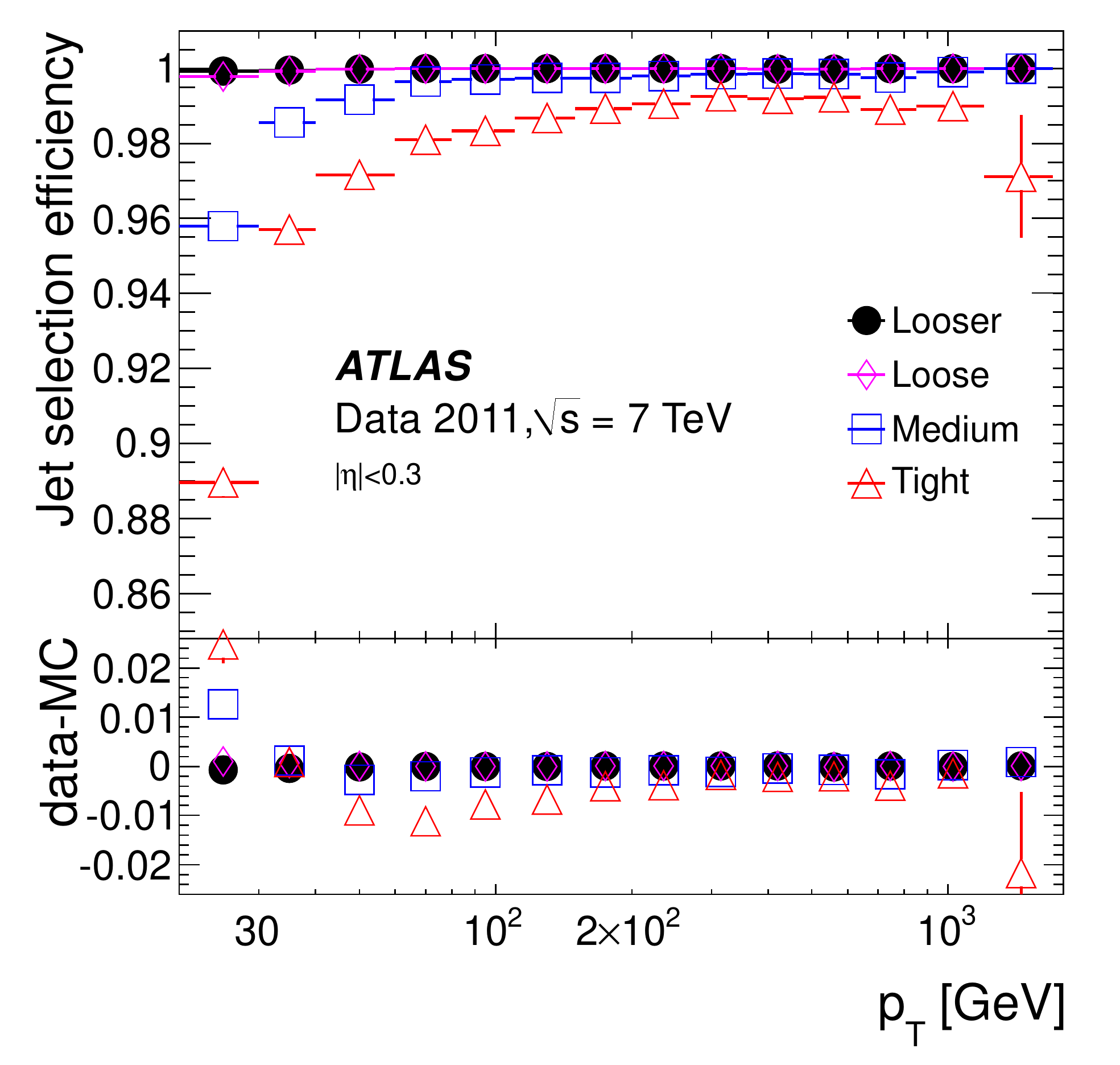}}
\subfloat[\etaRange{0.3}{0.8}]{\includegraphics[width=0.33\textwidth]{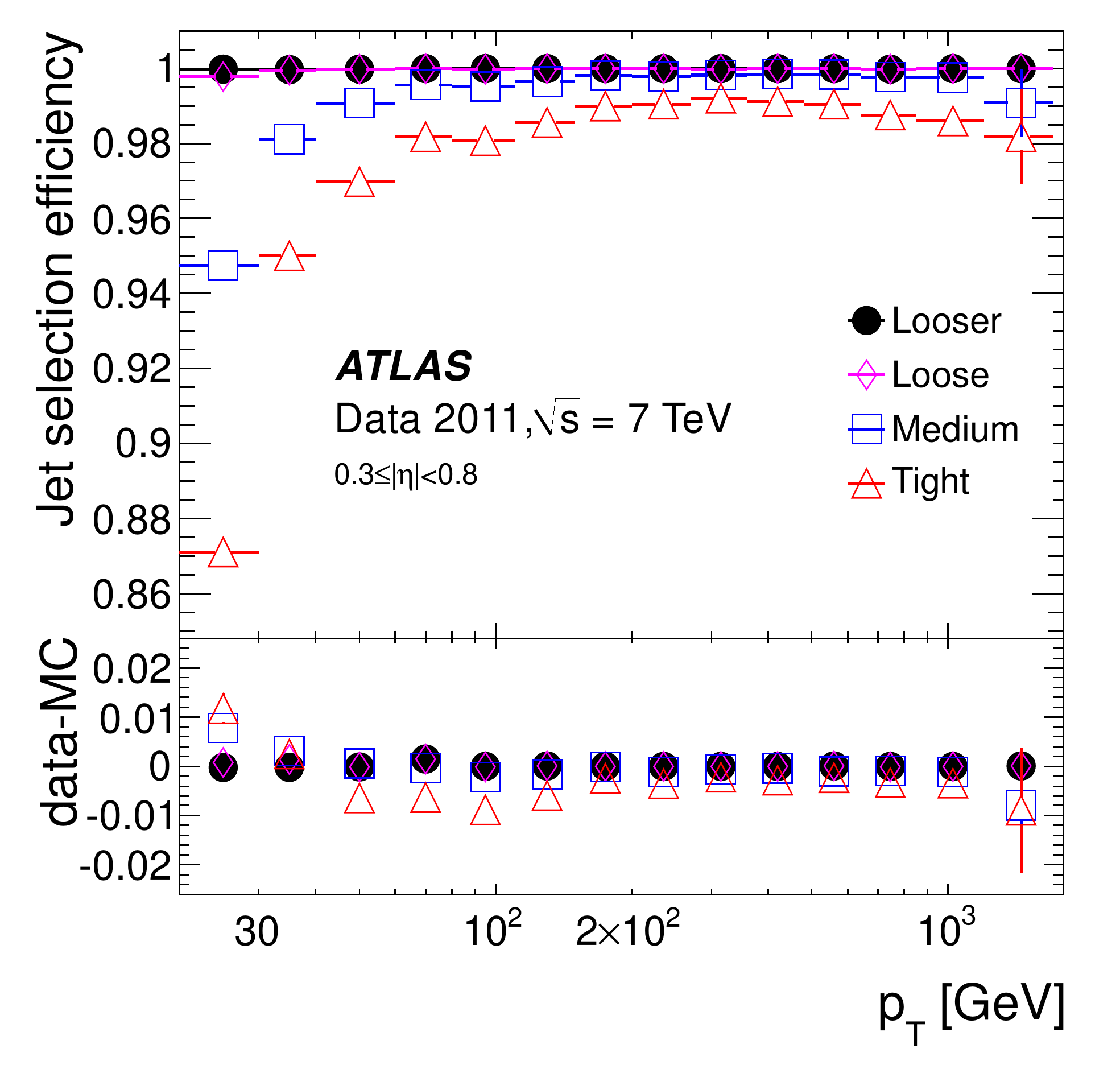}}
\subfloat[\etaRange{0.8}{1.2}]{\includegraphics[width=0.33\textwidth]{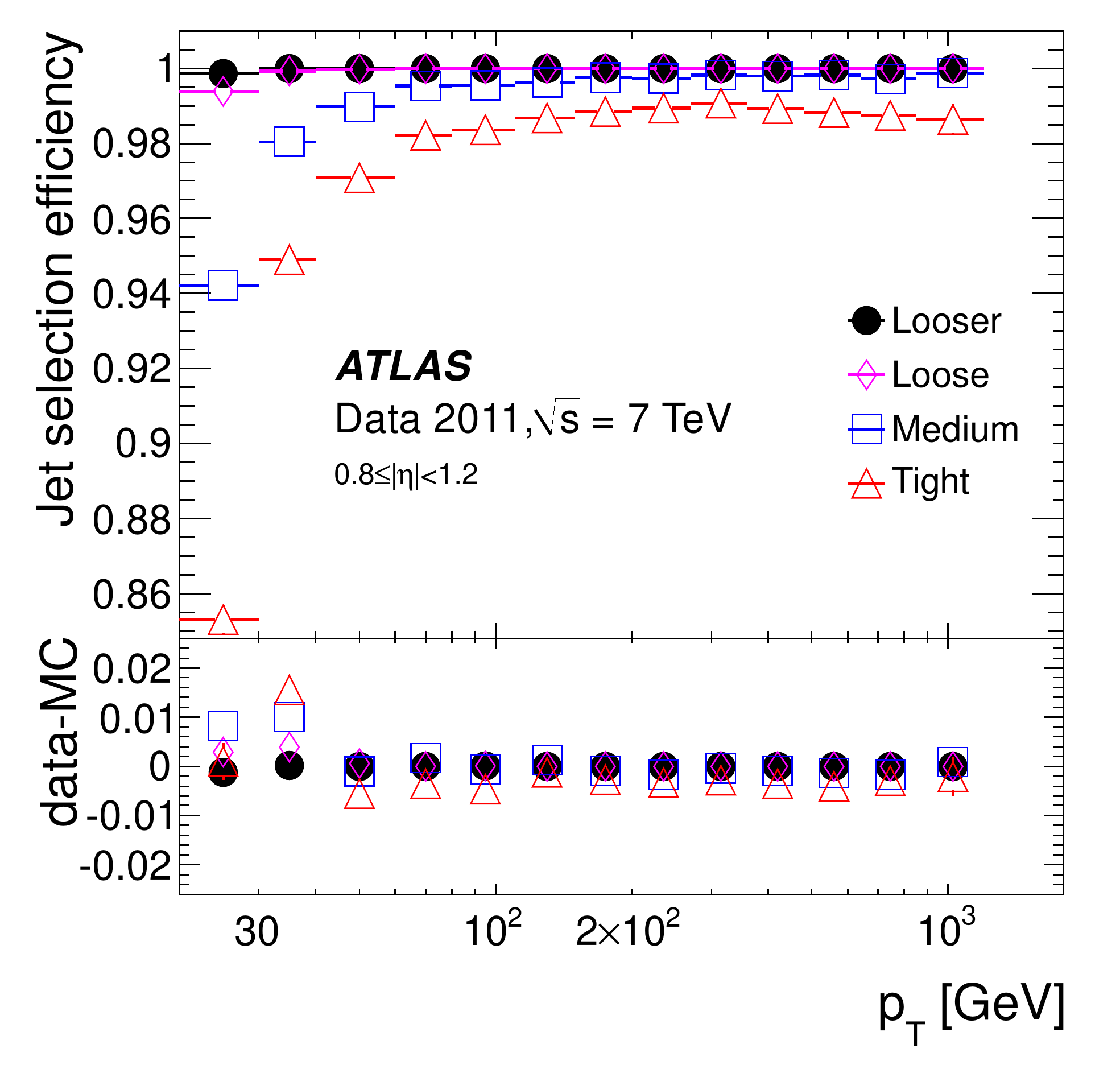}}\\
\subfloat[\etaRange{1.2}{2.0}]{\includegraphics[width=0.33\textwidth]{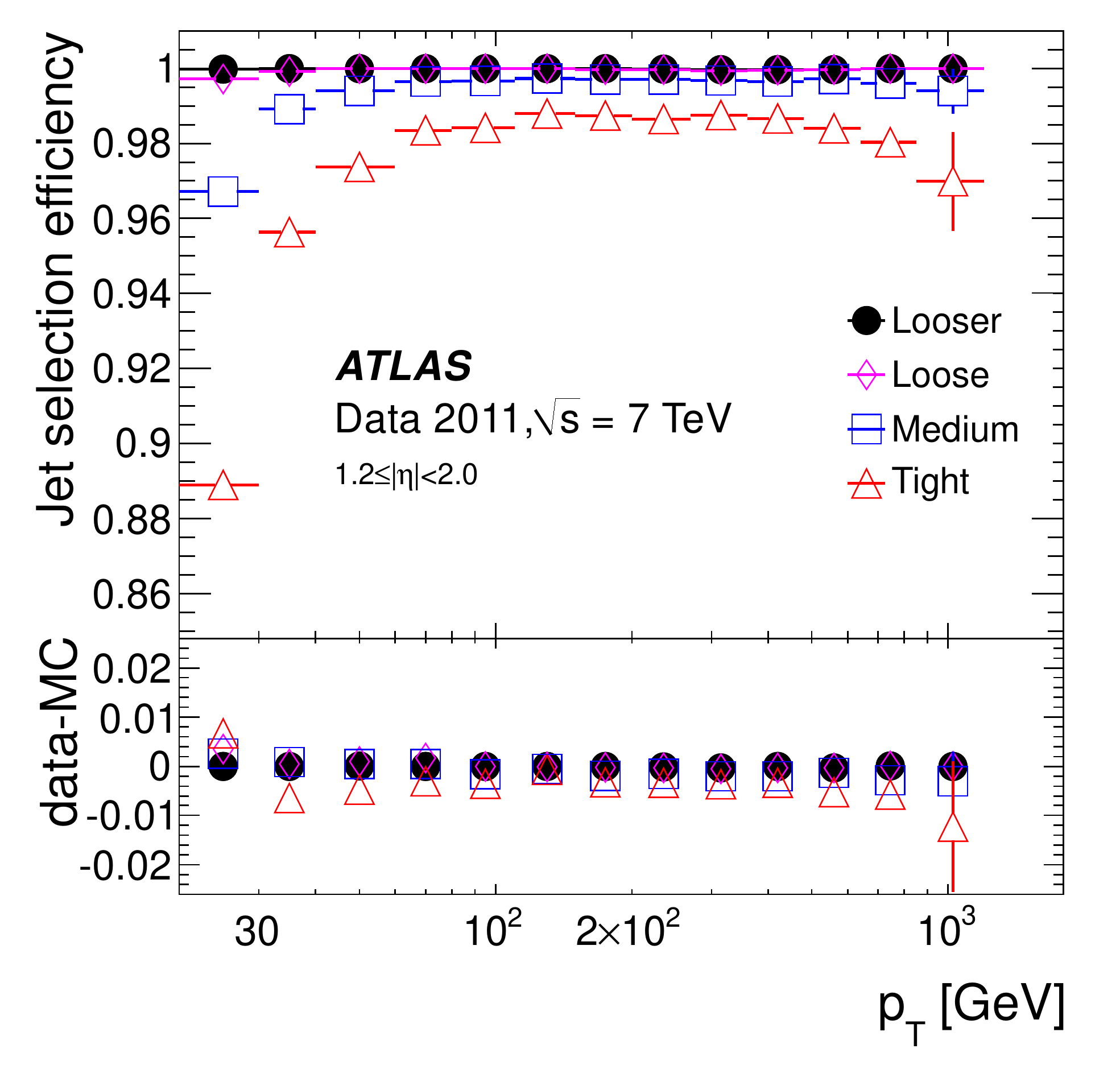}}
\subfloat[\etaRange{2.0}{2.5}]{\includegraphics[width=0.33\textwidth]{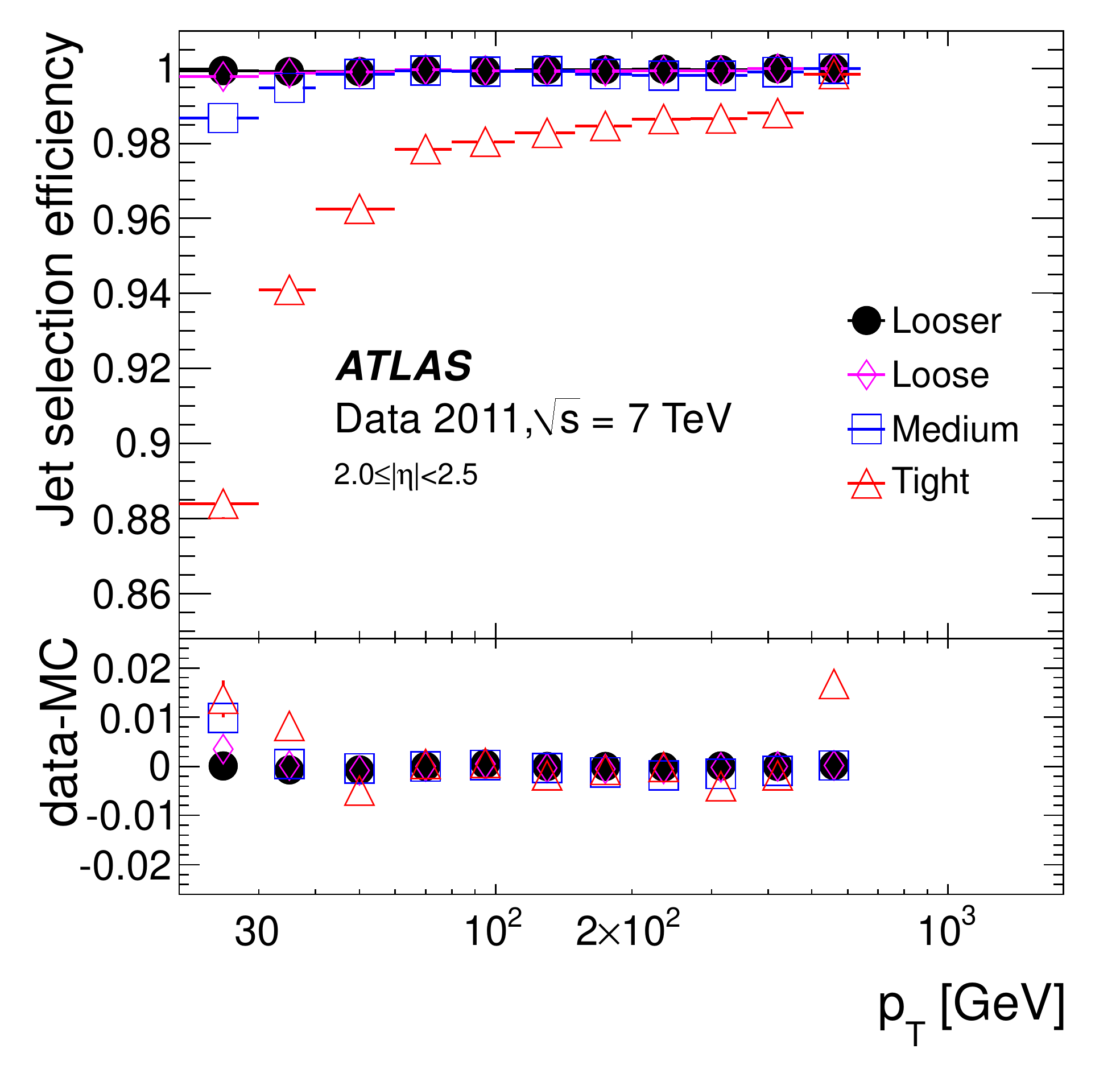}}
\subfloat[\etaRange{2.5}{2.8}]{\includegraphics[width=0.33\textwidth]{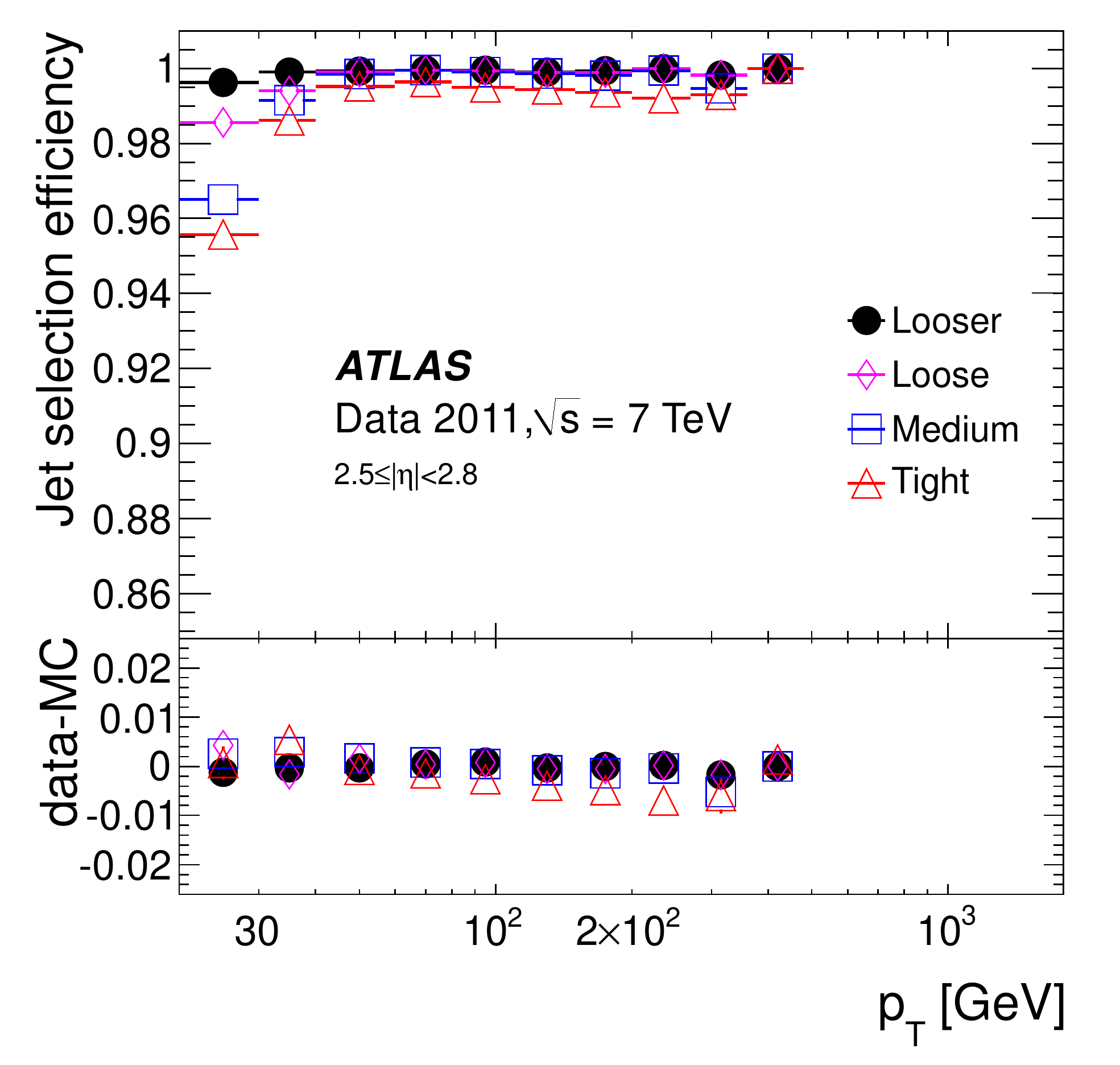}}\\
\subfloat[\etaRange{2.8}{3.6}]{\includegraphics[width=0.33\textwidth]{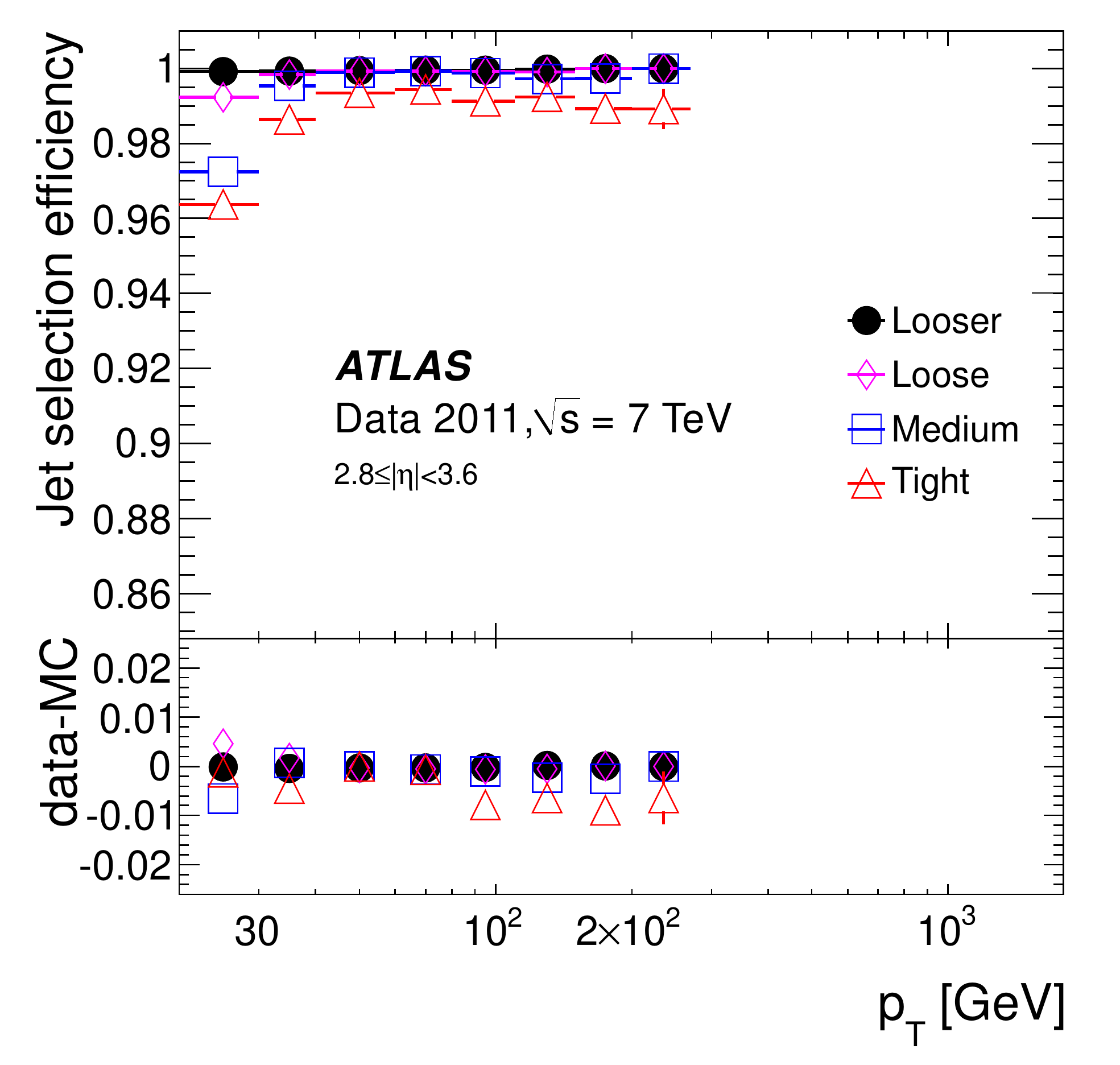}}
\subfloat[\etaRange{3.6}{4.5}]{\includegraphics[width=0.33\textwidth]{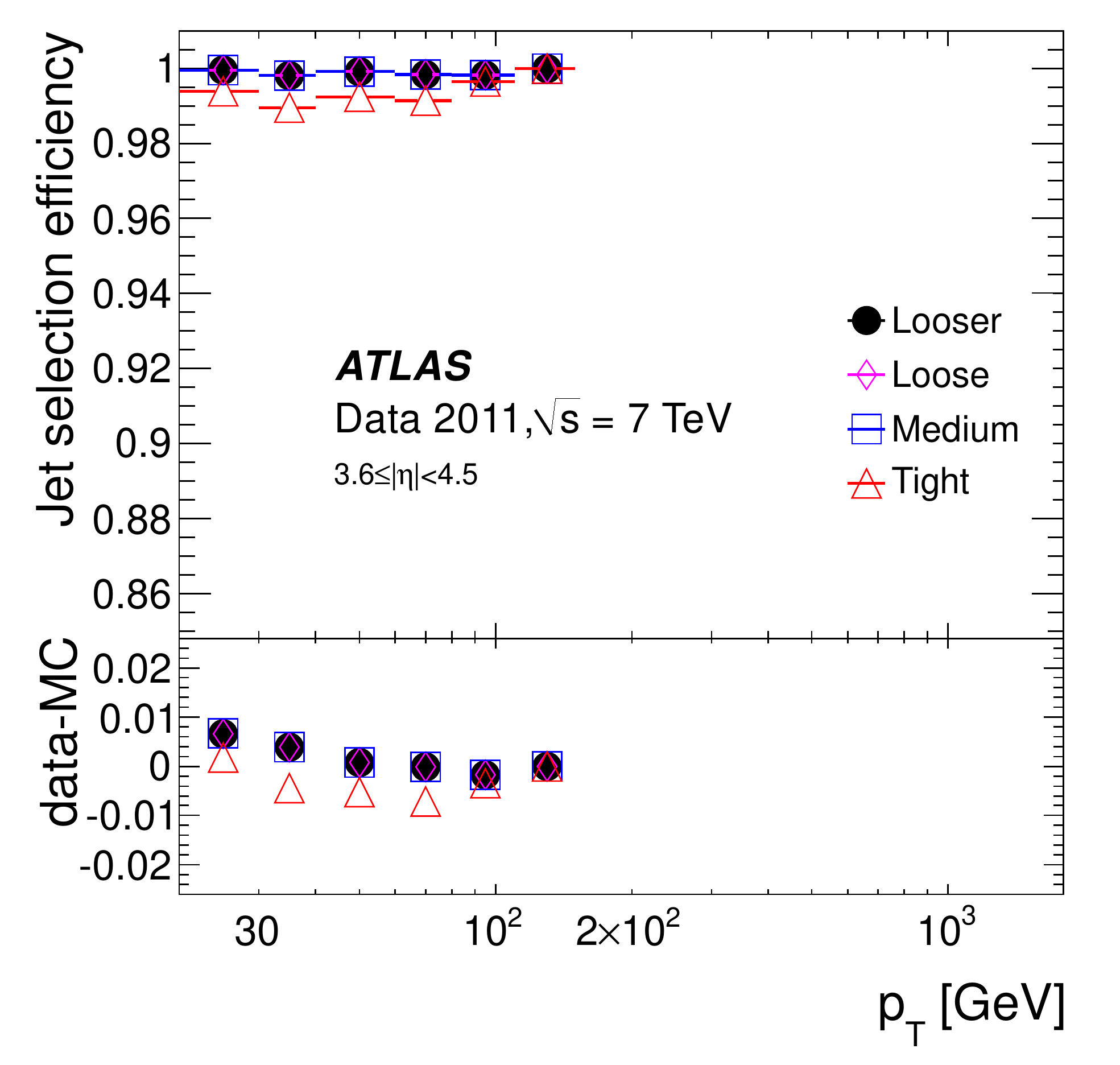}}
\caption{Jet quality selection efficiency for \antikt{} jets with $R = 0.4$ measured with a tag-and-probe technique
	 as a function of \ptjet{} in various $\etajet$ ranges, for the four sets of selection criteria.
	 Only statistical uncertainties are shown. Differences between data and \MC{} simulations are also shown.}
\label{fig:eff}
\end{figure*}

\subsection{Jet quality selection}
\label{sec:JetSel}
Jets with high transverse momenta produced in \pp{} collisions must be distinguished from background jet candidates 
not originating from hard-scattering events. A first strategy to select jets from collisions and to suppress background
is presented in Ref.~\cite{jespaper2010}.
 
The main sources of potential background are:
\noindent
\begin{enumerate}
\item Beam--gas events, where one proton of the beam collides with the residual gas within the beam pipe.
\item Beam-halo events, for example caused by interactions in the tertiary collimators in the beam-line far away from the \ATLAS{} detector.
\item Cosmic-ray muons overlapping in-time with collision e\-v\-ents.
\item Calorimeter noise.
\end{enumerate}
The jet quality selection criteria should efficiently reject jets from these background processes while maintaining high efficiency for selecting jets produced in \pp{} collisions. Since the level and composition of background depend on the event topology and the jet kinematics, 
four sets of criteria called \jclooser, \jcloose, \jcmedium{} and \jctight{} are introduced in Ref. \cite{beambackground2011}. They correspond to different levels of fake-jet rejection and jet selection efficiency, with the \jclooser{} criterion being the one with the highest jet selection efficiency while the \jctight{} criterion is the one with the best rejection. 
The discrimination between jets coming from the collisions and background jet candidates is based on several pieces of experimental 
information, including the quality of the energy reconstruction at the cell level, jet energy deposits in the direction 
of the shower development, and reconstructed tracks matched to the jets.

The efficiencies of the jet selection criteria are measured using the tag-and-probe method described in Ref.~\cite{jespaper2010}. 
The resulting efficiencies for \antikt{} jets with $R = 0.4$ for all selection criteria are shown in \figRef{fig:eff}. 
The jet selection efficiency of the \jclooser{} selection is greater than $99.8\%$ over all 
calibrated transverse jet momenta \ptjet{} and $\eta$ bins. 
A slightly lower efficiency of about $1$-$2$\%
is measured for the \jcloose{} selection, in particular at low \ptjet{} 
and for $2.5 <|\eta|< 3.6$. The \jcmedium{} and \jctight{} selections have lower jet selection efficiencies 
mainly due to cuts on the jet charged fraction, which is the ratio of the scalar sum of the \pT{} of all reconstructed tracks matching the jet, and the jet \pT{} itself, see Ref. \cite{beambackground2011} for more details.
For jets with $\ptjet \approx 25$~\GeV{}, the \jcmedium{} and \jctight{} selections have inefficiencies of $4\%$ and $15\%$, respectively.
For $\ptjet > 50$~\GeV, the \jcmedium{} and \jctight{} selections have efficiencies greater than $99\%$ and $98\%$, respectively.

The event selection is based on the azimuthal distance between the probe and tag jet \deltaphi{\mathrm{tag}}{\mathrm{probe}} and the significance of the 
missing transverse momentum \Etmiss{} 
\cite{Atlasetmiss} 
reconstructed for the event, which is measured by the ratio $\Etmiss/\sqrt{\sumet}$. Here \sumet{} is the scalar transverse  momentum
sum of all particles, jets, and soft signals in the event. The angle \deltaphi{\mathrm{tag}}{\mathrm{probe}}, $\Etmiss/\sqrt{\sumet}$,  and the \jctight{} 
selection of the 
reference (tag) jet are varied to study the systematic uncertainties. For the \jcloose{} and \jclooser{} selections, 
the jet selection efficiency is almost unchanged by varying the selection cuts, with variations of less than $0.05\%$. 
Slightly larger changes are observed for the two other selections, but they are not larger than $0.1\%$ for the \jcmedium{} and $0.5\%$ 
for the \jctight{} selection.

The jet selection efficiency is also measured using a \MC{} simulation sample. A very good agreement between data and simulation is observed for the \jclooser{} and \jcloose{} selections. Differences not larger than $0.2\%$ and $1\%$ are observed for the \jcmedium{} and \jctight{} selections, respectively, for $\ptjet > 40$~\GeV. Larger differences are observed at lower \ptjet, but they do not exceed $1\%$($2\%$) for the \jcmedium (\jctight) selection.

\subsection{Track jets}
\label{sec:trackjets}
In addition to the previously described calorimeter jets reconstructed from \topos, track jets in \ATLAS{} are built from reconstructed charged particle tracks associated  with the reconstructed primary collision vertex, which is defined by 
\begin{displaymath}
\sum (\pttrk)^{2} = \max .
\end{displaymath}
Here \pttrk{} is the transverse momentum of tracks pointing to a given vertex. The tracks associated with the primary vertex are required to have $\pttrk > 500$ \MeV{} and to be within $|\eta| < 2.5$. Additional reconstruction quality criteria are applied, including the number of hits in the pixel detector (at least one) and in the silicon microstrip detector (at least six) of the \ATLAS{} ID system. Further track selections are  based on the transverse ($d_{0}$, perpendicular to the beam axis) and longitudinal ($z_{0}$, along the beam axis) impact parameters of the tracks measured with respect to the primary vertex ($|d_{0}|<1.5$ \mm, $|z_{0}\sin\theta|<1.5$ \mm). Here $\theta$ is the polar angle of the track. 

Generally, track jets used in the studies presented in this paper are reconstructed with the same configurations as calorimeter jets, i.e. using the \antikt{} algorithm with $R = 0.4$ and $R = 0.6$. As only tracks originating from the hardest primary vertex in the collision event are used in the jet finding, the transverse momentum of any of these track jets provides a rather stable kinematic reference for matching calorimeter jets, as it is independent of the pile-up activity. Track jets can of course only be formed  within the tracking detector coverage (\AetaRange{2.5}), yielding an effective acceptance for track jets of $|\eta_{\mathrm{track jet}}| < 2.5 - R$.

Certain studies may require slight modifications of the track selection and the track-jet formation criteria and algorithms. Those are indicated in the respective descriptions of the applied methods. In particular, track jets may be further selected by requirements concerning the number of clustered tracks, the track-jet \pT, and the track-jet direction.

\subsection{Truth jets}
\label{sec:truthjets}
Truth jets can be formed from stable particles generated in \MC{} simulations. In general those are particles with a lifetime $\tau$ defined by $c \tau > 10$ \mm{} \cite{Beringer:1900zz}. 
The jet definitions applied are the same as the ones used for calorimeter and track jets (\antikt{} with distance parameters  $R = 0.4$ and $R = 0.6$, respectively). If truth jets are employed as a reference for calibrations purposes in \MC{} simulations, neither final-state muons nor neutrinos are included in the stable particles considered for its formation. The simulated calorimeter jets are calibrated with respect to truth jets consisting of stable particles leaving an observable signal (\emph{visible energy}) in the detector.\footnote{Muons can generate an observable signal in some of the \ATLAS{} calorimeters, but it is generally small and usually not proportional to the actual muon energy loss. Their contribution to the truth-jet energy, which can be large, is excluded to avoid biases and tails in the response function due to occasionally occurring high-\pt{} muons in the \MC-simulated calibration samples.} This is a particular useful strategy for inclusive jet measurements and the universal jet calibration discussed in this paper, but special truth-jet references including muons and/or neutrinos  may be utilised as well, in particular to understand the heavy-flavour jet response, as discussed in detail in \secRef{sec:bjets}.       

\subsection{Jet kinematics and directions}
\label{sec:jetdirections}
Kinematic properties of jets relevant for their use in final-state selections and final-state reconstruction are the transverse momentum \pt{} and the rapidity $y$. The full reconstruction of the jet kinematics including these variables takes into account the physics frame of reference, which in \ATLAS{} is defined event-by-event by the primary collision vertex discussed in \secRef{sec:trackjets}.
 
On the other hand, many effects corrected by the various \JES{} calibrations discussed in this paper are highly localised, i.e. they are due to specific detector features and inefficiencies at certain directions or ranges. The relevant directional variable to use as a basis for these corrections is then the detector pseudorapidity \etaDet, which is reconstructed in the nominal detector frame of reference in \ATLAS, and is centred at the nominal collision vertex $(x = 0,y = 0,z=0)$.

Directional relations to jets, and e.g. between the con\-sti\-tu\-ents of jet and its principal axis, can then be measured either in the physics or the detector reference frame, with the choice depending on the analysis. In the physics reference frame (\yphispace) the distance between any two objects is given by 
\begin{equation}
 \DeltaR = \sqrt{(\Delta y)^{2} + (\Delta \phi)^{2}} ,
\label{eq:deltaRphys}
\end{equation}    
where $\Delta y$ is the rapidity distance and $\Delta \phi$ is the azimuthal distance between them. The same distance measured in the detector frame of reference (\etaphispace) is calculated as
\begin{equation}
\DeltaR = \sqrt{(\Delta\eta)^{2} + (\Delta\phi)^{2}} ,
\label{eq:deltaRdet}
\end{equation}
where $\Delta\eta$ is the distance in pseudorapidity between any two objects. In case of jets and their constituents (\topos{} or tracks), $\eta = \etaDet$ is used. All jet clustering algorithms used in \ATLAS{} apply the physics frame distance in \eqRef{eq:deltaRphys} in their distance evaluations, as jets are considered to be massive physical objects, and the jet clustering is intended to follow energy flow patterns introduced by the physics of parton showers, fragmentation, and hadronisation from a common (particle) source. In this context  \topos{} and reconstructed tracks are considered  \emph{pseudo-particles} representing the true particle flow within the limitations introduced by the respective detector ac\-cep\-t\-an\-ces and resolutions.

\begin{figure*}[htp!]\centering
\subfloat[\pTrec{\EM}(\Npv), \antikt{} jets, $R = 0.4$ (MC)]{\includegraphics[width=0.45\textwidth]{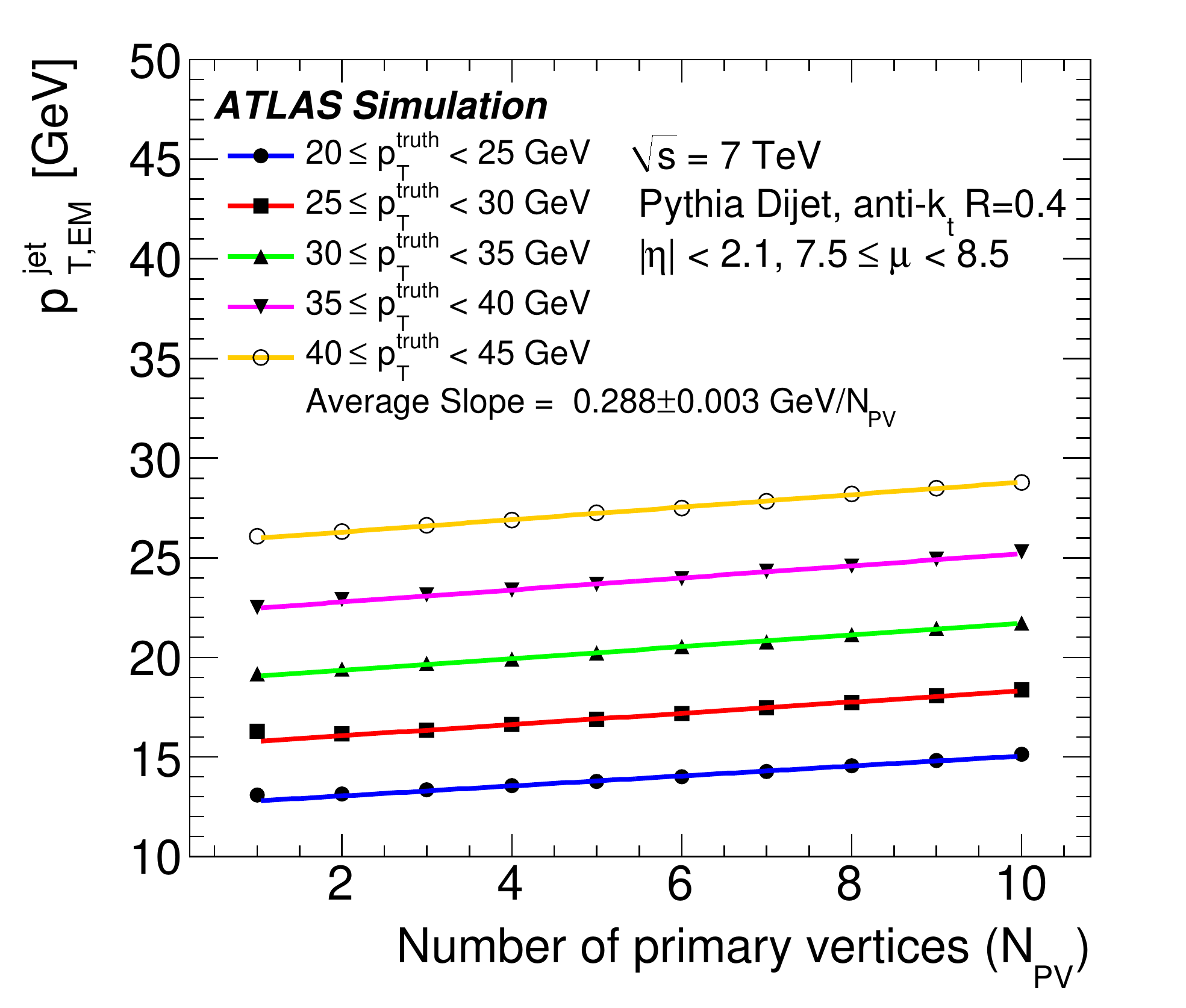} \label{fig:slope:akt4mc}}
\subfloat[\pTrec{\EM}(\Npv), \antikt{} jets, $R = 0.6$ (MC)]{\includegraphics[width=0.45\textwidth]{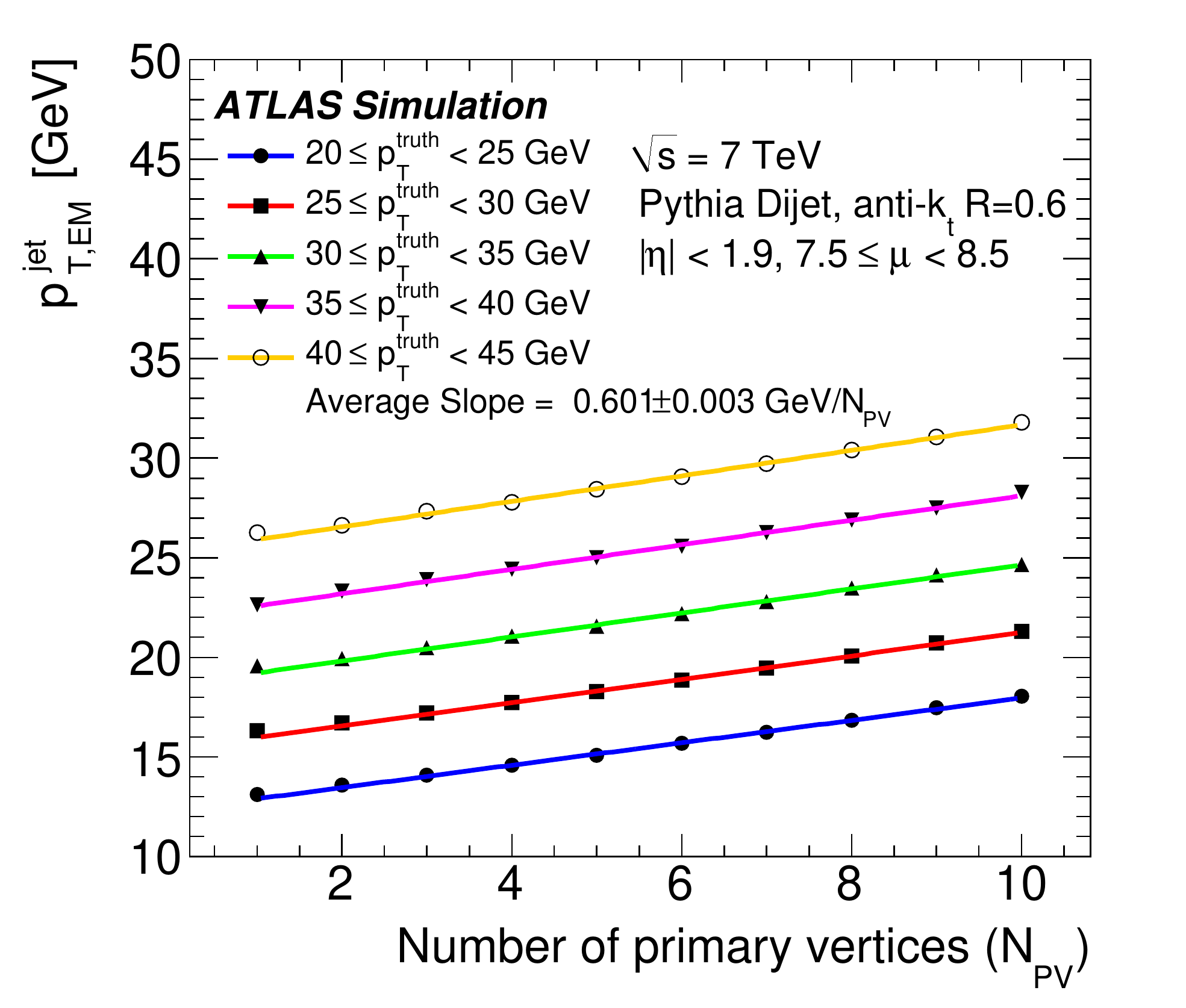} \label{fig:slope:akt6mc}}
\qquad
\subfloat[\pTrec{\EM}(\Npv), \antikt{} jets, $R = 0.4$ (Data)]{\includegraphics[width=0.45\textwidth]{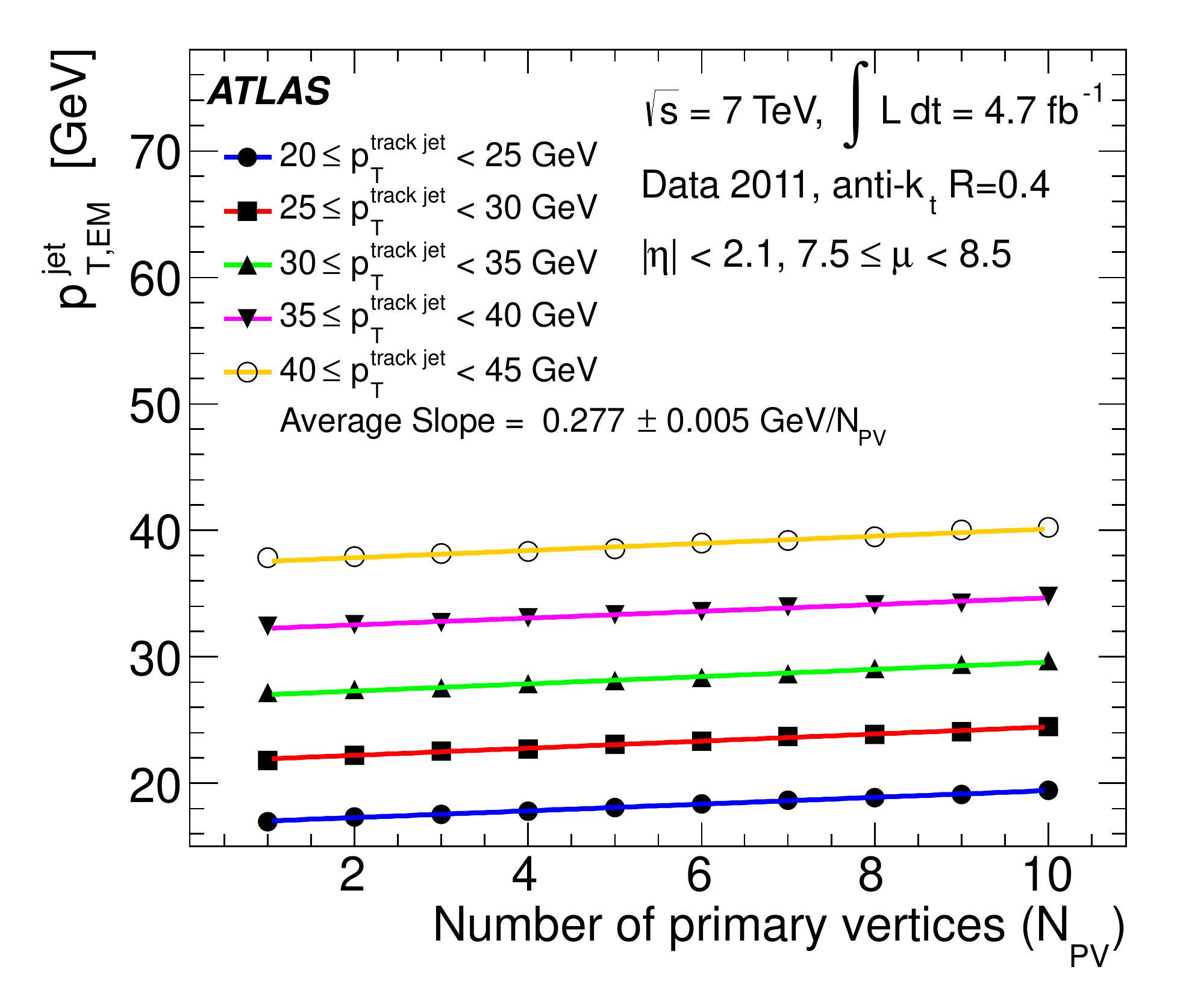}  \label{fig:slope:akt4trk}}
\subfloat[\pTrec{\EM}(\Npv), \antikt{} jets, $R = 0.6$ (Data)]{\includegraphics[width=0.45\textwidth]{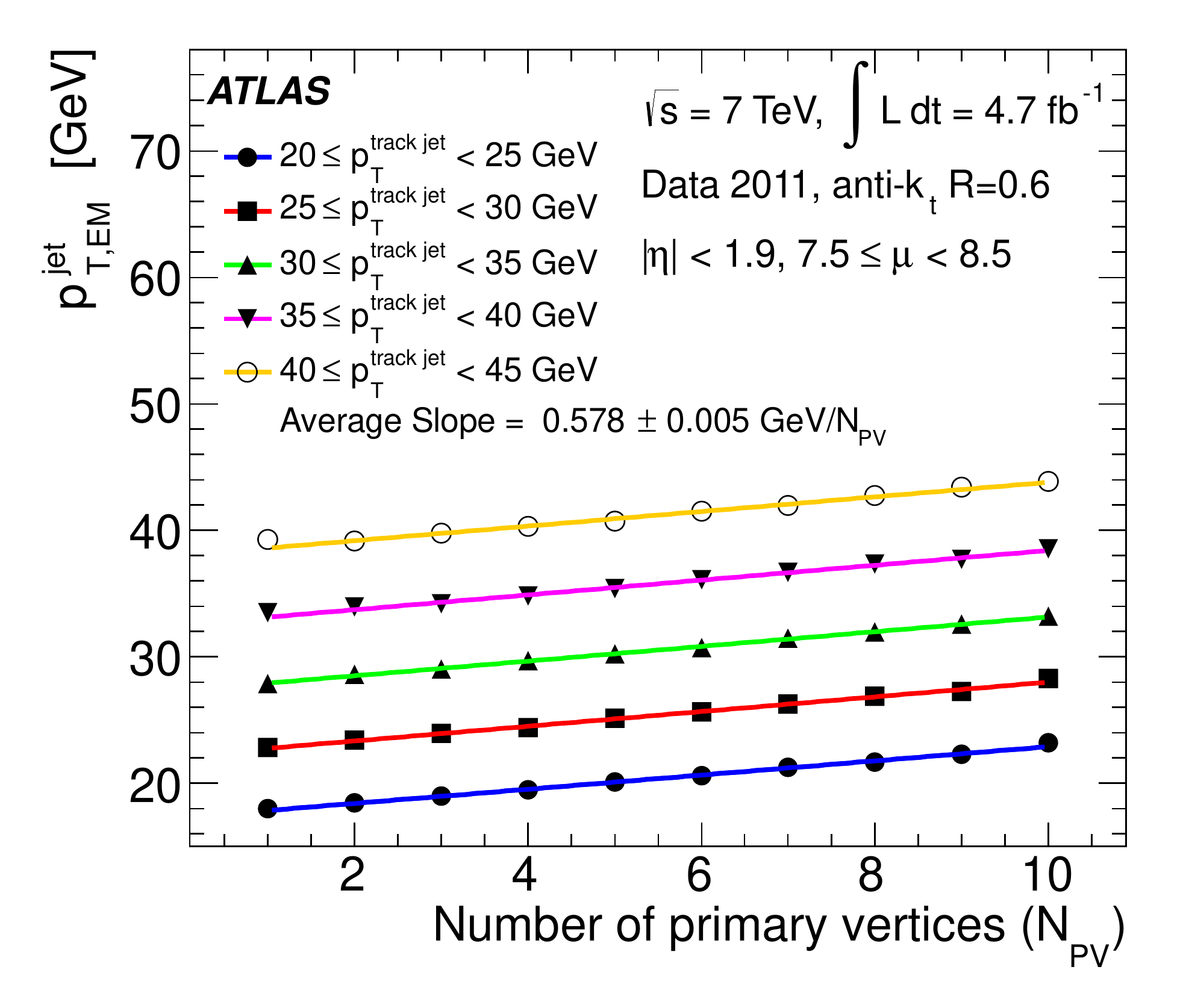}  \label{fig:slope:akt6trk}}
\caption[]{The average reconstructed transverse momentum \ptjetEM{} on \EM{} scale for jets in \MC{} simulations, as function of the number of reconstructed primary 
vertices \Npv{} and $7.5 \leq \axing < 8.5$, in various bins of truth-jet transverse momentum \pttrue, for jets with \subref{fig:slope:akt4mc} $R =0.4$  
and \subref{fig:slope:akt6mc} $R = 0.6$. The dependence of \ptjetEM{} on \Npv{} in data, in bins of
track-jet transverse momentum \pttrk{}, is shown in \subref{fig:slope:akt4trk}  for $R = 0.4$ jets, and in \subref{fig:slope:akt6trk} 
for $R = 0.6$ jets.\label{fig:slope}}
\end{figure*}

\begin{figure*}[htp!!]
\centering
\subfloat[\pTrec{\EM}(\axing), $20 \leq \pttrue < 25$~\GeV{} (MC)]{\includegraphics[width=0.45\textwidth]{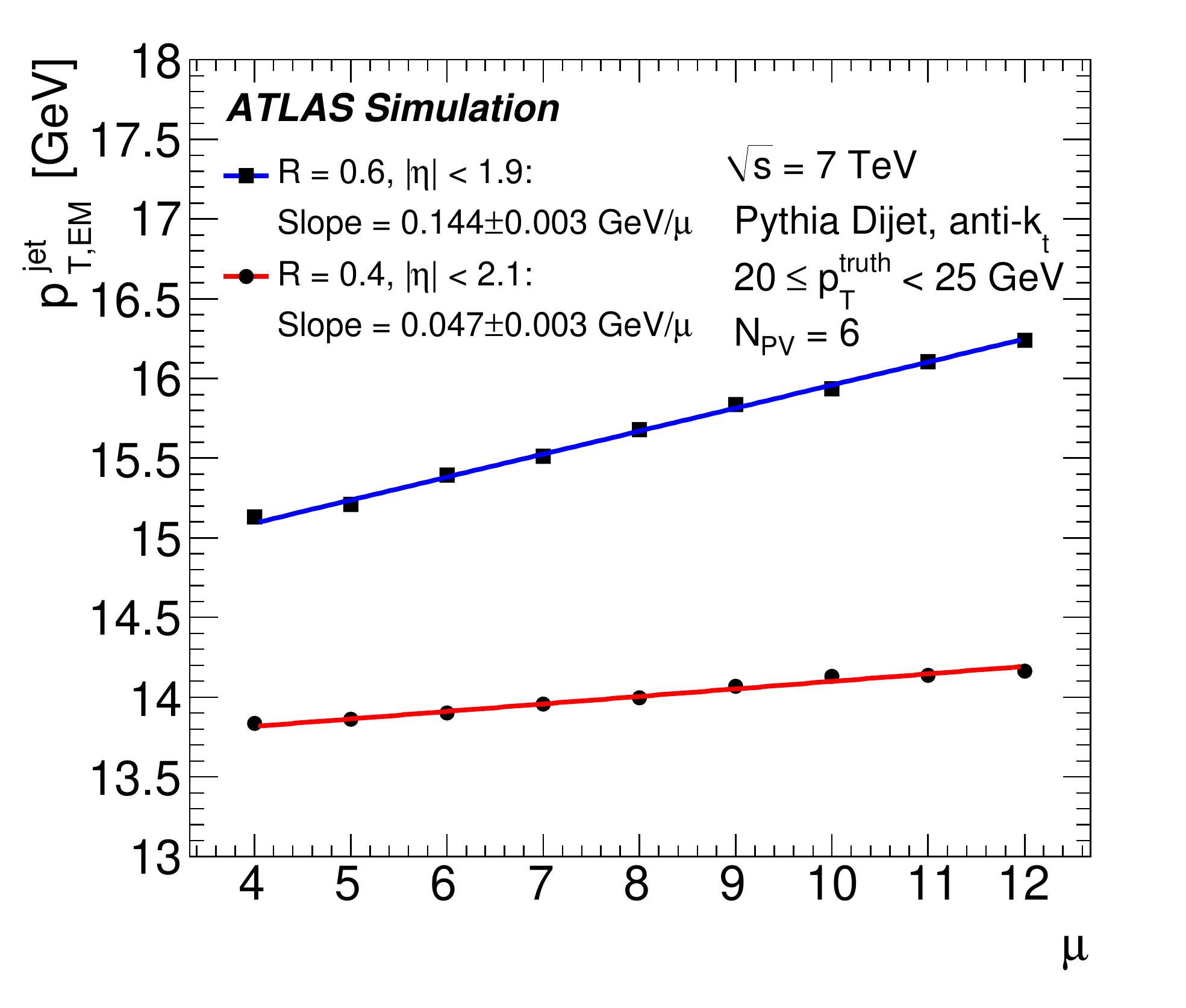} \label{fig:muslope:mc}}
\subfloat[\pTrec{\EM}(\axing), $20 \leq \ptjetTrk < 25$~\GeV{} (Data)]{\includegraphics[width=0.45\textwidth]{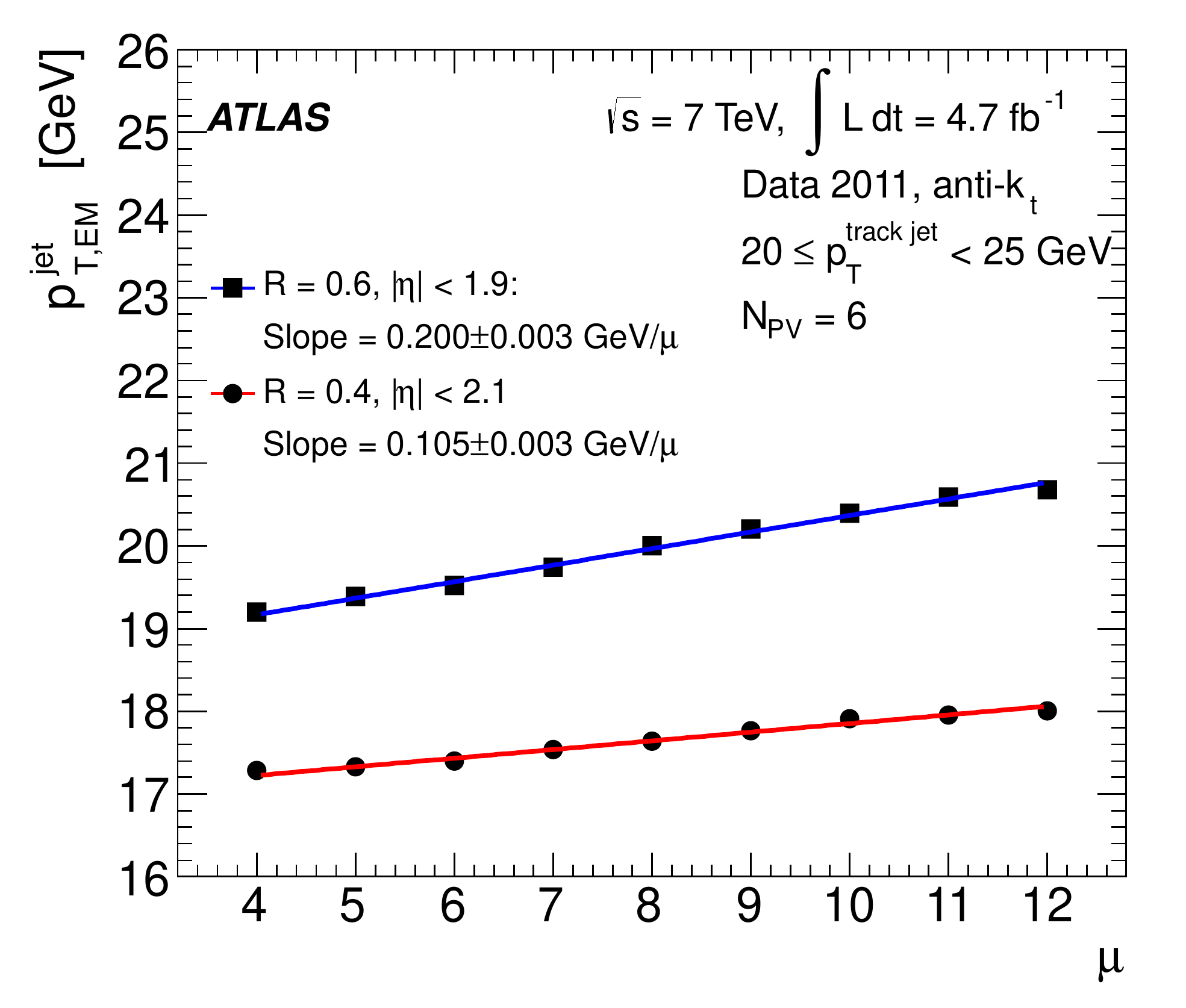} \label{fig:muslope:trk}}
\caption[]{The average reconstructed jet transverse momentum \ptjetEM{} on \EM{} scale as function of the average number of collisions \axing{} at
a fixed number of primary vertices 
$\Npv = 6$, for truth jets in MC simulation \subref{fig:muslope:mc} in the lowest bin of \pttrue{} and \subref{fig:muslope:trk} in the lowest bin of track jet transverse momentum \ptjetTrk{} considered in data.
\label{fig:muslope}}
\end{figure*}

\section{Jet energy correction for pile-up interactions}
\label{sec:pileupsection}

\subsection{Pile-up correction method}
\label{sec:pileupmethod}

The pile-up correction method applied to reconstructed jets in \ATLAS{} is derived from \MC{}
simulations and validated with \insitu{} and simulation based techniques. 
The 
approach is to calculate the amount of transverse momentum generated by pile-up in a jet in \MC{} simulation,
and subtract this offset \pToff{} from the reconstructed jet $\pT^{\mathrm{jet}}$ at any given
signal scale (\EM{} or \LCW{}). At least to first order, pile-up contributions to the jet signal can be considered 
stochastic and diffuse with respect to the true jet signal. Therefore, 
both in-time and out-of-time pile-up are expected to depend only
on the past and present pile-up activity, with linear relations between the
amount of activity and the pile-up signal.

\subsection{Principal pile-up correction strategy}
\label{sec:pileupstrategy}

To characterise the in-time pile-up activity, the number of reconstructed primary vertices (\Npv{}) is used.
The \ATLAS{} tracking detector timing resolution allows the reconstruction of only in-time tracks and vertices, so that \Npv{} provides
a good measure of the actual number of \pp{} collisions in a re\-cor\-ded event.

For the out-of-time pile-up activity, the average number
of interactions per bunch crossing (\axing) at the time of the re\-cor\-ded events provides a
good estimator. It is derived by averaging the actual number of interactions per bunch
crossing over a rather large window $\Delta t$ in time, which safely encompasses the time
interval during which the \ATLAS{} calorimeter signal is sensitive to the activity in
the collision history ($\Delta t \gg 600$~ns for the liquid-argon calorimeters). The observable \axing{}
can be reconstructed from the average luminosity $L$ over this period $\Delta t$, the
total inelastic \pp{} cross section ($\sigma_{\mathrm{inel}} = 71.5$~mb \cite{Aad:2011eu}),
the number of colliding bunches in LHC ($N_{\mathrm{bunch}}$) and the LHC revolution
frequency ($f_{\mathrm{LHC}}$) (see Ref.~\cite{Aad:2011dr} for details): 
\begin{displaymath}
        \axing = \frac{L \times \sigma_{\mathrm{inel}}}{N_{\mathrm{bunch}} \times f_{\mathrm{LHC}}} .
\end{displaymath}

The \MC-based jet calibration is derived for a given (reference) pile-up condition\footnote{The particular choice for a working point, here ($\NpvRef = 4.9, \axingRef = 5.4$),
is arbitrary and bears no consequence for the correction method and its uncertainty.}
$(\NpvRef,\axingRef)$ such that $\pToff(\Npv = \NpvRef,$ $\axing = \axingRef) = 0$. 
As the amount of energy scattered into a jet by pile-up and the signal modification imposed by 
the pile-up history determine \pToff, a general dependence on the distances from the 
reference point is expected. From the nature of pile-up discussed earlier, the linear 
scaling of \pToff{} in both \Npv{} and \axing{} provides the ansatz for a correction,
\begin{eqnarray}
\lefteqn{\pToff(\Npv,\axing,\etaDet) = \pT^{\mathrm{jet}}(\Npv,\axing,\etaDet) - \pttrue} \nonumber \\
              &  = & \frac{\partial\pT}{\partial\Npv}(\etaDet) \left(\Npv - \NpvRef\right) +
                  \frac{\partial\pT}{\partial{\axing}}(\etaDet) \left(\axing - \axingRef\right) \nonumber \\
              &  = & \alphaFct \cdot \left(\Npv - \NpvRef\right) + \betaFct \cdot \left(\axing - \axingRef\right)
\label{eq:corrfct}
\end{eqnarray}
Here, 
$\pT^{\mathrm{jet}}(\Npv,\axing,\etaDet) $
is the reconstructed transverse momentum of the jet (without the \JES{} correction described in Section
\ref{sec:jetrecocalibsequence} applied) in a given
pile-up condition (\Npv,\axing) and at a given direction \etaDet{} in the detector. The true transverse momentum of the
jet (\pttrue{}) is available from the generated particle jet matching a reconstructed jet in \MC{} simulations. 
The coefficients \alphaFct{} and \betaFct{} depend on \etaDet{}, as both in-time and out-of-time pile-up signal contributions manifest themselves 
differently in different calorimeter regions, according to the following influences:
\begin{enumerate}
\item The energy flow from collisions into that region. 
\item The calorimeter granularity and occupancy after \topo{} reconstruction, leading to
different acceptances at cluster level and different probabilities for multiple particle
showers to overlap in a single cluster.
\item The effective sensitivity to out-of-time pile-up introduced by different calorimeter
signal shapes.
\end{enumerate}
The offset \pToff{} can be determined in \MC{} simulation for jets on the \EM{} or the \LCW{} scale by using
the corresponding reconstructed transverse momentum on one of those scales,
i.e. $\pT^{\mathrm{jet}} = \pTrec{\EM}$ or $\pT^{\mathrm{jet}} = \pTrec{\LCW}$ in \eqRef{eq:corrfct}, and \pttrue.
The particular choice of scale affects the magnitude of the coefficients and, therefore, the transverse 
momentum offset itself,
\begin{eqnarray*}
  \pToff^{\EM}  & \mapsto & \left\{\alphaEMFct,\betaEMFct\right\} \\
  \pToff^{\LCW} & \mapsto & \left\{\alphaLCWFct,\betaLCWFct\right\} .
\end{eqnarray*}
The corrected transverse momentum of the jet at either of the two scales (\pTcor{\EM}{}
or \pTcor{\LCW}) is then given by
\begin{eqnarray}
  \pTcor{\EM} & = & \pTrec{\EM} - \pToff^{\EM}(\Npv,\axing,\etaDet) \\ 
  \pTcor{\LCW} & = & \pTrec{\LCW} - \pToff^{\LCW}(\Npv,\axing,\etaDet) .
\end{eqnarray}
After applying the correction, the original \pTrec{\EM}{} and \pTrec{\LCW}{}
dependence on \Npv{} and \axing{} is expected to vanish in the corresponding corrected
\pTcor{\EM}{} and \pTcor{\LCW}. 
\begin{figure*}[htp!]\centering
\subfloat[\alphaEMFct{}, \antikt{} jets, $R = 0.4$]{\includegraphics[width=0.49\textwidth]{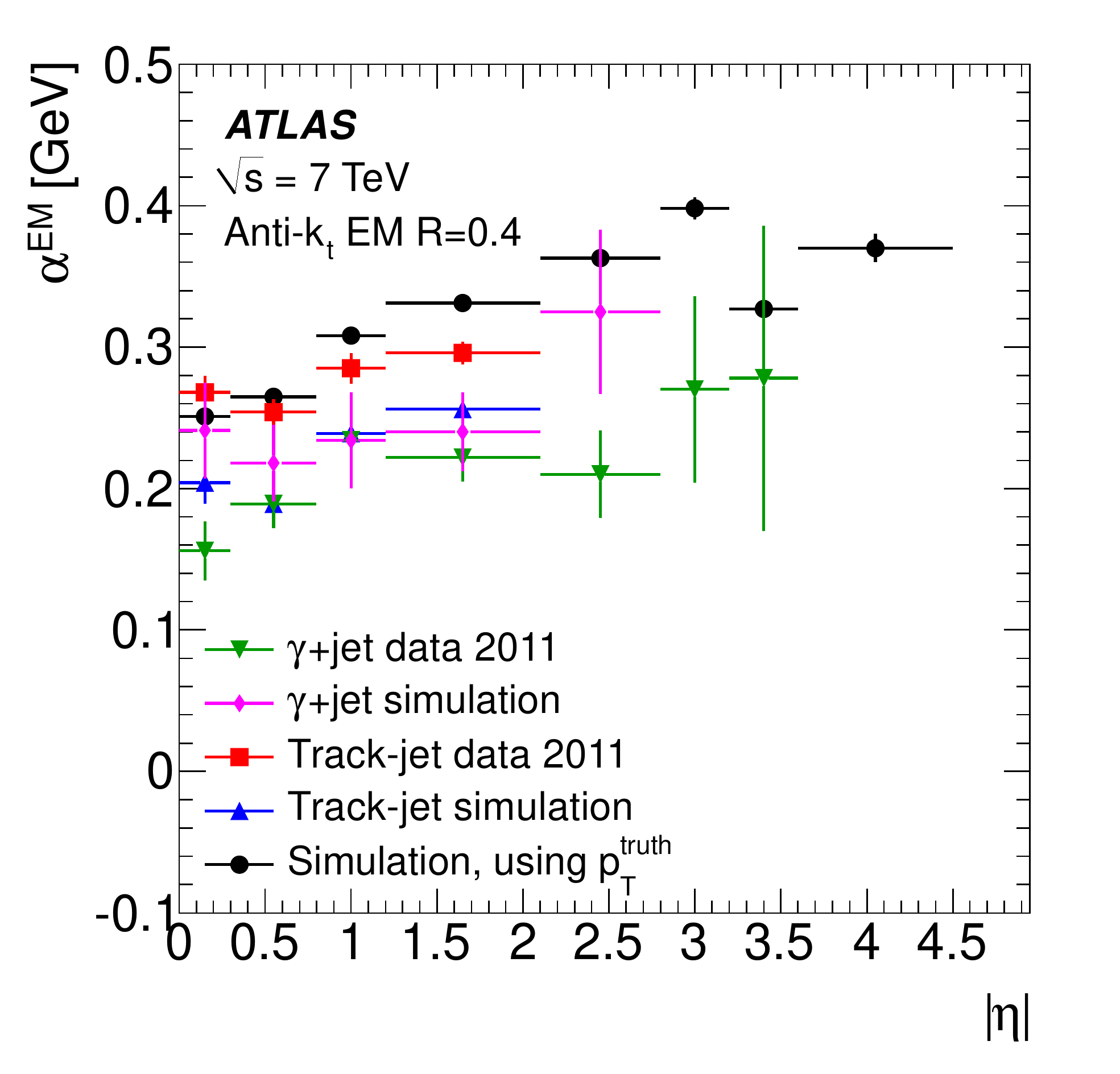} \label{fig:slope_summary:npv4}}
\subfloat[\alphaEMFct{}, \antikt{} jets, $R = 0.6$]{\includegraphics[width=0.49\textwidth]{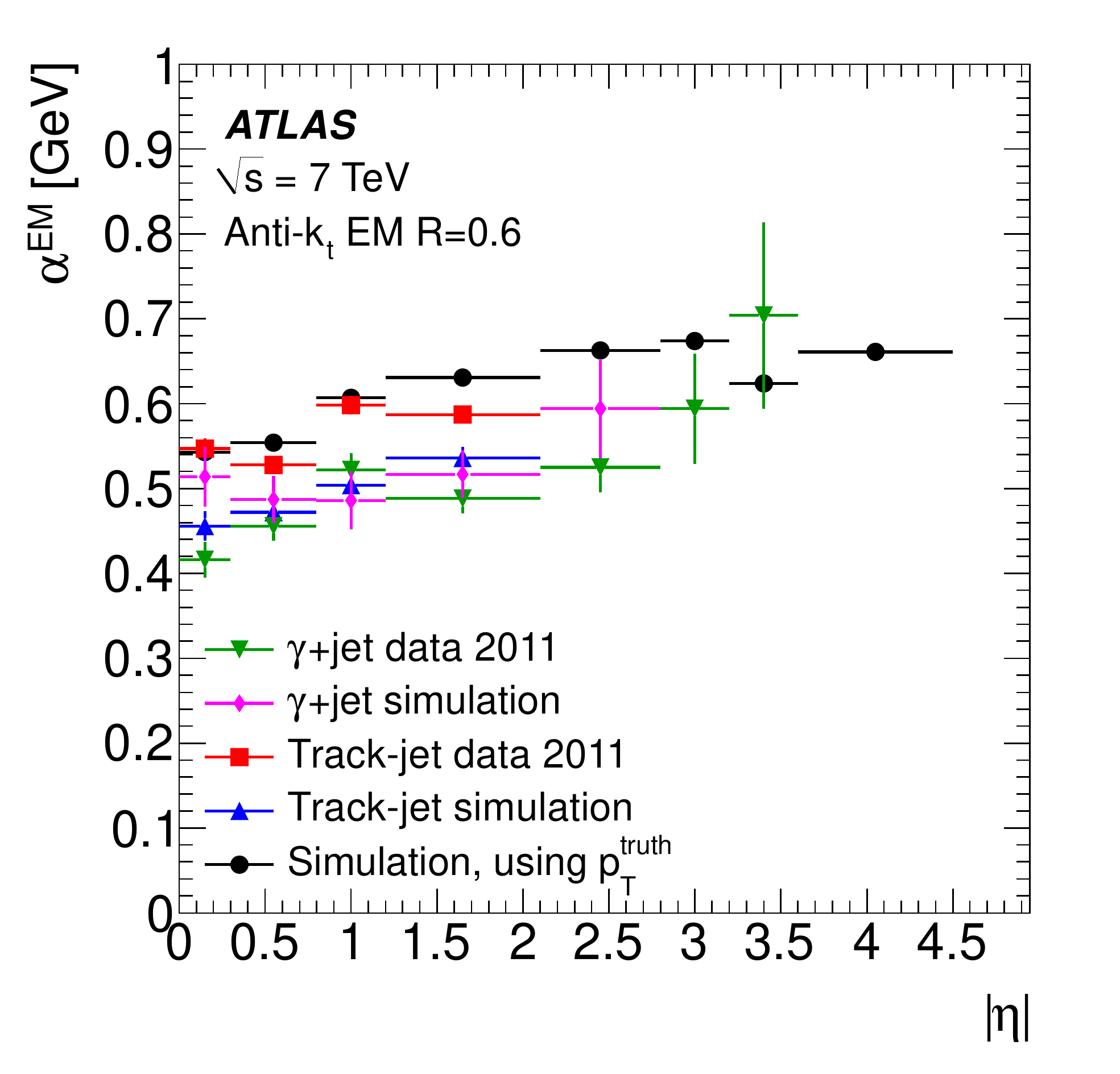} \label{fig:slope_summary:npv6}} \\
\subfloat[\betaEMFct{}, \antikt{} jets, $R = 0.4$]{\includegraphics[width=0.49\textwidth]{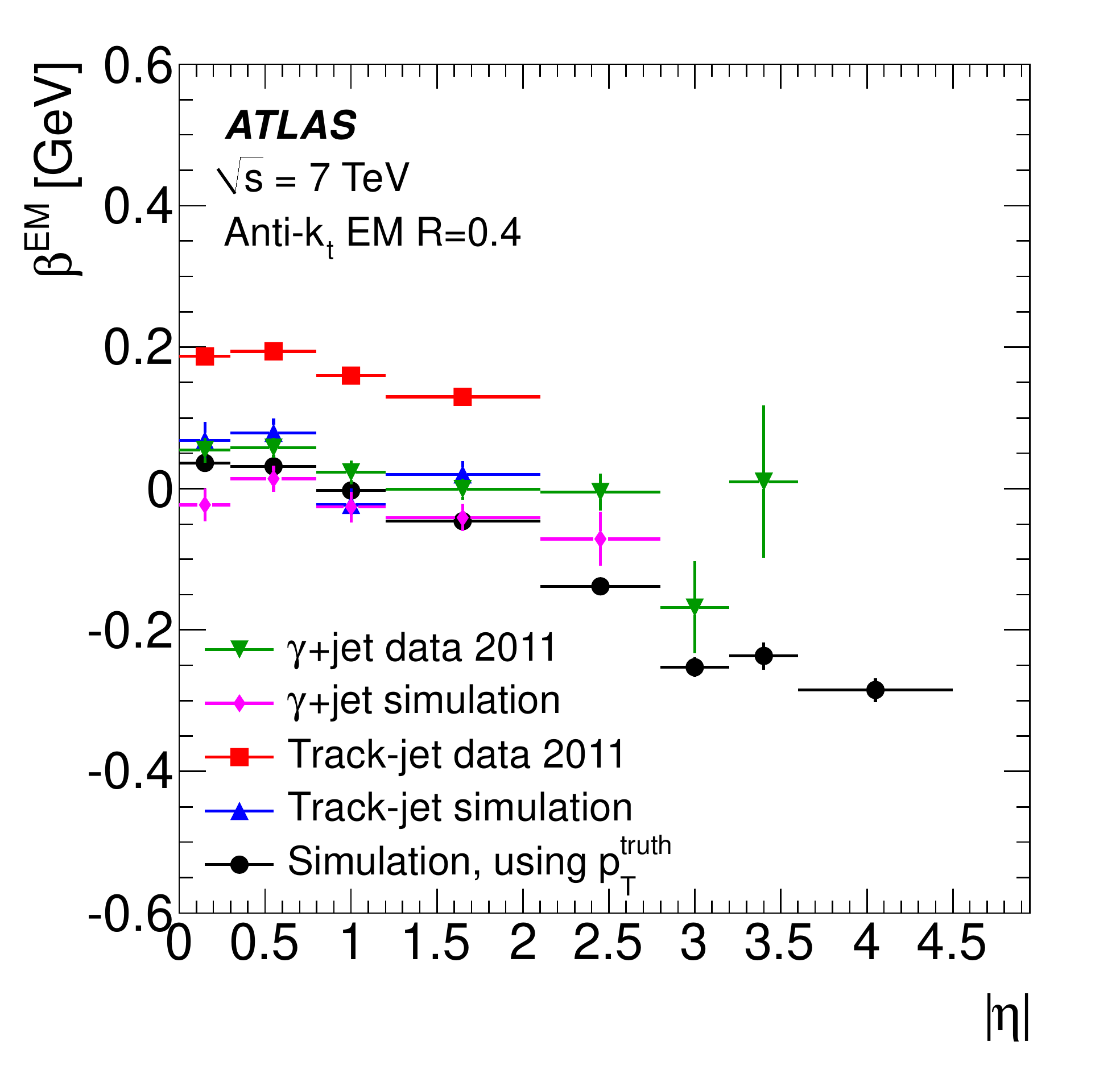}  \label{fig:slope_summary:mu4}}
\subfloat[\betaEMFct{}, \antikt{} jets, $R = 0.6$]{\includegraphics[width=0.49\textwidth]{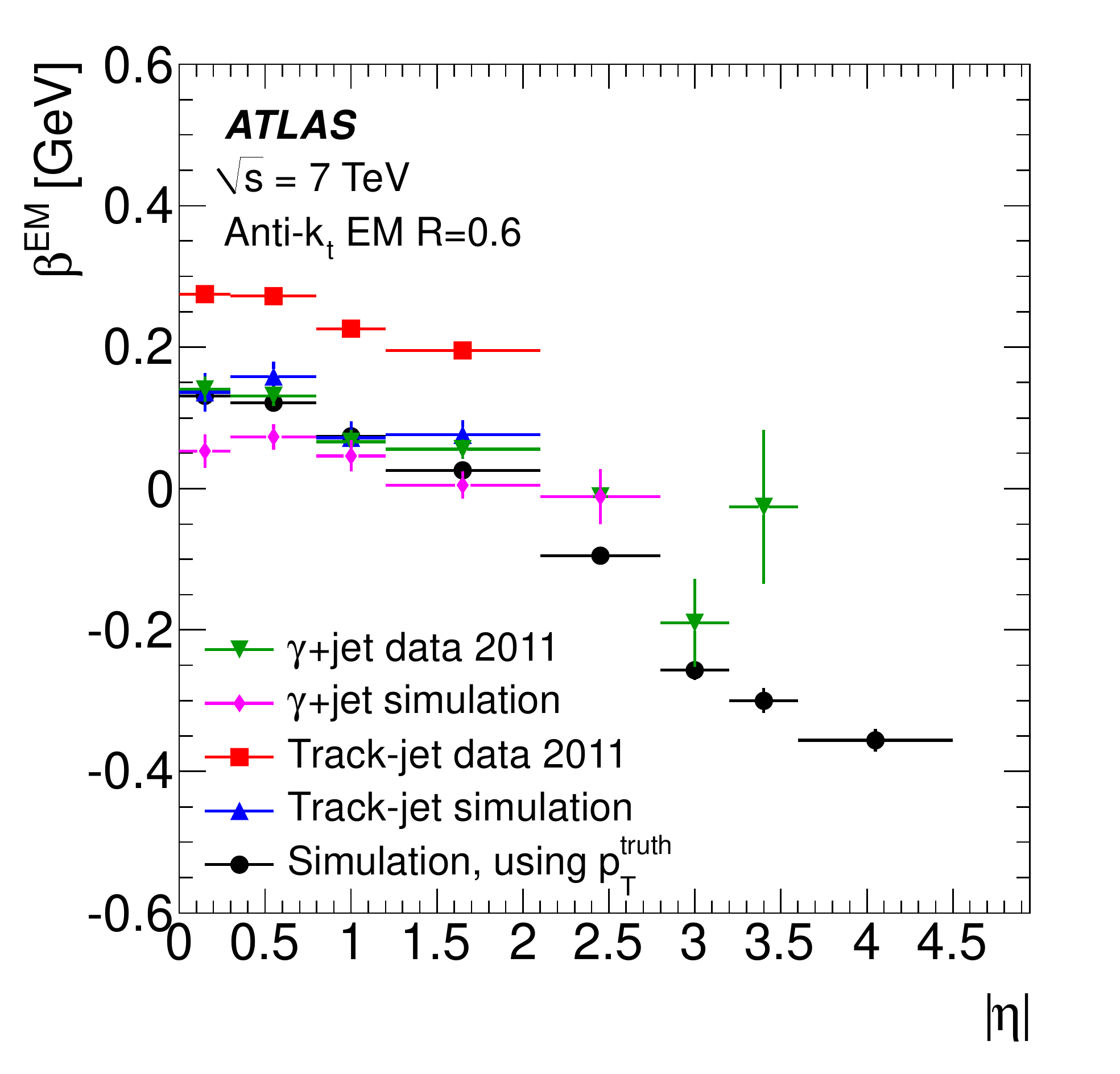}  \label{fig:slope_summary:mu6}}
\caption[]{The pile-up contribution per additional vertex, measured as $\alphaEM = \partial\ptjetEM/\partial\Npv$, as function of $\left|\etaDet\right|$, for the various methods 
discussed in the text, for \subref{fig:slope_summary:npv4} $R = 0.4$ and \subref{fig:slope_summary:npv6} $R = 0.6$ jets. 
The contribution from \axing{}, calculated as $\betaEM = \partial\ptjetEM/\partial\axing$ and displayed for the 
various methods as function of $\left|\etaDet\right|$, is shown for the two jet sizes in \subref{fig:slope_summary:mu4} and 
\subref{fig:slope_summary:mu6}, respectively. The points for the determination of \alphaEM{} and \betaEM{} from \MC{} simulations use the offset 
calculated from the reconstructed \pTrec{\EM} and the true (particle level) \pttrue, as indicated in Eq.~\ref{eq:corrfct}. \label{fig:slope_summary}}
\end{figure*}

\subsection{Derivation of pile-up correction parameters}
\label{sec:pileupparameters}

Figures~\ref{fig:slope}\subref{fig:slope:akt4mc} and \ref{fig:slope}\subref{fig:slope:akt6mc} show the 
dependence of \pTrec{\EM}, and thus $\pToff^{\EM}$, on \Npv{}. In this example, 
narrow ($R = 0.4$, $\left|\etaDet\right| < 2.1$) and wide ($R = 0.6$, $\left|\etaDet\right| < 1.9$) central jets reconstructed in \MC{} simulation
are shown for events within a given 
range $7.5 \leq \axing < 8.5$. 
The jet \pt{} varies by 
$0.277 \pm 0.005$ \GeV (in data) and 
$0.288 \pm 0.003$ \GeV (in \MC{} simulations) per primary vertex 
for jets with $R=0.4$ and by 
$0.578 \pm 0.005$ \GeV (in data) and 
$0.601 \pm 0.003$ \GeV (in \MC{} simulations) per primary vertex
for jets with $R=0.6$.
The slopes  \alphaEM{} are found to be independent of the true jet transverse momentum 
\pttrue, as expected from the diffuse character of in-time pile-up signal contributions. 

A qualitatively similar behaviour can be observed in collision data for calorimeter jets 
individually matched with track jets, the latter reconstructed as discussed in \secRef{sec:trackjets}. 
%
%
The \Npv{} dependence of \pTrec{\EM}{} can be measured in bins of the track-jet transverse momentum
\ptjetTrk. 
Jets formed from tracks are much less sensitive to pile-up and can be used
as a stable reference to investigate pile-up effects.
Figures~\ref{fig:slope}\subref{fig:slope:akt4trk} and 
\ref{fig:slope}\subref{fig:slope:akt6trk} show the results for the same calorimeter regions and out-of-time pile-up 
condition as for the \MC-simulated jets in Figs. \ref{fig:slope}\subref{fig:slope:akt4mc} and 
\ref{fig:slope}\subref{fig:slope:akt6mc}. The results shown in \figRef{fig:slope} also confirm the expectation 
that the contributions from in-time pile-up to the jet signal are larger for wider jets
($\alphaEM(R = 0.6) > \alphaEM(R = 0.4)$), but scale only approximately with the size of 
the jet catchment area \cite{Cacciari:2008jetarea}
determined by the choice of distance parameter $R$ in the \antikt{} 
algorithm.

The dependence of \ptjetEM{} on \axing{}, for a fixed $\Npv = 6$, is shown in 
\figRef{fig:muslope}\subref{fig:muslope:mc} for \MC{} simulations using truth jets, and in 
\figRef{fig:muslope}\subref{fig:muslope:trk} for collision data using track jets. 
The kinematic bins shown are the lowest bins considered, with $20 < \pttrue < 25$~\GeV{} and 
$20 < \ptjetTrk < 25$~\GeV{} for \MC{} simulations and data, respectively. 
The jet \pt{} varies by 
$0.047 \pm 0.003$~\GeV{} (in \MC{} simulations)
$0.105 \pm 0.003$~\GeV{} (in data) per primary vertex 
for jets with $R=0.4$ 

The result confirms 
the expectations that the dependence of \ptjetEM{} on the out-of-time pile-up is linear 
and significantly less than its dependence on the in-time pile-up contribution 
scaling with \Npv. Its magnitude is still different for jets with $R = 0.6$, as the size 
of the jet catchment area again determines the absolute contribution to \ptjetEM{}.

The correction coefficients for jets calibrated with the \EM+\JES{} scheme, \alphaEM{} and \betaEM{}, are both determined from \MC{} simulations as functions of the jet 
direction \etaDet. For this, the \Npv{} dependence of \pTrec{\EM}(\etaDet) reconstructed in various bins
of \axing{} in the simulation is fitted and then averaged, yielding \alphaEMFct. Accordingly and independently,
the dependence of \pTrec{\EM}{} on \axing{} is fitted in bins of \Npv, yielding the average
\betaEMFct, again using \MC{} simulations. An identical procedure is used to find the correction functions \alphaLCWFct{} and 
\betaLCWFct{} for jets calibrated with the \LCWJES{} scheme. 

The parameters \alphaEM(\alphaLCW) and \betaEM(\betaLCW) can be 
also measured 
with \insitu{} techniques. This is discussed in 
\secRef{sec:pileupineta}.

\subsection{Pile-up validation with \insitu{} techniques and effect of out-of-time pile-up in different calorimeter regions}
\label{sec:pileupineta}
The parameters \alphaEM(\alphaLCW) and \betaEM(\betaLCW) can be measured in data with respect to a
reference that is stable under pile-up
using track jets or photons in \gammajet{} events 
as kinematic reference that does not depend on pile-up. 

%
%
%
%

The variation of the \pt{}
balance $\pTrec{\EM} - \pT^{\gamma}$ ($\pTrec{\LCW} - \pT^{\gamma}$) in \gammajet{} events can be used in
data and \MC{} simulation (similarly to the strategy discussed in \secRef{sec:gammajetInSitu}),
as a function of \Npv{} and \axing{}. 
\FigRef{fig:slope_summary} summarises \alphaEMFct{} and \betaEMFct{} determined 
with track jets 
and \gammajet{} events, and their dependence on \etaDet. Both methods suffer from lack of statistics or large systematic uncertainties in the 2011 data, but are used in \datatomc{} comparisons to determine systematic uncertainties of the \MC-based method (see the corresponding discussion in \secRef{sec:pileupinsitusystematics}).

%

The decrease of \betaEMFct{} towards higher \etaDet{}, as shown in Figs.~\ref{fig:slope_summary}\subref{fig:slope_summary:mu4} and \ref{fig:slope_summary}\subref{fig:slope_summary:mu6}, 
indicates a decreasing signal contribution to \pTrec{\EM}{} per out-of-time pile-up interaction. For jets with 
$\left|\etaDet\right| > 1.5$, the offset is increasingly suppressed in the signal 
with increasing \axing{} ($\betaEMFct < 0$). This constitutes a qualitative 
departure from the behaviour of the pile-up history contribution in the central region of \ATLAS{}, where
this out-of-time pile-up leads to systematically increasing signal contributions
with increasing \axing.

This is a consequence of two effects. First, for $\left|\etaDet\right|$ larger than about $1.7$ the
hadronic calorimetry in \ATLAS{} changes from the \Tile{} calorimeter to the \LAr{}
end-cap (\HEC) calorimeter. The \Tile{} calorimeter has a unipolar and fast signal shape
\cite{TileReadiness}. It has little sensitivity to out-of-time pile-up, with an
approximate shape signal baseline of $150$~\ns. The out-of-time history manifests itself
in this calorimeter as a small positive increase of its contribution to 
the jet signal with increasing \axing. 

The \HEC, on the other hand, has the typical \ATLAS{} \LAr{} calorimeter bipolar pulse shape
with approximately $600$~ns baseline. This leads to an increasing suppression of the
contribution from this calorimeter to the jet signal with increasing \axing, as more
activity from the pile-up history increases the contribution weighted by the negative
pulse shape.

Second, for $\left|\etaDet\right|$ larger than approximately $3.2$, coverage is provided by the \ATLAS{}
forward calorimeter (\FCal). While still a liquid-argon calorimeter, the \FCal{} features 
a considerably faster signal due to very thin argon gaps. The shaping function for this 
signal is bipolar with a net zero integral and a similar positive shape as in other \ATLAS{} liquid-argon calorimeters, 
but with a shorter overall pulse baseline (approximately $400$~ns). Thus, the \FCal{} shaping 
function has larger negative weights for out-of-time pile-up of up to $70\%$ of the 
(positive) pulse peak height, as compared to typically $10\%$ to $20\%$ in the other \LAr{}
calorimeters \cite{LArReadiness_mod}. These larger negative weights lead to larger signal
suppression with increasing activity in the pile-up history and thus with increasing \axing.


%

\section{\Insitu{} transverse momentum balance techniques}
\label{sec:InSitu}
In this section an overview is given on how the \datatomc{} differences
are assessed using \insitu{} techniques exploiting the transverse momentum
balance between the jet and a well-mea\-su\-red reference object.

The calibration of jets in the forward region of the detector relative
to jets in the central regions is discussed in more detail in \secRef{sec:etaintercalibration}}.
Jets in the central region are calibrated using photons or \Zboson{} bosons
as reference objects  up to a transverse momentum of $800$ \GeV{}
(see \secRef{sec:ZjetInSitu} and \secRef{sec:gammajetInSitu}).
Jets with higher \pt{} are calibrated using a system of low-\pt{} jets recoiling
against a high-\pt{} jet (see \secRef{sec:multijet}).

\subsection{Relative \insitu{} calibration between the central and forward rapidity regions}
\label{sec:etaInterCalib}
Transverse momentum balance in dijet events is exploited to study the pseudorapidity dependence of the jet response. A relative \etaic{} is derived using the \emph{\etamm} described in Ref.~\cite{jespaper2010} 
to correct the jets in data for residual effects not captured by the initial calibration derived from \MC{} simulations and based on truth jets. 
This method is applied for jets with $20 \leq \ptjet<1500$~\GeV{} and $|\etaDet| \leq 4.5$.
Jets up to $|\etaDet| = 2.8$ are calibrated using $|\etaDet| < 0.8$ %
as a reference region.
For jets with $\etaDet>2.8$ ($\etaDet<-2.8$), for which the uncertainty on the derived calibration becomes large,
the calibration determined at $\etaDet=2.8$ ($\etaDet=-2.8$) is used.\footnote{The relative jet response is measured
independently for each \etaDet{} hemisphere of the detector to accommodate asymmetries introduced by the actual collision vertex position during data taking.}
Jets that fall in the reference region receive no additional correction on average. The \etaic{} is applied to all jets prior to deriving the absolute %
calibration of the central region.

The largest uncertainty of the dijet balance technique is due to the modelling of the additional parton radiation altering the \pt{} balance. This uncertainty is estimated using \MC{} simulations employing the \pythia{} and \herwigpp{} generators, respectively.

\subsection{\Insitu{} calibration methods for the central rapidity region}
\label{sec:absolute_insitu_methods}
The energy scale of jets is tested \insitu{} using a well-calibrated object as reference.
The following techniques are used for the central rapidity region $\etaDet < 1.2$: %

\begin{mylist}
\myitem{Direct transverse momentum balance between a photon or a $\textit{Z}$ boson and a jet}
  Events with a photon or a \Zboson{} boson and a recoiling jet are used to directly compare the transverse momentum of the jet to that 
  of the photon  or the \Zboson{} boson (direct balance, \DB). %
   The data are compared to \MC{} simulations in the jet pseudorapidity range $|\etaDet| < 1.2$.  The \gammajet{} analysis covers a range in    photon transverse momentum from $25$ to $800$~\GeV, while the \Zjet{} analysis covers a range in \Zboson{} transverse momentum from $15$ to $200$~\GeV.  However, only the direct transverse momentum balance between the \Zboson{} and the jet is used in the derivation of the residual \JES{} correction, as the method employing \pt{} balance between a photon and the full hadronic recoil, rather than the jet (see item \ref{item:MPF} below),  is used in place of the direct \gammajet{} balance, see \secRef{sec:mpddbgamma} for more details.

\myitem{Transverse momentum balance between a photon and the hadronic recoil \label{item:MPF}}
      The photon transverse momentum is balanced against the full hadronic recoil using the projection of the missing transverse momentum onto the photon direction. With this missing transverse momentum projection fraction (\MPF) technique,  the calorimeter response for the hadronic recoil is measured, which is independent of any jet definition. The comparison is done in the same kinematic region as the direct photon balance method. 
\myitem{Balance between a low-\textit{p}$_{\mathrm{\textbf{T}}}$ jet system and a high-\textit{p}$_{\mathrm{\textbf{T}}}$ jet}
      Jets at high \pt{} can be balanced against a recoil system of low \pt{} jets within $\etaDet < 2.8$ %
      if the low \pt{} jets are well calibrated using \gammajet{} or \Zjet{} \insitu{} techniques.
      The multijet balance can be iterated several times to increase the \nonleading{} (in terms of \pt) jets \pt{} range
      beyond the values covered by  \gammajet{} or \Zjet{} balance, and reaching higher \pt{} of
      the leading jet, until statistical limitations preclude a precise measurement.       
      This method can probe the jet energy scale up to the \TeV{} regime.
\end{mylist}

In addition to the methods mentioned above, the mean transverse momentum sum of tracks within a cone around the jet direction
provides an independent test of the calorimeter energy scale over the entire measured \ptjet{} range within the tracking acceptance.
This method, described in Ref.~\cite{jespaper2010}, is used for the 2010 dataset and is also studied for the inclusive jet data sample in 2011. It is also used for \bjets{} (see \secRef{sec:bjets}). However, because of the relatively large associated systematic uncertainties, it is not included in the \JES{} calibration derived from the combination of \insitu{} methods for inclusive jets in 2011. This  calibration can be constrained to much higher quality by applying the three methods described above.

\section{Relative forward-jet calibration using dijet events }
\label{sec:etaintercalibration}
The calibration of the forward detector can be performed by exploiting the
transverse momentum balance in events with two jets at high transverse momentum.
A well calibrated jet in the central part of the detector is balanced against a jet in the forward
region.
Thus the whole detector acceptance in $\eta$ can be equalised
as a function of \ptjet.
In addition to this simple approach, a matrix method is used where jets
in all regions (and not only the central one) are used for the $\eta$-intercalibration.

In the following the results for the \EMJES{} scheme are discussed
as an example.
While the measured relative response can deviate by a few percent 
between the \EMJES{} and the \LCWJES{} calibration schemes, 
the ratio between data and Monte Carlo simulation agrees within
a few permille.

\begin{figure}
  \centering
\includegraphics[width=.48\textwidth]{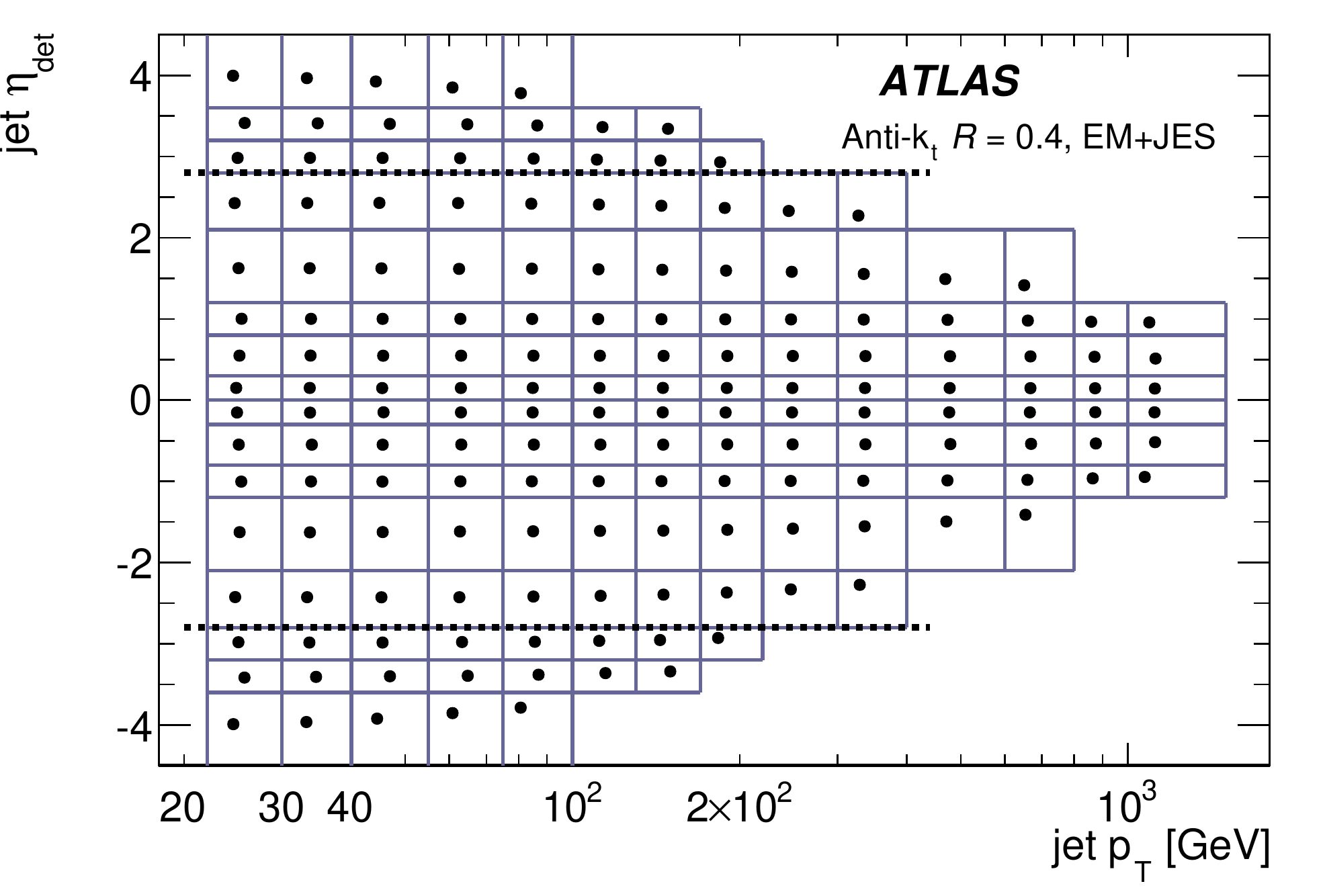}
  \caption{
    Overview of the $(\ptavg,\etaDet)$ bins of the dijet balance measurements for jets reconstructed with distance parameter $R=0.4$ 
calibrated using the \EMJES{} scheme. 
    The solid lines indicate the $( \ptavg,\etaDetProbe)$ bin edges, and the points show the average transverse momentum and 
    pseudorapidity of the probe jet 
    within each bin. The measurements within the $\etaDet$ range spanned by the two thick, dashed lines are used to derive the residual calibration.
    \label{fig:bins} 
  }
\end{figure}

\subsection{Intercalibration using events with dijet topologies}
\subsubsection{Intercalibration using a central reference region}
\label{sec:etaintercalibrationstandardmethod}
The standard approach for \etaic{} with dijet events is to use the central region of the calorimeters as the reference region, as described in Ref.~\cite{cscbook}. The relative calorimeter response of jets in other calorimeter regions is measured by the \pt{} balance between the reference jet (with \ptref) and the probe jet (with \ptprobe), exploiting the fact that these jets are expected to have equal \pt{} due to transverse momentum conservation. The \pt{} balance is expressed in terms of the asymmetry \asym,
\begin{equation}
  \label{eq:asym}
  \asym = \frac{\ptprobe - \ptref}{\ptavg},
\end{equation}
with $\ptavg = (\ptprobe + \ptref)/2$. The reference region is chosen as the central region of the barrel calorimeter, given by  $|\etaDet|<0.8$. 
If both jets fall into the reference region, each jet is used, in turn, to probe the other. As a consequence, the average asymmetry in the reference region will be zero by construction.

The asymmetry is then used to measure an \etaic{} factor $c$ of the probe jet, or its response relative to the reference jet $1/c$, using
the relation
\begin{equation}
  \label{eq:rr}
  \frac{\ptprobe}{\ptref}
  =\frac{2+\asym}{2-\asym}
  =1/c.
\end{equation}

The measurement of $c$ is performed in bins of jet \etaDet{} and \ptavg{}, where \etaDet{} is defined as discussed in  \secRef{sec:jetdirections}. 
Using the standard method outlined above, there is an asymmetry distribution $\asym_{ik}$ for each probe jet \etaDet{} bin~$i$ and each \ptavg{} bin~$k$ 
An overview of the binning is given in \figRef{fig:bins} for jets
with $R=0.4$ calibrated with the \EMJES{} scheme. 
The same bins are used for jets calibrated with the \EMJES{} or \LCWJES{} scheme.
However, the binning is changed for jets with $R=0.6$ to take the different trigger thresholds
into account.
Intercalibration factors are calculated for each bin according to \eqRef{eq:rr}, resulting in
\begin{displaymath}
  \label{eq:c}
  c_{ik}=\frac{2-\langle\asym_{ik}\rangle}{2+\langle\asym_{ik}\rangle}, 
\end{displaymath}
where the $\langle\asym_{ik}\rangle$ is the mean value of the asymmetry distribution in each bin. The uncertainty on $\langle\asym_{ik}\rangle$ is taken to be the RMS/$\sqrt{N}$ of each distribution. For the data, $N$ is the number of events in the bin, while for the \MC{} sample the number of effective events $N_{\mathrm{eff}}$ is used ($N = N_{\mathrm{eff}}$) to incorporate \MC{} event weights $w_{k}$,
\begin{displaymath}
N_{\mathrm{eff}} = \left(\sum{w_k}\right)^2/\sum{w_k^2}.
\end{displaymath}
Here the sums are running over all events of the \MC{} sample. 
The above procedure is referred to as the \emph{\etacref}.

\subsubsection{Intercalibration using the matrix method}
A disadvantage with the \etacref{} outlined a\-bove is that all events are required to have a jet in the central reference region. This results in a significant loss of event statistics, especially in the forward region, where the dijet cross section drops steeply as the rapidity interval between the jets increases. 
In order to use the full statistics, one can extend the \etacref{} by replacing the probe and reference jets by ``left'' and ``right'' jets, defined by $\etaDetLR{left}<\etaDetLR{right}$. Equations \eqref{eq:asym} and \eqref{eq:rr} then become
\begin{eqnarray*}
  \label{eq:frederik}
  \asym = \frac{\ptl - \ptr}{\ptavg}, & \mathrm{and} & 
  \mathcal{R}=
  \frac{\ptl}{\ptr}=
  \frac{c^{\rm right}}{c^{\rm left}}=
  \frac{2+\asym}{2-\asym},
\end{eqnarray*}
where the term $\mathcal{R}$ denotes the ratio of the responses, and $c^{\rm left}$ and $c^{\rm right}$ are the \etaic{} factors
for the left and right jet, respectively. 

This approach yields response ratio ($\mathcal{R}_{ijk}$) distributions with an average value $\langle R_{ijk}\rangle$, evaluated for each $\etaDetLR{left}$ bin~$i$, $\etaDetLR{right}$ bin~$j$, and \ptavg{} bin~$k$. The relative correction factor $c_{ik}$ for a given jet in $\etaDet$ bin $i$, 
with $i = 1 \ldots N$, 
and for a fixed $\ptavg$ bin $k$ is then obtained by a minimisation procedure using a set of $N$ 
equations,
\begin{eqnarray}
  \lefteqn{S(c_{1k},...,c_{Nk}) =}   \label{eq:MM} \\ \nonumber
  & & \sum_{j=1}^{N} \sum_{i=1}^{j-1}
  \left(\frac{1}{\Delta \langle\mathcal{R}_{ijk} \rangle }\left(c_{ik}
    \langle\mathcal{R}_{ijk} \rangle -c_{jk}\right)\right)^2
  + X(c_{1k},...,c_{Nk}) .
\end{eqnarray}
Here $\Delta\langle\mathcal{R}\rangle$ is the statistical uncertainty of $\langle\mathcal{R}\rangle$ and the function $X(c_{ik})$ is used to quadratically suppress deviations from unity of the average corrections,\footnote{This term prevents the minimisation from choosing the trivial solution, which is all $c_{ik} = 0$.}
\begin{displaymath}
  X(c_{1k},...,c_{Nk})=K\left( N^{-1}  \sum_{i=1}^{N} c_{ik} - 1\right)^{2}.
\end{displaymath}
The value of the constant  $K$ does not influence the solution as long as it is sufficiently large, e.g. $K \approx N_{\mathrm{bins}}$, where $N_{\mathrm{bins}}$ is the number of \etaDet{} bins. 
The minimisation according to \eqRef{eq:MM}{} is performed separately for each \pt{} bin~$k$, and the resulting calibration factors $c_{i}$ obtained in each $\etaDet$ bin $i$ are scaled such that the average calibration
factor in the reference region $|\etaDet|<0.8$ equals unity. This method is referred to as the \emph{\etamm}.

\subsection{Event selection for dijet analysis}
\label{sec:evsel}
\label{sec:selection}

\subsubsection{Trigger selection}
Events are retained from the calorimeter trigger stream using a combination of central ($|\etaDet|<3.2$) and forward ($|\etaDet|>3.1$) jet triggers \cite{triggerperformance}. 

The selection is designed such that the trigger efficiency for a specific region of $\ptavg{}$ 
is greater than $99\%$, and approximately flat as a function of the pseudorapidity of the probe jet. 
Due to the different prescales for the central and forward jet triggers, the data collected by each trigger correspond to different integrated luminosities.
To correctly normalise the data, events are assigned weights depending on the luminosity and the trigger 
decisions,
according to the exclusion method described in Ref.~\cite{Lendermann:2009ah}.

\subsubsection{\Ds{} and jet quality selection}
All \ATLAS{} sub-detectors are required to be operational and events are rejected if any data-quality issues are present. 
The leading two jets are required to fulfil the default set of jet quality criteria
(see \secRef{sec:JetSel}). 
A dead calorimeter region was present for a subset of the data. To remove any bias from this region, events are removed if any jets are reconstructed close to this region.

\subsubsection{Dijet topology selection}
\label{sec:dijetselection}
In order to use the momentum balance of dijet events to measure the jet response, it is important that the events
used have a relatively clean $2 \to 2$ topology. If a third jet is produced in the same hard-scatter \pp{} interaction, the balance between the 
leading (in \pt) two jets is affected. To enhance the number of events in the sample that have this $2 \to 2$ topology, 
selection criteria on the azimuthal angle  \deltaphi{\jetnum{1}}{\jetnum{2}}{} between the two leading jets, 
and \pt{} requirements on additional jets 
are applied. Table~\ref{tab:eventSelection} summarises these topology selection criteria.

In addition, all jets used for balancing and topology selection have to originate from the hard-scattering vertex, and not from a vertex reconstructed from a pile-up interaction. 
For this, each jet considered is evaluated with respect to its jet vertex fraction (\JVF), a likelihood measure estimating the vertex contribution to a jet \cite{jespaper2010}. 
To calculate \JVF, reconstructed tracks originating from reconstructed primary vertices $i = 1,\ldots,\Npv$ are matched to jets using an angular matching criterion in $(\eta,\phi)$ space of $\Delta R < 0.4$ with 
respect to the jet axis. 
The track parameters calculated at the distance of closest approach to the selected 
hard-scattering vertex are used for this matching.
For each jet, the scalar sum of the \pt{} of these matched tracks, $\Sigma_i$, is calculated for each vertex $i$ contributing to the jet. 
The \JVF{} variable is then defined as the \pt{} sum for the hard-scattering vertex, $\Sigma_0$, divided by the sum of $\Sigma_i$ over all primary vertices. 
Any jet that has $|\etaDet|<2.5$ and $\JVF>0.6$ is classified as ``vertex confirmed'' since it is likely to 
originate from the hard-scattering vertex.

\begin{table}[!ht]
 \renewcommand{\arraystretch}{\myarraystretch}
  \caption{Summary of the event topology selection criteria applied in this analysis. 
The symbols ``\jetnum{1}" and ``\jetnum{2}" refer to the leading two jets (two highest-\pT{} jets), 
while ``\jetnum{3}" indicates the highest-\pT{} sub-leading (third) jet in the 
event. %
} 
  \begin{center}
    \begin{tabular}{l|l}
      \hline \hline
      Variable & Selection \\
      \hline
      $\deltaphi{\jetnum{1}}{\jetnum{2}}$         & $ > 2.5~{\rm rad}  $  \\
      \ptjetn{3}, $|\etaDet(\jetnum{3})| < 2.5$   & $< \max(0.25\, \ptavg{}, 12 \GeV)$ \\
      \ptjetn{3}, $|\etaDet(\jetnum{3})| > 2.5$   & $< \max(0.20\, \ptavg{}, 10 \GeV)$ \\
      $\JVF(\jetnum{3})$, $|\etaDet(\jetnum{3})|<2.5$  & $> 0.6$ \\

      \hline \hline
    \end{tabular}
  \end{center}
 \label{tab:eventSelection}
\end{table}

This selection differs from that used in previous studies \cite{jespaper2010}
due to the much higher instantaneous luminosities experienced during data taking and the consequentially increasing pile-up.
In the forward region $|\etaDet|>2.5$, no tracking is available, and events containing any additional forward jet with significant $\pt$ are 
removed 
 (see the third criteria in Table~\ref{tab:eventSelection}).
%

\begin{figure*}[htp!]
  \centering
  \mbox{
    \subfloat[$40 \leq \ptavg{}<55$~\GeV]{ \includegraphics[width=.48\textwidth]{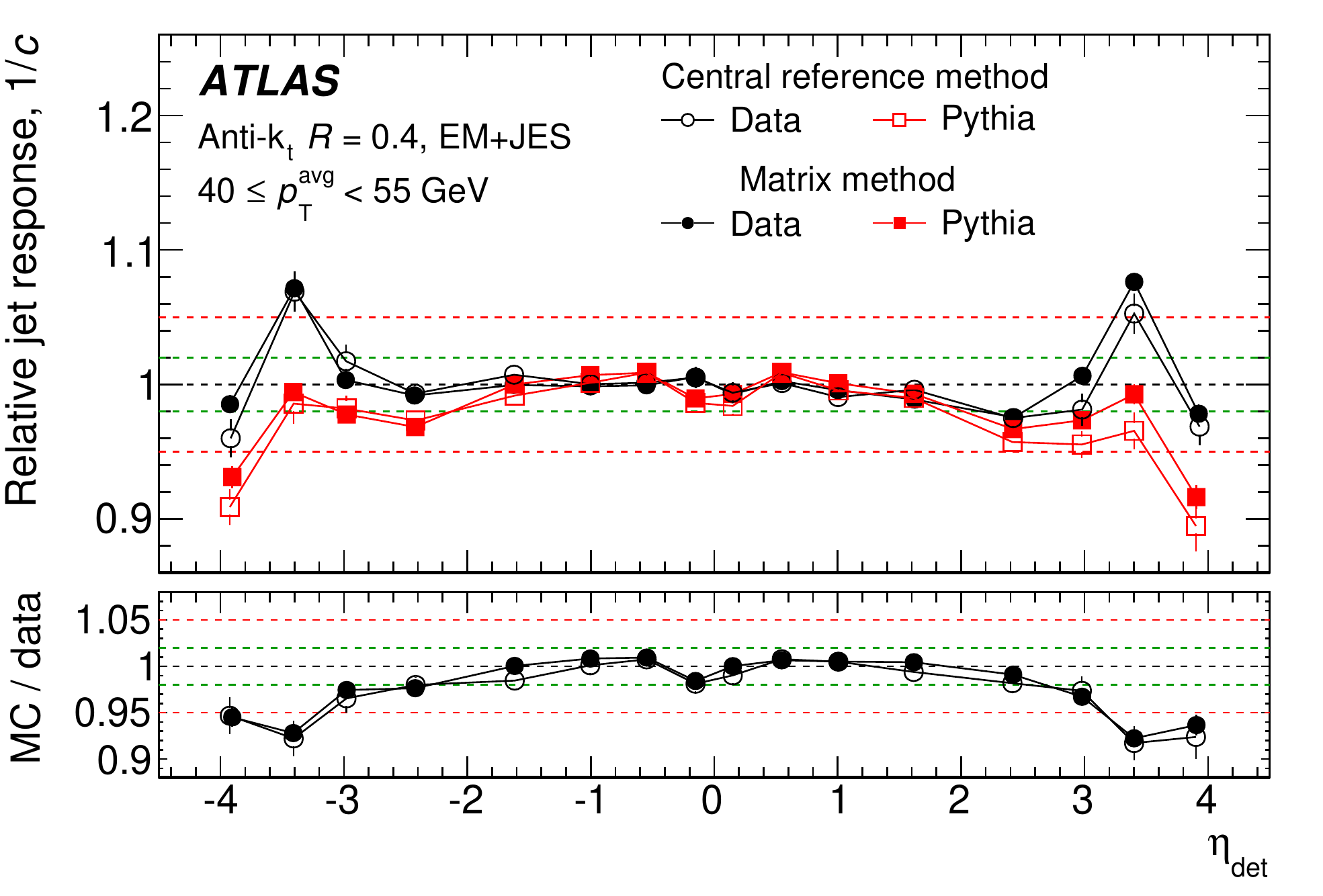}\label{fig:compLowPt}}\quad
    \subfloat[$220 \leq \ptavg{}<300$~\GeV]{ \includegraphics[width=.48\textwidth]{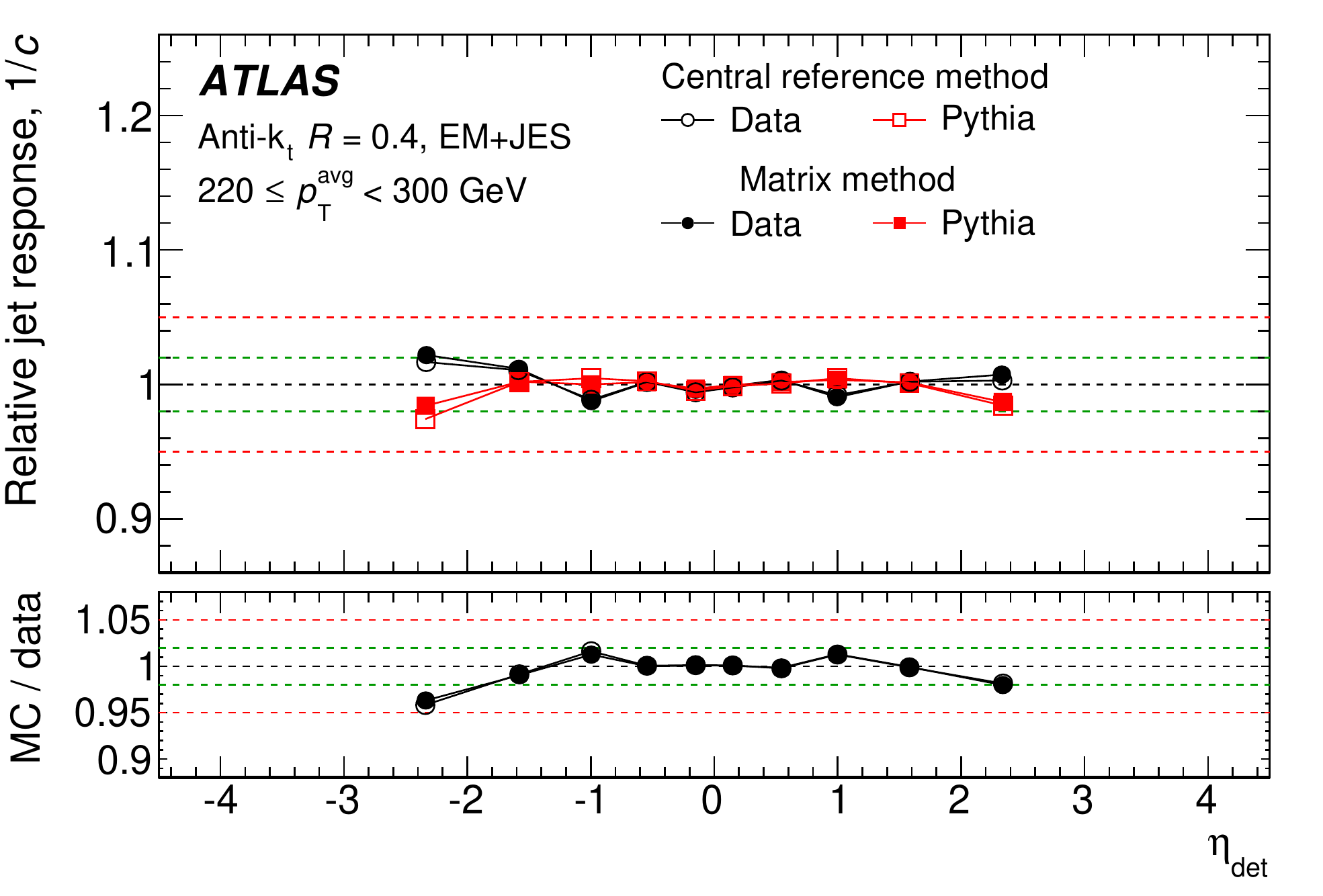}\label{fig:compHighPt}}
  }
  \caption[]{
    Relative jet response ($1/c$) for \antikt{} jets with $R=0.4$ calibrated with the \EMJES{} scheme
    as a function of the probe jet pseudorapidity measured using the 
    matrix and the central reference methods. Results are presented in \subref{fig:compLowPt} for 
    $40 \leq \ptavg<55$~\GeV{} and in \subref{fig:compHighPt} for $220 \leq \ptavg<300$~\GeV{}.
    The lower parts of the figures show the ratios between relative response in data and \MC.  
    \label{fig:comp} 
  }
\end{figure*}

\subsection{Dijet balance results}
\label{sec:res}
\subsubsection{Binning of the balance measurements}
An overview of the (\ptavg,\etaDet) bins used in the analysis is presented in \figRef{fig:bins}.
All events falling in a given \ptavg{} bin are collected using a dedicated central and forward trigger combination. 
The statistics in each \ptavg{} bin are similar, except for the highest and lowest bins which contain fewer events.
The loss of statistical precision of the measurements for the lower \ptavg{} bins is introduced by a larger sensitivity to the
inefficiency of the  pile-up suppression strategy, which rejects relatively more events due to the kinematic overlap of the hard-scatter
jets with jets from pile-up. In addition, the asymmetry distribution broadens due to worsening relative jet \pt{} resolution, leading to larger 
fluctuations in this observable. 

Each \ptavg{} bin is further divided into several \etaDet{} bins. The \etaDet{} binning is motivated by detector geometry and
statistics.

\begin{figure*}
  \centering
    \subfloat[$22<\ptavg{}<30$~\GeV]{ \includegraphics[width=.44\textwidth]{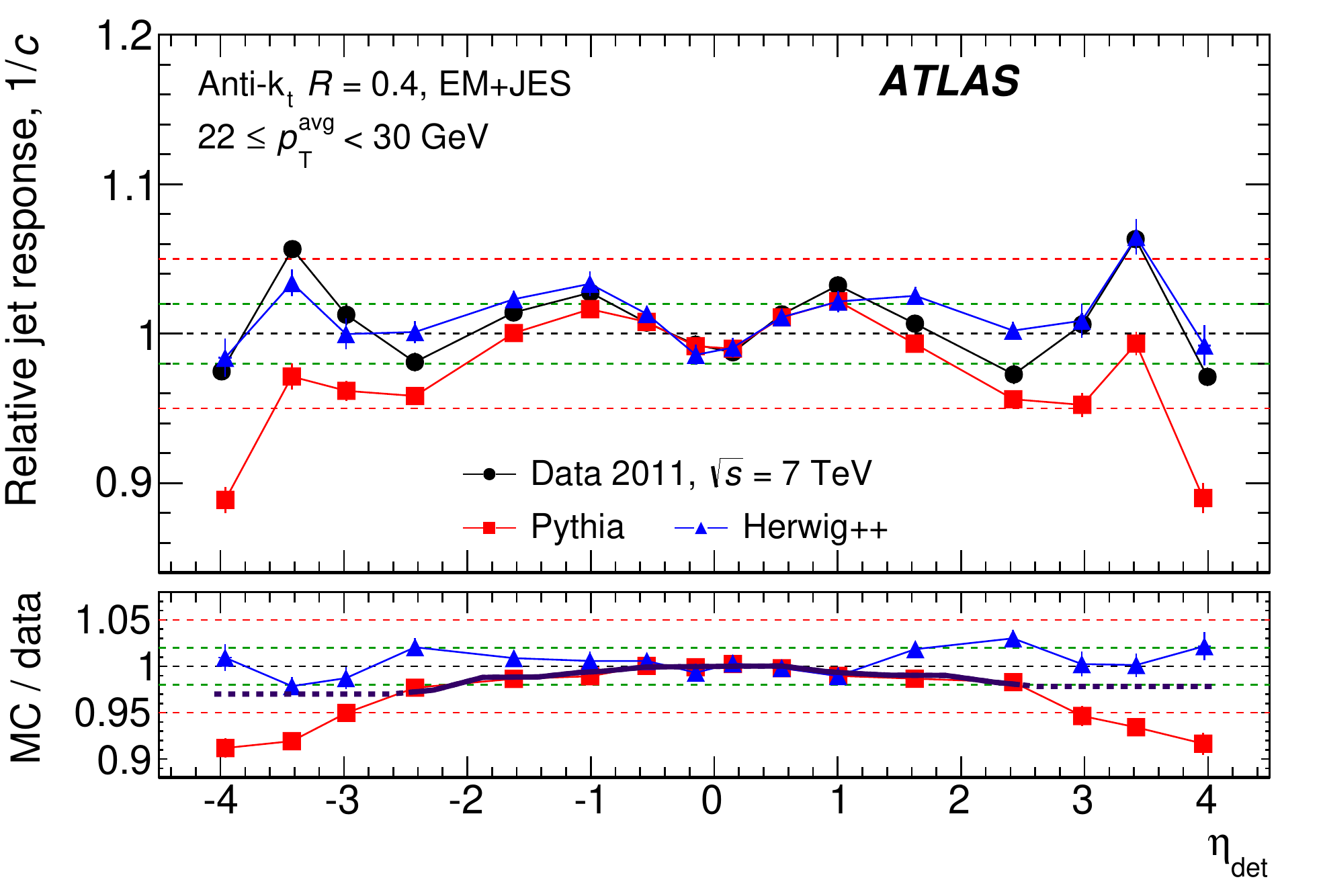}\label{fig:resp_vs_eta_pt0}}
    \subfloat[$55<\ptavg{}<75$~\GeV]{ \includegraphics[width=.44\textwidth]{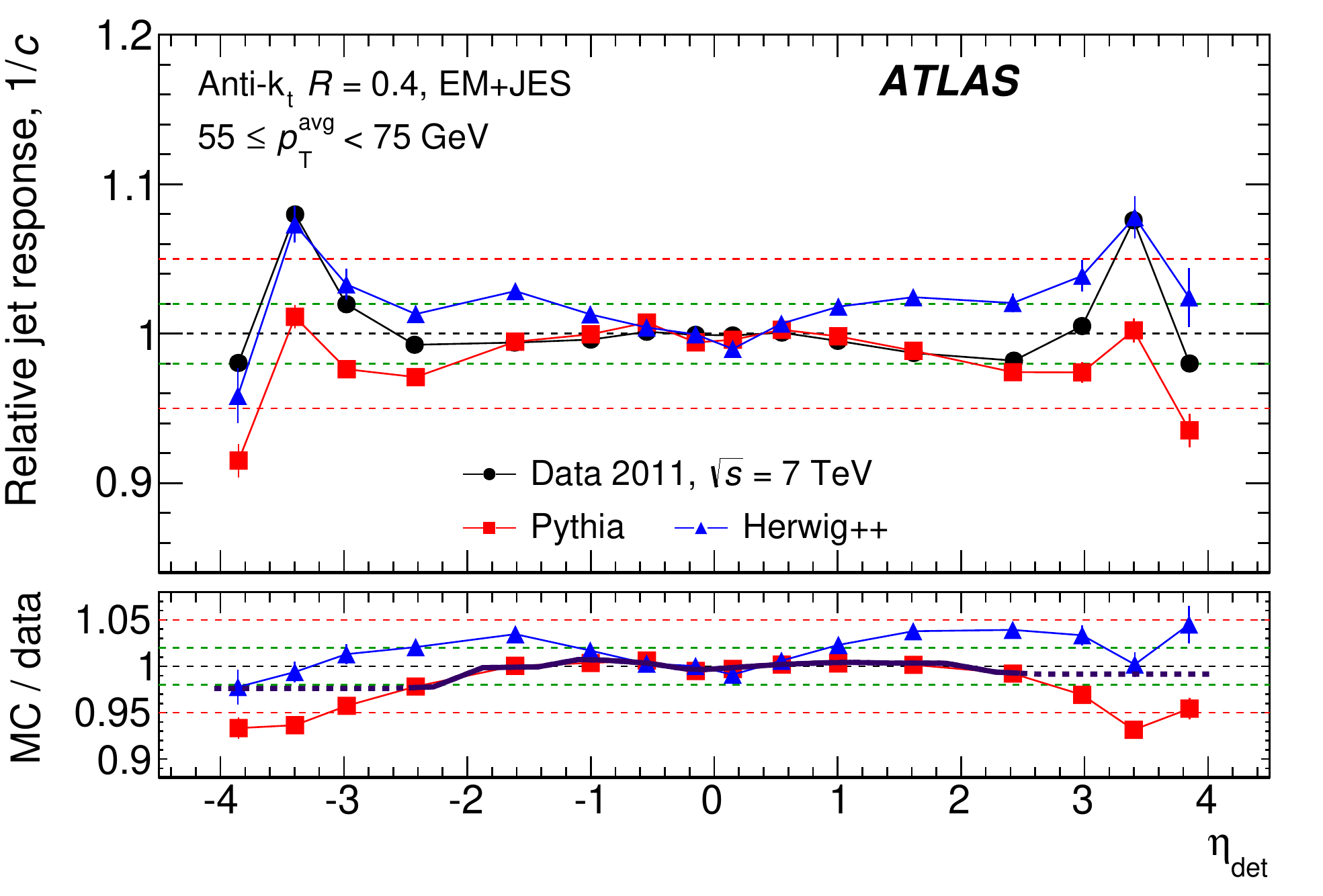}\label{fig:resp_vs_eta_pt1}} \\
    \subfloat[$170<\ptavg{}<220$~\GeV]{ \includegraphics[width=.44\textwidth]{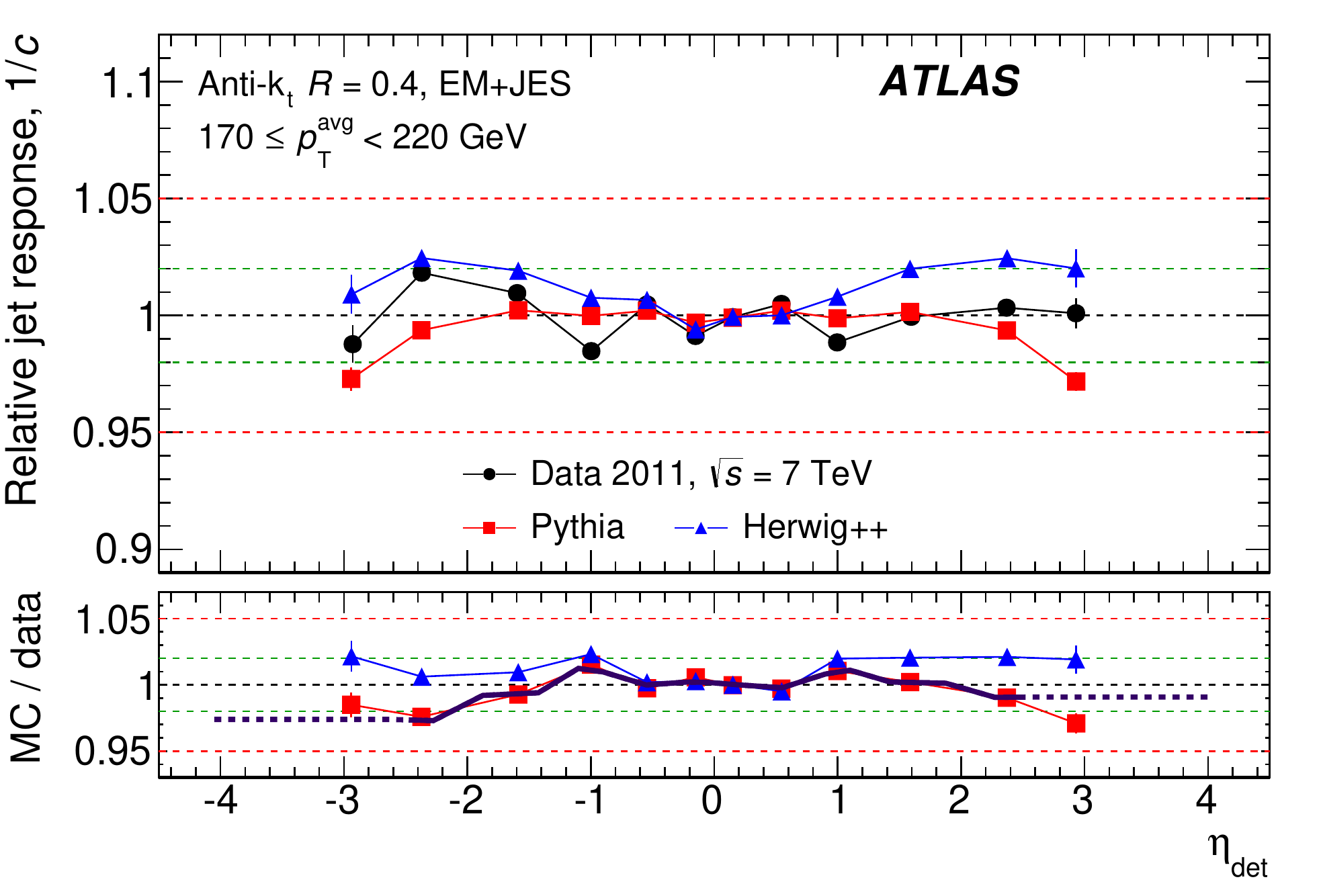}\label{fig:resp_vs_eta_pt2}}
    \subfloat[$600<\ptavg{}<800$~\GeV]{ \includegraphics[width=.44\textwidth]{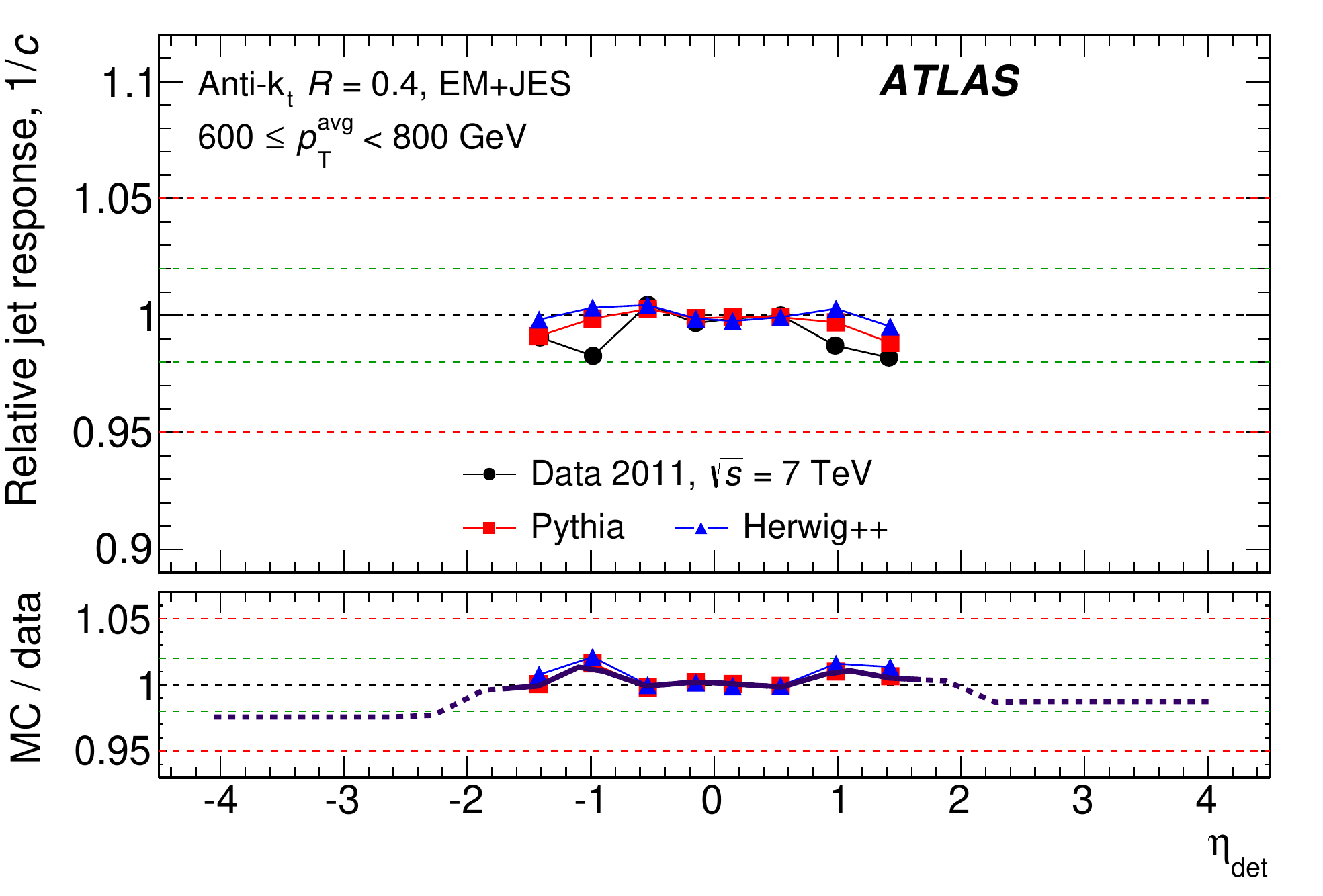}\label{fig:resp_vs_eta_pt3}}
  \caption[]{
    Relative jet response ($1/c$) as a function of the jet 
    pseudorapidity for \antikt{} jets with $R=0.4$ calibrated with the \EMJES{} scheme, separately for
    \subref{fig:resp_vs_eta_pt0} $22 \le \ptavg{}<30$~\GeV, \subref{fig:resp_vs_eta_pt1}  $55 \le \ptavg{}<75$~\GeV, 
    \subref{fig:resp_vs_eta_pt2} $170 \le \ptavg{}<220$~\GeV{} and \subref{fig:resp_vs_eta_pt3} $600 \le \ptavg{}<800$~\GeV.
    The lower parts of the figures show the ratios between the data and \MC{} relative response.  These measurements are performed using the matrix method.
The applied correction is shown as a thick line.
The line is solid over the range where the measurements is used to constrain the calibration, 
and dashed in the range where extrapolation is applied.
    \label{fig:resp_vs_eta}
  }
\end{figure*}

\begin{figure*}
  \centering
    \subfloat[$-1.2\leq\etaDet<-0.8$]{\includegraphics[width=.44\textwidth]{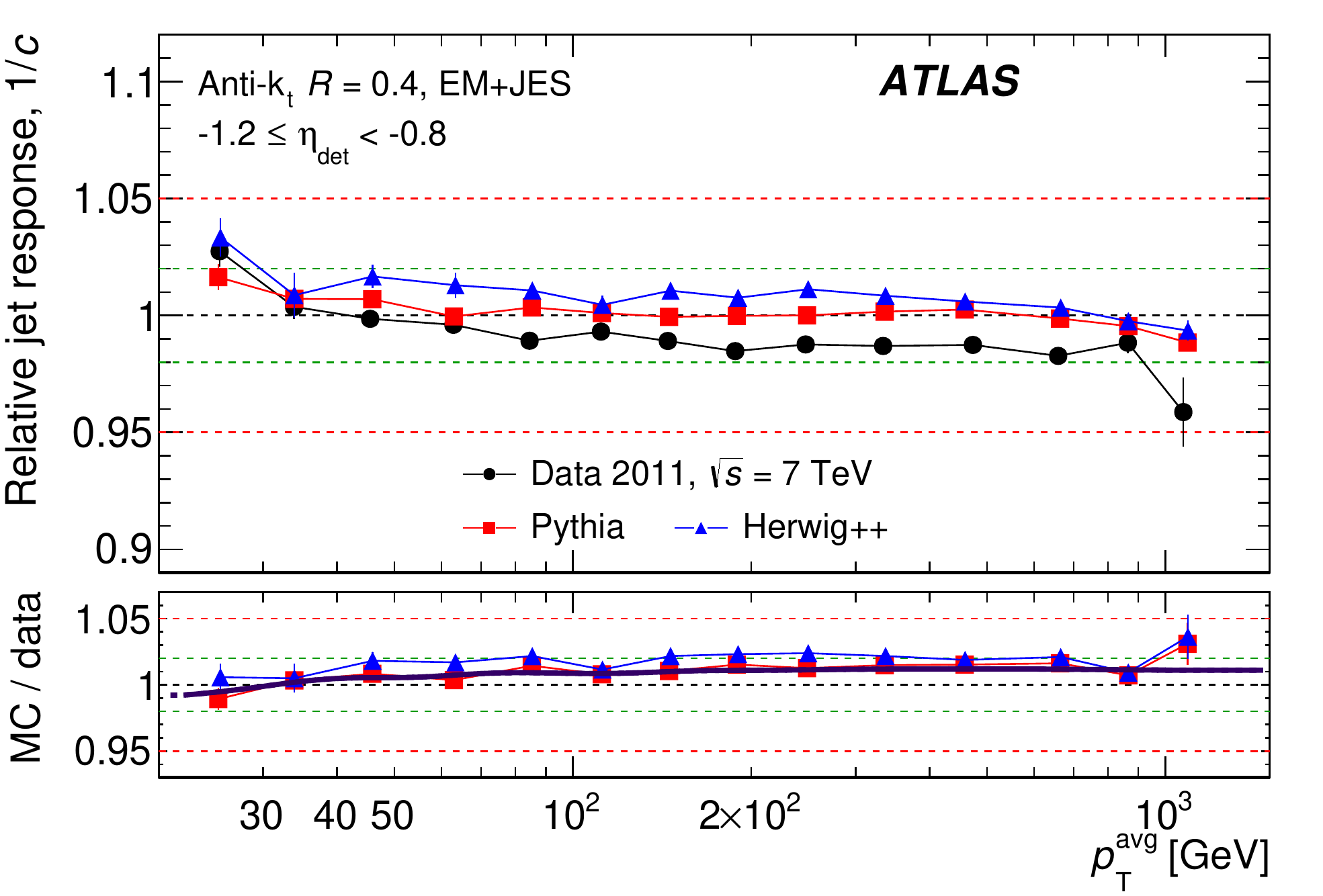}\label{fig:resp_vs_pt_eta0}}
    \subfloat[$2.1\leq\etaDet<2.8$]{ \includegraphics[width=.44\textwidth]{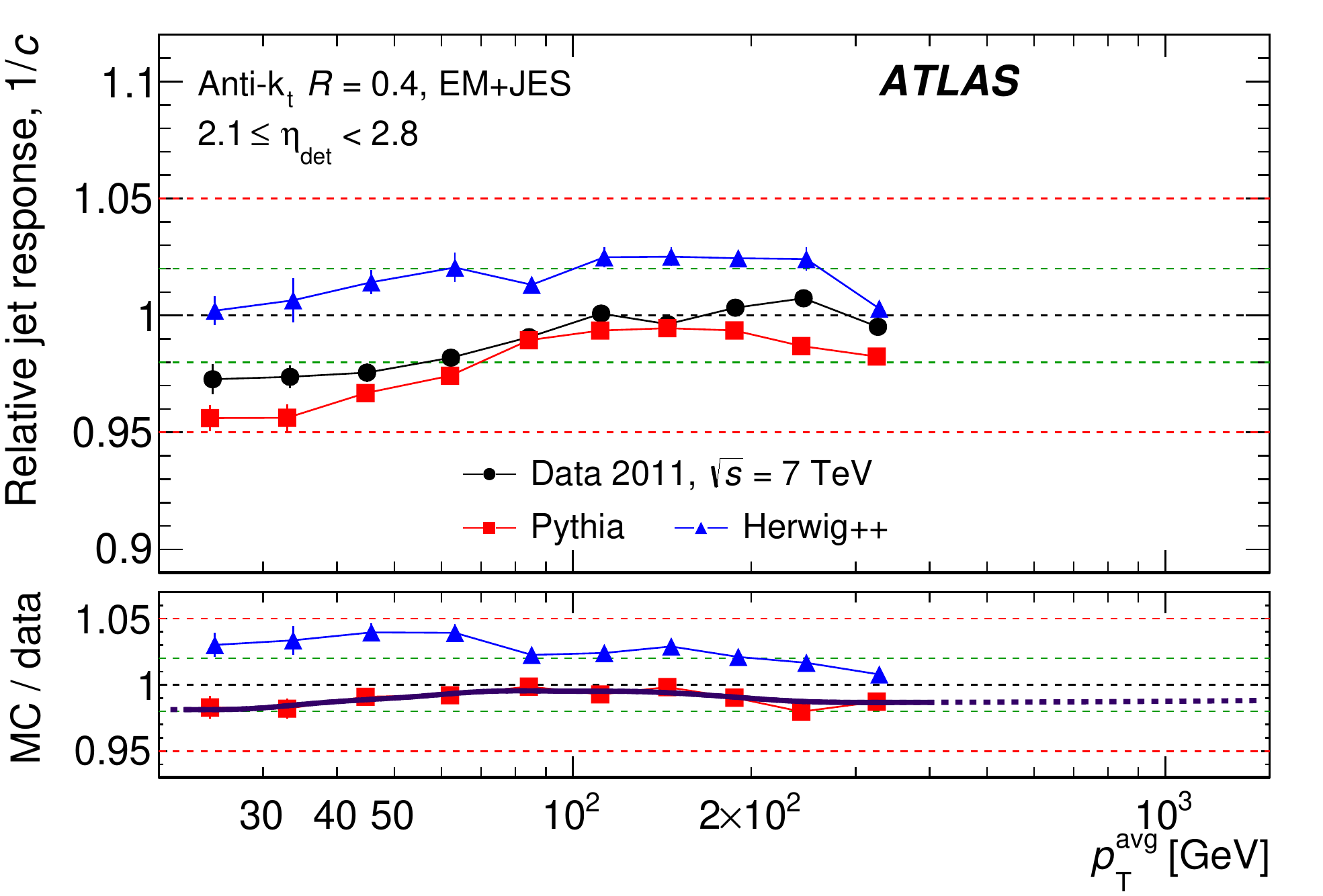}\label{fig:resp_vs_pt_eta1}}
  \caption[]{
    Relative jet response ($1/c$) as a function of the average jet \pt{} of the dijet system 
    for \antikt{} jets with $R=0.4$ calibrated with the \EMJES{} scheme, separately for \subref{fig:resp_vs_pt_eta0}    
    $-1.2\leq\etaDet<-0.8$  and  \subref{fig:resp_vs_pt_eta1} $2.1\leq\etaDet<2.8$. The lower parts of the figures show the ratios between the data and \MC{} relative response. 
The applied correction is shown as a thick line.
    \label{fig:resp_vs_pt}
  }
\end{figure*}

\subsubsection{Comparison of intercalibration methods}
The relative jet response obtained with the matrix
method is compared to the relative jet response obtained using the
central reference method. Figures \ref{fig:comp}\subref{fig:compLowPt}
and  \ref{fig:comp}\subref{fig:compHighPt} show the jet response relative
to central jets ($1/c$) for two \ptavg{} bins,  $40 \leq \ptavg{}<55$~\GeV{} and $220\leq \ptavg{}<300$~\GeV. 
In the most forward region at low \pt{}, the matrix method tends to give a slightly higher relative response compared to the central 
reference me\-thod (see \figRef{fig:comp}\subref{fig:compLowPt}). However, the same relative shift is observed both for data and \MC{} simulations, and
consequently the \datatomc{} ratios are consistent.
The matrix method is therefore used to measure the relative response 
as it has better statistical precision.

\subsubsection{Comparison of data with Monte Carlo simulation}
\label{sec:results}
\FigRef{fig:resp_vs_eta} shows the relative response obtained using the \etamm{} as a function of the jet pseudorapidity for
data and \MC{} simulations. Four different \ptavg{} regions are shown, $22\leq\ptavg<30$~\GeV, $55\leq\ptavg<75$~\GeV, $170\leq\ptavg<220$~\GeV, and $600\leq\ptavg<800$~\GeV. 
\FigRef{fig:resp_vs_pt} shows the relative response as a function of $\ptavg$ for two representative $\etaDet$ bins, namely
$-1.2\leq\etaDet<-0.8$ and  $2.1\leq\etaDet<2.8$.
The general features of the response in data are reasonably well reproduced by the \MC{} simulations. However, as observed in
previous studies \cite{jespaper2010}, the \herwigpp{} \MC{} generator predicts a higher relative response than \pythia{} for jets outside the central
reference region ($|\etaDet|>0.8$). Data tend to fall in-between the two predictions. This discrepancy was investigated and is observed both for truth jets built from stable particles (before any detector modelling), and also jets built from partons (before hadronisation). 
The differences therefore reflect a difference in physics modelling  between the event generators, most likely due to the parton showering.
The \pythia{} predictions are based upon a \ptordered{} parton shower 
whereas the \herwigpp{} predictions are based on an angular-ordered parton
shower.

For $\pt{}>40$~\GeV{} and $|\etaDet|<2$, \pythia{} tends to agree better with data than \herwigpp{} does. In the more forward region, the spread between the \pythia{} and \herwigpp{} response predictions increases and reaches approximately $5\%$ at $|\etaDet|=4$. 
In the more forward region ($|\etaDet|>3$) the relative response prediction of \herwigpp{} generally agrees better with data than \pythia{}.

\subsubsection{Derivation of the residual correction}
\label{sec:etaintercorrection}
The residual calibration is derived from the data/\pythia\ ratio $C_i = c^{\rm data}_i/c^{\textsc{Pythia}}_i$ 
of the measured \etaic{} factors. \pythia{} is used as the reference 
as it is also used to obtain the initial (main) calibration, see \secRef{sec:jetrecocalib}.
The correction is a function of jet $\pt$ and $\etaDet$ ($F_{\rm rel}(\pt,\etaDet)$) and is constructed by combining the $N_{\rm bins}$
measurements of the $(\ptavg,\etaDet)$ bins using a two-dimensional Gaussian kernel, like
\begin{displaymath}
F_{\rm rel}(\pt,\etaDet) =   \dfrac{\sum_{i=1}^{N_{\rm bins}} C_i w_i}{\sum_{i=1}^{N_{\rm bins}} w_i},
\end{displaymath}
with
\begin{eqnarray*}
\lefteqn{w_i =} \\
& & \dfrac{1}{\Delta C_i^2} \times {\rm Gaus}\left(\dfrac{\log \pt - \log \langle\ptprobe\rangle_i}{\sigma_{\log\pt}} \oplus \dfrac{\etaDet - \langle\etaDet\rangle_i}{\sigma_\eta}\right).
\end{eqnarray*}
Here $i$ denotes the index of a $(\ptavg,\etaDet)$-bin, $\Delta C_i$ is the statistical uncertainty of $C_i$,$\langle\ptprobe\rangle_i$ and $\langle\etaDet\rangle_i$ are the average $\pt$ and $\etaDet$ of the probe jets in the bin (see \figRef{fig:bins}). 
The Gaussian function has a central value of zero and a width controlled by $\sigma_{\log\pt}$ and $\sigma_\eta$.

Only the measurements with $|\etaDet|<2.8$ are included in the derivation of the correction function
because of the large discrepancy between the modelled response of the \MC{} simulation samples in the more forward region. 
This \etaDet{} boundary is indicated by a thick, dashed line in \figRef{fig:bins}. 
The residual correction is held fixed for pseudorapidities larger than those of
the most forward measurements included ($|\etaDet|\approx 2.4$). 
All jets with a given $\pt$ and $|\etaDet|>2.4$ will hence receive the same \etaic{} correction.
The kernel-width parameters used\footnote{A width of $\sigma_{\log\pt} = 0.25$ is used for the \pt{} interpolation and $\sigma_\eta = 0.18$ for the \etaDet{} interpolation.} 
are found to capture the shape
of the \datatomc{} ratio, but at the same time provide stability against
statistical fluctuations. This choice introduces a stronger constraint
across \pt{}. 
The resulting residual correction is shown as a thick line in the
lower sections of Figs.~\ref{fig:resp_vs_eta} and~\ref{fig:resp_vs_pt}. 
The line is solid over the range where the measurements is used to constrain the calibration, 
and dashed in the range where extrapolation is applied.


\subsection{Systematic uncertainty}
\label{sec:etaintercaliUncertainty}
The observed difference in the relative response between data and \MC{} simulations could be due to mis-modelling of physics or detector effects used in the simulation. 
Suppression and selection criteria used in the analysis (e.g. topology selection and radiation suppression) can also 
affect the response through their influence on the mean asymmetry. 
The systematic uncertainty is evaluated by considering the following effects:

\begin{enumerate} \itemsep1pt \parskip0pt \parsep0pt
\item Response modelling uncertainty.
\item Additional soft radiation.
\item Response dependence on the $\Delta \phi$ selection between the two leading jets.
\item Uncertainty due to trigger inefficiencies.
\item Influence of pile-up on the relative response.
\item Influence of the jet energy resolution (JER) on the response measurements.
\end{enumerate}

\noindent
All systematic uncertainties are derived as a function of \pt\ and  $|\etaDet|$. 
No statistically significant difference is observed for positive and negative $\etaDet$ for any of the uncertainties.

%

\begin{figure*}
  \centering
    \subfloat[$\pt{} = 35$~\GeV]{ \includegraphics[width=.48\textwidth]{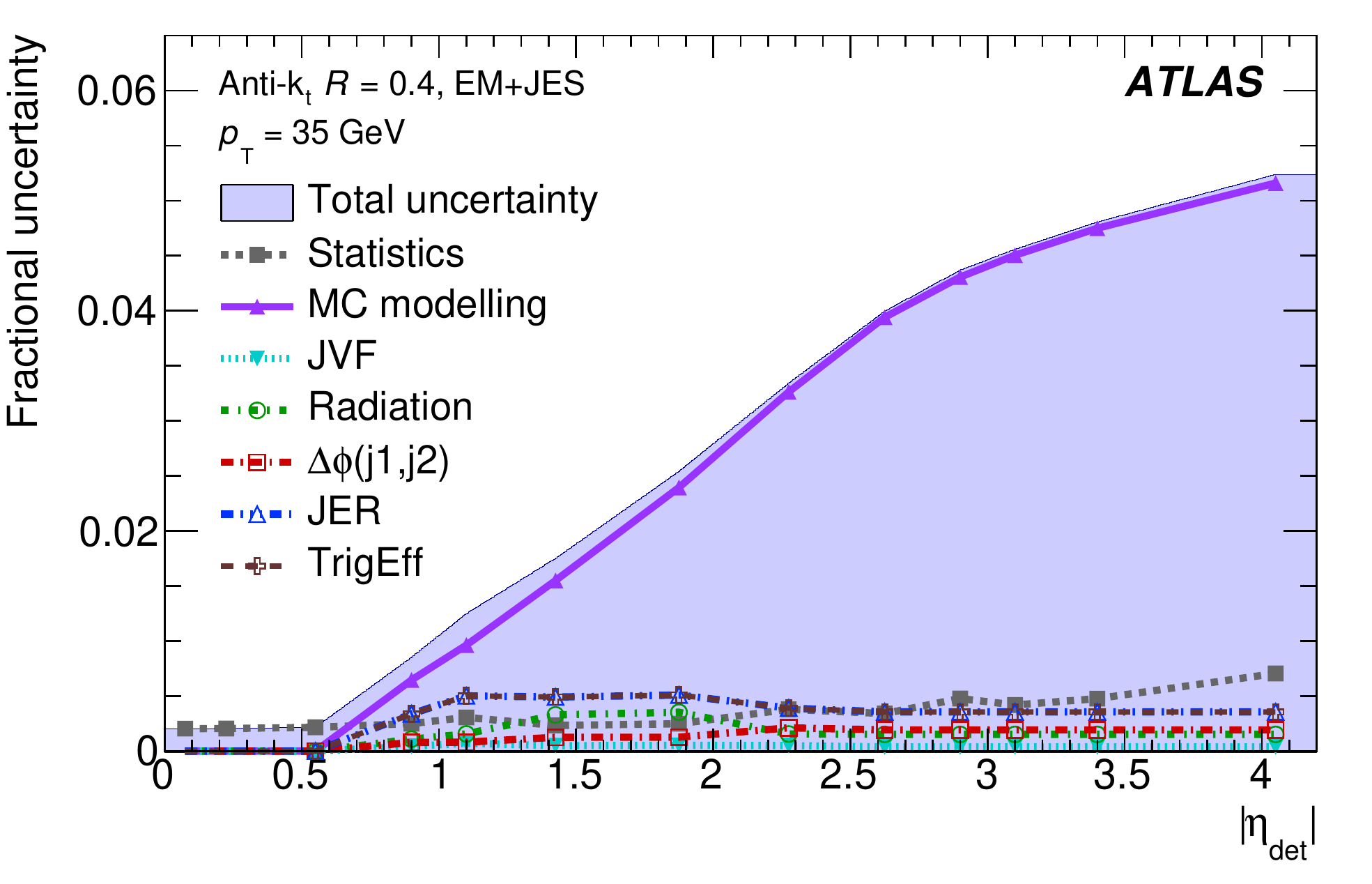}\label{fig:etaintercalibrationuncertainty_low}}
    \subfloat[$\pt = 350$~\GeV]{ \includegraphics[width=.48\textwidth]{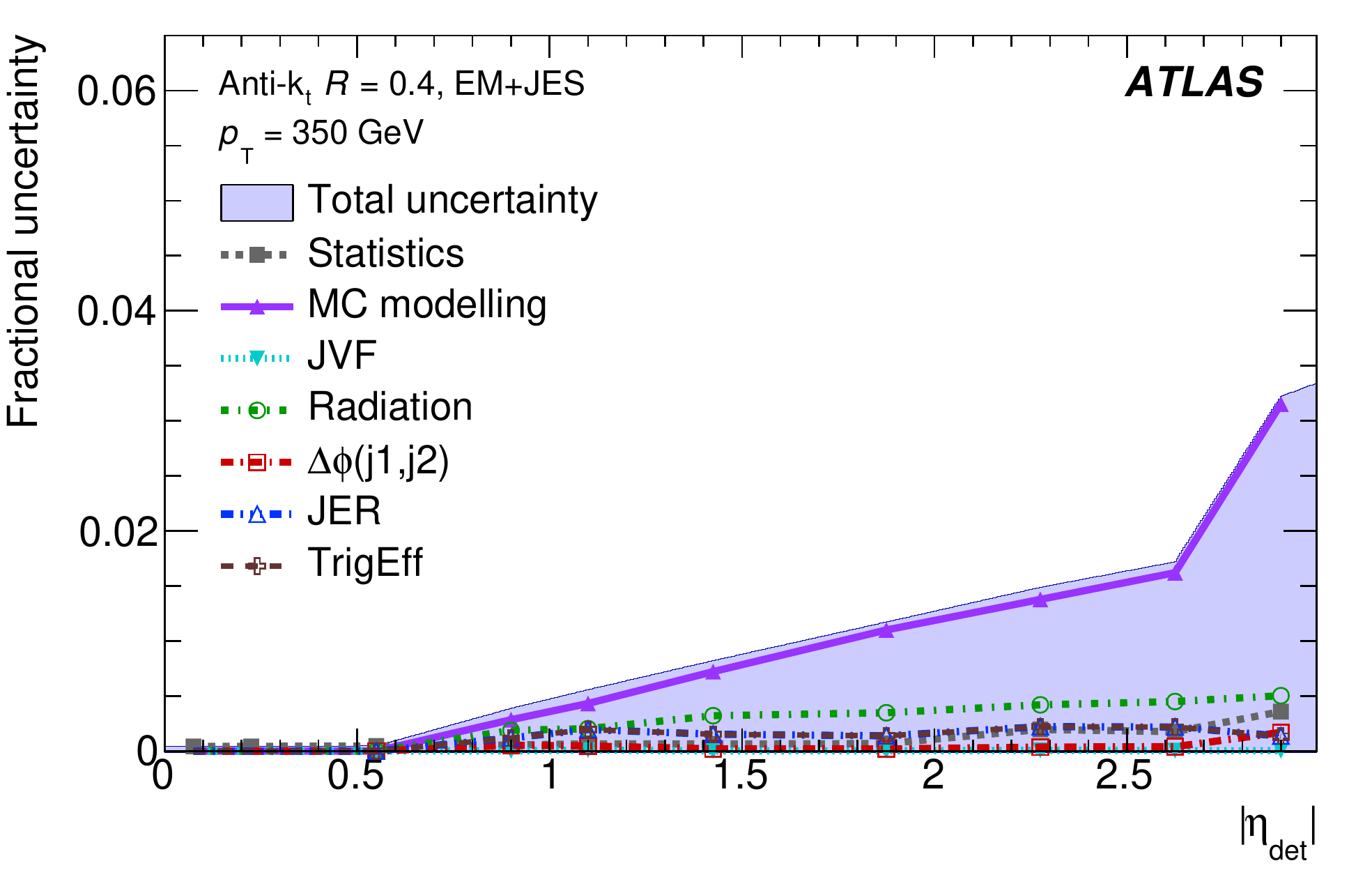}\label{fig:etaintercalibrationuncertainty_high}}
  \caption[]{
    Summary of uncertainties on the intercalibration  as a function of the jet \etaDet{}
    for \antikt{} jets with $R=0.4$ calibrated with the \EMJES{} scheme, separately for  \subref{fig:etaintercalibrationuncertainty_low}
    $\pt{} = 35$~\GeV{} and  \subref{fig:etaintercalibrationuncertainty_high} $\pt = 350$~\GeV. The individual components are added in quadrature 
    to obtain the total uncertainty. 
The \MC{} modelling uncertainty is the dominant component.
    \label{fig:etaintercalibrationuncertainty}
  }
\end{figure*}


\subsubsection{Modelling uncertainty}
\label{sec:modelUncert}
The two generators used for the \MC{} simulation deviate in their
predictions of the response for forward jets as discussed in \secRef{sec:results}.
Since there is no {\it a priori} reason to trust one generator over
the other, the full difference between the two predictions is 
used as the modelling uncertainty. This uncertainty is the
largest component of the intercalibration uncertainty. 
In the reference region ($|\etaDet|<0.8$), no uncertainty is assigned.
For $0.8\leq|\etaDet|<2.4$, where data are corrected to the \pythia{} \MC{} predictions, 
the full difference between \pythia\ and \herwig{} is taken as the
uncertainty. For $|\etaDet|>2.4$, where
the calibration is extrapolated, the uncertainty is taken as the 
difference between the calibrated data and either \pythia{} or \herwig, whichever is larger.

\subsubsection{Sub-leading jet radiation suppression}
Additional radiation from sub-leading jets can affect the dijet
balance. In order to mitigate these effects, selection criteria are
imposed on the \pt{} of any additional jets in an event as discussed
in  \secRef{sec:evsel}. To assess the uncertainties due to the
radiation suppression, the selection criteria are varied 
for both data and \MC{} simulations, and the calibration is re-evaluated. The
uncertainty is taken as the fractional difference between the varied and
nominal calibrations. Each of the three selection criteria are varied
independently. The \JVF{} requirement is changed by $\pm 0.2$ from its 
nominal value ($0.6$) for central jets, and the fractional amount of \pt{}
carried by the third jet relative to \ptavg{} is varied by $\pm 10\%$.
Finally, the minimum \pt{} cutoff is changed by $\pm 2$ $\mathrm{GeV}$.

\subsubsection{\deltaphi{\jetnum{1}}{\jetnum{2}}{} event selection}
The event topology selection requires that the two leading jets have a $\Delta \phi$ separation greater than 2.5~rad. In order to assess the influence of this selection on the \pt{} balance, the residual calibration is re-derived twice after shifting the selection criterion by $\pm 0.4$ rad ($\deltaphi{\jetnum{1}}{\jetnum{2}} < (2.5 \pm 0.4)$ rad), separately in either direction. The largest difference between the shifted and nominal calibrations is taken as the uncertainty.

\subsubsection{Trigger efficiencies}
Trigger biases can be introduced if the trigger selection, which is applied only to data, is not fully efficient.
To assess the uncertainty associated with the small inefficiency in the trigger, the measured efficiencies are applied to the \MC{} samples. 
The effect on the \MC{} response is found to be negligible in comparison to the other sources,
even when exaggerating the effect by shifting the measured efficiency curves to reach the plateau $10\%$ earlier in \pt{}. 
This uncertainty is hence neglected.

\subsubsection{Impact of pile-up interactions}
The influence of pile-up on the relative response is evaluated. To assess the magnitude of the effect, the differences between low and high pile-up subsets are investigated.
Two different selections are used, high and low $\axing$ subsets ($\axing < 7$ and $\axing \geq 7$), and high and low \Npv{} subsets ($\Npv<5$ and $ \Npv \geq 5$). 
The discrepancies observed are well within the systematic uncertainty for the pile-up correction itself (see  
\secRef{sec:pileupsystematics}). 
Therefore, no further contribution from pile-up is included in the evaluation of the full systematic uncertainty of the \etaic.

\subsubsection{Jet resolution uncertainty}
\label{sec:etaic_jer_uncertainty}
The jet energy resolution (\JER)~\cite{jerpaper2010} in the \MC{} simulation is comparable to the resolution observed in data. To assess the impact of the \JER{} on the \pt{} balance, a smearing factor is applied as a scale factor to the \MC{} jets, which results in an increased jet resolution consistent with the \JER{} measured in data plus its error. It is randomly sampled from a Gaussian with width
\begin{equation}
\sigma = \sqrt{(\sigma_{\mathrm{data}}+\Delta\sigma_{\mathrm{data}})^{2} - \sigma_{\mathrm{data}}^{2}} ,
\label{eq:jer_smearing}
\end{equation}
where $\sigma_{\mathrm{data}}$ is the measured jet resolution in data and $\Delta\sigma_{\mathrm{data}}$ is the corresponding uncertainty. 
The difference between the nominal and smeared \MC{} results is taken as the \JER{} systematic uncertainty.

\subsection[Summary of the \etaic{} and its uncertainties]{Summary of the \ETAIC{} and its uncertainties}
\label{sec:etaintercalibrationsummary}
The pseudorapidity dependence of the jet response is analysed using dijet pseudorapidity \etaic{} techniques.
A re\-sid\-u\-al $\pt$ and $\etaDet$ dependent response correction is derived with a \etamm{} for jets 
with $|\etaDet|<2.4$. The correction is applied to data to correct for effects not captured 
by the default \MC-derived calibration. 
The correction to the jet response 
is measured to be approximately $+1\%$ at $|\etaDet| = 1.0$ and falling to $-3\%$ 
and to $-1\%$ for $|\etaDet|=2.4$ and beyond.
The total systematic uncertainty is obtained as the quadratic sum of the various components mentioned. 
Figure~\ref{fig:etaintercalibrationuncertainty} presents a summary of the uncertainties as a function 
of $\etaDet$ for two representative values of jet transverse momentum, 
namely $\pt = 35$~\GeV{} and $\pt = 350$~\GeV{}. 
The uncertainty is parameterised in the same way as the correction as described in 
\secRef{sec:etaintercorrection}.
There is no strong variation of the uncertainties as a function of jet \pt.
For a $\pt = 25$~\GeV{} jet, the uncertainty is about $1\%$ at $|\etaDet|=1.0$, $3\%$ at $|\etaDet|=2.0$ and about $5\%$ for $|\etaDet|>3.0$.
The uncertainty is below $1\%$ for $\pt = 500$~\GeV{} jets with $|\etaDet|<2$.

\section[Jet energy calibration using \Zjet{} events]{Jet energy calibration using \ZJET{} events}
\label{sec:ZjetInSitu}
This section presents results based on events where a \Zboson{} boson 
decaying to an \ee{} pair is produced together with a jet, which balance each other in the transverse plane. 
The \pt{} balance is compared in data and in \MC{} simulations, and a study of systematic uncertainties 
on the \datatomc{} ratio is carried out. 
The results from a similar study with \gammajet{} events are discussed in \secRef{sec:gammajetInSitu}. 

The advantage of \Zjet{} events is the possibility of probing low-\pt{} jets, 
which are difficult to reach with \gammajet{} events due to trigger thresholds and 
background contamination in that region. On the other hand, \gammajet{} events 
benefit from larger statistics for \pt{} above $150 \GeV$. 
In the  \Zjet{} and \gammajet{} analyses, jets with 
a pseudorapidity $|\etaDet|<1.2$ are probed.

%
\begin{figure}[!ht]
\begin{center}
\begin{tabular}{lr}
\includegraphics[width=7.9cm]{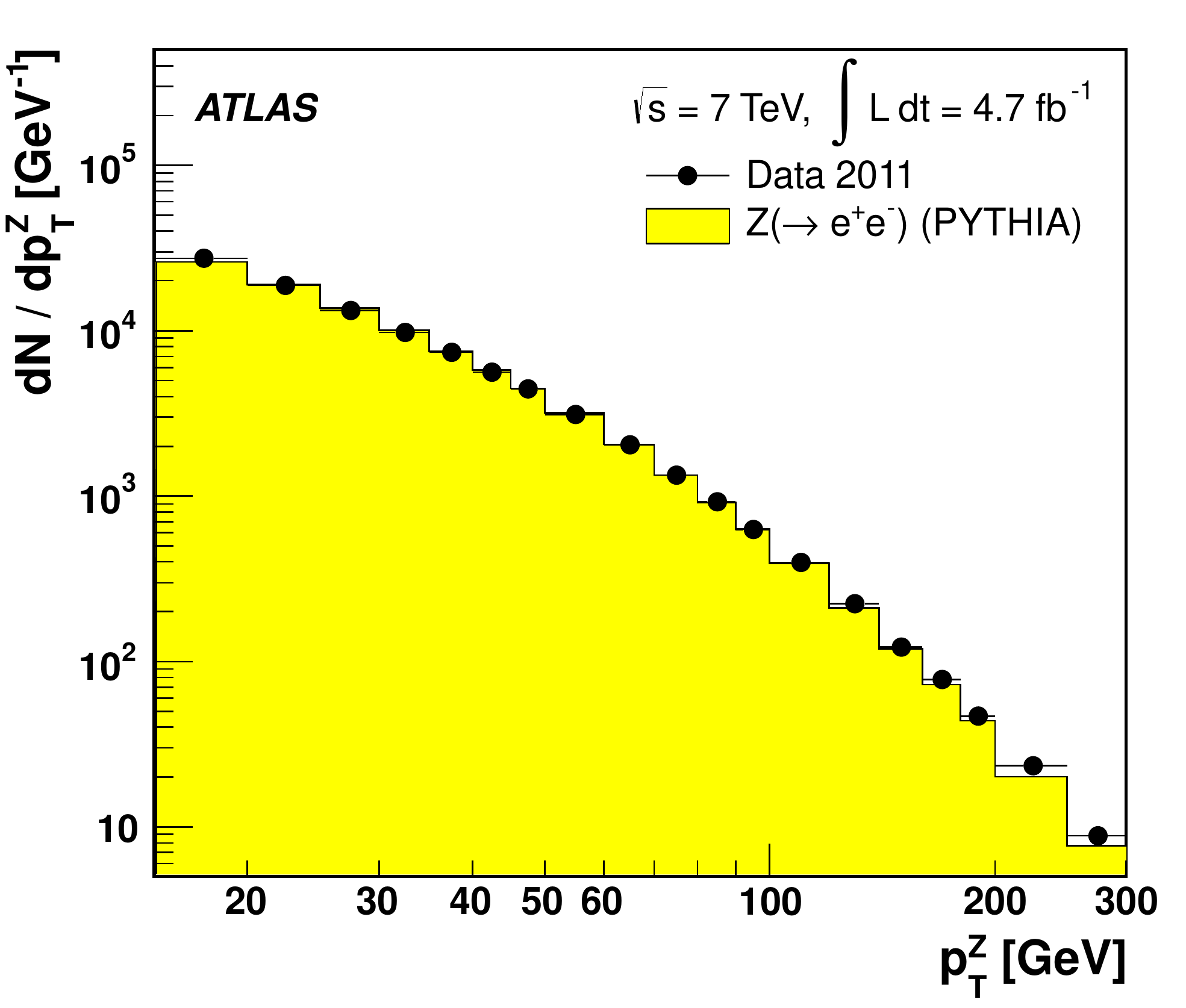}
\end{tabular}
\end{center}
\caption{The \Zboson{} boson \pt{} distribution in selected \Zboson{} events. Data and prediction from the \Zee{} \pythia{} simulation, normalised to the observed number of events, are compared. }
\label{fig:zDistrib}
\end{figure}

\subsection[Description of the \pt{} balance method]{Description of the \PTBF{} balance method}
\label{sec:ZjetMethod}
In events where one \Zboson{} boson and only one jet are produced, 
the jet recoils against the \Zboson{} boson, ensuring approximate momentum balance between 
them in the transverse plane. The direct \pt{} balance technique exploits this relationship 
in order to improve the jet energy calibration.

If the \Zboson{} boson decays into electrons, its four-momentum is reconstructed using the electrons,
which are accurately measured in the electromagnetic calorimeter and the inner 
detector~\cite{Atlaselectronpaper}. Ideally, if the jet includes all the particles that 
recoil against the \Zboson{} boson, and if the electron energies are perfectly measured, 
the response of the jet in the calorimeters can be determined by using $\pt^{\Zboson}$ as the 
reference truth-jet \pt{}. However, this measurement is affected by the following: 
\begin{enumerate} 
  \item Uncertainty on the electron energy measurements.
  \item Particles contributing to the \pt{} balance
        that are not included in the jet cone (out-of-cone radiation). 
  \item Additional parton radiation contributing to the recoil against the \Zboson{} boson.
  \item Contribution from the underlying event.
  \item In-time and out-of-time pile-up. 
\end{enumerate}

Therefore, the direct \pt{} balance between a \Zboson{} boson and a jet ($\pt^{\text{jet}}/\pt^{\text{ref}}$) 
is not used to estimate the jet response, but only to assess how well the \MC{} simulation 
can reproduce the data. 

To at least partly reduce the effect of additional parton radiation perpendicular to the jet axis in the transverse plane, a reference $\pt^{\text{ref}}=\pt^{\Zboson}\times|\cos(\deltaphi{\mathrm{jet}}{\Zboson})|$ is constructed from the azimuthal angle \deltaphi{\mathrm{jet}}{\Zboson} between the \Zboson{} boson and the jet, and the \Zboson{} boson transverse momentum $\pt^{\Zboson}$. 
 
The jet 
calibration in the data is then adjusted using the \datatomc{} comparison of the $\pt^{\text{jet}}/\pt^{\text{ref}}$ ratio for 
the two jet calibration schemes \EMJES{} and \LCWJES{} described in \secRef{sec:jetrecocalib}. The effects altering
this ratio are evaluated by changing kinematic and topological selections and \MC{} event generators and other modelling parameters. In particular the extrapolation of the $\Delta\phi(\text{jet},\Zboson)$ dependence of $\pt^{\text{jet}}/\pt^{\text{ref}}$ to the least radiation-biased regime ($\Delta\phi(\text{jet},\Zboson) = \pi$) is sensitive to the \MC-mo\-del\-ling quality and is investigated with \datatomc{} comparisons. 
%
%

\begin{figure*}[!ht]
\begin{center}
\begin{tabular}{lr}
  \subfloat[Balance distribution and fit]{\label{fig:balanceDistribution}
\includegraphics[width=0.45\textwidth]{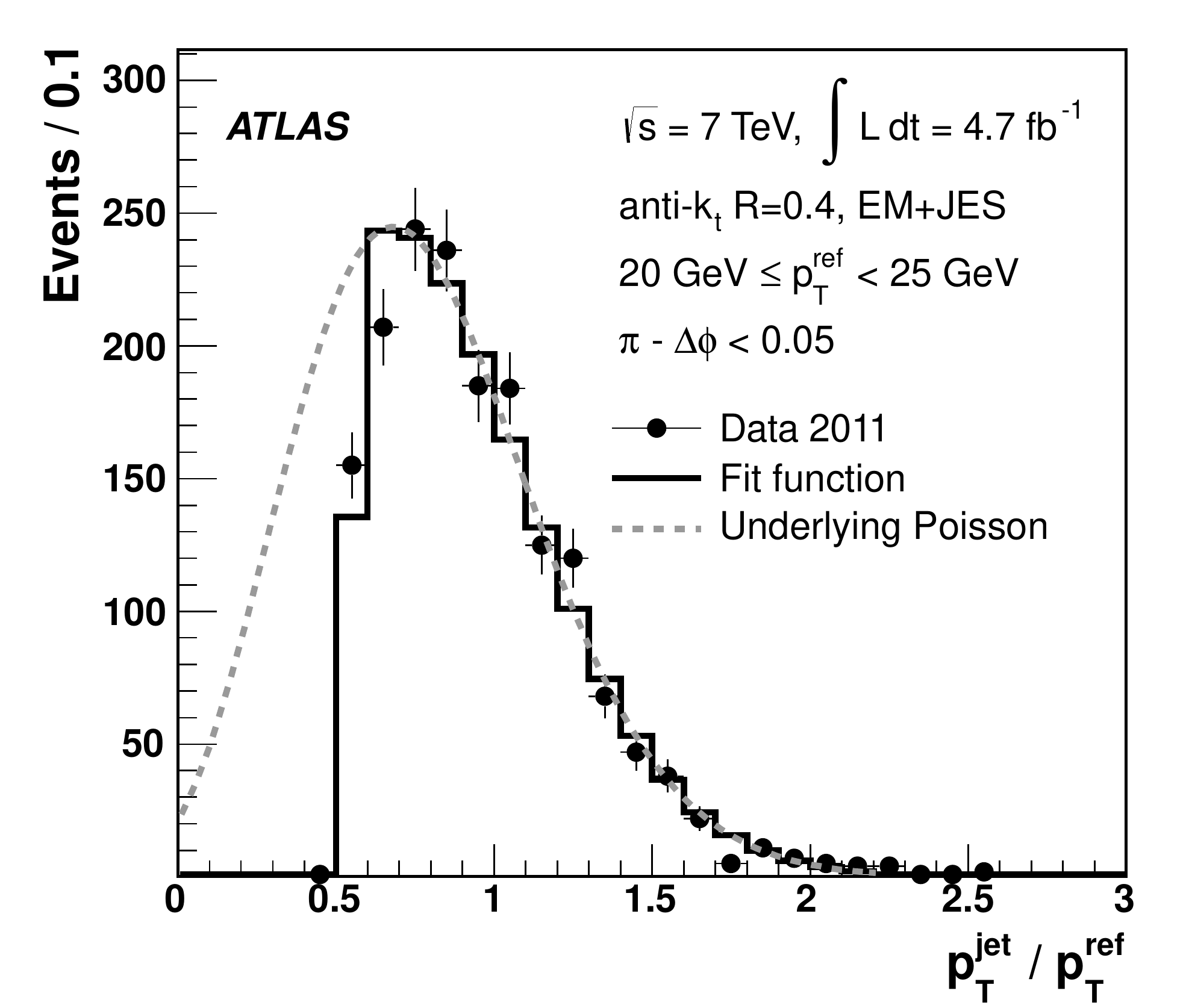}}
  \subfloat[Balance width as a function of $\pt^{\text{ref}}$]{\label{fig:resolution}
\includegraphics[width=0.45\textwidth]{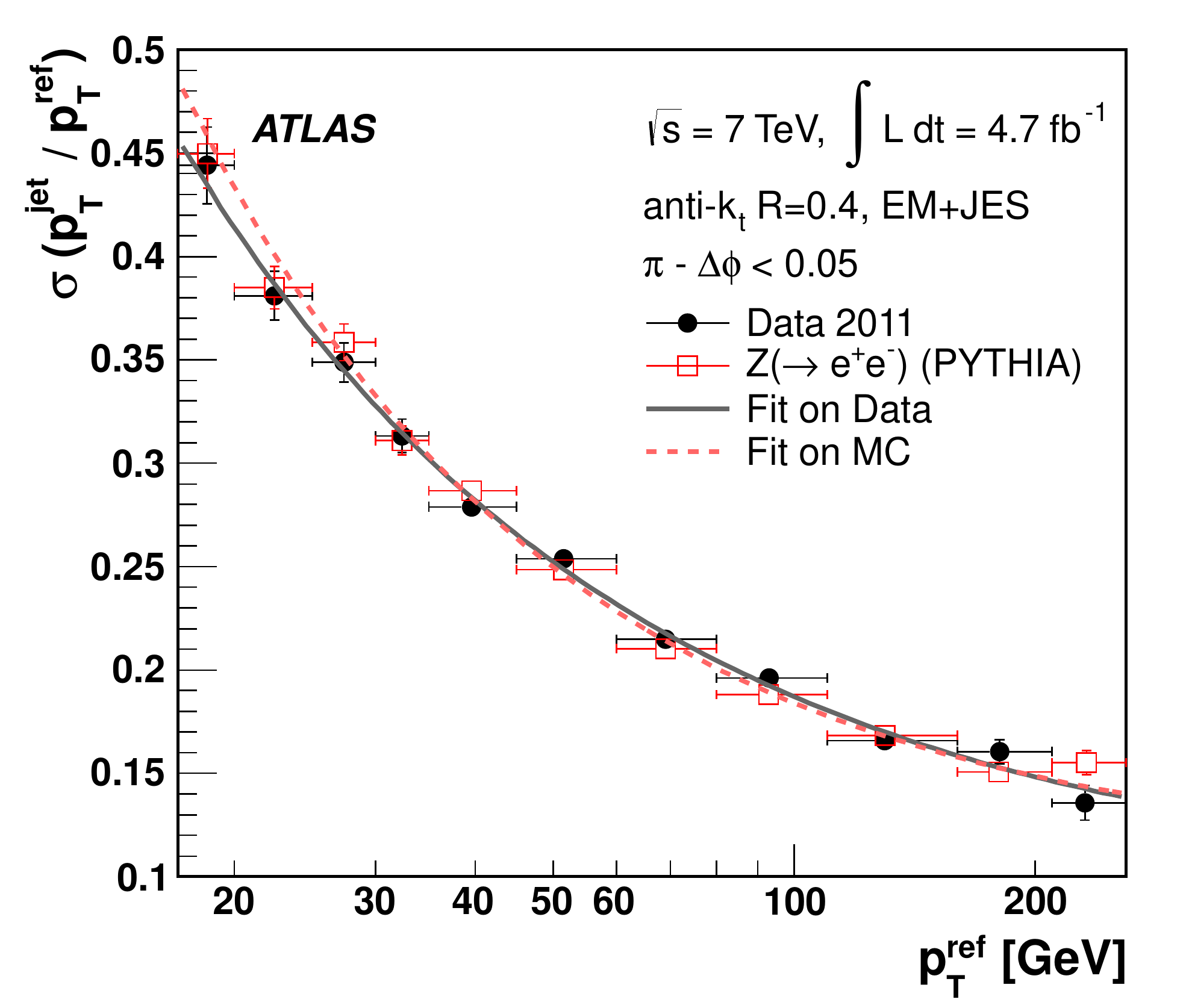}}
\end{tabular}
\end{center}
\caption[]{The $\pt^{\text{jet}}/\pt^{\text{ref}}$ distribution in the data for 
$20 \leq \pt^{\text{ref}} < 25\GeV$ and $\pi-\deltaphi{\mathrm{jet}}{\Zboson}<0.05$ is shown in \subref{fig:balanceDistribution}. 
The black solid histogram represents the fit function, a Poisson distribution extended to 
non-integer values, multiplied by a turn-on curve. The value used in each bin is the mean 
value of that function in the bin. The gray dashed line shows the underlying Poisson distribution, 
from which the mean is taken as the measurement of the \pt{} balance. 
In \subref{fig:resolution} the measured widths of the $\pt^{\text{jet}}/\pt^{\text{ref}}$ distributions as a 
function of $\pt^{\text{ref}}$ is shown for data and \MC{} simulations, for events
 with $\pi-\deltaphi{\mathrm{jet}}{\Zboson}<0.05$. The fitted functional form defined 
by \eqRef{eq:resolution} is superimposed. In both figures, \antikt{} jets with 
distance parameter $R=0.4$ calibrated with the \EMJES{} scheme are used. 
Only statistical uncertainties are shown.}
\label{fig:balanceEstimation}
\end{figure*}

\begin{figure}[!ht]
\begin{center}
\begin{tabular}{lr}
  \includegraphics[width=0.45\textwidth]{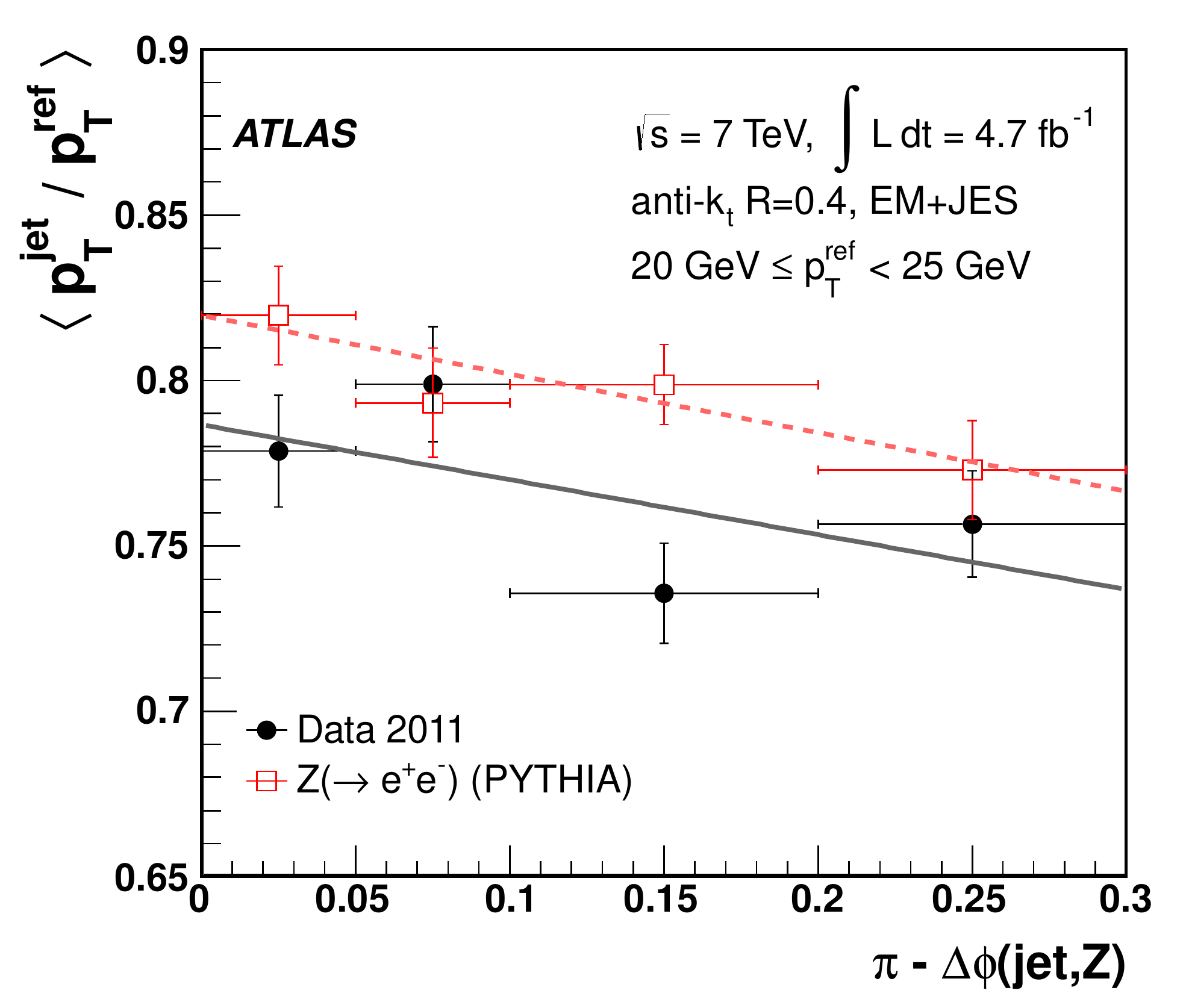}
\end{tabular}
\end{center}
\caption{Mean $\ptjet/\pt^{\text{ref}}$ balance vs \deltaphi{\mathrm{jet}}{\Zboson}{} 
for $20 \leq \pt^{\text{ref}} < 25\GeV$ in the data and in the simulation. A linear 
function used to extrapolate the balance to $\Delta\phi = \pi$ is superimposed. \Antikt{} jets 
with distance parameter $R=0.4$ calibrated with the \EMJES{} scheme are used. 
Only statistical uncertainties are shown.}
\label{fig:extrapolation}
\end{figure}

\subsection[Selection of \Zjet{} events]{Selection of \ZJET{} events}
\label{sec:jetSelection}
Events are selected online using a trigger logic that requires the presence of at least one 
well-identified electron with transverse energy ($E^e_{\text{T}}$) above $20\GeV$ (or $22\GeV$, 
depending on the data-taking period) or two well-identified electrons with $E^e_{\text{T}} > 12\GeV$, 
in the region $|\eta|<2.5$~\cite{egammaTrigger2}.
Events are also required to have a primary hard-scattering vertex, as defined in \secRef{sec:trackjets}, with at least three tracks 
associated to it. 
This renders the contribution from fake vertices due to beam backgrounds negligible. 

Details of electron reconstruction and identification can be found in Ref.~\cite{Atlaselectronpaper}.
Three levels of electron identification quality are defined, based on different requirements on 
shower shapes, track quality, and track--cluster matching. 
The intermediate one (``medium'') is used in this analysis. 

Events are required to contain exactly two such electron candidates with $E^e_{\text{T}}>20\GeV$ 
and pseudorapidity in the range $|\eta^e|<2.47$, where the transition region between calorimeter 
sections $1.37<|\eta^e|<1.52$ is excluded, as well as small regions where an accurate energy 
measurement is not possible due to temporary hardware failures. 
If these electrons have op\-po\-si\-te-sign charge, and yield a combined invariant mass 
in the range $66 < M_{\ee} < 116\GeV$, the event is kept and the four-momentum of the \Zboson{} boson 
candidate is reconstructed from the four-momenta of the two electrons. 
The transverse momentum distribution of these \Zboson{} boson candidates 
is shown in \figRef{fig:zDistrib}.

All jets within the full calorimeter acceptance and with a \JES-corrected transverse 
momentum $\pt^{\text{jet}}>12\GeV$ are considered. 
For each jet the \JVF{} (see \secRef{sec:dijetselection}) is used to estimate the degree of pile-up contamination of a jet based 
on the vertex information.
The highest-\pt{} (leading) jet must pass the quality criteria described in \secRef{sec:JetSel}, 
have a $\JVF>0.5$, and be in the fiducial region $|\etajet|<1.2$. 

Furthermore, the leading jet is required to be isolated from the two electrons 
stemming from the \Zboson{} boson. The distance \DeltaR{} between the jet and each of the two electrons in \etaphispace, measured according to \eqRef{eq:deltaRdet}{} in \secRef{sec:jetdirections}, is required to be $\DeltaR > 0.35\,(0.5)$ 
for \antikt{} jets with $R=0.4\,(0.6)$.

The presence of additional high-\pt{} parton radiation altering the balance between 
the \Zboson{} boson and the leading jet is suppressed by requiring that the next-highest-\pt{} (sub-leading) 
jet has a calibrated \pt{} less than $20\%$ of the \pt{} of the \Zboson{} boson, 
with a minimal \pt{} of $12\GeV$. For sub-leading jets within the tracking acceptance, 
this cut is only applied if the jet has a $\JVF>0.75$. 
A summary of the event selection is presented in Table~\ref{tab:ZjeteventSelection}.

\begin{table}[!ht]
  \caption{Summary of the event selection criteria applied in the \Zjet{} analysis.}
  \renewcommand{\arraystretch}{1.25}
  \begin{center}
   \begin{tabular}{l|ll}
   \hline \hline
   Variable & \multicolumn{2}{|l}{Selection} \\
   \hline
   $E^e_{\text{T}}$      & \multicolumn{2}{|l}{$>20\GeV$}  \\
   $|\eta^e|$         & \multicolumn{2}{|l}{$<2.47$} \\
                      & (excluding calorimeter transition regions)  \\
   $\pt^{\text{jet}}$    & \multicolumn{2}{|l}{$>12\GeV$}  \\
   $|\eta^{\text{jet}}|$ & \multicolumn{2}{|l}{$<1.2$}  \\
   $M_{\ee}$           & $66 <M_{\ee}<116\GeV$ \\
   $\Delta R(\text{jet,electrons})$ & \multicolumn{2}{|l}{$>0.35\,(0.5)$, \antikt{} jets with $R=0.4\,(0.6)$}\\
   $\pt^{\text{jet2}}/\pt^{\Zboson}$      & \multicolumn{2}{|l}{$<0.2$} \\
   \hline \hline
  \end{tabular}
 \end{center}
 \label{tab:ZjeteventSelection}
\end{table}

%
\subsection[Measurement of the \pt{} balance]{Measurement of the \PTBF{} balance}
\label{sec:balanceMeasurement}
\begin{figure*}[!ht]
\begin{center}
\begin{tabular}{lr}
  \subfloat[\Antikt{} $R=0.4$]{\label{fig:compareBalance_akt4}\includegraphics[width=0.44\textwidth]{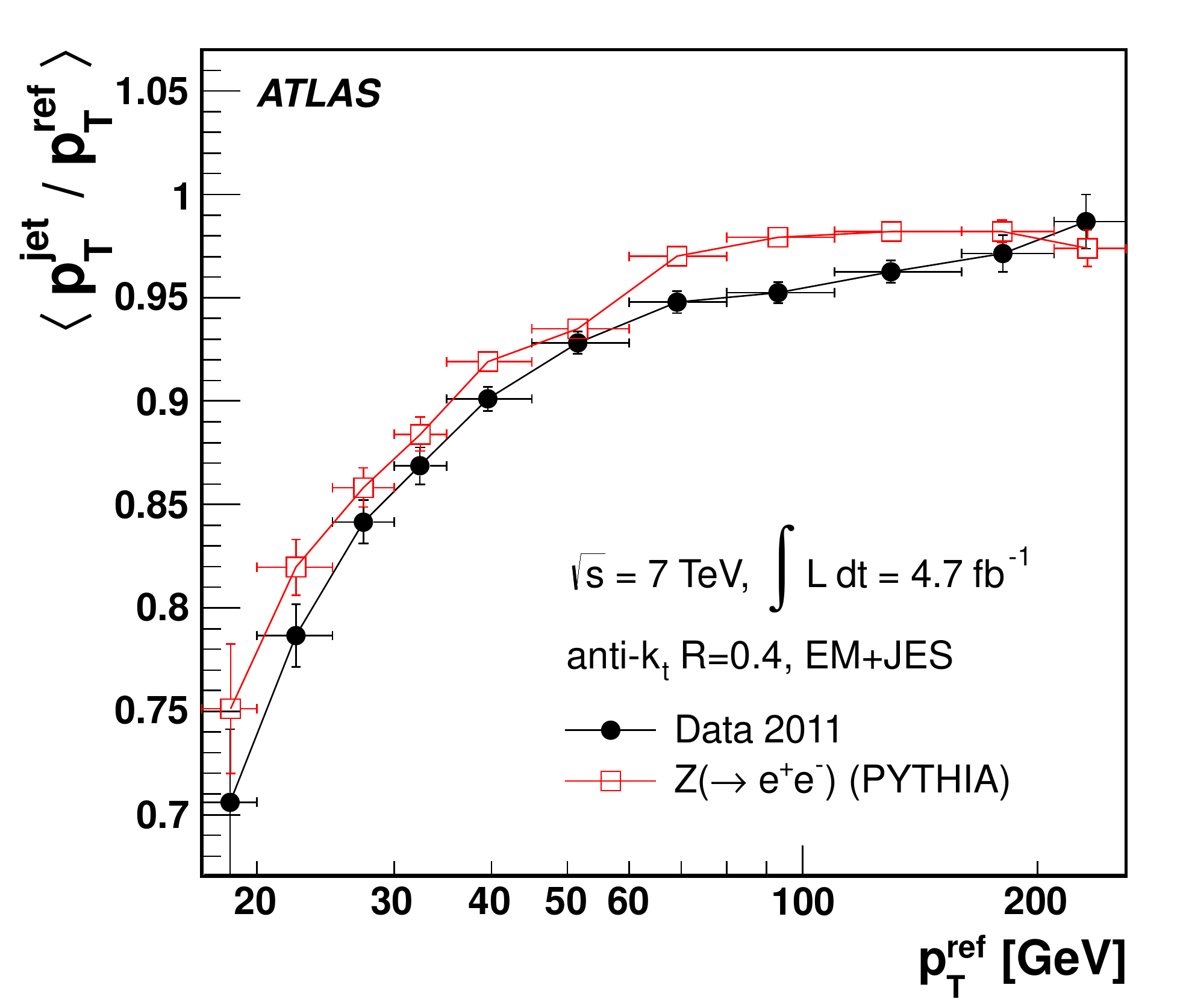}\label{fig:compareBalanceNarrow}}
\subfloat[\Antikt{} $R=0.6$]{\label{fig:compareBalance_akt6}\includegraphics[width=0.44\textwidth]{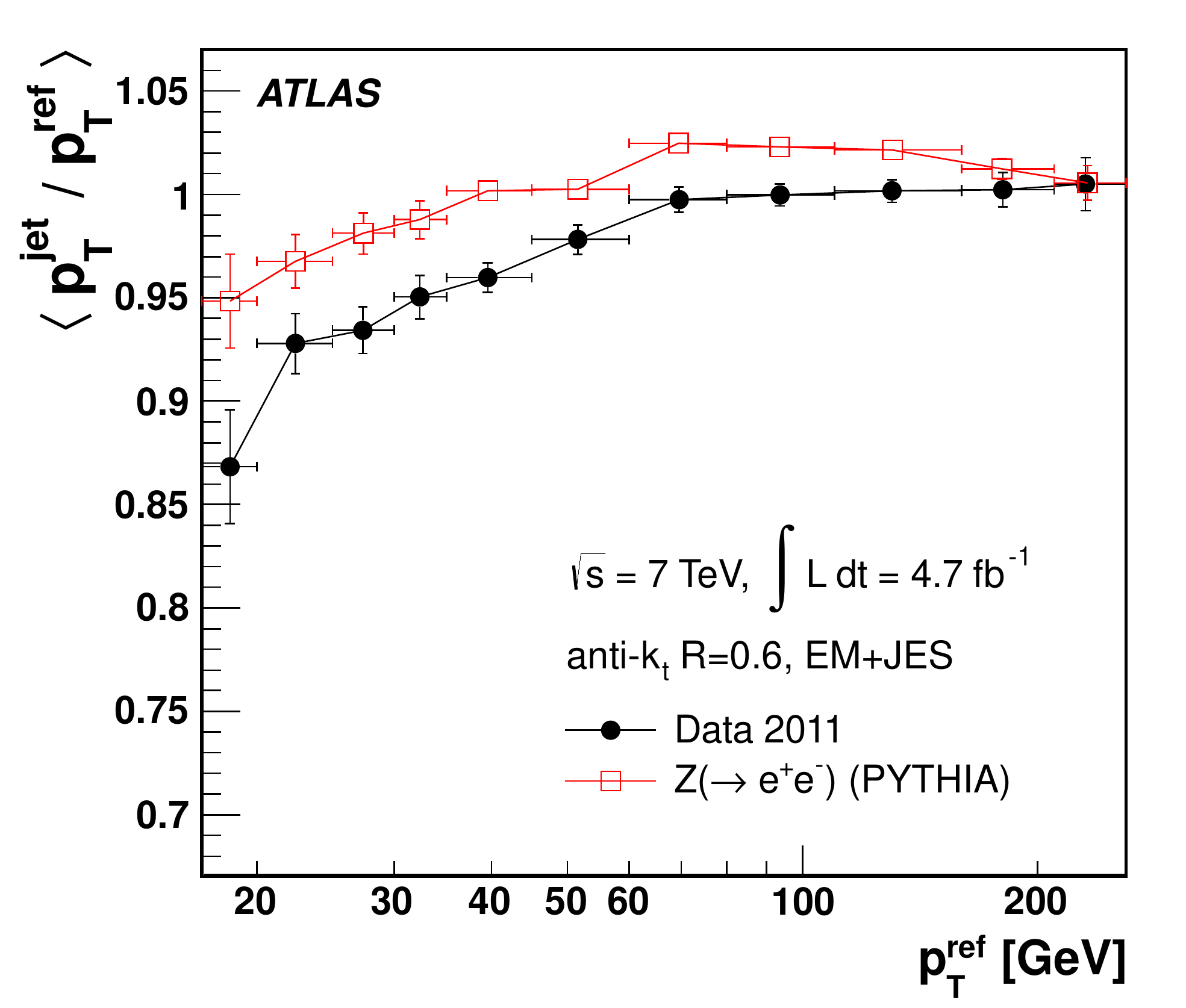}\label{fig:compareBalanceWide}}
\end{tabular}
\end{center}
\caption[]{Mean \pt{} balance obtained in the data and with the \pythia{} simulation. Results for \antikt{} jets with distance parameter \subref{fig:compareBalanceNarrow} $R=0.4$ and \subref{fig:compareBalanceWide} $R=0.6$ calibrated with the \EMJES{} scheme are shown. Only statistical uncertainties are shown.}
\label{fig:compareBalance}
\end{figure*}

The mean value of the $\ptjet/\pt^{\text{ref}}$ ratio distribution is computed in 
bins of $\pt^{\text{ref}}$ and \deltaphi{\mathrm{jet}}{\Zboson}. 
This mean value is obtained with two methods, depending on $\pt^{\text{ref}}$. 
\begin{enumerate}
  \item In the low-$\pt^{\text{ref}}$ region ($17 \leq \pt^{\text{ref}}<35\GeV$), 
it is obtained by a maximum likelihood fit applied to the distribution 
of the $\ptjet/\pt^{\text{ref}}$ ratio. 
The function used, hereafter denoted as the ``fit function'', is a Poisson distribution 
extended to non-integer values\footnote{This continuous Poisson function is obtained by extending the 
discrete Poisson distribution to real values by replacing the factorials in the discrete
Poisson function with Euler's Gamma function.
This function has only one free parameter ($\lambda$). 
A linear transformation of the $x$-scale ($x' = a*x$) is introduced 
and the mean and width of this function are expressed in terms
of $\lambda$ and $a$.
} 
and multiplied by a turn-on curve to model the effect 
of the \ptjet{} threshold, as depicted in \figRef{fig:balanceEstimation}\subref{fig:balanceDistribution}. 
For a given $[\pt^{\text{ref,min}}, \pt^{\text{ref,max}}]$ bin, the turn-on curve  is equal to 1 
above $12\GeV/\pt^{\text{ref,min}}$ and equal to 0 below $12\GeV/\pt^{\text{ref,max}}$. 
A linear function is used to interpolate the turn-on between these two values. 
The mean value of the underlying Poisson distribution is taken as the mean \pt{} balance. 
A fit is preferred to an arithmetic mean calculation because of the jet \pt{} cut, which 
biases the mean value of the balance distribution at low $\pt^{\text{ref}}$ due to the jet energy 
resolution~\cite{Resolution2010}. 
\item For larger $\pt^{\text{ref}}$ ($\pt^{\text{ref}}\geq35\GeV$), the arithmetic mean calculation is 
not sensitive to the jet threshold, and it gives results equivalent to those obtained with a fit. 
In this $\pt^{\text{ref}}$ region, an arithmetic mean is therefore used as it leads to smaller 
uncertainties.  
\end{enumerate}

In the region where the fit is used, $17 \leq\pt^{\text{ref}}<35\GeV$, the fit is actually 
performed twice, in order to reduce the impact of statistical fluctuations:
\begin{enumerate}
  \item In each bin of $\pt^{\text{ref}}$ and $\Delta\phi$, the mean and the width of the Poisson
 distribution are fitted simultaneously. 
  \item The distribution of the widths is parameterised as a function of $\pt^{\text{ref}}$ 
in each $\Delta\phi$ bin according to: 
\begin{equation}
  w(\pt^{\text{ref}}) = \frac{a}{\pt^{\text{ref}}} \oplus \frac{b}{\sqrt{\pt^{\text{ref}}}} \oplus c .  
  \label{eq:resolution}
\end{equation}
The parameters $a$, $b$, and $c$ are obtained from a fit to the widths of the fitted Poisson 
distributions for $\pt<35\GeV$ and to the arithmetic RMS for larger \pt{} 
(see \figRef{fig:balanceEstimation}\subref{fig:resolution}). It is emphasised that this measured width  can not 
directly be compared to the resolution determined in Ref.~\cite{Resolution2010}, since no extrapolation
 to a topology without radiation is per\-for\-m\-ed here.
  \item The fits to the $\ptjet/\pt^{\text{ref}}$ distributions are repeated, but now with 
the widths fixed to the values resulting from the parameterisations.
\end{enumerate}

In order to estimate the mean balance for a topology where the jet and the \Zboson{} boson 
are back-to-back, the mean balances in $\Delta\phi$ bins are extrapolated to $\Delta\phi = \pi$
for each $\pt^{\text{ref}}$ bin, using a linear function (see \figRef{fig:extrapolation}). 
This extrapolation reduces the sensitivity of the mean balance to additional parton radiation 
transverse to the leading jet axis, as discussed earlier in \secRef{sec:ZjetMethod}.
The extrapolated mean balances for the data and \MC-simulated samples generated by \pythia{} 
are shown in \figRef{fig:compareBalance} for \antikt{} jets with distance parameters 
of $R=0.4$ and $R=0.6$, calibrated with the \EMJES{} scheme. 
The mean balance obtained for jets with $R=0.6$ is larger compared to jets with $R=0.4$, 
which is a direct consequence of the larger jet size, and has smaller variations with the 
transverse momentum.

%
\begin{figure}[!ht]
\begin{center}
\begin{tabular}{lr}
  \includegraphics[width=0.45\textwidth]{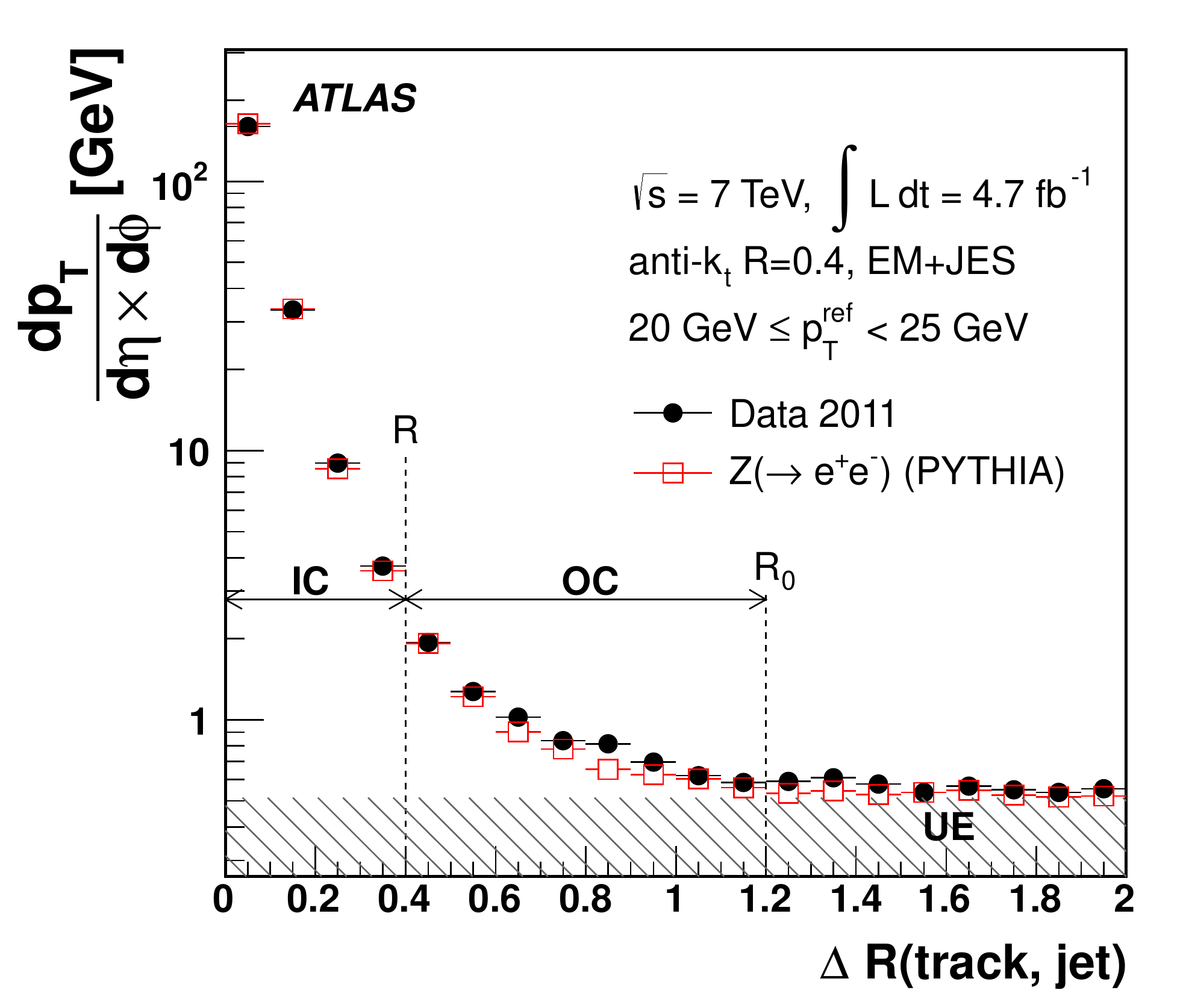}
\end{tabular}
\end{center}
\caption{Transverse momentum profile of tracks around the leading jet axis for events with 
$20 \leq p_{\text{T}}^{\text{ref}} < 25\GeV$ in \Zjet{} events. The radii $R$ and $R_0$ are those used in 
\eqRef{eq:kterm} to define the IC and IC+OC regions. The hatched area indicates the
 contribution from the underlying event (UE). \Antikt{} jets with $R=0.4$ calibrated with 
the \EMJES{} scheme are considered.}
\label{fig:trackFlow}
\end{figure}

%
\subsection{Measuring out-of-cone radiation and underlying event contributions}
\label{sec:outofcone}
The transverse momentum of the \Zboson{} boson is only approximately equal to the transverse momentum
of the truth jet, because of out-of-cone radiation and contributions from the underlying event:
\begin{enumerate}
  \item The \Zboson{} boson balances against all particles inside and outside the jet cone, whereas the truth jet clusters particles inside the jet cone only.
  \item The truth jet's \pt{} includes any UE particles that are clustered in the jet, whereas the \Zboson{} boson's \pt{} receives almost no such contribution.
\end{enumerate}

These two contributions are estimated by measuring the transverse momentum profile of tracks around 
the leading jet axis (see \figRef{fig:trackFlow}). Tracks associated to the hard-scattering 
vertex are used instead of clusters of calorimeter cells in order to reduce the sensitivity to pile-up 
interactions. Tracks associated to the two electrons stemming from the \Zboson{} boson are 
removed when computing the transverse momentum profiles.

A factor is calculated from the out-of-cone and underlying event contributions:
\begin{equation}
  \kooc{} = \frac{\pt^{\rm IC, ALL}}{\pt^{\rm IC+OC, ALL} - \pt^{\rm IC+OC, UE}}\, ,
  \label{eq:kterm}
\end{equation}
where $\pt^{\rm IC, ALL}$ is the average scalar \pt{} sum of all the tracks inside the jet cone with 
radius $R$, $\pt^{\rm IC+OC, ALL}$ is the average scalar \pt{} sum of all the tracks inside and 
outside the jet cone, and $\pt^{\rm IC+OC, UE}$ is the average contribution of the underlying event 
to $\pt^{\text{IC+OC, ALL}}$. The transverse momenta $\pt^{\rm IC+OC, ALL}$ and $\pt^{\rm IC+OC, UE}$ 
are estimated in a cone of radius $R_0$, above which only the UE contributes 
to $\pt^{\rm IC+OC, ALL}$, and from where the transverse momentum density is constant 
(see \figRef{fig:trackFlow}). In practice, $R_0$ is the value where the logarithmic 
derivative of \kooc{} with respect to $R_0$ is equal to $0.05$.

\begin{figure*}[!ht]
\begin{center}
\begin{tabular}{lr}
\subfloat[\Antikt{} $R=0.4$]{\label{fig:UncertEMJES_akt4}\includegraphics[width=0.45\textwidth]{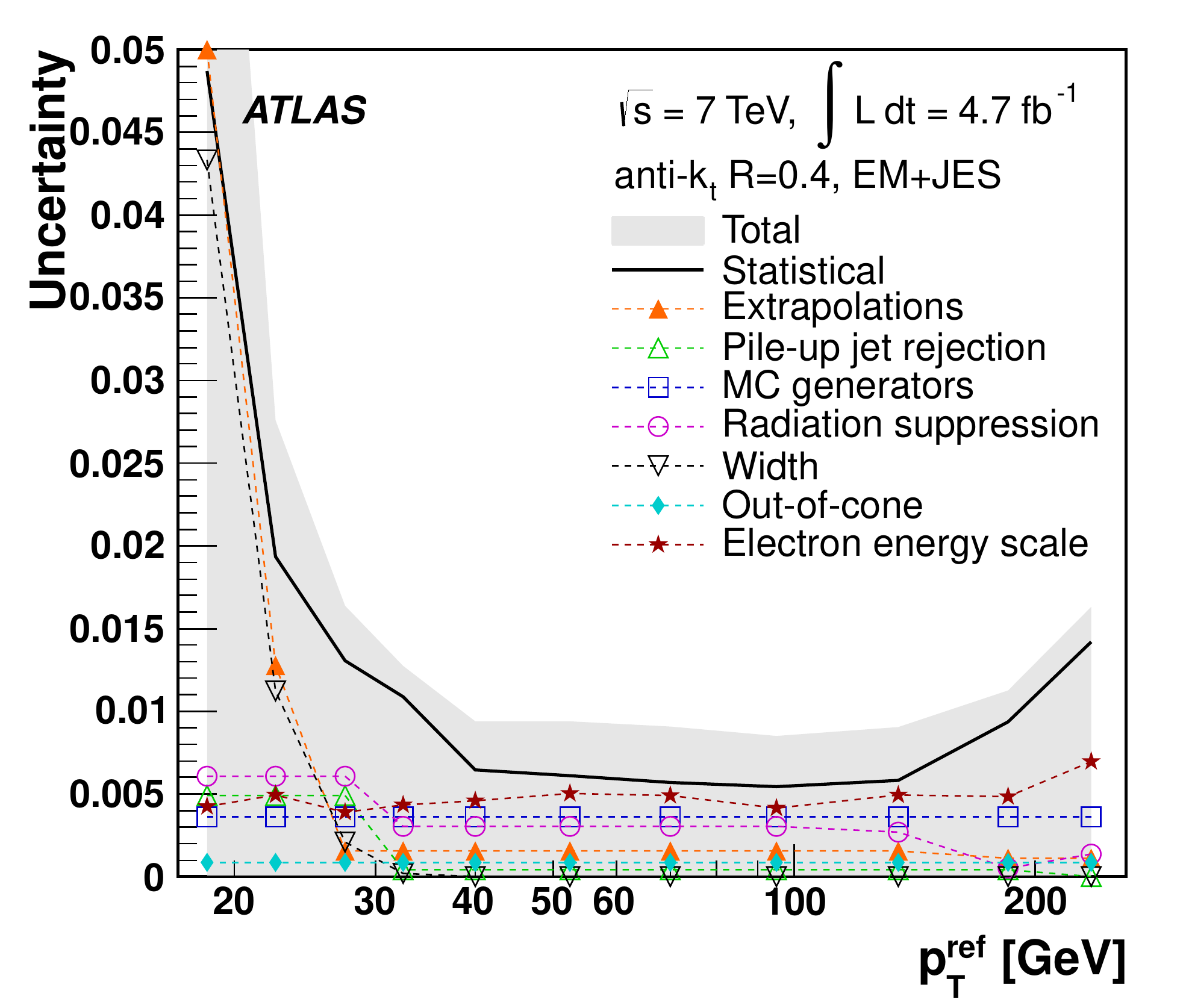}\label{fig:UncertEMJESNarrow}}
\subfloat[\Antikt{} $R=0.6$]{\label{fig:UncertEMJES_akt6}\includegraphics[width=0.45\textwidth]{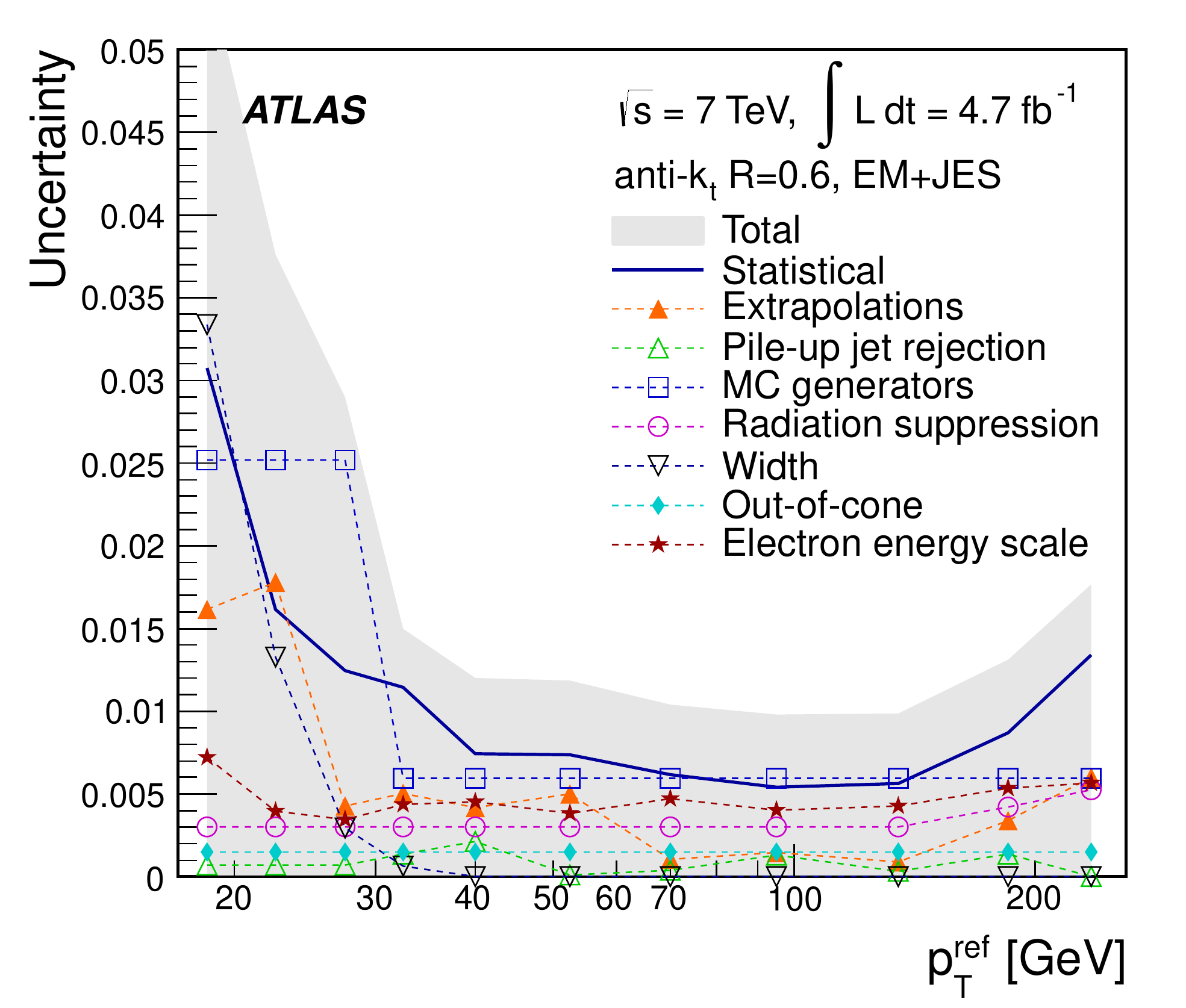}\label{fig:UncertEMJESWide}}
\end{tabular}
\end{center}
\caption[]{Summary of the \Zjet{} uncertainties on the \datatomc{} ratio of the mean \pt{} balance, for \antikt{} jets with distance parameter \subref{fig:UncertEMJESNarrow} $R=0.4$ and \subref{fig:UncertEMJESWide} $R=0.6$ calibrated with the \EMJES{} scheme.}
\label{fig:UncertEMJES}
\end{figure*}

\begin{figure*}[!ht]
\begin{center}
\begin{tabular}{lr}
\subfloat[\Antikt{} $R=0.4$ \EMJES]{\includegraphics[width=0.45\textwidth]{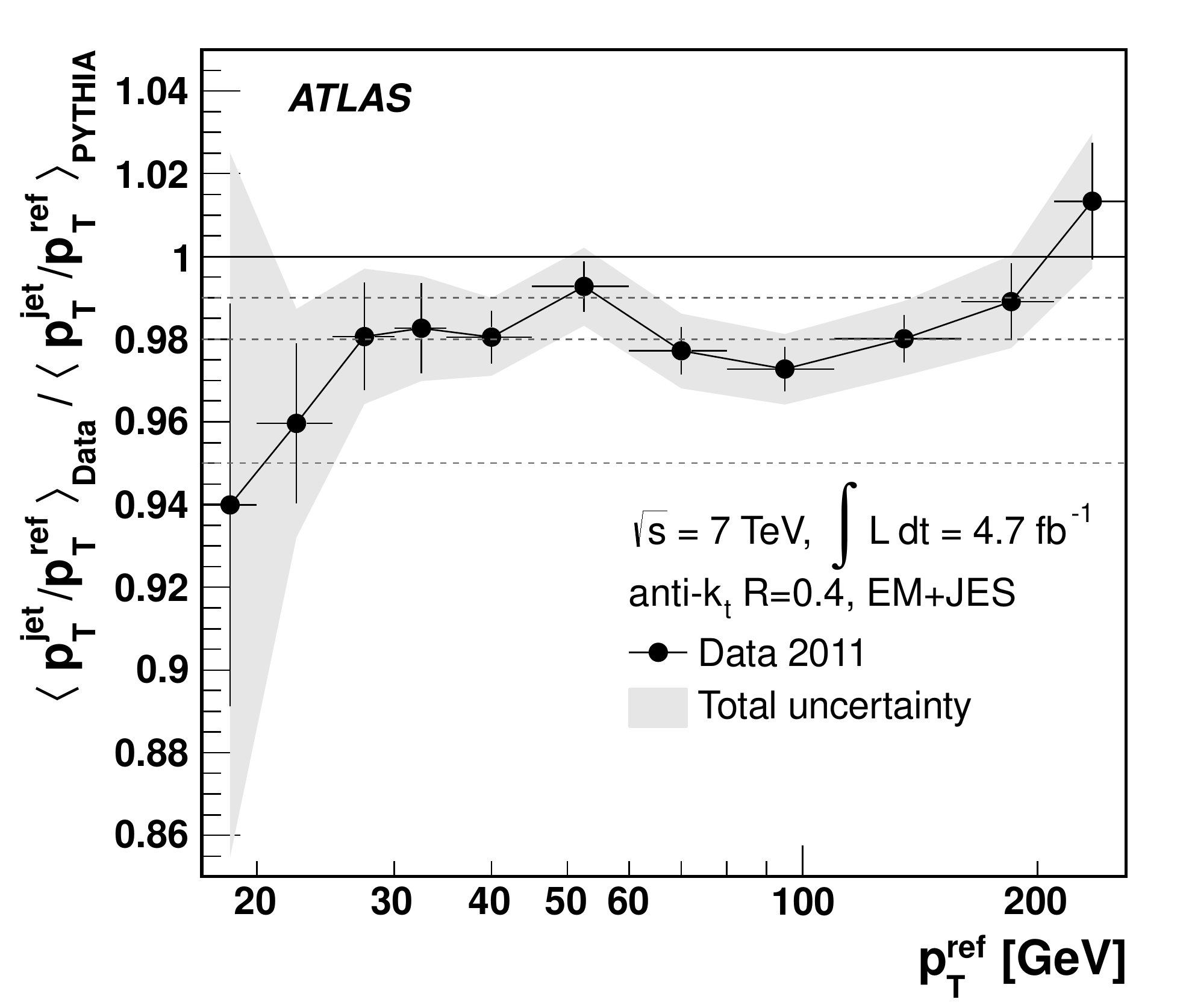}\label{fig:ResultsZjetJESEMJESNarrow}}
\subfloat[\Antikt{} $R=0.6$ \EMJES]{\includegraphics[width=0.45\textwidth]{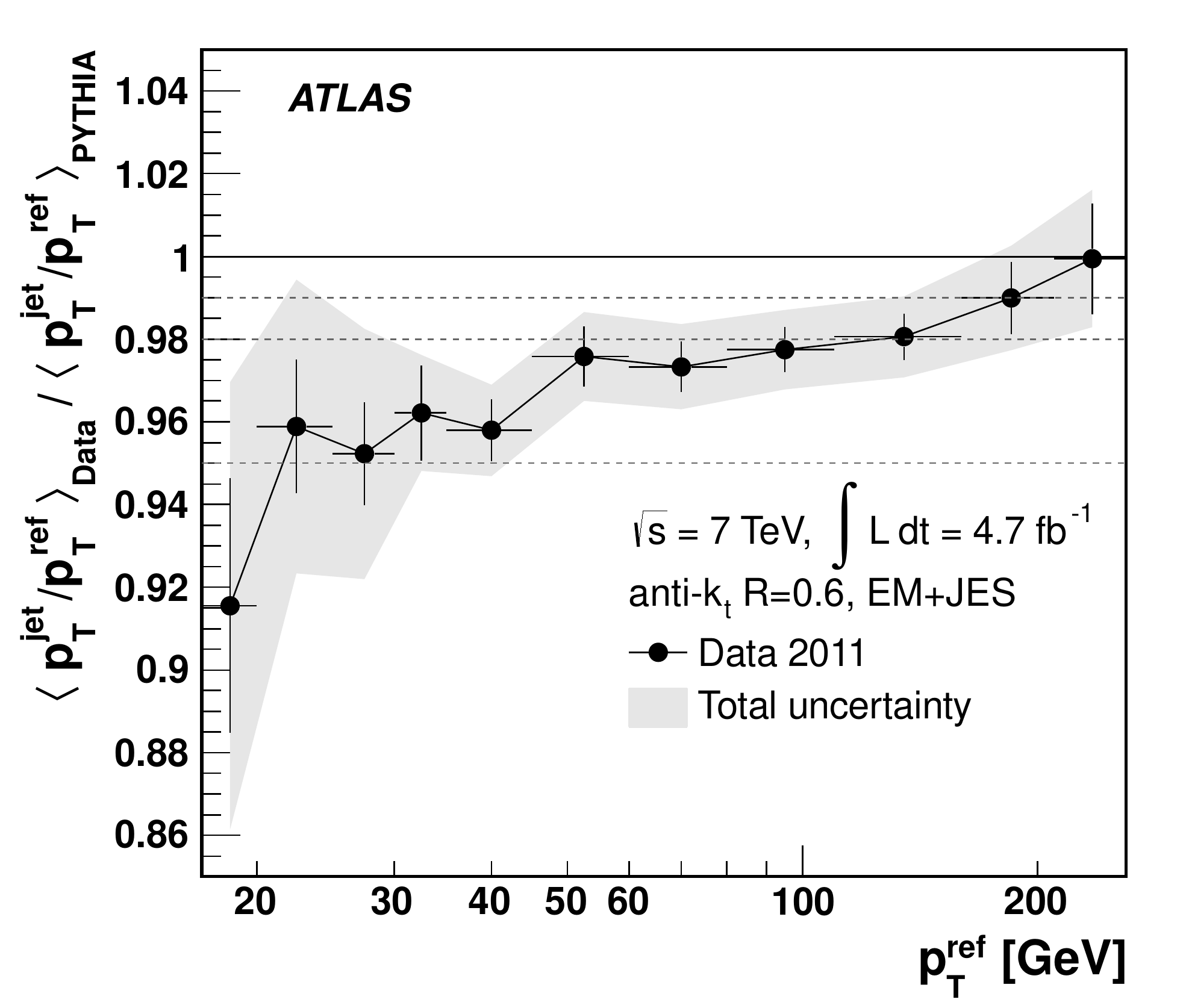}\label{fig:ResultsZjetJESEMJESWide}}\\
\subfloat[\Antikt{} $R=0.4$ \LCWJES]{\includegraphics[width=0.45\textwidth]{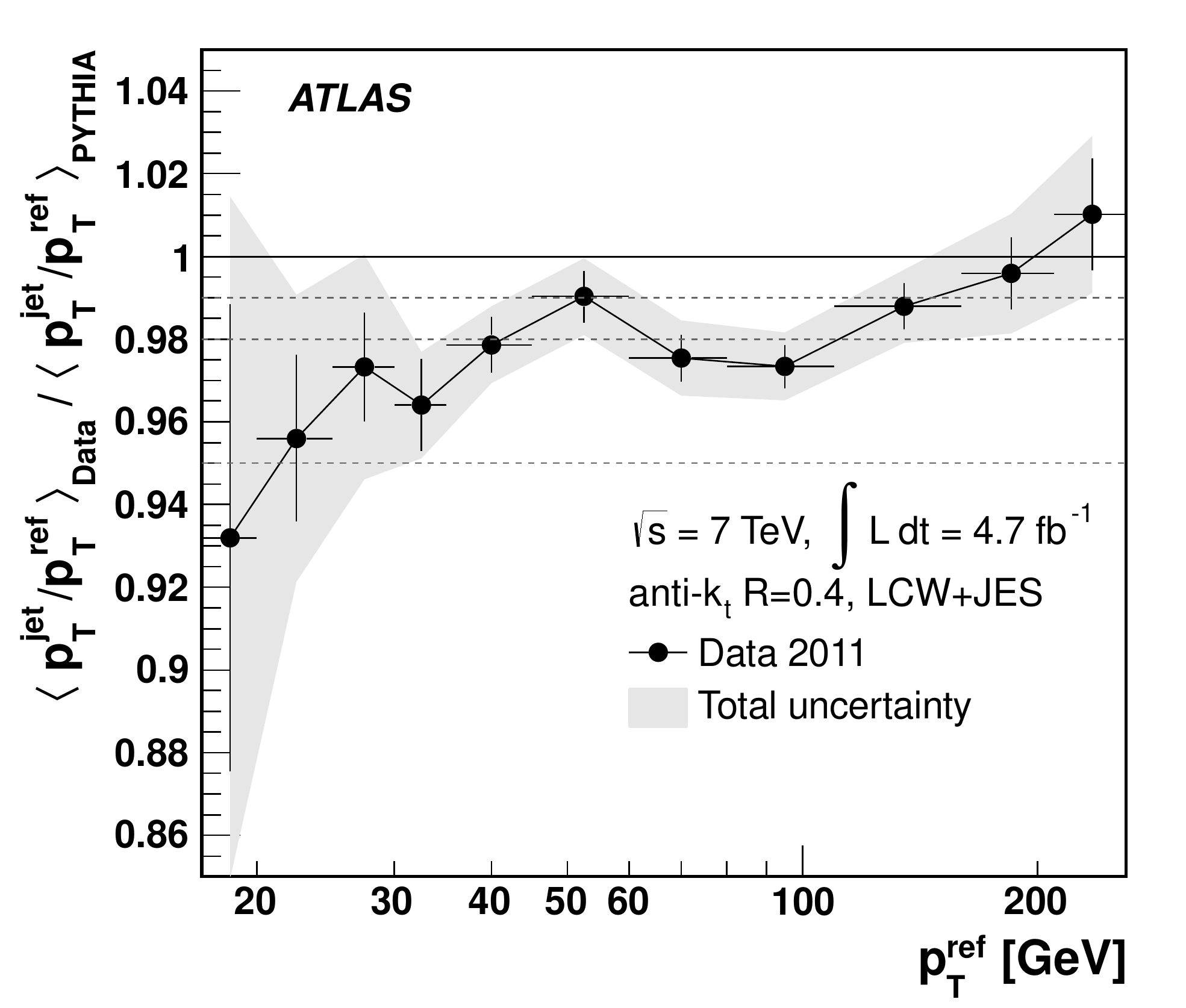}\label{fig:ResultsZjetJESLCWJESNarrow}}
\subfloat[\Antikt{} $R=0.6$ \LCWJES]{\includegraphics[width=0.45\textwidth]{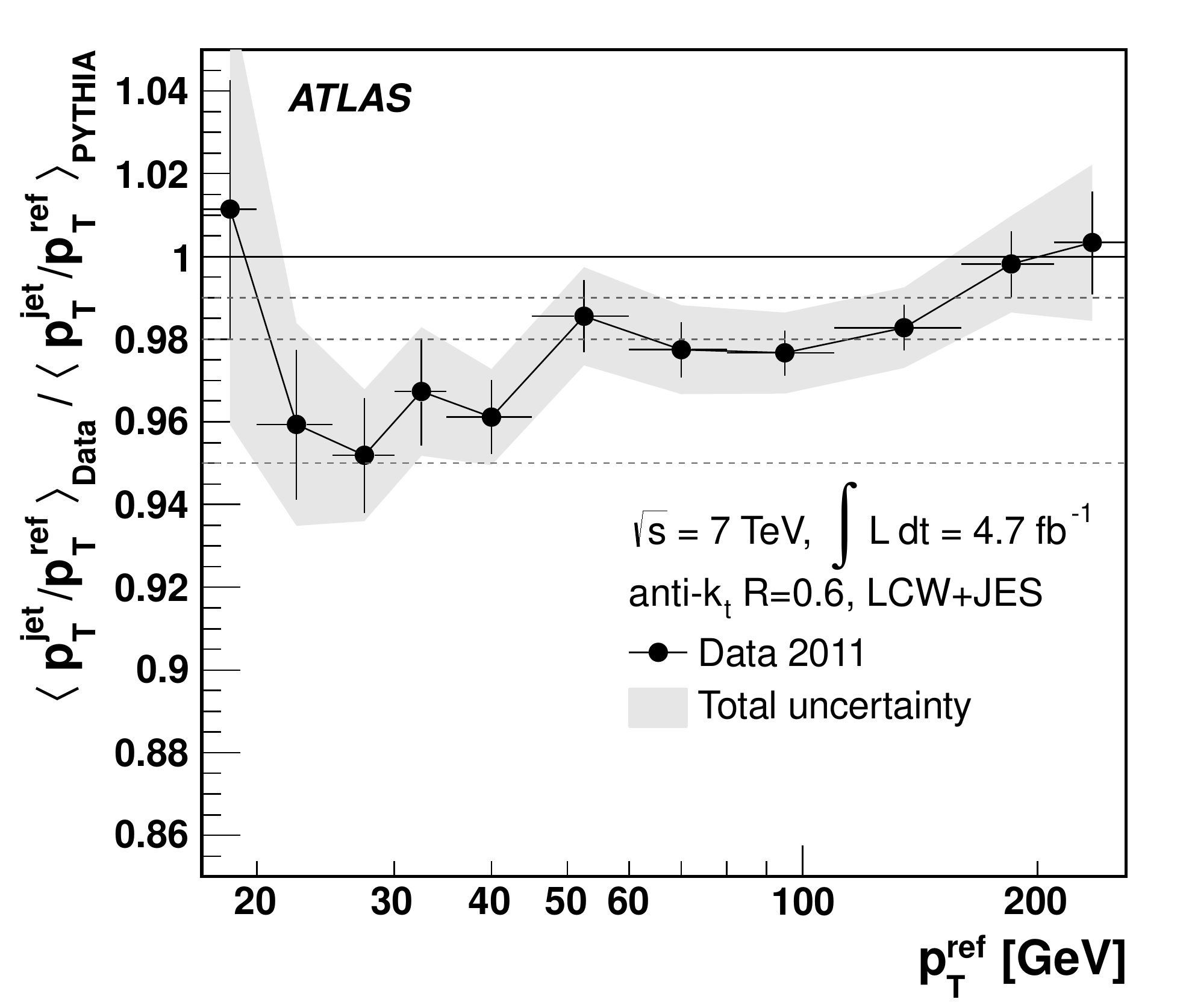}\label{fig:ResultsZjetJESLCWJESWide}}

\end{tabular}
\end{center}
\caption[]{\Datatomc{} ratio of the mean \pt{} balance for \Zjet{} events as a function of $\pt^{\text{ref}}$ 
for \antikt{} jets with distance parameter (\subref{fig:ResultsZjetJESEMJESNarrow}, \subref{fig:ResultsZjetJESLCWJESNarrow}) $R=0.4$  and  (\subref{fig:ResultsZjetJESEMJESWide}, \subref{fig:ResultsZjetJESLCWJESWide})  $R=0.6$ calibrated with 
the (\subref{fig:ResultsZjetJESEMJESNarrow}, \subref{fig:ResultsZjetJESEMJESWide}) \EMJES{} and the (\subref{fig:ResultsZjetJESLCWJESNarrow}, \subref{fig:ResultsZjetJESLCWJESWide}) \LCWJES{}  schemes. 
The total uncertainty on this ratio is depicted by grey bands. 
Dashed lines show the $-1\%$, $-2\%$, and $-5\%$ shifts.}
\label{fig:ResultsZjetJES}
\end{figure*}
%
\subsection{Systematic uncertainties}
\label{sec:Zjetsystematics}

The differences between the balances observed in data and tho\-se observed in \MC{} simulations 
may be due to physics or detector effects directly influencing the calorimeter response to jets 
(e.g. fragmentation or material in front of the calorimeter), which may not be correctly modelled 
in the simulation. They can also be due to effects that have an influence on the direct 
\pt{} balance method itself, e.g. the estimation of the mean balance or higher-order parton emissions. 
For a more detailed evaluation of the systematic uncertainties, the following steps are performed:
\begin{enumerate}
  \item The uncertainty on the width parameterisation is propagated to the mean estimation.
  \item The fit range used for the $\Delta\phi$ extrapolation is varied.
  \item The cut on sub-leading jets is varied to assess the effect of additional high-\pt{} parton radiation altering the balance.
  \item The effect of soft particles produced outside the jet cone and the underlying event contribution to the jet energy is compared in data and simulation.
  \item The impact of additional (pile-up) interactions is studied.
  \item The uncertainty in the electron energy measurement is propagated to the \pt{} balance.
  \item The results obtained with \pythia{} and \alpgen+\herwig{} are compared.
\end{enumerate}

\subsubsection{Fitting procedure}
For $\pt^{\text{ref}}<35\GeV$, the mean balance in a given bin of $\pt^{\text{ref}}$ and $\Delta\phi$ 
is first obtained using the nominal parameterised width given in \eqRef{eq:resolution}. 
The fit is then performed again with a larger and a smaller width according to the uncertainty 
on the parameterisation. 
The four differences obtained in the resulting mean balances for the up and down variations in data and 
Monte Carlo simulation are propagated independently
, after $\Delta\phi$ extrapolation, to the \datatomc{} ratio. 
The two positive and two negative deviations are both summed in quadrature and the final uncertainty
is taken as the average of the absolute values of the two deviations.

\subsubsection{Extrapolation procedure}
The nominal extrapolated balance is determined with a linear fit from $\Delta\phi = \pi-0.3$ 
to $\Delta\phi = \pi$. The lower limit is decreased to $\pi-0.4$ and increased to $\pi-0.2$, 
and the average of the absolute values of the two deviations is taken as a systematic uncertainty 
on the \datatomc{} ratio.

\subsubsection{Additional radiation suppression}\label{sec:sysRadiationSup}
While the extrapolation of the \pT{} balance in $\Delta\phi$ attempts to reduce the effect of radiation 
perpendicular to the jet axis at angular scales within the range from $[\pi-0.3,\pi]$, additional
radiation not reflected by the $\Delta\phi$ measurement and extrapolation can still occur and modify 
the \pt{} balance between the \Zboson{} boson and the leading jet with respect to expectations for true back-to-back topologies. Therefore, events with energetic sub-leading jets are vetoed. 
Systematic uncertainties associated with this second jet veto are studied, and the mean 
\pt{} balances in the data and the simulation are compared when applying different second jet vetoes.
The nominal
\begin{displaymath} 
\ptjetn{2,\text{nom}}=\text{max}\{ 12\GeV,\, 0.2\times\pt^{\Zboson} \}
\end{displaymath}
 is varied up and down to
\begin{displaymath}
\begin{array}{ll} 
	\ptjetn{2,\text{nom}\uparrow} = \ptjetn{2,\text{nom}} + 0.1\times\pt^{\Zboson}  & \;\;\text{(down)} \\
        \ptjetn{2,\text{nom}\downarrow} = \max\{ 10\GeV,\, 0.1\times\pt^{\Zboson} \}   & \;\;\text{(up)} .
\end{array}
\end{displaymath}
The average of the absolute values of the two deviations is taken as a systematic uncertainty on the 
\datatomc{} ratio.

\subsubsection{Out-of-cone radiation and underlying event}
\label{sec:outofconesyst}
This \kooc{} factor defined in \eqRef{eq:kterm} and measured as described in \secRef{sec:outofcone} indicates how the \Zboson{} boson's \pt{} differs from the 
truth jet's \pt{}. 
In order to evaluate the systematic uncertainties coming from out-of-cone radiation and the 
underlying event, this factor is applied to the \Zboson{} boson's \pt{}. It is measured in the 
data and in the simulation in bins of $\pt^{\text{ref}}$. Its value depends on the \pt{} as well 
as on the jet size. For jets with $R=0.4$, \kooc{} increases from about $0.93$ at low \pt{} 
to about $0.99$ at high \pt{}. For jets with $R=0.6$, it varies between $1$ and $1.02$ without 
any systematic \pt{} dependence. A modified \datatomc{} ratio of the balance is calculated using 
the \kooc{} factors and the difference with respect to the nominal ratio is taken 
as a systematic uncertainty.

 \subsubsection{Impact of additional pile-up interactions}
The impact of in-time and out-of-time pile-up is studied by comparing the \pt{} balance in 
two samples with different numbers of primary vertices ($\Npv\leq 5$ and $\Npv>5$), 
and two samples with different average number of interactions per bunch crossing ($\axing<8$ 
and $\axing>8$). The differences observed between the samples are small compared to the uncertainty 
on the pile-up offset correction (see \secRef{sec:pileupineta}). 
Therefore, they are not taken into 
account in this analysis in order to avoid double-counting between the different steps of the 
jet calibration procedure.

The direct impact of additional interactions on the leading jet is also studied by relaxing the \JVF{} cut, 
introduced in \secRef{sec:jetSelection}, for the leading jet. The difference with respect to the nominal result is taken as an additional uncertainty.

\subsubsection{Electron energy scale}
The \pt{} of the \Zboson{} boson, measured from the energy of the electrons, is used as a 
reference to probe the jet energy scale. The electron energy is shifted upwards and 
downwards according to the uncertainty on its measurement~\cite{Atlaselectronpaper}, 
updated using data recorded in $2011$. 

\subsubsection{Impact of the Monte Carlo generator}
The mean balances are obtained from \pythia{} and \alpgen{} samples, using the procedure 
described in \secRef{sec:balanceMeasurement}. The difference between the data-to-\pythia{} 
and the data-to-\alpgen{} ratios is taken as a systematic uncertainty. 
The \alpgen{} \MC{} generator uses different theoretical models for all steps of the event 
generation and therefore gives a reasonable estimate of the systematic variations. 
However, the possible compensation of modelling effects that shift the jet response in 
opposite directions cannot be excluded. 
To reduce the impact of statistical fluctuation the first three bins are merged,
since they give the same result within their statistical uncertainties.

\subsubsection{Summary of systematic uncertainties}
Additional sources of uncertainties are considered: 
\begin{enumerate}
  \item The main background to \Zjet{} events is from multijet e\-v\-ents, 
and its fraction in the selected events is only of the order of $3\%$~\cite{ZJetPaper}. 
Furthermore, jets passing the electron identification cuts contain an important electromagnetic 
component and the detector response should therefore be similar to the response for true electrons. 
No additional systematic uncertainty is considered for the contamination of \Zjet{} events 
with background events.
  \item As already mentioned, the uncertainty on the pile-up offset correction is treated
        as extra uncertainty (see \secRef{sec:pileupineta}) 
  \item The uncertainty induced by quark and gluon response differences as well as different 
quark and gluon compositions in data and in the simulation is addressed 
in \secRef{sec:FlavorTopology}.
\end{enumerate}

In the final evaluation of systematic uncertainties, only effects that are significant with 
respect to their statistical uncertainties are taken into account~\cite{asymErrors}. 
The systematic effects and their statistical uncertainties are first evaluated using the 
initial binning. Then the results in neighbouring bins are iteratively combined until the 
observed effects become significant. 
The quadratic sum of all the components previously described is taken as the overall 
systematic uncertainty. \FigRef{fig:UncertEMJES} summarises the different contributions 
to the total uncertainty, for \EMJES{} jets, in the whole \pt{} range. 
For $R=0.4$ jets and $25\GeV<\pt^{\text{ref}}<260 \GeV$, uncertainties are typically 
between $1\%$ and $2\%$, and increase up to $10\%$ for low transverse momenta.

\subsection[Summary of the \Zjet{} analysis]{Summary of the \ZJET{} analysis}
\label{sec:Zjetsummary}
The two \ATLAS{} jet energy calibration schemes \EMJES{} and \LCWJES{} are probed using the direct \pt{} balance 
between a central jet and a \Zboson{} boson. 
The responses measured in the data and in the simulation are compared for jets defined 
by the \antikt{} clustering algorithm with distance parameters of $R=0.4$ and $R=0.6$. 

\FigRef{fig:ResultsZjetJES} shows the \datatomc{} ratio of the mean \pt{} balance for jets 
calibrated with the \EMJES{} and the \LCWJES{} schemes, with statistical and systematic uncertainties. 
For $R=0.4$ jets and $\pt^{\text{ref}}>25\GeV$, this ratio is shifted by at most $-4\%$ from unity, 
and typically by $-2\%$ over most of the \Zboson{} boson \pt{} range. 
Uncertainties are typically between $1\%$ and $2\%$ for $25 <\pt^{\text{ref}}<260\GeV$, 
and increase up to $10\%$ for low transverse momenta. 

%

\begin{figure*}[ht!p]
\begin{center}
\subfloat[$25 \leq \ptgamma < 45$~\GeV{}]
{\includegraphics[width=0.45\textwidth]{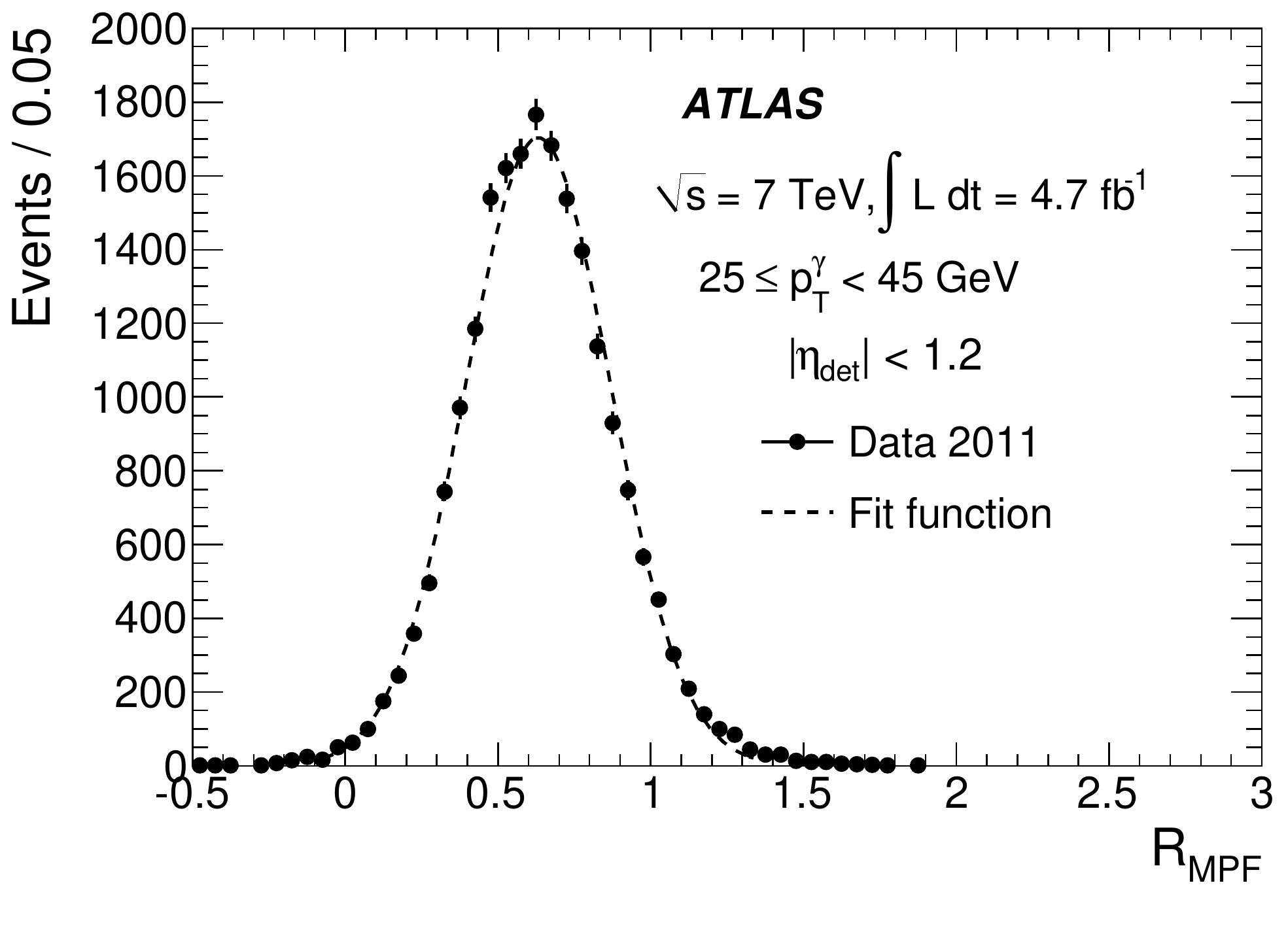}\label{fig:MPF_DistributionsLowPt}}
\subfloat[$160 \leq \ptgamma < 210$~\GeV{}]
{\includegraphics[width=0.45\textwidth]{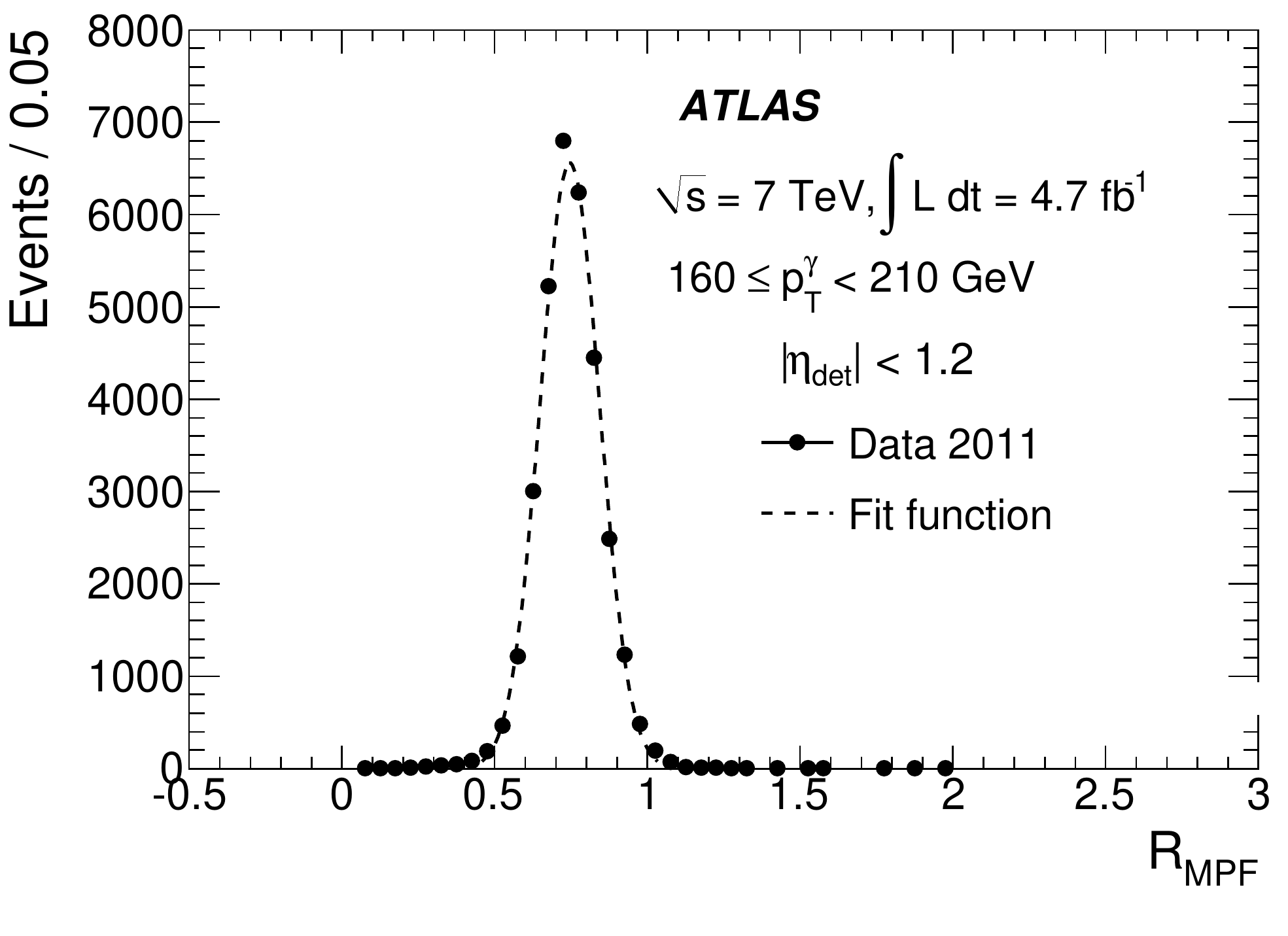}\label{fig:MPF_DistributionsHighPt}}
\end{center}
\caption[]{\MPF{} response distributions in the \gammajet{} data for \subref{fig:MPF_DistributionsLowPt}  $25 \leq \ptgamma < 45$~\GeV{} and  \subref{fig:MPF_DistributionsHighPt} $160 \leq \ptgamma < 210$ \GeV{} when using \topos{} at the \EM{} scale. The dashed lines 
represent the fits with a Gaussian function. The mean value from the fit in each 
\ptgamma{} bin is the value used as the measured average \MPF{} response.}
\label{fig:MPF_Distributions}
\end{figure*}

\begin{figure*}[ht!p]
\begin{center}
\subfloat[$25 \leq \ptgamma < 45$~\GeV{}]
{\includegraphics[width=0.45\textwidth]{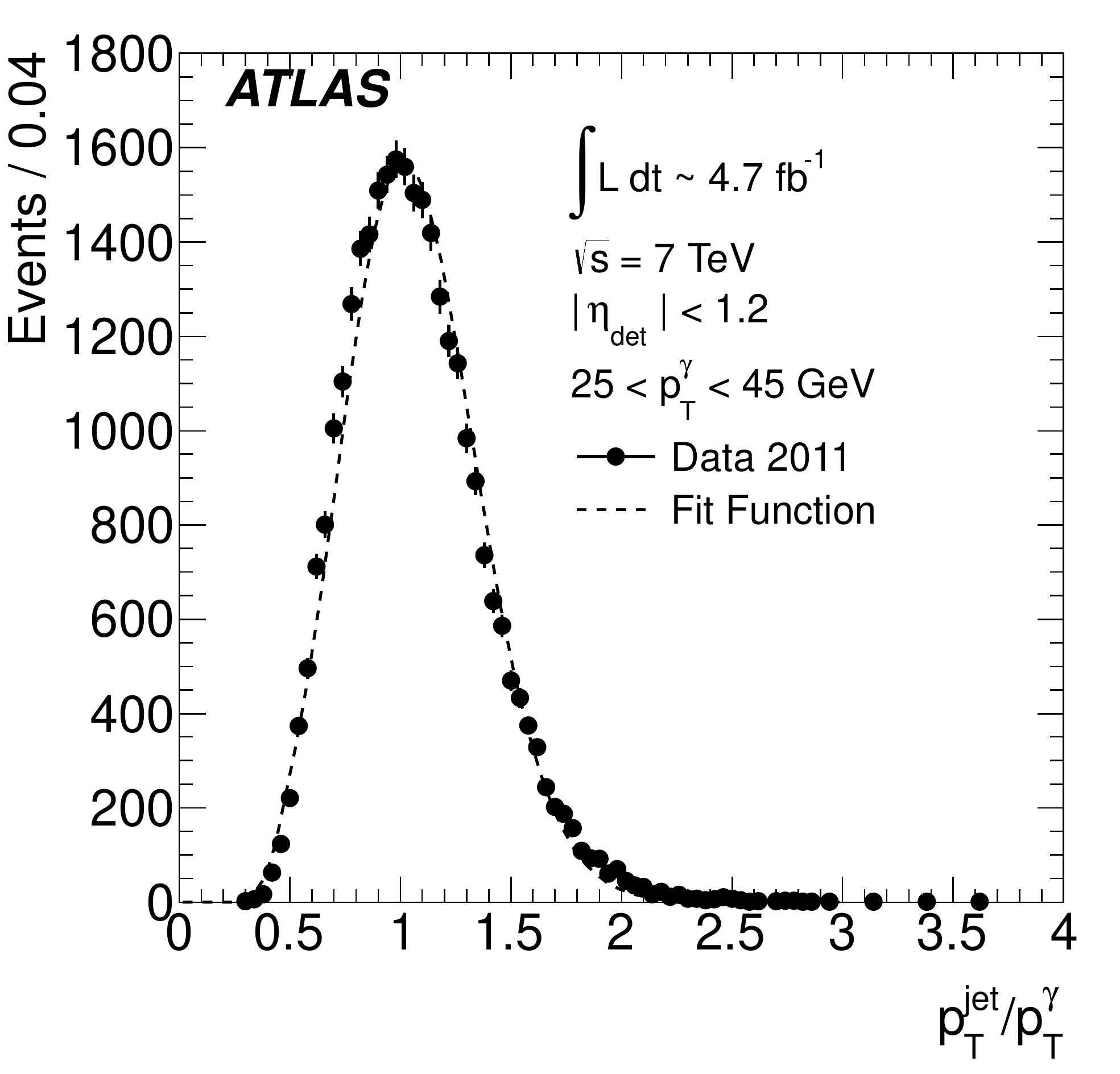}\label{fig:DB_DistributionsLowPt}}
\subfloat[$160 \leq \ptgamma < 210$~\GeV{}]
{\includegraphics[width=0.45\textwidth]{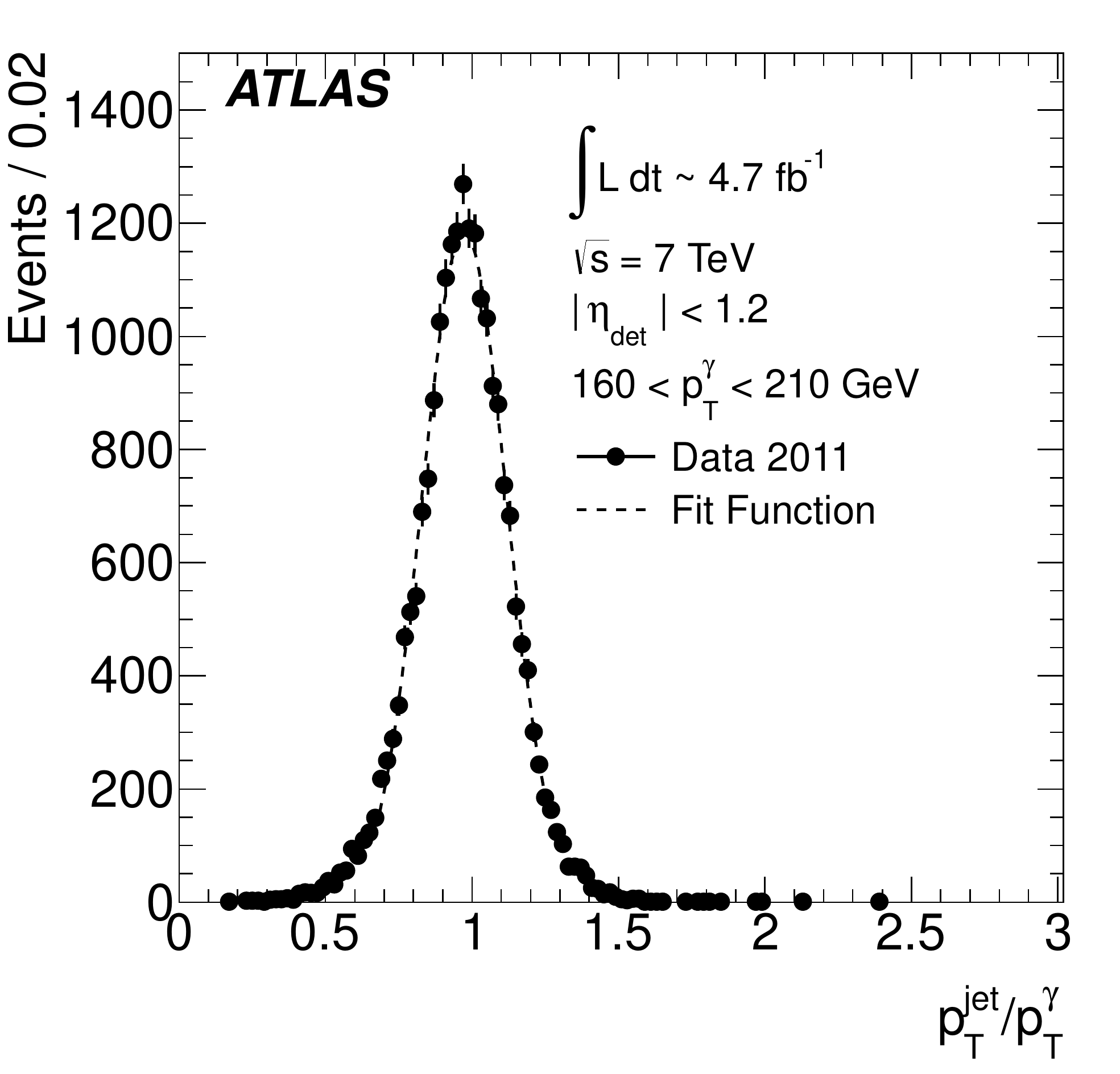}\label{fig:DB_DistributionsHighPt}}
\end{center}
\caption[]{Jet response distributions in the \gammajet{} data for \subref{fig:DB_DistributionsLowPt} $25 \leq \ptgamma < 45$~\GeV{} and  \subref{fig:DB_DistributionsHighPt} 
$160 \leq \ptgamma < 210$~\GeV{} as measured by the \DB{} technique 
for \antikt{} jets 
with $R = 0.6$ at the \EMJES{} scale. The dashed lines represent fits of Gaussian functions, except in the lowest bin ($25 \leq \ptgamma < 45$~\GeV{}), where the fit function is a Poisson distribution. 
The mean value from the fit in each \ptgamma{} bin is the value used as the measured average jet response 
in \DB{}.}
\label{fig:DB_Distributions}
\end{figure*}

\begin{figure*}[ht!p]
\begin{center}
\subfloat[\EM]
{\includegraphics[width=0.45\textwidth]{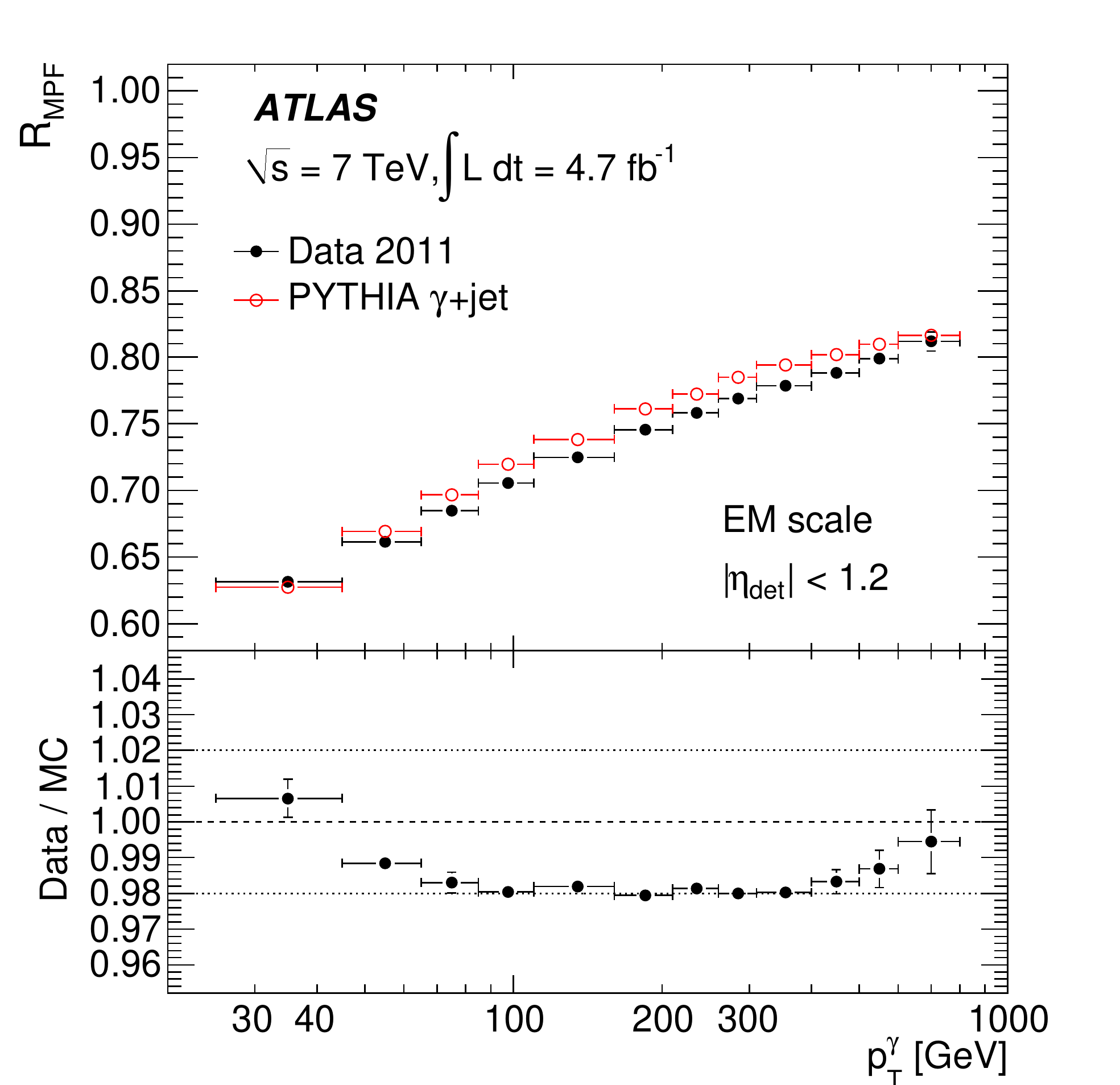}\label{fig:MPF_DataMCEM}}
\subfloat[\LCW]
{\includegraphics[width=0.45\textwidth]{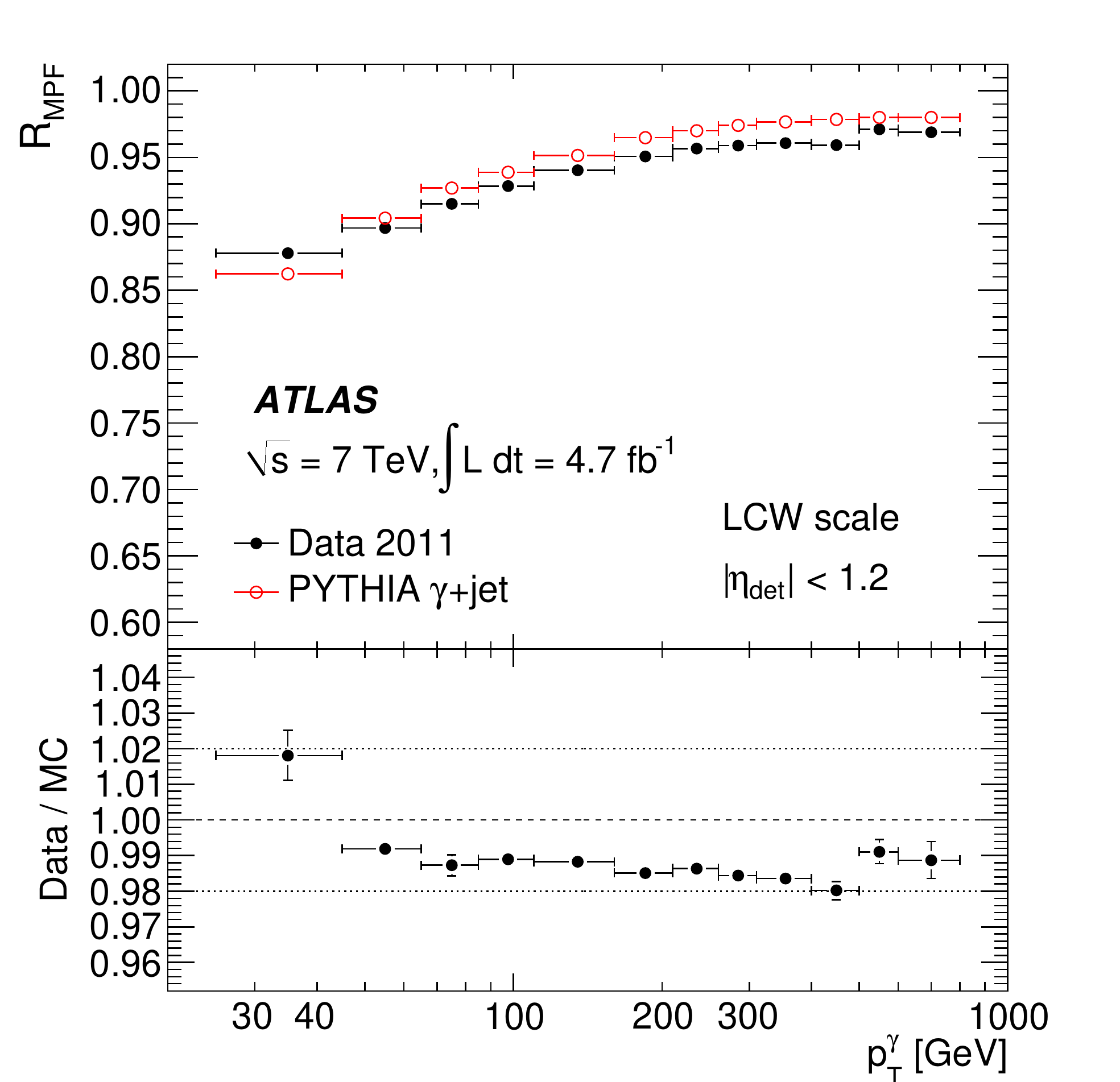}\label{fig:MPF_DataMCLCW}}
\end{center}
\caption[]{Average jet response as determined by the \MPF{} technique in \gammajet{} events using \topos{} at the \subref{fig:MPF_DataMCEM}  \EM{} and \subref{fig:MPF_DataMCLCW} \LCW{}
energy scales, for both data and \MC{} simulations, as a function of the photon transverse momentum. 
The \datatomc{} response ratio is shown in the bottom inset of each figure. Only the statistical uncertainties are shown.}
\label{fig:MPF_DataMC}
\end{figure*}

\begin{figure*}[ht!p]
\begin{center}
\subfloat[$R = 0.4$, \EMJES]
{\includegraphics[width=0.45\textwidth]{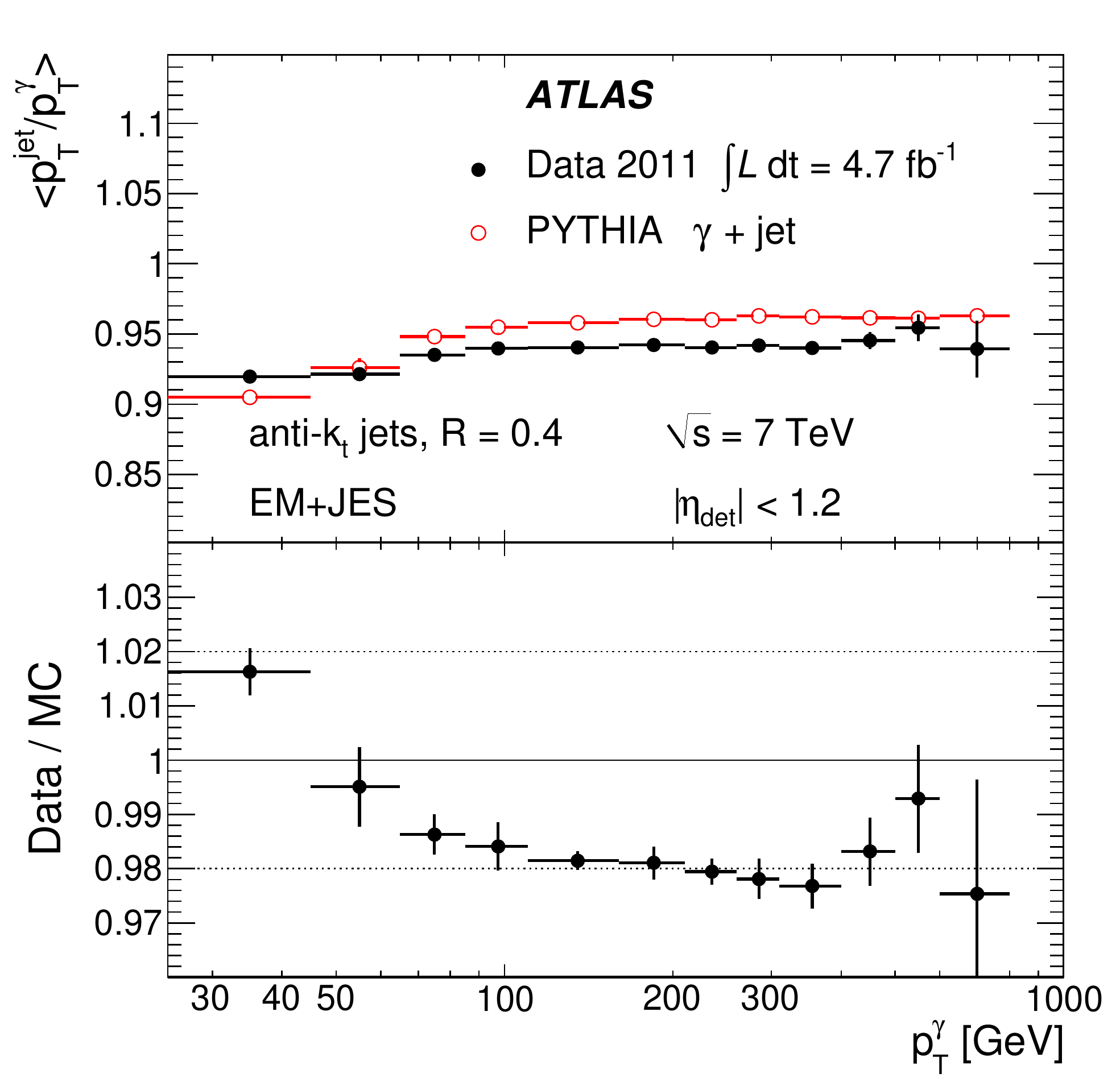}\label{fig:DirectBalance_DataMC_EM_akt4}}
\subfloat[$R = 0.4$, \LCWJES]
{\includegraphics[width=0.45\textwidth]{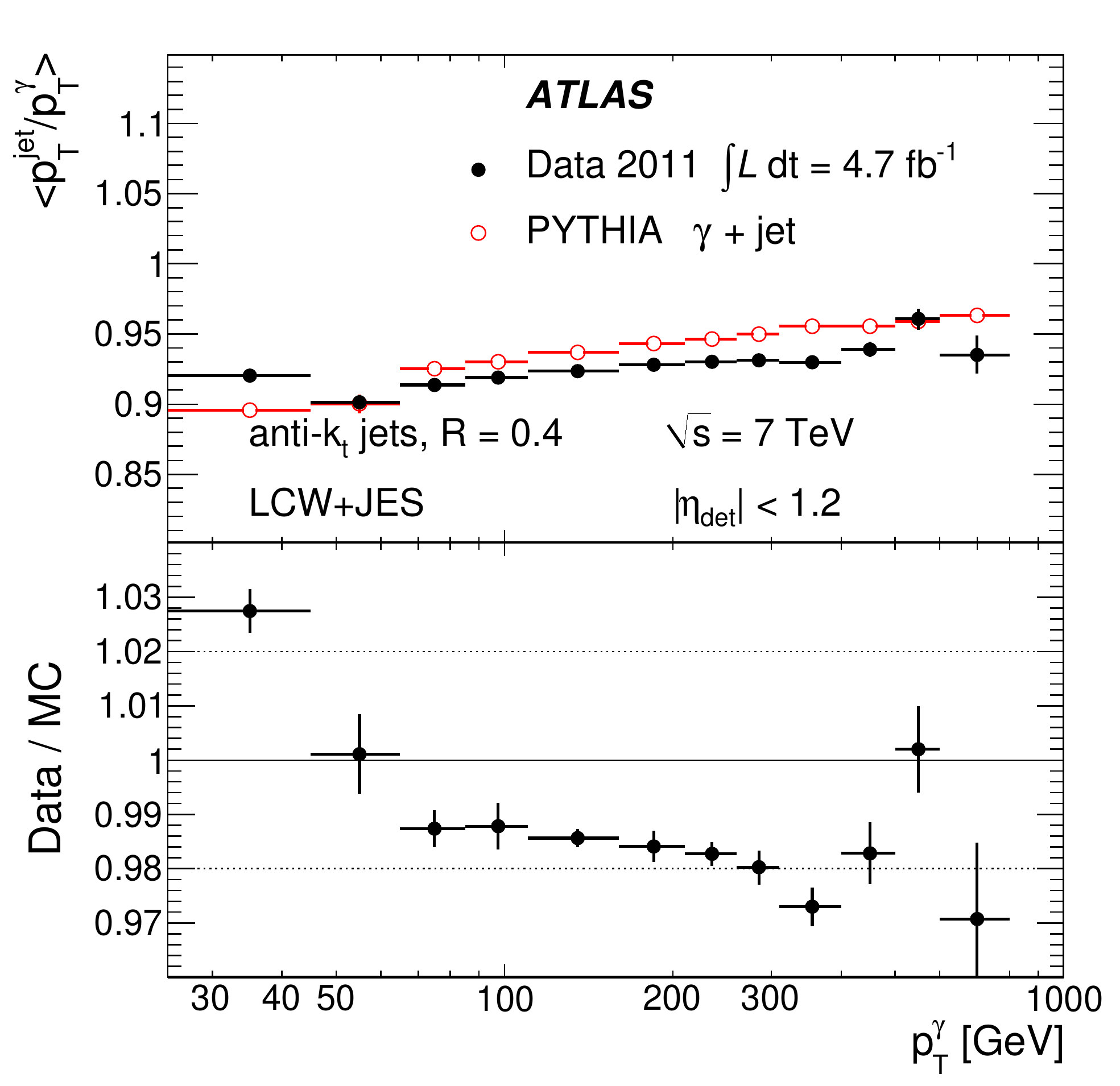}\label{fig:DirectBalance_DataMC_LCW_akt4}}\\
\subfloat[$R = 0.6$, \EMJES]
{\includegraphics[width=0.45\textwidth]{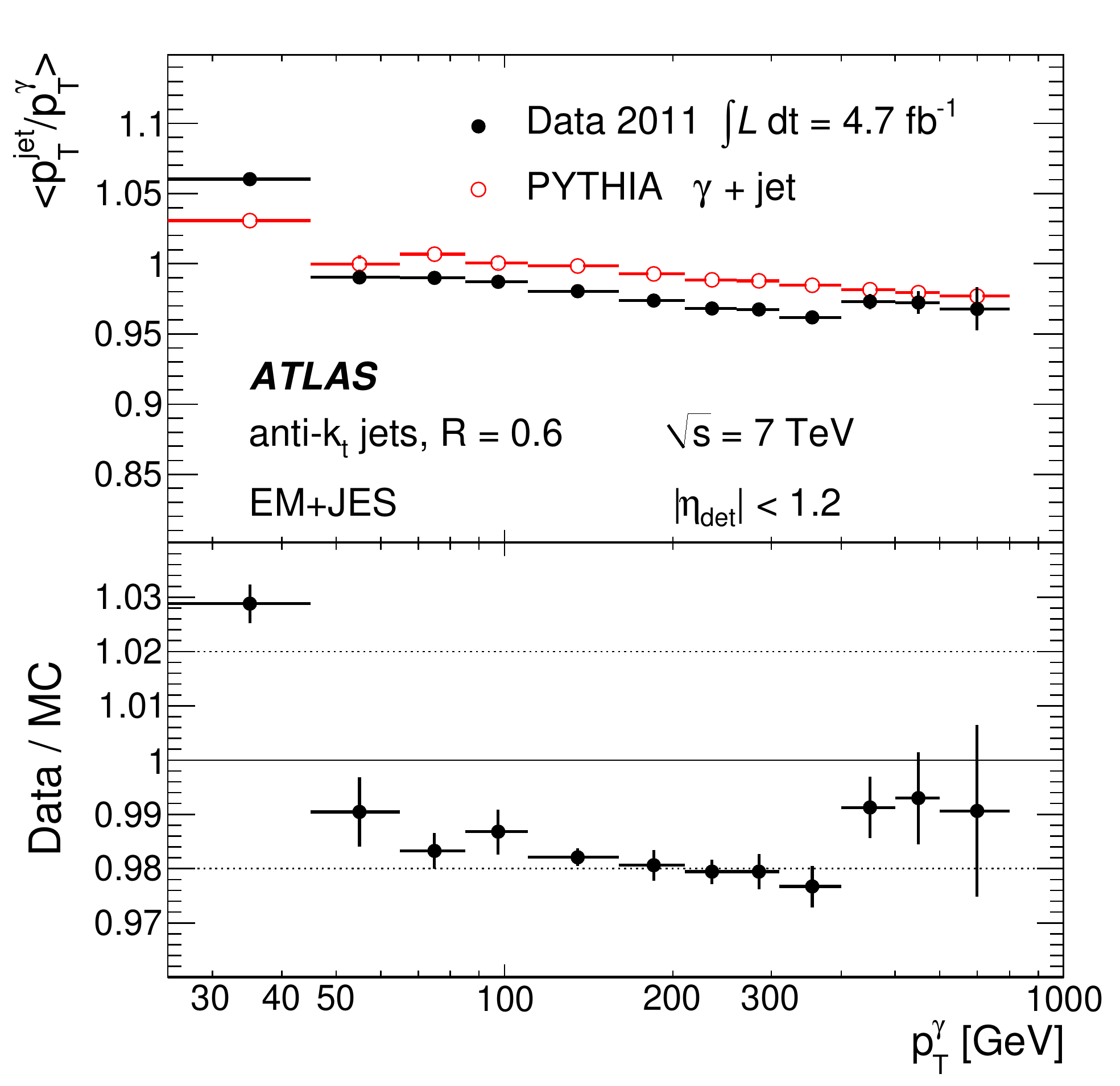}\label{fig:DirectBalance_DataMC_EM_akt6}}
\subfloat[$R = 0.6$, \LCWJES]
{\includegraphics[width=0.45\textwidth]{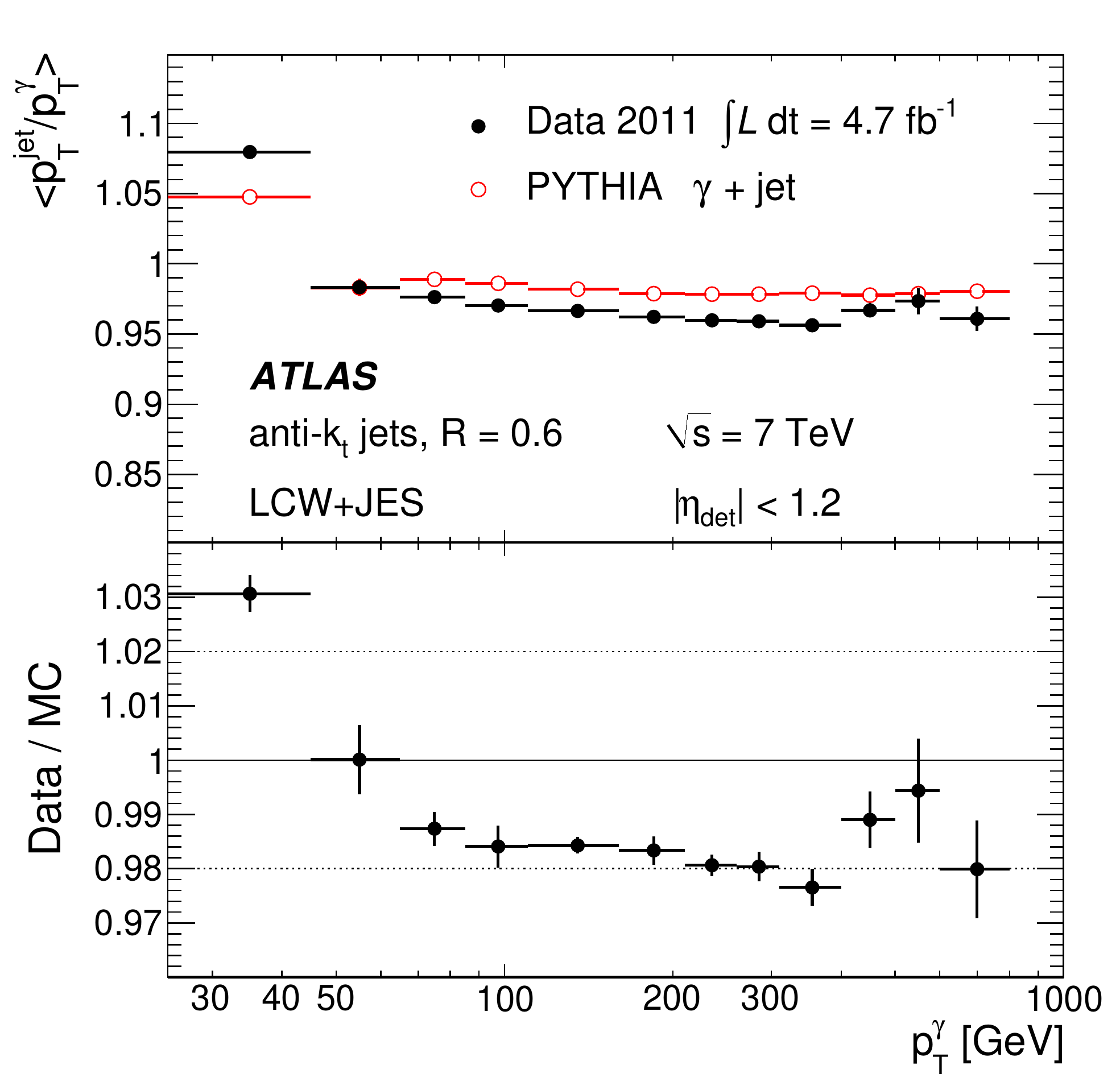}\label{fig:DirectBalance_DataMC_LCW_akt6}}
\end{center}
\caption[]{Average jet response as determined by the \DB{} technique 
in \gammajet{} events for \antikt{} jets with (\subref{fig:DirectBalance_DataMC_EM_akt4}, \subref{fig:DirectBalance_DataMC_LCW_akt4}) $R = 0.4$ 
and  (\subref{fig:DirectBalance_DataMC_EM_akt6}, \subref{fig:DirectBalance_DataMC_LCW_akt6}) $R = 0.6$, calibrated with the (\subref{fig:DirectBalance_DataMC_EM_akt4}, \subref{fig:DirectBalance_DataMC_EM_akt6}) \EMJES{} scheme and with the  (\subref{fig:DirectBalance_DataMC_LCW_akt4}, \subref{fig:DirectBalance_DataMC_LCW_akt6}) \LCWJES{} scheme, for both 
data and \MC{} simulations, as a function of the photon transverse momentum. 
}
\label{fig:DirectBalance_DataMC}
\end{figure*}

\section[Jet energy calibration using \gammajet{} events]{Jet energy calibration using \GAMMAJET{} events}
\label{sec:gammajetInSitu}
\subsection{\Insitu{} jet calibration techniques}
\label{sec:insitugammajetmethods}

Two \insitu{} techniques probing the calorimeter response to the jet balancing the photon 
are employed in this analysis:
\begin{mylist}
\myitem{Direct \ptbf{} balance (\DBbf)}
The transverse momentum of the jet with the highest \pt{} 
is compared to the transverse momentum of the reference photon (\ptgamma). The response is then 
computed as the ratio $\ptjet/\ptgamma$.

\myitem{Missing transverse momentum projection fraction \\(\MPFbf)}
The total hadronic recoil is
used to estimate the calorimeter response to jets. The hadronic recoil is reconstructed from the vectorial 
sum of the transverse projections of the energy deposits in the calorimeter projected onto the 
photon direction. As in the direct \pt{} balance, the photon \pt{} serves as reference. The \MPF{} 
response is defined as
\begin{displaymath}
\RMPF{} = 1 +
\frac{ \vec{p}_{\rm T}^{\gamma} \cdot \vecEtmiss }{|\ptgamma|^2},
\label{eq:JetResponse-MPF}
\end{displaymath}
where the \vecEtmiss{} is computed with \topos{} at the \EM{} or \LCW{} scales. A more detailed 
description of these two techniques can be found in Ref.~\cite{jespaper2010}.
\end{mylist}

Each technique has different sensitivities to additional soft-parton radiation, as well as to 
pile-up. The \MPF{} is in general less sensitive to additional particle activity that is symmetric 
in the transverse plane, like for example pile-up and the underlying event. 

The explicit use of jets in the jet response measurement from \DB{} makes this technique clearly
dependent on the jet reconstruction algorithm. Conversely, the dependence of the \MPF{} technique 
on the jet algorithm is relegated to a second-order effect.\footnote{Any dependence of 
the \MPF{} response on the jet reconstruction algorithm is introduced solely by the event selection.}
Thus, in the following, when presenting the results from the \MPF{} technique, no jet algorithm 
is in general explicitly mentioned.

\begin{table}[h!]
\renewcommand{\arraystretch}{\myarraystretch}
\caption{%
Summary table of the criteria to select \gammajet{} events.}
\begin{center}
\begin{tabular}{l|l}
\hline \hline
 Variable & Selection \\
\hline
$N_{\mathrm{tracks}}^{\mathrm{vertex}}$ & $> 4$ \\
\ptgamma{} & $> 25$~\GeV{} \\
$|\eta^{\gamma}|$ & $< 1.37$ \\
$\ptjet$ & $> 12$~\GeV{} \\
$|\eta^{\mathrm{jet}}|$ & $< 1.2$ \\
$E_{\mathrm{T}}^{\gamma~\mathrm{Iso}}$ & $< 3$~\GeV{} \\
\deltaphijetgamma{} & $> 2.9$ rad \\
$\ptsecondjet/\ptgamma$ & $< 0.2$ for \DB{} ($< 0.3$ for \MPF{}) \\
\hline \hline
\end{tabular}
\label{tab:EvtSel}
\end{center}
\end{table}

\begin{table}[h!]
\renewcommand{\arraystretch}{\myarraystretch}
\caption{Table with the approximate number of selected events in each \ptgamma{} bin.}
\begin{center}
\begin{tabular}{r@{--}l|r|r@{--}l|r}
\hline \hline
 \multicolumn{2}{c|}{\ptgamma{} [\GeV]} & Events & \multicolumn{2}{|c|}{\ptgamma{} [\GeV]} & Events \\
\hline 
$25$ & $45$   &  $20\,480$ & $210$ & $260$ & $10\,210$ \\
$45$ & $65$   &  $61\,220$ & $260$ & $310$ &  $4\,650$ \\
$65$ & $85$   & $125\,040$ & $310$ & $400$ &  $2\,770$ \\
$85$ & $110$  & $262\,220$ & $400$ & $500$ &    $800$ \\
$110$ & $160$ & $143\,180$ & $500$ & $600$ &    $240$ \\
$160$ & $210$ &  $32\,300$ & $600$ & $800$ &    $100$ \\
\hline \hline
\end{tabular}
\label{tab:NumberOfEvents}
\end{center}
\end{table}

\subsection[Event selection of \gammajet{} events]{Event selection of \GAMMAJET{} events}
\label{sec:gammajeteventselection}
The event selection used in this analysis is basically the same as that described in Ref.~\cite{jespaper2010} 
for the $2010$ analysis, except for changes that are either to adapt to the higher instantaneous luminosity 
of the $2011$ \ds{} or to the different detector conditions. The event selection proceeds as follows:
\begin{enumerate}

\item Events are required to have a primary vertex, as defined in \secRef{sec:trackjets}, with at least five associated tracks 
($N_{\mathrm{vertex}}^{\mathrm{tracks}} \geq 5$). 

\item There must be at least one reconstructed photon; the highest-\pt{} (leading) photon 
is taken as the hard-process photon and must have $\ptgamma > 25$~\GeV.

\item The event is required to pass a single-photon trigger, with trigger \pt{} threshold 
depending on the \pt{} of the leading photon. %

\item The leading photon must pass strict identification criteria \cite{photon-isolation}, meaning 
that the pattern of energy deposition in the calorimeter is consistent with the expected photon 
showering behaviour.

\item The leading photon must lie in the pseudorapidity range $|\eta^{\gamma}| < 1.37$, 
meaning it is fully contained within the electromagnetic barrel calorimeter.

\item Jets with high electromagnetic content (e.g., jets fluctuating to a leading $\pi^{0}$, with 
$\pi^{0} \rightarrow \gamma\gamma$) may be misidentified as photons. In order to reduce this 
background, the leading photon is required to be isolated from other activity in the calorimeter. The 
isolation variable ($E_{\mathrm{T}}^{\gamma~\mathrm{Iso}}$)~\cite{photon-isolation} is computed in a cone of size $R = 0.4$ around 
the photon, and corrected for pile-up energy inside the isolation cone. 
Only photons with $E_{\mathrm{T}}^{\gamma~\mathrm{Iso}} < 3$~\GeV{} are selected.

\item The photon reconstruction algorithm attempts to retain photons that have converted into an 
electron-positron pair. Whi\-le clusters without matching tracks are directly classified as ``unconverted'' 
photon candidates, clusters matched to pairs of tracks originating from reconstructed conversion vertices 
are considered as ``converted'' photon candidates (double-track conversions). To increase the reconstruction 
efficiency of converted photons, conversion candidates whe\-re only one of the two tracks is reconstructed 
(single-track conversions) are also retained. Jets that are misidentified as photons fall more often in the 
category of converted photons, because fake photons produce wider showers and have tracks associated to them. 
To suppress this background further, the ratio of the transverse energy of the photon candidate 
cluster to the scalar sum of the \pt{} of the matching tracks 
($E_{\mathrm{T}}^{\gamma~{\mathrm{cluster}}}/(\sum p_{\mathrm{T}}^{\mathrm{tracks}})$) is required 
to be in the range from $0$ to $2$ for single-track conversions, and from $0.5$ to $1.5$ for double-track 
conversions. The fraction of converted photons is $\sim 30\%$ throughout the $\ptgamma$ range under consideration. 

\item Only jets with $\ptjet > 12$~\GeV{} are considered. From those, only jets that pass quality 
criteria designed to reject fake jets originating from noise bursts in the calorimeters or from 
non-collision background or cosmics (see \secRef{sec:JetSel}), are used. After these jet selections, each event is required 
to have at least one jet. 

\item The highest-\pt{} (leading) jet must be in the region $|\eta^{\mathrm{jet}}| < 1.2$. 
This choice is motivated by the small \etaic{} correction below $1.5\%$ in this region.
\item To suppress soft radiation that would affect the \pt{} balance between the jet and the photon, 
the following two conditions are required: 
\begin{alphalist}
\item The leading jet must be back-to-back to the photon in the transverse plane ($\deltaphi{\mathrm{jet}}{\gamma} > 2.9$~rad).
\item The \pt{} of the \subleading{} 
jet from the hard process ($\ptsecondjet$) must be less than $20\%$ ($30\%$) of the \pt{} of the photon 
for \DB{} (\MPF{}\footnote{For \MPF{}, a less strict criterion can be used, since this technique is less 
sensitive to soft radiation.}). In order to distinguish jets from the hard process against jets 
from pile-up, the \subleading{} jet is defined as the highest-\pt{} jet from the subset of \nonleading{} 
jets that either have $\JVF > 0.75$ or for which \JVF{} could not be computed because they are 
outside the region covered by the tracking system. See \secRef{sec:dijetselection} for the explanation of \JVF.
\end{alphalist}

\item In the case of \DB{}, the event is rejected if either the leading jet or the \subleading{} 
jet falls in a region where, for a certain period, the read-out of the \EM{} calorimeter was not functioning. 
For \MPF{}, the condition is extended to all jets with $\ptjet > 20$~\GeV{} in the event. A similar 
condition is imposed on the reference photon.

\end{enumerate}
A summary of the event selection criteria is given in Table~\ref{tab:EvtSel}. 
Table~\ref{tab:NumberOfEvents} shows the approximate number of selected events per \ptgamma{} bin.

\subsection{Jet response measurement}
\label{sec:jetenergymeasurement}

The calorimeter response to jets is measured in bins of the photon transverse momentum. Distributions 
of the \MPF{} and the jet responses in the data are shown in Figs.~\ref{fig:MPF_Distributions} 
and~\ref{fig:DB_Distributions}, respectively, for $25 \leq \ptgamma{} < 45$~\GeV{} and for 
$160 \leq \ptgamma{} < 210$~\GeV{}. The distributions are fitted with a Gaussian function, except in 
the lowest \ptgamma{} bin for \DB{} where a Poisson distribution is used to address the issues introduced by the jet reconstruction \pt{} threshold, as discussed in \secRef{sec:balanceMeasurement}.
The mean values from the fits 
define the average \MPF{} and \DB{} jet responses for each \ptgamma{} bin. 
\FigRef{fig:MPF_DataMC} 
presents the results obtained in data and \MC{} simulations for \MPF{} when the \Etmiss{} is calculated from 
\topos{} at the \subref{fig:MPF_DataMCEM} \EM{} and \subref{fig:MPF_DataMCLCW} \LCW{} scales. \FigRef{fig:DirectBalance_DataMC} shows the results 
for \DB{} for \antikt{} jets with radius parameter $R = 0.4$ and $R = 0.6$ for the 
\EMJES{}  and \LCWJES{} calibration schemes. 

With increasing jet energies, the particles inside the jet get more energetic as well. Higher incident energies for hadrons in non-compensating
calorimeters, like the ones 
in \ATLAS, increase the amount of energy invested in intrinsically induced electromagnetic showers, thus leading to 
an increase of the calorimeter response \cite{Gabriel:1993ai}.   
This increase is clearly observed for \MPF{}, especially when 
\topos{} at the \EM{} scale are used as  for the observations shown in \figRef{fig:MPF_DataMC}\subref{fig:MPF_DataMCEM}. 
For \DB{}, the effect is masked, because the jets used are already calibrated. \DB{} is, in this case, measuring 
calibration residuals only.

Furthermore, a comparison of the \MPF{} responses at \EM{} scale in \figRef{fig:MPF_DataMC}\subref{fig:MPF_DataMCEM} and \LCW{} scale in \figRef{fig:MPF_DataMC}\subref{fig:MPF_DataMCLCW} shows the effect of having applied the \LCW{} calibration 
to the \topos. The response for jets built from \LCW{} \topos{} is much closer to unity, because the response differences between 
electromagnetic and hadronic particles in the jet are largely corrected by \LCW{} at the level of the \topos.

The lower part in Figs.~\ref{fig:MPF_DataMC} and~\ref{fig:DirectBalance_DataMC}
shows the ratio of the response in data to that in \MC{} simulations. 
The \MC{} simulation features a response that 
is $1\%$ to $2\%$ higher than that in data for $\ptgamma > 110$~\GeV{}. For lower values of 
\ptgamma{}, the \datatomc{} ratio tends to increase. Systematic studies have shown that 
the increase at low \pt{}
is due to the presence of contamination from multijet background events in the data, 
the different out-of-cone energy observed in data and in \MC{} simulations, and the different effect of the 
$12$~\GeV{} jet \pt{} reconstruction threshold (due to differences in the jet \pt{} spectrum) on the response 
in data and in \MC{} simulations.

\subsection{Systematic uncertainties of photon--jet balance}
\label{sec:systematicgammajetuncertainties}

The following sections briefly describe the procedure to estimate the systematic uncertainties 
of the \gammajet{} \insitu{} techniques. The dominant sources of systematic uncertainties, for 
$\ptgamma \lesssim 75$~\GeV{}, are the purity of the \gammajet{} data sample and for \DB{} also the 
out-of-cone correction (see \secRef{sec:gammajetkterm}) in the case of $R = 0.4$ jets. For 
$\ptgamma \gtrsim 75$~\GeV{}, the uncertainty on the energy scale of the photon dominates.

\subsubsection{Influence of pile-up interactions}
\label{sec:gammajetpileup}
The influence of in-time pile-up is evaluated by comparing the response in events with six or more 
reconstructed primary vertices ($\Npv \geq 6$) to the response in events with one or two reconstructed primary vertices, 
inclusively in $\axing$. Similarly, the effect of out-of-time pile-up is estimated comparing the response 
in events with $\axing > 7$ to the response in events with $3.5 < \axing < 5.5$, inclusively in \Npv{}. 
Since these two comparisons are highly correlated, the pile-up uncertainty is estimated in each \ptgamma{} 
bin as the maximum difference between the two high pile-up responses and the two low pile-up responses.
For \MPF{}, the uncertainty due to pile-up is typically about $\sim 0.5\%$ or smaller.

In the case of \DB{} however, the jet \pt{} is already corrected for the additional energy from 
pile-up interactions, as detailed in \secRef{sec:jetrecocalib}. 
The variations in the \datatomc{} response ratio obtained with the procedure explained above are found to be much 
smaller than other uncertainties on the measurement. They are also well contained within the variations 
obtained by propagating the uncertainty on the pile-up offset correction (see \secRef{sec:pileupinsitusystematics}).

\subsubsection{Soft-radiation suppression}
\label{sec:gammajetsoftradiation}

The stability of the \datatomc{} response ratio under soft radiation is evaluated in two steps. 
First, the cut on the \pt{} of the \subleading{} jet is varied, while keeping \deltaphi{\mathrm{jet}}{\gamma}{} 
fixed to its nominal cut value, and second, the cut on \deltaphi{\mathrm{jet}}{\gamma}{} is varied, with the cut 
on the \subleading{} jet fixed to its nominal value. The cut on the \subleading{} jet is varied 
to looser or tighter values as follows: 
\begin{enumerate}
\item Tight: 
\begin{displaymath}
\begin{array}{ll}
\ptsecondjet < \max\{ 10~\GeV, 0.2\times \ptgamma \}& \mathrm{for\ \MPF{},\ and}  \\
\ptsecondjet < \max\{ 10~\GeV, 0.1\times \ptgamma \}& \mathrm{for\ \DB{}.}
\end{array}
\end{displaymath}
\item Loose:
\begin{displaymath}
\begin{array}{ll} 
\ptsecondjet < \max\{ 12~\GeV, 0.3\times \ptgamma\} + 0.1\times \ptgamma & \mathrm{for\ \MPF{},\ and} \\ 
\ptsecondjet < \max\{ 12~\GeV, 0.2\times \ptgamma\} + 0.1\times \ptgamma & \mathrm{for\ \DB{}.} 
\end{array}
\end{displaymath}
\end{enumerate}
The typical variation on the \datatomc{} response ratio is of the order of $0.5\%$ for \DB{} and smaller for \MPF{}. 
Similar variations are observed when the \deltaphi{\mathrm{jet}}{\gamma}{} cut is relaxed to be $\deltaphi{\mathrm{jet}}{\gamma} > 2.8$ 
or tightened to be $\deltaphi{\mathrm{jet}}{\gamma} > 3.0$.
Other tests of the stability of the \datatomc{} response ratio under soft radiation are explored, 
such as relaxing and tightening the \deltaphi{\mathrm{jet}}{\gamma}{} and \ptsecondjet{} selection criteria at the same time, and lead to similar results.

\subsubsection{Background from jet events}
\label{sec:photonbackground}

The uncertainty on the response due to the presence of jets that are identified as photons (fakes) in the data
can be estimated, to first order, as $(1-P)\times(\Response_{\mathrm{dijet}} - \Response_{\gammajet})/\Response_{\gammajet}$, 
where $P$ is the purity of the \gammajet{} sample, and $\Response_{\gammajet}$ and $\Response_{\mathrm{dijet}}$ 
are the responses in signal and background events, respectively. 

The difference in response is estimated from \MC{} simulations as in the $2010$ analysis~\cite{jespaper2010}, using the 
nominal signal \pythia{} sample, and an inclusive jet \pythia{} sample (see \secRef{sec:MC}) enriched 
in events with narrow jets, which are more likely to be misidentified as photons. The comparisons indicate 
that the relative response differences are below $5\%$ for both 
techniques, which is taken as a 
conservative estimate. This is also confirmed by studying the response variation after relaxing the photon 
identification criterion.

The determination of the purity of the \gammajet{} data sample is done in the data using a sideband technique 
which is described in detail in Refs.~\cite{photon-isolation,jespaper2010}. 
The purity is about $60\%$ at $\ptgamma = 40$~\GeV{}, rises with \ptgamma{}, and becomes 
greater than $95\%$ for $\ptgamma \gtrsim 200$~\GeV{}. This purity is lower than that measured in the $2010$ 
analysis~\cite{jespaper2010}, due to the larger number of pile-up events in the 2011 data. The effect 
of pile-up is tested by measuring the purity under the same high and low pile-up conditions used 
to estimate the uncertainty on the response due to pile-up (see \secRef{sec:pileupsection}). Variations in the purity 
of the order of $5\%$ to $10\%$ are found. 
The systematic uncertainty on the purity measurement is not taken into account in the estimation of the uncertainty 
due to background events, because it becomes negligible when multiplied by the relative response 
difference between the signal and background events. 

The same purity estimate is used for \MPF{} and \DB{}, since both techniques have the same photon 
selection. The uncertainty due to background from jet events is $\sim 2.5\%$ at low \ptgamma{}, and decreases 
to about $0.1$\% towards high \ptgamma.

\subsubsection{Photon energy scale}
\label{sec:phes}

The electron energy is calibrated \insitu{} using the measurements of the \Zboson{} mass 
in $e^{-}e^{+}$ decays~\cite{Atlaselectronpaper}. The main sources of the electron energy scale uncertainty are the energy loss 
in the interactions with the material in front of the calorimeter and the leakage of energy 
transversely to the \topos{} axis. The calibration factors obtained from the $Z \rightarrow e^{-}e^{+}$ 
measurements are also applied to photons, with a
corresponding increase in the systematic uncertainty (the difference between the electron and 
the photon energy scales is caused mainly by the different interaction of electrons and photons 
with the material in front of the calorimeter). The photon calibration and its uncertainty are 
propagated to the jet response measurement, leading in both techniques to an uncertainty of 
approximately $+0.8\%$ and $-0.5\%$, independent of \ptgamma{}.

\subsubsection{Jet energy resolution}
\label{sec:gammjetjer}

The energy resolution for jets~\cite{jerpaper2010} in the \MC{} simulation is very close to the resolution 
observed in data. The uncertainty on the jet energy resolution measurement in data is propagated as 
an uncertainty in the response in \MC{} simulations. This is done as described in \secRef{sec:etaic_jer_uncertainty} and \eqRef{eq:jer_smearing} therein. 
The 
observed difference in response between the varied and the nominal results is defined as the 
systematic uncertainty due to jet energy resolution.

\subsubsection{Monte Carlo generator}
\label{sec:gammajetMCgenerator}

Uncertainties due to different modelling of the parton shower, jet fragmentation and multiple parton 
interactions affecting the \pt{} balance between the photon and the jet, can be estimated using 
different \MC{} generators which implement different models. The jet response derived with \pythia{} 
is compared to the response derived using \herwig{}. The results are shown in 
\figRef{fig:DB_MPF_PythiaVsHerwig}. The central value for the jet response in \MC{} simulations is taken from
\pythia, since this is the generator used to derive the JES corrections, and the observed full 
difference between \pythia{} and \herwig{} is taken as a (symmetric) systematic uncertainty. 
The difference in the responses between \herwig{} and \pythia{} is 
maximally about $ 1\%$.

\begin{figure*}[ht!p]
\begin{center}
\subfloat[\MPF{}, \EM]
{\includegraphics[width=0.495\textwidth]{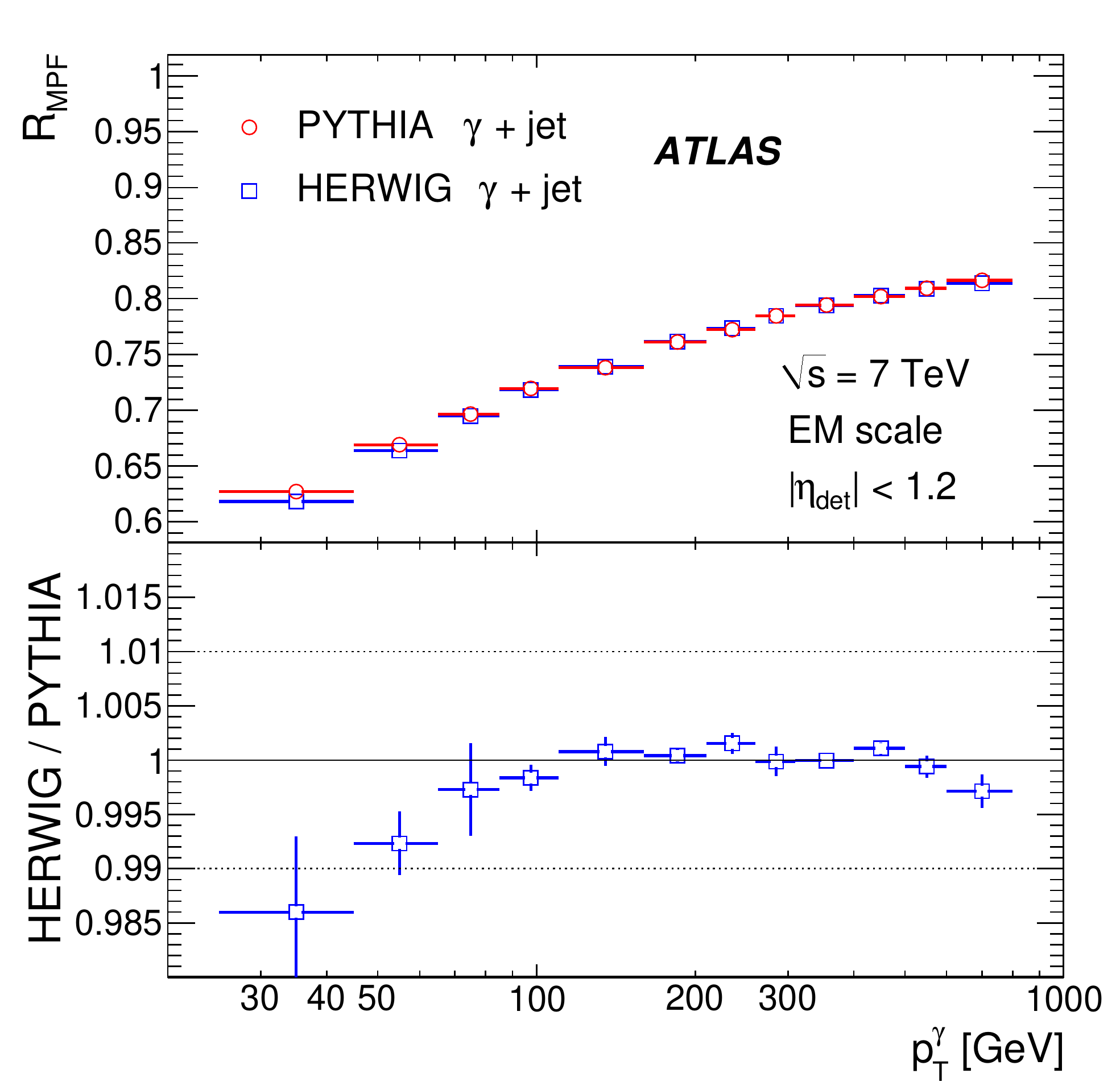}\label{fig:DB_MPF_PythiaVsHerwig_EM}}
\subfloat[\DB{}, $R = 0.4$, \EMJES]
{\includegraphics[width=0.495\textwidth]{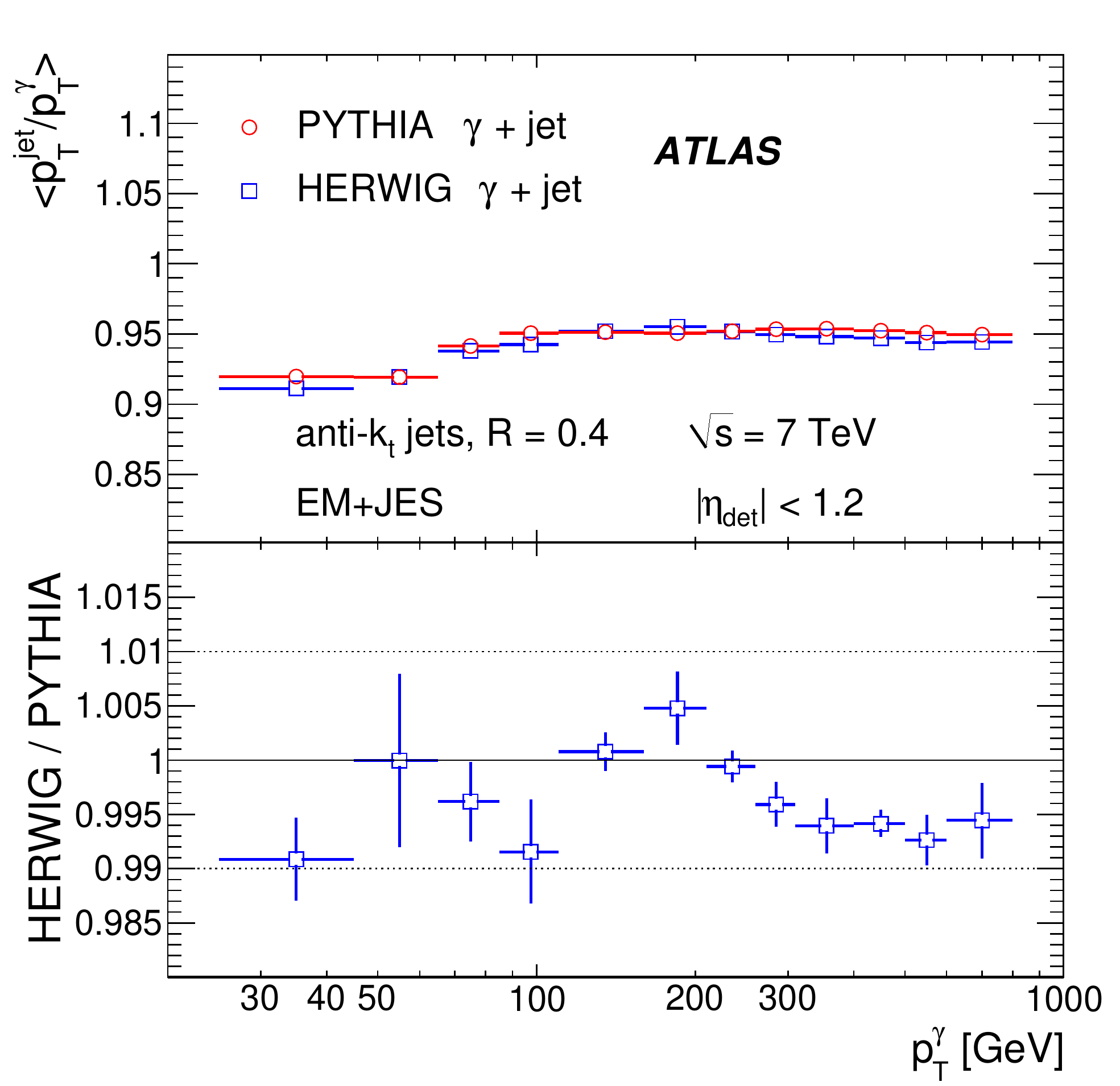}\label{fig:DB_MPF_PythiaVsHerwig_EMJES}}
\end{center}
\caption[]{Average jet response as determined by the  \subref{fig:DB_MPF_PythiaVsHerwig_EM} \MPF{} and \subref{fig:DB_MPF_PythiaVsHerwig_EMJES}  \DB{} techniques, using \antikt{} 
jets with $R = 0.4$ at the \EM{} and \EMJES{} energy scales respectively, for \pythia{} (circles) and \herwig{}
(squares) \MC{} simulations, as a function of the photon transverse momentum. 
The HERWIG-to-PYTHIA response ratio is shown in the bottom inset of each figure. Only the statistical uncertainties are shown.}
\label{fig:DB_MPF_PythiaVsHerwig}
\end{figure*}

\subsubsection{Out-of-cone radiation and underlying event}
\label{sec:gammajetkterm}

Even in a $2 \to 2$ \gammajet{} event, where the outgoing photon and parton (quark or gluon) 
perfectly balance each other in transverse momentum, 
the transverse momentum of the photon is 
only approximately equal to the transverse momentum of the truth jet, formed as described in \secRef{sec:truthjets}, originating 
from the parton. The two main reasons for this are the same already described for the \Zjet{} events in \secRef{sec:outofcone}, namely the fact that the jet does not capture all particles recoiling from the photon, and the contribution to the jet from the underlying event. 
The amount of momentum carried by particles outside the jet and by particles coming from soft 
interactions not contributing to the \pt{} balance needs to be compared in data and \MC{} simulation. 

When averaging over many events, particles not associated to the hard 
scattering are distributed isotropically, and therefore they do not 
contribute to the hadronic recoil vector constructed in the \MPF{} method. Thus, their contribution to the \MPF{} response is zero. 
This is also supported by studies in the \MC{} simulation using the particles produced by the 
underlying event model.
Moreover, in the \MPF{} technique the photon 
is balanced against the full hadronic recoil, not only against the leading jet.
For the \DB{} method the out-of-cone radiation is computed as explained in \secRef{sec:outofcone}.

The measured \kooc{} factor (\eqRef{eq:kterm}) is shown as a function of \ptgamma{} in \figRef{fig:gammajetkterm} for \antikt{} 
jets with $R=0.4$ (\figRef{fig:gammajetkterm}\subref{fig:gammajetkterm_akt4}) and with $R=0.6$ (\figRef{fig:gammajetkterm}\subref{fig:gammajetkterm_akt6}), for both data and \MC{} simulations. 
Systematic uncertainties obtained by varying the parameters in the \kooc{} factor definition are added 
in quadrature to the statistical uncertainties. The \kooc{} varies from $0.92$ 
($0.97$) at low \pt{} to $0.99$ ($1.01$) at high \pt{} for $R = 0.4(0.6)$, respectively. 
The data are described by the \MC{} simulation within $1\%$ to $2\%$ at low \pt{}. 
This deviation is taken as a systematic uncertainty in the \DB{} technique.

\begin{figure*}[ht!p]
\begin{center}
\subfloat[$R=0.4$]
{\includegraphics[width=0.495\textwidth]{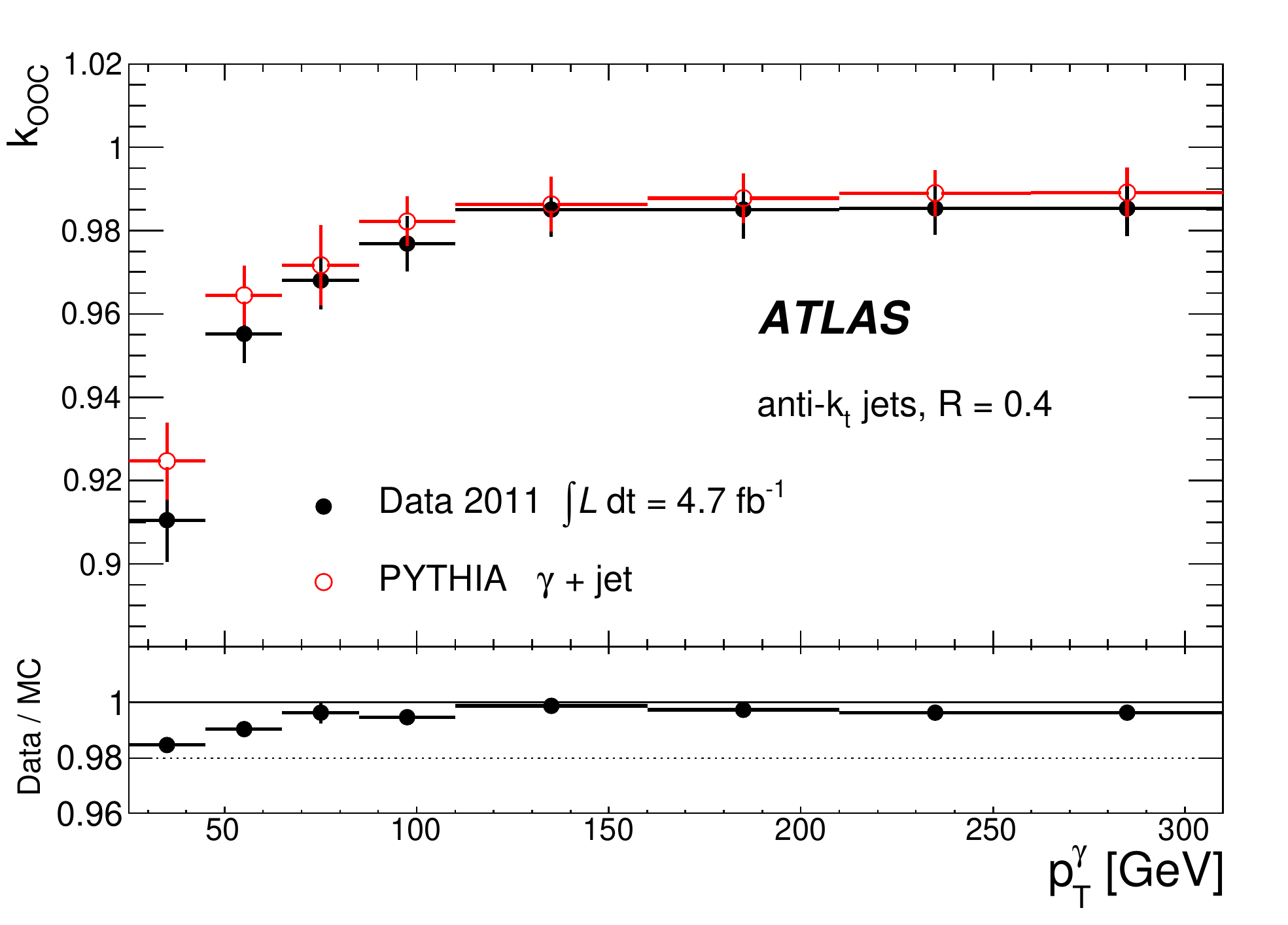}\label{fig:gammajetkterm_akt4}}
\subfloat[$R=0.6$]
{\includegraphics[width=0.495\textwidth]{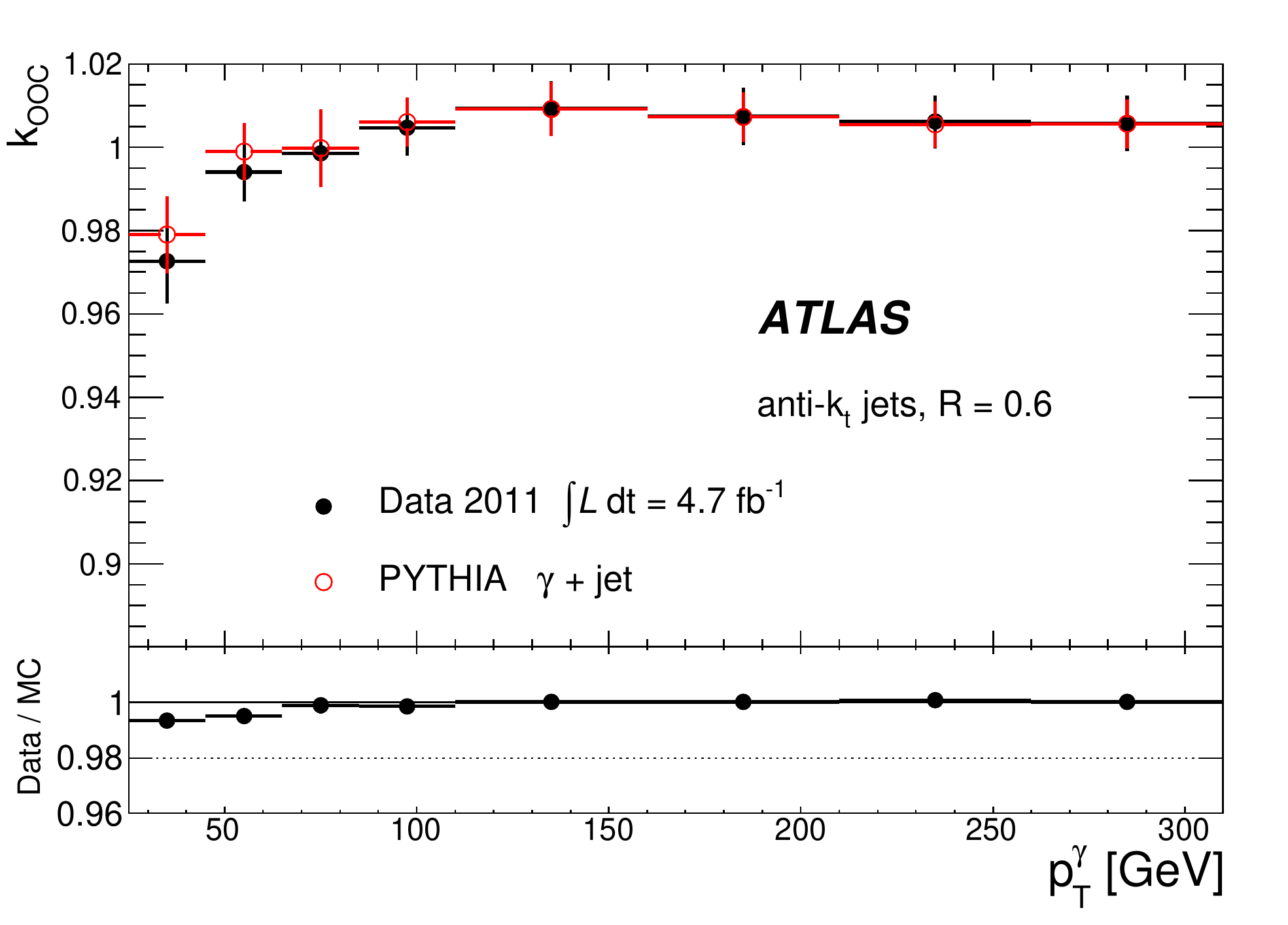}\label{fig:gammajetkterm_akt6}}
\end{center}
\caption[]{%
Out-of-cone radiation factor \kooc{} 
relating the \pt{} of the photon with the \pt{} of the truth jet as a function of the 
photon transverse momentum, measured using charged particles,
for \antikt{} jets with \subref{fig:gammajetkterm_akt4}  $R=0.4$ and \subref{fig:gammajetkterm_akt6} $R=0.6$, in data and in \MC{} simulations. 
The \datatomc{} response ratio is shown in the bottom inset of each plot. Statistical and systematic 
uncertainties are added in quadrature.}
\label{fig:gammajetkterm}
\end{figure*}

\subsubsection{Summary of systematic uncertainties}

A summary of the systematic uncertainties for the \MPF{} and the \DB{} techniques as a function 
of the photon \pt{} are presented in Figs.~\ref{fig:MPF_Systematics} and~\ref{fig:DirectBalance_Systematics}, 
respectively. The systematic uncertainties are shown for jets calibrated with the \EM{} and \LCW{} schemes for \MPF{}, and 
with the \EMJES{} and \LCWJES{} schemes for \DB{} where also jets with $R=0.4$ 
and $R=0.6$ are considered. The figures also show the statistical uncertainty, and the total uncertainty, 
which corresponds to the quadratic sum of all individual components (statistical and systematic).
Table~\ref{tab:SystSummary} shows the components of the systematic uncertainty for both methods 
in two representative $\pt^{\gamma}$ bins.

For the \DB{} technique, the total uncertainty is as large as $2\%$ to $3\%$ at very low and very high \pt{} values, 
and it is around $0.9\%$ in the \pt{} range from $100$~\GeV{} to $500$~\GeV{}. The uncertainties are smaller 
for \MPF{}; the total uncertainty is $\sim 0.7\%$ in the range $100$~\GeV{} to $500$~\GeV{} and it is 
dominated by the photon energy scale uncertainty. 

\begin{table}[ht!p]
\caption{Systematic uncertainties on the \datatomc{} ratio of the jet response on the \EM{} scale for both \DB{} and \MPF{} 
in two representative $\pt^{\gamma}$ bins. 
}
\renewcommand{\arraystretch}{\myarraystretch}
\begin{center}
\begin{tabular}{l|cc|cc}
\hline \hline
  & \multicolumn{2}{|c}{\DB{}, $R=0.6$ [\%]} & \multicolumn{2}{|c}{\MPF{} [\%]} \\
\hline
$\pt^{\gamma}$ range [\GeV] & $45 - 65$ & $310 - 400$ & $45 - 65$ & $310 - 400$ \\
\hline
\emph{Event} & & & &  \\
Pile-up                                 & $   --   $        & $   --   $      & $\pm 0.21$       & $\pm 0.16$ \\ \hline
\emph{Radiation} & & & &  \\ 
$\ptsecondjet$              & $\pm 0.43$        & $\pm 0.28$      & $\pm 0.09$       & $\pm 0.10$ \\
\deltaphi{\mathrm{jet}}{\gamma}         & $\pm 0.35$        & $\pm 0.20$      & $\pm 0.03$       & $\pm 0.03$ \\ \hline
\emph{Photon} & & & & \\
Purity                           & $\pm 1.18$        & $\pm 0.15$      & $\pm 1.18$       & $\pm 0.15$ \\
Energy                           & $\pm 0.46$        & $\pm 0.71$      & $\pm 0.57$       & $\pm 0.61$ \\ \hline
\emph{Jet} & & & & \\ 
\JER                          & $\pm 0.01$        & $\pm 0.11$      & $\pm 0.04$       & $\pm 0.01$ \\
Out-of-cone                             & $\pm 0.60$        & $\pm 0.00$      & $   --   $       & $   --   $ \\ \hline
\emph{Modelling} & & & & \\
\MC{} generator                            & $\pm 0.48$        & $\pm 0.44$      & $\pm 0.38$       & $\pm 0.00$ \\
\hline\hline
\end{tabular}
\label{tab:SystSummary}
\end{center}
\end{table}

\subsection[Summary of the \gammajet{} analysis]{Summary of the \GAMMAJET{} %
analysis}
\label{sec:gammajetsummary}

The average jet response in events with an isolated photon and a jet at high transverse momentum 
is computed using the $2011$ \ds, and compared to the average jet response obtained using \MC{} simulations. 
Two different %
techniques are used, the direct \pt{} balance and the missing-\pt{} projection fraction methods. 
Both techniques are highly correlated and show consistent results within systematic uncertainties. The \datatomc{} 
response ratio is close to $98\%$ for $\ptgamma > 85$~\GeV{}. Systematic uncertainties are evaluated for both 
methods to be of the order of $1\%$ or smaller in most of the \ptgamma{} range under consideration.

\begin{figure*}[ht!p]
\begin{center}
\subfloat[\EM]
{\includegraphics[width=0.45\textwidth]{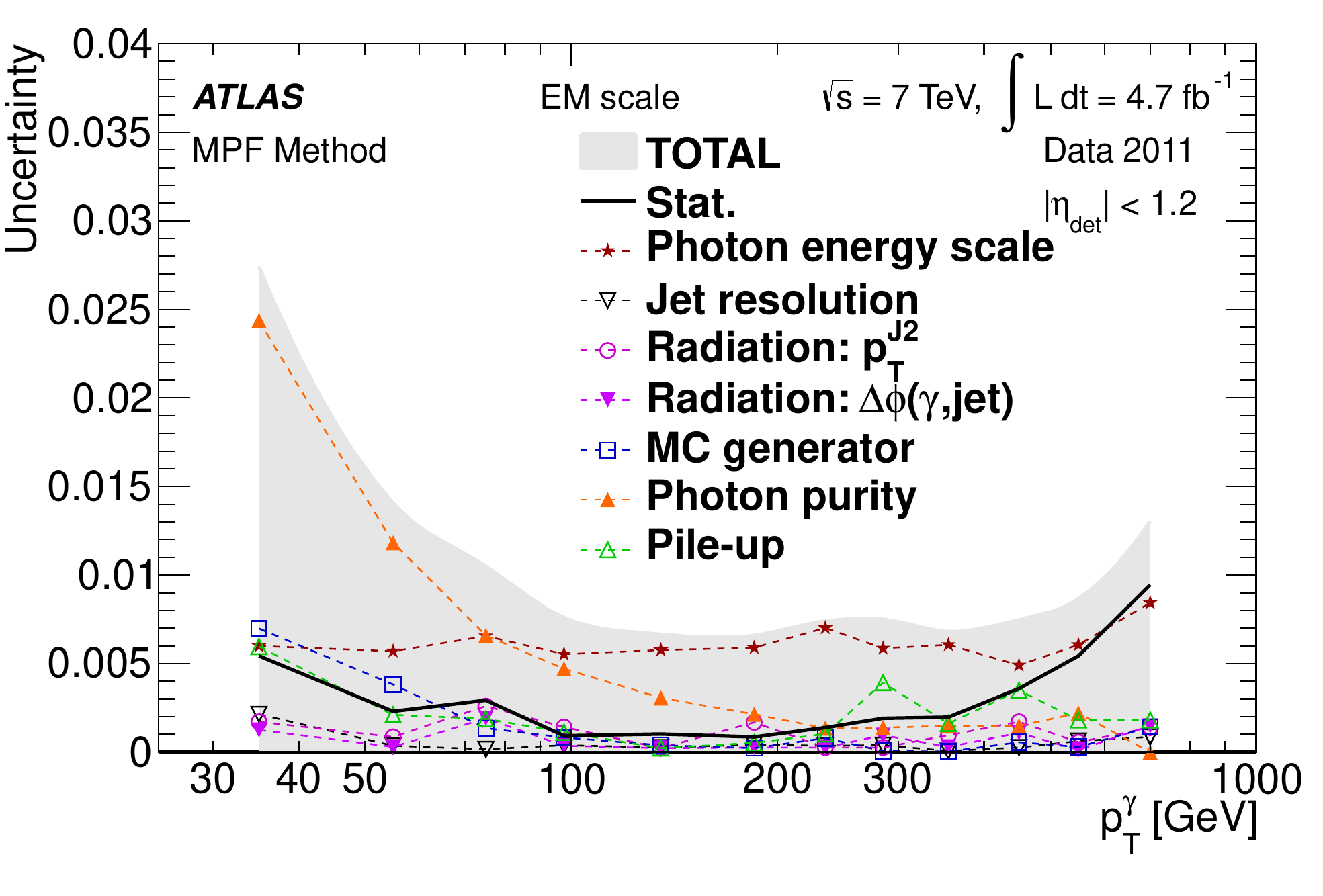}\label{fig:MPF_systematicsEM}}
\subfloat[\LCW]
{\includegraphics[width=0.45\textwidth]{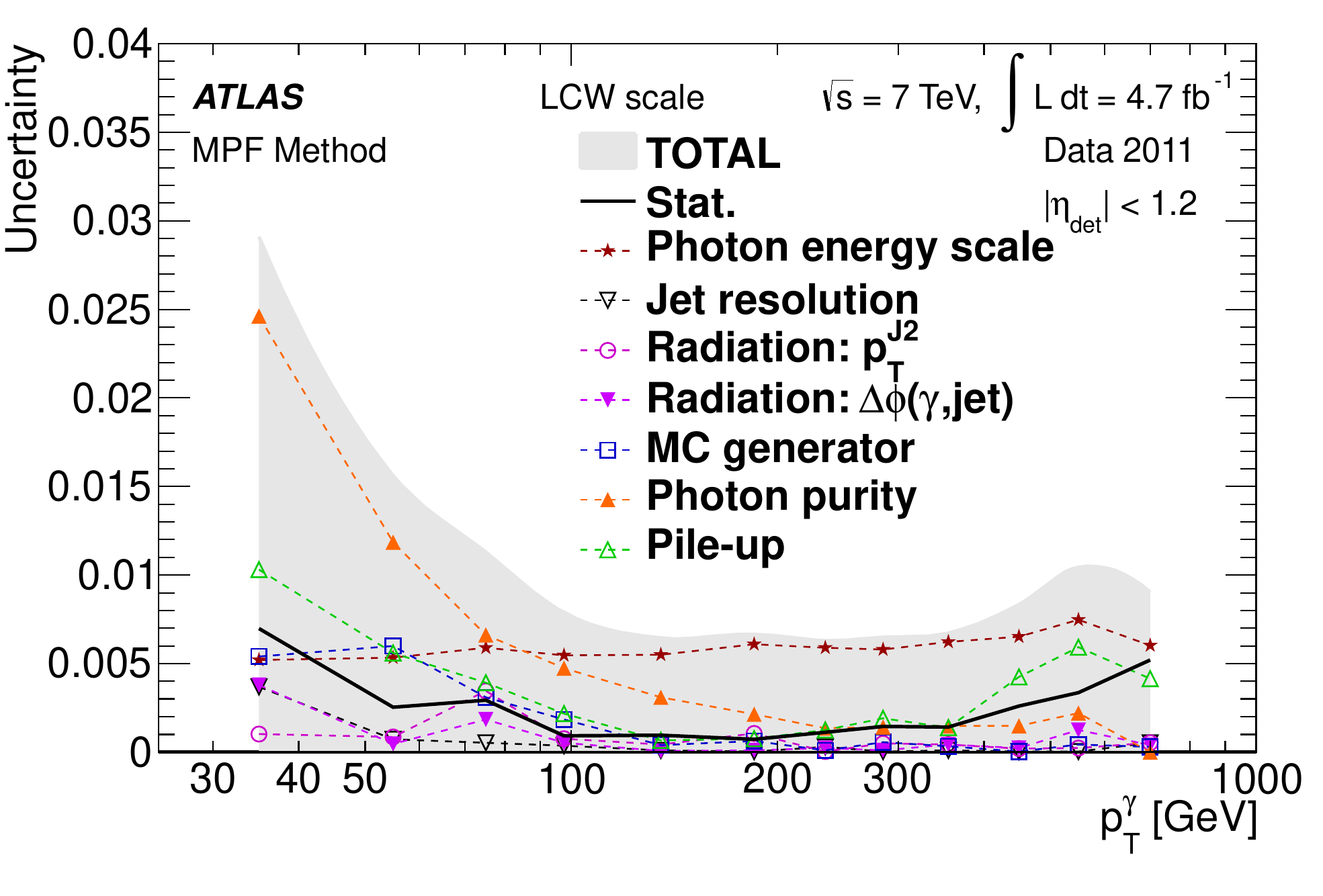}\label{fig:MPF_systematicsLCW}}
\end{center}
\caption[]{Systematic uncertainties on the \datatomc{} ratio of the jet response, as determined 
by the \MPF{} technique for \gammajet{} events 
using \topos{} at the \subref{fig:MPF_systematicsEM} \EM{} and  \subref{fig:MPF_systematicsLCW} \LCW{} energy scales, as a function of 
the photon transverse momentum.} 
\label{fig:MPF_Systematics}
\end{figure*}

\begin{figure*}[ht!p]
\begin{center} 
\subfloat[$R = 0.4$, \EMJES]
{\includegraphics[width=0.45\textwidth]{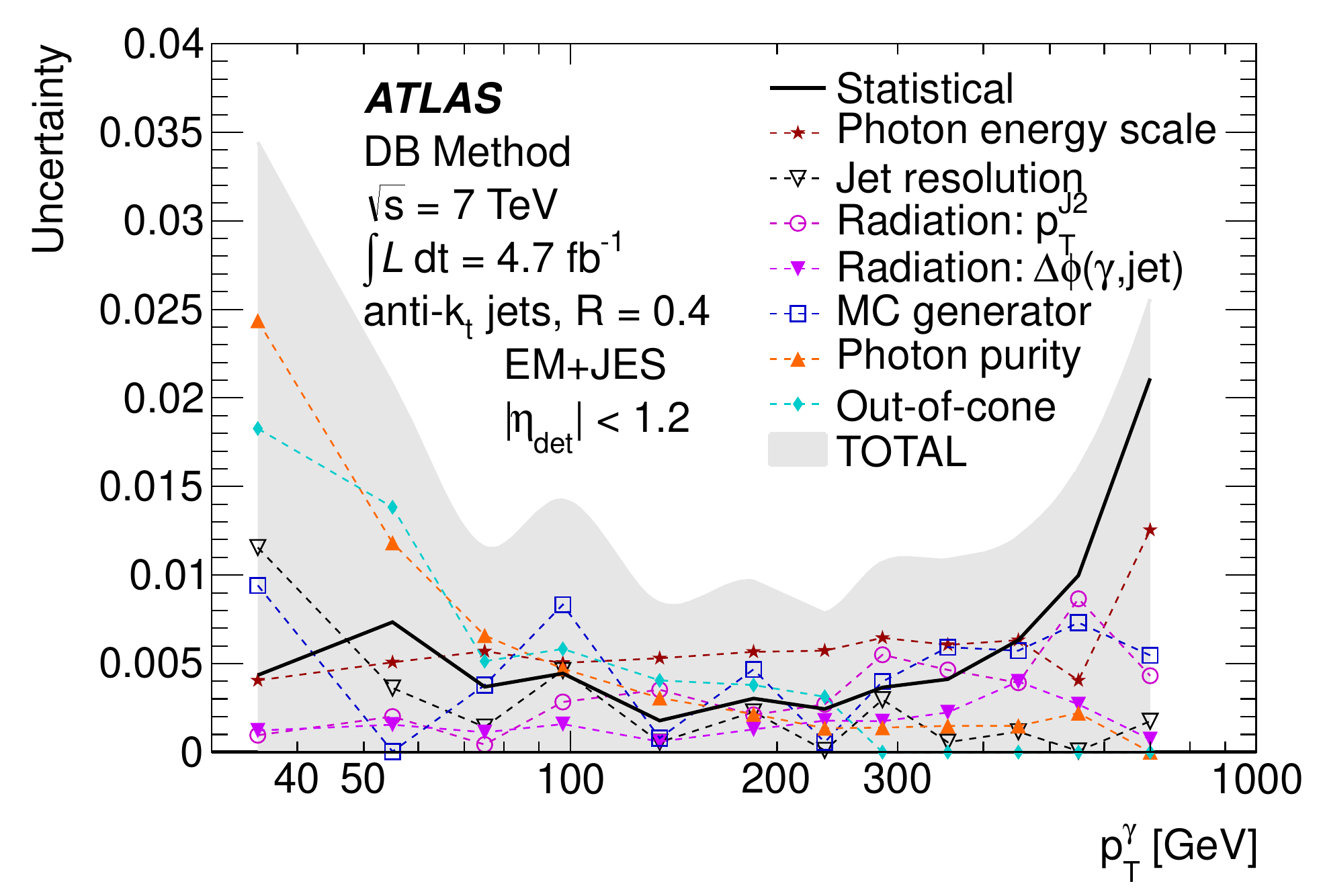}\label{fig:DirectBalanceSystematicsEMNarrow}}
\subfloat[$R = 0.4$, \LCWJES]
{\includegraphics[width=0.45\textwidth]{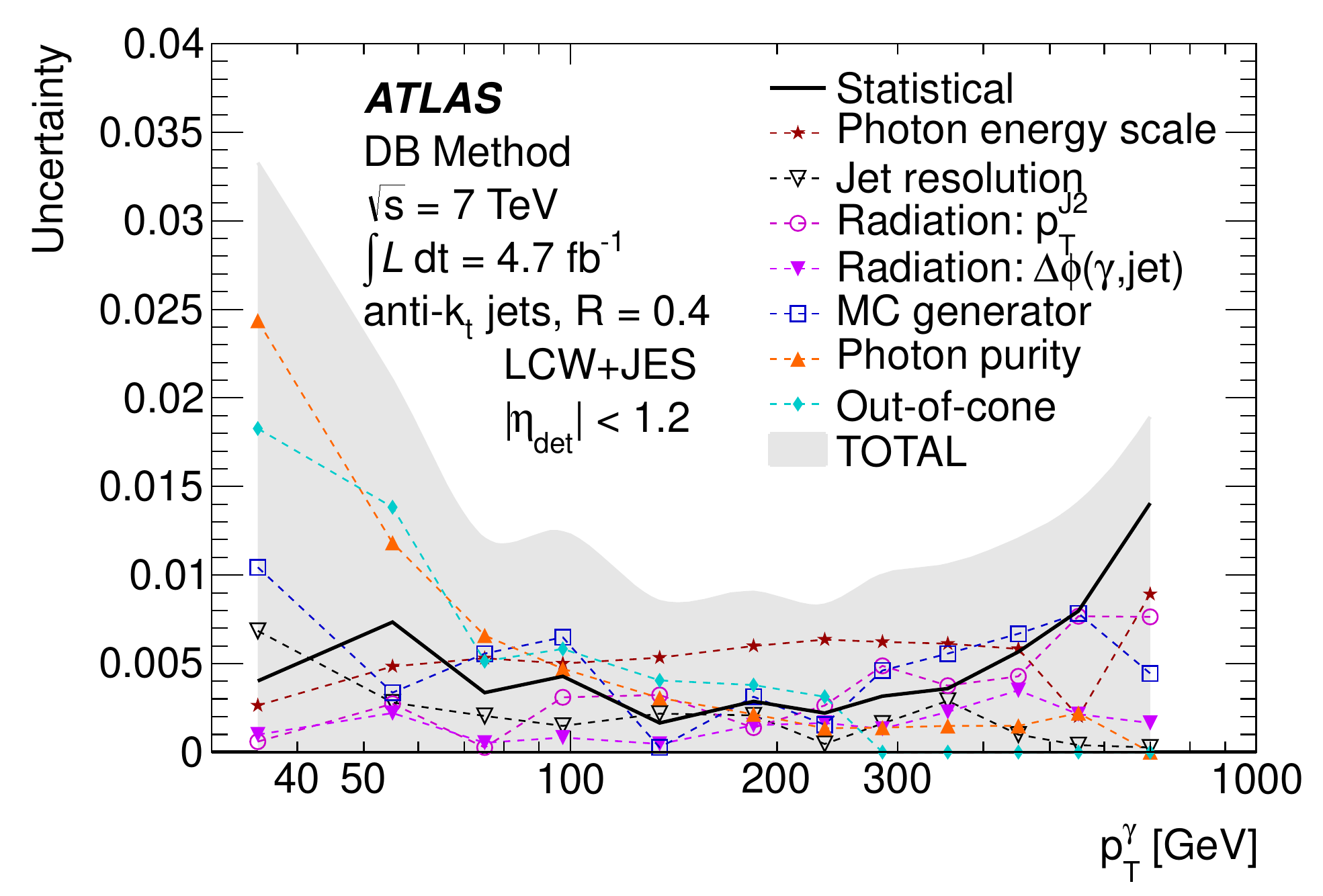}\label{fig:DirectBalanceSystematicsLCWNarrow}}\\
\subfloat[$R = 0.6$, \EMJES]
{\includegraphics[width=0.45\textwidth]{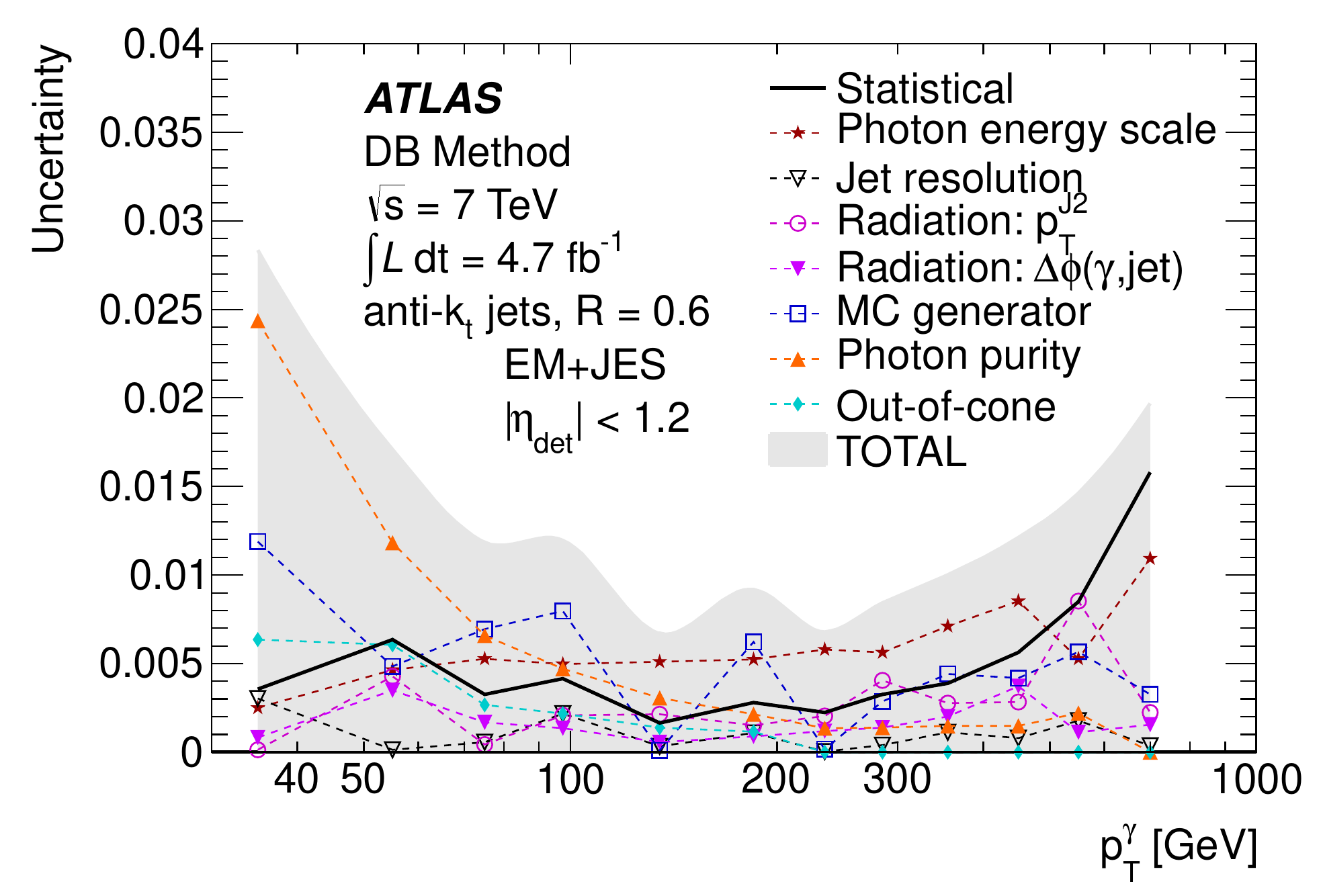}\label{fig:DirectBalanceSystematicsEMWide}}
\subfloat[$R = 0.6$, \LCWJES]
{\includegraphics[width=0.45\textwidth]{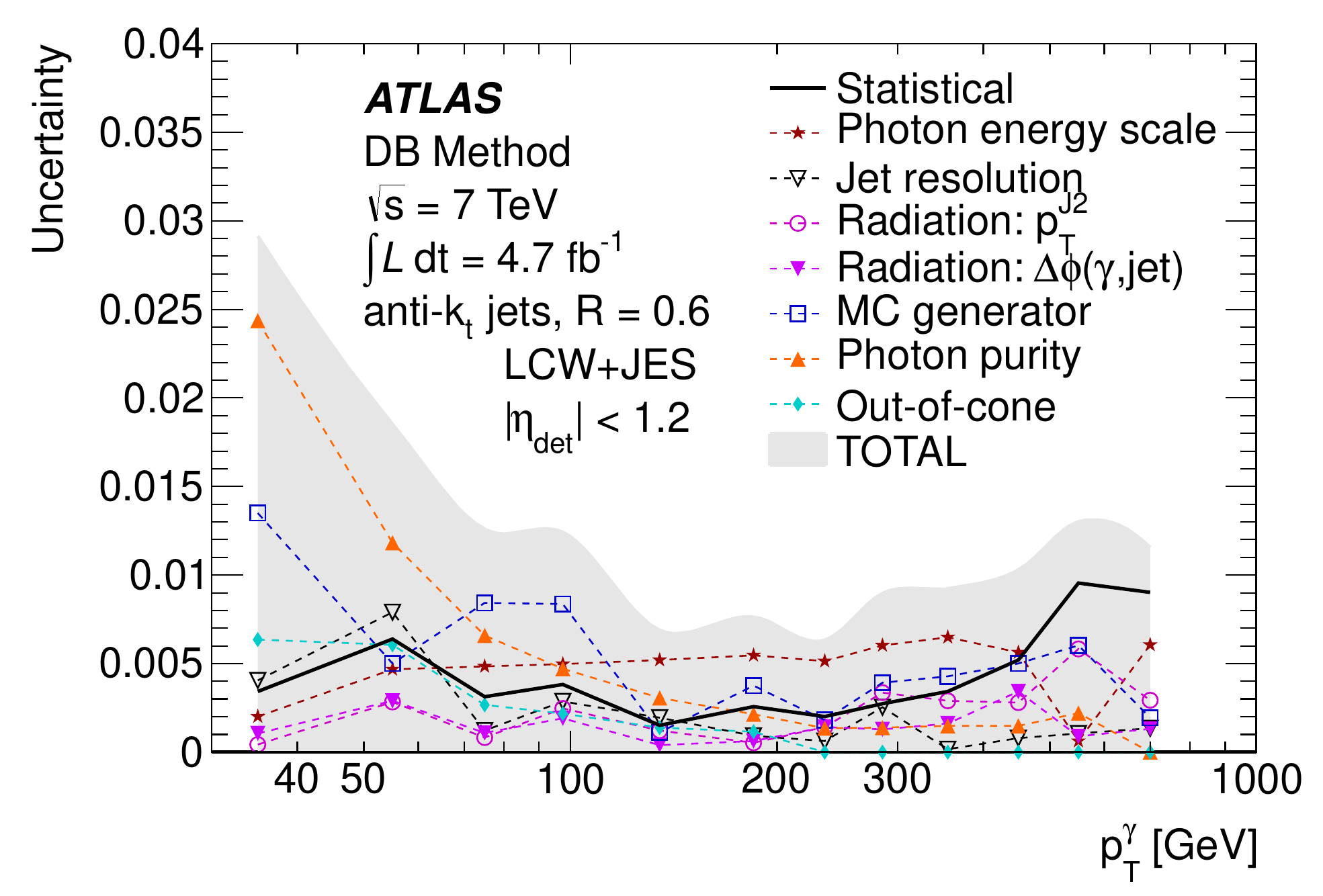}\label{fig:DirectBalanceSystematicsLCWWide}}
\end{center}
\caption[]{Systematic uncertainties on the \datatomc{} ratio of the jet response, as determined by the \DB{} technique in \gammajet{} events, for \antikt{} jets with (\subref{fig:DirectBalanceSystematicsEMNarrow},\subref{fig:DirectBalanceSystematicsLCWNarrow}) $R = 0.4$ and  (\subref{fig:DirectBalanceSystematicsEMWide}, \subref{fig:DirectBalanceSystematicsLCWWide}) $R = 0.6$, calibrated 
with the (\subref{fig:DirectBalanceSystematicsEMNarrow}, \subref{fig:DirectBalanceSystematicsEMWide}) \EMJES{} scheme and with the  (\subref{fig:DirectBalanceSystematicsLCWNarrow}, \subref{fig:DirectBalanceSystematicsLCWWide}) \LCWJES{} scheme, as a function of the photon 
transverse momentum.}
\label{fig:DirectBalance_Systematics}
\end{figure*}


\begin{figure*}[htp!]
 \begin{center}
    \subfloat[$R = 0.4$, \EMJES]
    {\includegraphics[width=0.45\textwidth]{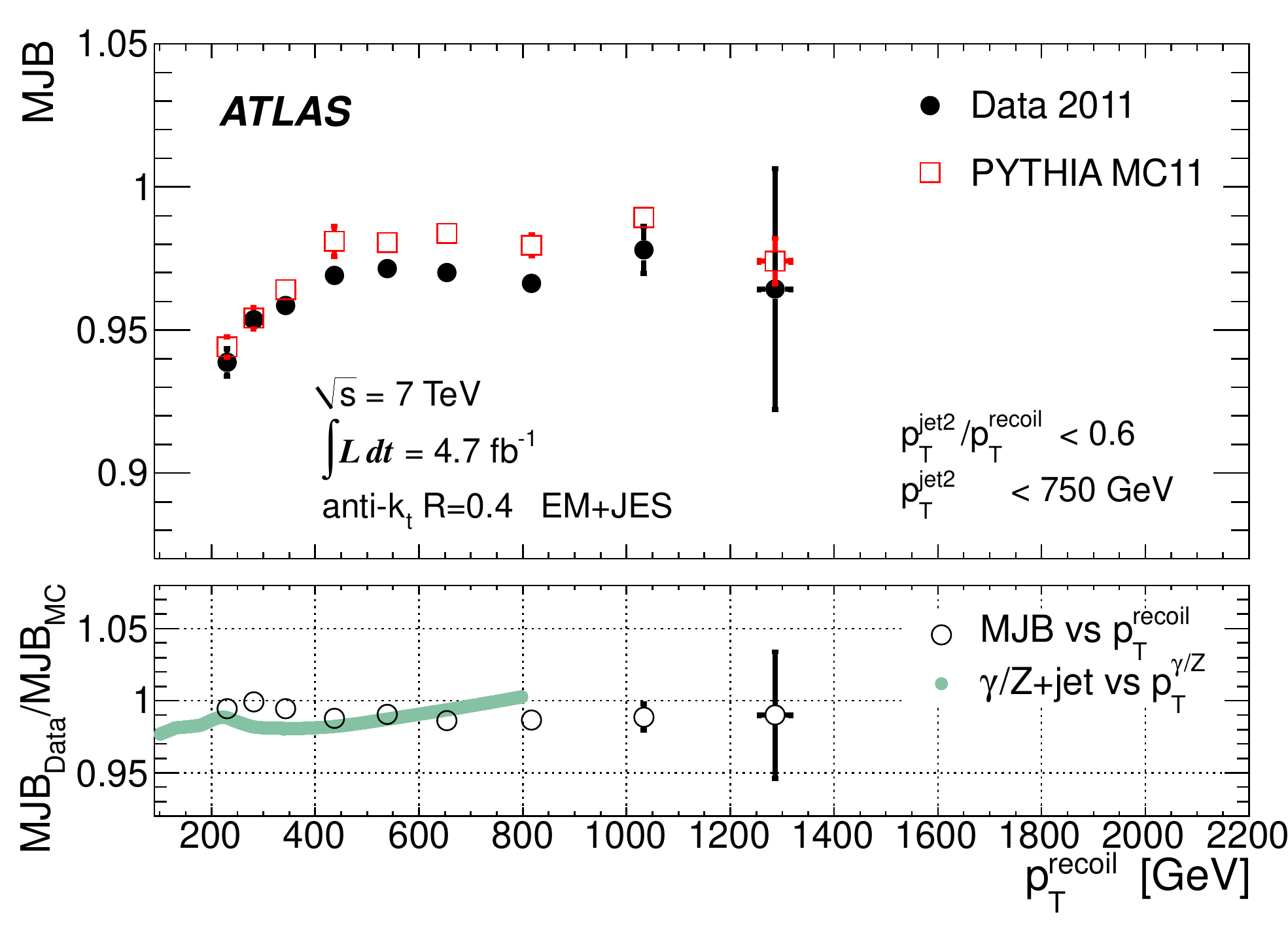}\label{fig:PtBalance_MJB_EMJES_akt4}}
    \subfloat[$R = 0.4$, \LCWJES]
    {\includegraphics[width=0.45\textwidth]{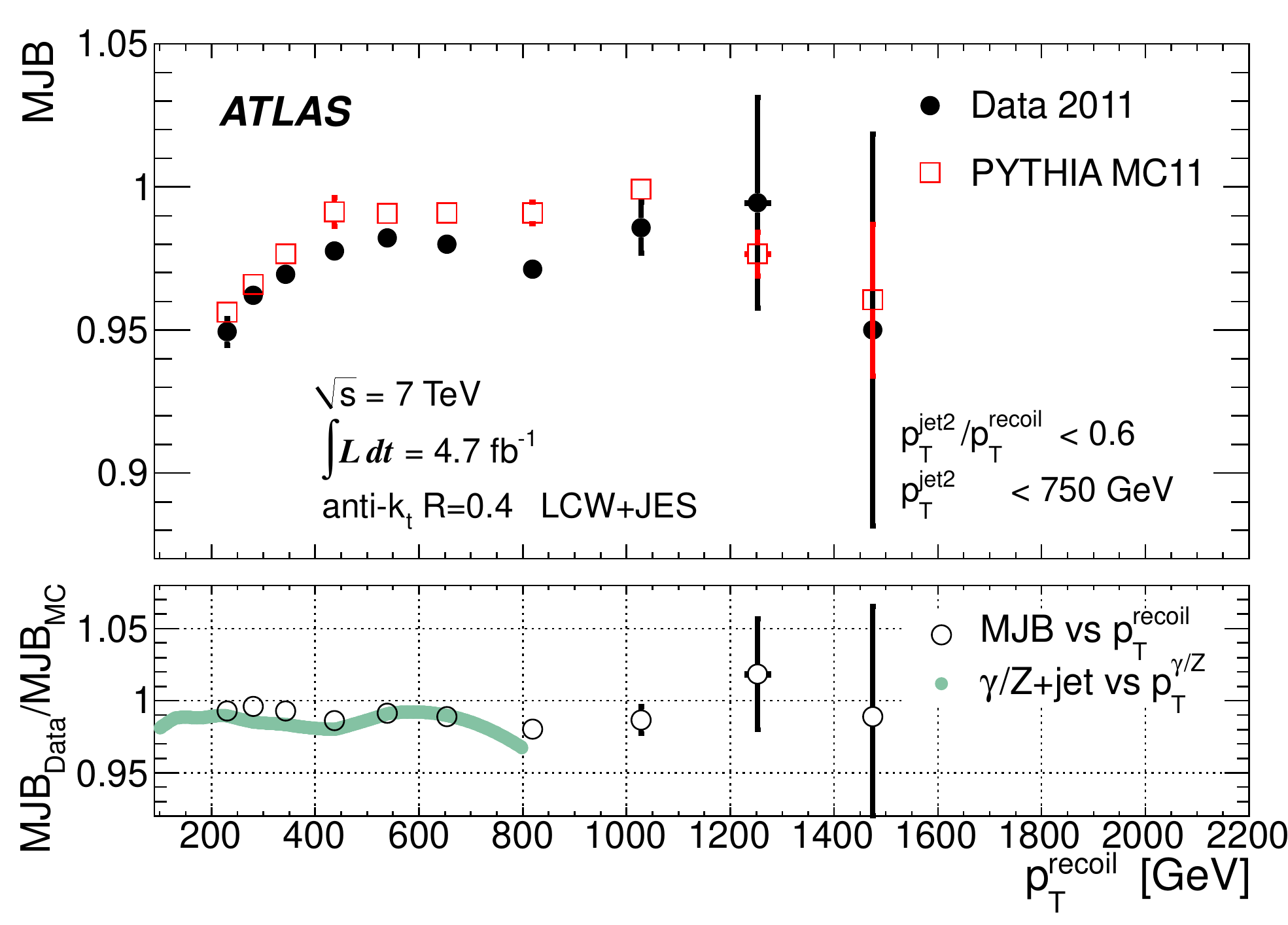}\label{fig:PtBalance_MJB_LCWJES_akt4}}\\
    \subfloat[$R = 0.6$, \EMJES]
    {\includegraphics[width=0.45\textwidth]{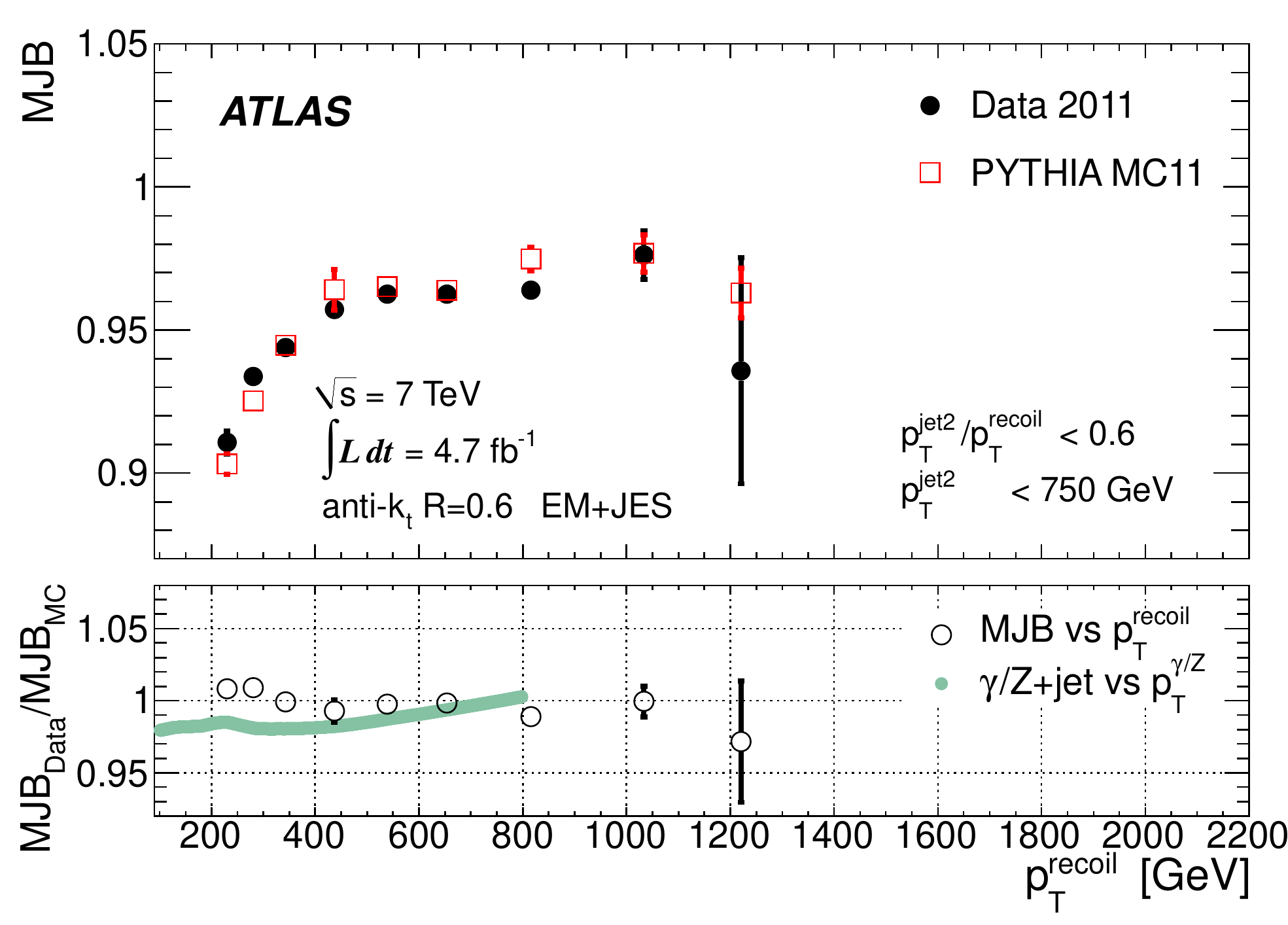}\label{fig:PtBalance_MJB_EMJES_akt6}}
    \subfloat[$R = 0.6$, \LCWJES]
    {\includegraphics[width=0.45\textwidth]{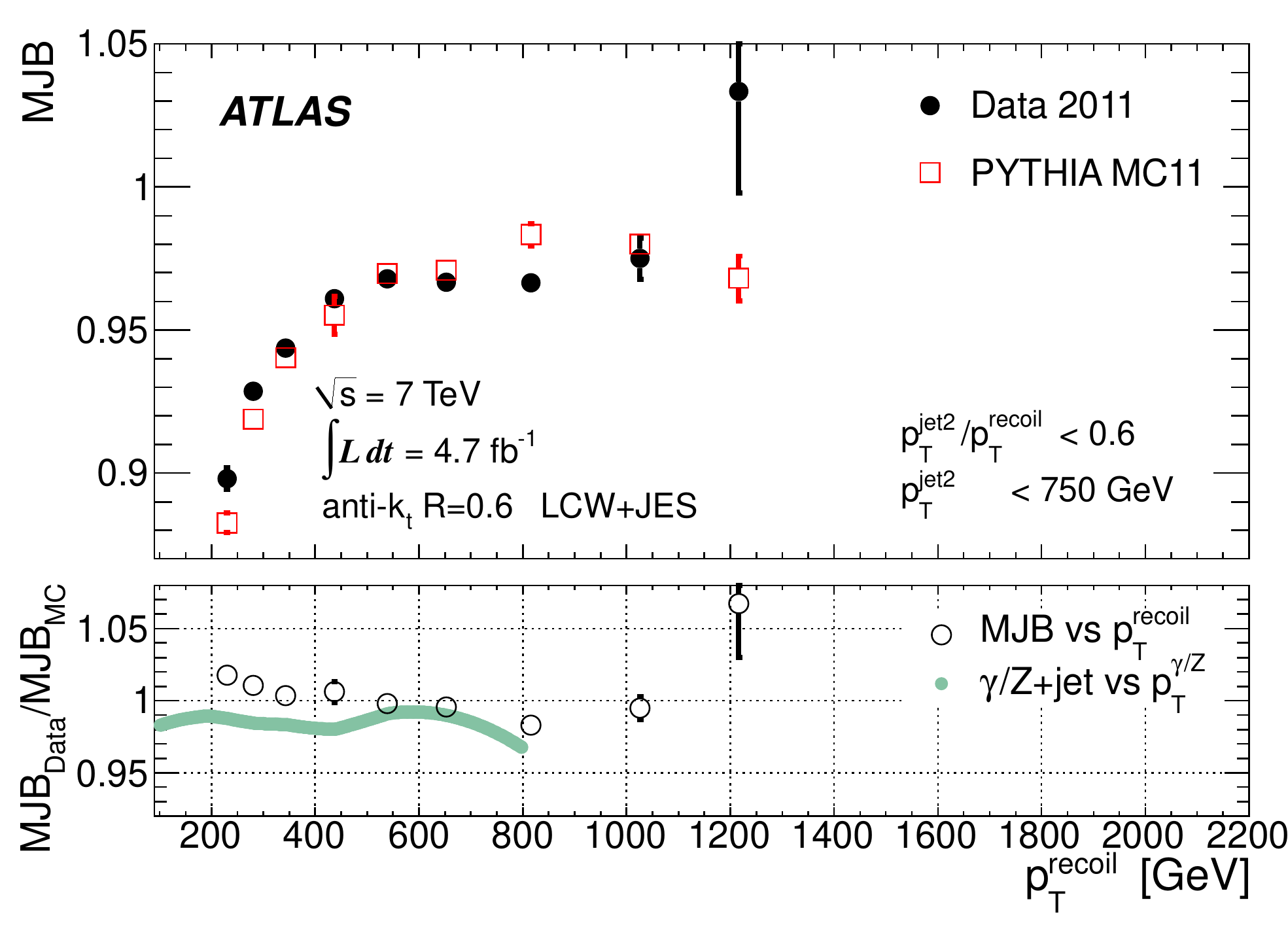}\label{fig:PtBalance_MJB_LCWJES_akt6}}
 \end{center}
 \caption[]{Multijet balance as a function of the recoil system \ptrecoil{} for \antikt{} jets with (\subref{fig:PtBalance_MJB_EMJES_akt4}, \subref{fig:PtBalance_MJB_LCWJES_akt4}) $R = 0.4$  and (\subref{fig:PtBalance_MJB_EMJES_akt6}, \subref{fig:PtBalance_MJB_LCWJES_akt6})
$R = 0.6$, calibrated with the (\subref{fig:PtBalance_MJB_EMJES_akt4}, \subref{fig:PtBalance_MJB_EMJES_akt6}) \EMJES{} scheme and with the (\subref{fig:PtBalance_MJB_LCWJES_akt4}, \subref{fig:PtBalance_MJB_LCWJES_akt6}) \LCWJES{} scheme, for both data and \MC{} simulations. 
The \nonleading{} jets in the data with $\pt<750$~\GeV{} are corrected by the combination of \gammajet{} and \Zjet{} \insitu{} calibrations as described in \secRef{sec:method}. 
The open points in the bottom panel show the ratio of the \MJB{} 
values between data and \MC{} simulations. The curve in the same panel shows the \datatomc{} ratio of the jet \pt{} relative
to the \pt{} of a photon (\ptgamma) or a \Zboson{} boson ($\pt^{\Zboson}$) as a function of the $\pt^{\gamma}$ or $\pt^{Z}$
in \gammajet{} or \Zjet{} events, obtained in the combination mentioned above.
Only the statistical uncertainties are shown.}
 \label{fig:PtBalance_MJB}
\end{figure*}

\section[High-\pt{} jet energy calibration using multijet events]{High-\PTBF{} jet energy calibration using multijet events}
\label{sec:multijet}
\subsection{Multijet balance technique and uncertainty propagation}
\label{sec:method}
The multijet balance (\MJB) technique described in Ref.~\cite{jespaper2010} can be used to verify the energy scale of jets and obtain correction factors that can correct for any non-linearity at very high \pt{}. The method exploits the \pt{} balance in events where the highest-\pt{} jet (leading jet) is produced back-to-back to a system composed of \nonleading{} jets, referred to as a ``recoil system". The leading jet is required to have significantly larger \pt{} than the jets in the recoil system in order to ensure that \MJB{} is testing the absolute high-\pt{} jet energy scale.

The vectorial sum of the \pt{} of all non-leading jets defines the transverse momentum of the recoil system (\ptrecoil) that is expected to approximately balance the \pt{} of the leading jet. The ratio 
\begin{displaymath}
\MJB = \frac{|\ptleadvec|}{|\ptrecoilvec|}
\end{displaymath}
thus allows the verification of the \JES{} of the leading jet using the properly calibrated non-leading jets at a lower \pt{} scale. The asymmetry in the \pt{} scale between the leading jet and \nonleading{} jets is established by introducing a maximum limit on the ratio between the \pt{} of the \subleading{} (second-highest \pt) jet (\ptjetn{2}) and \ptrecoil. The calibration for the \nonleading{} jets in the recoil system is provided by the combination 
of the %
\JES{} corrections derived from the \pt{} balance in events 
with a jet and a \Zboson{} boson (see \secRef{sec:ZjetInSitu})
or a photon (see \secRef{sec:gammajetInSitu}) for the absolute jet energy calibration, in addition to the \pt{} balance in dijet events (see \secRef{sec:etaintercalibration})
for the relative (\etaDet{} dependent) jet energy correction. See later \secRef{sec:insitucombination} for detailed descriptions of the combination strategies in various \pt{} ranges. 

The \MJB{} measured in data with the corrected \nonleading{} jets (${\rm MJB}^{\rm Data}$) is compared with that in the simulation ($\MJB^{\rm MC}$) to evaluate the \JES{} calibration for the leading jet and assess the systematic uncertainty for high-\pt{} jets. 
The statistical and systematic uncertainties of the \gammajet{} and \Zjet{} measurements are propagated through the combination. They are taken into account, together with the systematic uncertainty of the \etaic, by fluctuating each sub-leading jet four momentum within its uncertainties individually, and propagating those to higher \pt{} as a variation in the \MJB{} measurement. This whole procedure is repeated by increasing the \subleading{} jet \pt{} in steps, and applying the \JES{} calibration derived in the previous step to the new event sample with harder \nonleading{} jets. The \MJB-based calibration is then calculated for the specific \pt{} range and applied in the following increase of the \subleading{} jet \pt. The procedure terminates once the number of events available for the next step becomes too low for a precise evaluation of \MJB{} with the corresponding sample.

A cut on the ratio between \ptjetn{2}{} and \ptrecoil, which defines the hard scale for the \subleading{} jets, is also relaxed in the repetition sequences to effectively increase the statistics available in the calibration. The convolution of the propagated uncertainties from the \JES{} calibrations applied to the \nonleading{} jets with systematic uncertainties associated with the \MJB{} method itself, as described in \secRef{sec:multijetsystematics}, gives rise to a \JES{} systematic uncertainty across the whole jet \pt{} range accessible in 2011 data.

\subsection{Selection of multijet events}
\label{sec:multijetselection}
In order to cover a wide \pt{} range with enough event statistics, the analysis uses four single-jet triggers, each with a different jet-\pt{} threshold. The highest \pt-threshold trigger that is active for the full dataset requires at least one jet with $\pt>240$~\GeV{} at the \EM{} scale. The other three triggers are pre-scaled, i.e. only a defined fraction of them are recorded, 
and they require respective jet-\pt{} thresholds of $55$, $100$, and $135$~\GeV. As shown below, the analysis is not limited by the statistical accuracy even with these pre-scaled jet triggers.
In the offline analysis the data collected by a given trigger are used in non-overlapping \ptrecoil{} ranges where the trigger is $>99\%$ efficient. 

Only events containing at least one primary vertex, defined as described in \secRef{sec:trackjets} and associated with at least five tracks, are considered. Events are rejected if they contain either an identified lepton (electron or muon) or a photon. Events are also rejected if they contain at least one jet which has $\pt>20$~\GeV{} that does not pass the jet cleaning criteria discussed in \secRef{sec:JetSel} to suppress noise or detector problems and mismeasured jets. 
For a certain period of time the read-out of a part of the \EM{} calorimeter was not functioning, and events containing jets pointing to the affected region are also rejected. At the last stage of the event pre-selection, events are required to have at least three good-quality jets that have $\pt>25$~\GeV{} and $|\eta|<2.8$. The leading jet is required to be within $|\eta|<1.2$.

In order to select events having one jet produced against a well-defined recoil system, a selection is applied using two angular variables,
\begin{enumerate}
\item $\alpha = |\Delta\phi - \pi| < 0.3$ rad, where $\Delta\phi$ is the azimuthal opening angle between the highest-\pt{} jet and the recoil system, and
\item The azimuthal opening angle between the leading jet and the \nonleading{} jet that is closest in $\phi$ ($\beta$) is required to be $\beta > 1$ rad.
\end{enumerate}
Two more selection criteria ensure that the \subleading{} jets have a \pt{} in the range where the \insitu{} \gammajet{} and \Zjet{} calibrations are available and the leading jet is well above this range. The former is achieved by requiring the \subleading{} jet  \ptjetn{2}{} to be less than 750~\GeV{} and the latter by requiring that the ratio $A$ between \ptjetn{2}{} and \ptrecoil{} satisfies $\ptjetn{2}/\ptrecoil < 0.6$. These two initial selections are modified when the analysis procedure is repeated as described above. 

A summary of all cuts used in the analysis is given in Table~\ref{tab:AnalysisCuts}.

\begin{table}
\renewcommand{\arraystretch}{\myarraystretch}
 \caption{Summary of the event selection cuts used in the analysis. The first (second, third) values for \ptjetn{2}{} and $\ptjetn{2}/\ptrecoil$ cuts are used in the first (second, subsequent) repetition of the \MJB{} calibration procedure as described in Sec.~\ref{sec:method}.}
  \begin{center}
    \begin{tabular}{lc}
      \hline \hline
     Variable     & \multicolumn{1}{c}{Cut value}\\ 
      \hline
      Jet \pt{}             & $>25$~\GeV  \\
      Jet rapidity          & $|\eta|<2.8$  \\
      Leading jet rapidity  & $|\eta|<1.2$  \\
      Number of good jets   & $\geq 3$ \\ 
      \ptRecoil             & $>210$~\GeV \\
      $\alpha$              & $<0.3$~rad \\
      $\beta$               & $>1$~rad  \\
      \ptjetn{2}                  & $<750$ ($1200$, $1450$)~\GeV \\      
      $\ptjetn{2}/\ptrecoil$ & $<0.6$ ($0.8$, $0.8$) \\      
      \hline \hline
    \end{tabular}
   \label{tab:AnalysisCuts}
  \end{center}
\end{table}

\begin{figure*}
 \begin{center}
    \subfloat[$R = 0.4$, \EMJES]
    {\includegraphics[width=0.48\textwidth]{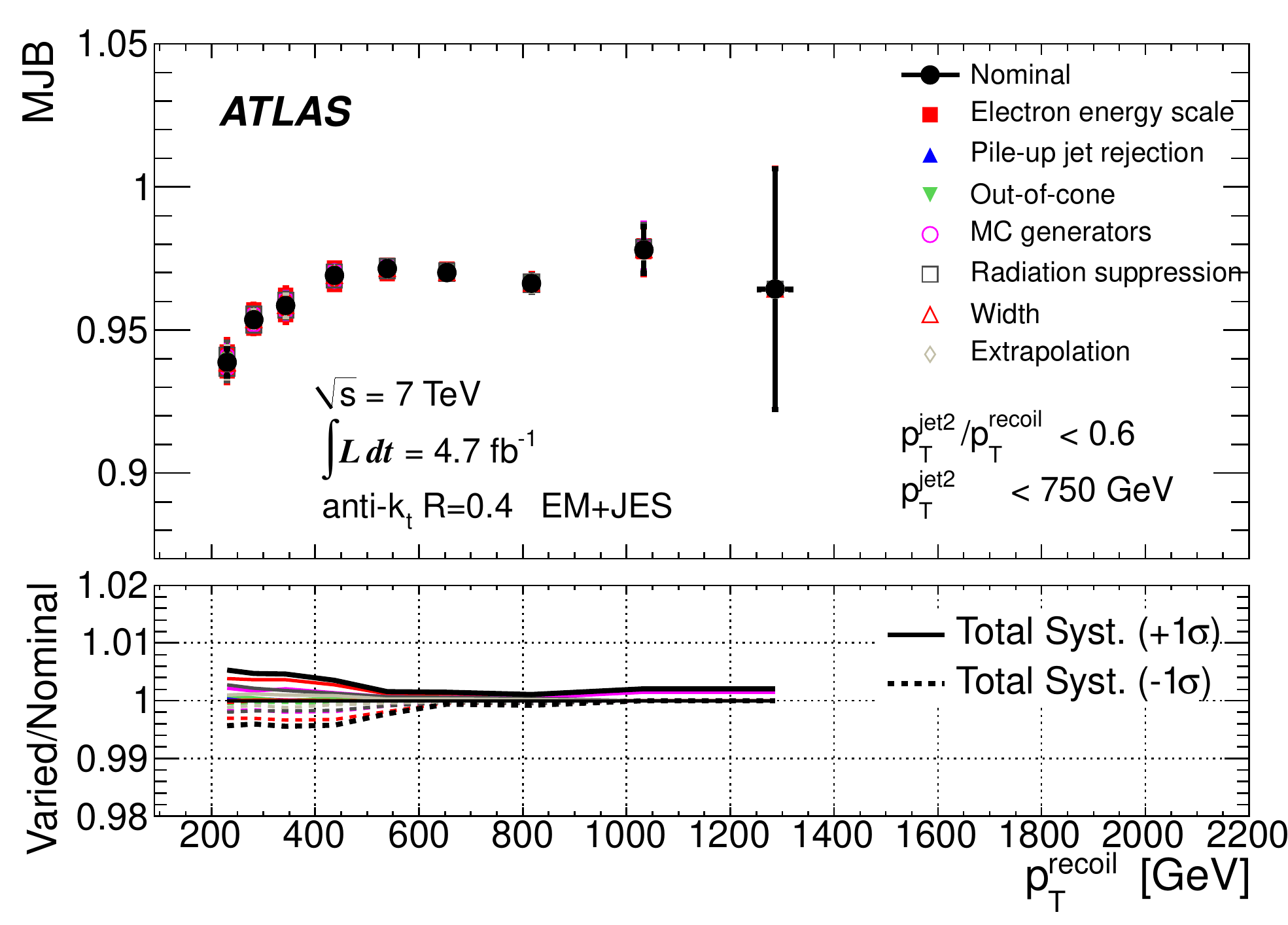}\label{fig:PtBalance_ZJsyst_EMJES_akt4}}
    \subfloat[$R = 0.4$, \LCWJES]
    {\includegraphics[width=0.48\textwidth]{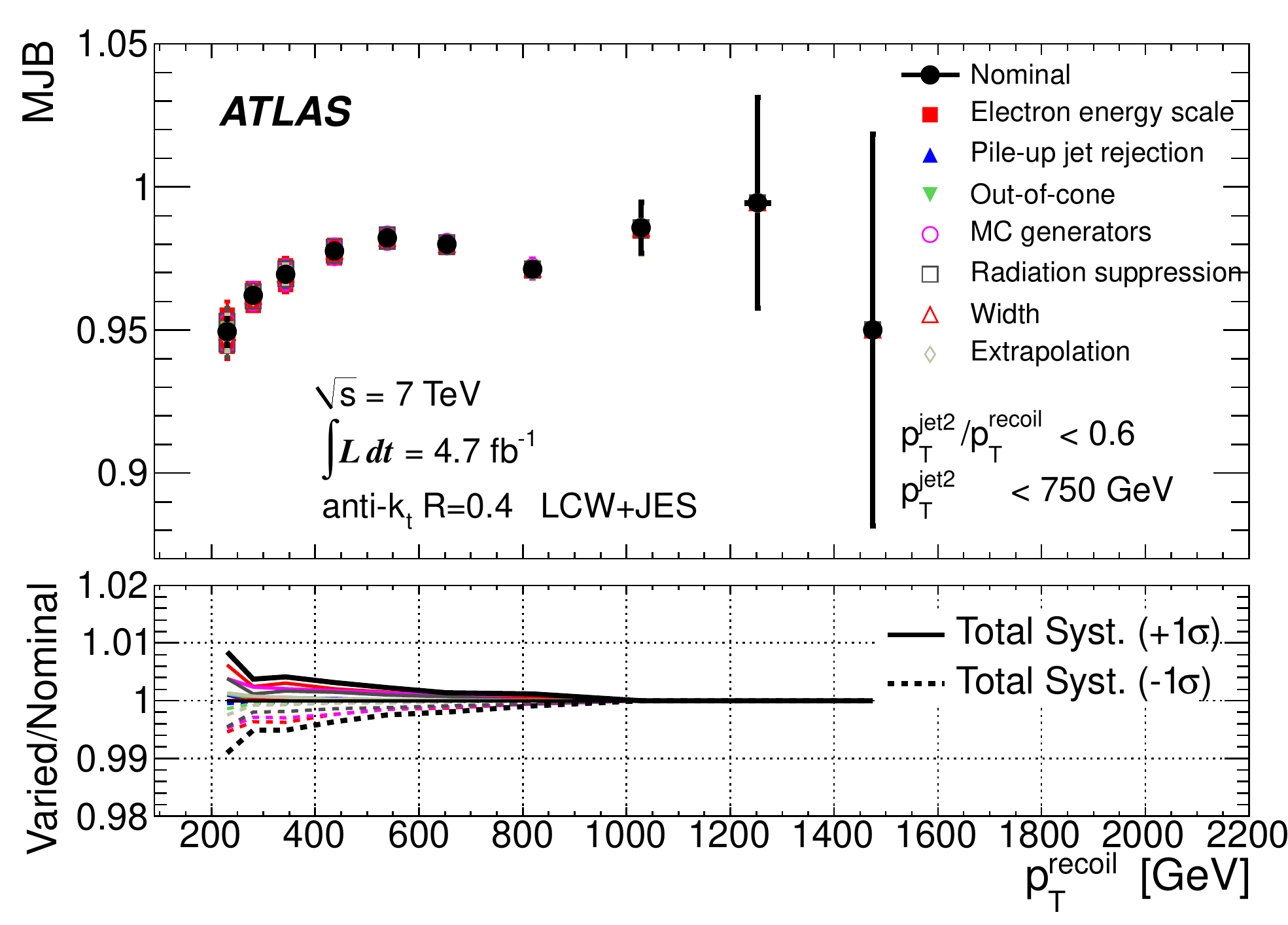}\label{fig:PtBalance_ZJsyst_LCWJES_akt4}}\\
    \subfloat[$R = 0.6$, \EMJES]
    {\includegraphics[width=0.48\textwidth]{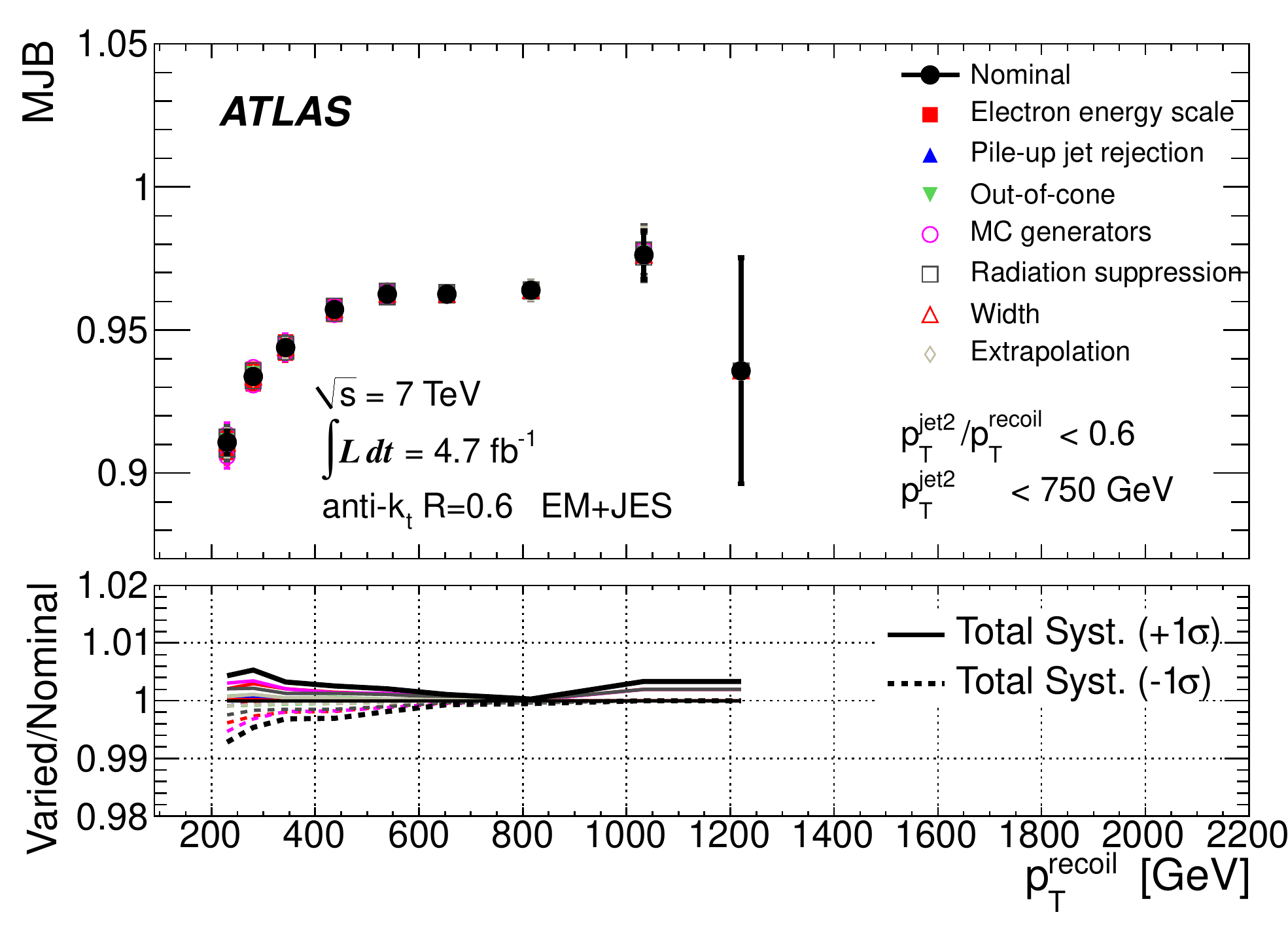}\label{fig:PtBalance_ZJsyst_EMJES_akt6}}
    \subfloat[$R = 0.6$, \LCWJES]
    {\includegraphics[width=0.48\textwidth]{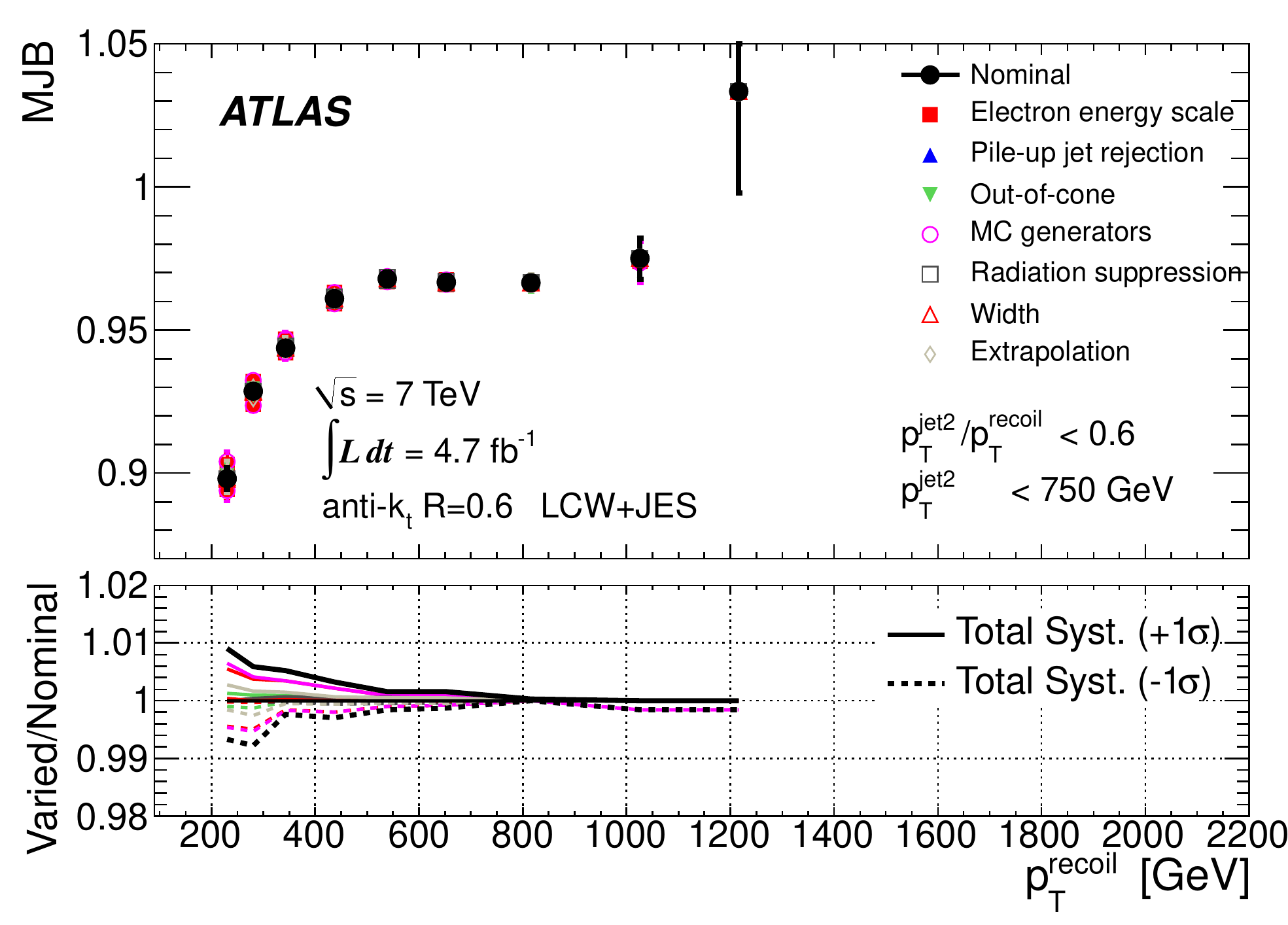}\label{fig:PtBalance_ZJsyst_LCWJES_akt6}}
 \end{center}
 \caption[]{Multijet balance with the nominal and varied \Zjet{} \insitu{} calibrations
as a function of the recoil system \ptrecoil{} for \antikt{} jets with (\subref{fig:PtBalance_ZJsyst_EMJES_akt4}, \subref{fig:PtBalance_ZJsyst_LCWJES_akt4})  $R = 0.4$ and  (\subref{fig:PtBalance_ZJsyst_EMJES_akt6}, \subref{fig:PtBalance_ZJsyst_LCWJES_akt6}) 
$R = 0.6$, calibrated with the  (\subref{fig:PtBalance_ZJsyst_EMJES_akt4}, \subref{fig:PtBalance_ZJsyst_EMJES_akt6})  \EMJES{} scheme and with the (\subref{fig:PtBalance_ZJsyst_LCWJES_akt4}, \subref{fig:PtBalance_ZJsyst_LCWJES_akt6}) \LCWJES{} scheme.
The varied distributions are obtained by fluctuating the jet energy scale for the \nonleading{} jets by $\pm 1\sigma$ for each of
the systematic uncertainties for the \Zjet{} calibration and repeating the analysis over the 
data sample. The bottom panel shows the relative variations of the \MJB{} with respect to the nominal case. The uppermost (lowermost) thick line
in the bottom panel shows the total variation obtained by adding all the positive (negative) variations in quadrature.
 The colour coding used in the lower part of the figure is the same as that used in the upper one.
}
 \label{fig:PtBalance_ZJsyst}
\end{figure*}

\begin{figure*}
 \begin{center}
    \subfloat[$R = 0.4$, \EMJES]
    {\includegraphics[width=0.48\textwidth]{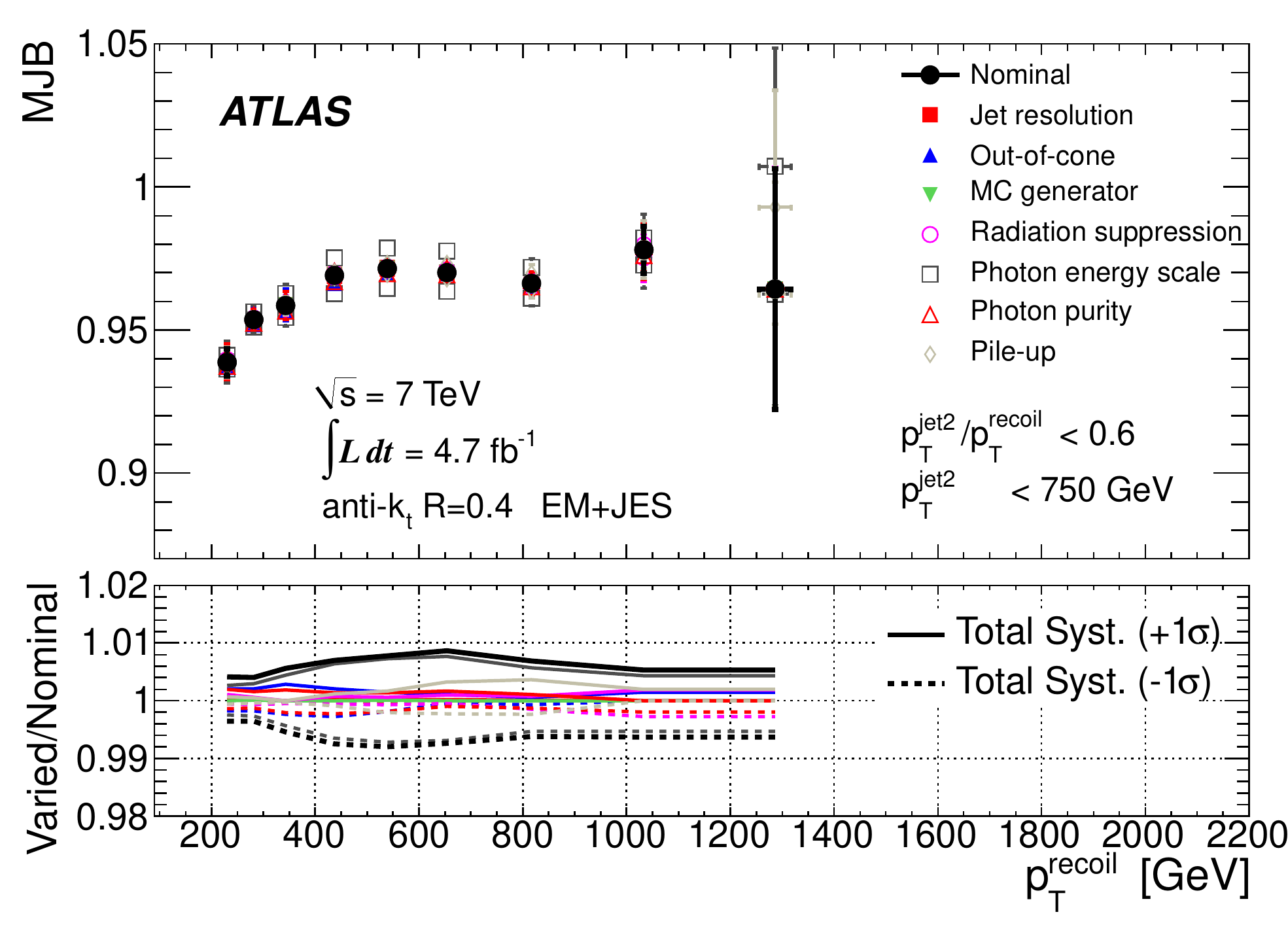}\label{fig:PtBalance_MPFsyst_EMJES_akt4}}
    \subfloat[$R = 0.4$, \LCWJES]
    {\includegraphics[width=0.48\textwidth]{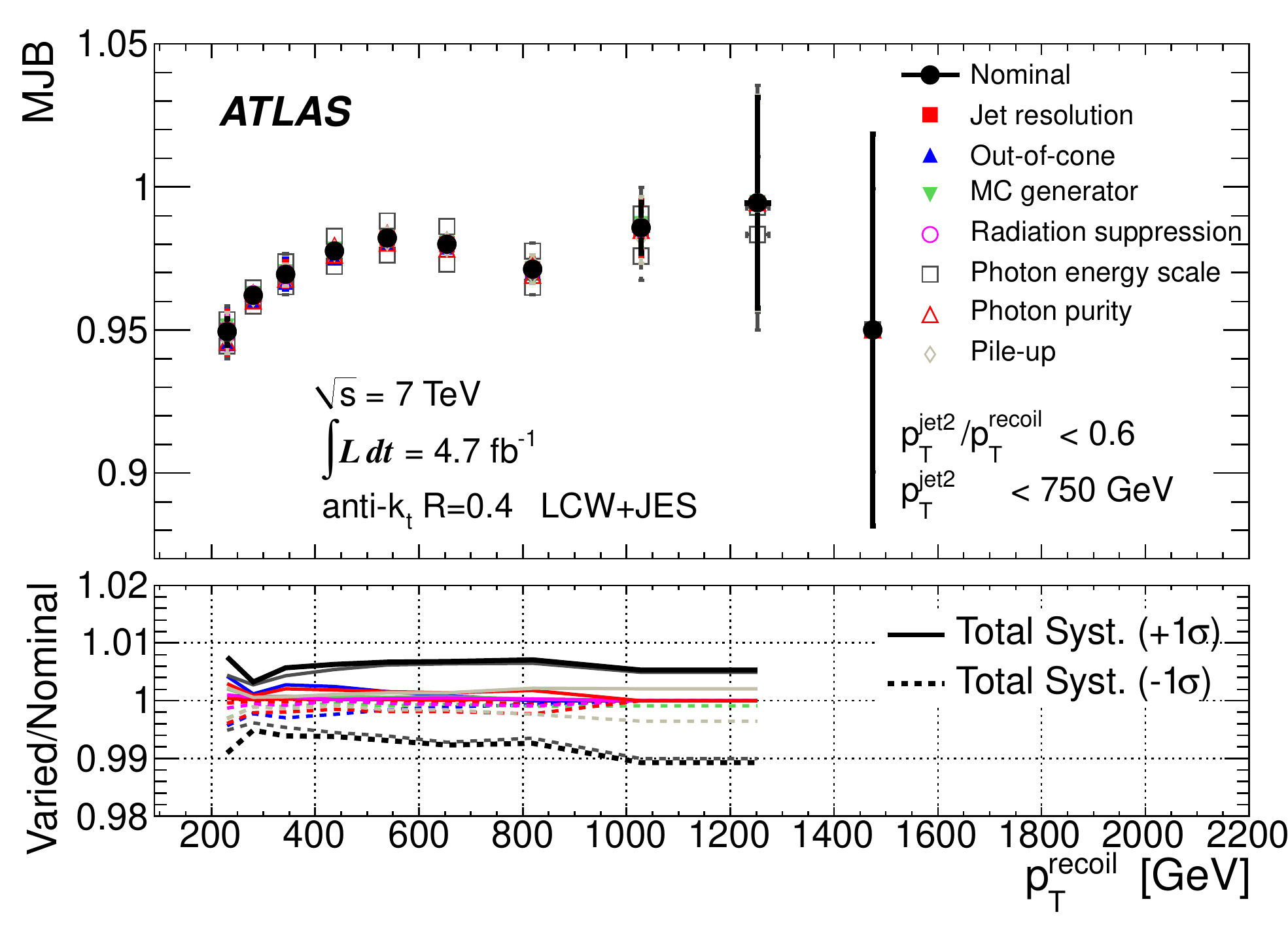}\label{fig:PtBalance_MPFsyst_LCWJES_akt4}}\\
    \subfloat[$R = 0.6$, \EMJES]
    {\includegraphics[width=0.48\textwidth]{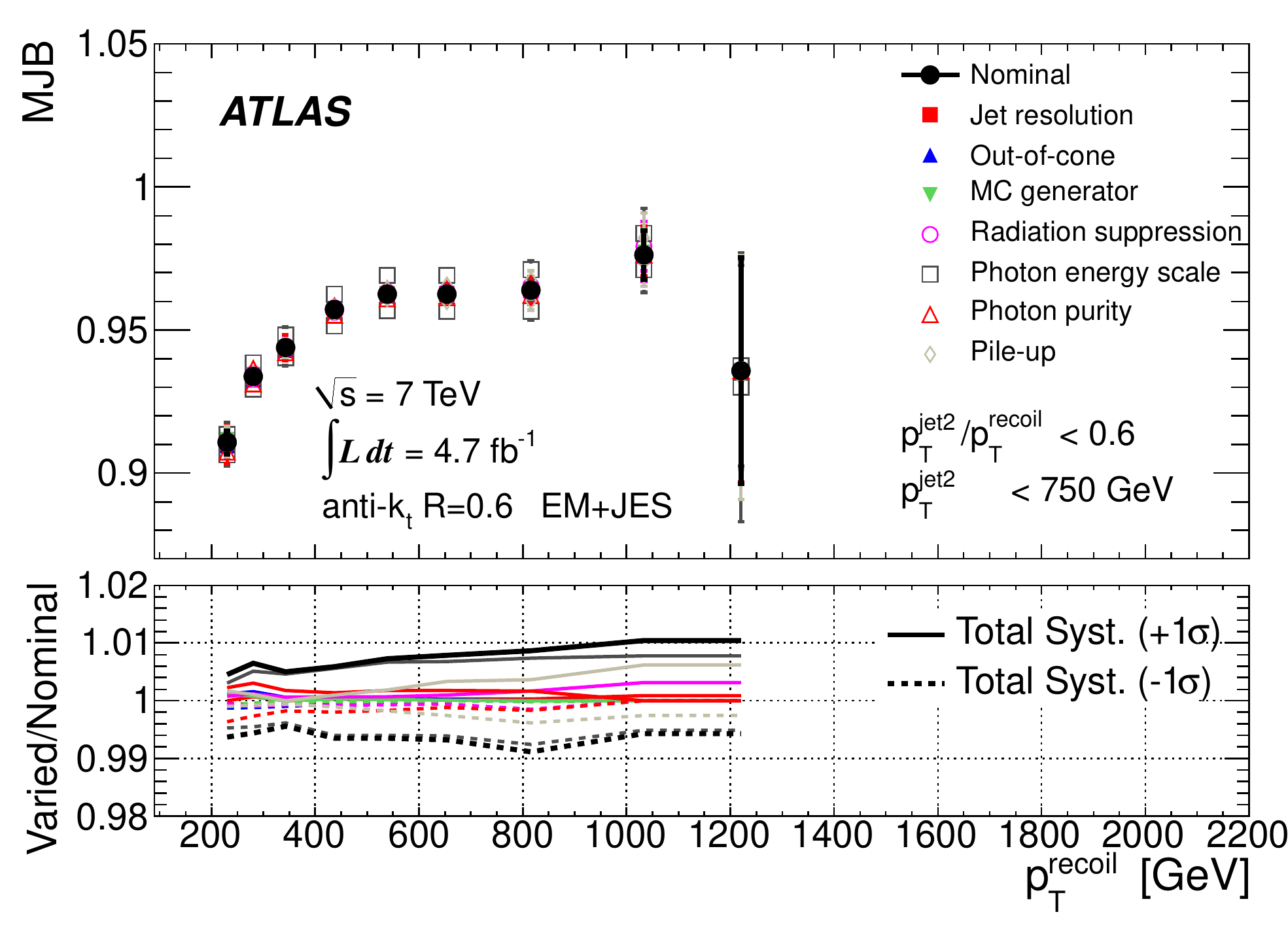}\label{fig:PtBalance_MPFsyst_EMJES_akt6}}
    \subfloat[$R = 0.6$, \LCWJES]
    {\includegraphics[width=0.48\textwidth]{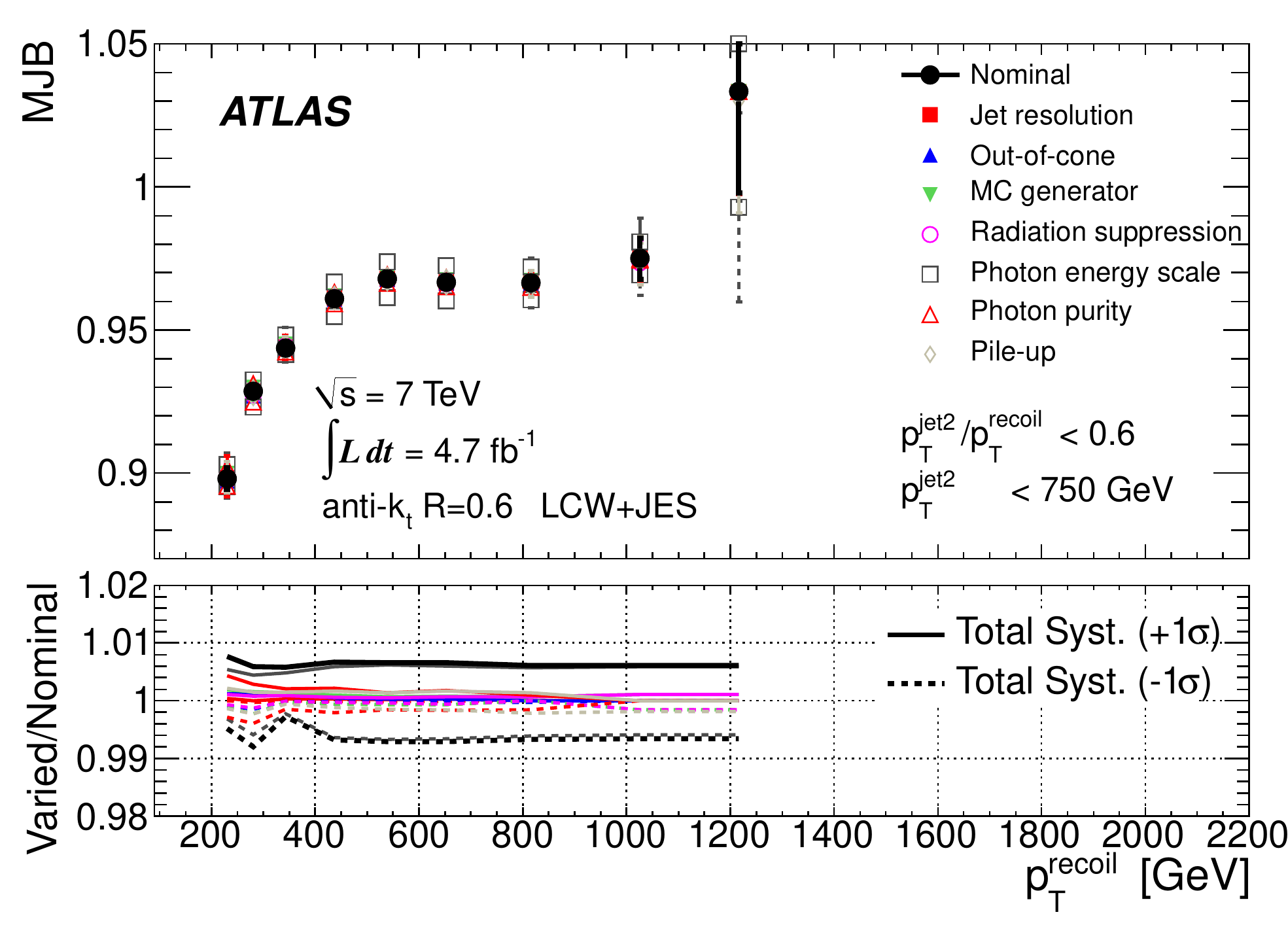}\label{fig:PtBalance_MPFsyst_LCWJES_akt6}}
 \end{center}
 \caption[]{Multijet balance with the nominal and varied \gammajet{} \insitu{} calibrations
as a function of the recoil system \ptrecoil{} for \antikt{} jets with (\subref{fig:PtBalance_MPFsyst_EMJES_akt4}, \subref{fig:PtBalance_MPFsyst_LCWJES_akt4}) $R = 0.4$ and  (\subref{fig:PtBalance_MPFsyst_EMJES_akt6}, \subref{fig:PtBalance_MPFsyst_LCWJES_akt6})
$R = 0.6$, calibrated with the (\subref{fig:PtBalance_MPFsyst_EMJES_akt4}, \subref{fig:PtBalance_MPFsyst_EMJES_akt6}) \EMJES{} scheme and with the (\subref{fig:PtBalance_MPFsyst_LCWJES_akt4}, \subref{fig:PtBalance_MPFsyst_LCWJES_akt6}) \LCWJES{} scheme.
The varied distributions are obtained by fluctuating the jet energy scale for the \nonleading{} jets by $\pm1\sigma$ for each of
the systematic uncertainties and repeating the analysis over the 
data sample. The bottom panel shows the relative variations of the \MJB{} with respect to the nominal case. The uppermost (lowermost) thick line
in the bottom panel shows the total variation obtained by adding all the positive (negative) variations in quadrature.
 The colour coding used in the lower part of the figure is the same as that used in the upper one.
}
 \label{fig:PtBalance_MPFsyst}
\end{figure*}

\begin{figure*}
 \begin{center}
    \subfloat[$R = 0.4$, \EMJES]
    {\includegraphics[width=0.44\textwidth]{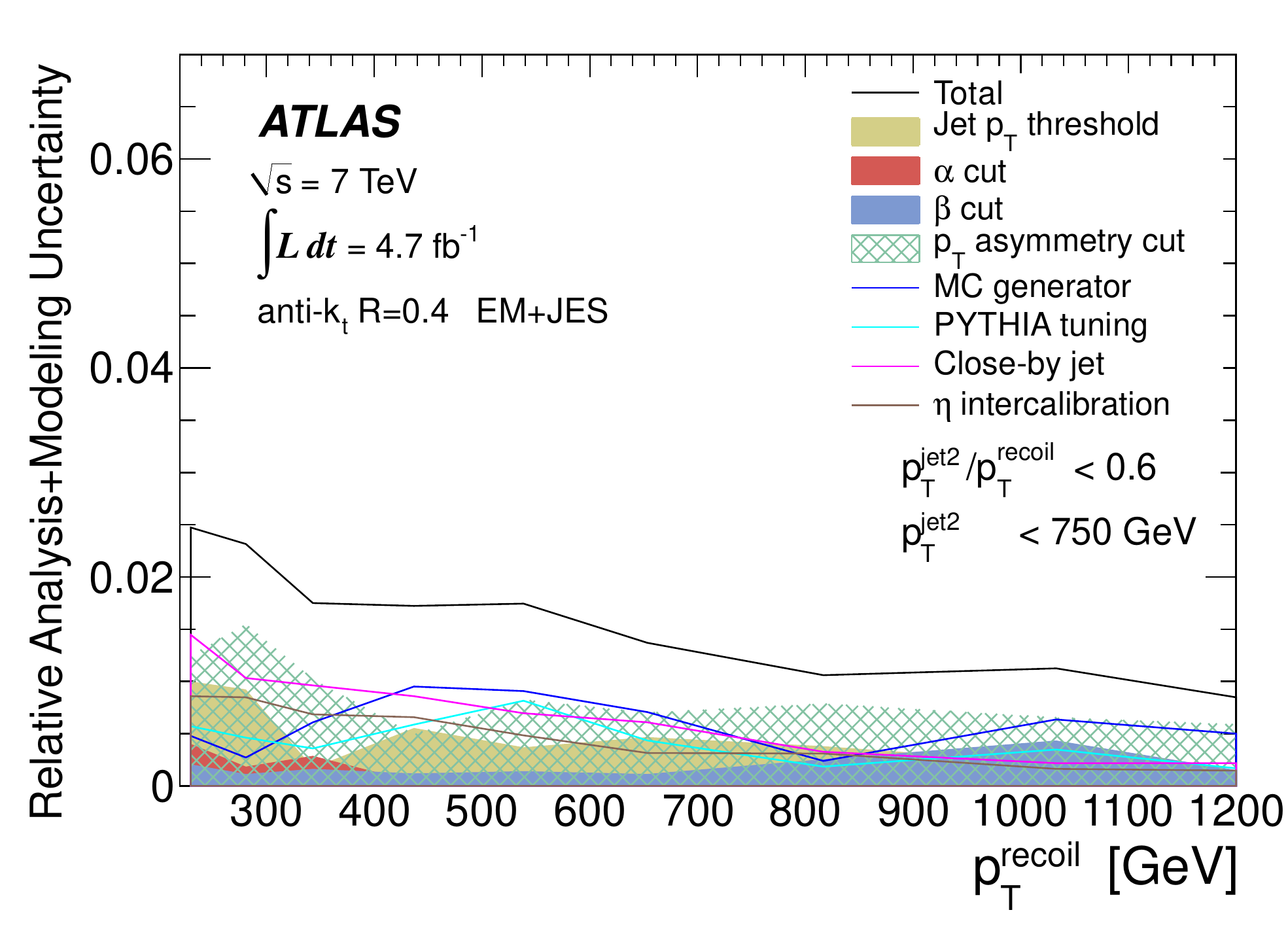}\label{fig:MJBsyst_Summary_EMJES_akt4}}
    \subfloat[$R = 0.4$, \LCWJES] 
    {\includegraphics[width=0.44\textwidth]{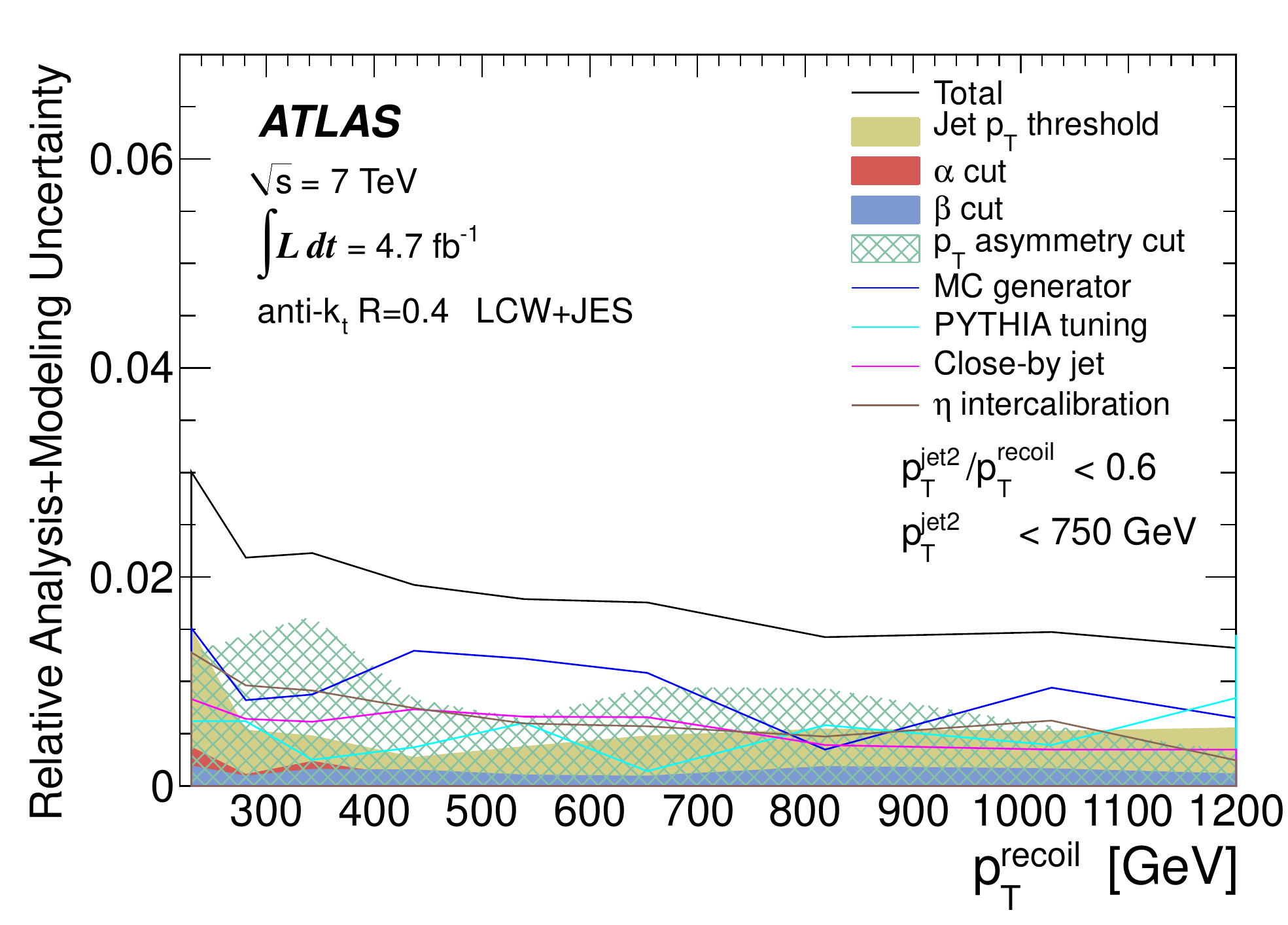}\label{fig:MJBsyst_Summary_LCWJES_akt4}}\\
    \subfloat[$R = 0.6$, \EMJES]
    {\includegraphics[width=0.44\textwidth]{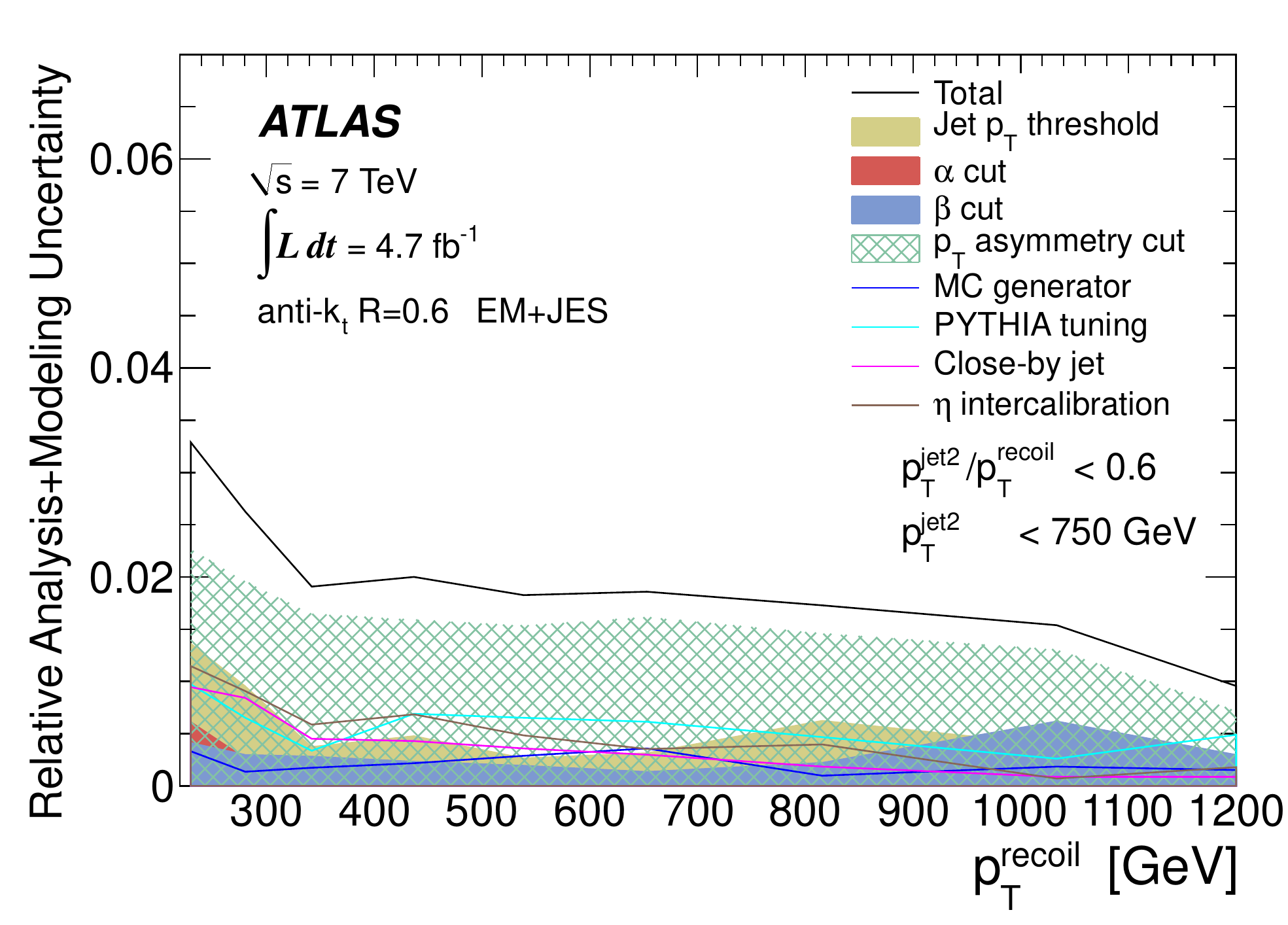}\label{fig:MJBsyst_Summary_EMJES_akt6}}
    \subfloat[$R = 0.6$, \LCWJES]
    {\includegraphics[width=0.44\textwidth]{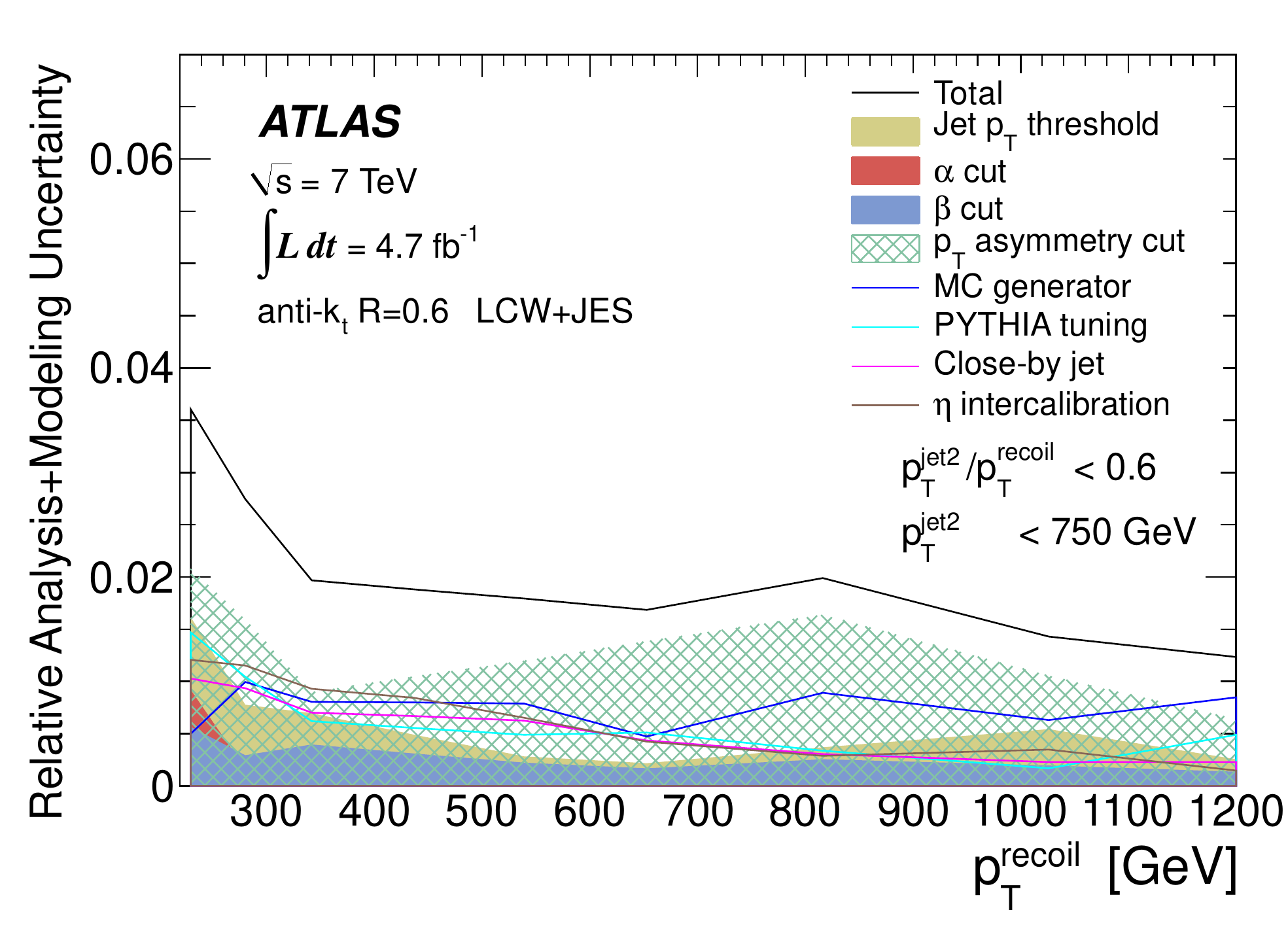}\label{fig:MJBsyst_Summary_LCWJES_akt6}}
 \end{center}
 \caption[]{Relative uncertainties on the \MJB{} due to the systematic uncertainty sources considered in the analysis as a function of the recoil system \pt{}
for \antikt{} jets with (\subref{fig:MJBsyst_Summary_EMJES_akt4}, \subref{fig:MJBsyst_Summary_LCWJES_akt4}) $R = 0.4$ and (\subref{fig:MJBsyst_Summary_EMJES_akt6}, \subref{fig:MJBsyst_Summary_LCWJES_akt6}) $R = 0.6$, calibrated with the (\subref{fig:MJBsyst_Summary_EMJES_akt4}, \subref{fig:MJBsyst_Summary_EMJES_akt6}) \EMJES{} scheme and with the (\subref{fig:MJBsyst_Summary_LCWJES_akt4}, \subref{fig:MJBsyst_Summary_LCWJES_akt6}) \LCWJES{} scheme.
The black line shows the total uncertainty obtained as a sum of all uncertainties in quadrature.}
 \label{fig:MJBsyst_Summary}
\end{figure*}

\begin{figure*}
 \begin{center}
    \subfloat[$R = 0.4$, \EMJES]
    {\includegraphics[width=0.44\textwidth]{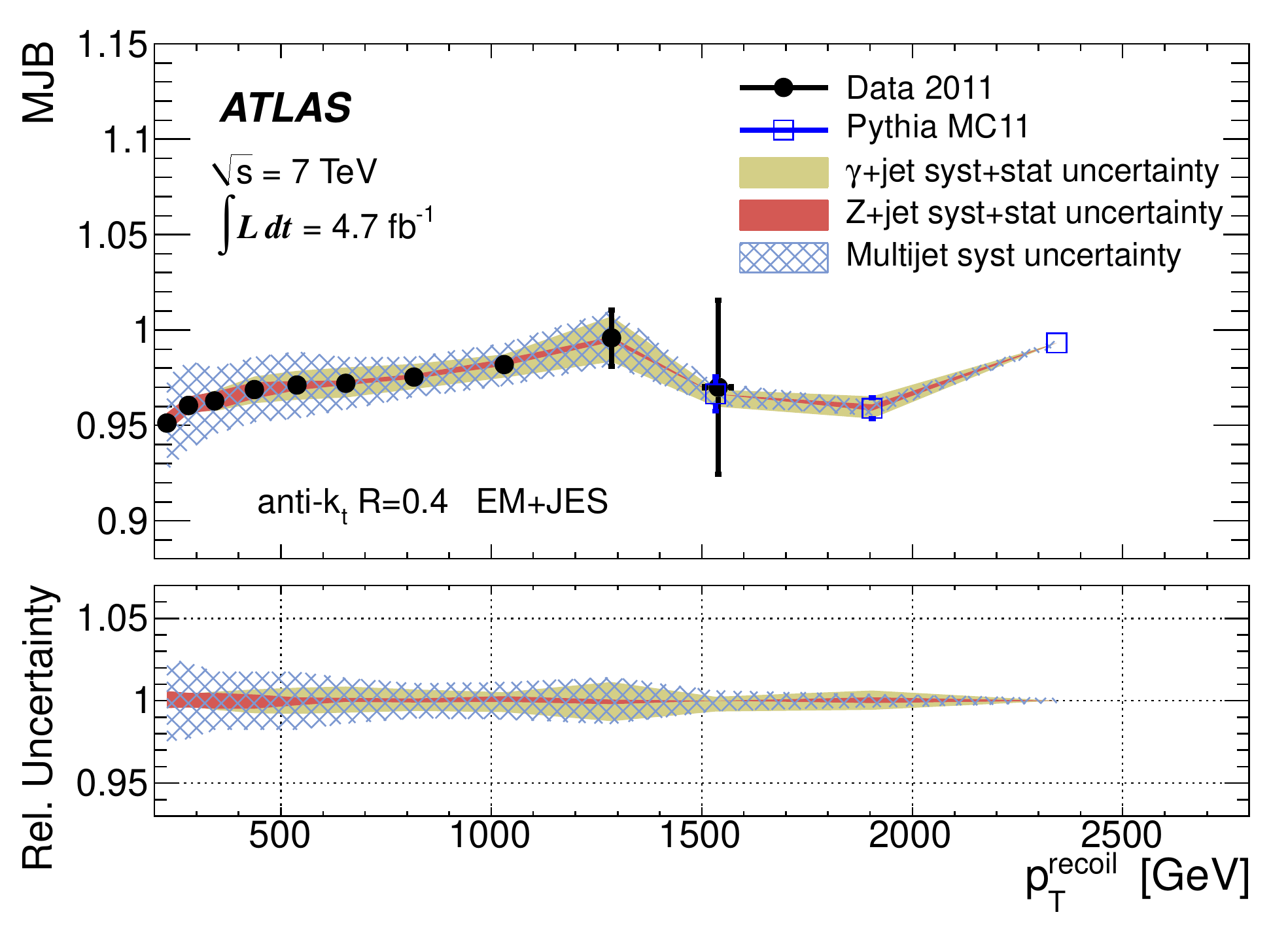}\label{fig:JES_SummaryEMNarrow}}
    \subfloat[$R = 0.4$, \LCWJES] 
    {\includegraphics[width=0.44\textwidth]{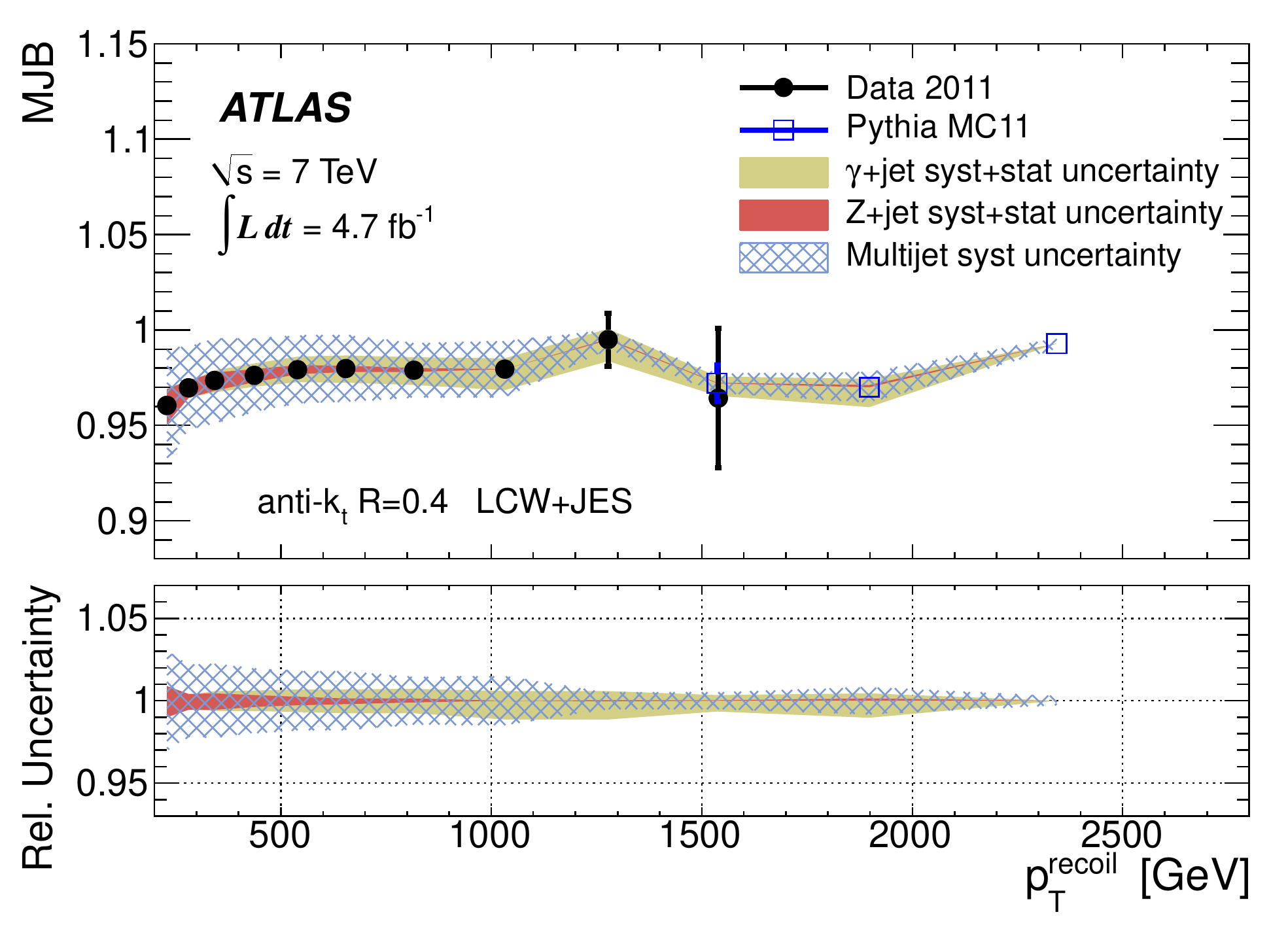}\label{fig:JES_SummaryLCWNarrow}}\\
    \subfloat[$R = 0.6$, \EMJES] 
    {\includegraphics[width=0.44\textwidth]{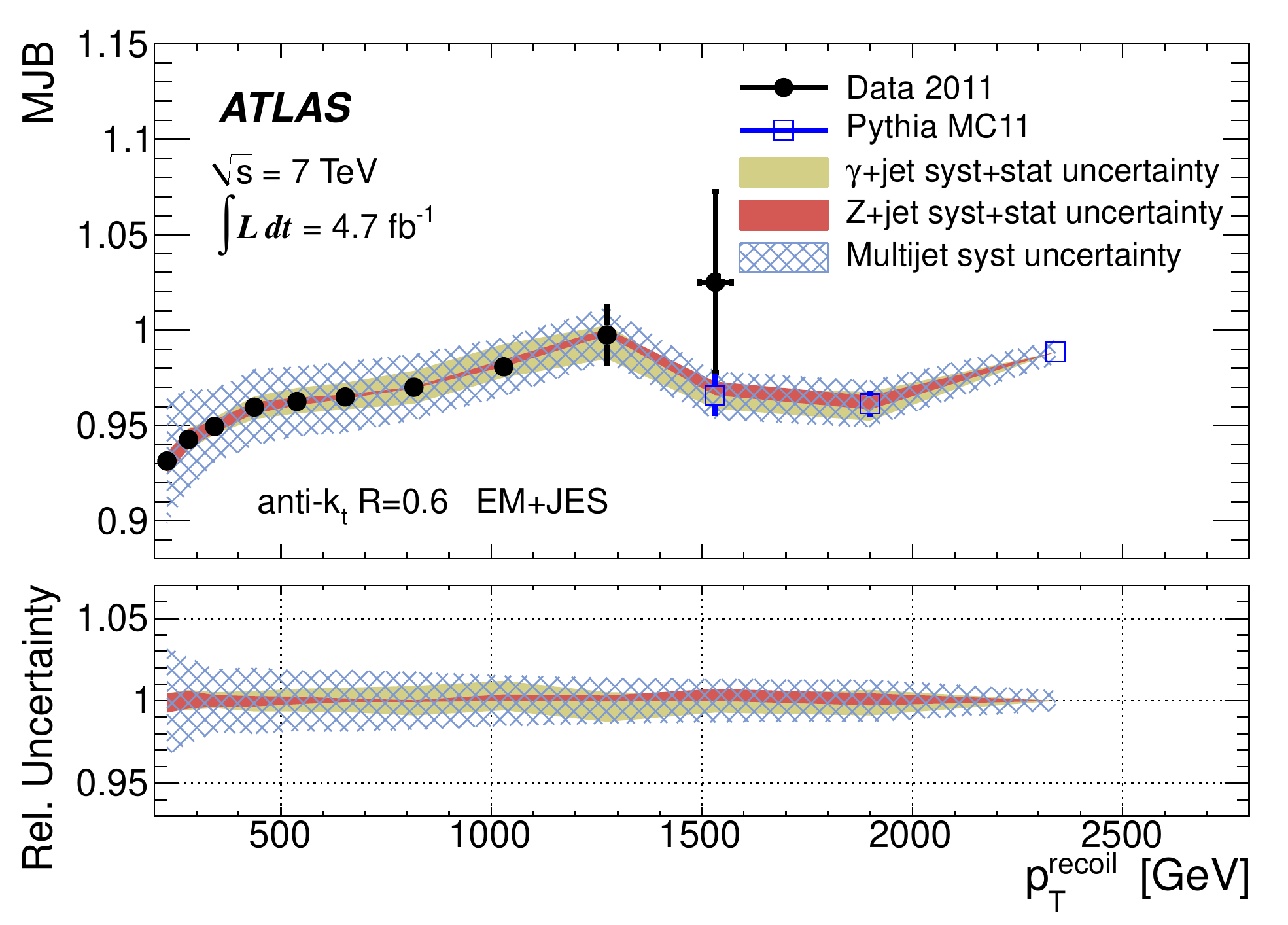}\label{fig:JES_SummaryEMWide}}
    \subfloat[$R = 0.6$, \LCWJES]
    {\includegraphics[width=0.44\textwidth]{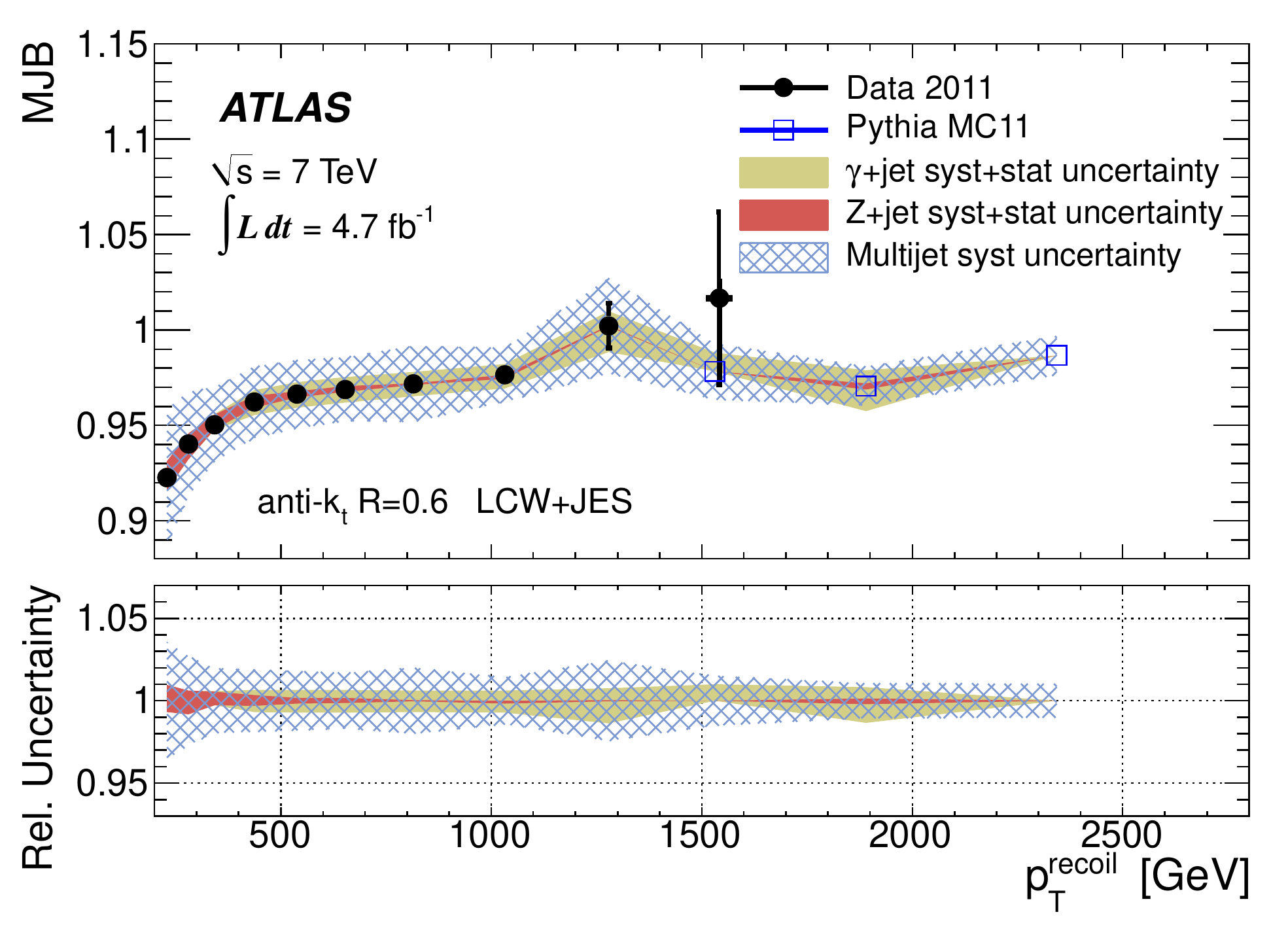}\label{fig:JES_SummaryLCWWide}}
 \end{center}
 \caption[]{
Multijet balance and systematic uncertainties related to the multijet balance technique and 
to the propagated uncertainties from the \gammajet{} and \Zjet{} balance
as a function of the recoil system \ptrecoil{} for \antikt{} jets with  (\subref{fig:JES_SummaryEMNarrow}, \subref{fig:JES_SummaryLCWNarrow}) $R = 0.4$ and (\subref{fig:JES_SummaryEMWide}, \subref{fig:JES_SummaryLCWWide})
$R = 0.6$, calibrated with the (\subref{fig:JES_SummaryEMNarrow}, \subref{fig:JES_SummaryEMWide}) \EMJES{} scheme and with the (\subref{fig:JES_SummaryLCWNarrow}, \subref{fig:JES_SummaryLCWWide}) \LCWJES{} scheme. 
The subleading jets in the data are corrected by the combination of \gammajet{} and \Zjet{} \insitu{} calibrations at $\pt<750$~\GeV{} and 
\MJB{} calibration at higher \pt{} as described in \secRef{sec:method}. The three systematic uncertainty bands are obtained
by adding individual systematic uncertainties for each calibration technique in quadrature.
Also shown are predictions of the \MJB{} from \MC{} simulations for the three highest \ptrecoil-values, together with their systematic uncertainties propagated by using distribution from \MC{} simulations. The bottom panel shows the relative variations of the \MJB{} with respect to the nominal case.}
 \label{fig:JES_Summary}
\end{figure*}

\subsection{Multijet balance measurement}
\label{sec:multijetmeasurement}
The multijet balance obtained from the selected events for the \EMJES{} and \LCWJES{} calibrated jets with the \antikt{} jet algorithm with $R = 0.4$ or $R = 0.6$ is shown in Fig.~\ref{fig:PtBalance_MJB} for data and 
the \MC{} simulations with \pythia. 

The \MJB{} decreases slightly at \ptrecoil{} below $400$ \GeV, which is a consequence of the broadening of the \ptrecoil{} distribution that can already be observed for jets formed from truth particles.
The ratio between the distributions obtained from the data to the corresponding ones from \MC{} simulations is shown in the lower part of each figure. It is compared with the \datatomc{} ratio observed in the \gammajet{} and \Zjet{} \insitu{} measurements. The agreement between data and \MC{} simulations in the \pt{} range covered by the \gammajet{} and \Zjet{} calibration, evaluated as the average value of the \datatomc{} ratio, is within $2\%$ ($3\%$) for jets with $R = 0.4$ (0.6). 

\begin{table}
\renewcommand{\arraystretch}{\myarraystretch}
  \caption{Default values and the range of variation used to evaluate the systematic uncertainty on the analysis cuts.}
  \begin{center}
    \begin{tabular}{lcr@{--}l}
      \hline \hline
     Variable     & \multicolumn{1}{c}{Default}&\multicolumn{2}{c}{Range}\\ 
      \hline
      Jet \pt{}                              & $25$~\GeV & $20$     & $30$~\GeV \\
      $\alpha$                              & $0.3$~rad  & $0.1$   & $0.4$~rad \\
      $\beta$                               & $1.0$~rad  & $0.50$ & $1.50$~rad \\
      $\ptjetn{2}/\ptrecoil$ & $0.6$          & $0.4$   & $0.7$ \\      
      \hline \hline
    \end{tabular}
  \label{tab:AnalysisSys}
  \end{center}
\end{table}

\begin{table}
 \renewcommand{\arraystretch}{\myarraystretch}
 \caption{Representative values of systematic uncertainties in the \ptRecoil{} range 
 $500 \GeV<\ptRecoil<1.2$~\TeV{} for all effects considered in the analysis.}
  \begin{center}
    \begin{tabular}{lcccc}
      \hline \hline
      Source   & \multicolumn{2}{c}{\EMJES} & \multicolumn{2}{c}{\LCWJES}\\ 
      \hline
      Jet size & $R = 0.4$ & $R = 0.6$ & $R = 0.4$ & $R = 0.6$ \\ 
      \hline
      Absolute \JES   & $0.8\%$ & $0.7\%$ & $0.7\%$ & $0.7\%$ \\
      Relative \JES   & $0.3\%$ & $0.4\%$ & $0.5\%$ & $0.4\%$ \\
      Close-by jet    & $0.6\%$ & $0.3\%$ & $0.6\%$ & $0.4\%$ \\
       \hline
      Jet \pt{} threshold       & \multicolumn{4}{c}{$<0.4\%$} \\
      $\alpha$ cut              & \multicolumn{4}{c}{$<0.1\%$} \\
      $\beta$ cut               & \multicolumn{4}{c}{$<0.2\%$} \\
      $\ptjetn{2}/\ptrecoil$ cut & $<0.1\%$ & $1.5\%$ & $<0.1\%$ & $1.2\%$ \\ 
      UE/radiation model & \multicolumn{4}{c}{$<0.5\%$} \\
      Fragmentation model    & $1.0\%$ & $0.3\%$ & $1.0\%$ & $0.5\%$ \\
      \hline \hline
    \end{tabular} 
  \label{tab:SystSumm}
  \end{center}
\end{table}

\begin{figure*}
  \centering
    \subfloat[$55\leq\ptavg{}<75$~\GeV]{ \includegraphics[width=0.45\textwidth]{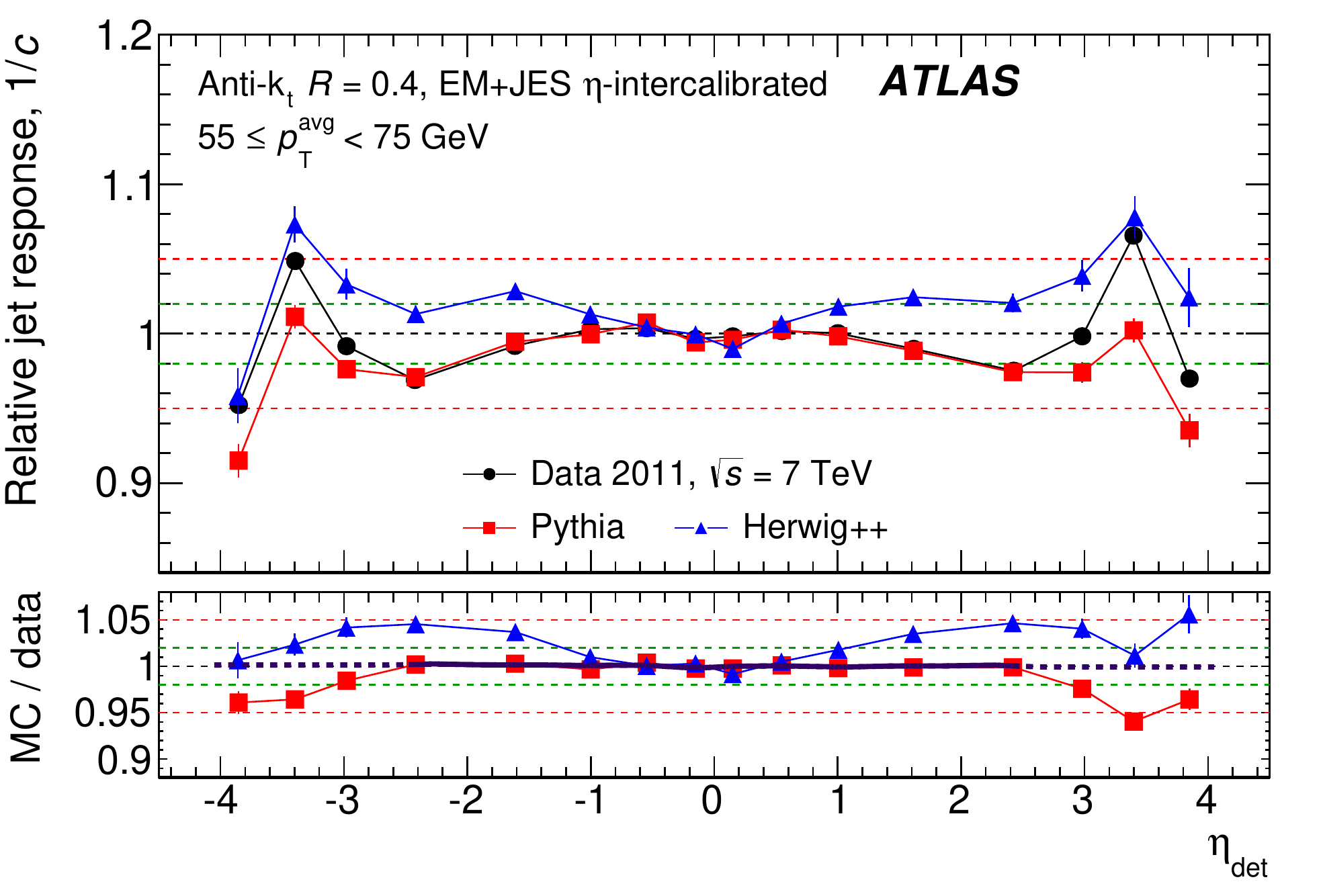}\label{fig:closure_low}}  
    \subfloat[$300\leq\ptavg{}<400$~\GeV]{ \includegraphics[width=0.45\textwidth]{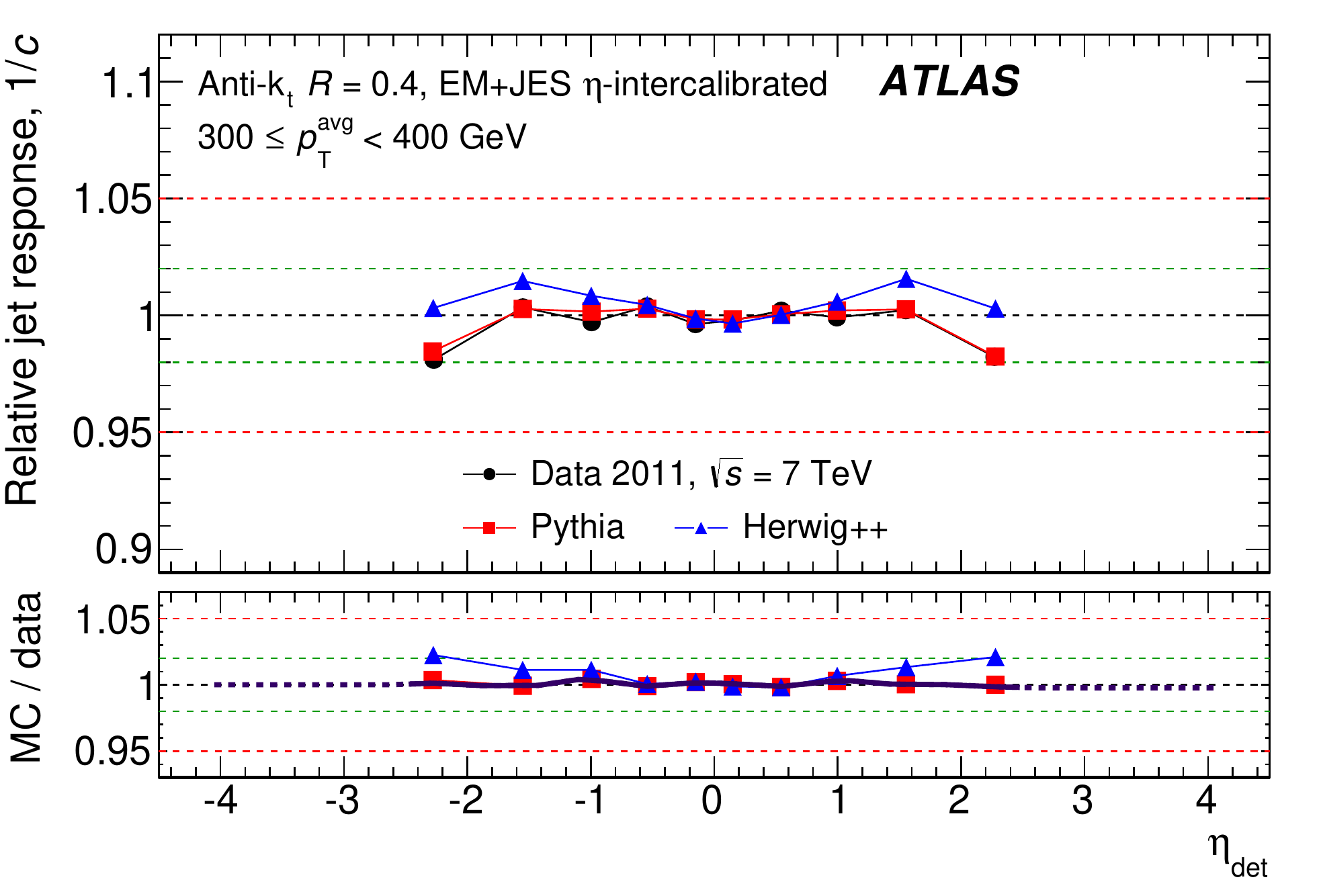}\label{fig:closure_high}}
  \caption[]{
    Relative jet response, $1/c$, as a function of the jet \etaDet{} 
    for \antikt{} jets with $R=0.4$ calibrated with the \EMJES{} scheme and in addition
    the derived \etaic. Results are shown separately for  \subref{fig:closure_low} 
    $55\leq\ptavg{}<75$~\GeV{} and  \subref{fig:closure_high} $300\leq\ptavg{}<400$~\GeV{}.
    For all points included in the original calibration ($|\etaDet|<2.8$), the data are corrected 
    to be consistent with the response in \MC{} simulations using  \pythia, as intended. 
    The resulting calibration derived from the already calibrated data is shown as a thick
    line and is consistent with unity.
    The lower parts of the figures show the ratios between the relative jet response in data and \MC{}.  
    \label{fig:closure}
  }
\end{figure*}

\begin{figure*}
  \centering
    \subfloat[$25\leq\ptref<35$ \GeV]{ \includegraphics[width=.45\textwidth]{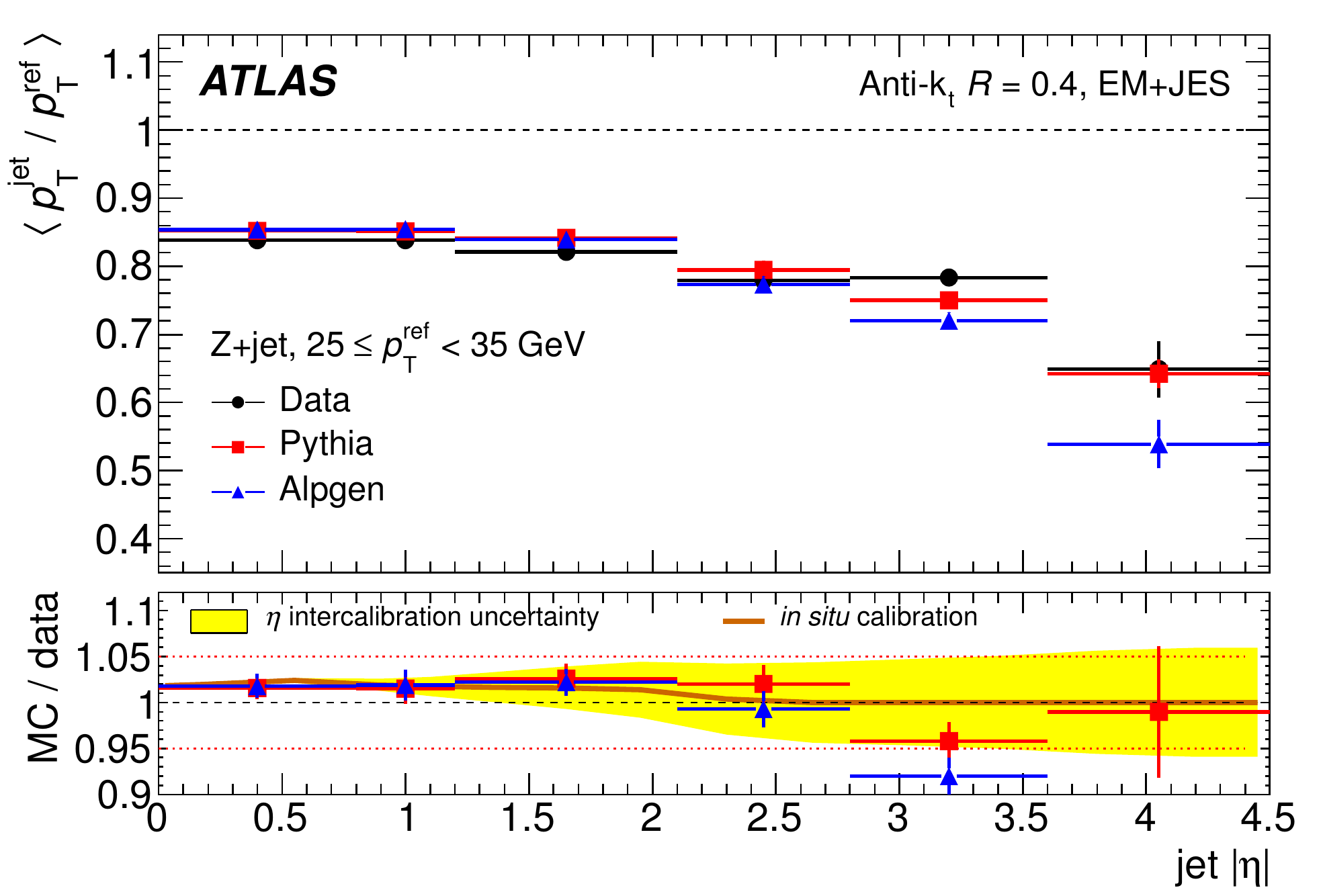}\label{fig:forwardclosure_0}}\quad
    \subfloat[$50\leq\ptref<80$ \GeV]{ \includegraphics[width=.45\textwidth]{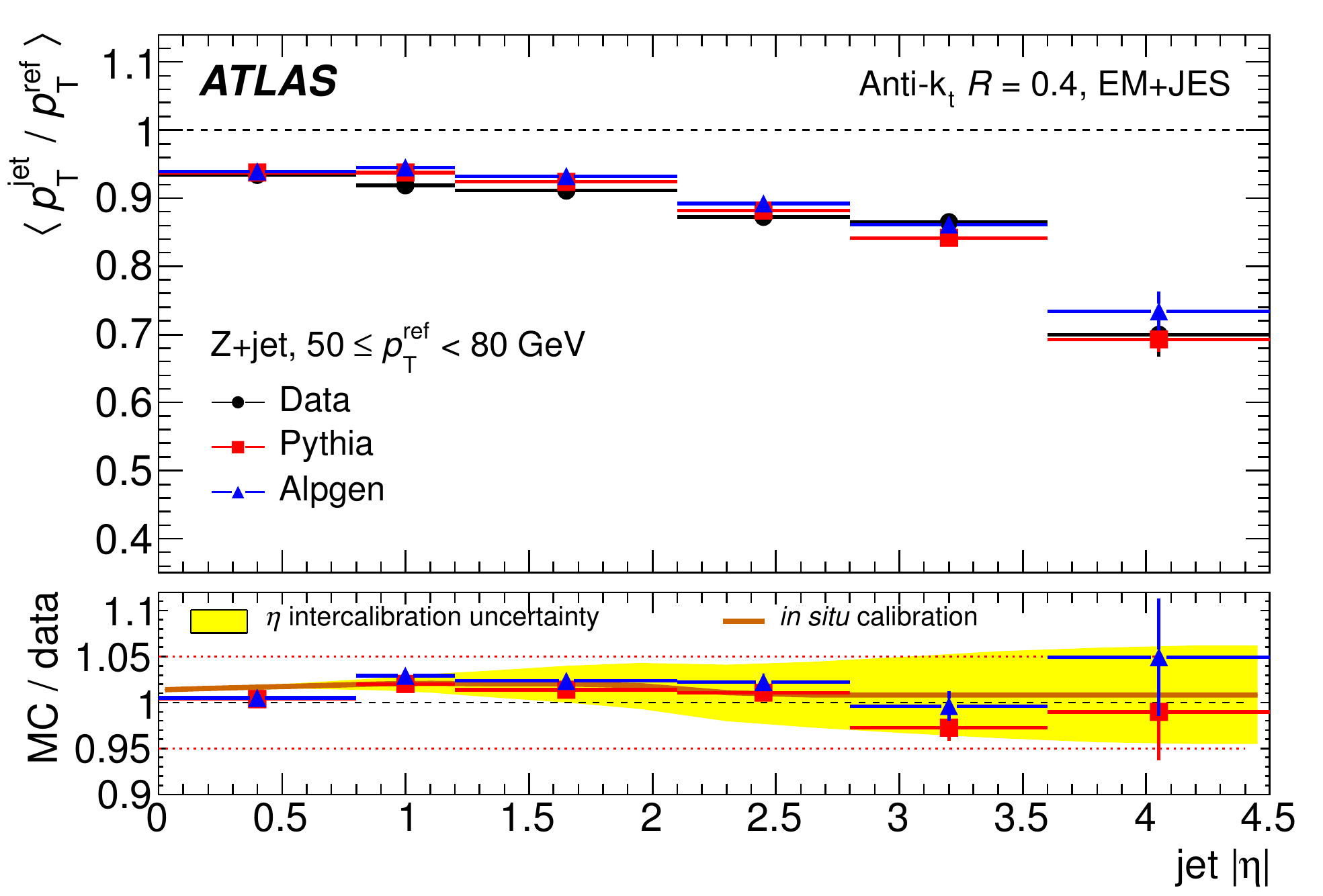}\label{fig:forwardclosure_1}}\\
    \subfloat[$85\leq\pt^\gamma<110$ \GeV]{ \includegraphics[width=.45\textwidth]{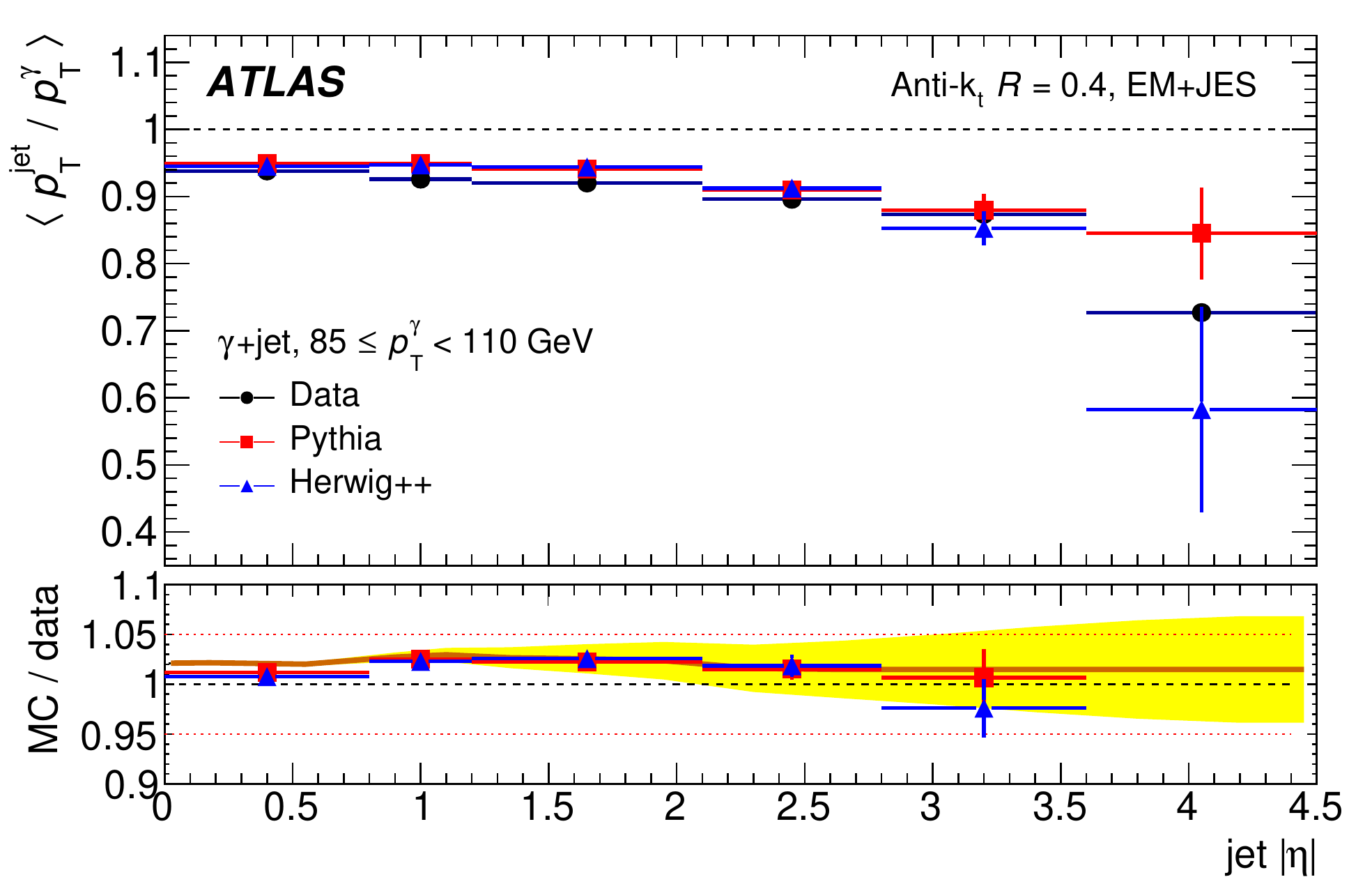}\label{fig:forwardclosure_2}}
    \subfloat[$210\leq\pt^\gamma<260$ \GeV]{ \includegraphics[width=.45\textwidth]{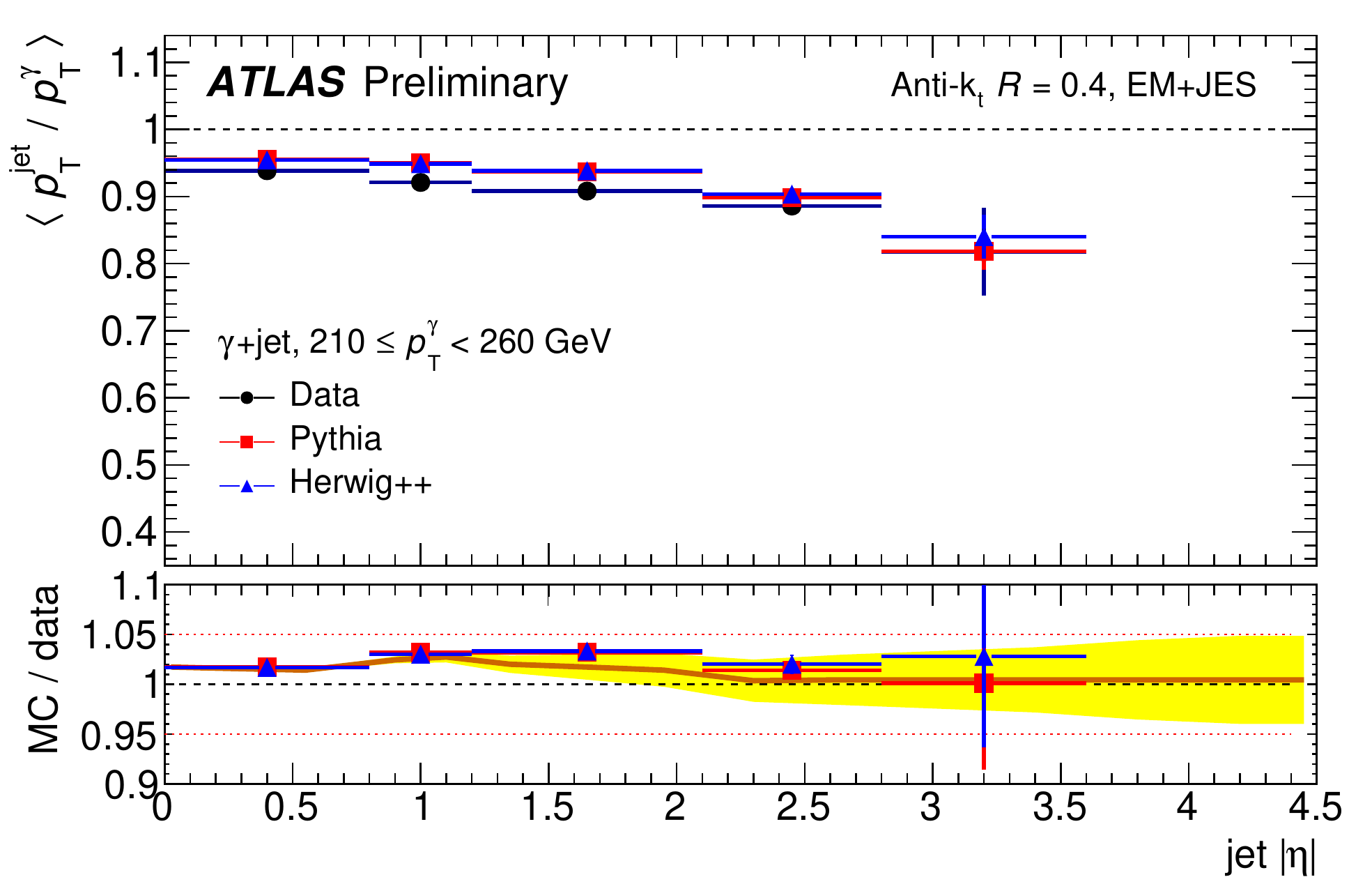}\label{fig:forwardclosure_3}}
  \caption[]{
    The (\subref{fig:forwardclosure_0}, \subref{fig:forwardclosure_1}) \Zjet{} and  (\subref{fig:forwardclosure_2}, \subref{fig:forwardclosure_3}) \gammajet{} balance for \antikt{} jets with $R=0.4$, calibrated with the \EMJES{} scheme. In \subref{fig:forwardclosure_0} and \subref{fig:forwardclosure_1} the results for events with $25\leq\ptref<35$ \GeV and  $50\leq\ptref<80$ \GeV{} are shown, respectively. Here \ptref{} is the \pt{} of the reconstructed \Zboson{} boson projected onto the axis of the balancing jet. The \pT{} balance for \gammajet{} events with photons with transverse momenta \ptgamma{} within $85 \leq \ptgamma < 100$~\GeV{} is shown in \subref{fig:forwardclosure_2}, while \subref{fig:forwardclosure_3} shows the \pT{} balance for higher photon transverse momenta ($210 \leq \ptgamma < 260$~\GeV). As no \insitu{} calibration is applied to these measurements, it is expected that data and \MC{} simulations using \pythia{} are shifted relative to each other by the absolute correction  multiplied by the relative ($\etaDet$ dependent) correction presented herein. The resulting 
\JES{} calibration is shown as a solid line in the lower part of the figures. The dijet modelling uncertainty is shown as a filled band around the \insitu{} correction.
\label{fig:forwardclosure}
  }
\end{figure*}

\subsection{Systematic uncertainties on the multijet balance}
\label{sec:multijetsystematics}
Two main categories of systematic uncertainties are considered. The first category contains those which affect the reference \pt{} of the recoil system. The second category includes those that affect the \MJB{} variables used to probe the leading jet \pt{}, introduced mostly by effects from analysis cuts and imperfect \MC{} modelling of the event.  

The systematic uncertainty on the recoil system includes the following contributions:
\begin{mylist}
\myitem{Absolute JES uncertainty} The standard absolute \JES{} uncertainties obtained from the combination of %
\gammajet{} and \Zjet{} techniques (see \secRef{sec:insitucombination}) are included for each jet composing the recoil system. Figures~\ref{fig:PtBalance_ZJsyst} and \ref{fig:PtBalance_MPFsyst}  show the \MJB{} variations obtained by scaling the \nonleading{} jet energy and momentum scale by $\pm 1 \sigma$ for each of the individual systematic uncertainties in the \gammajet{} and \Zjet{} calibrations, for the four jet calibration schemes. Each source of systematic uncertainty is described in \secRef{sec:Zjetsystematics} and \secRef{sec:systematicgammajetuncertainties}, respectively.
%
In case there are fewer than 10 events in a bin,
the uncertainty is taken to be the RMS of the last bin with more than 10 events divided by the square root of the number of events in that bin. The central value of the ratio is unchanged.

This uncertainty ranges from $0.2\%$ to $0.4\%$ for \Zjet{} and $0.6\%$ to $1.0\%$ for \gammajet{} in the jet \pt{} range of $0.5$--$1.2$~\TeV{} for the two jet sizes of $R = 0.4$ and $0.6$.
\myitem{Relative JES uncertainty} Relative jet response uncertainties evaluated in the 
dijet \etaic{} (\secRef{sec:etaintercaliUncertainty}) are included in a similar manner for each jet with $|\eta|<2.8$ in the recoil system. 
\myitem{Close-by jet uncertainty} The jet response is known to depend on the angular distance to the closest jet in \etaphispace{} \cite{jespaper2010}, and the response variation is expected to be more significant for jets belonging to the recoil system. Any discrepancy between \MC{} simulations and data in describing the jet response with close-by jets therefore results in an additional systematic uncertainty. 
The measurement performed to evaluate the effect and the resulting systematic uncertainty are described in \secRef{sec:close-by}. The close-by jet effect on \MJB, shown in \figRef{fig:MJBsyst_Summary}, is obtained by scaling the jet energy and momentum for each recoil jet using the results in \secRef{sec:close-by}.
\end{mylist}

The flavour composition of the jets could affect the agreement between \MC{} simulations and data, and in principle cause an additional contribution to the JES uncertainty. Previous studies with $2010$ data \cite{jespaper2010}, however, show that the resulting uncertainty on \MJB{} is less than $1\%$, and is therefore ignored in this evaluation of systematic uncertainties.

The jet response is corrected for energy deposited by additional \pp{} collisions in the same bunch crossings using the pile-up offset correction described in \secRef{sec:pileupsection}. The residual pile-up effect on \MJB{} is checked by comparing the \MJB{} values using sub-samples of data and \MC{} simulations with different \Npv{} and $\axing$ values. The result shows that the agreement between \MC{} simulations and data is stable within its statistical uncertainty, and 
therefore an uncertainty due to pile-up is not considered.

The second systematic uncertainty category includes sour\-ces that affect the \MJB{} variable which is used to probe the high-\pt{} jet energy scale. As said earlier,  those are mainly due to effects from analysis cuts or imperfect \MC{} modelling with the following considerations: %
\begin{mylist}
\myitem{Analysis cuts} A systematic uncertainty might be induced by event selection cuts on physical quantities that are not perfectly described by the \MC{} simulation. In order to evaluate this systematic uncertainty, all relevant analysis cuts are varied in a range where the corresponding kinematic variables are not strongly biased and can be examined with small statistical fluctuations (see Table \ref{tab:AnalysisSys} for the range of variation). For each value of the cuts, the ratio of the value of \MJB{} 
in data and simulation is evaluated. The maximum relative deviation of this ratio from the default value is taken as the systematic uncertainty from the source under consideration.  
\myitem{Jet rapidity acceptance} The analysis uses only jets with $|y|<2.8$ in order to reduce the impact of the large \JES{} uncertainties in the forward region. This selection, however, can cause additional systematic uncertainty because the fraction of jets produced outside the rapidity range can be different in the data and \MC{} simulations, and hence affect the \MJB{} values. This effect is checked, as is done in Ref.~\cite{jespaper2010}, by looking at the \MJB{} for events with $\ptrecoil>210$~\GeV, as a function of the total transverse energy (\sumet) summed over all jets with $|y|>2.8$. The majority of events have a very small \sumet{} and the effect turns out to be negligible.
\myitem {Underlying event, fragmentation and ISR/FSR modelling} Imperfect modelling of the UE, fragmentation and ISR/FSR may influence the multijet balance by affecting variables used to select events and kinematic properties of the leading jet and the recoil system. 
The systematic uncertainty for each of the mentioned sources is estimated by evaluating the \datatomc{} ratio of the \MJB, measured using the default simulation based on \pythia{} and simulations using alternative \MC{} generators. For the systematic uncertainty contribution from fragmentation, the \herwigpp{} samples are used as an alternative. For the underlying event and radiation modelling systematics, the \pythia{} \Perugia{} $2011$ \cite{Perugia2010} samples are used. The systematic uncertainty introduced by these effects is $2~\%$ or smaller in all cases except the lowest \ptrecoil{} bins below $300$~\GeV.
\end{mylist}

All systematic uncertainties due to the analysis cuts and event modelling, and the total uncertainty obtained by summing them in quadrature, are shown as a function of \ptrecoil{} in \figRef{fig:MJBsyst_Summary} for jets with $R = 0.4$ and $R = 0.6$, calibrated with the \EMJES{} and \LCWJES{} schemes. The uncertainties due to dijet \etaic{} and close-by jet effects are also included in the figure as well as the total uncertainty. 
Representative values of the uncertainties in the \ptrecoil{} range between $0.5$ and $1.2$~\TeV{} are summarised in Table \ref{tab:SystSumm}. 

The summary of all systematic uncertainties associated with
the multijet balance technique and the propagated uncertainties from the \gammajet{} and \Zjet{}
\insitu{} techniques
overlaid on the \datatomc{} ratio of the multijet balance, is shown in \figRef{fig:JES_Summary}, for \antikt{} jets with the distance parameters $R = 0.4$ and $0.6$.
The \JES{} uncertainty is
determined more precisely at jet \pt{} below $\sim0.6$~\TeV{} by the \gammajet{} and \Zjet{} calibrations than the \MJB{} calibration.

\subsection{Summary of multijet analysis}
\label{sec:multijetsummary}
The multijet balance technique is used to probe the jet energy scale in the \TeV{} region for \antikt{} jets with 
distance parameters $R = 0.4$ and $R = 0.6$. 
Exploiting the \pt{} balance between the highest-\pt{} jet and the recoil system 
composed of jets corrected by the \gammajet{} and \Zjet{} calibrations allows the extension of the \insitu{} \JES{} determination to 
higher \pt{}, beyond the range covered by the \gammajet{} calibration.
Propagating systematic uncertainties associated with the \gammajet{}, \Zjet{} and dijet calibrations as well as the systematic uncertainty 
due to the knowledge of the recoil system transverse momentum in the \MJB{} method (including the close-by jet uncertainty), the total systematic uncertainties 
for the \gammajet{}, \Zjet{} and \MJB{} calibration methods 
are obtained to be about $0.6 \%$, $0.3 \%$ and $1.5 \%$ respectively, for jets with $\pt = 1$~\TeV. 
At high transverse momentum, the main contribution to the systematic uncertainty is due to the uncertainty on the \MJB{} calibration. 
Considering the statistical uncertainty of the \MJB{} calibration based on the $2011$ data, the high-\pt{} jet energy scale is 
validated at $\pt>500$~\GeV{} within 
$2.4 \%$ ($2.0 \%$) and $2.2 \%$ ($3.0 \%$) up to $1.2$~\TeV{} for \antikt{} jets with $R = 0.4$ and $R = 0.6$, both calibrated with the \EMJES{} (\LCWJES) scheme.

\section[Forward-jet energy measurement validation using \Zjet{} and \gammajet{} data]{Forward-jet energy measurement validation using \ZJET{} and \GAMMAJET{} data}
\label{sec:closure}
To test the performance of the forward-jet calibration derived in \secRef{sec:etaintercalibration}, 
this calibration
is applied to all jets in the original dataset and the full analysis is repeated.
The resulting intercalibration results are within $0.3\%$ of unity 
across the full $(\ptavg,\etaDet)$ phase space 
in which the calibration is derived, both for jets with $R=0.4$ and $R=0.6$, 
and for the \EMJES{} and \LCWJES{} calibrations.
The measured relative response for two representative bins of $\ptavg$ is shown in \figRef{fig:closure}.

Similar to the analyses described in \secRef{sec:ZjetInSitu} and \secRef{sec:gammajetInSitu}, the balance between a \Zboson{} boson decaying to an electron--positron pair and a recoiling forward jet,
and the balance between a photon and a forward jet, are used to study the jet response in the forward direction.
The results for \Zjet{} and \gammajet{}, as shown in \figRef{fig:forwardclosure}, agree with the calibrations
and uncertainty derived from the dijet analysis.

The \Zjet{} study also includes predictions from the \alpgen{} generator, which uses 
\herwig{} for parton shower and fragmentation into particles (see \secRef{sec:MC} for generator configuration details). 
The \alpgen+\herwig{} response predictions generally agree with the expectations within the modelling uncertainty
of this analysis (see \secRef{sec:modelUncert}).
The \gammajet{} results include comparisons with \pythia{} events, 
generated with the same tune and version as the \pythia{} dijet samples used in this analysis,
and a sample produced with \herwig{}, using the already mentioned \ATLAS{} {\sc AUET2B MRST LO**} tune and the {\sc MRST LO**} PDF set (see \secRef{sec:MC}).

\begin{table*}[htp!]
\renewcommand{\arraystretch}{\myarraystretch}
\begin{center}
\caption{Summary of the number of events available for various \insitu{} techniques after all selection cuts.
The numbers are given for illustration in specific \ptjet{} ranges for \antikt{} jets with $R=0.4$ reconstructed
with the \EMJES{} scheme. The \gammajet{} results are based on the MPF method. 
}
\vskip10pt
\begin{tabular}{c|ccccc}
\hline
\hline
 \Zjet{} method        &              &                 &                  &                  &  \\ 
 \ptjet & $20$--$25$ \GeV & $35$--$45$ \GeV & $210$--$260$ \GeV  &                  &  \\
 Number of events      &     $8\,530$   &     $8\,640$      &     $309$        &                  &  \\ 
\hline
\gammajet{} method     &              &                 &                  &                  &  \\ 
 \ptjet & $25$--$45$ \GeV & $45$--$65$ \GeV & $210$--$260$ \GeV  & $600$--$800$ \GeV   &  \\
Number of events       &  $20\,480$     &     $61\,220$     &     $10\,210$       &      $100$        &  \\ 
\hline
 multijet method       &              &                 &                  &                  &     \\ 
 \ptjet                &              &                 & $210$--$260$ \GeV & $750$--$950$ \GeV & $1.45$--$1.8$ \TeV\\
 Number of events      &              &                 &     $2\,638$       &      $3\,965$      &         $48$ \\ 
\hline
\hline
\end{tabular}
\label{table:numberofevents}
\end{center}
\end{table*}

\section{Jet energy calibration and uncertainty combination}
\label{sec:strategy}

\begin{figure*}[!tpb]
\begin{center}
\subfloat[\EMJES]{\includegraphics[width=0.49\textwidth]{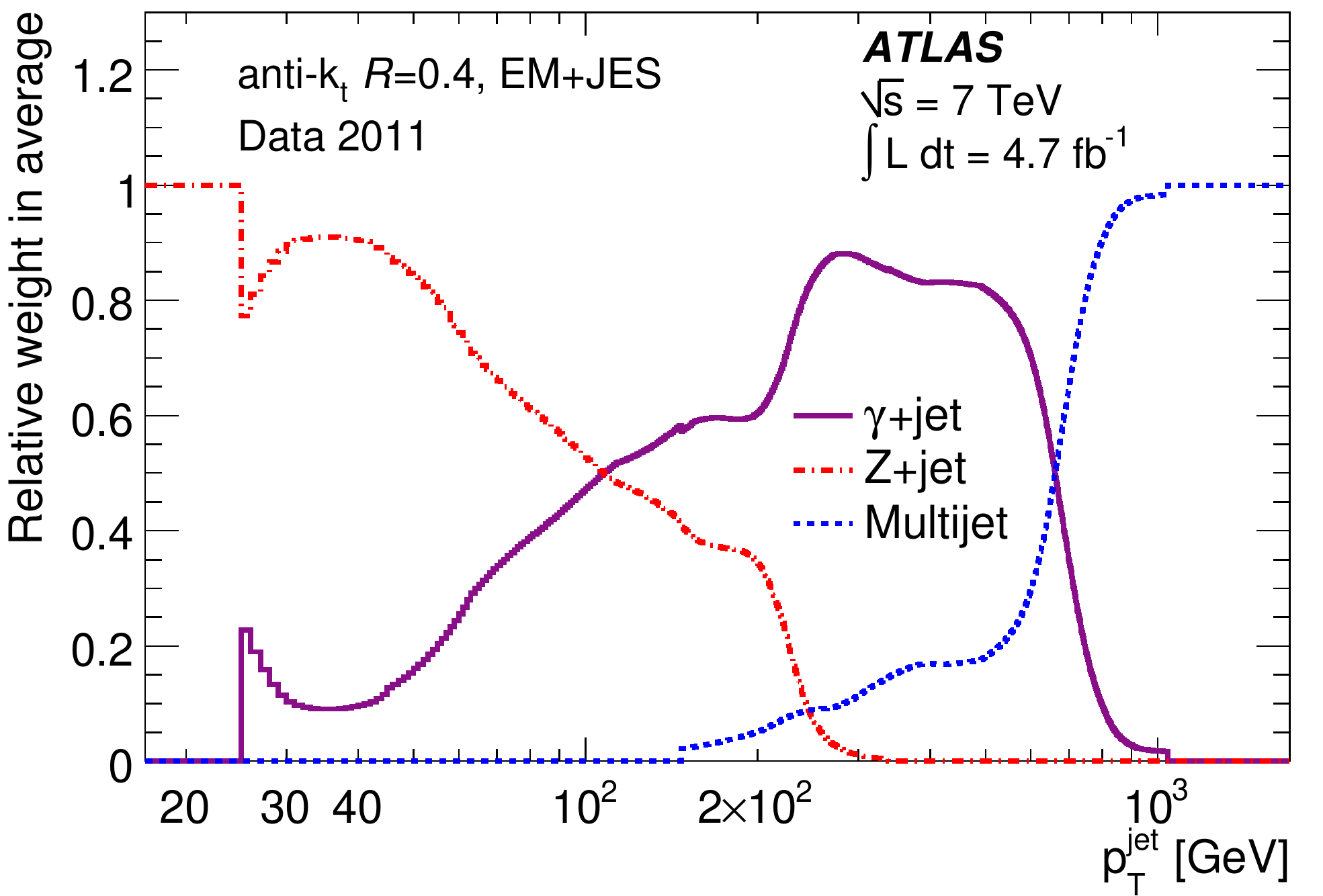}\label{fig:weightinsituEM}}
\subfloat[\LCWJES]{\includegraphics[width=0.49\textwidth]{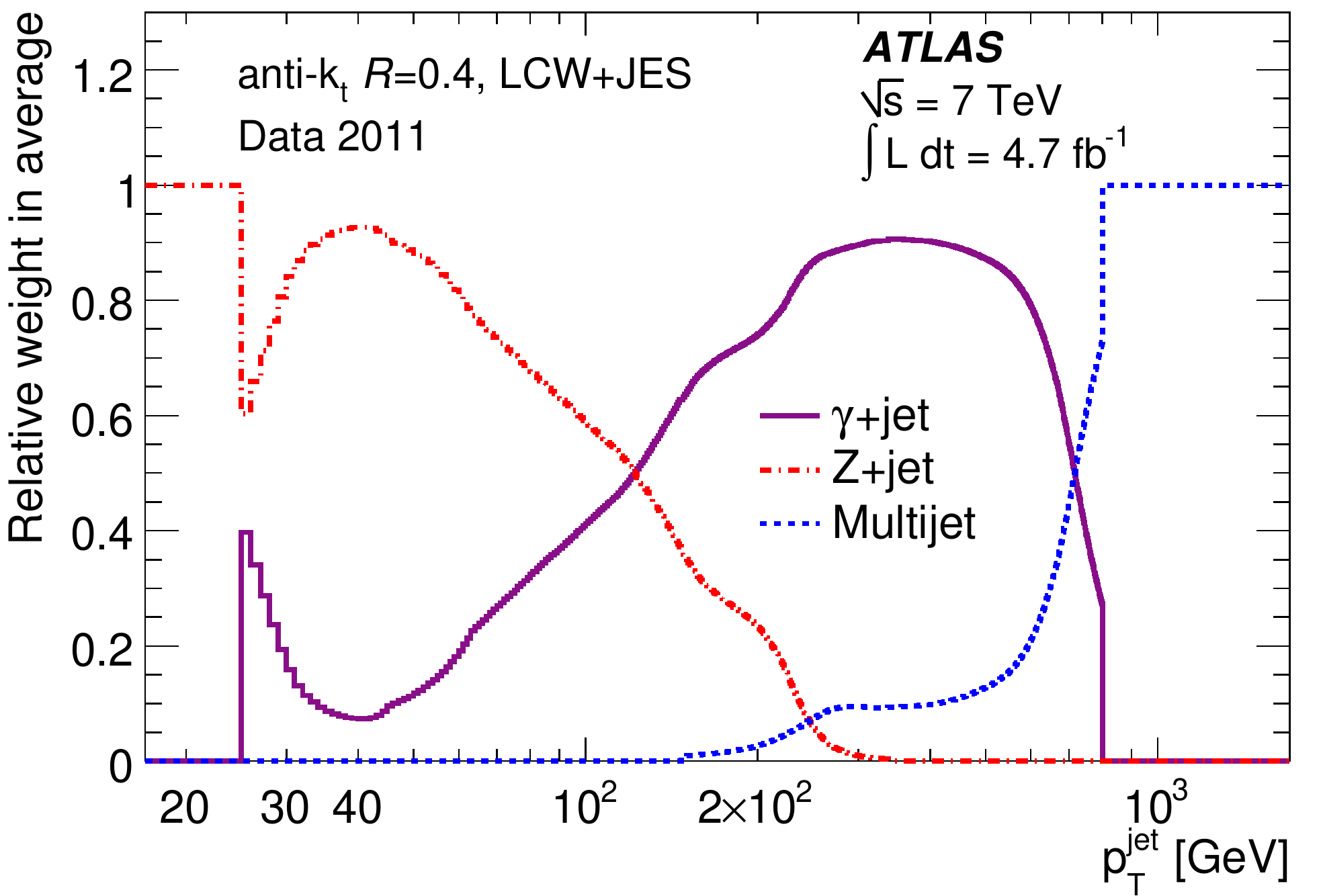}\label{fig:weightinsituLCW}}
\caption[]{%
Weight carried by each \insitu{} technique in the combination to derive the residual jet energy scale calibration
as a function of the jet transverse momentum \ptjet{} for \antikt{} jets with $R = 0.4$ calibrated with the  \subref{fig:weightinsituEM} \EMJES{} and the \subref{fig:weightinsituLCW} \LCWJES{} scheme.
The \ptjet{} dependence of the weights is discussed in \secRef{sec:combinationresults}.
\label{fig:weightinsitu}}
\end{center}
\end{figure*}

\begin{figure*}[!tpb]
\begin{center}
 \subfloat[\EMJES{}  \Zjet~DB   ]{\includegraphics[width=0.49\textwidth]{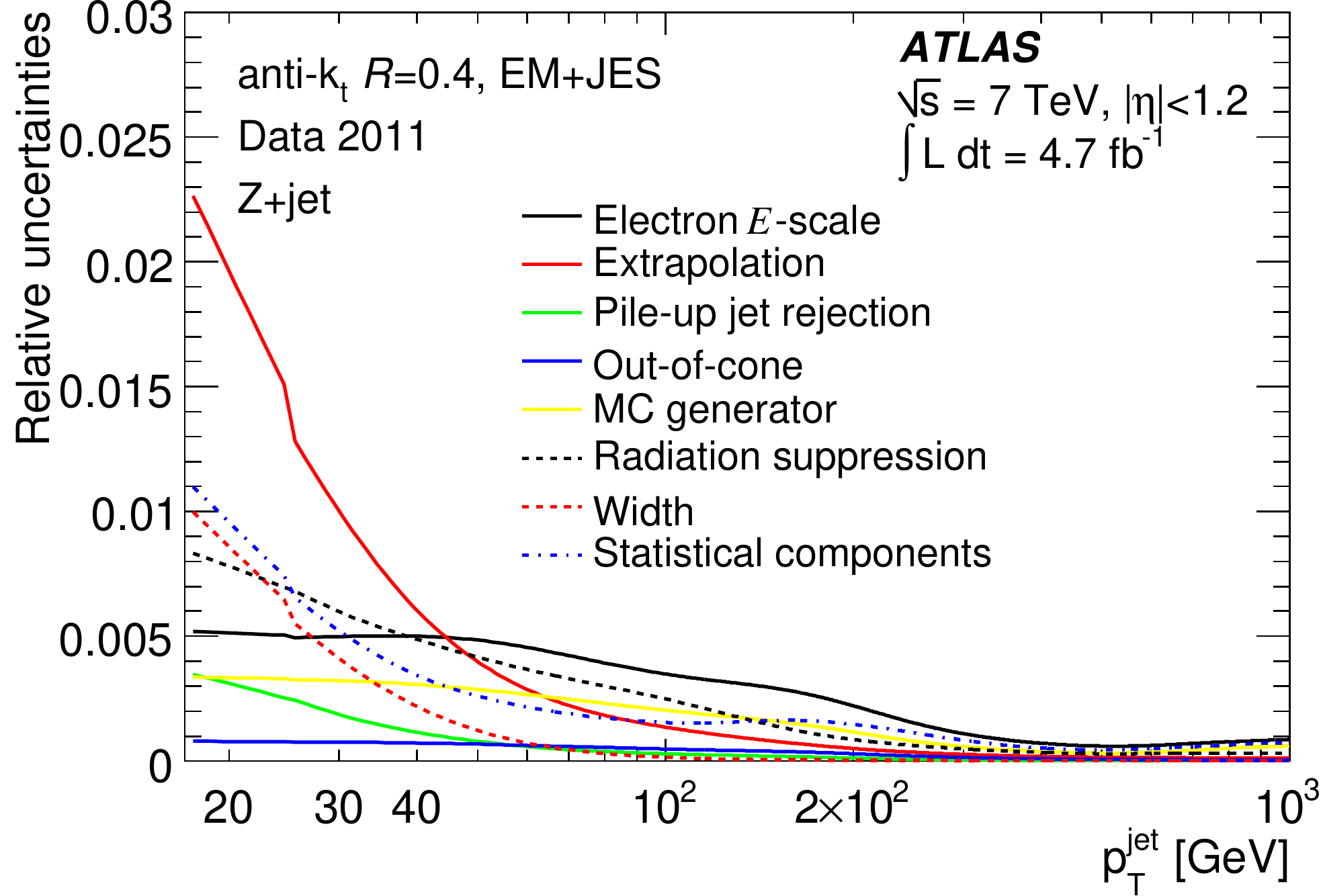} \label{fig:np_EMJES_R4_ZDB}}
 \subfloat[\LCWJES{} \Zjet~DB]{\includegraphics[width=0.49\textwidth]{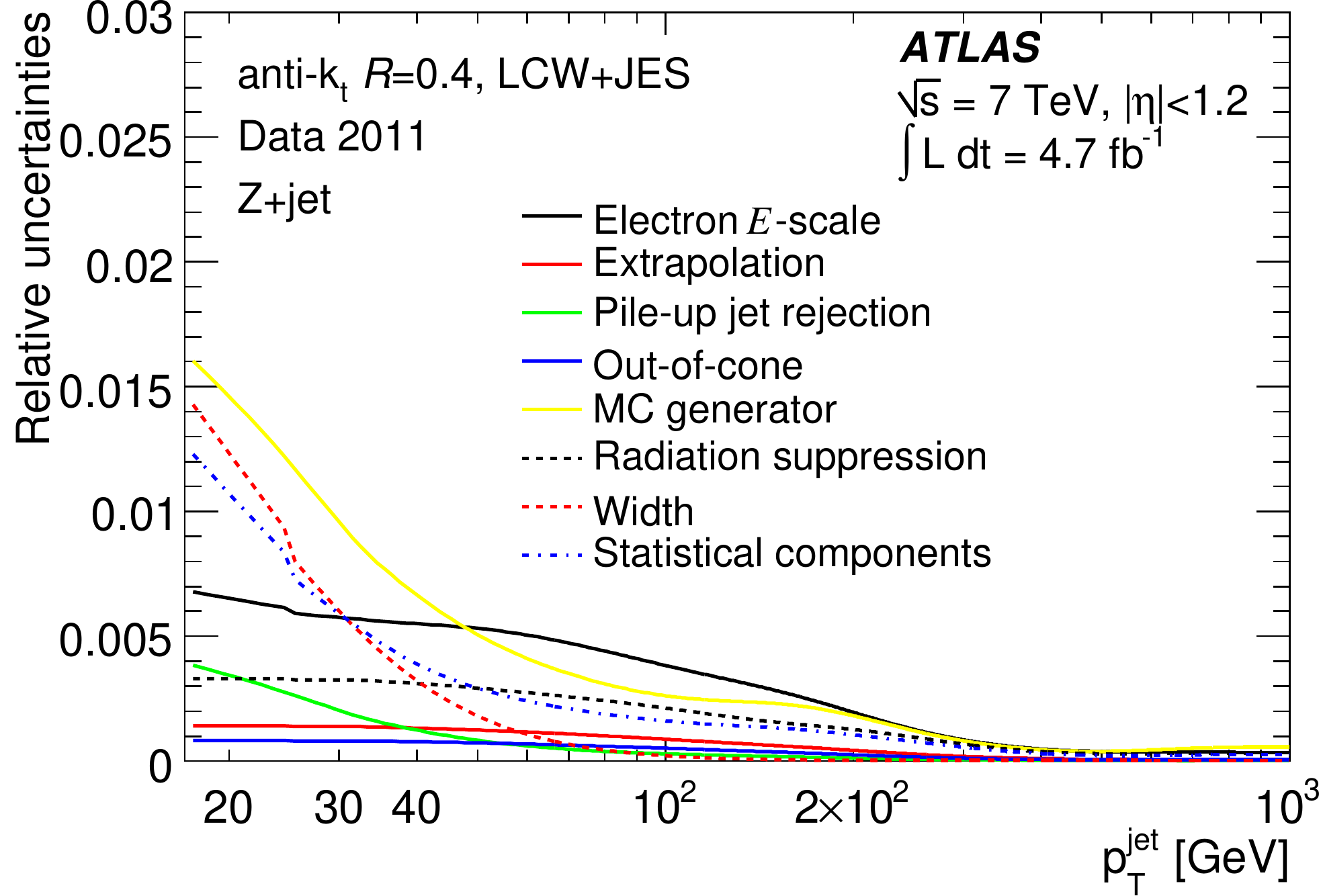} \label{fig:np_LCWJES_R4_ZDB}} \\
 \subfloat[\EMJES{}  \gammajet~\MPF     ]{\includegraphics[width=0.49\textwidth]{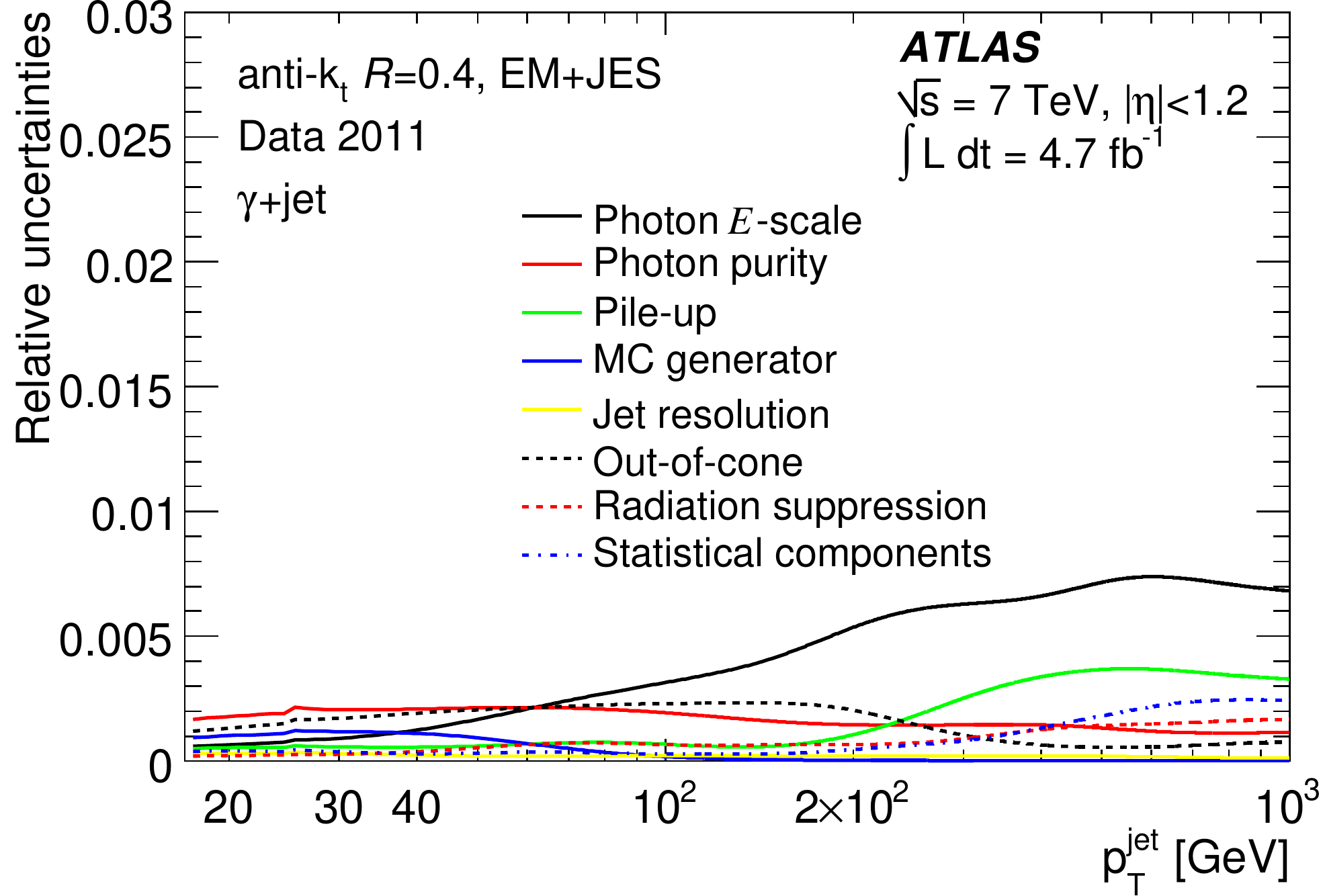} \label{fig:np_EMJES_R4_MPF}}
 \subfloat[\LCWJES{} \gammajet~\MPF]{\includegraphics[width=0.49\textwidth]{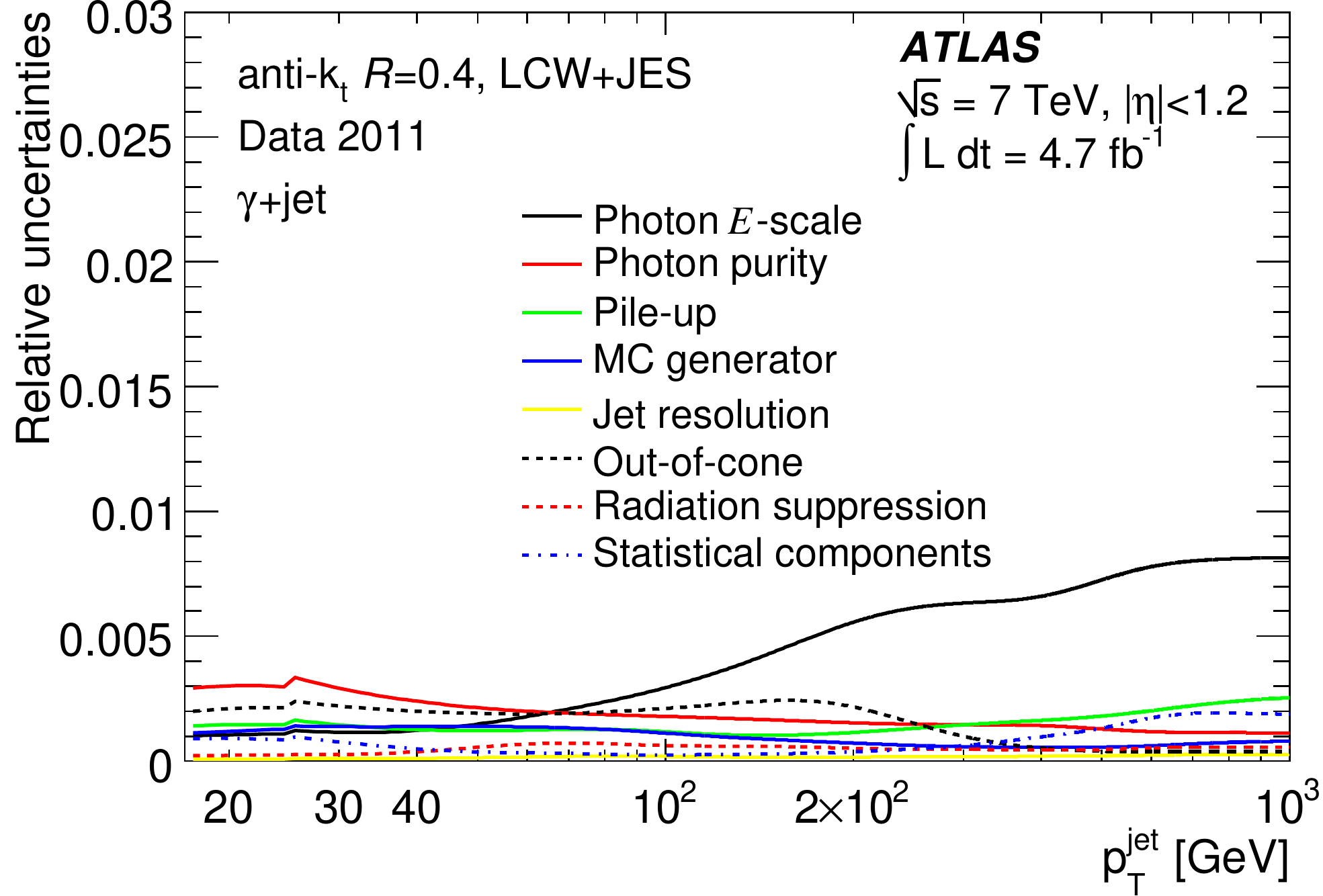} \label{fig:np_LCWJES_R4_MPF}}\\
 \subfloat[\EMJES{} Multijet]{\includegraphics[width=0.49\textwidth]
{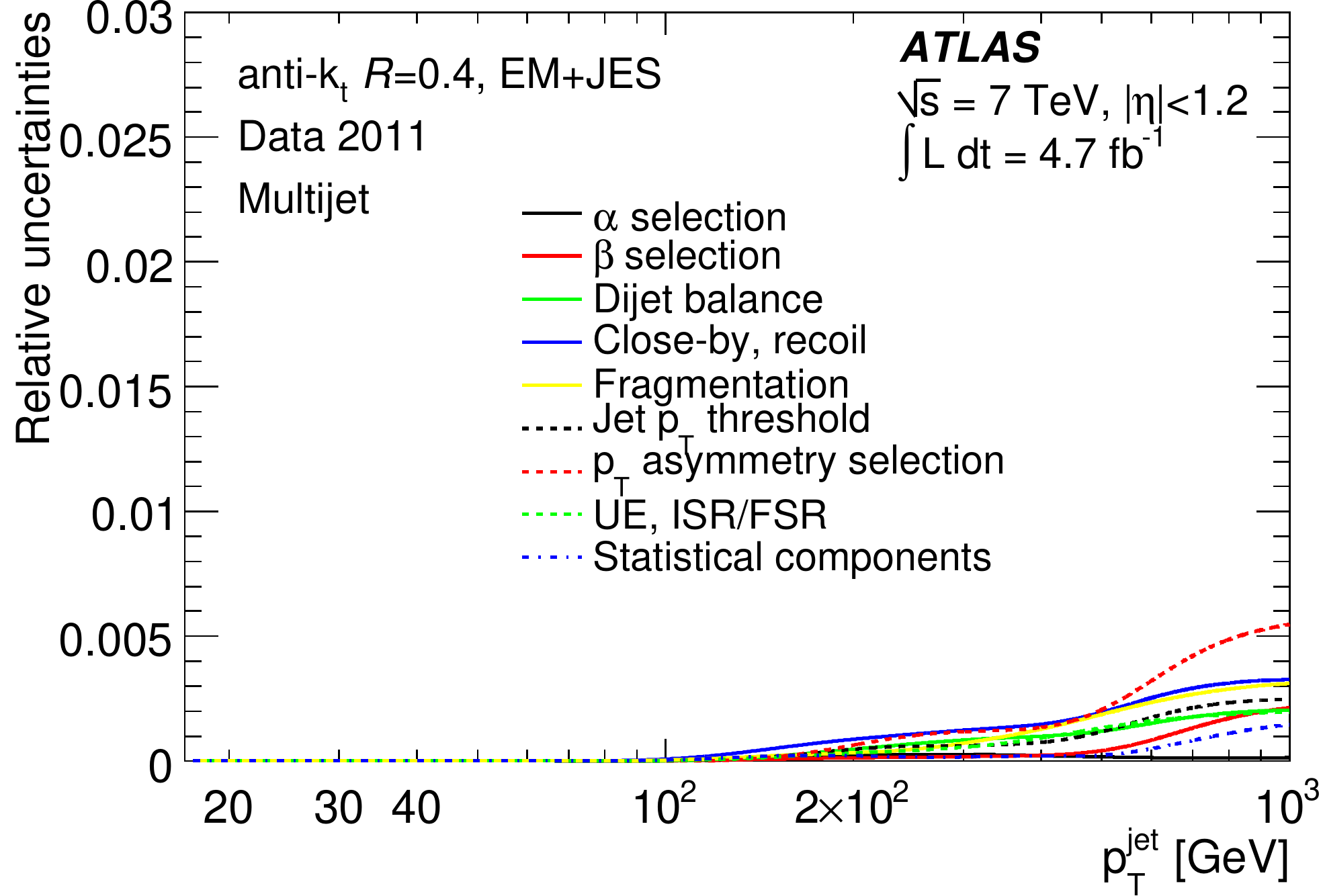} \label{fig:np_EMJES_R4_MJB}} 
 \subfloat[\LCWJES{} Multijet]{\includegraphics[width=0.49\textwidth]{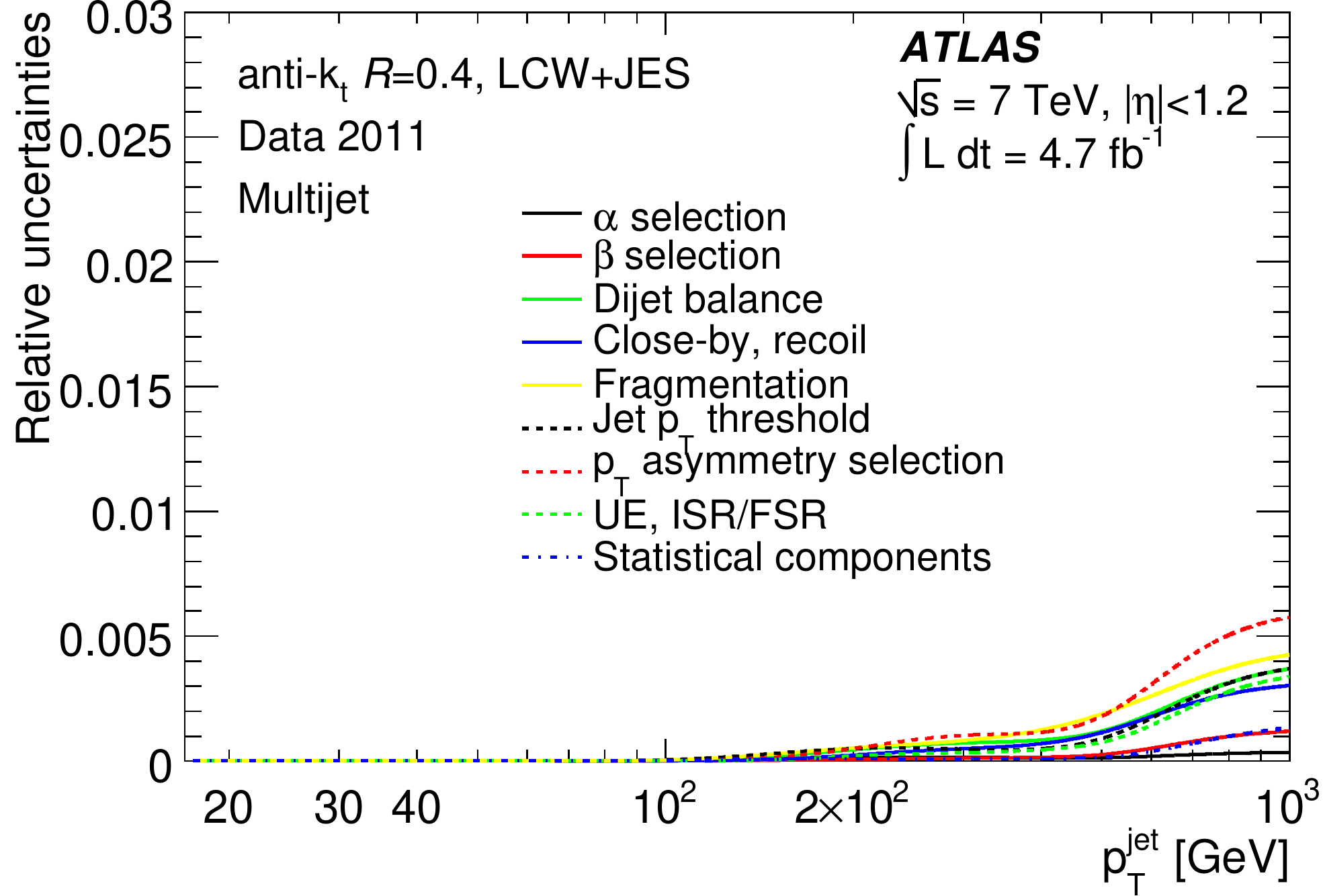} \label{fig:np_LCWJES_R4_MJB}}
 \caption[]{%
   Individual uncertainty sources applicable to the combined response ratio
   as a function of the jet \pt{} for the three \insitu{} techniques:
    (\subref{fig:np_EMJES_R4_ZDB},\subref{fig:np_LCWJES_R4_ZDB}) \Zjet{} direct balance,  (\subref{fig:np_EMJES_R4_MPF},\subref{fig:np_LCWJES_R4_MPF}) \gammajet{} \MPF{} and (\subref{fig:np_EMJES_R4_MJB},\subref{fig:np_LCWJES_R4_MJB}) multijet balance for
   \antikt{} jets with $R=0.4$ calibrated with the (\subref{fig:np_EMJES_R4_ZDB},\subref{fig:np_EMJES_R4_MPF},\subref{fig:np_EMJES_R4_MJB}) \EMJES{} and the (\subref{fig:np_LCWJES_R4_ZDB},\subref{fig:np_LCWJES_R4_MPF},\subref{fig:np_LCWJES_R4_MJB}) \LCWJES{}
  scheme. The systematic uncertainties displayed here correspond to the 
   components listed in Table~\ref{tab:nuissanceparameters}.
   \label{fig:nuisanceparameters}}
\end{center}
\end{figure*}

\begin{figure*}[!tpb]
\begin{center}
 \subfloat[\EMJES]{\includegraphics[width=0.49\textwidth]{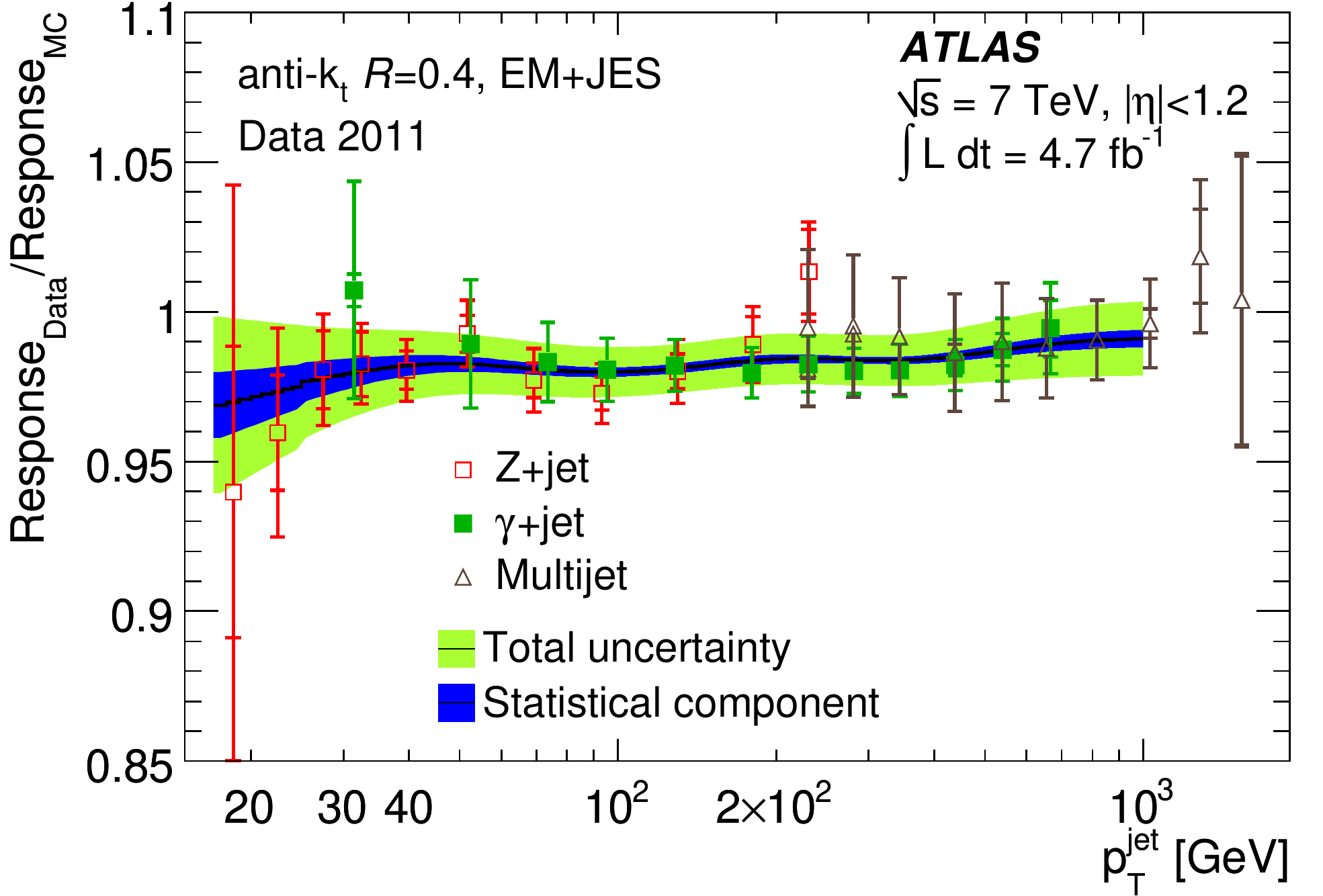}\label{fig:responseratioinsituEM}}
 \subfloat[\LCWJES]{\includegraphics[width=0.49\textwidth]{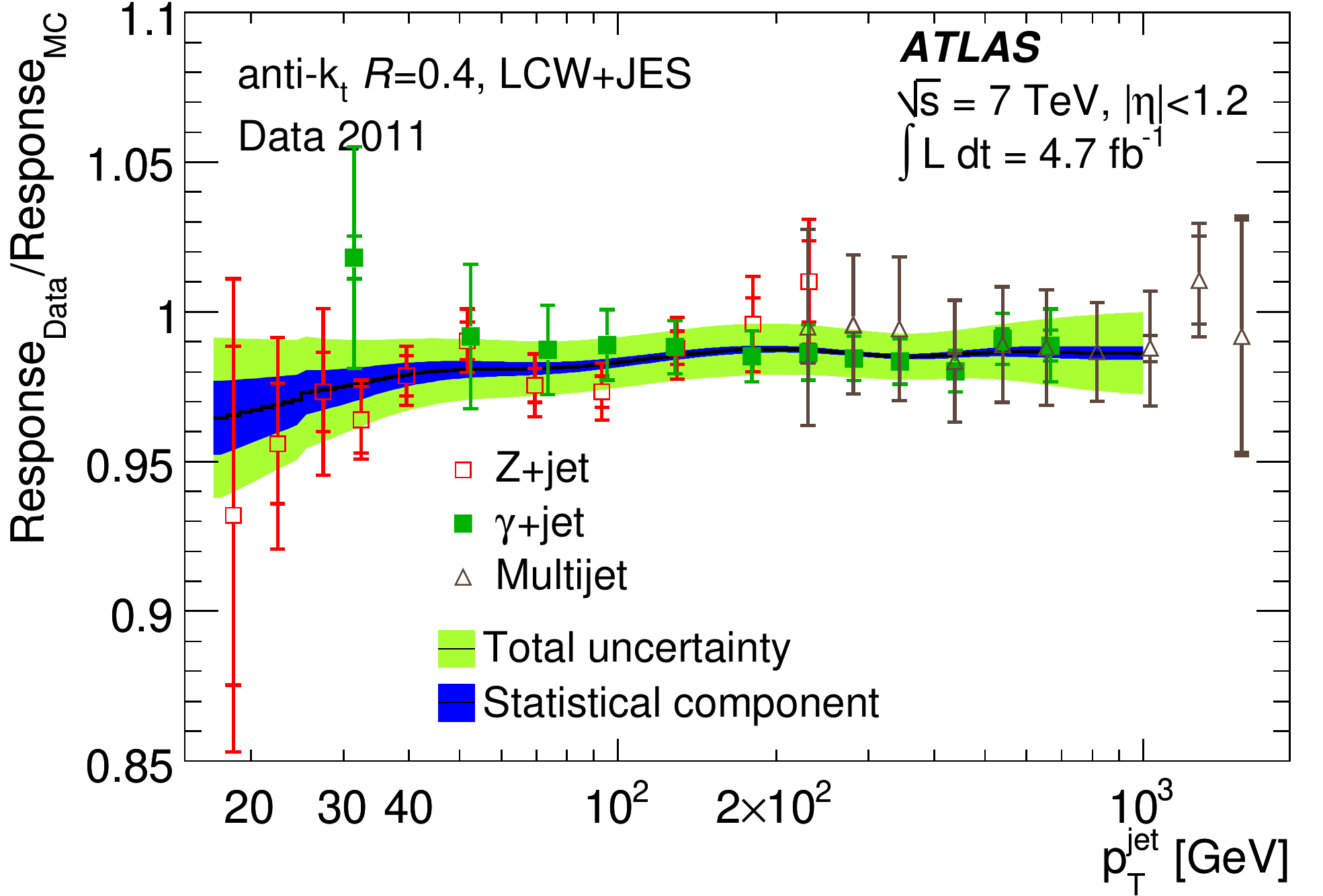}\label{fig:responseratioinsituLCW}}
\caption[]{Ratio of the average jet response $\langle \ptjet/\ptref \rangle$ measured in data to that measured 
in \MC{} simulations
for jets within $|\etajet| < 1.2$ as a function of the transverse jet momentum \ptjet{}.
The \datatomc{} jet response ratios are shown 
 separately for the three \insitu{} techniques used in the combined calibration:
direct balance in \Zjet{} events, MPF in \gammajet{} events, and multijet \pt{} balance in inclusive jet events.
The error bars indicate the statistical and the total uncertainties~(adding in quadrature statistical and systematic uncertainties).
Results are shown for \antikt{} jets with $R = 0.4$
calibrated with the \subref{fig:responseratioinsituEM} \EMJES{} and the \subref{fig:responseratioinsituLCW} \LCWJES{} scheme.
The light band indicates the total uncertainty
from the combination of the \insitu{} techniques.
The inner dark band indicates the statistical component only.
\label{fig:responseratioinsitu}}
\end{center}
\end{figure*}

\begin{table*}[htp!]
\renewcommand{\arraystretch}{\myarraystretch}
\caption{
  Summary table of the uncertainty components for each \insitu{} technique 
(\Zjet{} (see Section~\ref{sec:ZjetInSitu}), \gammajet{} (see Section~\ref{sec:gammajetInSitu}),
and multijet \pt{} balance (see Section~\ref{sec:multijet}) used to derive the jet energy scale 
uncertainty.
Shown are the $21$ systematic uncertainty components together with the $11$, $12$ and $10$
statistical uncertainty components for each \insitu{} technique.
  Each uncertainty component is categorised depending on its source as either detector (\npdet), physics modelling (\npmodel), mixed detector and modelling (\npmixed), or as statistics and method (\npstatmeth).
}\begin{center}
\begin{tabular}{|l|l|l|}
\hline
Name                  & Description & Category\\
\hline
\textbf{Common sources}        & &\\
Electron/photon $E$ scale & electron or photon energy scale & \npdet \\ 
\hline
\DBbf{} \Zjet{} \pt{}  balance& &\\
\MC{} generator	      &  \MC{} generator difference between \alpgen/\herwig{} and \pythia{} & \npmodel \\
Radiation suppression & radiation suppression due to second jet cut  & \npmodel \\
Extrapolation         & extrapolation in \deltaphijetZ{} between jet and Z boson& \npmodel \\
Pile-up jet rejection & jet selection using jet vertex fraction  & \npmixed \\
Out-of-cone           & contribution of particles outside the jet cone & \npmodel \\
Width	              & width variation in Poisson fits to determine jet response& \npstatmeth \\
Statistical components & statistical uncertainty for each of the $11$ bins & \npstatmeth \\
\hline 
\MPFbf{} \gammajet{} \pt{} balance (\MPF)   & &\\
\MC{} generator          & \MC{} generator difference \herwig{} and \pythia{} & \npmodel \\
Radiation suppression & radiation suppression due to second jet cut  & \npmodel \\
Jet resolution        & variation of jet resolution within uncertainty & \npdet \\
Photon Purity         & background response uncertainty and photon purity estimation & \npdet \\
Pile-up               & sensitivity to pile-up interactions & \npmixed \\
Out-of-cone           & contribution of particles outside the jet cone & \npmodel \\
Statistical components & statistical uncertainty for each of the $12$ bins & \npstatmeth \\
\hline 
\MJBbf{} Multijet \pt{} balance  & &\\
$\alpha$ selection    & angle between leading jet and recoil system & \npmodel \\
$\beta$  selection    & angle between leading jet and closest sub-leading jet & \npmodel \\
Dijet balance         & dijet balance correction applied for \AetaRange{2.8}& \npmixed \\
Close-by, recoil      & \JES{} uncertainty due to close-by jets in the recoil system& \npmixed \\
Fragmentation         & jet fragmentation modelling uncertainty & \npmixed \\
Jet \pt{} threshold   & jet \pt{} threshold & \npmixed \\
\pt{} asymmetry selection & \pt{} asymmetry selection between leading jet and sub-leading jet& \npmodel \\
UE,ISR/FSR            & soft physics effects modelling: underlying event and soft radiation & \npmixed\\
Statistical components & statistical uncertainty for each of the $10$ bins & \npstatmeth \\
\hline
\end{tabular}
\label{tab:nuissanceparameters}
\end{center}
\end{table*}

%
\subsection[Overview of the combined \JES{} calibration procedure]{Overwiew of the combined JES calibration procedure}
After the first JES calibration step described in \secRef{sec:jetrecocalib}, 
the jet transverse momenta \ptjet{} in data and \MC{} simulation are compared 
using \insitu{} techniques that exploit the balance\footnote{As for all \pt{} balance evaluations between a reference and a probe object,  the expectation value of this balance is not unity, due to physics effects (e.g., ISR) and jet reconstruction inefficiencies (e.g., out-of-cone energy losses). The ability of the \MC{} simulation to reproduce all of these effects is further discussed in the context of the evaluation of the systematic uncertainties in  \secRef{sec:insitutechniquesuncertainties}.}  between
\ptjet{} and the \pt{} of a reference object (\ptref):
\begin{equation}
  \label{eq:response_ratio}
 \mathcal{R}(\ptjet,\eta) = \dfrac{\langle\ptjet/\ptref\rangle_{\mathrm{data}}}{\langle\ptjet/\ptref\rangle_{\mathrm{MC}}}
\end{equation}
The inverse of this quantity is the residual %
\JES{} correction factor for jets measured in data, and thus reflects the final \JES{} calibration in \ATLAS. It is derived from corrections individually described in \secRef{sec:InSitu}. The sequence of these corrections is briefly summarised again below, with references to the corresponding more detailed descriptions:
\begin{enumerate}
\item Apply \etaic{} to remove the \etaDet{} dependence of the detector response to jets within \etaRange{0.8}{4.5} by equalising it with the one for jets within $|\etaDet|<0.8$ (see \secRef{sec:etaInterCalib}).
\item Apply the absolute %
correction, as derived using a combination of the \Zjet{} (\secRef{sec:ZjetInSitu}) and the \gammajet{} (\secRef{sec:gammajetInSitu}) methods, to the the central jet response ($|\etaDet|<1.2$). The slightly larger  \etaDet{} range used here, compared to the one used in \etaic, provides more statistics while keeping systematic uncertainties small. The corresponding combined \JES{} uncertainty is determined  from the uncertainties of each of these techniques, as presented in detail in \secRef{sec:insitutechniquesuncertainties}. The absolute scale correction, together with its systematic uncertainties, is also evaluated for jets in the end-cap and forward detector region ($|\etaDet| \ge 1.2$), and accordingly applied to those as well.
\item Jets with energies in the \TeV{} regime are calibrated using the multijet transverse momentum balance technique (\MJB{} in \secRef{sec:multijet}). The lower-\pt{} jets are within $|\etaDet|<2.8$, while the leading jet is required to be within $|\etaDet|<1.2$. The uncertainties derived from \gammajet{}, \Zjet{} and dijet \pt{} balance for the lower-\pt{} jets are propagated to the higher-\pt{} jets (\secRef{sec:multijetsystematics}). 
\end{enumerate}
The \insitu{} \JES{} calibration and the corresponding \JES{} uncertainty for central jets ($|\etaDet|<1.2|$) are hence derived by  a combination of the \datatomc{} ratios $\mathcal{R}$, individually determined as given in \eqRef{eq:response_ratio}, obtained from the \gammajet, \Zjet{} and \MJB{} correction methods.  The \JES{} uncertainties for forward jets 
$1.2<|\etaDet|<4.5$  are then derived from those for central jets using the dijet \etaic{} technique. 

Table \ref{table:numberofevents} summarises the number of events available for each correction method in various kinematic bins. Details of the combination method, including  the full evaluation of the systematic uncertainties and its underlying components (nuisance parameters), are further explained in the remainder of this section. 

\label{sec:insitucombination}
\subsection{Combination technique}
\label{sec:combinationtechnique}
The \datatomc{} response ratios (see Eq.~\ref{eq:response_ratio}) of the various \insitu{} methods are combined 
using the procedure described in Ref. \cite{jespaper2010}.
The \insitu{} jet response measurements are made in bins of \ptref{} and within $|\etaDet|<1.2$, and are evaluated at the barycentre $\langle\ptref\rangle$ of each \ptref{} bin, for each \etaDet{} range.\footnote{Since $\langle\ptjet/\ptref\rangle$ is close to unity for all $\ptref$ bins, the bin barycentre $\langle\ptref\rangle$ is a good approximation of $\langle\ptjet\rangle$. In the following \ptjet{} is used.}

First, a common, fine \pT{} binning is introduced for the combination of methods. In each of these \pT{} bins, and for each \insitu{} method that contributes to that bin, the \datatomc{} response ratio is determined using interpolating splines based on second-order polynomials. The combined \datatomc{} ratio  \rrextrapfct{\langle\ptjet\rangle}{\etaDet}{} is then determined by the weighted average of 
the interpolated contributions from the various methods. The weights are obtained by a $\chi^{2}$ minimisation of the response ratios in each \pt{} bin, and are therefore proportional to the inverse of the square of the uncertainties of the input measurements.
The local $\chi^{2}$ is also used to test the level of agreement between the \insitu{} methods.

Each uncertainty source of the \insitu{} methods is treated as fully correlated across \pt{} and \etaDet, while the individual uncertainty sources 
inside a method and between the methods
are assumed to be independent of each other. The full set of uncertainties is propagated from the \insitu{} methods to the combined result in each \pt{} bin using pseudo-experiments \cite{jespaper2010}.
For some applications like the combination and comparison of several experimental measurements using jets, it is necessary to understand the contribution of each uncertainty component to the final total uncertainty.
For this purpose, each uncertainty component is propagated separately from each \insitu{} method to the combined result.
This is achieved by coherently shifting all the correction factors obtained by the \insitu{} methods by one standard deviation
of a given uncertainty component, and redoing the combination using the same set of averaging weights as in the nominal combination.
The comparison of the shifted average correction factors with the nominal ones provides the propagated 
systematic uncertainty.

To account for potential disagreement between \insitu{} measurements constraining the same term 
(referred to as measurements which are \emph{in tension}), each uncertainty source is 
rescaled by the factor $\sqrt{\chi^{2}/{\rm dof}}$, if this factor is larger than $1$.
This is conservative, as values of $\sqrt{\chi^{2}/{\rm dof}}$ larger than $1$ can also be reached due to statistical fluctuations.

$\rrextrapfct{\langle\ptjet\rangle}{\etaDet} = 1/c$ is used as the \insitu{} correction calibration factor and its inverse $c$ is applied
to data. The correction factor still contains part of the statistical fluctuations of the \insitu{} measurements.
The influence of the statistical fluctuations is reduced by applying a minimal amount of smoothing using a sliding Gaussian 
kernel to the combined correction factors \cite{jespaper2010}.

Each uncertainty component from the \insitu{} methods is also propagated 
through the smoothing procedure. Propagating information between close-by \pt{} regions, 
the smoothing procedure changes the amplitude of the uncertainties 
(e.g. reducing them at low \pt).

\subsection{Uncertainty sources of the \insitu{} calibration techniques}
\label{sec:insitutechniquesuncertainties}
The \insitu{} techniques usually rely on assumptions that are only approximately fulfilled. One example is the assumption that the calibrated jet and the reference object are balanced in transverse momentum, while this balance can be altered by the presence of additional high-\pt{} particles. In order to determine the \JES{} uncertainties, the modelling of physics effects 
has to be disentangled from detector effects.
These effects can be studied by looking at the changes of the \datatomc{} response ratios introduced by systematic variation of the event selection criteria.
The ability of the \MC{} simulation to describe these changes under large variations of the selection criteria 
determines the systematic uncertainty in the \insitu{} methods, since physics effects can be suppressed 
or amplified by these variations. 
In addition, systematic uncertainties related to the selection, calibration and 
modelling of the reference object need to be considered.

When performing the variations of the selection criteria, only statistically significant variations of the response ratios
are propagated to the systematic uncertainties.
This is achieved by evaluating the systematic uncertainties in intervals which can be larger than 
the bins used for the measurement of the response ratios, meaning that several bins are iteratively combined until the observed deviations are significant.
By doing so, one avoids multiple counting of the statistical uncertainties in the systematics that are evaluated.
Using this approach, it is found that the radiation suppression uncertainty for the \deltaphi{\mathrm{jet}}{\gamma}{} cut on the \MPF{} 
method (see \secRef{sec:gammajetsoftradiation}) %
can be dropped.\footnote{This uncertainty is very small, and the corresponding variations 
are not significant, even when the evaluation is performed on the full \pt{} range.}

For the relative \etaic{}  described in \secRef{sec:etaInterCalib}  
the dominant uncertainty source is due to \MC{} modelling of jets at 
forward rapidities, where properties differ significantly for the generators under consideration (\pythia{} and \herwig{}). 
Other systematic uncertainty sources arise due to the modelling of the jet resolution,
the trigger, and dijet topology selection. However, these components are negligible 
when compared to the \MC{} modelling uncertainty.

The \datatomc{} response ratio given in \eqRef{eq:response_ratio} for the direct balance in \Zjet{} events, the MPF technique in 
\gammajet{} events, and the multijet balance method are combined as described in the previous \secRef{sec:combinationtechnique}. 
In this combination, the ability of the \MC{} simulation to describe the data, 
the individual uncertainties of the \insitu{} techniques and their compatibility, are considered.
The uncertainties of the three central \insitu{} methods combined here are described by a set of $54$
systematic uncertainty sources listed in Table \ref{tab:nuissanceparameters}.
The photon and electron energy scale uncertainties are treated as being fully correlated at this level.
Components directly related to the dijet balance technique are $\eta$ dependent quantities, and are thus treated differently.
Such parameters are not included in the list of the $54$ components, although uncertainties related to their propagation
through other methods are included.

In Table~\ref{tab:nuissanceparameters}, each uncertainty component is assigned to one of four categories, based on its source and correlations:
\begin{enumerate}
\item Detector description (\npdet)
\item Physics modelling (\npmodel)
\item Statistics and method (\npstatmeth)
\item Mixed detector and modelling (\npmixed). 
\end{enumerate}
The motivation for these categories, and to some extend the guidance for assigning the $54$ individual components to them, are given by considerations concerning the comparability of jet measurements and their uncertainties in different experiments. For example, the \npdet{} and \npstatmeth{} categories can be considered 
largely
uncorrelated between experiments, while the \npmodel{} category is 
likely correlated.
%

%
\begin{figure}[!htbp]
\begin{center}
   \includegraphics[width=0.5\textwidth]{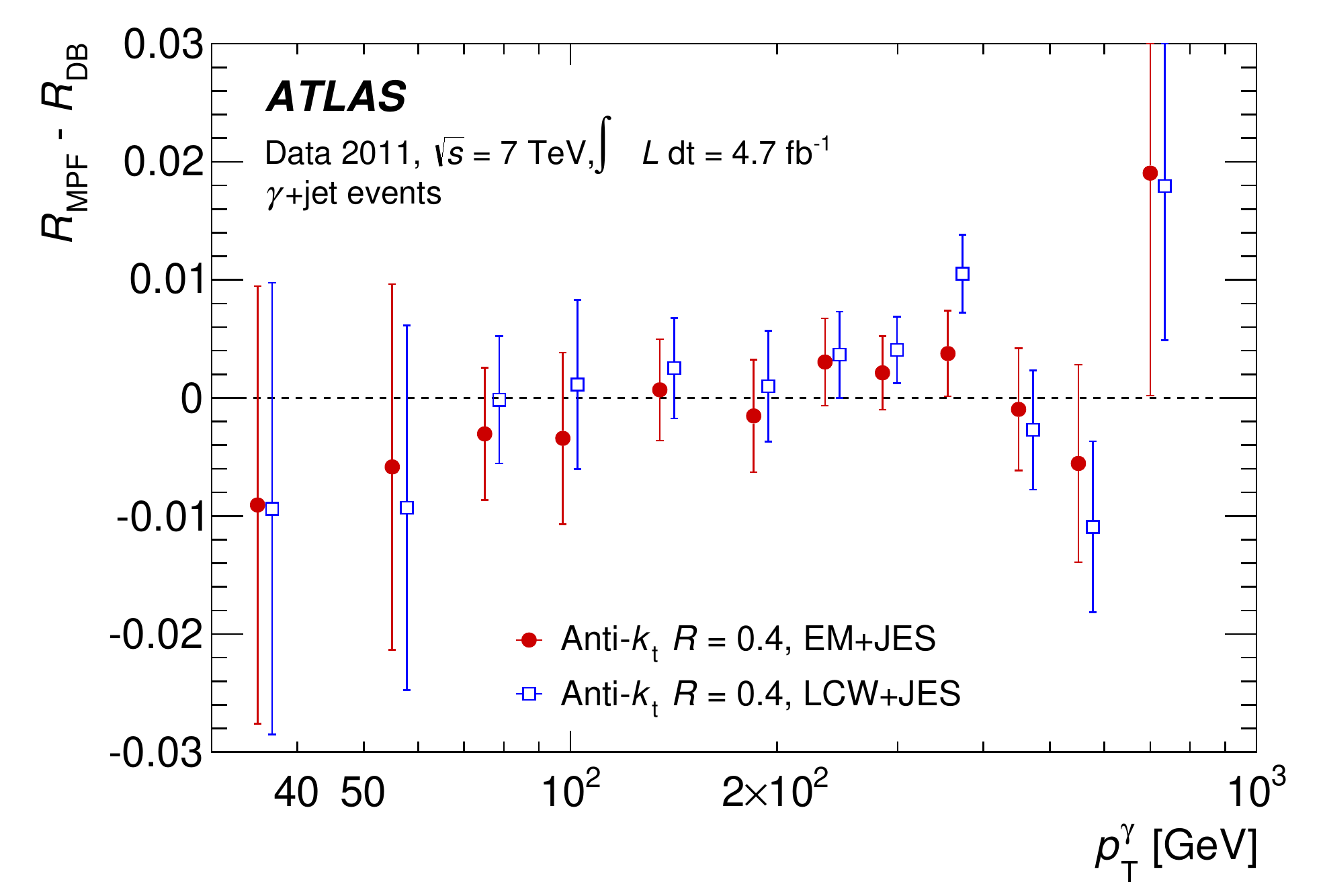}
  \caption{%
    Difference between the \datatomc{} response ratio $R$ measured using the direct balance (\DB) and the missing momentum fraction (\MPF) methods for
    jets reconstructed with the \antikt{} algorithm with $R=0.4$ calibrated with the \EMJES{} and \LCWJES{} schemes.
    The error bars shown only contain the uncorrelated uncertainties.
    \label{fig:MPVvsDB}}
\end{center}
\end{figure}

\subsection{Combination results}
\label{sec:combinationresults}
\FigRef{fig:weightinsitu} shows the contribution of each \insitu{} technique to the \JES{} residual calibration, defined to be the fractional weight carried in the combination. In the region $\ptjet \lesssim 100$~\GeV{}, the \Zjet{} method has the highest
contribution to the overall \JES{} average. The contribution is $100\%$ for $\ptjet$ below $25$ \GeV{}, the region covered only by \Zjet,
about $90\%$ at $\ptjet = 40$~\GeV{}, and decreases to about $50\%$ at  $\ptjet = 100$~\GeV.
In order to prevent the uncertainties specific to the low-\ptjet{} region from propagating to higher $\ptjet$ in the combination,
the \Zjet{} measurements below and above $\ptjet = 25$ \GeV{} are treated separately, meaning no interpolation is performed across $\ptjet = 25$ \GeV, although the magnitude of the original systematic uncertainty sources is used, separately, in both regions.

The weaker correlations between the uncertainties of the \Zjet{} measurements, compared to ones from \gammajet{},
lead to a faster increase of the extrapolated uncertainties, hence to the reduction of the \Zjet{} weight in the region between $25$ and 40$ \GeV{}$.
In the region  $100 \lesssim \ptjet \lesssim 600$ \GeV,  the \gammajet{} method dominates
with a weight increasing from $50\%$ at  $\ptjet = 100$~\GeV{} to
about $80\%$ at  $\ptjet = 500$ \GeV.
For $\ptjet \gtrsim 600$ \GeV{} the measurement based on multijet balance becomes increasingly
important and for  $\ptjet \gtrsim 800$~\GeV{} it is the only method 
contributing to the \JES{} residual calibration.
The combination results and the relative uncertainties are considered 
in the \pt{} range from 17.5~\GeV{} to  1~\TeV, where sufficient statistics are available.

The individual uncertainty components for the final combination results,\footnote{The uncertainties 
apply to the overall result of the combination of the \insitu{} techniques and differ from the 
original uncertainties of the \insitu{} methods, as they are convoluted with the corresponding weights.} 
are shown in \figRef{fig:nuisanceparameters}
for \antikt{} jets with $R=0.4$ for the \EMJES{} and the \LCWJES{} calibration scheme and
for each \insitu{} technique. 

The agreement between the \insitu{} methods is good, with $\chi^{2}/{\rm dof} < 1$ 
for most \pt{} bins, and values up to $\chi^{2}/{\rm dof} = 1.5$ in only a few bins.
The largest  $\chi^{2}/{\rm dof} = 2$ is found for \antikt{} jets with $R=0.6$ calibrated
with the \LCWJES{} scheme for $\ptjet = 25$ \GeV. 

The final \JES{} residual calibration obtained from the combination of the \insitu{}
techniques is shown in \figRef{fig:responseratioinsitu}, together with
statistical and systematic uncertainties.
A general offset of about $-2\%$ is observed in the \datatomc{} response ratios for jet transverse momenta below $100$ \GeV.
The offset decreases to about $-1\%$ at higher \pT{} ($\ptjet \gtrsim 200$).
The \JES{} uncertainty from the combination of the \insitu{} techniques
is about $2.5\%$ at $\ptjet = 25$ \GeV{}, and decreases to below $1\%$ for $55 \leq \ptjet < 500$ \GeV.
The multijet balance method is used up to $1$ \TeV{}, as at higher \pt{} values it has large statistical uncertainties.
At $1$~\TeV{} the total uncertainty is about $1.5\%$.

The results for the \EMJES{} and the \LCWJES{} calibration schemes for jets with $R=0.6$ are similar to those for $R=0.4$. 

\subsection[Comparison of the %
\gammajet{} calibration methods]{Comparison of the %
\GAMMAJET{} calibration methods}
\label{sec:mpddbgamma}

As discussed in \secRef{sec:gammajetInSitu}, two different %
techniques exploiting the transverse momentum balance in \gammajet{} events are used to probe the jet response, the direct balance (\DB) and the missing momentum fraction (\MPF) method. 
These methods have different sensitivities to parton radiation, pile-up interactions and 
photon background contamination, and hence different systematic uncertainties, as explored in \secRef{sec:systematicgammajetuncertainties}. %

Since the \MPF{} method uses the full hadronic recoil and not only the jet, a systematic uncertainty due to the possible difference in data and \MC{} simulation of the calorimeter response to particles inside and outside
of the jet needs to be taken into account. This systematic uncertainty
contribution is estimated to be small compared to other considered uncertainties.
However, in the absence of a more quantitative estimation, the full
energy of all particles produced outside of the jet as estimated
in the \DB{} technique is taken as the systematic uncertainty.
A comparison between the two results is shown in \figRef{fig:MPVvsDB}. 
The results are compatible within their uncorrelated uncertainties.

As the methods use similar datasets, the measurements are highly correlated and cannot easily be included together
in the combination of the \insitu{} techniques. 
In order to judge which method results in the most precise calibration, the 
combination described 
in \secRef{sec:combinationtechnique} is performed twice, 
both for \Zjet{}, \gammajet{} \DB{} and multijet balance, and 
separately for \Zjet{}, \gammajet{} \MPF{} and multijet balance. 
The resulting combined calibration that includes the \MPF{} method has slightly smaller uncertainties, by up to about $0.1\%$, and is therefore used as the main result.

%
%

\begin{figure*}[ht!]
\begin{center}
 \subfloat[\EMJES{} ]{\includegraphics[width=0.49\textwidth]{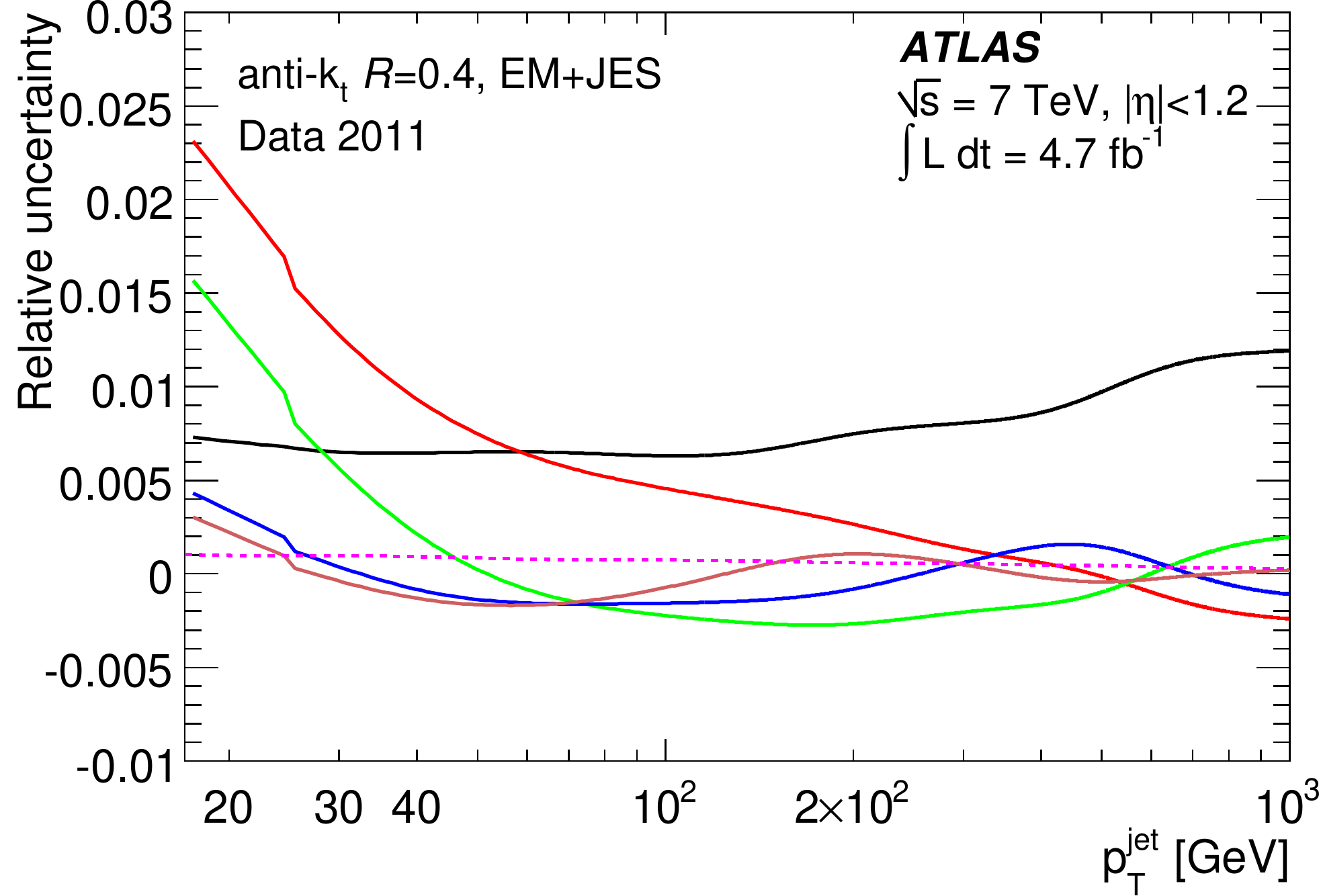}\label{fig:usedeigenvectorEM}}
 \subfloat[\LCWJES{}]{\includegraphics[width=0.49\textwidth]{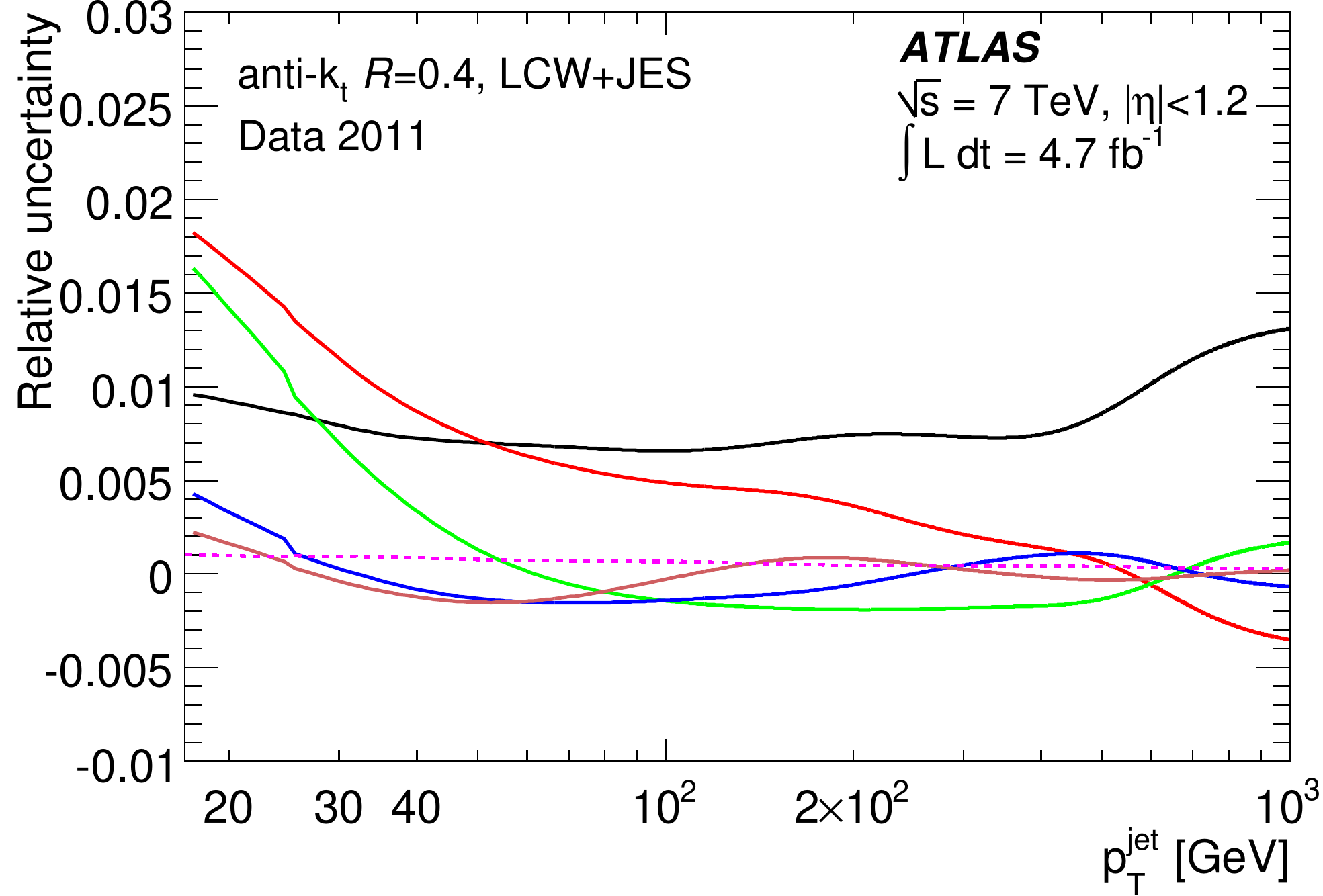}\label{fig:usedeigenvectorLCW}}
\caption[]{
  Systematic (effective)~relative uncertainties displayed as a function of jet \pt{} for \antikt{} jets with $R=0.4$ 
  calibrated with the  \subref{fig:usedeigenvectorEM} \EMJES{} and the \subref{fig:usedeigenvectorLCW} \LCWJES{} calibration schemes
  for the reduced scheme with six nuisance parameters. 
  Each curve can be interpreted as a $1\sigma$ \JES{} systematic nuisance parameter, symmetric around zero.
  They represent eigenvectors of the covariance matrix~(continuous lines) and the residual component~(dashed line).
  \label{fig:usedeigenvector}
}
\end{center}
\end{figure*}

\begin{figure*}[htp!]
  \centering
  \subfloat[Nominal correlation]{\includegraphics[width=0.49\textwidth]{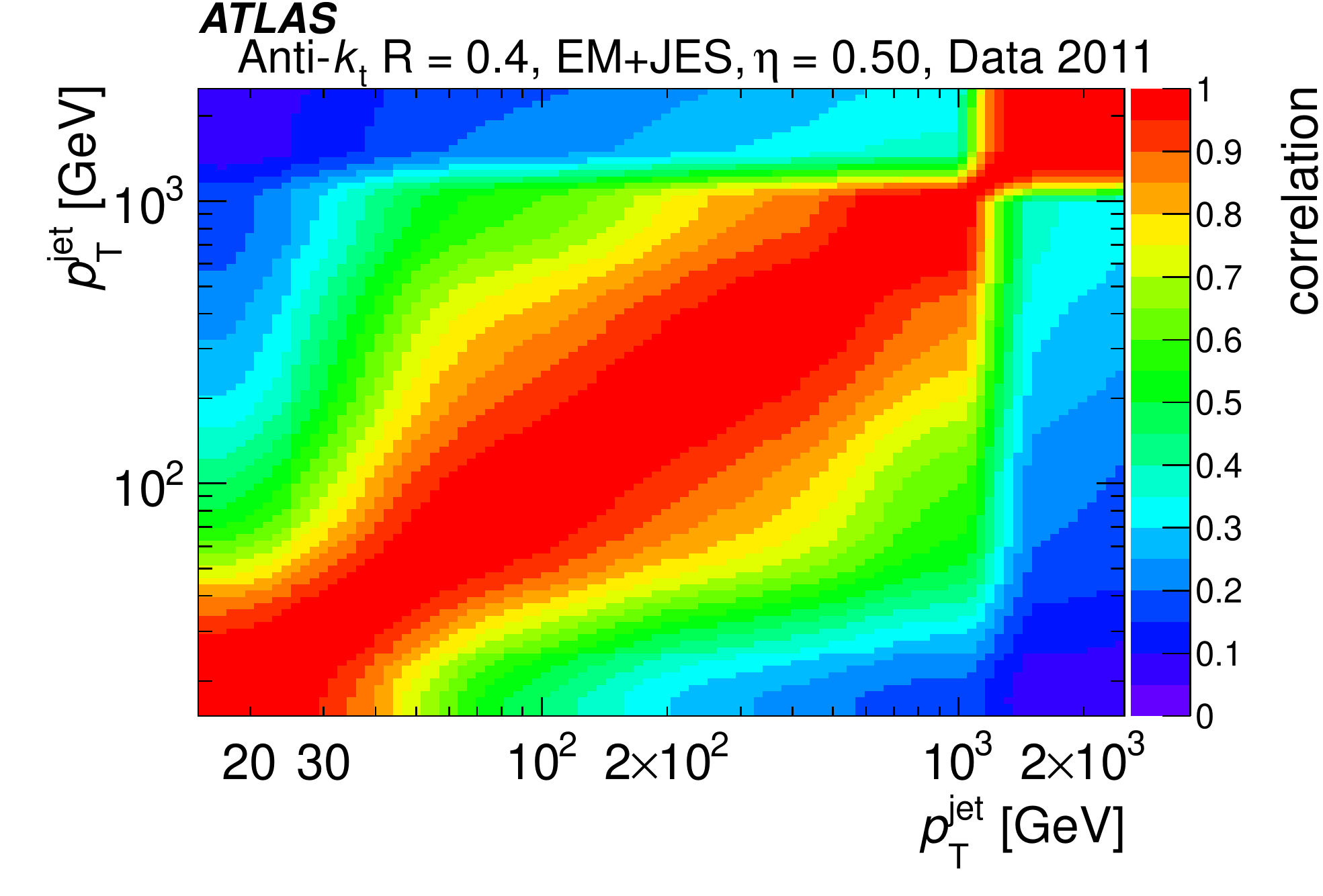}\label{fig:CorrelationMatricesFull}}
  \subfloat[Difference between weaker and stronger correlation]{\includegraphics[width=0.49\textwidth]{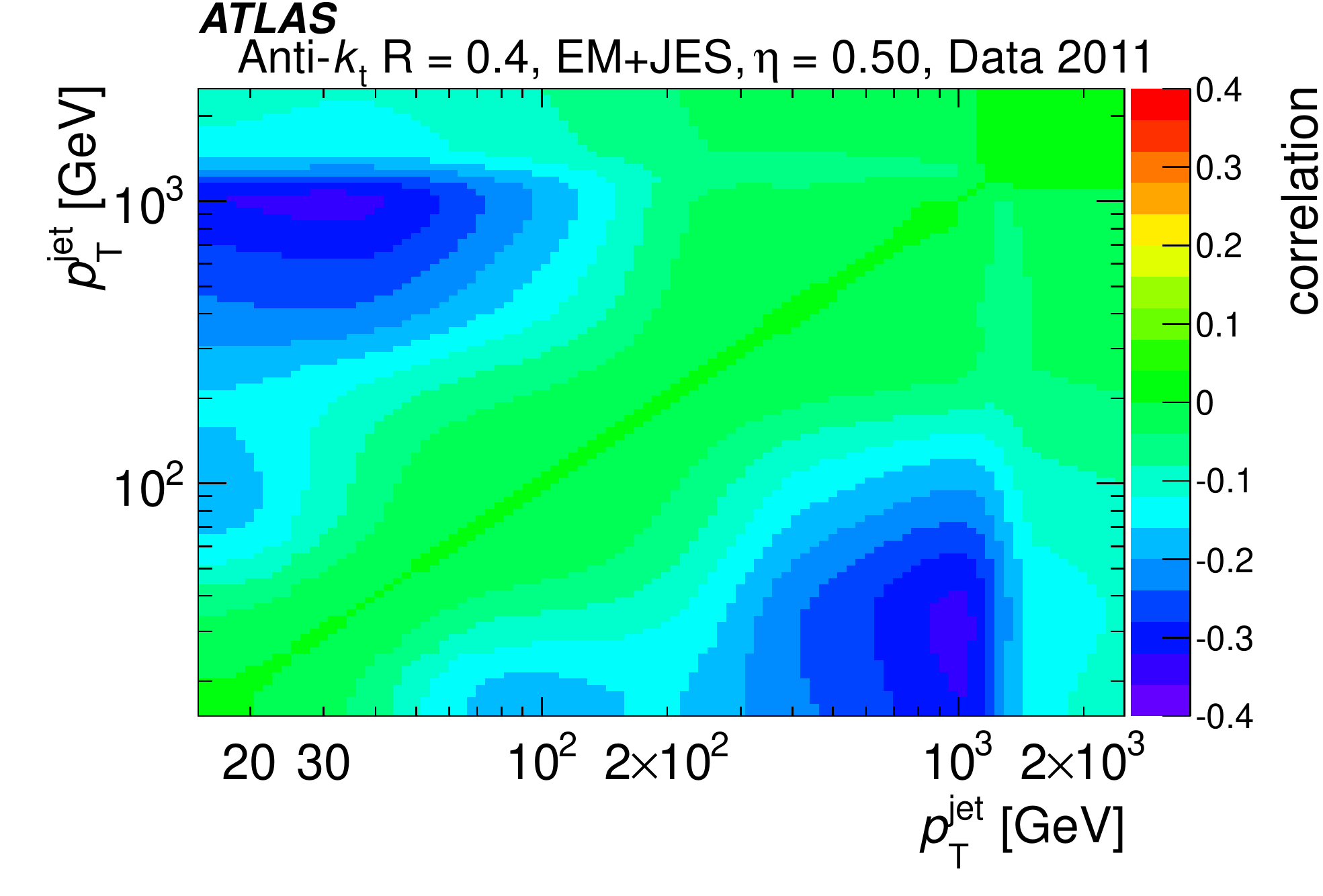}\label{fig:CorrelationMatricesDiff}}
  \caption[]{
    In  \subref{fig:CorrelationMatricesFull}, the nominal \JES{} correlation matrix is shown. The difference between the correlation matrices of the interpretations resulting in stronger and weaker 
    correlations for \antikt{} jets with $R=0.4$ calibrated using the \EMJES{} calibration scheme in the central calorimeter region ($\etajet=0.5$) is depicted
    in  \subref{fig:CorrelationMatricesDiff} . 
    \label{fig:CorrelationMatrices}
  }
\end{figure*}

%
%
\begin{figure*}[!htp]
  \centering
  \subfloat[Statistical and method components]{\includegraphics[width=0.48\textwidth]{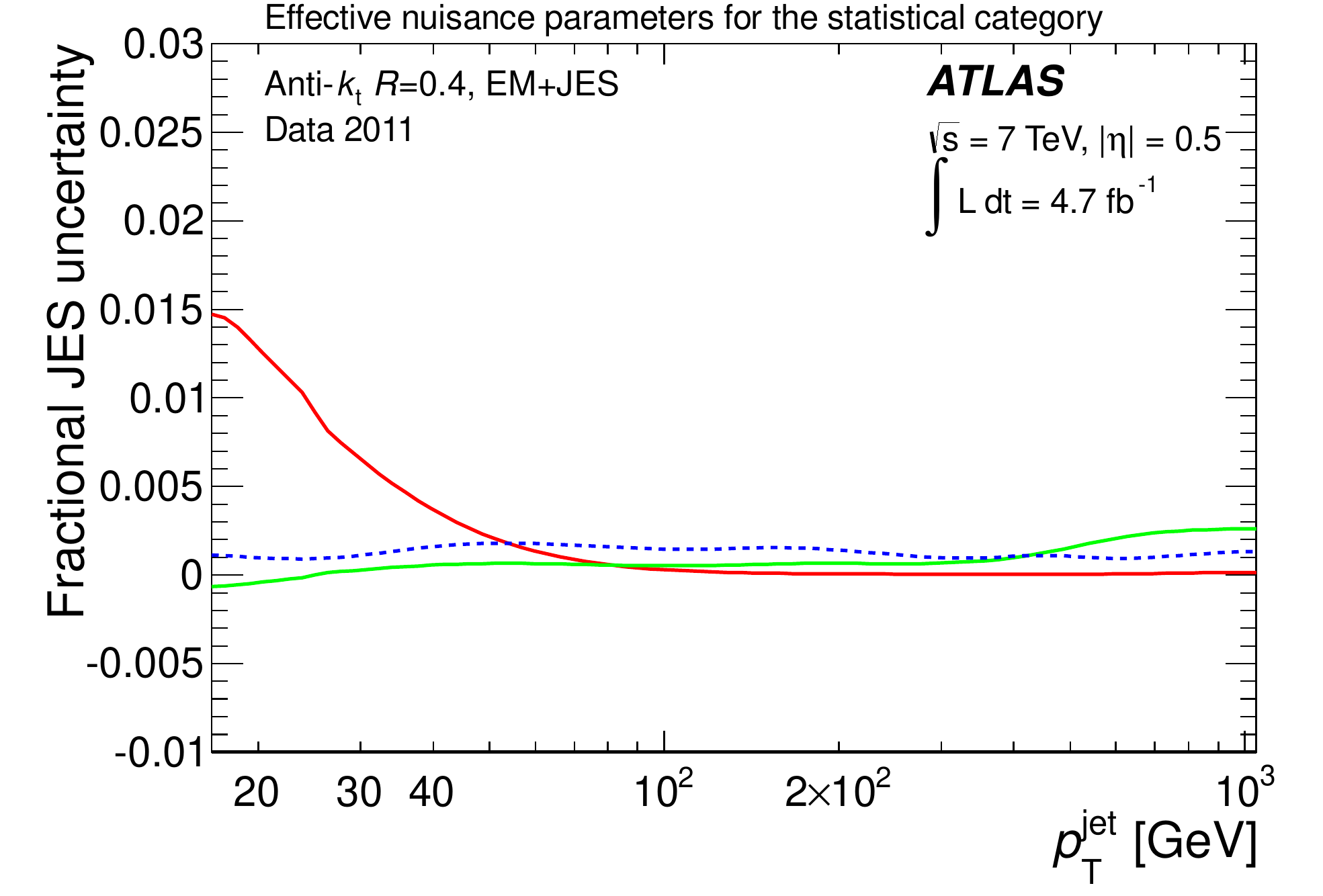}\label{fig:CategoryReductionStat}}
  \subfloat[Detector components]{\includegraphics[width=0.48\textwidth]{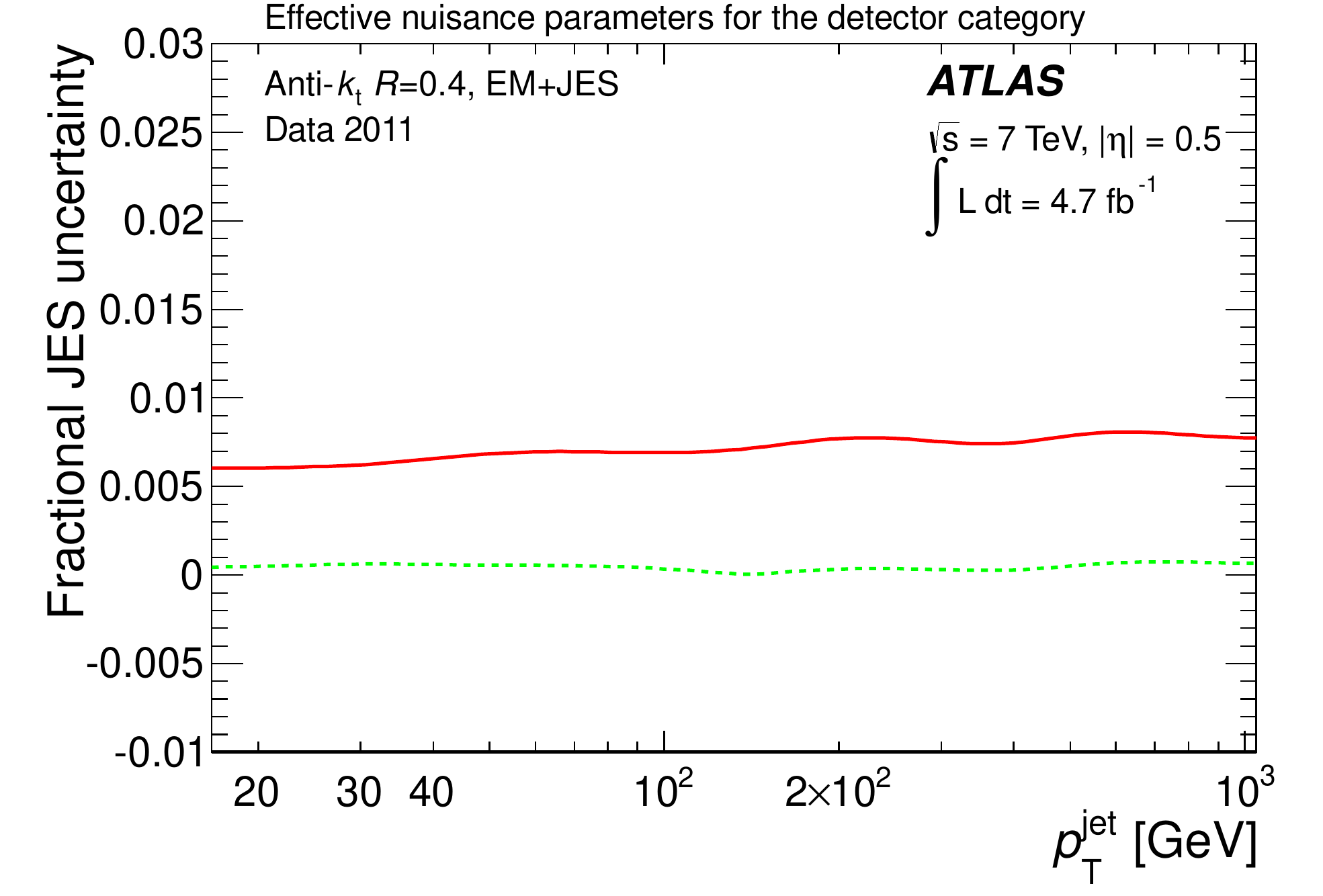}\label{fig:CategoryReductionDet}}\\
  \subfloat[Modelling components]{\includegraphics[width=0.48\textwidth]{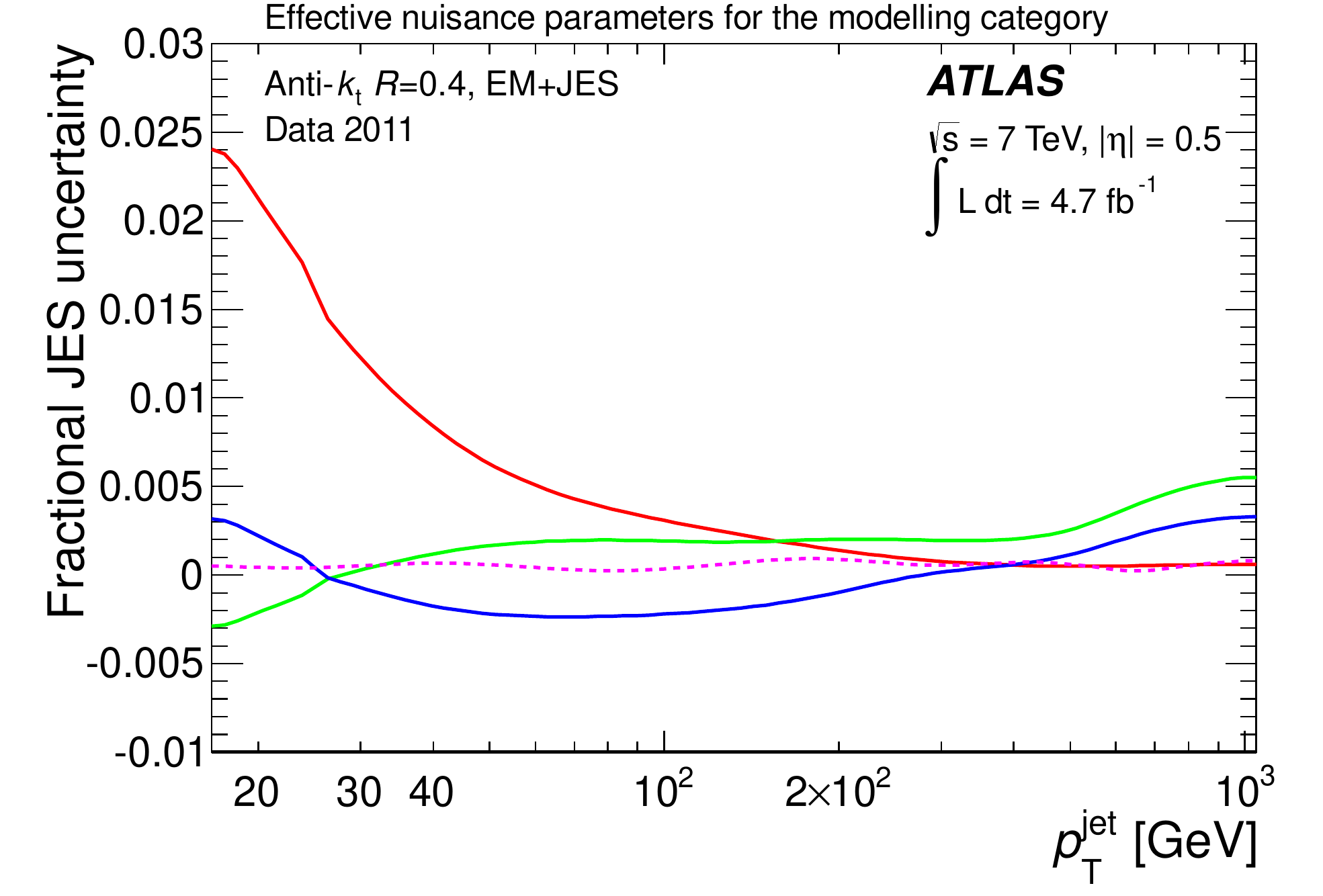}\label{fig:CategoryReductionModel}}
  \subfloat[Mixed detector and modelling components]{\includegraphics[width=0.48\textwidth]{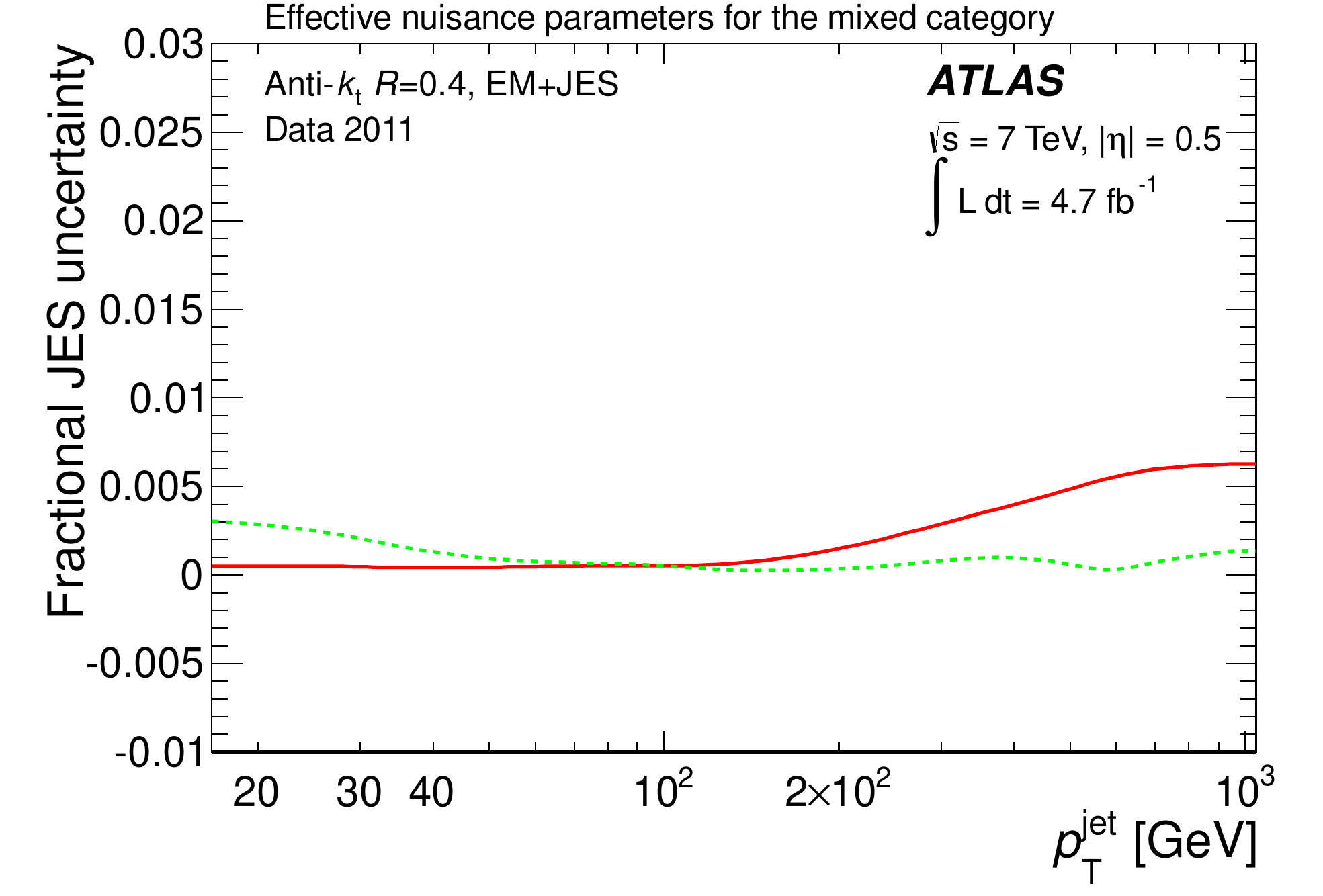}\label{fig:CategoryReductionMixed}}
  \caption[]{
    Relative uncertainties for reduced (effective)~components within a single category displayed as a fraction of jet \pT{} for \antikt{} jets with $R=0.4$, calibrated with the \EMJES{} scheme. The convention from Fig.~\ref{fig:usedeigenvector} is followed here.  The 54 nuisance parameters that are input to the reduction for each of the categories are listed in Table \ref{tab:nuissanceparameters}. The reduction is performed for all nuisance parameters belonging to any given category, which are statistical and method components (a), detector components~(b), modelling components~(c), and mixed detector and modelling components~(d). 
Each of the curves can be interpreted as an effective $1\sigma$ \JES{} systematic nuisance parameter, symmetric around zero.  They represent eigenvectors of the covariance matrix~(continuous lines) and the residual component~(dashed line), for the specified category.
    \label{fig:CategoryReduction}
  }
\end{figure*}

\subsection{Simplified description of the correlations}
\label{sec:simplifiednuisanceParameters}

For some applications like parameterised likelihood fits it is pre\-fe\-ra\-ble to have the \JES{} uncertainties 
and correlations described by a reduced set of uncertainty components. This can be a\-chie\-ved by combining the least significant (weakest) nuisance parameters into one component while maintaining a sufficient accuracy for the \JES{} uncertainty correlations. 

The total covariance matrix  $C^{\rm tot}$ of the \JES{} correction factors can be derived from the individual components of the statistical and systematic uncertainties:
\begin{equation}
\label{Eq:totCovMat}
   C^{\rm tot} = \sum_{k=1}^{N_{\rm sources}} C^{k},
\end{equation}
where the sum goes over the covariance matrices of the individual uncertainty components $C^{k}$.
Each uncertainty component $s^{k}$ is treated as fully correlated in \pT{} and
the covariance of the \pt{} bins $i$ and $j$ is given by $C^{k}_{ij} = s^{k}_{i} s^{k}_{j}$.
All the uncertainty components are treated as independent of one another, 
except for the photon and electron energy scales which are treated as correlated.\footnote{A single systematic uncertainty source is assigned to account for both the photon and electron energy scales by first adding the photon and electron scales linearly, deriving the full covariance matrix, and add it linearly to the covariance matrix of the other uncertainty components.}

A reduction of the number of nuisance parameters while retaining the information on the correlations 
can be achieved by deriving the total covariance matrix in \eqRef{Eq:totCovMat} and diagonalising it:
\begin{displaymath}
\label{Eq:diagonalization}
   C^{\rm tot} = S^{T} D~S.
\end{displaymath}
Here $D$ is a (positive definite) diagonal matrix, containing the eigenvalues  $\sigma_{k}^{2}$ of the total 
covariance matrix, while the $S$ matrix contains on its columns the corresponding
(orthogonal) unitary eigenvectors  $V^k$.
A new set of independent uncertainty sources can then be obtained by multiplying each eigenvector 
by the corresponding eigenvalue.
The covariance matrix can be re-derived from these uncertainty sources using:
\begin{displaymath}
   C^{\rm tot}_{ij} = \sum_{k=1}^{N_{\rm bins}} \sigma_{k}^{2}~V^k_i~V^k_j,
\end{displaymath}
where $N_{\rm bins}$ is the number of bins used in the combination.

A good approximation of the covariance matrix can be obtained by separating out only a small subset 
of $N_{\rm eff}$ eigenvectors that have the largest corresponding eigenvalues. 
From the remaining $N_{\rm bins}-N_{\rm eff}$ components, a residual,
left-over uncertainty source is determined, with an associated covariance matrix $C'$.
The initial covariance matrix can now be approximated as:
\begin{displaymath}
   C^{\rm tot}_{ij} \approx \sum_{k=1}^{N_{\rm eff}} \sigma_{k}^{2}~V^k_i~V^k_j +  C'.
\end{displaymath}
This approximation conserves the total uncertainty, while the precision on the description of the correlations 
can be directly determined by comparing the original full correlation matrix and the approximate one.
The last residual uncertainty could in principle be treated either as correlated or as uncorrelated 
between the \pt{} bins.
It is observed that treating this uncertainty source as uncorrelated in \pt{} provides a better 
approximation of the correlation matrix.
This is expected, as this residual uncertainty source includes many orthogonal eigenvectors with small amplitudes and
many oscillations, hence the small correlations.
The original exact covariance matrix is thus decomposed into a part with strong correlations and 
another one with much smaller correlations. It is this residual uncertainty source that incorporates the part with small correlations.

\FigRef{fig:usedeigenvector} shows the obtained five eigenvectors $\sigma_kV^k$ and the residual sixth component, as a function of the jet \pt{}. 
The \pt-dependent sign of these eigenvectors allows to keep track of the (anti-)correlations of each component 
in different phase-space regions. 
This is necessary for a good description of the correlations of the total \JES{} uncertainty.
These six nuisance parameters are enough to describe the correlation matrix with sufficient precision at the level of percent.
As explained above, the quadratic sum of these six components is identical to the quadratic sum of the original uncertainties shown in \figRef{fig:nuisanceparameters}. In the high-\pt{} region above $300$~\GeV, one eigenvector has a significantly larger amplitude than all the others, 
see the black  
curve in \figRef{fig:usedeigenvector}, hence the strong correlations between the bins.
Approximately 60\% to 80\% of this component is due to the photon and electron energy scale uncertainties up to about $700$ \GeV{} (see Figs. \ref{fig:nuisanceparameters}\subref{fig:np_EMJES_R4_MPF} and  \ref{fig:nuisanceparameters}\subref{fig:np_LCWJES_R4_MPF}), while some other uncertainties contribute to it at higher \pt{}.

%
\begin{table}[h!] \centering
\caption[\JES{} configuration summary]{A summary of the various explored \JES{} configurations. The precision of the reduction is defined by the largest deviation in correlations between the original full set of parameters and the reduced version. The full \pT{} phase space is considered in this determination.
The values quoted are for jets in the region $|\etajet|<1.2$ 
for \antikt{} with $R=0.4$ calibrated with \EMJES{} scheme. The number of nuisance parameters quoted refers only to the parameters entering the reduction procedure, which are relevant to the \insitu{} techniques.}
  \label{table:AllConfigurations}
\renewcommand{\arraystretch}{\myarraystretch}
\begin{tabular}{l|c|c|c}
  \hline \hline
  Configuration & Reduction & $N_{\rm params}$ & Reduction  \\
  type          &           &               & precision (\%) 
  \\\hline
  All parameters & none & $54$ & $100$
  \\
                          & global & $6$ & $97$ 
  \\
                         & category & $11$ & $95$
  \\\hline
  Stronger correlations & none & $45$ & $100$
  \\
                                    & global & $6$ & $97$
  \\
                                   & category & $12$ & $96$
  \\\hline
  Weaker correlations & none & $56$ & $100$
  \\
                                   & global & $6$ & $97$
  \\
                                   & category & $12$ & $95$
  \\\hline\hline
  \end{tabular}
\end{table}

%

\subsection{Jet energy scale correlation scenarios}

The \JES{} uncertainty and its correlations discussed so far can play a crucial role in physics analyses.  
In order to quantify these correlations, knowledge of the interdependence of the
systematic uncertainty sources is needed. The limitations in this knowledge lead to uncertainties on the correlations. 

The variation of the systematic uncertainty sources as a function of \pt{} and \etajet{} can be described as a nuisance parameter, as explained before.
The total set of correlations can be expressed in the form of a correlation matrix
calculated from the full set of nuisance parameters as presented in \secRef{sec:combinationresults}.
The correlation matrix, derived assuming that the nuisance parameters are
independent from each other, is shown in \figRef{fig:CorrelationMatrices} \subref{fig:CorrelationMatricesFull}. 
The nuisance parameters are affected by the strength of the correlations between uncertainty components, which can be difficult to estimate. The investigation of alternative correlation 
scenarios for the components thus allows to determine the uncertainty on the global correlations shown in \figRef{fig:CorrelationMatrices} \subref{fig:CorrelationMatricesFull}.

Two additional configurations are specifically designed to weaken and to strengthen the global correlations. They cover the space of reasonable \JES{} component dependencies.
In a given physics analysis these scenarios can be used to examine how the final results are affected by variations of the correlation strengths. This allows propagation of the uncertainties on the correlations.
The difference between the weaker and stronger correlation matrices is shown in Figure \figRef{fig:CorrelationMatrices}\subref{fig:CorrelationMatricesDiff}.
\subsection{Alternative reduced configurations}

A global reduction of nuisance parameters, irrespective of the uncertainty source, 
is performed in order to reduce the number of these parameters required to represent the full correlation matrix,
see \secRef{sec:simplifiednuisanceParameters}.  
However, it is also useful to keep track of the physical meaning of the uncertainty components,
e.g. for a proper combination of measurements from different experiments.
In \secRef{sec:insitutechniquesuncertainties} each \JES{} systematic uncertainty component
is assigned to a representative category, as given in Table \ref{tab:nuissanceparameters}.

The same reduction technique discussed in \secRef{sec:simplifiednuisanceParameters} is applied independently to each set of uncertainty components within each individual category.
The resulting reduced set of uncertainty components for the nominal configuration are shown in \figRef{fig:CategoryReduction}. This category reduction approach generally results in a larger 
number of nuisance parameters than the global reduction. This is because two components 
from different categories with very similar shapes can be globally combined without significant loss of information for the correlations. 
However, when the reduction is performed in categories, components may require a nuisance parameter not lose 
significant precision for the description of the global correlation.

This technique is applied to each of the correlation scenarios. Category reduction configurations 
are derived for the set of all parameters, the stronger correlation scenario, and the weaker correlation scenario.  
In each case, correlation matrices are compared to ensure that the reduction preserved correlation information 
to within a few percent.  Table \ref{table:AllConfigurations} lists the various configurations evaluated,
together with the accuracy achieved with the reduction procedure.

%
%
%
\begin{figure}[!!!h]
  \centering
\subfloat[]{\includegraphics[width=0.49\textwidth]{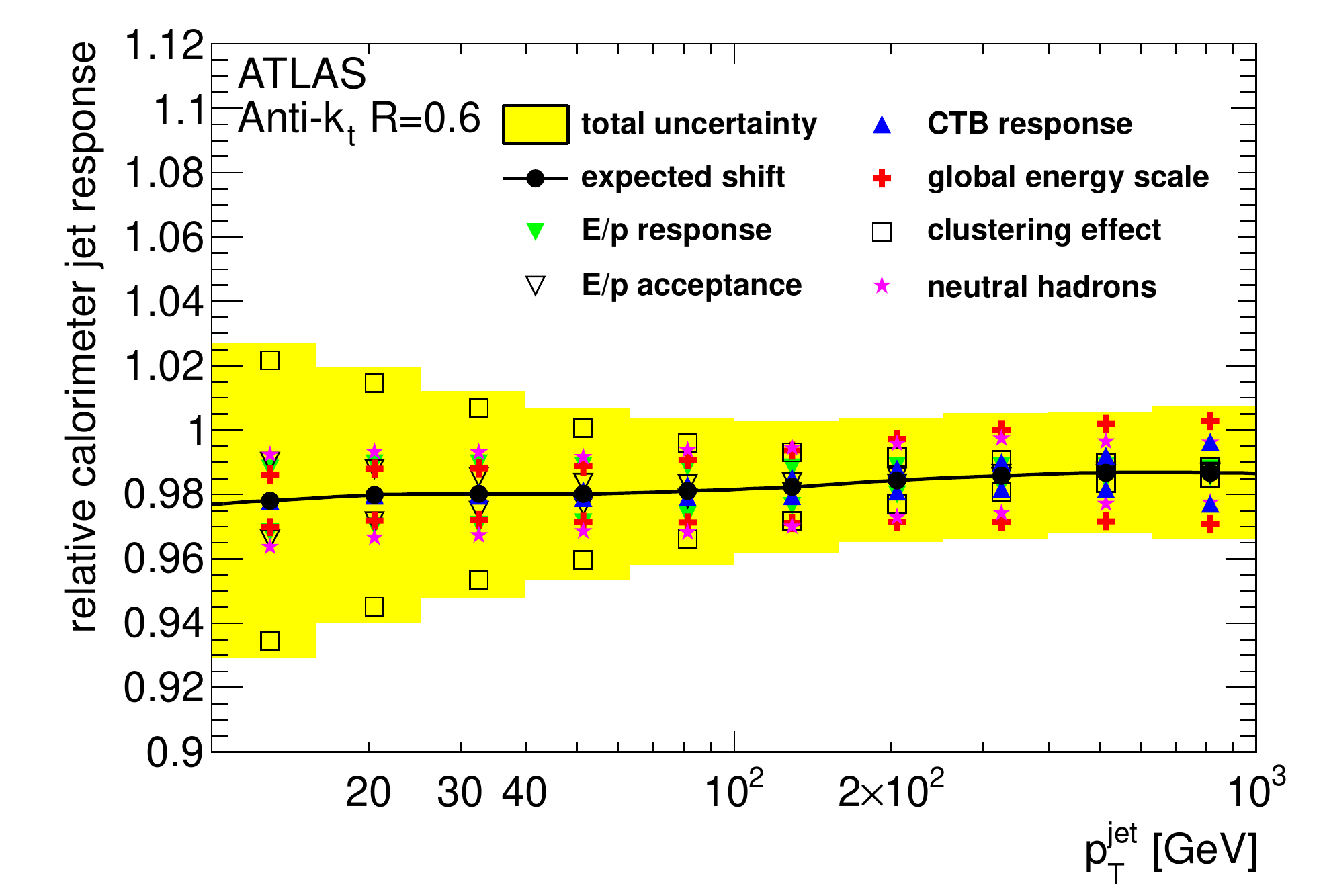} \label{fig:ep_response}}\\
\subfloat[]{\includegraphics[width=0.49\textwidth]{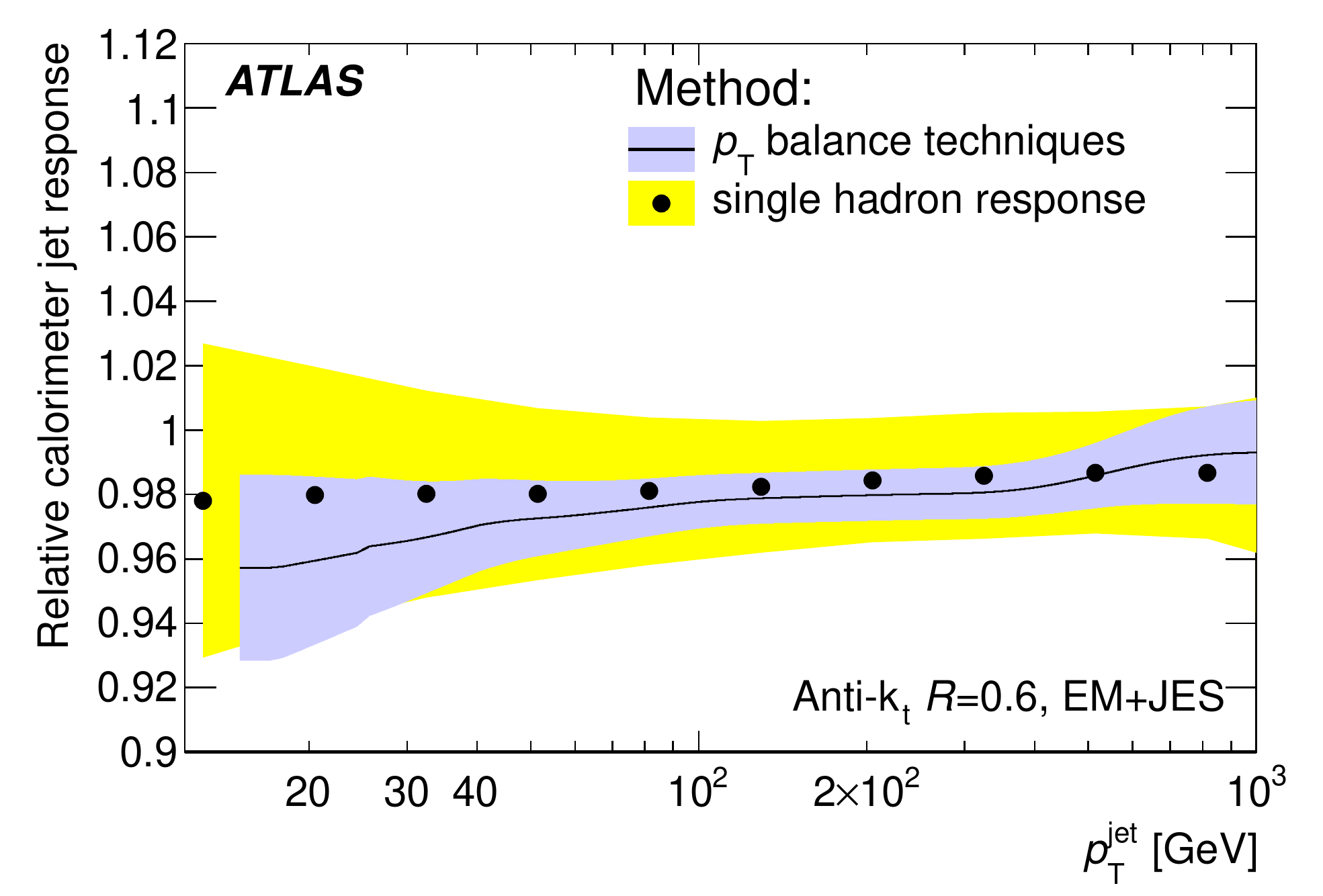} \label{fig:ep_jes}}
 \caption[]{Relative calorimeter jet response ratio between data and \MC{} simulations,
as estimated from the single-hadron response measurements as a function of the jet transverse momentum, is shown in \subref{fig:ep_response}. 
The total systematic uncertainty together with the uncertainty from the individual components
is shown as a lighter band.
The black circles denote the 
estimated mean shift of the calorimeter response to jets in
data over the one in \MC{} simulations.  In \subref{fig:ep_jes},  the uncertainty from the single-hadron response measurements is shown as a lighter (yellow) band, while
the \JES{} uncertainty, as derived from the \insitu{} methods based on \pt{} balance, is shown as a dark (gray) band. 
The closed markers denote the 
estimated shift of the calorimeter response to jets in data over the one 
in \MC{} simulations, and
the line shows the \JES{} correction derived from the \pt{} balance \insitu{} methods.
\label{fig:EOverP}
}
\end{figure}

\section{Comparison to jet energy scale uncertainty from single-hadron response measurements}
\label{sec:SingleParticle}
The \JES{} correction and uncertainty  derived from \insitu{} techniques exploiting the \pt{} balance between
a jet and a reference object can be compared to the method where the jet energy scale is estimated
from single-hadron response measurements, as described in Ref.\cite{jespaper2010}.
In this method, jets are treated as a superposition of energy deposits of single particles.
For each calorimeter energy deposition within the jet cone, 
the type of the particle inside the jet is determined,
and the expected mean shift and the systematic uncertainty of the calorimeter response 
between data and \MC{} simulation is evaluated. 
The corresponding uncertainty is derived from \insitu{} measurements or systematic \MC{} variations.
This deconvolution method is described in Ref. \cite{jespaper2010,eppaper2010} and is used for the derivation of the \JES{} uncertainty for the \ATLAS{} $2010$ data analysis.

Measurements of the calorimeter response to pions in the combined test-beam \cite{CTB04pion}
are used for pions with momenta between $20$ and $350$ \GeV{}.\footnote{The \MC{} simulation was updated from the version used
for the combined test-beam studies to the version used for the collision data simulation.} 
Single isolated hadrons with momenta up to $20$ \GeV{} are selected in a minimum bias sample 
produced in \pp{} collisions at $\sqrt{s} = 7$~\TeV{} taken in $2011$
and the calorimeter energy ($E$) in a narrow cone around an isolated track is compared
to the track momentum ($p$) (see Refs. \cite{eppaper2010,TauEnergyUncertainty2012} for more details).
Effects from the  noise thresholds and from the calorimeter acceptance are 
estimated by comparing the energy measured in calorimeter cells to the one measured in \topos.
In addition, the uncertainty on the absolute electromagnetic energy scale is considered
and the response uncertainty of protons, anti-protons and neutral hadrons 
is evaluated using different hadronic shower models, again as described 
in Refs. \cite{eppaper2010,TauEnergyUncertainty2012}. 
For hadrons with $p > 400$ \GeV{}, for which no measurements are available in the combined test-beam, 
the uncertainty is conservatively estimated as $10\%$ to account for possible calorimeter non-linearities 
or longitudinal leakage. 

The mean $E/p$ is well described by the \MC{} simulation for $p>6$ \GeV.
However, for lower momenta ($1 \lesssim  p < 6$ \GeV) the data are shifted down with respect 
to the \MC{} simulation by about $4\%$.
This is in contrast to the $2010$ measurement, where an agreement within $3\%$ is found \cite{eppaper2010}. 
The worse \datatomc{} agreement is due to the new corrections in the absolute electromagnetic
energy scale obtained \insitu{} using the $Z$ boson mass constraint reconstructed from $Z \to e^+ e^-$,
the increased \topo{} thresholds, and the use of a new \geant{} version.

\FigRef{fig:EOverP}\subref{fig:ep_response} shows the estimated calorimeter jet response ratio 
between data and \MC{} simulation as estimated from the single-hadron response
measurements as a function of the jet transverse momentum. 
A lower calorimeter response to jets 
in data than in the \MC{} simulations is observed (black circles),
consistent with that obtained using \insitu{} techniques. 
The uncertainty on this ratio is about $4\%$ at very low and very high \pt. 
It decreases to about $2\%$ between $100 \le \pt < 600$ \GeV. 
The individual uncertainty components are also shown. The dominant uncertainties
at low \pt{} are those from noise threshold effects, which can be different for single isolated
hadrons and hadrons inside jets. At high \pt{} the response
differences between data and \MC{} simulation as measured in the \ATLAS{} combined
test-beam and the uncertainty for hadrons with $p>400$ \GeV{} are largest.
The uncertainty on the global electromagnetic energy scale and the response
uncertainty for neutral hadrons contribute about $1\%$.

\FigRef{fig:EOverP}\subref{fig:ep_jes} compares the \JES{} uncertainty as obtained from single hadron
response measurements to the one obtained from the \insitu{} method based on 
the \pt{} balance between a jet and a well-measured reference object.
For both methods the mean jet calorimeter response in data is observed to be shifted down 
by about $2\%$ with respect to the one in the \MC-simulated events. 
However, the \pt{} balance methods give a considerably smaller uncertainty.

 \begin{figure*}[htbp]
   \centering
 \subfloat[]{\includegraphics[width=0.45\textwidth]{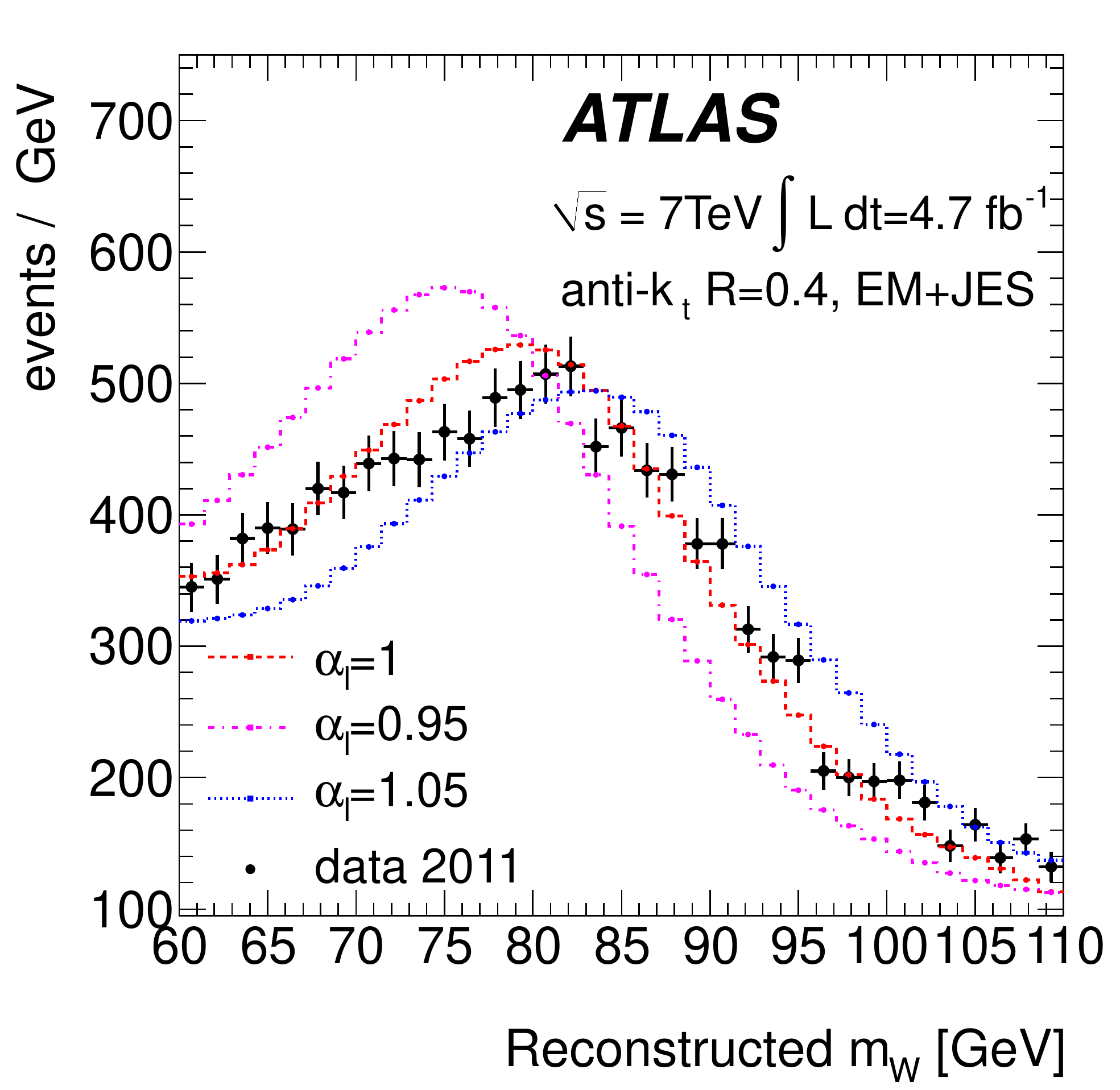}
\label{fig:wtd_templates}}
 \subfloat[]{\includegraphics[width=0.45\textwidth]{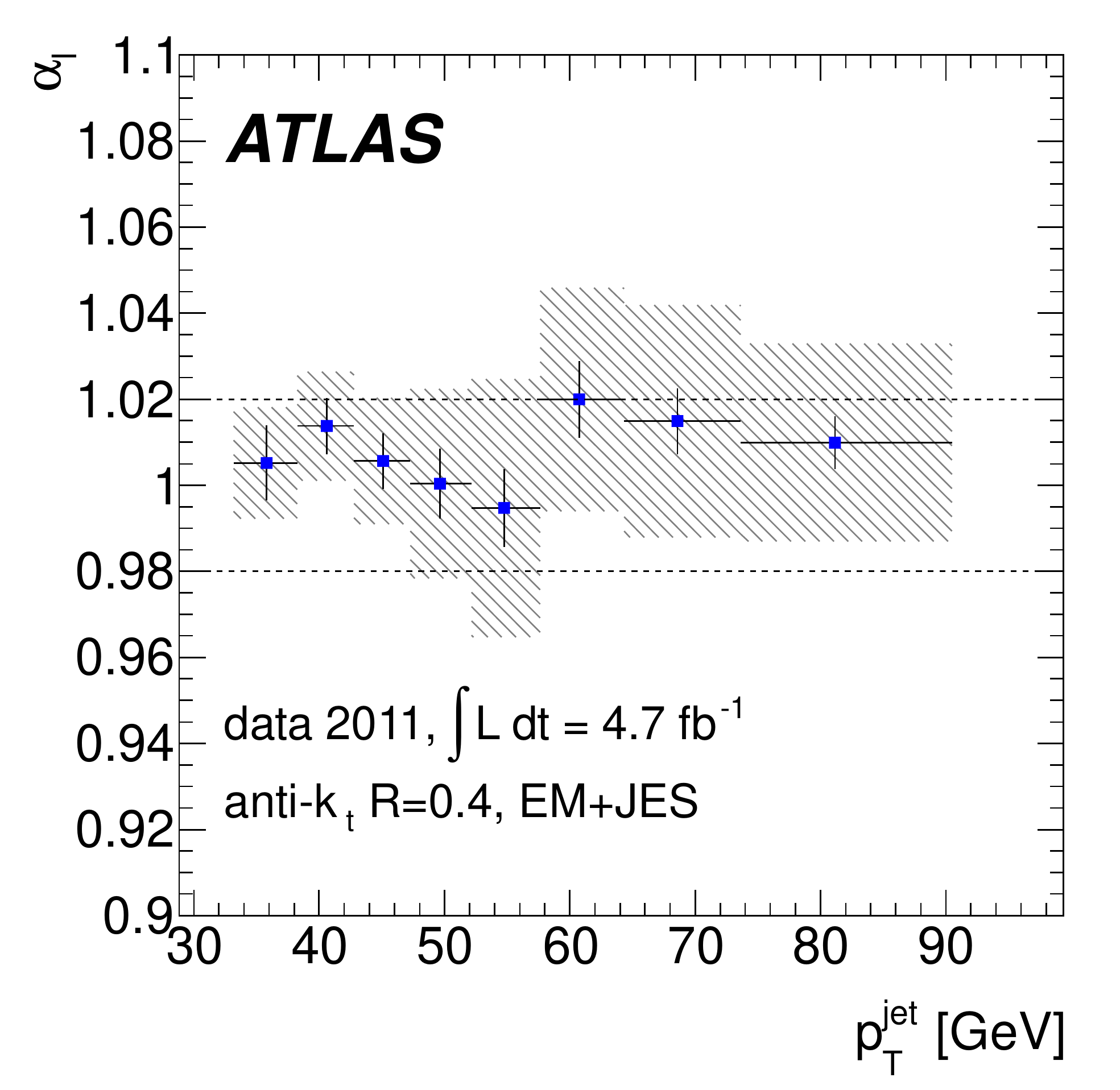}\label{fig:wtd_measure}} 
  \caption[]{The three templates distributions for the reconstructed \Wboson{}
mass in \ttbar{} events obtained by shifting the jet energy
by a factor $\Rjpesl = 0.95$, $1$ and $1.05$ in the Monte Carlo simulation with respect
to the one in data \subref{fig:wtd_templates}. 
The Monte Carlo simulation templates are also compared to the data distribution.
The \Rjpesl{} measurement as a
   function of mean jet \pt{} using the maximum \pt{} reconstruction approach, is shown in \subref{fig:wtd_measure}. Error bars are statistical while %
 hashed    
 rectangles represent the total uncertainties.}
  \label{fig:template_data}
 \end{figure*}  

\section{Jet energy scale uncertainty from the W boson mass constraint}
\label{sec:wmass}

The mass of the \Wboson{} boson (\mw) provides a stable reference for the determination
jet energy scale uncertainty.
In events where a top pair (\ttbar) is produced,
the hadronically decaying \Wboson{} bosons  give rise to two jets that can be well identified. 
A dedicated event reconstruction is developed in order to 
find the jets from the \Wboson{} decay.
The jet energy measurement can be assessed by measuring the residual difference between the 
observed and the simulated invariant \Wboson{} mass spectrum.

\Wboson{} provide a pure source of jets induced by quarks. A sizeable
fraction of these jets are induced by charm quarks and contain charm hadrons.
Given that an unbiased sample of charm jets can not be selected in data,
all jets from \Wboson{} decays are treated in the same way.

\subsection{Event samples}

The \ds{} is selected using %
single-electron or  single-muon triggers. 
Jets are reconstructed with the \antikt{} algorithm  with $R = 0.4$ starting from 
\topos{} and are calibrated with the \EMJES{} scheme.
Jets from the decay of heavy-flavour hadrons are selected by the so-called {\sc MV1} algorithm, a neural-network-based $b$-tagging algorithm described in Ref. \cite{mv1}. 
It is used at an operating point with $70\%$ efficiency for \bjets, and a mistag rate of less
than $1\%$, as determined from simulated \ttbar{} events.

Events with leptonically decaying \Wboson{} bosons are selected as follows:

Candidate electrons with transverse momenta $\pT > 25$~\GeV{} are
required to pass the tight \ATLAS{} electron quality
cuts \cite{Atlaselectronpaper}. Muons with transverse momentum $\pt > 20$ \GeV{} are
required to pass \ATLAS{} standard muon quality
cuts \cite{muon}. Events with an electron (muon) are required to be
triggered by an electron (muon) trigger with a threshold of $20$ ($18$) \GeV, thus
ensuring the trigger is fully efficient.
Events are required to have a missing transverse momentum 
$\Etmiss > 30$ \GeV{} ($\Etmiss> 20$ \GeV) in the electron
(muon) channel. 
The signal region for this analysis requires exactly one char\-ged lepton and four or more jets.
Two \btagged{} jets are required in each event.  
\Etmiss{} is calculated from the vector sum of the energy in the calorimeter
cells associated to \topos \cite{MET2011}. 
Additionally, the transverse mass of the reconstructed 
leptonic \Wboson{} boson is required to pass $m_{\mathrm{T}}^{\Wboson} > 30$~\GeV{} in the electron channel, or $\Etmiss + m_{\mathrm{T}}^{\Wboson} > 60$~\GeV{} in the muon channel. Here $m_{\mathrm{T}}^{\Wboson}$ 
is defined as:
\begin{displaymath}
m_{\mathrm{T}}^{\Wboson} = \sqrt{2 p_{\mathrm{T}}^{\ell} \Etmiss (1 - \cos(\Delta\Phi(\ell,\met))} ,
\end{displaymath}
with the lepton transverse momentum $p_{\mathrm{T}}^{\ell}$ and the azimuthal angle $\Delta\Phi$ between the lepton and the missing transverse energy. 

A cut is applied on each event to have fewer than seven reconstructed jets, to
 significantly reduce the number of possible jet pair combinations per event.
The main background processes to \ttbar{} are single-top production, multijet and \Wboson{} boson
production in association with jets. The \ttbar{} signal purity is greater than $90\%$ after this selection.

\begin{table}[htbp]\centering
\caption{Systematic uncertainties on the \Rjpesl{} measurement. Uncertainties lower than $0.05\%$ are not listed. The two different jet selection strategies for the \Wboson{} boson reconstruction discussed in the text are topological proximity (``topo. prox.") and \pt-maximisation (``\pt-max."). }
\renewcommand{\arraystretch}{\myarraystretch}
  \centering \small
  \begin{tabular}[]{l | c | c}
    \hline \hline
      \multicolumn{1}{c|}{Effects}   & $\Delta$\Rjpesl{} topo. prox. [\%] & $\Delta$\Rjpesl{} \pt--max.[\%]   \\ \hline 
      Multijet background                  & $\pm 0.12$ & $\pm 0.18$  \\ 
      Jet resolution                       & $\pm 0.39$ & $\pm 0.80$  \\        
      MC generator                         & $\pm 0.41$ & $\pm 0.25$  \\      
      Fragmentation                        & $\pm 0.65$ & $\pm 0.68$  \\    
      Parton radiation                     & $\pm 2.48$ & $\pm 2.42$  \\       
      Total                                & $\pm 2.63$ & $\pm 2.65$  \\ \hline \hline
  \end{tabular}
 \label{tab:syste} 
\end{table}

\subsection {Reconstruction of the W boson}

The reconstruction efficiency for hadronically decaying \Wboson{} bo\-sons is measured by the fraction of reconstructed 
jet pairs matching the same \Wboson{} boson.  This can be done by forming all possible light-quark jet pairs consisting of jets which are not \btagged, and calculating their invariant mass $m_{\mathrm{jj}}$. Then, only pairs with $|m_{\text{jj}} - m_{\Wboson}^{\mathrm{MC}}| < 4 \sigma_{\Wboson}$ are considered as originating from \Wboson{} boson decays. Here $m_{\Wboson}^{\mathrm{MC}}$ is the \Wboson{} mass and $\sigma_{\Wboson}$ is the expected \mw{} resolution, both taken from \MC-simulation samples. This relatively large window of about $11$ \GeV{} avoids biases in the reconstructed \Wboson{} mass peak, and only about $3\%$ of true \Wboson{} bosons are rejected by this mass cut 

Two methods are used to select one jet pair per event. The first method is based 
on topological proximity in the detector, where the jet pair which minimises the distance 
between the two jets \DRjj{}, calculated in \etaphispace{} as defined in \eqRef{eq:deltaRdet} 
in  \secRef{sec:jetdirections}, is selected.  This reconstruction has an efficiency of $51\%$ 
in finding the signal jet pair at the level of the selection for reconstructible events. 
The second jet selection method is based on 
transverse momentum maximisation such 
that the two light-quark jets maximising the \pT{} of the reconstructed \Wboson{} are taken 
as the two jets from the hadronic decay. This reconstruction has an efficiency of $55\%$.

 Jet pairs with $\DRjj < 0.7$ are rejected in order to avoid geometrically overlapping jets 
and to reduce the sensitivity to parton radiation in the \Wboson{} mass spectrum. 

In order not to be sensitive to the jet mass the reconstructed \Wboson{} mass \mwrec{} 
is calculated as:
\begin{displaymath}
\mwrec = \sqrt{2 E_{1}E_{2} \left( 1 - \cos{\theta_{1,2}} \right)} ,
\end{displaymath}
 where $E_1$, $E_2$ are the respective energies of the paired jets, and $\theta_{1,2}$ is the opening angle between them. 

\subsection {Extraction of the relative light jet scale}
 \label{sec:measure} 

  The relative light-quark jet calibration \Rjpesl{} is defined by
 \begin{displaymath}
  \Rjpesl = \frac{\alpha_{l}^{\rm data}}{\alpha_{l}^{\rm MC}}, 
 \end{displaymath}
 where $\alpha_{l}^{\rm data}$ ($\alpha_{l}^{\rm MC}$) is the jet energy scale in the data (simulation). This analysis uses the expected dependency of the \Wboson{} mass distribution on the \Rjpesl{} parameter.
 Templates for the \mw{} distributions are derived from \MC{} simulations, where $\alpha_{l}^{\rm MC}$
 is varied. This rescaling of $\alpha_{l}^{\rm MC}$  is applied before the event
 selection and the \Wboson{} reconstruction steps. A set of \mw{} distributions are
  produced for different \Rjpesl{} values. In order to obtain the \mw{} 
  distribution of an arbitrary \Rjpesl{} value, a bin-by-bin interpolation is
  performed using the two generated and adjacent \Rjpesl{} values.

 A binned likelihood maximisation with a Poisson law is used. It identifies
 the \Rjpesl{} values whose associated \mw{} distribution fits the best to the
 observed \mw{} distribution. 
The analysis templates are defined for  \Rjpesl{} values ranging from
\Rjpesl = $0.85$ to \Rjpesl = $1.15$.

 In order to test the consistency of the extraction method, an arbitrary jet energy
 scale is applied to one pseudo-experiment of arbitrary luminosity.
The comparison is then done between the applied scale and the
 measured one. The difference between both is compatible with zero for a wide
 range of \Rjpesl{} hypotheses.  

 The expected statistical precision on \Rjpesl{} is determined using
 pseudo-experiments each one containing a number of events corresponding to the
 luminosity recorded in $2011$. A pull variable is computed, reflecting the differences between the measured and the
expected mean values scaled with the observed uncertainties.
The mean pull is compatible with zero and its standard deviation with unity. The mean value of
 the uncertainties obtained from the different pseudo-experiments is taken as
 the expected statistical precision. It is $0.28\%$ for the maximum \pt{}
 reconstruction method and $0.29\%$ for the topological proximity
 reconstruction method.  

\subsection{Systematic uncertainties}

The main sources of systematic uncertainties on the \Rjpesl{} measurement are
summarised in Table~\ref{tab:syste} and presented for the topological proximity and 
the \pt{} maximisation reconstruction methods. 

A variety of potential systematic effects are evaluated. The uncertainty from
the shape of the multijet background, %
the uncertainty on the jet energy resolution, the jet reconstruction efficiency 
and the $b$-tagging efficiency and mistag rate. 
The uncertainties on the Monte Carlo simulation model are estimated in
ter\-ms of generator variations, fragmentation uncertainty and parton radiation variation rate.
In particular the parton radiation rate can alter the \ttbar{} final states,
inducing distortions in the reconstructed \mw{} distribution.

\subsection{Results}

 \FigRef{fig:template_data}\subref{fig:wtd_templates} shows the observed \mw{} distribution from the maximum \pt{} reconstruction compared to three different templates. The relative scale correction \Rjpesl{} is extracted for electron and muon channels together as
 well as for the two channels separately. Results are summarised in
 Table~\ref{tab:rjpesl}. 

 In order to test the stability of the measurement, cross-checks are performed by relaxing the \DRjj{} cut and by changing the \mw{} reconstruction definition. None of these changes affects the measured \Rjpesl{} by more than
 $0.15\%$. Since the definition of \mw{} depends on \DRjj, a cross-check is done by
 an event re-weighting in \MC{} simulation in order to reproduce the observed
 \DRjj{} distribution in data. The effect on \Rjpesl{} is about $0.12\%$ for the two
 reconstruction methods.

 The relative scale \Rjpesl{} is studied as a function of the mean \pt, see \figRef{fig:template_data}, as well as a function of $\etajet$ of the two jets coming from the \Wboson{} boson decay. The tested \pt{} values range  
 from $33$ to $90$~\GeV. Templates of the \mw{} are produced for each bin of
 \pt{} or \etajet. Taking into account systematical uncertainties, no
 significant dependence is observed with respect to the average \pt{} or \etajet{} of
 the two jets. 
The mean \Rjpesl{} is 
measured as \Rjpesl$ = 1.0130 \pm 0.0028 \pm 0.027$. 

The agreement between the jet energy scale in data and Monte Carlo simulation is found
to be in agreement within the estimated uncertainties.
The main systematic uncertainty is related to the modelling of additional parton radiation
(see Table~\ref{tab:syste}).

\begin{table}[ht!]
\renewcommand{\arraystretch}{\myarraystretch}
 \caption{The measurement of \Rjpesl{} using the closest proximity ($\Delta R_{\mathrm{jj}}^{\mathrm{min}}$) and the maximum \pt{} ($p_{\mathrm{T}}^{\mathrm{max}}$) approach, respectively, for the electron channel, the muon channel and
 both together. Uncertainties are statistical only.} 
  \centering \small
  \begin{tabular}[]{c | c | c | c}
    \hline \hline
        {\Rjpesl} & $e$ channel & $\mu$ channel & e + $\mu$ channels \\   \hline
 $\Delta R_{\mathrm{jj}}^{\mathrm{min}}$  & $ 1.0130 \pm 0.0048 $ &  $ 1.0143 \pm 0.0038 $  & $ 1.0137 \pm 0.0031 $ \\ 
 $p_{\mathrm{T}}^{\mathrm{max}}$    & $ 1.0105 \pm 0.0045 $ &  $ 1.0141 \pm 0.0038 $  & $ 1.0130 \pm 0.0028 $ \\ \hline \hline
  \end{tabular}
 \label{tab:rjpesl} 
\end{table}

\section{Systematic uncertainties on corrections for pile-up interactions}
\label{sec:pileupsystematics}
\subsection{Event and object selection}
\label{sec:pileupeventselection}
The pile-up corrections for jets derived from \MC{} simulation, 
as described in \secRef{sec:pileupsection}, 
can be validated with data samples of collisions events where a
stable reference that is insensitive to pile-up can be used
to assess the agreement of the Monte Carlo simulation with data.
Of particular interest here are
\gammajet{} events in prompt photon production, as the reconstructed photon kinematics 
are not affected by pile-up, and its transverse momentum 
\pTgam{} provides the stable reference for the pile-up dependent response of the 
balancing jet in the ratio $\ptjet/\ptref = \ptjet/\pTgam$. 
The \gammajet{} sample is selected as detailed in \secRef{sec:gammajeteventselection}.

Another per jet kinematic reference is provided by the track jets from the primary collision 
vertex introduced in \secRef{sec:trackjets}. These are matched with calorimeter jets, 
and the transverse momentum ratio $\ptjet/\ptref = \ptjet/\ptjetTrk$ is evaluated. 
%
%
The jet event sample needed for this evaluation can be extracted from samples with central jets in the final state. Both this and the \gammajet{} data samples are mostly useful for validation of the pile-up correction methods, as the limited statistics and phase space coverage in $2011$ data do not allow direct determination of the 
pile-up
corrections from these final states in data. 

To evaluate the pile-up corrections based on track jets, events with a calorimeter jet matching a \TrkJet{} with $\pttrk > 20$ \GeV{}
are extracted from an event sample triggered by high-\pT{} muons, thus avoiding potential jet-trigger biases. A \TrkJet{} is only associated 
with a calorimeter jet not overlapping with any reconstructed muon with $\pT^{\mu} > 5$~\GeV, to avoid potential biases from
heavy-flavour jets containing semi-leptonic decays. The general matching criterion for \TrkJet s to calorimeter jets is  based on the distance between the two jets \DeltaR{} in \etaphispace{},
as defined in \eqRef{eq:deltaRdet} in  \secRef{sec:jetdirections}.
Only uniquely matched track-jet--calorimeter-jet pairs with distances $\DeltaR < 0.3$ are considered. Outside of the imposed requirement for calorimeter jet reconstruction in \ATLAS{} in 2011 ($\ptjet > 10$~\GeV), no further cuts are applied on \ptjet, to avoid biases in the $\ptjet/\pttrk$ ratio, in particular at low \pttrk.

\begin{figure*}\centering
\subfloat[$\left|\etaDet\right| < 0.3$ (MC)]{\includegraphics[width=0.45\textwidth]{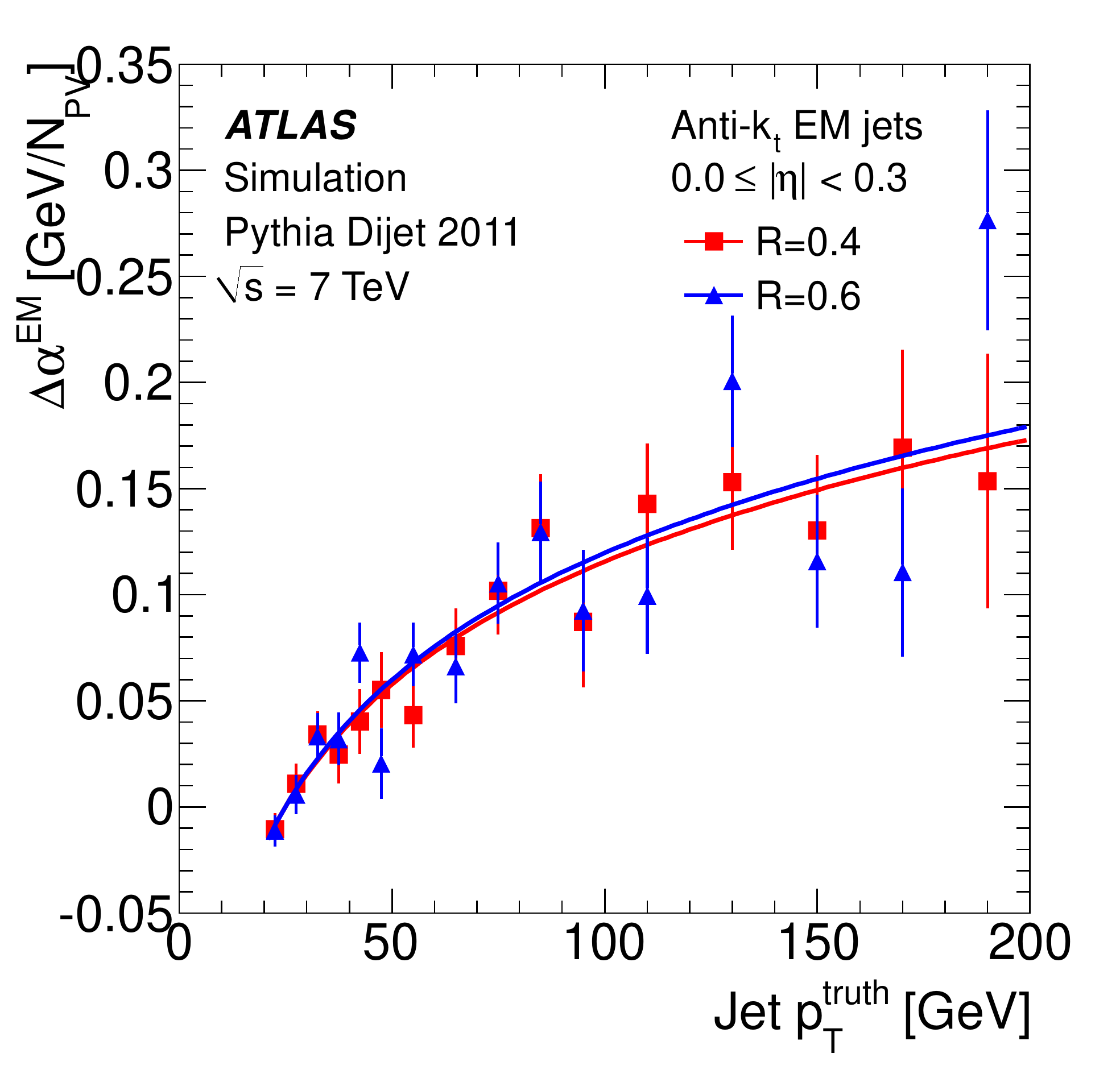} \label{fig:ptdep:npv_eta0}}
\subfloat[$1.2 \leq \left|\etaDet\right| < 2.1$ (MC)]{\includegraphics[width=0.45\textwidth]{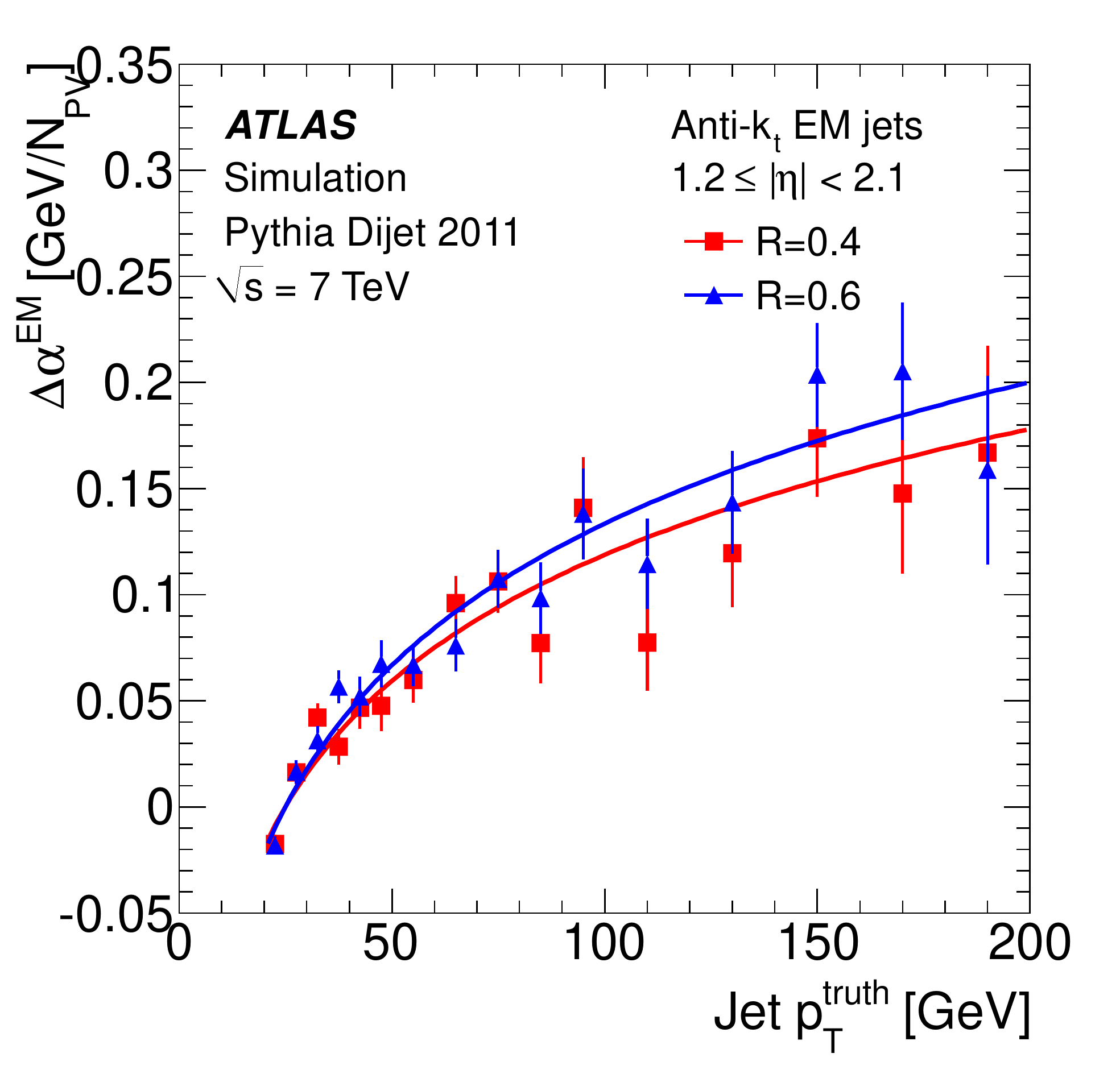} \label{fig:ptdep:npv_eta1}}
\qquad
\subfloat[$\left|\etaDet\right| < 0.3$ (MC)]{\includegraphics[width=0.45\textwidth]{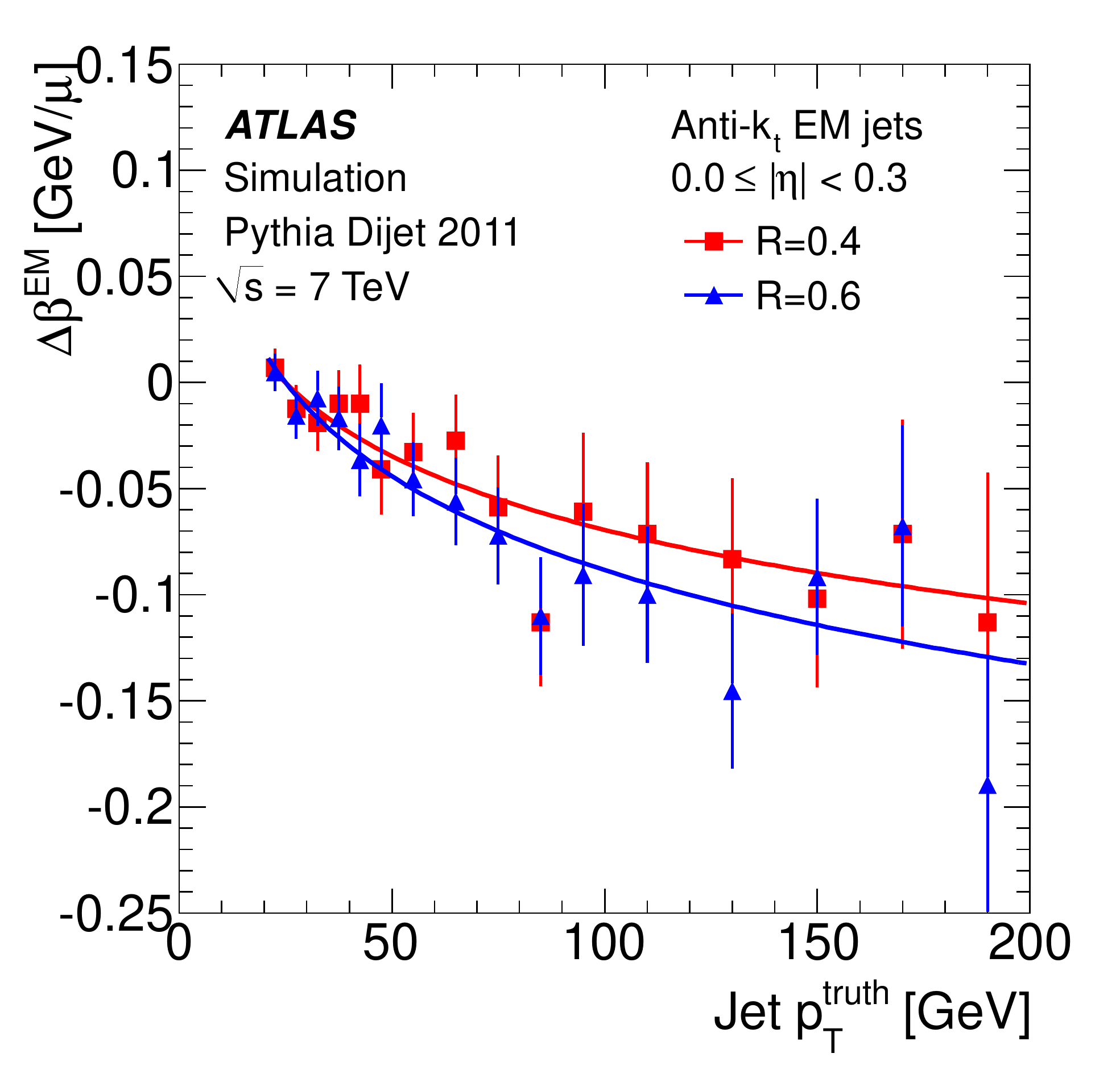} \label{fig:ptdep:mu_eta0}}
\subfloat[$1.2 \leq \left|\etaDet\right| < 2.1$ (MC)]{\includegraphics[width=0.45\textwidth]{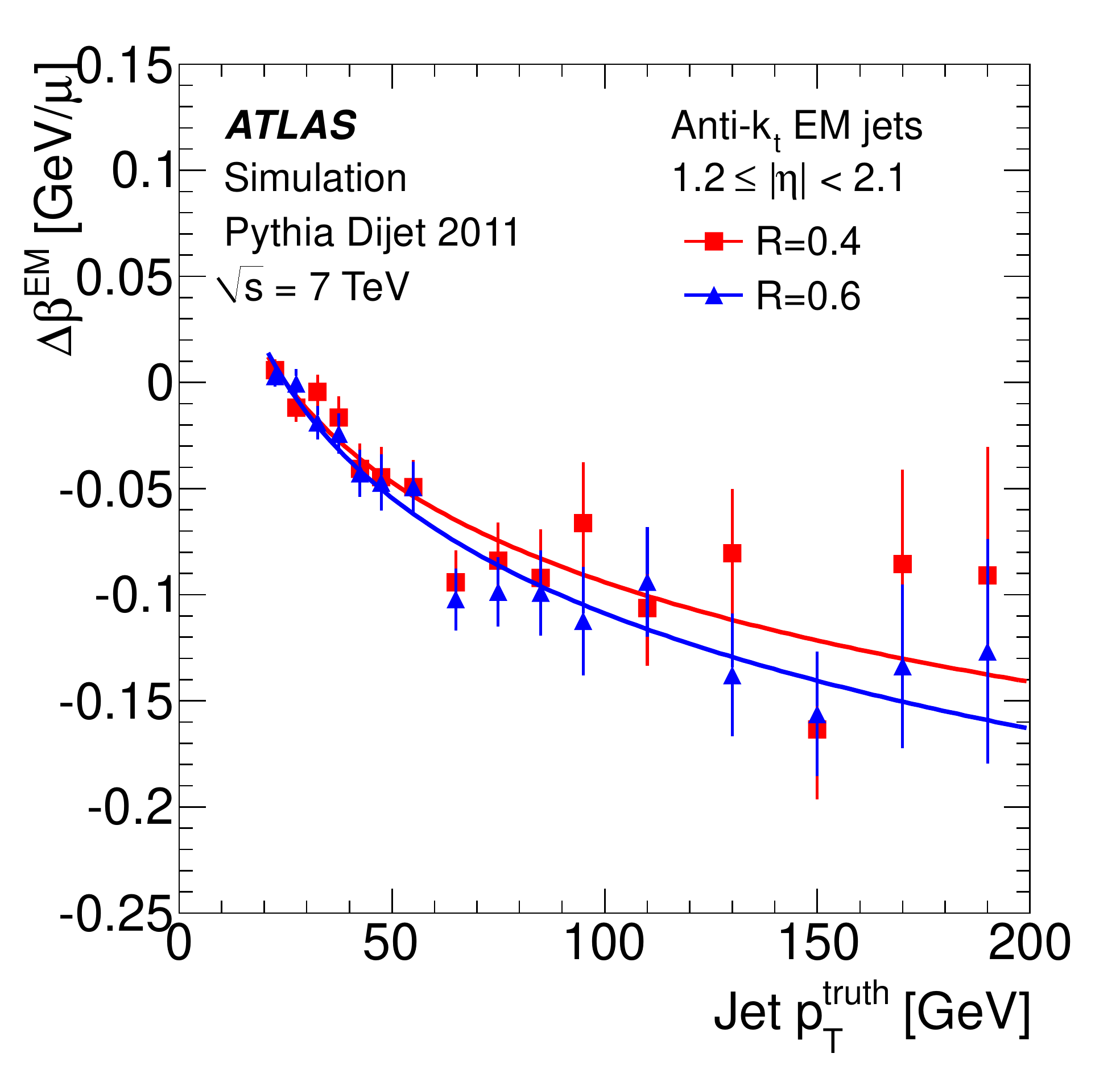} \label{fig:ptdep:mu_eta1}}
\caption[]{The difference from the average \alphaEMFct{} of the in-time pile-up signal
contribution per reconstructed primary vertex ($\Delta(\partial\pTrec{\EM}/\partial\Npv)(\pttrue)$)
as a function of the true jet transverse momentum \pttrue, for \MC-simulated jets reconstructed 
with \antikt{} $R = 0.4$ and $R = 0.6$ 
at the \EM{} scale, 
in two different regions  \subref{fig:ptdep:npv_eta0} 
$\left|\etaDet\right| < 0.3$ and  \subref{fig:ptdep:npv_eta1}  $1.2 \leq \left|\etaDet\right| < 2.1$
of the \ATLAS{} calorimeter. In \subref{fig:ptdep:mu_eta0}  and  \subref{fig:ptdep:mu_eta1}, the variations of the
out-of-time pile-up signal contribution per interaction with \pttrue{} around its average
\betaEMFct{} ($\Delta(\partial\pTrec{\EM}/\partial\axing{})(\pttrue)$) are shown for the same jet
samples and the same respective \etaDet{} regions. Logarithmic functions of \pttrue{} 
are fitted to the points obtained from \MC{} simulations.
\label{fig:ptdep}}   
\end{figure*}
\begin{figure*}\centering
\hspace*{\fill} \subfloat[\EMJES, $R = 0.4$]{\includegraphics[width=0.4\textwidth]{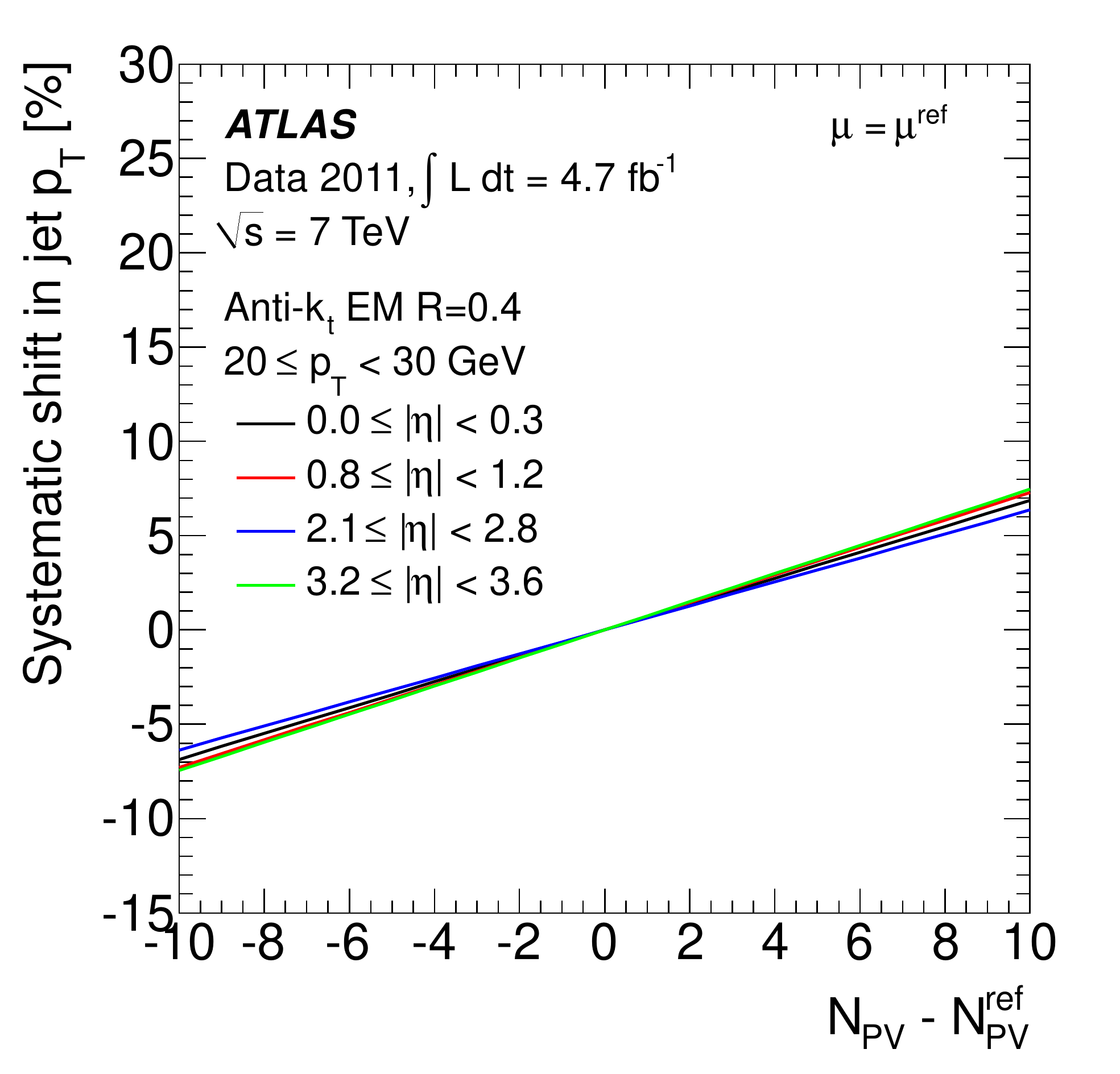} \label{fig:syst4n:npv_emjes_pt0}}  \hspace*{\fill}
\subfloat[\LCWJES{}, $R = 0.4$]{\includegraphics[width=0.4\textwidth]{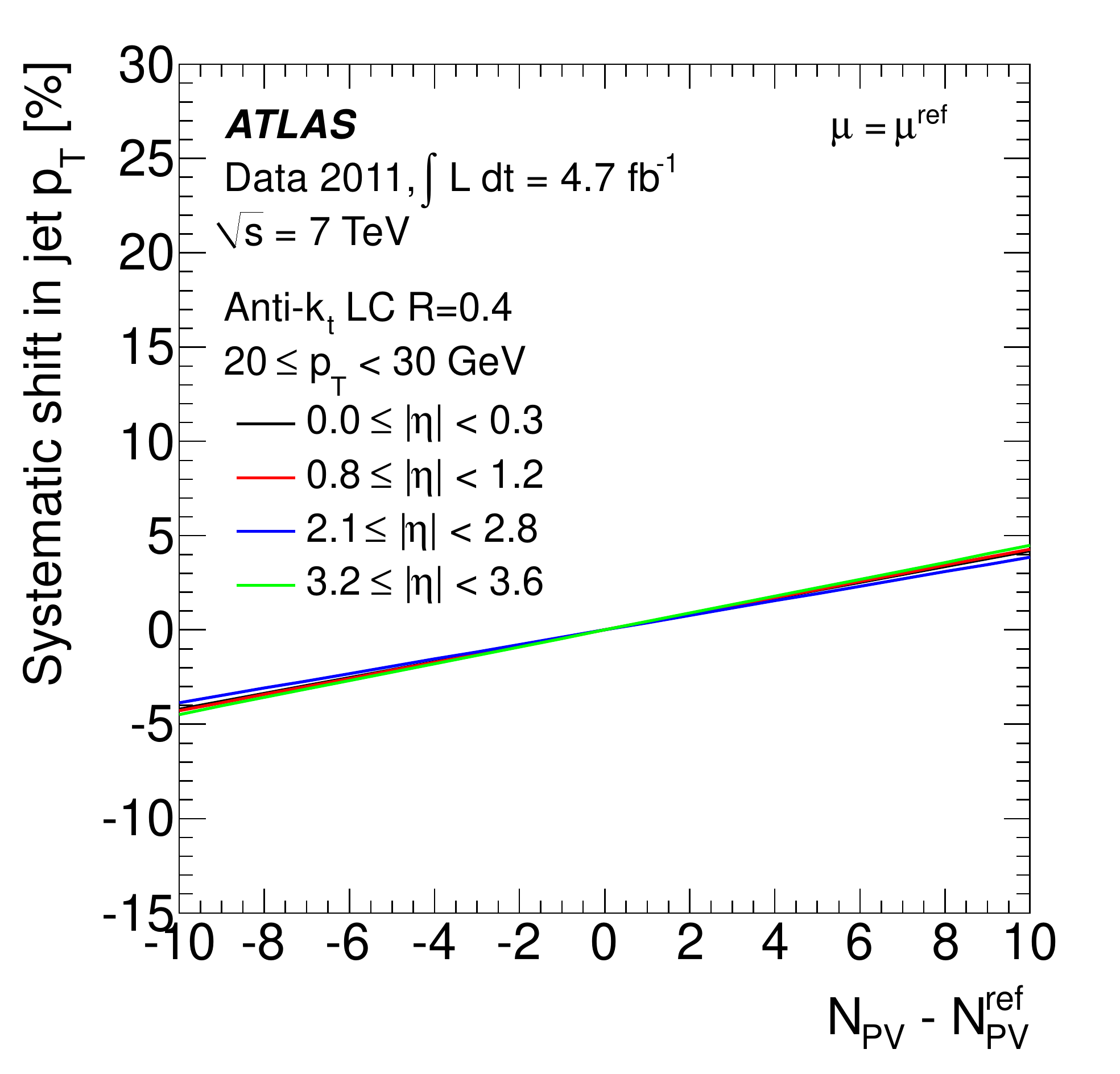} \label{fig:syst4n:npv_lcjes_pt0}}  \hspace*{\fill} \\
\hspace*{\fill} \subfloat[\EMJES, $R = 0.4$]{\includegraphics[width=0.4\textwidth]{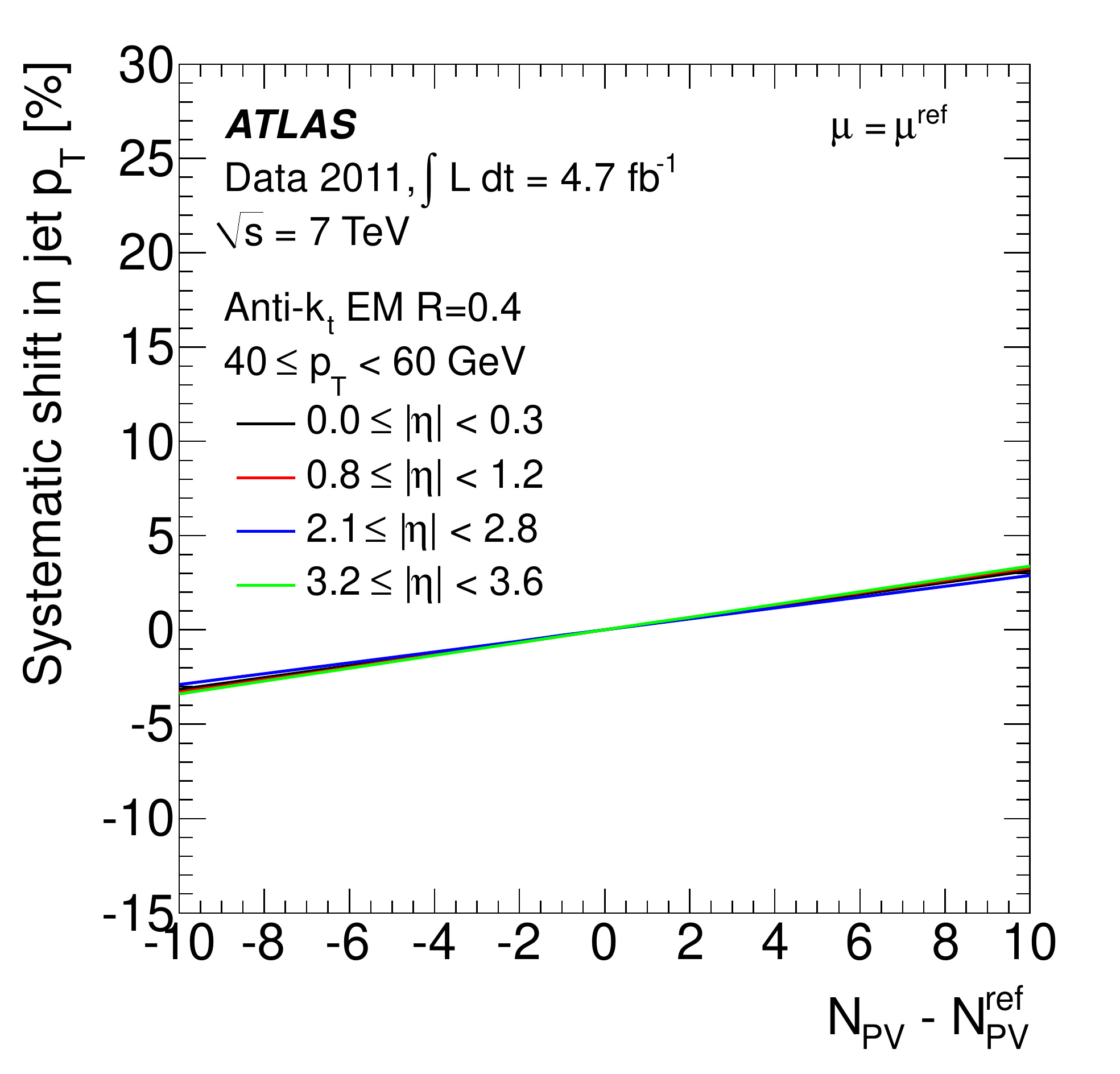} \label{fig:syst4n:npv_emjes_pt1}}  \hspace*{\fill}
\subfloat[\LCWJES, $R = 0.4$]{\includegraphics[width=0.4\textwidth]{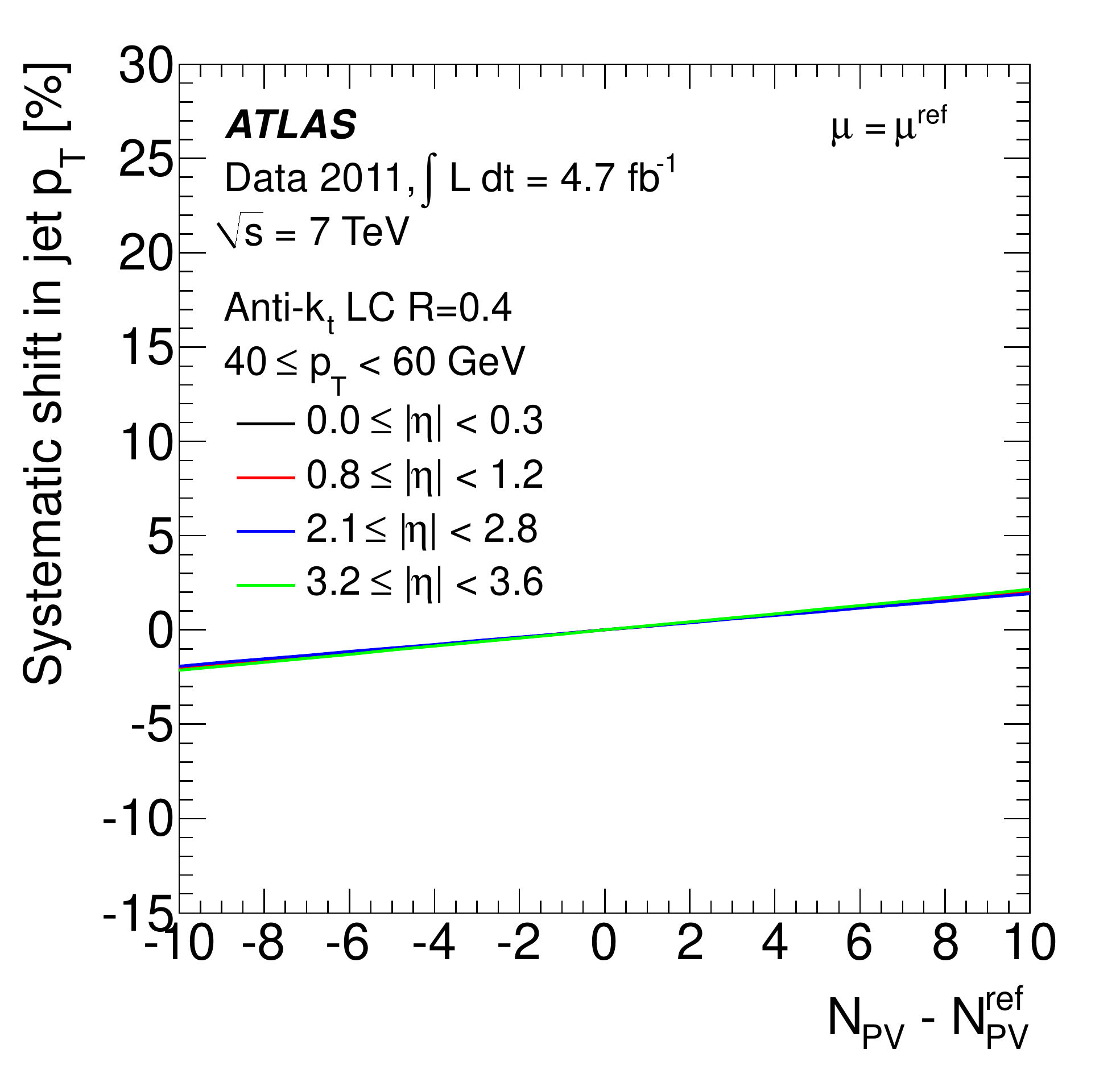} \label{fig:syst4n:npv_lcjes_pt1}}  \hspace*{\fill} \\
\hspace*{\fill} \subfloat[\EMJES, $R = 0.4$]{\includegraphics[width=0.4\textwidth]{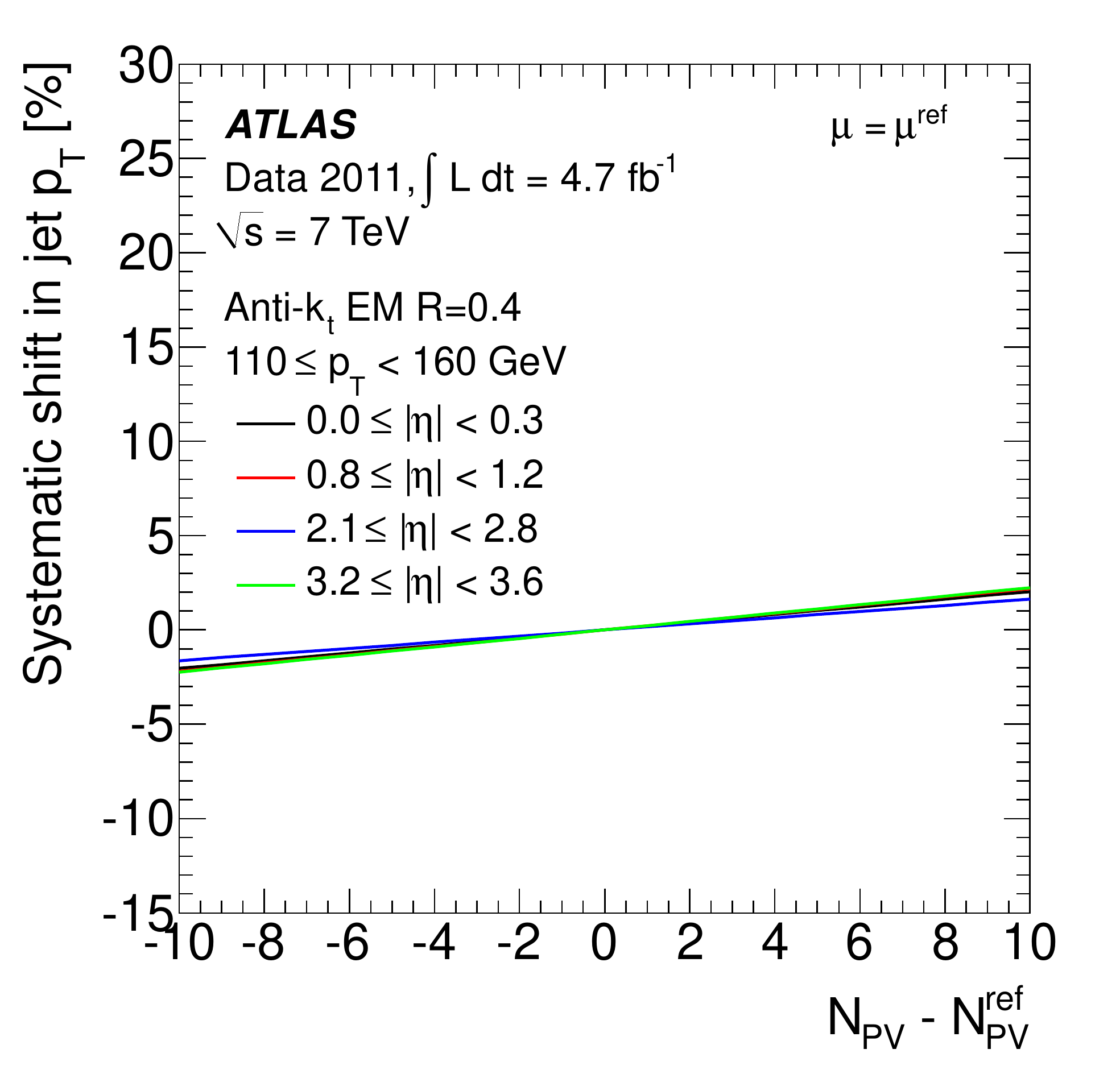} \label{fig:syst4n:npv_emjes_pt2}}  \hspace*{\fill}
\subfloat[\LCWJES, $R = 0.4$]{\includegraphics[width=0.4\textwidth]{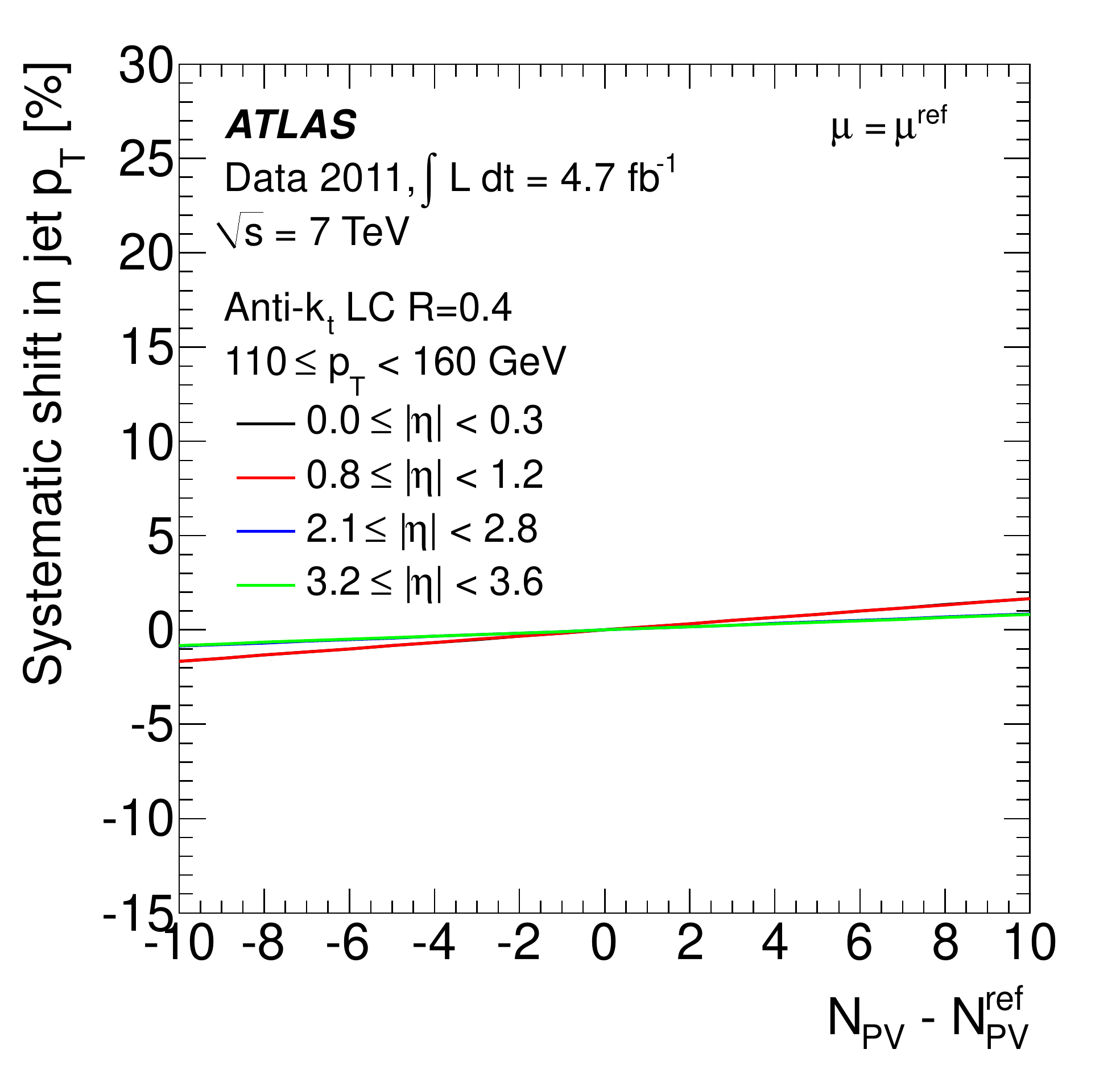} \label{fig:syst4n:npv_lcjes_pt2}}   \hspace*{\fill}
\caption[]{The fractional systematic shift due to mis-modelling of the effect of in-time 
pile-up on the transverse momentum \pTrec{\EMJES} of jets reconstructed with the 
\antikt{} algorithm with $R = 0.4$, and calibrated with the \EMJES{} scheme, is shown as a function 
of $\left(\Npv-\NpvRef\right)$ in \subref{fig:syst4n:npv_emjes_pt0}, \subref{fig:syst4n:npv_emjes_pt1}, and \subref{fig:syst4n:npv_emjes_pt2} for various \pTrec{\EMJES}{} bins. The same systematic shift is shown in 
\subref{fig:syst4n:npv_lcjes_pt0}, \subref{fig:syst4n:npv_lcjes_pt1}, and \subref{fig:syst4n:npv_lcjes_pt2} for jets calibrated with the \LCWJES{} scheme, now in bins of 
\pTrec{\LCWJES}. \label{fig:syst4n}}
\end{figure*}
\begin{figure*}\centering
\hspace*{\fill} \subfloat[\EMJES, $R = 0.6$]{\includegraphics[width=0.4\textwidth]{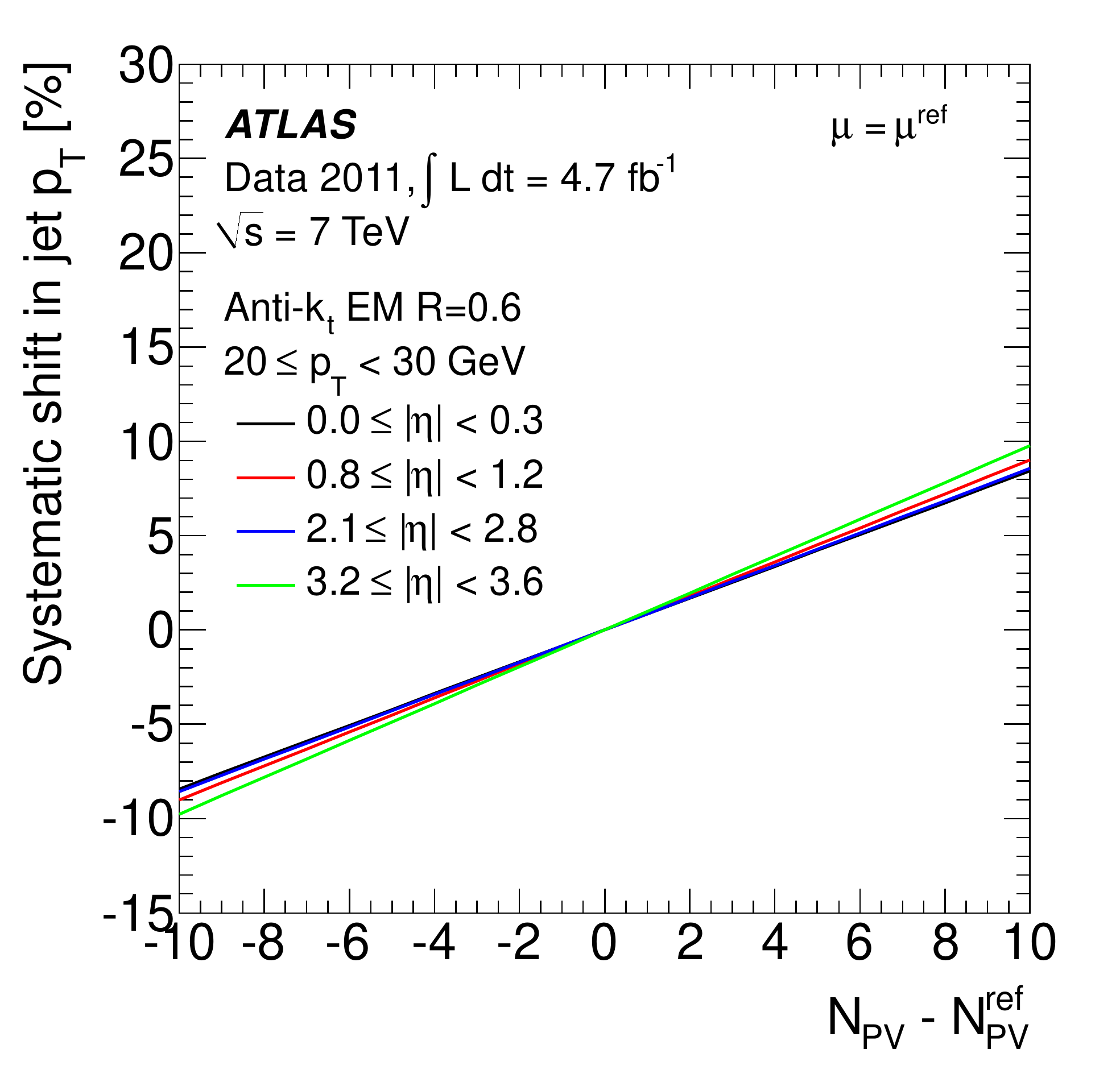} \label{fig:syst6n:npv_emjes_pt0}}  \hspace*{\fill}
\subfloat[\LCWJES, $R = 0.6$]{\includegraphics[width=0.4\textwidth]{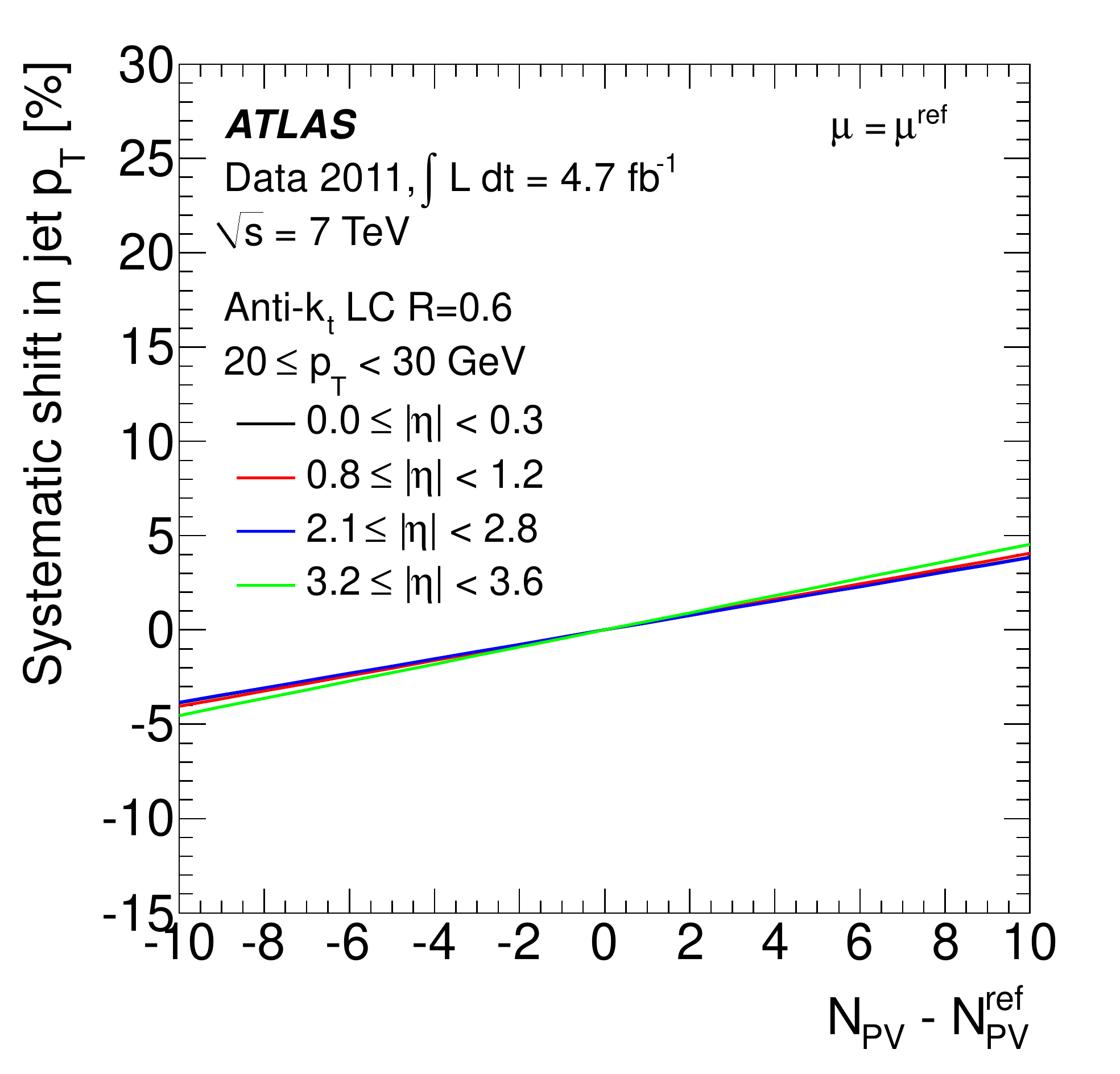} \label{fig:syst6n:npv_lcjes_pt0}}  \hspace*{\fill} \\
\hspace*{\fill} \subfloat[\EMJES, $R = 0.6$]{\includegraphics[width=0.4\textwidth]{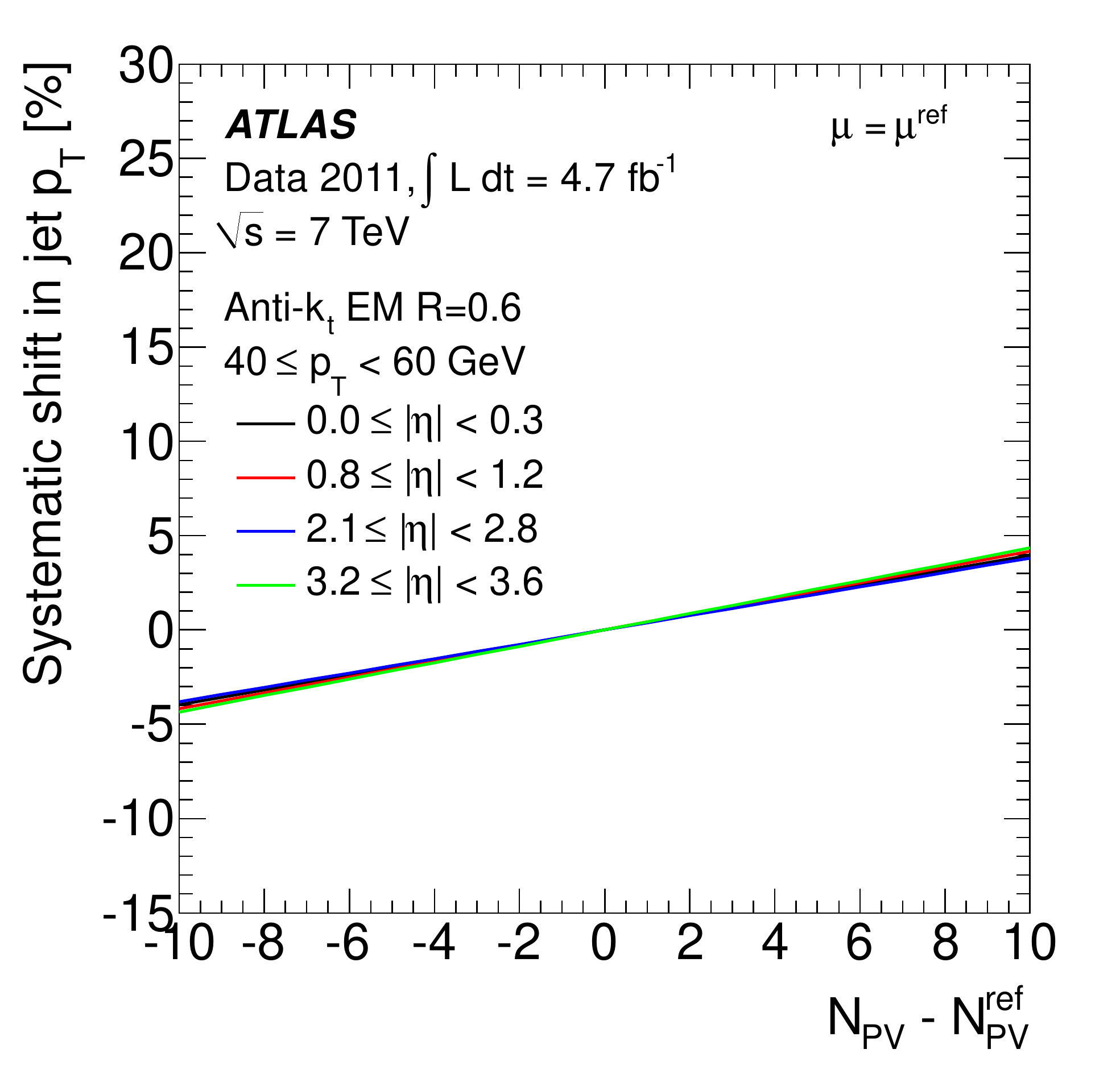} \label{fig:syst6n:npv_emjes_pt1}}  \hspace*{\fill}
\subfloat[\LCWJES, $R = 0.6$]{\includegraphics[width=0.4\textwidth]{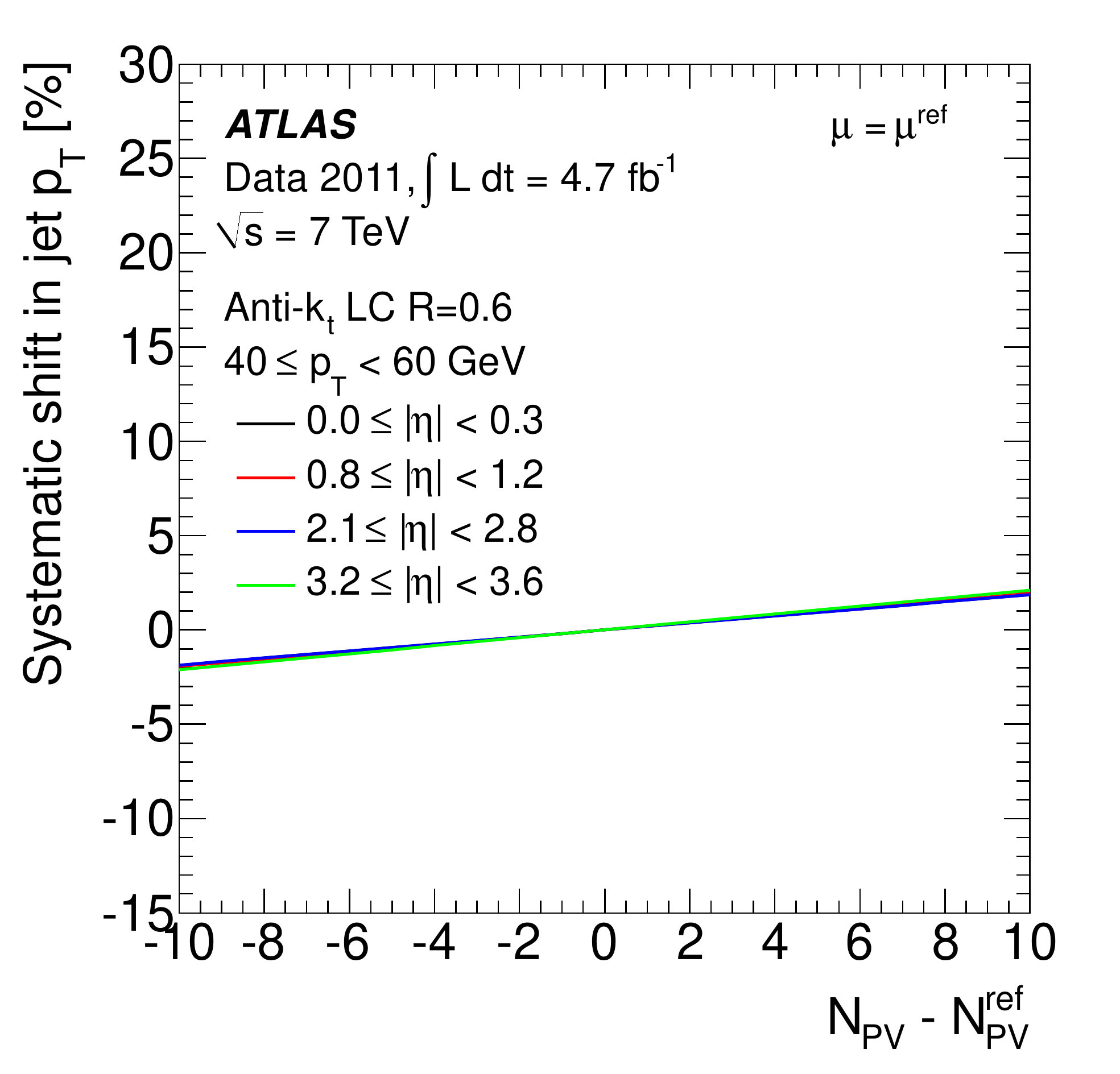} \label{fig:syst6n:npv_lcjes_pt1}}  \hspace*{\fill} \\
\hspace*{\fill} \subfloat[\EMJES, $R = 0.6$]{\includegraphics[width=0.4\textwidth]{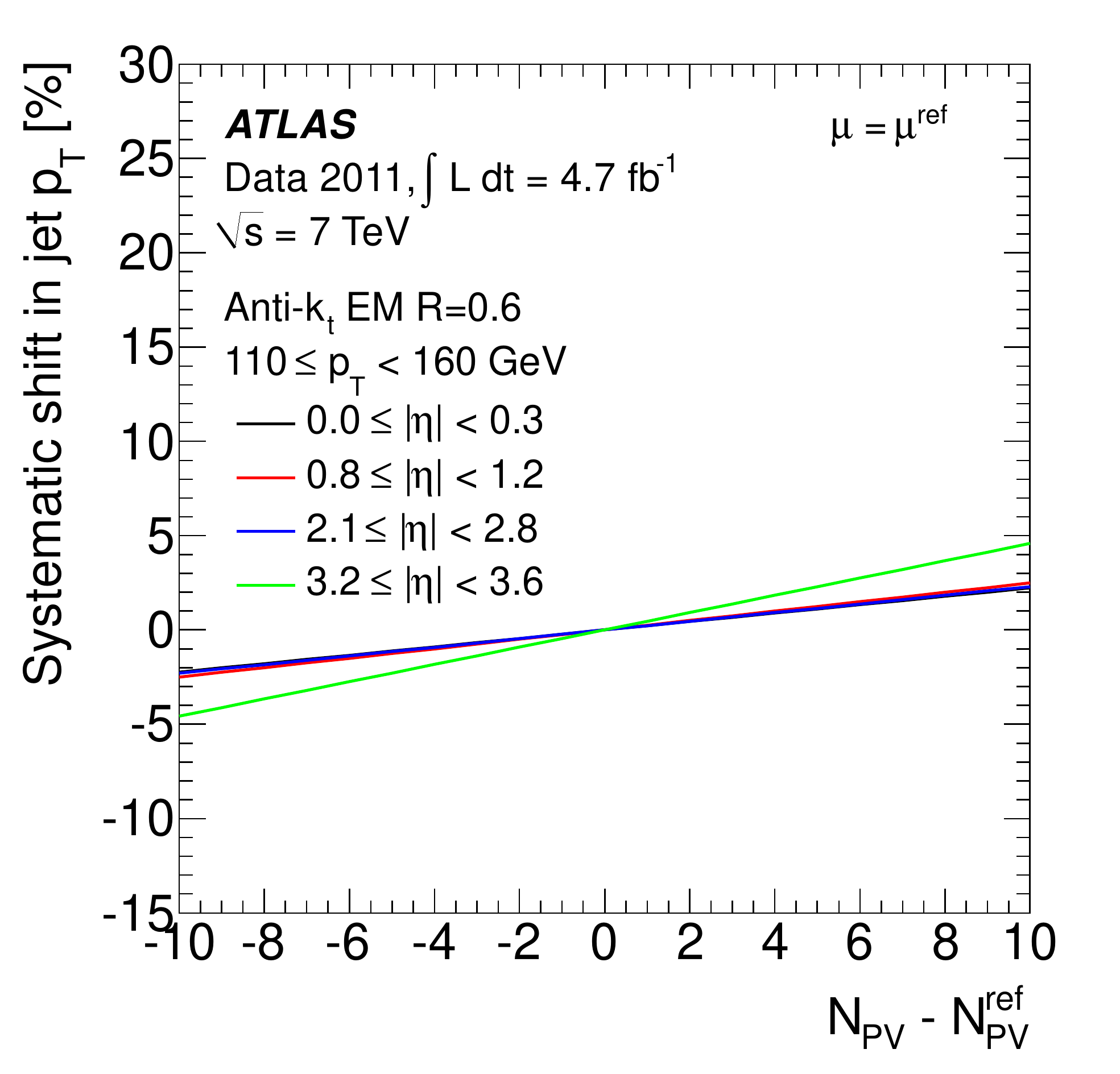} \label{fig:syst6n:npv_emjes_pt2}}  \hspace*{\fill}
\subfloat[\LCWJES, $R = 0.6$]{\includegraphics[width=0.4\textwidth]{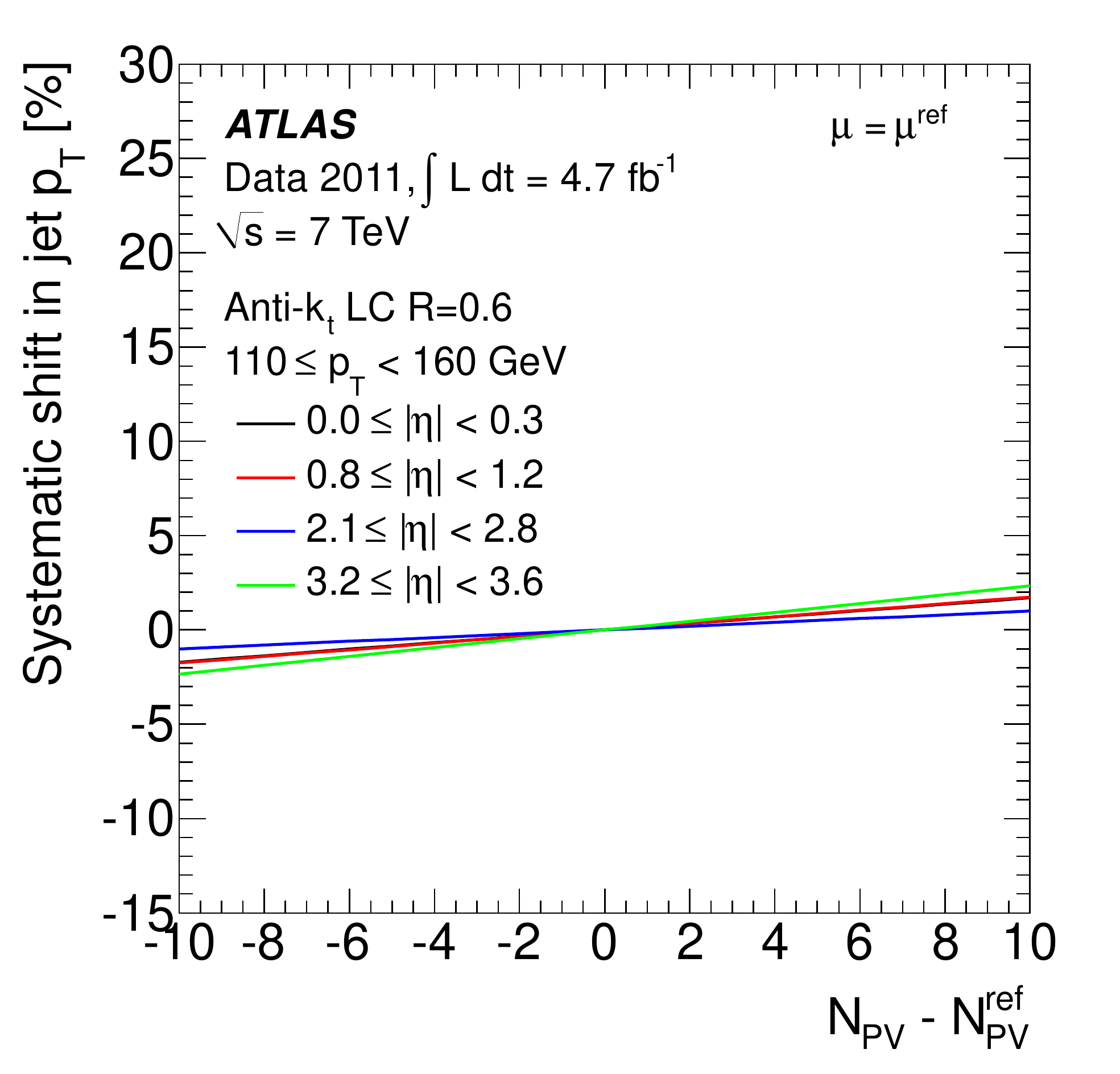} \label{fig:syst6n:npv_lcjes_pt2}}   \hspace*{\fill}
\caption[]{The fractional systematic shift due to mis-modelling of the effect of in-time 
pile-up on the transverse momentum \pTrec{\EMJES} of jets reconstructed with the 
\antikt{} algorithm with $R = 0.6$ and calibrated with the \EMJES{} scheme, is shown as a function 
of $\left(\Npv-\NpvRef\right)$ in \subref{fig:syst6n:npv_emjes_pt0}, \subref{fig:syst6n:npv_emjes_pt1}, and \subref{fig:syst6n:npv_emjes_pt2} for various \pTrec{\EMJES}{} bins. The same systematic shift is shown in 
\subref{fig:syst6n:npv_lcjes_pt0}, \subref{fig:syst6n:npv_lcjes_pt1}, and \subref{fig:syst6n:npv_lcjes_pt2} for jets calibrated with the \LCWJES{} scheme, now in bins of 
\pTrec{\LCWJES}. \label{fig:syst6n}}
\end{figure*}
\begin{figure*}\centering
\hspace*{\fill} \subfloat[\EMJES, $R = 0.4$]{\includegraphics[width=0.4\textwidth]{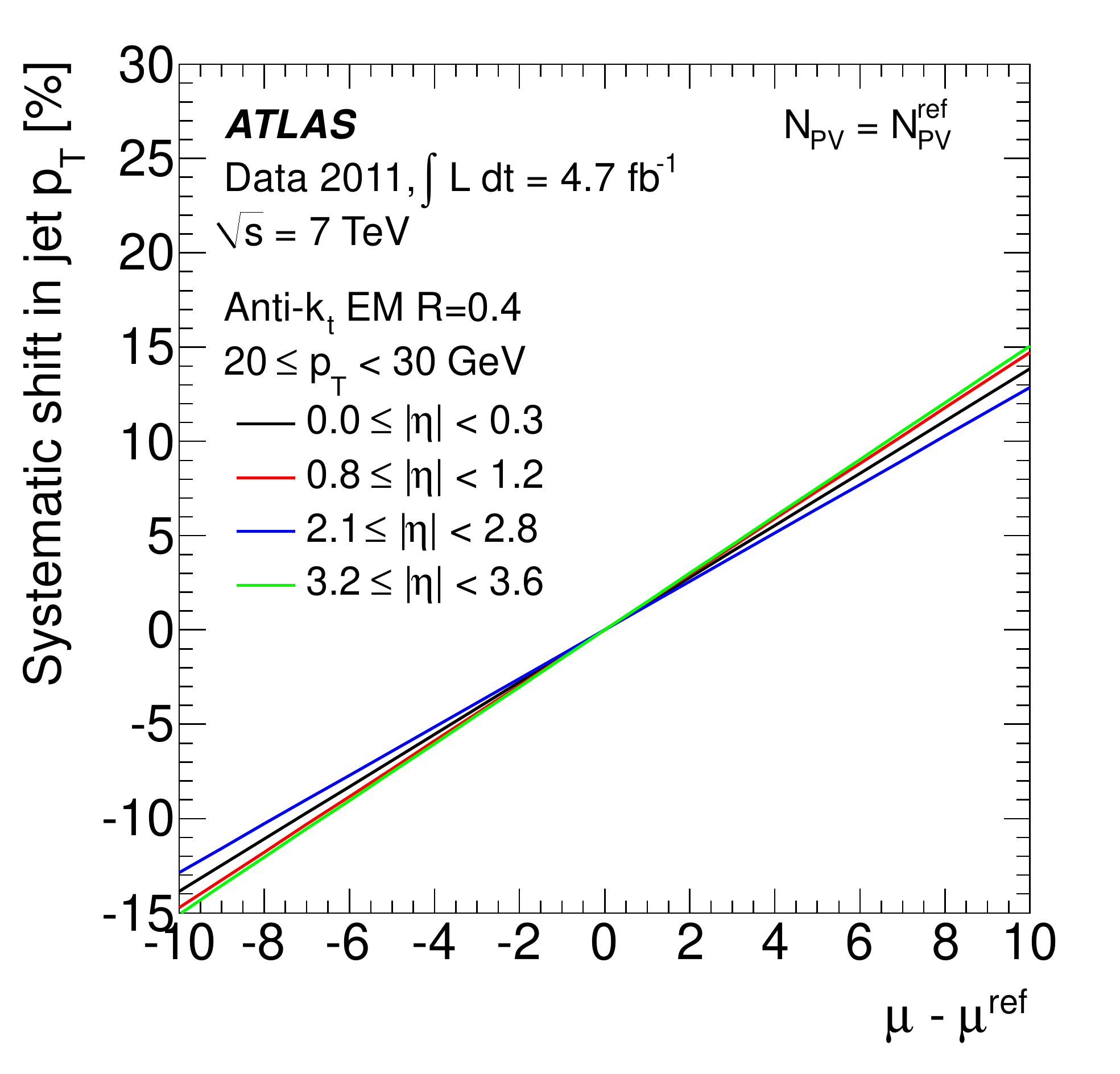} \label{fig:syst4m:mu_emjes_pt0}}  \hspace*{\fill}
\subfloat[\LCWJES, $R = 0.4$]{\includegraphics[width=0.4\textwidth]{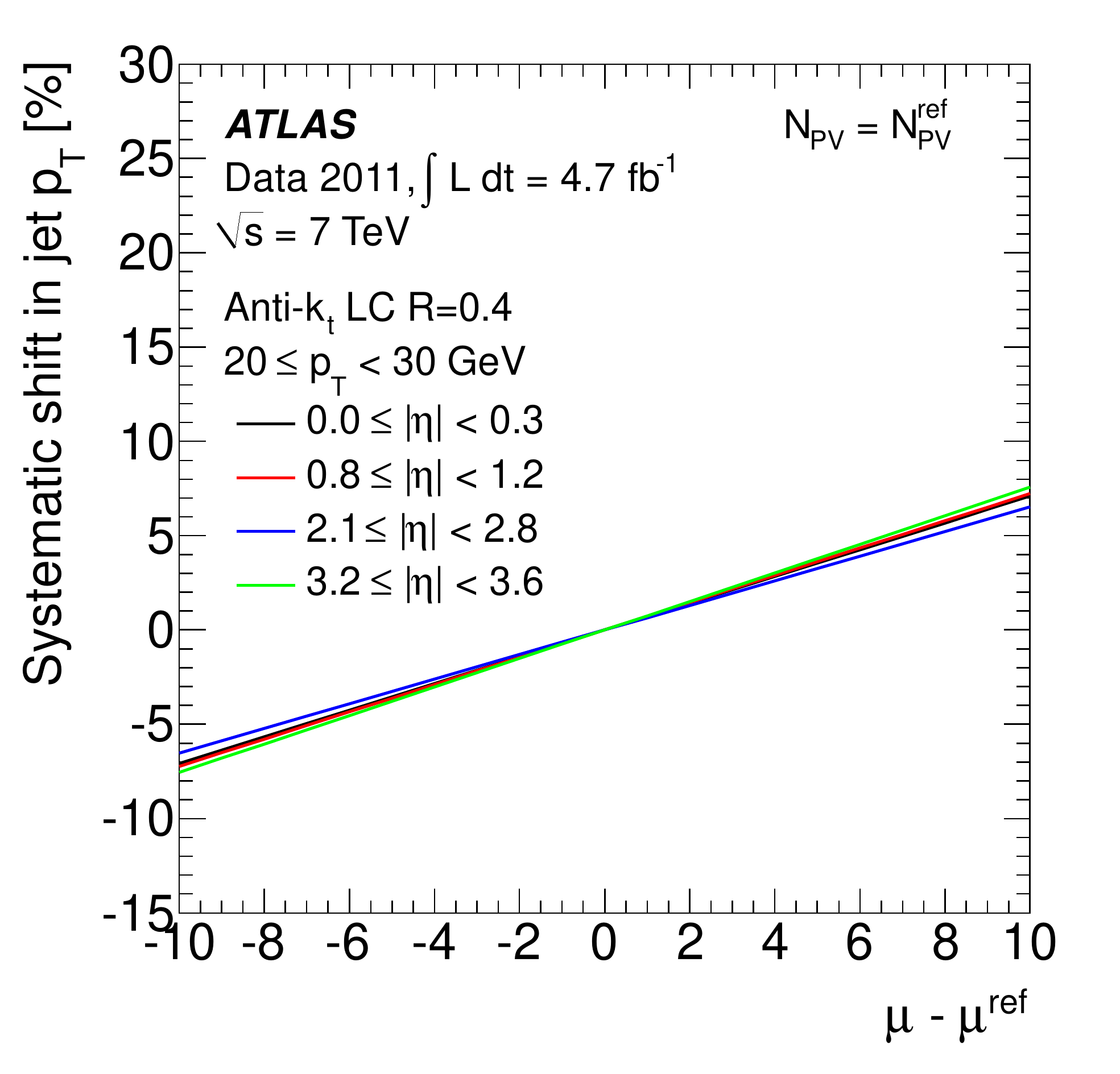} \label{fig:syst4m:mu_lcjes_pt0}}  \hspace*{\fill} \\
\hspace*{\fill} \subfloat[\EMJES, $R = 0.4$]{\includegraphics[width=0.4\textwidth]{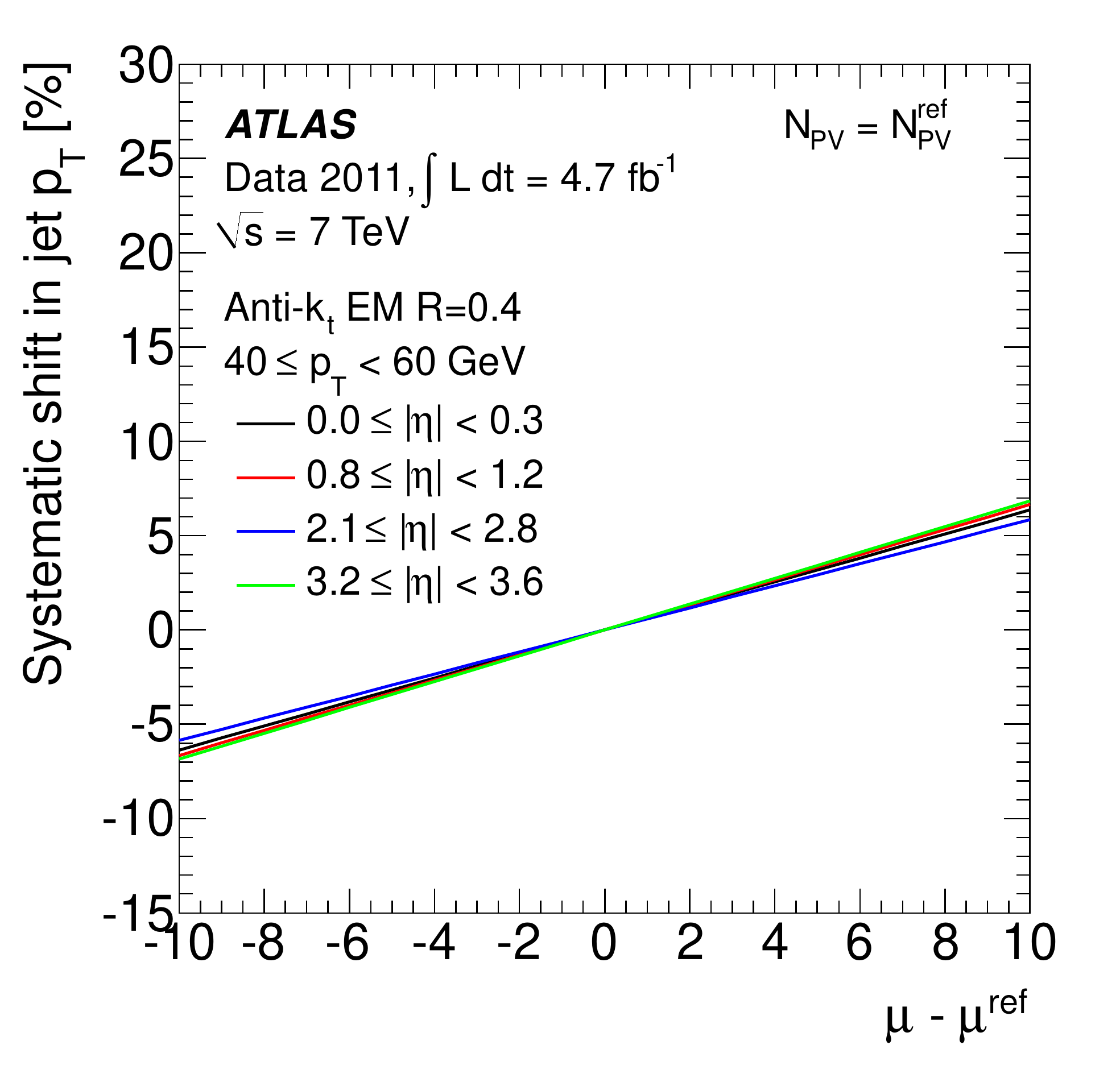} \label{fig:syst4m:mu_emjes_pt1}}  \hspace*{\fill}
\subfloat[\LCWJES, $R = 0.4$]{\includegraphics[width=0.4\textwidth]{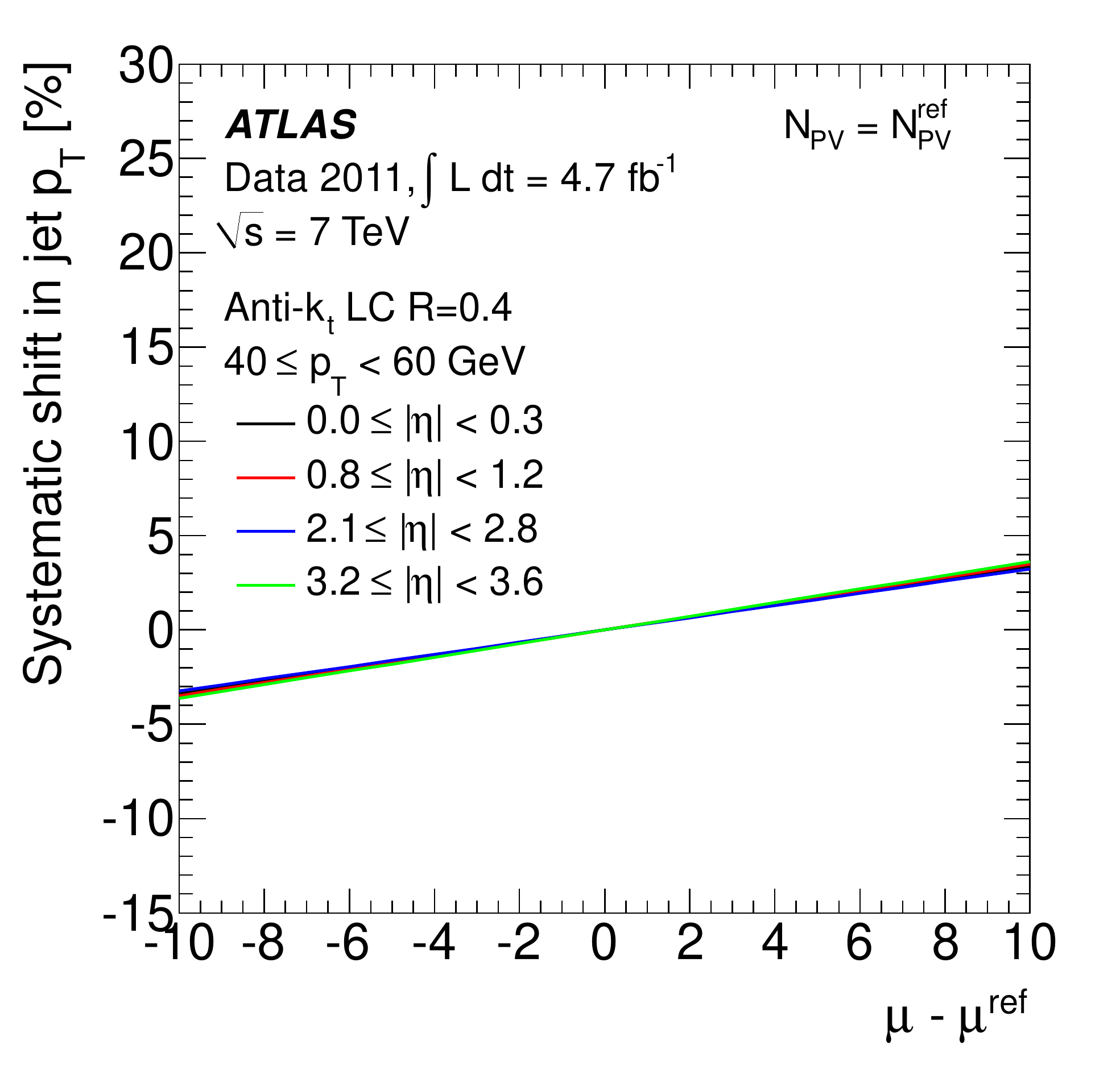} \label{fig:syst4m:mu_lcjes_pt1}}  \hspace*{\fill} \\
\hspace*{\fill} \subfloat[\EMJES, $R = 0.4$]{\includegraphics[width=0.4\textwidth]{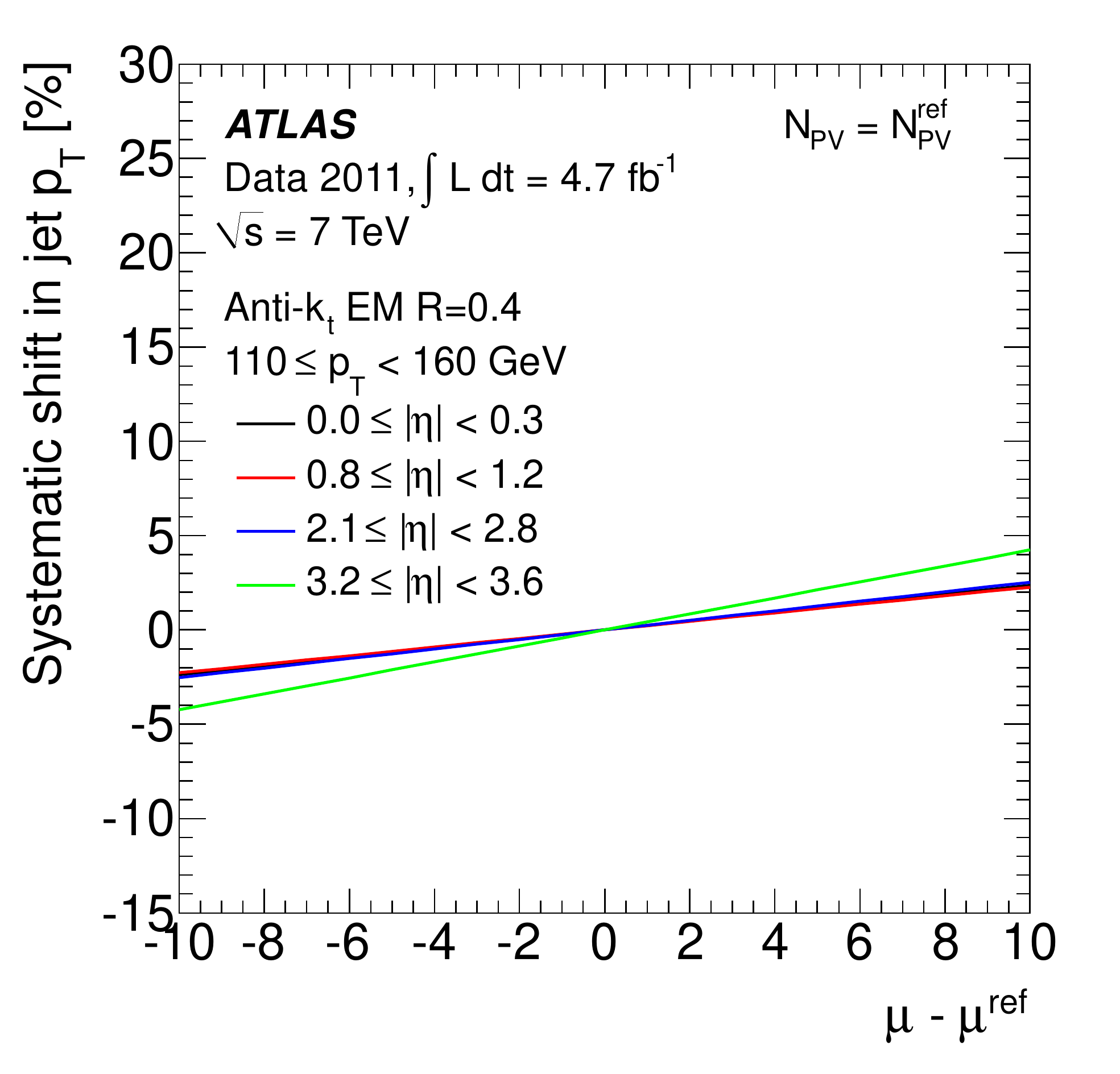} \label{fig:syst4m:mu_emjes_pt2}}  \hspace*{\fill}
\subfloat[\LCWJES, $R = 0.4$]{\includegraphics[width=0.4\textwidth]{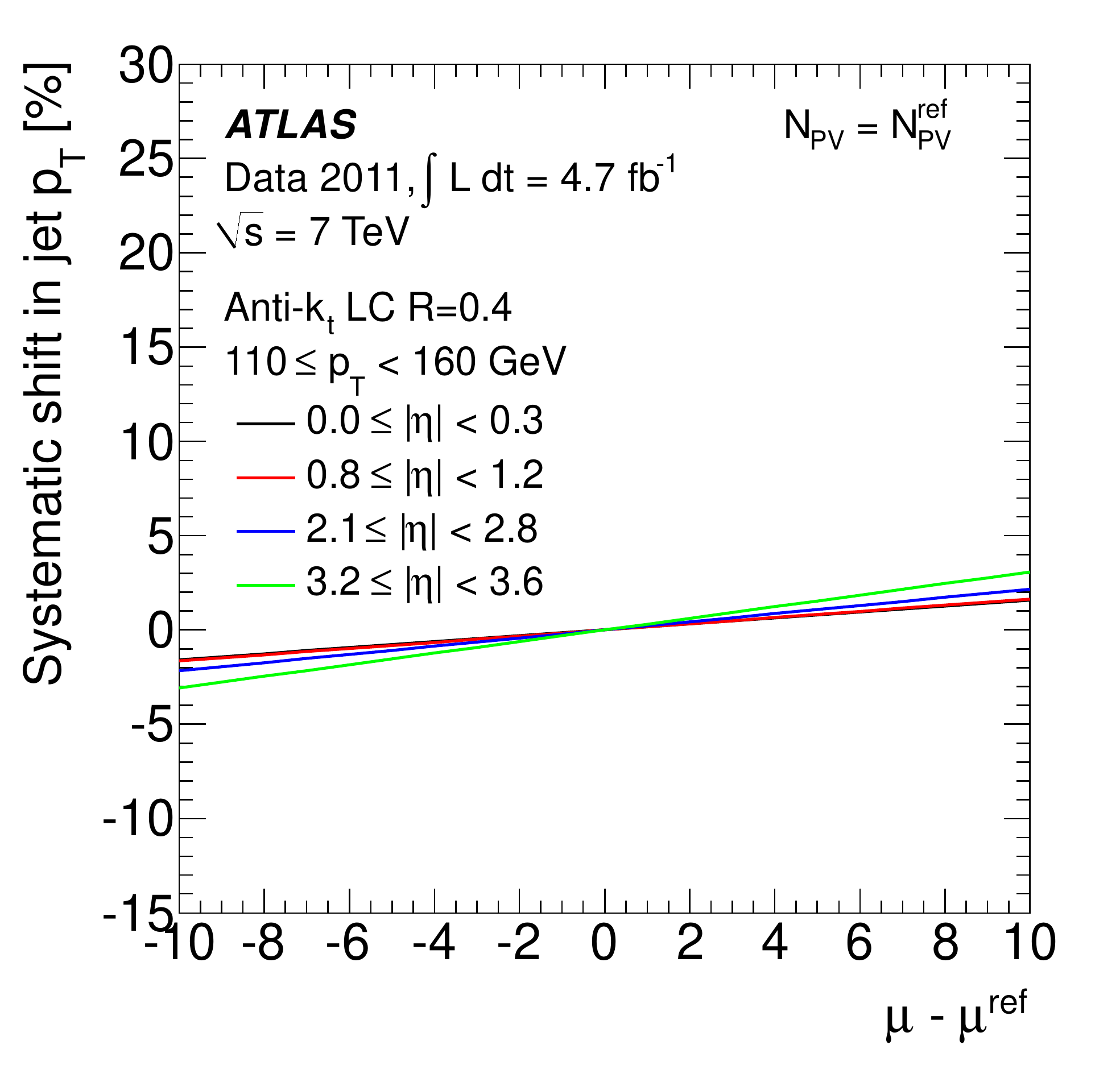} \label{fig:syst4m:mu_lcjes_pt2}}   \hspace*{\fill}
\caption[]{The fractional systematic shift due to mis-modelling of the effect of out-of-time 
pile-up on the transverse momentum \pTrec{\EMJES} of jets reconstructed with the 
\antikt{} algorithm with $R = 0.4$ and calibrated with the \EMJES{} scheme, is shown as a function 
of $\left(\axing-\axingRef\right)$ in \subref{fig:syst4m:mu_emjes_pt0}, \subref{fig:syst4m:mu_emjes_pt1}, and \subref{fig:syst4m:mu_emjes_pt2} for various \pTrec{\EMJES}{} bins. The same systematic shift is shown in 
\subref{fig:syst4m:mu_lcjes_pt0}, \subref{fig:syst4m:mu_lcjes_pt1}, and \subref{fig:syst4m:mu_lcjes_pt2} for jets calibrated with the \LCWJES{} scheme, now in bins of 
\pTrec{\LCWJES}. \label{fig:syst4m}}
\end{figure*}
\begin{figure*}\centering
\hspace*{\fill} \subfloat[\EMJES, $R = 0.6$]{\includegraphics[width=0.4\textwidth]{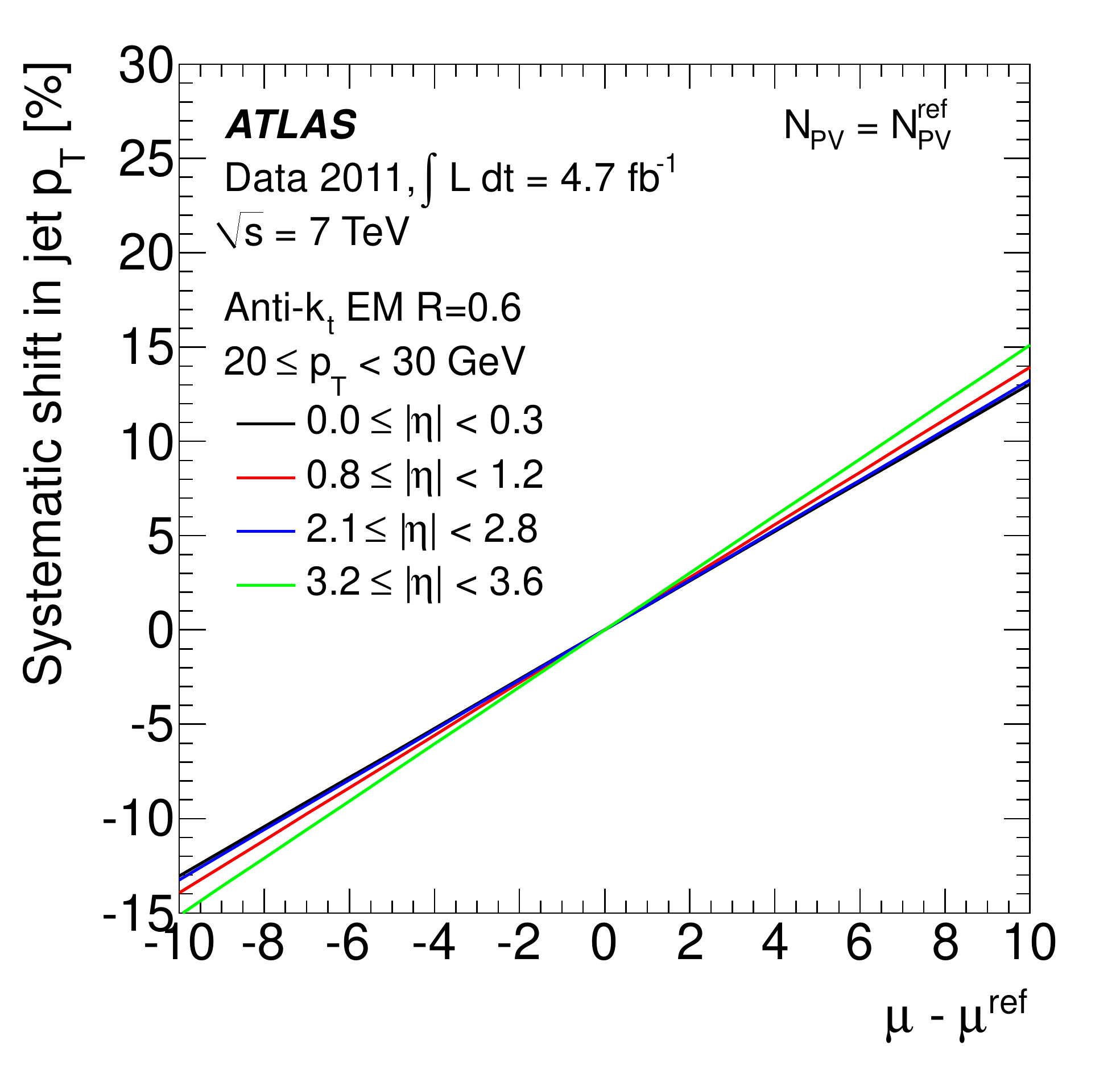} \label{fig:syst6m:mu_emjes_pt0}}  \hspace*{\fill}
\subfloat[\LCWJES, $R = 0.6$]{\includegraphics[width=0.4\textwidth]{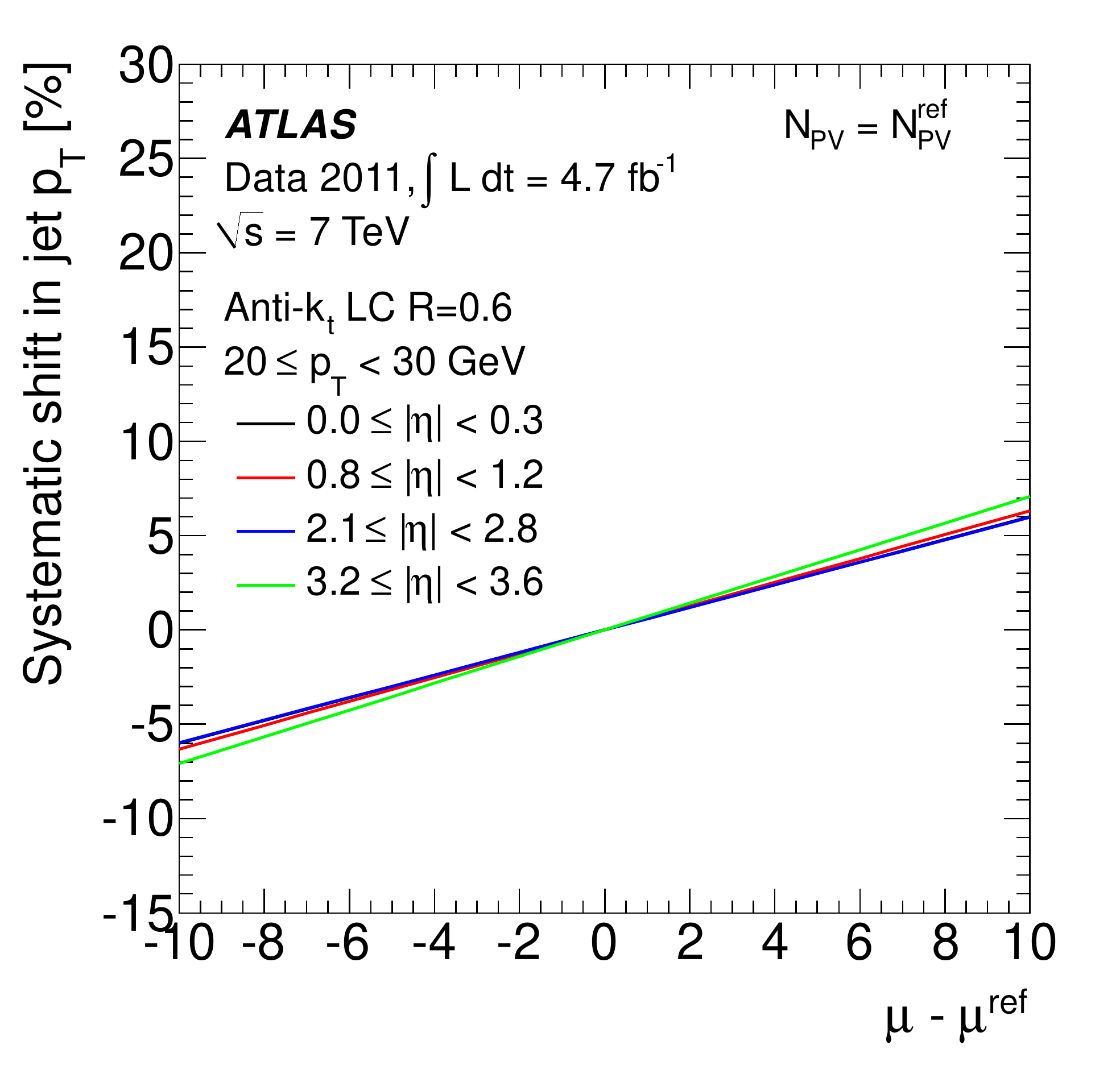} \label{fig:syst6m:mu_lcjes_pt0}}  \hspace*{\fill} \\
\hspace*{\fill} \subfloat[\EMJES, $R = 0.6$]{\includegraphics[width=0.4\textwidth]{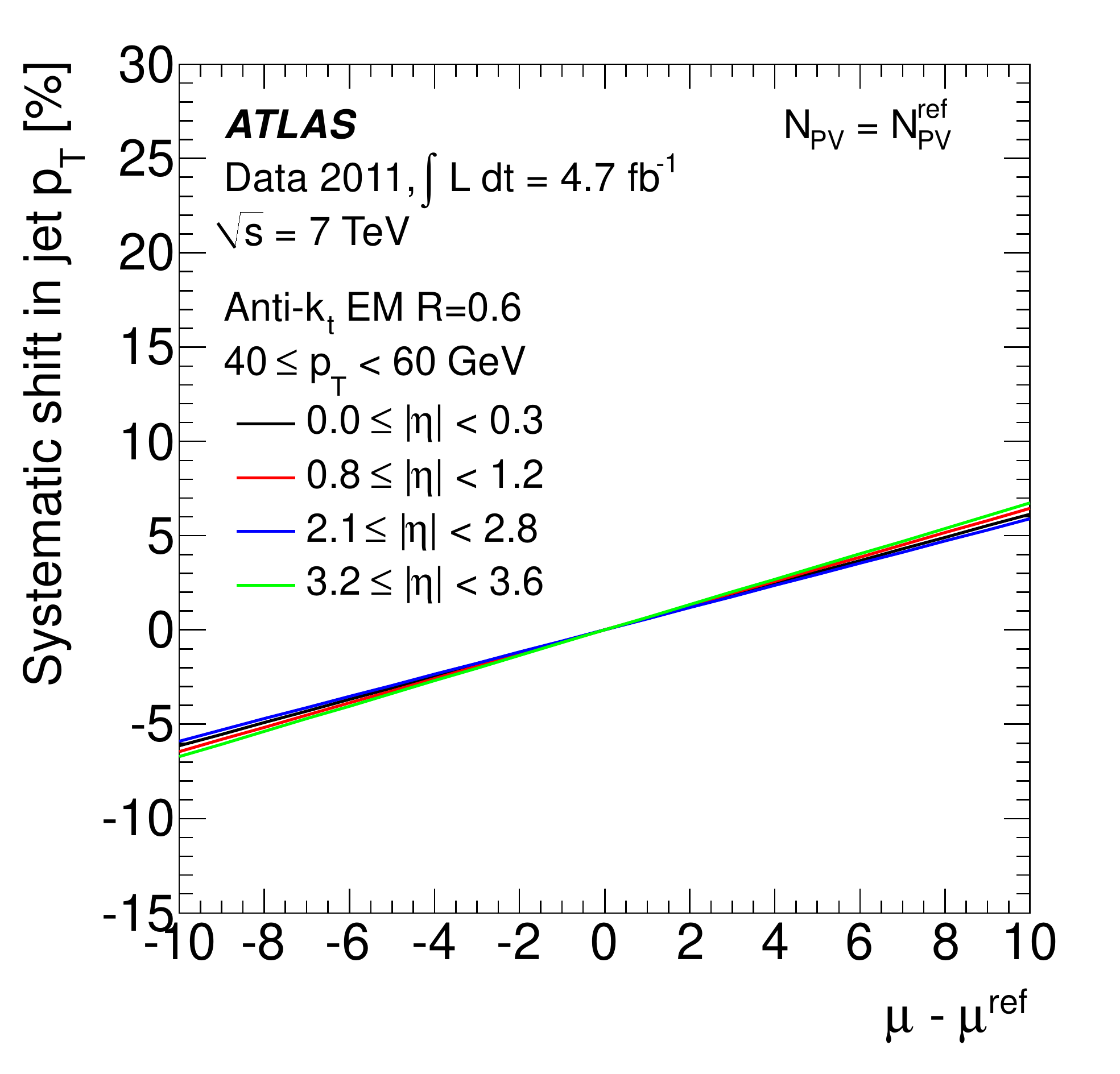} \label{fig:syst6m:mu_emjes_pt1}}  \hspace*{\fill}
\subfloat[\LCWJES, $R = 0.6$]{\includegraphics[width=0.4\textwidth]{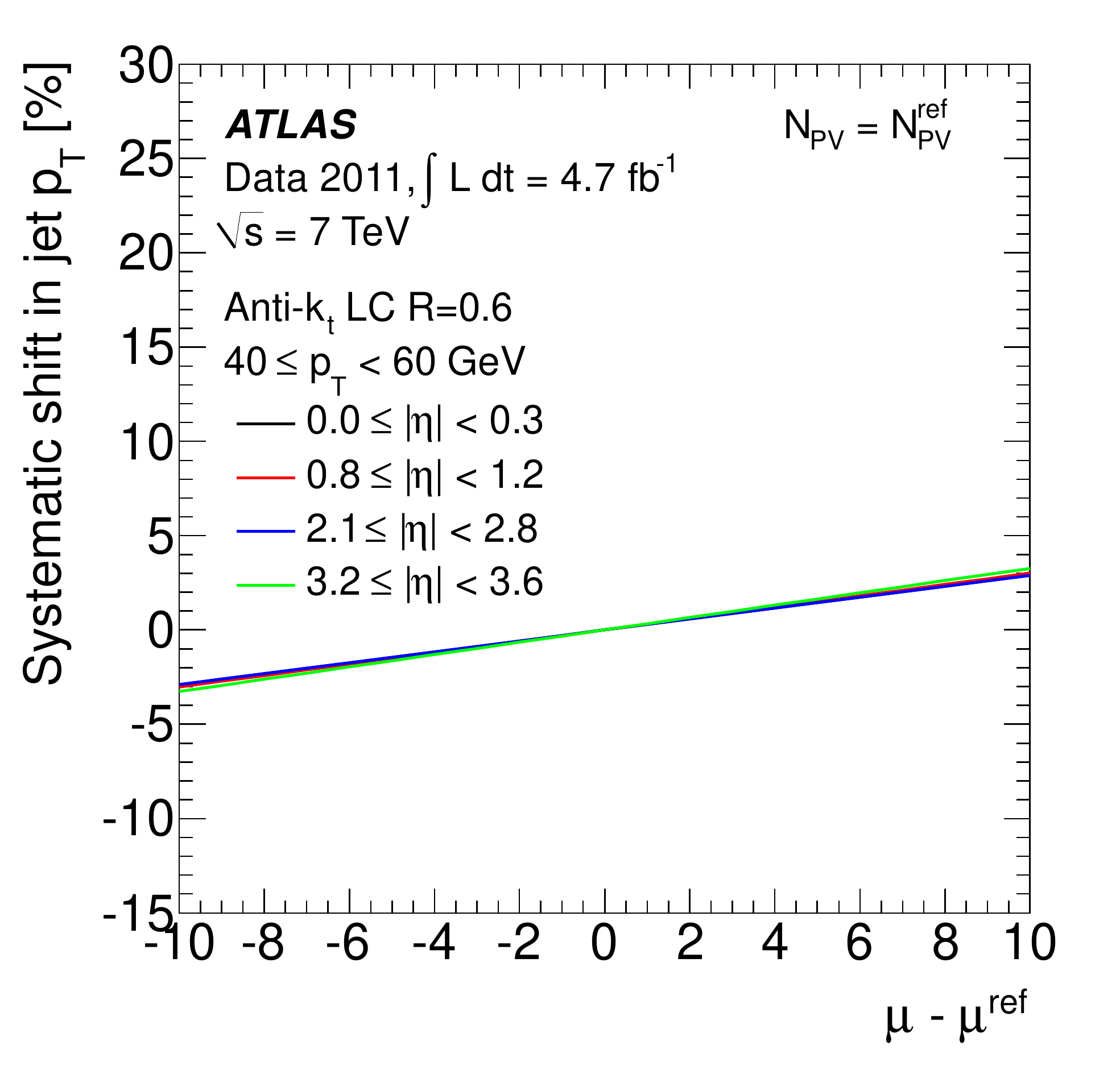} \label{fig:syst6m:mu_lcjes_pt1}}  \hspace*{\fill} \\
\hspace*{\fill} \subfloat[\EMJES, $R = 0.6$]{\includegraphics[width=0.4\textwidth]{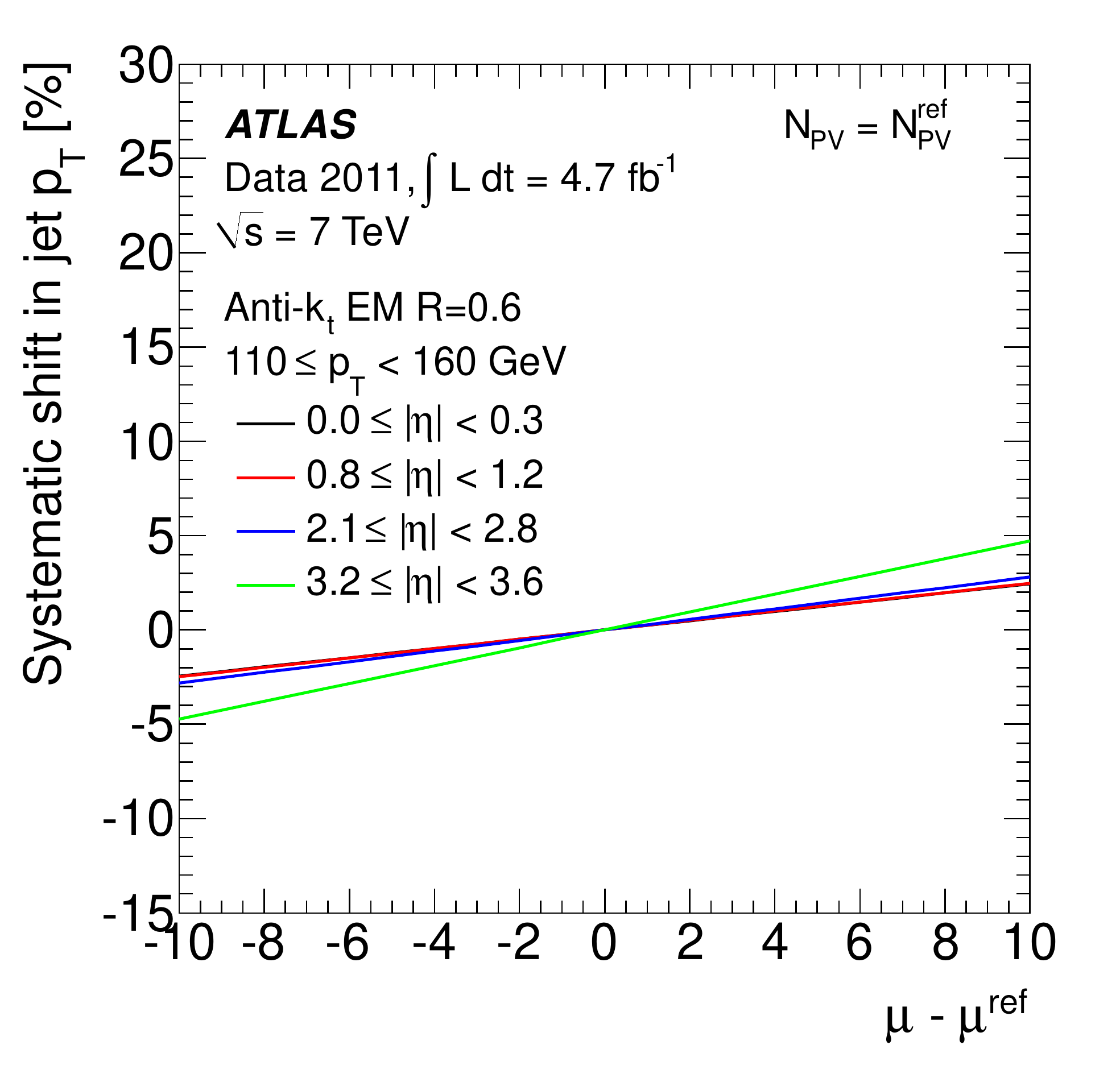} \label{fig:syst6m:mu_emjes_pt2}}  \hspace*{\fill}
\subfloat[\LCWJES, $R = 0.6$]{\includegraphics[width=0.4\textwidth]{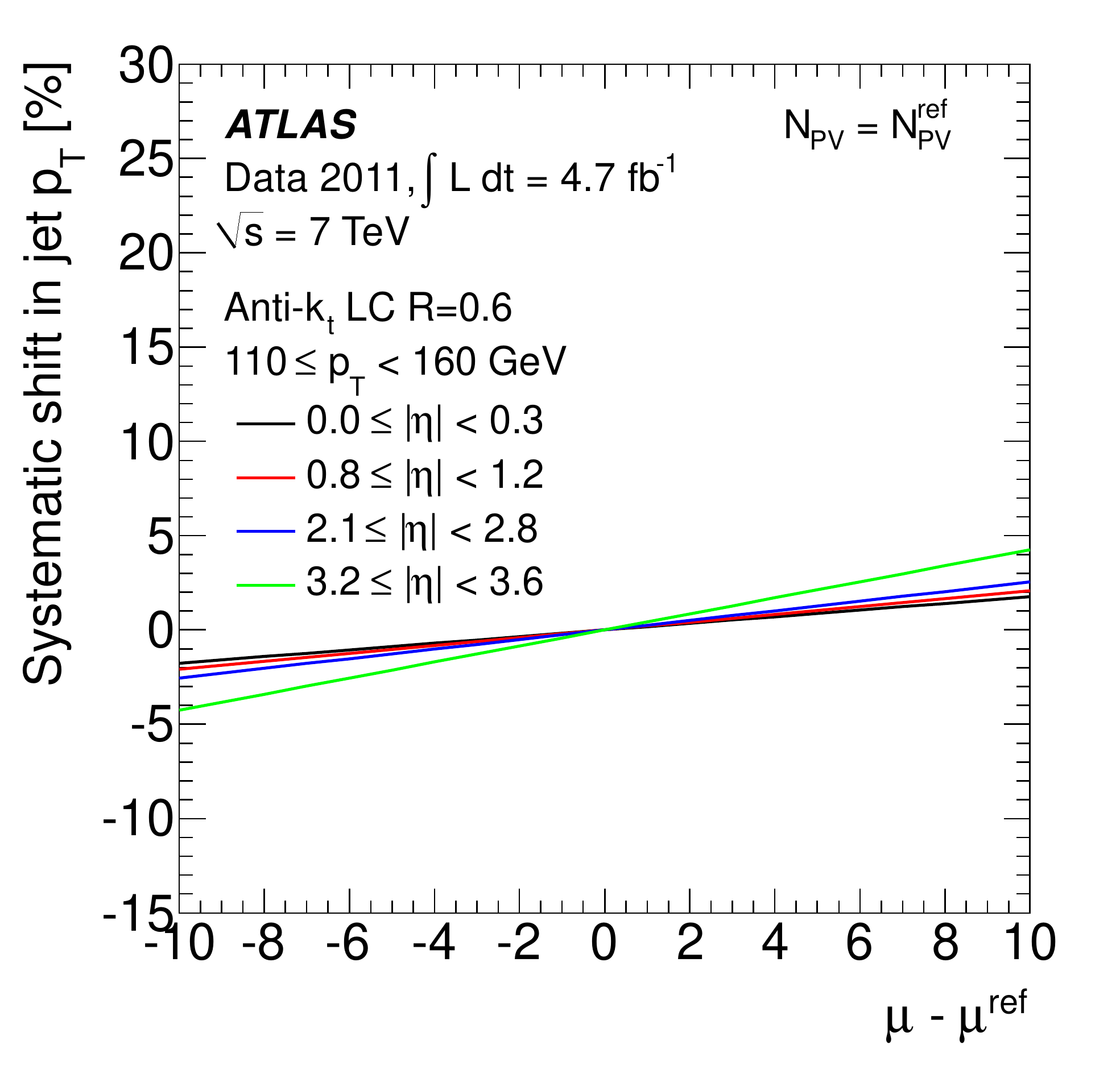} \label{fig:syst6m:mu_lcjes_pt2}}   \hspace*{\fill}
\caption[]{The fractional systematic shift due to mis-modelling of the effect of out-of-time 
pile-up on the transverse momentum \pTrec{\EMJES} of jets reconstructed with the 
\antikt{} algorithm with $R = 0.6$ and calibrated with the \EMJES{} scheme, is shown as a function 
of $\left(\axing-\axingRef\right)$ in \subref{fig:syst6m:mu_emjes_pt0}, \subref{fig:syst6m:mu_emjes_pt1}, and \subref{fig:syst6m:mu_emjes_pt2} for various \pTrec{\EMJES}{} bins. The same systematic shift is shown in 
\subref{fig:syst6m:mu_lcjes_pt0}, \subref{fig:syst6m:mu_lcjes_pt1}, and \subref{fig:syst6m:mu_lcjes_pt2} for jets calibrated with the \LCWJES{} scheme, now in bins of 
\pTrec{\LCWJES}. \label{fig:syst6m}}
\end{figure*}

\subsection{Derivation of the systematic uncertainty}
\label{sec:pileupinsitusystematics}
The systematic uncertainties introduced by applying the \MC-simulation-based pile-up 
correction to the reconstructed \pTrec{\EM}{} and \pTrec{\LCW}{} for jets in
collision data include the variation of the slopes $\alpha = \partial\pT/\partial\Npv$ and
$\beta = \partial\pT/\partial\axing$ with changing jet \pT. While the immediate 
expectation from the stochastic and diffuse nature of the (transverse) energy flow in 
pile-up events is that all slopes in \Npv{} (\alphaEM, \alphaLCW) and \axing{} (\betaEM, 
\betaLCW) are independent of this jet \pT, \figRef{fig:ptdep} clearly shows a 
\pttrue{} dependence of the signal contributions from in-time and out-of-time pile-up for jets
reconstructed on \EM{} scale. A similar \pttrue{} dependence can be observed for jets reconstructed on 
\LCW{} scale. 

The fact that the variations $\Delta(\partial\pT/\partial\Npv)$ with \pttrue{} are very
similar for narrow ($R = 0.4$) and wide ($R = 0.6$) \antikt{} jets indicates that this 
\pT{} dependence is associated with the signal core of the jet. The presence of dense 
signals from the jet increases the likelihood that small pile-up signals survive the 
noise suppression applied in the topological clustering algorithm, see \secRef{sec:topos}. 
As the core signal density of jets increases with \pT, the acceptance for small pile-up signals thus increases as well. 
Consequently, the pile-up signal contribution to the jet increases. This jet \pT{} dependence is 
expected to approach a plateau as the cluster occupancy in the core of the jet approaches 
saturation, which means that all calorimeter cells in the jet core survive the selection 
imposed by the noise thresholds in the \topo{} formation, and therefore all pile-up 
scattered into these same cells contributes to the reconstructed jet \pT{}. The jet \pT{} dependent pile-up contribution is not explicitly corrected for, and thus is 
implicitly included
in the systematic uncertainty discussed below.   

Since the pile-up correction is derived from \MC{} simulations, it explicitly does not
correct for systematic shifts due to mis-modelling of the effects of pile-up on simulated 
jets. The sizes of these shifts may be estimated from
the differences between the offsets obtained from data and from \MC{} simulations:
\begin{eqnarray*}
  \Delta\pToff^{\EM} & = & \left.\pToff^{\EM}(\Npv,\axing)\right|_{\mathrm{data}} - \left.\pToff^{\EM}(\Npv,\axing)\right|_{\mathrm{MC}} \\
  \Delta\pToff^{\LCW} & = & \left.\pToff^{\LCW}(\Npv,\axing)\right|_{\mathrm{data}} - \left.\pToff^{\LCW}(\Npv,\axing)\right|_{\mathrm{MC}} 
\end{eqnarray*}
To assign uncertainties that can cover these shifts, and to incorporate the results from 
each \insitu{} method, combined uncertainties are calculated as a weighted RMS of 
$\Delta\pToff(\Npv,\axing)$ from the offset 
measurements based on \gammajet{} and on track jets. The weight of each contribution is the inverse squared uncertainty of the 
corresponding $\Delta\pToff(\Npv,\axing)$. This yields absolute uncertainties in $\alpha$ 
and $\beta$, which are then translated to fractional systematic shifts in the fully 
calibrated and corrected jet \pT{} that depend on the pile-up environment, as described
by \Npv{} and \axing.

\FigRef{fig:syst4n} shows the fractional systematic shift in the \pT{} measurement for
\antikt{} jets with $R = 0.4$, as a function of the in-time pile-up activity measured by
the displacement $(\Npv-\NpvRef)$. The shifts are shown for various regions of the
\ATLAS{} calorimeters, indicated by \etaDet, and in bins of the reconstructed transverse
jet momentum \pTrec{\EMJES}{} for jets calibrated with the \EMJES{} scheme 
(Figs.~\ref{fig:syst4n}\subref{fig:syst4n:npv_emjes_pt0}, 
\ref{fig:syst4n}\subref{fig:syst4n:npv_emjes_pt1}, and \ref{fig:syst4n}\subref{fig:syst4n:npv_emjes_pt2}). 
Figures~\ref{fig:syst4n}\subref{fig:syst4n:npv_lcjes_pt0}, 
\ref{fig:syst4n}\subref{fig:syst4n:npv_lcjes_pt1} and \ref{fig:syst4n}\subref{fig:syst4n:npv_lcjes_pt2} show the shifts
for jets reconstructed with the \LCWJES{} scheme in the same regions of \ATLAS, in 
bins of \pTrec{\LCWJES}. The same uncertainty contributions from wider jets 
re\-con\-struc\-ted with the \antikt{} algorithm with $R = 0.6$ are shown in 
\figRef{fig:syst6n}.

Both the \EMJES{} and \LCWJES{} calibrations are nor\-ma\-li\-sed such that the pile-up
signal contribution is $0$ for $\Npv = \NpvRef$ and $\axing = \axingRef$, so 
the fractional systematic shifts associated with pile-up scale linearly with the displacement
from this reference. In general, jets reconstructed with \EMJES{} show a larger systematic
shift from in-time pile-up than \LCWJES{} jets, together with a larger dependence 
on the jet catchment area defined by $R$, and the jet direction \etaDet. In particular, 
the shift per reconstructed vertex for \LCWJES{} jets in the two lowest \pTrec{\LCWJES}{}
bins shows essentially no dependence on $R$ or \etaDet, as can be seen comparing 
Figs.~\ref{fig:syst4n}\subref{fig:syst4n:npv_lcjes_pt0} and 
\ref{fig:syst6n}\subref{fig:syst6n:npv_lcjes_pt0} to 
Figs.~\ref{fig:syst4n}\subref{fig:syst4n:npv_lcjes_pt1} and 
\ref{fig:syst6n}\subref{fig:syst6n:npv_lcjes_pt1}.

The systematic shift associated with out-of-time pile-up, on the other hand, is
independent of the chosen jet size, as shown in \figRef{fig:syst4m} for $R = 0.4$ and 
\figRef{fig:syst6m} for $R = 0.6$. Similar to the shift from in-time pile-up, 
the jets reconstructed with the \LCWJES{} scheme show smaller systematic shifts 
from out-of-time pile-up. The results shown in these figures also indicate that the 
shift from out-of-time pile-up is independent of the jet size. Note that both 
shifts contribute to the jet \pT{} reconstruction uncertainty in an 
uncorrelated fashion, which is justified as while \Npv{} and \axing{} are correlated in a
given sample, the corrections depending on them are derived independently.

\subsection{Summary on pile-up interaction corrections}
\label{sec:pileupsummary}
Dedicated correction methods addressing the signal contributions from in-time and
out-of-time pile-up to the jet energy measurement 
with the \ATLAS{} calori\-met\-ers were developed using \MC{} simulations 
to measure the change of the jet signal as function of the
characteristic variables measuring the pile-up activity, which are the number of
reconstructed primary vertices \Npv{} (in-time pile-up) and the average number of pile-up
interactions per bunch crossing \axing{} (out-of-time pile-up). The input to these
corrections are the slopes $\alpha = \partial\pT/\partial\Npv$ and
$\beta = \partial\pT/\partial\axing$, which are determined in the simulation for two jet
signal scales, the \EM{} scale (\ptjetEM) and the hadronic \LCW{} scale (\ptjetLCW),
both as functions of the truth-jet \pttrue{} and the direction of the jet in the 
detector \etaDet. 

As an alternative to the approach based on MC simulation, the cha\-nge 
of the reconstructed
(calorimeter) jet \pT{} with \Npv{} and \axing{} can be measured in data using the
matching track jet's \ptjetTrk{} as a kinematic reference independent of the pile-up 
activity. Furthermore, \gammajet{} events can be used in the same manner, with the photon 
\pT{} providing the reference in this case. These experimental methods are restricted by
the coverage of the \ATLAS{} tracking detector (track jets), and the lack of significant
statistics for events with jets at higher \etaDet{} in \gammajet{} events in 2011. 

Comparing the \insitu{} measurements of $\alpha$ and $\beta$ with the corresponding
simulation and the findings from the approach solely based on \MC{} simulations allows the
determination of systematic biases due to mis-modelling of the effects of pile-up on
simulated jets. To cover these biases, uncertainties are
assessed as functions of \Npv{} and \axing. These uncertainties amount to less than
$0.3\%(0.5\%)$ of the calibrated jet \pT{} per reconstructed vertex for central \antikt{}
jets with $R = 0.4 (0.6)$ with $20 < \pT < 30$~\GeV{} and for $\axing = \axingRef$, and about
$0.7\%$ per interaction for jets in the same phase space at $\Npv = \NpvRef$, independent 
of the jet size. The uncertainty contribution in the forward direction can be
significantly larger, by up to a factor of two, especially at higher jet \pT, where the 
uncertainty in the central detector is smaller than $0.1\%(0.2)\%$ per vertex and $0.2\%$
per interaction. These generally small uncertainties can be added in quadrature to give a
total fractional uncertainty for each pile-up condition (\Npv,\axing).  

A residual jet \pT{} dependence of the pile-up correction is observed in \MC{} simulation 
(see Fig.~\ref{fig:ptdep}), but
not yet fully confirmed in data due to limited size of the data set. 
It is therefore not explicitly addressed in
the correction procedure, rather it is implicitly included into the systematic uncertainties.
This dependence, which is not expected for a purely stochastic and diffuse signal
contribution from both in-time and out-of-time pile-up, is introduced by the \topos{}
formation in the calorimeter, which enhances the survivability of small (pile-up) signals
if higher density signals such as those in the core of a jet are close by. At very high
jet \pT{}, this dependence reaches a plateau, since the jet core gets so dense that all
calorimeter cells contribute to the jet signal, and therefore all signal generated by
pile-up in these cells is directly included in the jet signal.

In summary, the pile-up signal contribution to jets in the \ATLAS{} detector is well understood. The
correction based on \MC{} simulations controls this contribution to a high precision with 
uncertainties of less than $1\%$ per reconstructed primary vertex and additional \pp{}
collision per bunch crossing, yielding a small fractional contribution to the overall
jet energy scale uncertainty over the whole phase space, except for the very forward
region, where this uncertainty can be more significant.

\begin{figure*}
 \begin{center}
    \subfloat[$A_{\rm close-by}^{\trkjetid} - 1$ vs jet \pt]
    {\includegraphics[width=0.48\textwidth]{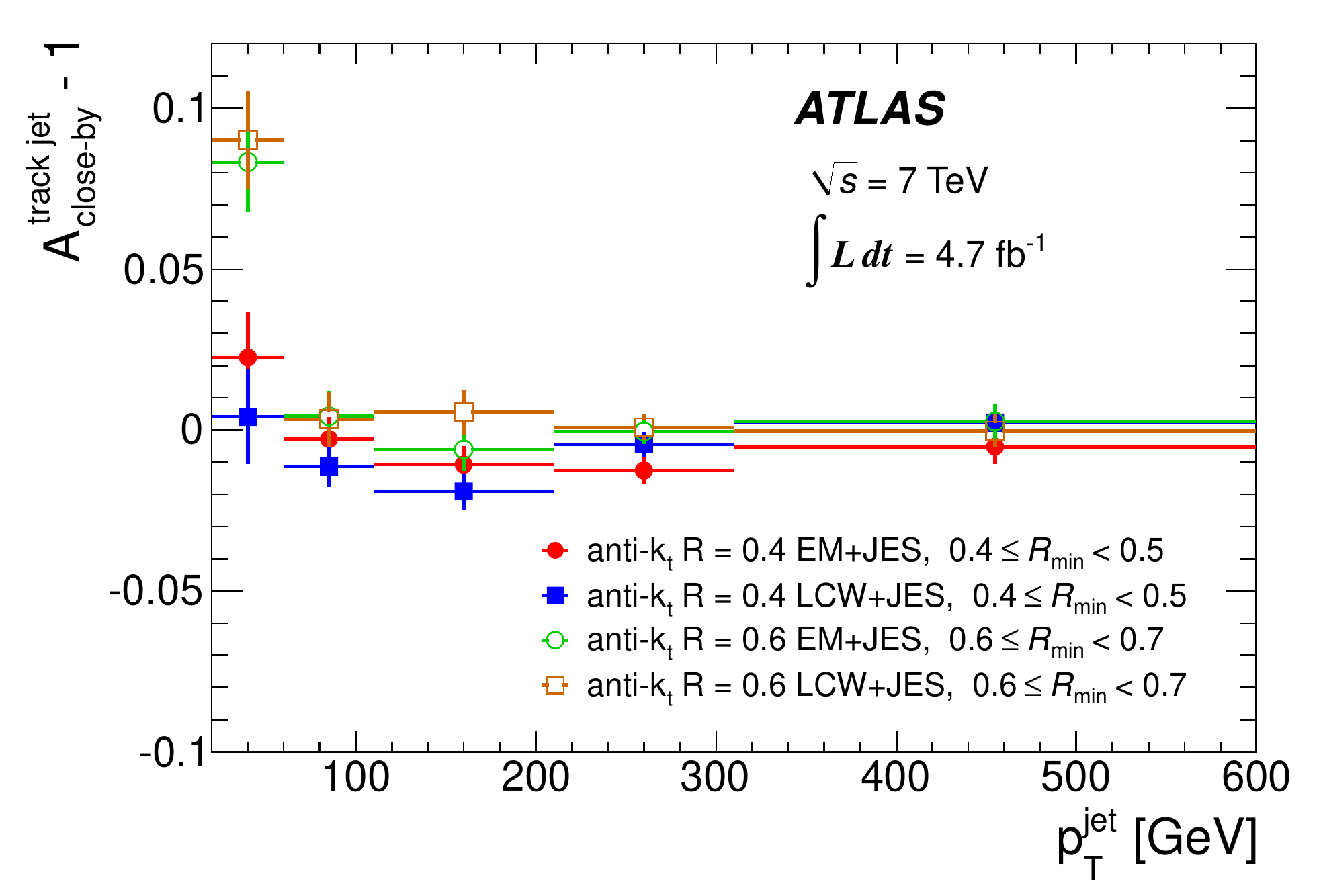}\label{fig:ClosebyTrkIso}}
    \subfloat[$A_{\rm close-by} - 1$ vs jet \pt]
    {\includegraphics[width=0.48\textwidth]{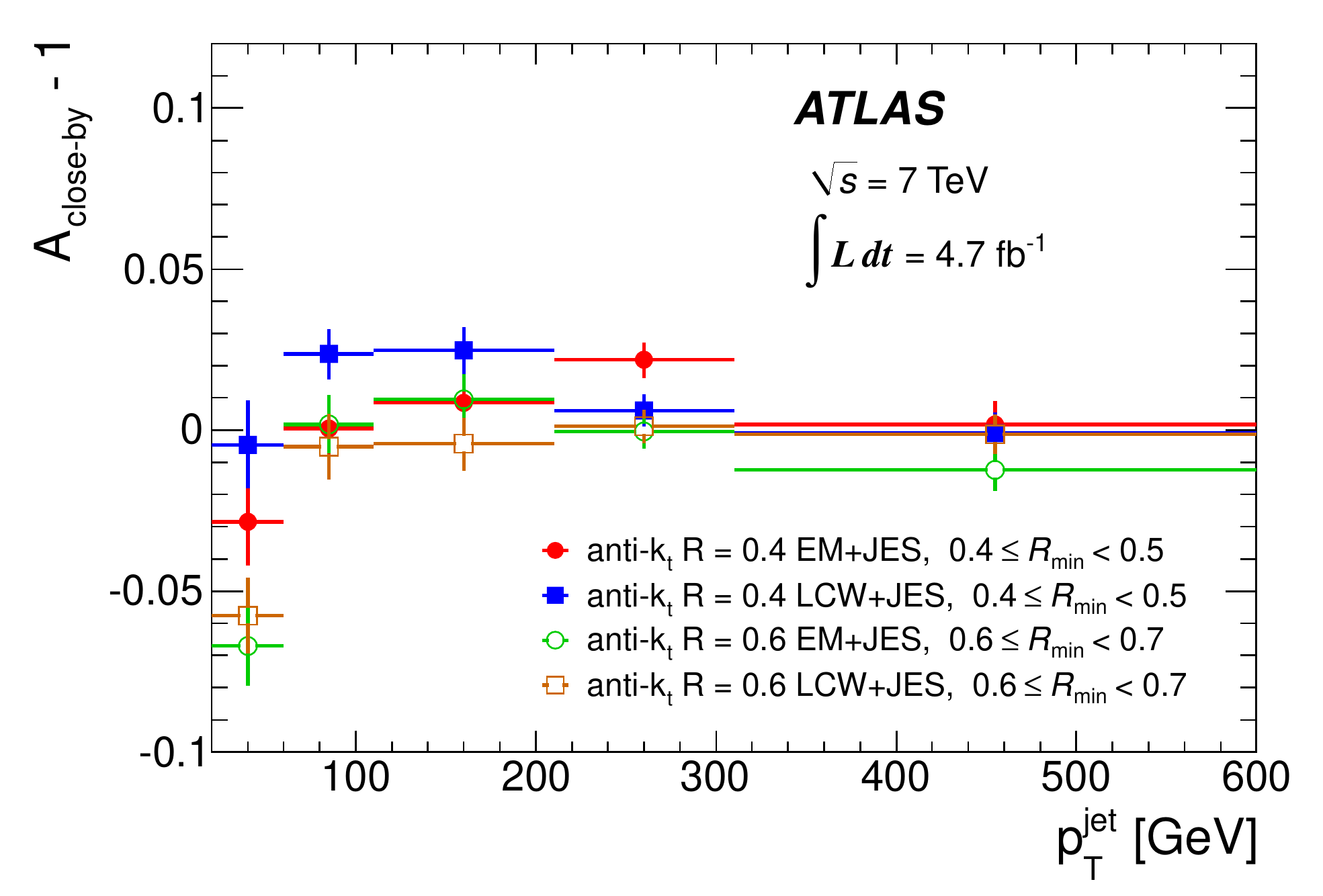}\label{fig:ClosebyTrkIsoNonIso}}
 \end{center}
 \caption[]{In  \subref{fig:ClosebyTrkIso}, the deviation from unity of the \datatomc{} ratio of the track-jet \pt{} for non-isolated jets
divided by the track-jet \pt{} for isolated jets, is shown as a function of the jet \pt{}. 
The deviation from unity of the \datatomc{} ratio of the relative response of non-isolated jets with respect
to that of isolated jets as a function of the jet \pt{} is shown in \subref{fig:ClosebyTrkIsoNonIso}. As described in the text, the distributions show the ratios given in \subref{fig:ClosebyTrkIso} \eqRef{eq:A_closeby_Track} and \subref{fig:ClosebyTrkIsoNonIso} \eqRef{eq:A_multijetcloseby} for the four jet calibration schemes. Only statistical uncertainties are shown.}
 \label{fig:Closeby}
\end{figure*}

\section{Close-by jet effects on jet energy scale}
\label{sec:close-by}
The variation of the jet energy response due to nearby jets and the associated systematic uncertainty 
are reported in
Ref. \cite{jespaper2010}, using the data collected in $2010$. The same analysis is performed to reassess this 
uncertainty for the $2011$ data.

The analysis uses track jets from the primary vertex, as defined in \secRef{sec:trackjets}, as a kinematic reference.  
The calorimeter jet's transverse momentum \ptjet{} relative to the track-jet transverse momentum \ptjetTrk{} provides an \insitu{} validation of the calorimeter jet 
response and the evaluation of the systematic uncertainty. The relative response measurement is performed in bins of 
$R_{\rm min}$, the distance in \etaphispace{} from the jet to the closest other jet with $\pt>7$~\GeV{} at the \EM{} scale. 
The response to track jets is also evaluated for the non-isolated condition $R_{\rm min} < 2.5 \times R$, 
where $R$ is the distance parameter used in the \antikt{} jet reconstruction, and the associated 
systematic uncertainty is assessed.
In the relative response measurement, the track jet is matched to the calorimeter jet with the distance requirement $\DeltaR<0.3$, where \DeltaR{} is measured according to 
\eqRef{eq:deltaRdet} (\secRef{sec:jetdirections}) in \etaphispace. 
When two or more jets are matched within the \DeltaR{} range, the closest matched jet  is taken. 

The calorimeter jet response relative to the matched track jet, defined as 
the $\pt$ ratio of the calorimeter to the track jet as a function of \ptjet, 
\begin{displaymath}
r^{\rm calo/\trkjetid}=\ptjet/\ptjetTrk , 
\end{displaymath}
is examined 
for different $R_{\rm min}$ values measured for the two close-by calorimeter jets.\footnote{Unless otherwise stated, 
both calorimeter jets are used in the jet response measurement if each of them can be matched to a track jet.}
The ratio of calorimeter jet response between non-isolated (i.e, small $R_{\rm min}$) and 
isolated (large $R_{\rm min}$)
jets, given by 
\begin{displaymath}
r^{\rm calo/\trkjetid}_{\rm non-iso/iso} = \dfrac{r^{\rm calo/\trkjetid}_{\rm non-iso}}{r^{\rm calo/\trkjetid}_{\rm iso}} , 
\end{displaymath}
is compared between data and \MC{} simulation. The relative difference between them,
\begin{equation}
  A_{\rm close-by} = \dfrac{\left.r^{\rm calo/\trkjetid}_{\rm non-iso/iso}\right|_{\rm Data}}{\left.r^{\rm calo/\trkjetid}_{\rm non-iso/iso}\right|_{\rm MC}}  ,
  \label{eq:A_multijetcloseby}
\end{equation}
is assumed to represent the calorimeter \JES{} uncertainty due to close-by jets. This uncertainty, 
convolved with the systematic uncertainty of the response to a track jet with a nearby jet, and evaluated in a similar way 
as the \datatomc{} difference between the average \pt{} ratio of the non-isolated to isolated track jets, 
provides the total JES systematic uncertainty due to the close-by jet effect.

\subsection{Samples and event selection}
Data collected with four single-jet, pre-scaled triggers with jet-\pt{} thresholds of $10$, $30$, $55$ and $135$~\GeV{} 
are used in the analysis. As in the \MJB{} analysis discussed in \secRef{sec:multijet},
the data from a given trigger are used in a certain 
non-overlapping jet-\pt{} range where the trigger is greater than $99\%$ efficient. 
For \MC{} simulation, 
the baseline \pythia{} samples described in \secRef{sec:MC} are used.

Events passing the trigger selections are required to satisfy the same primary vertex and event cleaning 
criteria for jets due to noise and detector problems as those used in the \MJB{} analysis (see \secRef{sec:multijetselection}). 
Finally, events that contain at least two jets with
calibrated $\pt>20$ \GeV{} and rapidity $|y|<2.8$ are selected for the analysis.

The track jets are reconstructed from the selected tracks by using the \antikt{} algorithm with 
$R = 0.4$ and $R = 0.6$, as described in \secRef{sec:trackjets}. In the analysis presented below, 
track jets with $\pt>10$~\GeV{} and $|\eta|<2.0$, composed of at least two tracks, are used. 
The close-by jet energy scale uncertainty is therefore assessed in the region of $|\eta|<2.0$ where
the calorimeter jets and track jets can be matched in $\eta$ and $\phi$.

\subsection{Non-isolated jet energy scale uncertainty}\label{sec:CloseByJES}
The average track-jet transverse momentum is examined as a function of the calorimeter jet \pt{}
for different $R_{\rm min}$ values starting from the jet radius in bins of $\DeltaR_{\rm min}=0.1$.
The ratio of the average track-jet \pt{} between the non-isolated and isolated track jets
$\pt^{\rm non-iso}/\pt^{\rm iso}$
in bin of the calorimeter \pt,
is used to quantify the uncertainty in the response to track jets.
This comparison
is shown in \figRef{fig:Closeby}\subref{fig:ClosebyTrkIso} as a deviation from unity of the \datatomc{} ratio:
\begin{equation}
  A_{\rm close-by}^{\rm track\,jet} = \dfrac{\left.\pt^{\rm non-iso}/\pt^{\rm iso}\right|_{\rm Data}}{\left.\pt^{\rm non-iso}/\pt^{\rm iso}\right|_{\rm MC}}
  \label{eq:A_closeby_Track}
\end{equation}
to quantify the uncertainty in the response to track jets in the small $R_{\rm min}$ range of $R\leq R_{\rm min}<R+0.1$. 
The $A_{\rm close-by}^{\trkjetid}$ has a strong $R_{\rm min}$ dependence, especially at small $\DeltaR_{\rm min}$ range
where the close-by jet overlaps the probe jet, and the dependence is more significant for jets with $R = 0.6$.
The agreement between data and \MC{} simulations improves with increasing $R_{\rm min}$.

The calorimeter jet response relative to the matched track jet ($r^{\rm calo/\trkjetid}$) is 
investigated as a function of \pt, in terms of the non-isolated jet response relative to the 
isolated jet response $r^{\rm calo/\trkjetid}_{\rm non-iso/iso}$, for data and \MC{} simulations.
The \datatomc{} ratio $A_{\rm close-by}$ of $r^{\rm calo/\trkjetid}_{\rm non-iso/iso}$ is shown 
in \figRef{fig:Closeby}\subref{fig:ClosebyTrkIsoNonIso} as the deviation from unity for the range of $R\leq R_{\rm min}<R+0.1$. 
As already seen in the track-jet response in \figRef{fig:Closeby}\subref{fig:ClosebyTrkIso}, 
there is a strong $R_{\rm min}$ dependence on $A_{\rm close-by}$ within the small $R_{\rm min}$ range mentioned above.
The deviation of $A_{\rm close-by}$ from unity is added in quadrature with the track-jet response uncertainties obtained
above to get the overall \JES{} uncertainties due to close-by jet effects. The convoluted uncertainty is about $3.5\%$ ($10\%$) 
at $R_{\rm min}<0.5$ ($0.7$) for $R = 0.4$ ($0.6$) jets with $\pt=30$~\GeV, and becomes smaller than $1\%$ at $R_{\rm min}$ above $0.8$
for both sizes of jets. The uncertainty decreases with increasing jet \pt{} and becomes about $2\%$ ($4\%$) at 
$R_{\rm min}<0.5$ ($0.7$) for $R = 0.4$ ($0.6$) jets with $\pt=100$~\GeV.

\section{Jet response difference for quark and gluon induced jets and associated uncertainty}
\label{sec:FlavorTopology}
%

All jet calibration schemes developed in \ATLAS{} achieve an average response of the calorimeter to jets near unity for jets 
in the inclusive jet sample. However, the calorimeter response to jets also exhibits variations that can be correlated to the 
flavour of the partons (i.e., light or heavy quarks, or gluons) produced in the sample under study. 
This dependence is to a large extent due to differences in fragmentation and showering 
properties of jets loosely labelled as originating from a light quark or a gluon.

In this section, the dependence of the jet energy scale on whether a jet originates from a
light quark or a gluon is studied. Also, a systematic uncertainty that accounts for the
sample dependence of the jet energy scale is established using different \MC{}
simulations. In addition, 
jet properties that can be shown to 
discriminate between jets initiated by light quarks and gluons
are used to build a light-quark/gluon tagger 
\cite{jespaper2010,Schwartz2}.
The focus in this section is on understanding how the \JES{} is affected by a selection based on the light-quark/gluon tagger, and the 
implications for the sample-dependent systematic uncertainty described if jets are tagged using this tagger. 
Details of the procedure to built a quark-gluon tagger can be found in Ref.~\cite{ATLASquarkgluon}. 

\subsection{Event selection}

\subsubsection{Jet and track selection}
Calorimeter jets with transverse momentum $\pt>20\GeV$ and $|\eta|<4.5$ are reconstructed 
using the \antikt{} jet algorithm with $R=0.4$.

The variables described in \secRef{sec:likelihoodDiscr} are constructed to describe the properties of jets. They are based on tracks with $\pttrk>1$ \GeV{} that are associated to jets if they are within a distance $\DeltaR = R$ (equal to the distance parameter $R$ used to build the jet) of the jet axis. The tracks are further selected as described in \secRef{sec:trackjets}, with slightly modified quality requirements in order to provide an even stronger association to the primary vertex (impact parameters $z_0 \sin(\theta)<1$ \mm{} and $d_0<1$ \mm).  

\subsubsection{Jet flavour definition}
Jets are labelled by partonic flavour, if they have $\pt>40\GeV$ and $|\eta|<2.1$. 
They are matched to the highest-energy parton found inside the cone of the jet.
This parton can be produced directly off the hard scatter, or by radiation.

This definition of partonic jet flavour is not theoretically sound, and that may have implications when attempting to apply this labelling to physics analyses.
However, several studies with {\sc MadGraph} \cite{MadGraph} have demonstrated that this definition is not changed by the parton shower model choices, and is equivalent to 
a matrix-element-based labelling for over $95\%$ of jets.
Since the partonic flavour of a jet can only be easily defined in leading order, 
and since only a labelling indicating differences in jet properties is required for the 
performance evaluations presented in this paper, this definition is sufficient.

\subsubsection{\Ds{} for flavour studies}
\label{sec:data-selection}
Two main event samples are used.
The first selects inclusive jet events (dijet sample). 
The second selects jets with a high-transverse momentum photon back-to-back with a jet (\gammajet{} sample). 
Both samples are defined using standard data-quality criteria and the requirement of a primary vertex 
with at least three associated tracks.

Central jet triggers are used for the dijet sample selection.
These triggers provide a fully efficient jet selection for $\pT > 40$~\GeV. 
Jet triggers with \pt{} thresholds less than $500$ \GeV{} are pre-scaled, so that only a fraction of the events in this kinematic regime are recorded.

The \gammajet{} sample is selected as described in \secRef{sec:gammajetInSitu}.
In addition, a photon with $\pt>45$ \GeV{} in the event is required to be back-to-back 
(azimuthal distance $\Delta\phi>2.8$ rad) to the leading jet. The 
sub-leading jet is required to have no more than $30\%$ of the photon $\pt$. 
%

%
\begin{figure*}[htbp]
\begin{center}
\subfloat[\EMJES]{
\includegraphics[width=0.32\textwidth]{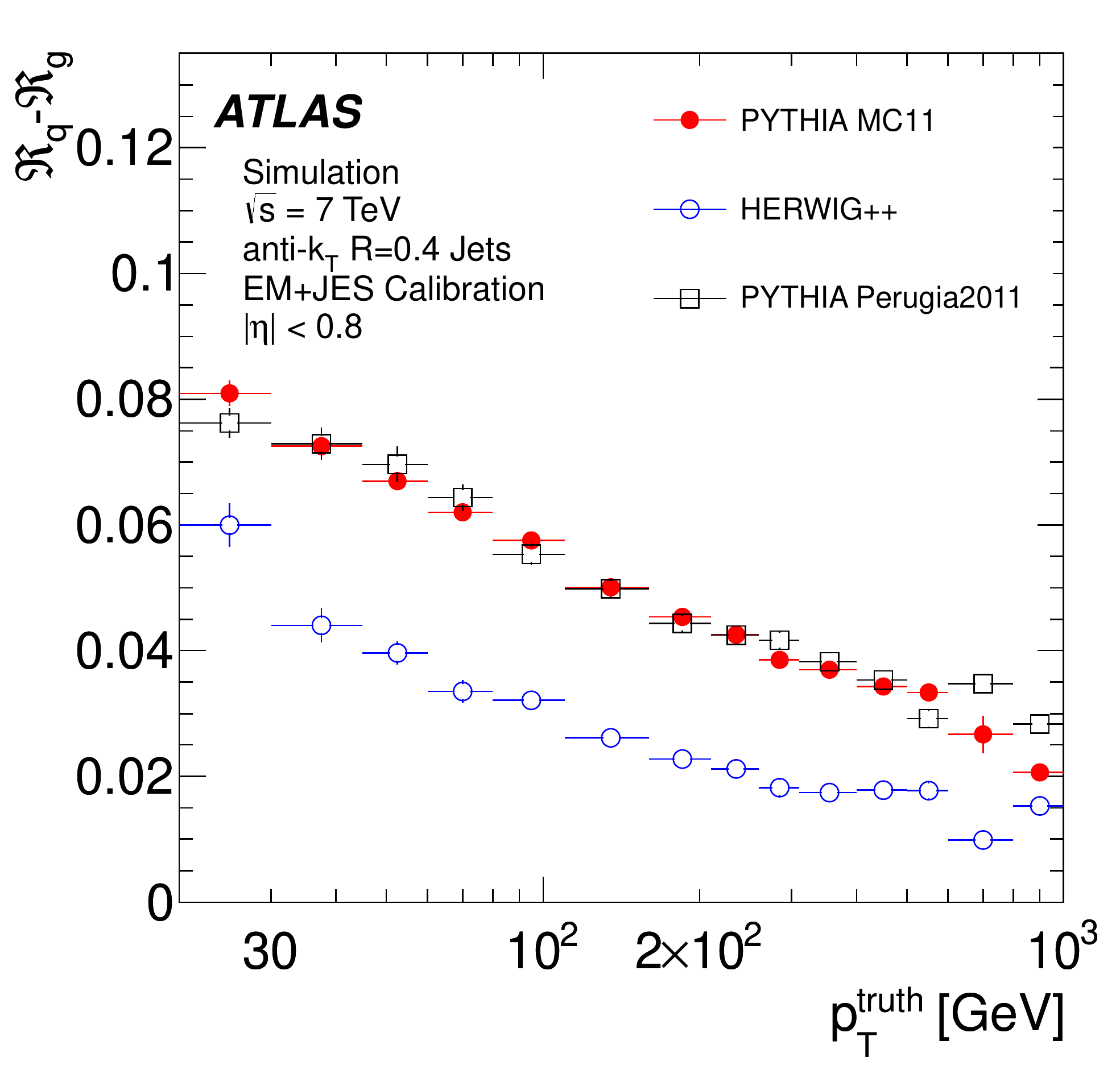} \label{fig:respLightGluEM}} 
\subfloat[\LCWJES]{
\includegraphics[width=0.32\textwidth]{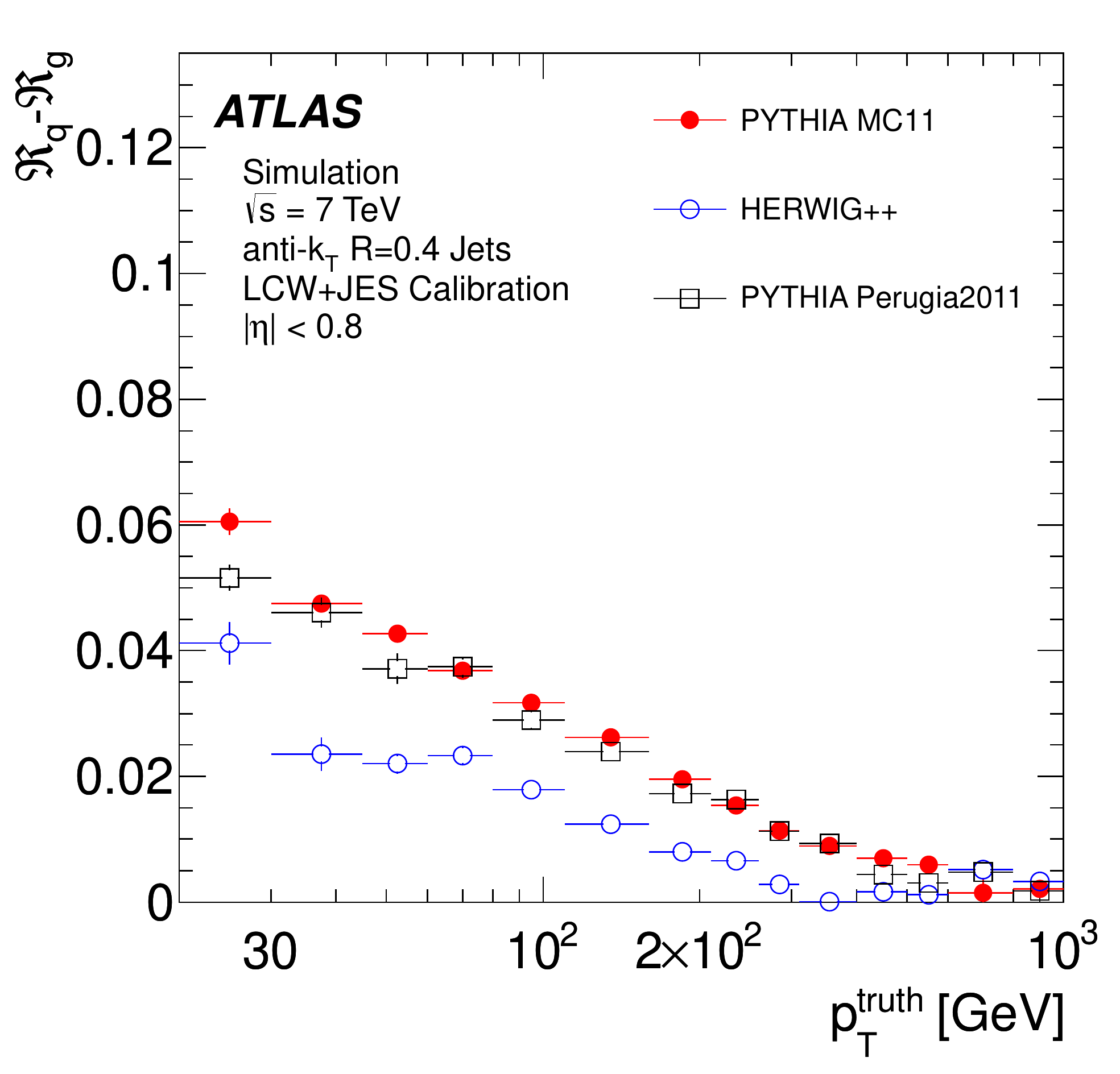} \label{fig:respLightGluLC}} 
\subfloat[\GS]{
\includegraphics[width=0.32\textwidth]{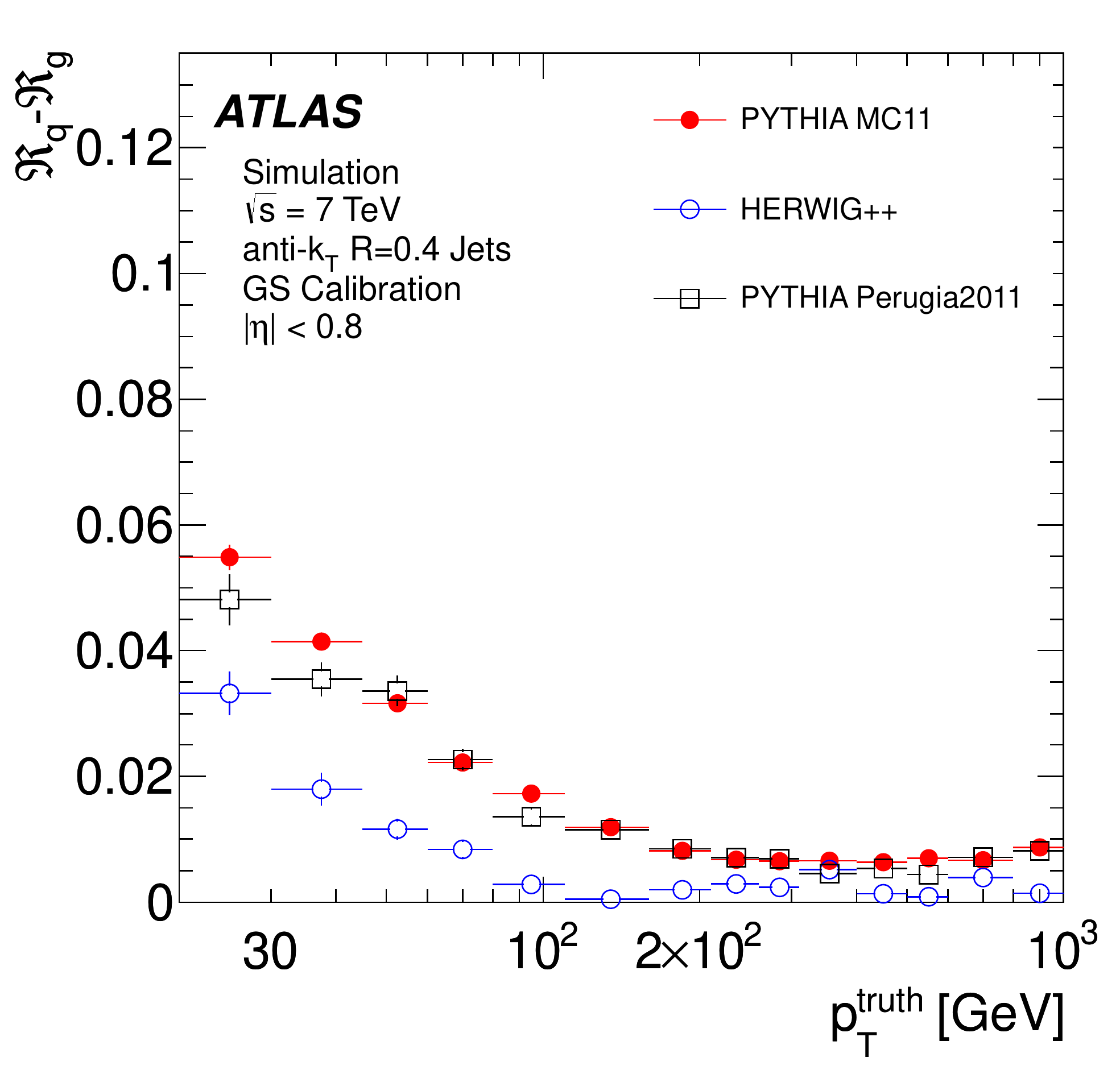}\label{fig:respLightGluGS}}
\caption[]{Difference in jet response $\mathcal{R} = \ptjet/\pttrue$ of isolated jets initiated by light quarks and gluons 
as a function of the true jet $\pt$, for \antikt{} jets with $R=0.4$ in the barrel calorimeter. 
Three different calibration schemes are shown for  \subref{fig:respLightGluEM} the \EMJES{} calibration, \subref{fig:respLightGluLC} 
the \LCWJES{} calibration, and  \subref{fig:respLightGluGS} the alternative Global Sequential (\GS) \cite{jespaper2010} scheme.
Three different \MC{} simulation samples are also shown, \pythia{} (solid red circles), 
\herwigpp{} (open blue circles) and \pythia{} \Perugia 2011 (open black squares).
\label{fig:respLightGlu}
}
\end{center}
\end{figure*}

%
\subsection{Calorimeter response to quark and gluon induced jets}
\label{sec:calorResponse}

Jets labelled as originating from light quarks have significantly different response 
($\ptjet/\pttrue$) from those labelled as originating from gluons 
in the \MC{} simulation. 
This difference is a result of a difference in fragmentation that can be correlated to differences 
in observable properties of the two types of jets. 
Gluon jets tend to have more particles, and as a result, those particles 
tend to have lower \pT{} than in the case of light-quark jets. 
Additionally, gluon jets tend to have a wider angular energy profile before interacting with 
the detector. 

The harder particles in light-quark jets have a higher probability of penetrating further 
into the calorimeter, and thus more often reaching the hadronic calorimeter layers. 
The lower response of the calorimeter for low-\pT{} particles combined with threshold 
and response effects related to the energy density inside the jet suggest that gluon jets 
should have a lower response than light-quark jets. 
The difference in calorimeter response in \MC{} simulations between isolated light-quark and gluon jets 
is shown in \figRef{fig:respLightGlu}, for \antikt{} jets with $R = 0.4$ in the 
barrel calorimeter ($|\etaDet|<0.8$).

Independent of the calibration scheme,
the flavour-de\-pen\-dent response difference is largest at low \pT{} (up to $8\%$ for \EMJES), 
and decreases to a few percent at high $\pt$. 
A more sophisticated calibration scheme like \LCWJES{} reduces the differences, 
because it exploits signal features of individual particle showers in the calorimeter for calibration, and thus partly compensates for 
variations in jet fragmentation and directional energy flow in the jet. Even more so, the Global Sequential (\GS) calibration introduced 
in Ref. \cite{jespaper2010}, which can be applied on top of the (standard) \EMJES{} or \LCWJES{} calibration, or just to jets at the \EM{} 
scale as done for the studies discussed here,   
shows the best performance at low \pT. This is due to its explicit use of a jet width variable which is strongly related 
with the transverse structure of the jet and is thus sensitive to differences between jets initiated by light-quarks 
and gluons. 
The response difference between light-quark- and gluon-initiated jets is reduced by roughly $1\%$ 
for \antikt{} jets with $R=0.6$, because the larger jet area diminishes the effect of the energy loss of the broader jet.

The differences in response between jets initiated by light quarks and gluons can impact analyses 
in which the flavour composition of the sample is not well known.
The corresponding \JES{} uncertainties can be reduced if the flavour composition of the analysis sample is known 
and the accuracy of the \MC{} description of the data can be established. 
This uncertainty can be extracted directly from \figRef{fig:respLightGlu} and 
amounts to about $2\%$ at low \pT{} and $0.5\%$ at high \pT{} for the \EMJES{} calibration, if the flavour composition 
of the sample is known within $25\%$. %
It can be reduced by a factor of two
at low \pt{} and even more at high $\pT$ 
through the use of one of the more sophisticated calibration schemes. 

These response differences between jets initiated by light quarks and gluons result in a 
sample dependence of the energy scale and suggests that the \JES{} calibration determined 
from \insitu{} techniques might only be applicable within a larger systematic uncertainty 
to different jet samples. 
With the techniques commissioned up to date, the 2011 dataset only allows for a coarse 
validation of the differences in the jet energy scale between light-quark- and gluon-initiated jets. 
\MC{} simulations are instead used to understand the impact of systematic effects in the response 
differences between light-quark and gluon jets.

\FigRef{fig:respLightGlu} shows 
the jet response difference between jets initiated by light quarks and 
gluons in the central $|\etaDet|$ region of \ATLAS{} for \pythia{} (standard \ATLAS{} MC11 tune), \pythia{} (\Perugia 2011 tune) 
and \herwigpp. Comparisons between the first two simulations show the impact of the underlying event tune 
on the response differences. 
Comparisons between \pythia{} and \herwigpp{} provide an estimate of the impact of differences 
in the modelling of the parton shower, fragmentation and hadronisation for generators 
modelling the jet fragmentation well within the constraints 
provided by data. The differences in the response between these two models are large, while the 
effect of the underlying event tune is small, as can be seen by comparing 
the standard \pythia{} MC11 tune with the \Perugia 2011 tune. 

Further analysis of the large differences between \pythia{} and \herwigpp{} indicate that the cause is almost exclusively the difference 
in the response to gluon jets. This leads to a sizable response difference for the inclusive jet sample, which in the lower-\pT{} region has mainly gluon-initiated jets in the final state. Significantly smaller differences are observed in the samples used to calibrate the absolute jet response in the lower-\pT{} regime, like \gammajet{} and \Zjet, which have a dominant contribution from light-quark jets.  

The systematic effect illustrated by the difference between the two \MC{} simulations can be included as an additional systematic uncertainty. For this, the response variation $\Delta\mathcal{R}_{\mathcal{S}}$ for a given event sample $\mathcal{S}$ can be written as 
\begin{eqnarray}
\Delta\mathcal{R}_\mathcal{S} & = & \Delta f_g(\mathcal{R}_g-1)+\Delta f_{uds}(\mathcal{R}_{uds}-1)  \nonumber \\
& + & f_g\Delta \mathcal{R}_g + f_{uds}\Delta \mathcal{R}_{uds} + f_b\Delta \mathcal{R}_b + f_c\Delta \mathcal{R}_c ,
\label{eq:flavcomp}
\end{eqnarray}
where $\mathcal{R}_{g}$, $\mathcal{R}_{uds}$, $\mathcal{R}_{c}$, and $\mathcal{R}_{b}$ refer to the response to jets initiated by gluons, light ($u$, $d$, $s$) quarks, \cquarks, and \bquarks, with $\Delta$ denoting the uncertainty on the respective variable. The fractions $f_{x}$ refer to the fractions of jets with a given partonic flavour $x \in \{ g, uds, c, b \}$ in the sample $s$.
Under the simplifying assumption that the jet energy scale uncertainty is established \insitu{}
for light-quark jets and that it is the same for jets from \bquarks{} and \cquarks, \eqRef{eq:flavcomp} can be simplified to 
\begin{equation}
\Delta \mathcal{R}_{\mathcal{S}}=\Delta f_q(\mathcal{R}_q-\mathcal{R}_g)+\Delta \mathcal{R}_q+f_g\Delta \mathcal{R}^{\rm ex}_{g} ,
\label{eq:flavSyst}
\end{equation}
where $\Delta\mathcal{R}_q\equiv\Delta\mathcal{R}_{uds}\equiv\Delta\mathcal{R}_b\equiv\Delta \mathcal{R}_c$ and 
$f_{q} = f_{uds} + f_{c} + f_{b} = 1 - f_{g}$. The additional term $\Delta \mathcal{R}^{\rm ex}_{g}$ reflects 
an additional variation that represents the uncertainty on the response of gluon 
jets that arises from the systematic effects captured by the different \MC{} simulations. 
Note that the first term of this equation is used to estimate the effect of the results shown in 
\figRef{fig:respLightGlu} on
the systematic uncertainty of the jet energy scale in a sample of imprecisely known 
flavour composition. 

The additional term $\mathcal{R}^{\rm ex}_{g}$ 
was
not added to the $2010$ \ATLAS{} jet energy scale uncertainty for simplicity, 
since it 
was
much smaller than the dominant contributing effects. 
The improvements in the jet energy measurement achieved with the $2011$ dataset require this more careful treatment.  
Using the response difference $\mathcal{R}_{q} - \mathcal{R}_{g}$ with the \EMJES{} calibration at low \pT{} shown in \figRef{fig:respLightGlu},
the uncertainty on $\mathcal{R}^{\rm ex}_{g}$ amounts to about $3\%$ in a sample with $75\%$ gluon content, which is close to the inclusive jet sample.   It is reduced to about $1\%$ in a sample with $25\%$ gluon content, as expected for \ttbar{} with radiation. 
The uncertainty at high $\pt$ is smaller than $1\%$. This term in the uncertainty can also
be reduced by 
a factor of $2$ 
or more when using the more evolved calibration schemes \LCWJES{} or \GS.

The \insitu{} jet energy scale uncertainty is derived using \gammajet{} and \Zjet{} samples, 
which at low $\pt$ are dominated by light-quark jets.  
The expression for the total uncertainty presented here could be generalised to account for the fact that there is 
some gluon-initiated jet contamination, and that the uncertainty on the light-quark jet response $\Delta \mathcal{R}_q$ cannot be established using these samples alone.
However, the approximation that the \gammajet{} and \Zjet{} sample are pure 
light-quark jet samples is most accurate at low \pt, where the gluon jet response uncertainty
is largest. Thus, this approximation leads to inaccuracies that are significantly smaller 
than other systematic uncertainties in the average jet response.

\subsection{Discrimination of light-quark and gluon induced jets}
\label{sec:likelihoodDiscr}
As indicated before, the differences between light-quark and gluon jets lead to (average) 
differences in observable final-state jet properties. 
Jets initiated by gluons are expected to be broader, with more low-\pt{} particles than those 
initiated by light quarks. Relevant observables like the jet width $w_{\mathrm{jet}}$, as reconstructed using the \pT{} flow of 
tracks associated with the jet, 
and the number of those tracks $n_{\mathrm{trk}}$, are already used to 
measure the average flavour fractions in different data samples \cite{jespaper2010}. They are identified as powerful
discriminators for the purpose of understanding partonic flavour in previous studies \cite{Schwartz2}.
More details on the quark-gluon tagger performance in the ATLAS detector can be found in Ref.~\cite{ATLASquarkgluon}. 

These jet properties, reconstructed using selected high-qua\-li\-ty tracks, are further exploited to build a 
likelihood discriminator or a light-quark/gluon tagger. Two-dimensional $(n_{\mathrm{trk}},w_{\mathrm{jet}})$ 
distributions are determined for data and \MC{} simulations using the inclusive jet and \gammajet{} event samples. 
The different fractions of light-quarks and gluons in these samples, which in \MC{} simulation are extracted from \pythia{} with the \ATLAS{} MC11 
tune,
are then reflected by variations in the $(n_{\mathrm{trk}},w_{\mathrm{jet}})$ distributions, and the expected ``pure'' jet sample properties can
be extracted.
This procedure is applied both in data and \MC{} simulations, and both data-driven and \MC-based taggers are built. 
Operating points are defined at fixed light-quark jet efficiencies of 30\%, 50\%, 70\% and 90\%, using the same 
extracted $(n_{\mathrm{trk}},w_{\mathrm{jet}})$ distributions.

The quark/gluon tagger essentially selects jets with both decreasing $n_{\mathrm{trk}}$ and $w_{\mathrm{jet}}$
as the operating point tightens, to achieve a higher gluon jet rejection at the expense
of a lower light-quark jet efficiency. It can then be expected that jets selected with different 
operating points of the tagger have different jet energy scales. This is shown in
\figRef{fig:respVsLLFlavs}, where the response as a function of the operating
point used to select jets in an inclusive \MC-simulation sample is shown for
two \pT{} bins for jets calibrated with the \EMJES{} calibration. 
\begin{figure*}[!htp]
	\centering
	\subfloat[$40 < \pT < 60$ \GeV]{
	\includegraphics[width=0.45\textwidth]{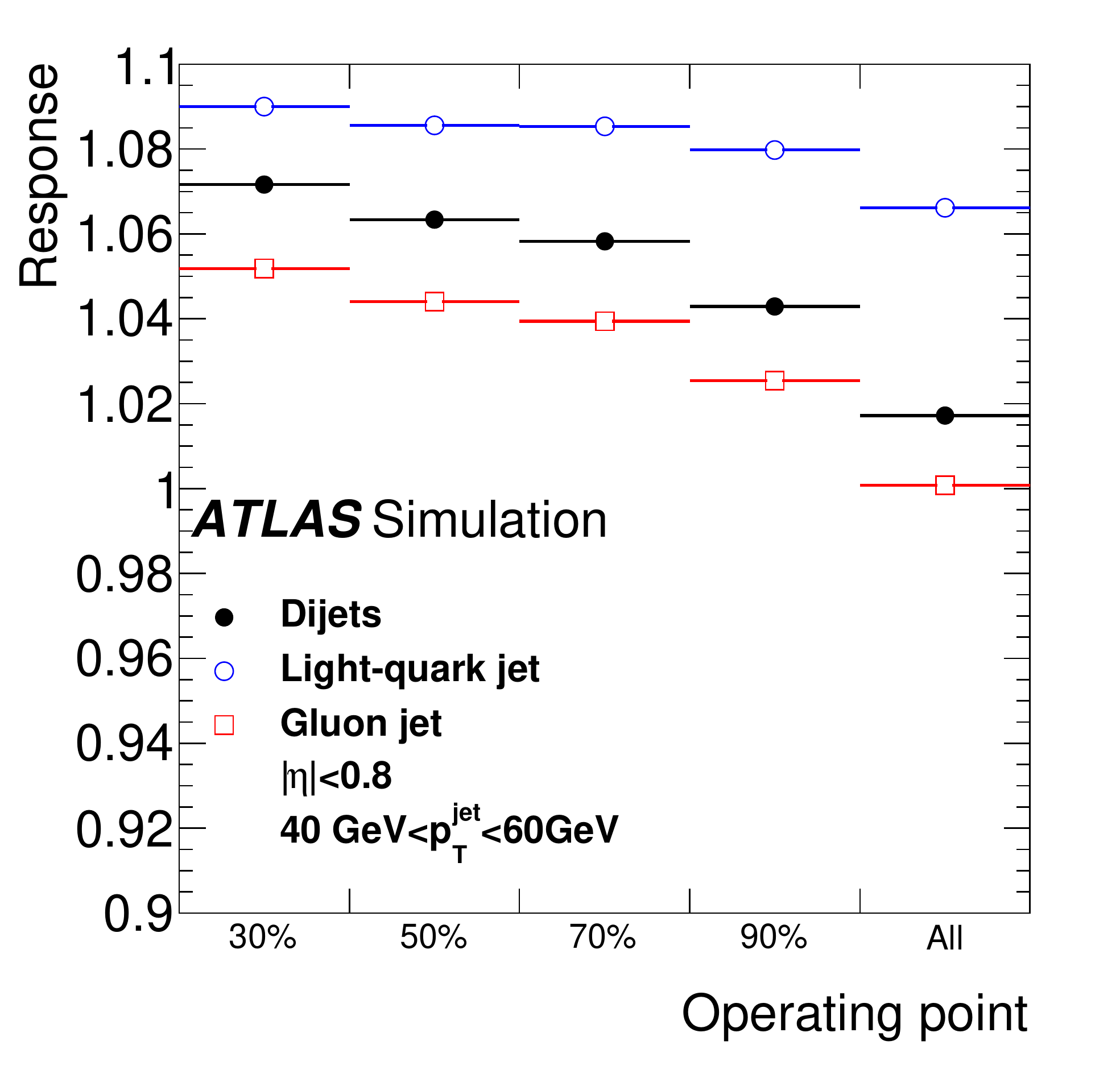}
	\label{fig:respVsLLFlavs:low}
	} \qquad
	\subfloat[$260 < \pT < 310$ \GeV]{
	\includegraphics[width=0.45\textwidth]{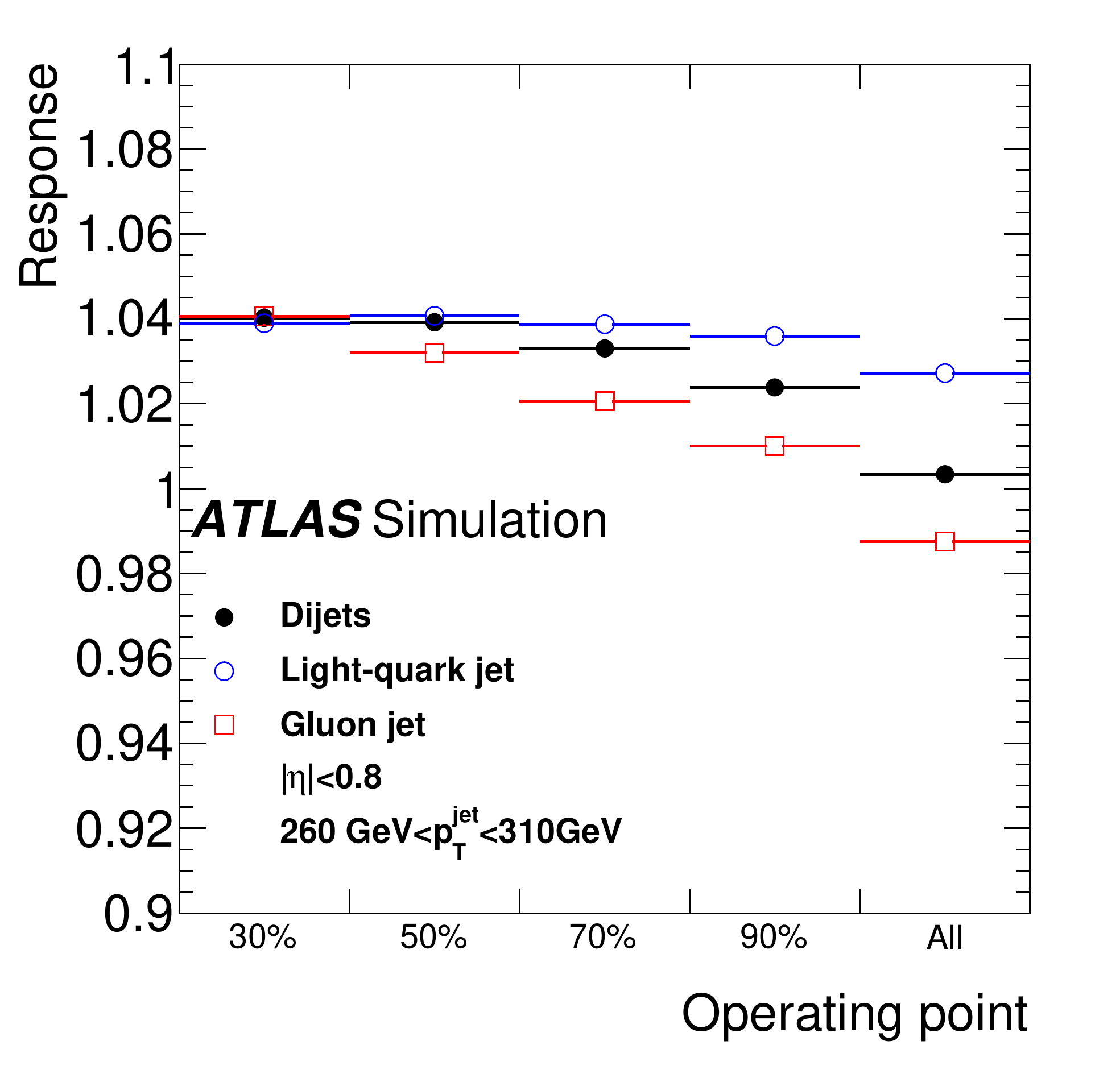}
	\label{fig:respVsLLFlavs:high}
	}
	\caption[]{Jet $\pt$ response for the two leading jets in the dijet sample for different
	tagger light-quark operating points for jets with  \subref{fig:respVsLLFlavs:low} $40\GeV<\pt<60\GeV$ and  \subref{fig:respVsLLFlavs:high}
	$260\GeV<\pt<310\GeV$, and $|\etaDet| <0.8$. Jets are labelled as light quark or gluon 
	using the \MC-simulation record and are further required to be isolated. }
	\label{fig:respVsLLFlavs}
\end{figure*}
Even choosing a high efficiency operating point increases the sample response significantly, particularly at low \pT, compared to the 
inclusive sample.  
The difference in response between light-quark and gluon jets is largest for the inclusive sample, and
basically vanishes for the tightest operating point at high jet $\pt$. This is expected, since it is shown in 
\figRef{fig:respLightGlu} that applying a $n_{\mathrm{trk}}$- and $w_{\mathrm{jet}}$-based \JES{} correction like \GS{} removes the
response differences between light-quark and gluon jets at high $\pt$. 
In addition, these jets are selected by the likelihood because they have quite similar (quark-jet-like) observable properties.

To gain confidence that the change in jet response does not affect analyses using the tagger, it is necessary to demonstrate that 
the agreement of the jet energy scale between \MC{} simulations and data does not change when the likelihood cut corresponding to each 
operating point is applied. 
This is verified using the \gammajet{} balance technique described in \secRef{sec:gammajetInSitu}, which finds changes of the \datatomc{} 
agreement to be below 1\%. 

The same \pT-balancing technique allows for a study of the dependence of the \JES{} on the tagger operating point 
in a specific sample, but not for an investigation of the light-quark and gluon jet responses directly. 
This is controlled through the sam\-ple- and flavour-dependent systematic uncertainties de\-scri\-bed in the 
previous section and summarised in \eqRef{eq:flavSyst}. The first term in this equation is 
based on the differences between light-quark and gluon \JES, which become smaller when the tagger is used, 
as shown in \figRef{fig:respVsLLFlavs}.  The second term is calculated comparing \herwigpp{} and \pythia{} in the dijet 
sample. Both comparisons are performed for tagged jets, and they
demonstrate that these uncertainties are actually smaller after the application
of the tagger than before. The use of the uncertainties derived in the previous section is thus
conservative for tagged jets, and the validation in the gluon jet sample is sufficient.

\subsection{Summary of the jet flavour dependence analysis}
\label{sec:flavoursummary}

The dependence of the jet energy scale on the flavour of the originating parton of the jet 
is evaluated in \MC{} simulations. This difference, which enters
the \JES{} systematic uncertainty, is shown to be sensitive
to certain details of the modelling of the decay and fragmentation of jets 
in the \MC{} generators. An additional term is derived that needs to be added 
to the \JES{} uncertainty to account for this dependence.
It amounts to about $3\%$ in a
sample with a $75\%$ gluon content (close to the inclusive jet sample) and is reduced 
to about $1\%$ in a sample with $25\%$ gluon content at low $\pt$ when using the \EMJES{} 
calibration scheme. The uncertainty at high \pT{} is smaller than $1\%$. 
This contribution to the \JES{} uncertainty can also be reduced by a factor of two or more when using the more 
sophisticated calibration schemes and is included as a part of the combined \ATLAS{} jet energy scale uncertainty. 

The flavour dependence of the \JES{} arises to a great extent from differences in observable properties of jets, such as the number of tracks 
and the jet width measured with tracks. These properties can be used to reduce this dependence, 
as well as to discriminate between light-quark and gluon jets. 
The properties are used in \ATLAS{} to build a quark/gluon jet tagger exploiting the differences in 
flavour composition between an inclusive jet 
and a \gammajet\ sample, in data as well as in \MC{} simulations. The \JES{} dependence on the choice of operating
point used in the tagger yields a \datatomc{} difference of less than $1\%$.
Furthermore, the sample
dependent uncertainties become smaller once jets are tagged, since the fragmentation
is constrained to a specific phase space for which differences between light-quark and
gluon jets between different \MC{} generator models are smaller.

\section{Jets with heavy-flavour content}
\label{sec:bjets}
In this section the measurement of the
jet energy is studied for jets from heavy-flavour decays. The main observable used in the corresponding analysis based both on \MC{} simulations and \insitu{} techniques is
the ratio \rtrk{}  of the sum of transverse momentum vectors \pttrkvec{} from all tracks in the jet cone to the calorimeter jet transverse momentum \ptjet,
%
\begin{equation}
\rtrk = \dfrac{| \sum \pttrkvec |}{\ptjet}.
\label{eq:def-rtrk}
\end{equation}
These studies assess the jet energy measurement in the calorimeter in
light-jet-enriched samples as well as for \bjet-enriched samples in
an inclusive jet sample and in an event sample where a top-quark pair
is produced (\ttbar). The uncertainty on the \bjet{} energy measurement 
is thus evaluated over a wide range of \pt{} and under different background 
conditions. Furthermore, the \pt{} imbalance in a dijet system is 
used to validate the description of the kinematics of the neutrino
coming from \bquarks{} decaying semileptonically in the \MC{} simulation.

In the following jets originating from a \bquark{} (\bjets) and identified by means of $b$-tagging techniques are referred to as ``$b$-tagged jets". 
%
The notation ``inclusive jets" is used to denote a mixture of jets initiated by light quarks, \bquarks, and gluons. All types of jets originating from \bquarks, including those containing semileptonic \bquark{} decays, are referred to as
``inclusive \bjets".

Since an unbiased sample of jets induced by charm quarks can not be selected in the
data, no dedicated studies for charm jets have been performed.
Charm jets are considered to be light jets and are treated as described 
in Section~\ref{sec:FlavorTopology}.

%
\subsection{Jet selection and response definition}\label{sec:bjetsjetreco}
Jets with a calibrated transverse momentum $\ptjet > 20$~\GeV{} and a pseudorapidity 
$|\etajet| < 2.5$ are used in this study. 

Two aspects of the jet energy scale are studied separately: the response to particles
absorbed in the calorimeter and the detector response to all produced particles including muons
and neutrinos. The former is characterised
by the calorimeter response $\Response^{\rm calo}=\ptjet/\pttruth$, where \pttruth{}
is the $\pt$ of a matched truth jet built from stable final-state particles, as defined in \secRef{sec:truthjets}, with the exclusion of 
muons and neutrinos. The latter is characterised by the all-particle response 
$\Response^{\rm all}=\pt^{\rm jet+\mu}/\pt^{\rm truth, all}$, where $\pt^{\rm jet+\mu}$ includes selected  
reconstructed muons inside the jet and $\pt^{\rm truth, all}$ is the \pt{} of a matched truth jet built from all stable final-state particles. 

The jet energy scale of $b$-tagged jets in the dijet sample is studied
using different $b$-tagging algorithms.  For each algorithm, different
operating points resulting in different efficiencies and purities are
studied, as detailed in Sect. \ref{sec:evtsel}. In the \MC{}
simulation, the flavour of jets is determined as described in
Ref. \cite{mv1}, by the presence of a heavy-flavour quark matched
geometrically to the reconstructed jet, using the distance \DeltaR{} in \etaphispace, see \eqRef{eq:deltaRdet} in \secRef{sec:jetdirections}.

In the \ttbar{} sample $b$-tagged jets are selected by means of the
\mvone{} tagger \cite{mv1}.  The \mvone{} tagger uses the results from
three $b$-tagging algorithms exploiting secondary-vertex and
track im\-pact-parameter information, which are input to a neural
network to derive a likelihood discriminant to select \bjets. 
In this analysis, a jet is experimentally identified as a \bjet{} if the
\mvone{} tagger weight (\wmvone) exceeds a threshold value of $0.6$.
This corresponds to 70\% per-jet efficiency for selecting \bjets{}
from $\ttbar$ decays, and a per-jet rejection factor for light-quark
jets of about $130$. To adjust the \MC{} simulations to the
$b$-tagging performance in data, a dedicated $b$-tagging efficiency
correction \cite{mv1} is applied to the simulation and the
related systematic uncertainties are evaluated.

The influence of nearby jets on the measurements is studied by
applying an isolation requirement which rejects jets that are separated from
the nearest other jet by a distance
$\DeltaR< 2 R$.  The influence of this
requirement is found to be negligible in the analyses presented,
so the requirement is omitted in the results shown.

The jet vertex fraction \JVF{} introduced in \secRef{sec:dijetselection} is 
used to quantify the amount of energy in a jet coming from pile-up interactions. 

\subsection{Track selection}
Tracks are associated to jets by requiring that the opening angle between the track and the jet 
direction be $\DeltaR({\rm jet,track})<0.4$, measured in \etaphispace. 
Tracks are required to pass the track selection criteria presented in \secRef{sec:trackjets} in the context of track jets. This assures an appropriate reconstruction quality and that the selected tracks come from the primary hard-scattering vertex.

\subsection{Event selection}\label{sec:evtsel}
Events are initially selected by means of single-jet and
single-lepton triggers. A primary vertex reconstructed from at least five tracks, 
which is consistent with the position and transverse size of the beam, is required. 
Analysis specific selections are described below.

\subsubsection{Jet sample selection}
Four complementary event selections are used for studies in the dijet sample: 
\begin{enumerate}
\item An inclusive selection is used to study the energy calibration in the inclusive jet sample,
and uses $11$ single-jet triggers to cover the full \pt{} range, to cope with the reduced data
rate allowed for lower-\pt{} triggers. 
\item Two $b$-tagged jet selections are used to study the energy calibration of \bjets. 
\begin{enumerate}
\item An inclusive $b$-tagged sample is selected using five different single-jet triggers, since
the range of \pt{} for \bjet{} studies is limited by the low trigger rates at low \pt{} and
by the measurements of $b$-tagging efficiencies at high \pt. 
\item A semileptonic $b$-tagged sample is selected using a single muon--jet trigger, 
requiring a muon candidate inside a jet, which is less heavily pre-scaled, increasing the size of
the sample collected with respect to a sample collected with a single-jet trigger. 
\end{enumerate}
\item A dijet selection is used to study the impact of semileptonic decays into muons and neutrinos. 
\end{enumerate}
Only one trigger is used to collect events in a specific \pt{} bin. This procedure is found to 
be compatible within statistical uncertainties with a procedure that combines all jet triggers
in each \pt{} bin by weighting contributing events according to the integrated luminosity collected by the trigger 
that allowed the event to be recorded. 

The measurement in the dijet sample is performed as a function of the average \pt{} 
(\ptavg) of the two leading jets, including the muon candidate if one is reconstructed 
inside the jet. The
estimated muon energy loss in the 
active layers of the calorimeter
is subtracted to avoid double counting.
 
The measurement in the inclusive samples is performed as a function of \ptjet. The dijet event selection further requires:
\begin{enumerate}
\item At least two jets are reconstructed with $\ptjet>20 \GeV$, $|\etajet|<1.2$ and $|\JVF|>0.75$.
\item The two leading (in \pt) jets are $b$-tagged with the \mvone{} algorithm ($\wmvone>0.6$).
\item At least one of the jets with a muon candidate within $\Delta R<0.4$ passes the selection
described in Ref.~\cite{mv1}.
\item No third-leading jet reconstructed in the event with $|\JVF|>0.6$ 
and $\ptjet > \max(12\GeV, 0.25 \cdot \ptavg)$.
\item The azimuthal distance between the two leading jets is $\Delta\phi_{jj}$ $> 2.5$. 
\end{enumerate}
The selection 
on the inclusive samples requires at least one jet with $\ptjet>25$ \GeV{} and $|\etajet|<2.5$, 
and the $|\JVF|>0.75$ cut. The muon selection is unchanged and different $b$-tagging algorithms and
operating points are studied, since the neutrino energy is expected to be largely independent
of the tagging algorithm, while \JES{} is not. 

The \bjet{} purity of these samples is measured with \MC{} simulations to vary from $50\%$ to $70\%$ for the inclusive selection, 
$60\%$ to $80\%$ for the semileptonic selection, and to be above $80\%$ for the dijet selection for the operating points studied. 
Observations at high $\pt\gtrsim 200$ \GeV{} suggest that the purity might be underestimated by as much as $10\%$ \cite{dijetFlav}. Uncertainties 
on the efficiency of the tagging algorithm to identify \bjets{} and \cjets{} can also impact these purity estimates by up to about $10\%$ \cite{mv1}. Despite
these systematic effects, the purity of these samples remains sufficiently large for the validation purposes of this study. 

\subsubsection{Top-quark pair sample selection}
Top-quark pair events where one of the $W$ bosons produced by the top-quark decays to
an electron or a muon are selected by the following requirements (see Ref.~\cite{mtop2011paper} 
for further details)
\begin{enumerate}
\item A single-lepton trigger is present.
\item Exactly one electron with transverse energy above $25 \GeV$, within pseudorapidity range of $|\eta|$ less than $2.47$, and
  outside the region of transition between the barrel and the endcap calorimeters, 
  $1.37 \leq |\eta| < 1.52$ is reconstructed; or, exactly one
  muon with transverse momentum above $20 \GeV$ is reconstructed within
  $|\eta|<2.5$. The reconstructed charged lepton has to match the
  trigger object corresponding to the required triggers that passed.
\item For the $\ttbar\to\ejets$ channel the transverse \Wboson{} boson mass \mtw, reconstructed from the electron and \MET, should
  be $\mtw > 25\GeV$, with $\MET>35$ \GeV. Alternatively, for the
  $\ttbar\to\mjets$ channel, $\MET >25$ \GeV{} and $\MET+\mtw>60$ \GeV{}
  are required.
\item At least four jets with $\ptjet> 25\GeV$, $|\JVF|>0.75$,
  and $|\etajet|<2.5$ are required.
  Among these, at least two jets should
  be $b$-tagged using the \mvone{}
  $b$-tagging algorithm ($\wmvone>0.6$). 
\end{enumerate}

After this selection the background contamination in the $\ttbar$
sample is expected to be of order $10\%$ and to mainly consist of
events from \Wboson/\Zboson+jets and single top-quark production. The
contribution from multijet background after the requirement of two
$b$-tagged jets is expected to be about $4\%$. 
The background contamination in the selected data sample has no
sizable impact in the studies performed, and it is considered as an
additional systematic uncertainty.

\subsection[\MC-based systematic uncertainties on the calorimeter \bjet{} energy scale]{\MC-based systematic uncertainties on the calorimeter \BJET{} energy scale}
\label{sec:mcBasedStudies}

The uncertainties on the \bjet{} transverse momentum measurement are
studied using systematic variations in the \MC{} simulation. The \bjet{} can be either
reconstructed using a calibration with respect to all stable particles
to study the all-particle energy scale, or excluding muons and
neutrinos to study the calorimeter energy scale, as described in
\secRef{sec:jetrecocalib}.  The former definition is currently most
relevant for $b$-tagging calibration analyses \cite{mv1}, and 
further discussed in \secRef{sec:semilCorrMC}. 

The uncertainty in the calorimeter response to \bjets{} can be
estimated using a combination of different \MC{} simulations as reported in
Ref. \cite{jespaper2010}. 
\FigRef{fig:systMC}\subref{fig:systMCEscale} shows the calorimeter response to 
\bjets{} for various \MC{} simulations. 
\begin{figure*}[htbp]
\begin{center}
\subfloat[Response]{
\includegraphics[width=0.48\textwidth]{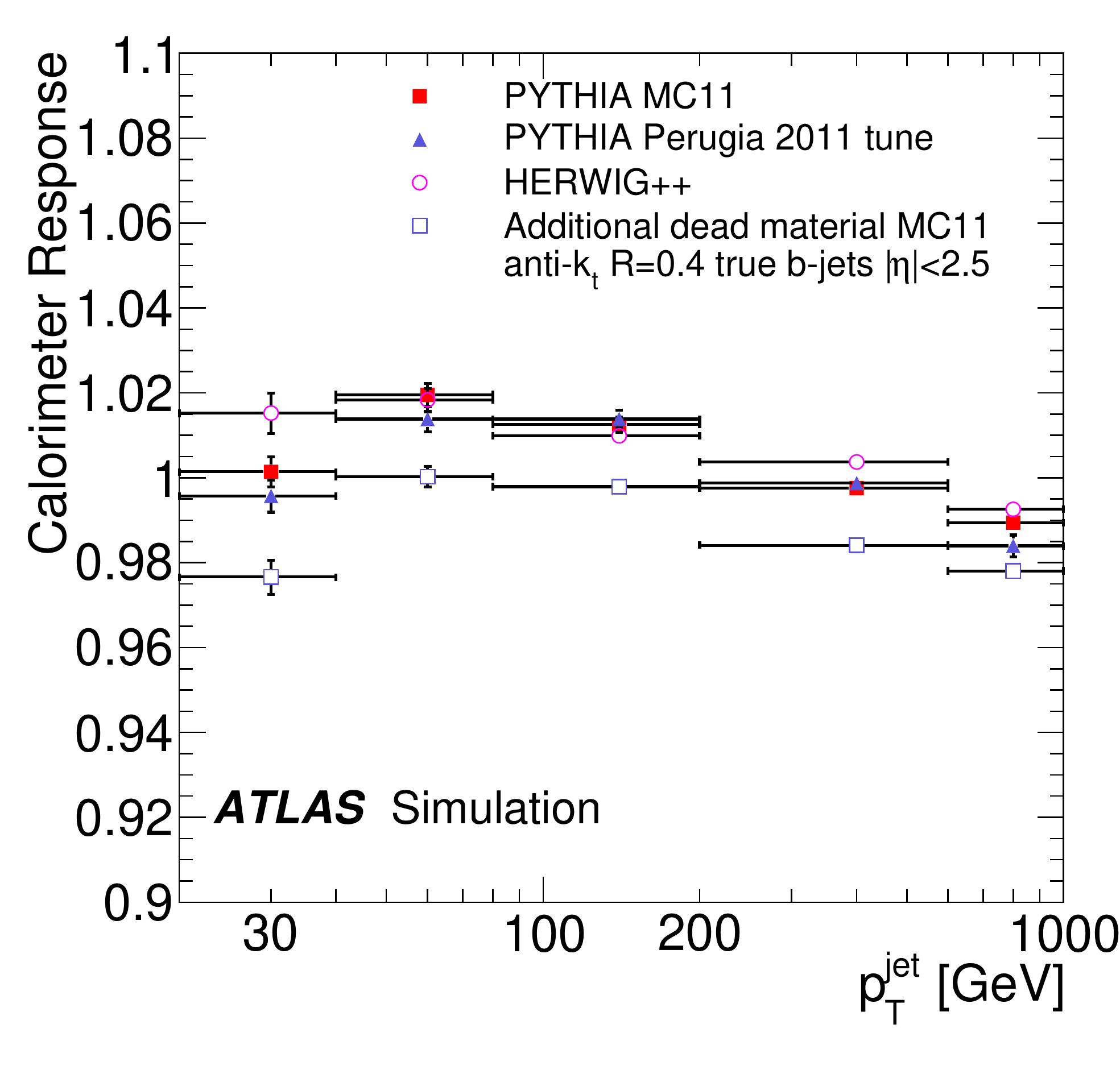} 
\label{fig:systMCEscale}}
\subfloat[Uncertainty]{
\includegraphics[width=0.48\textwidth]{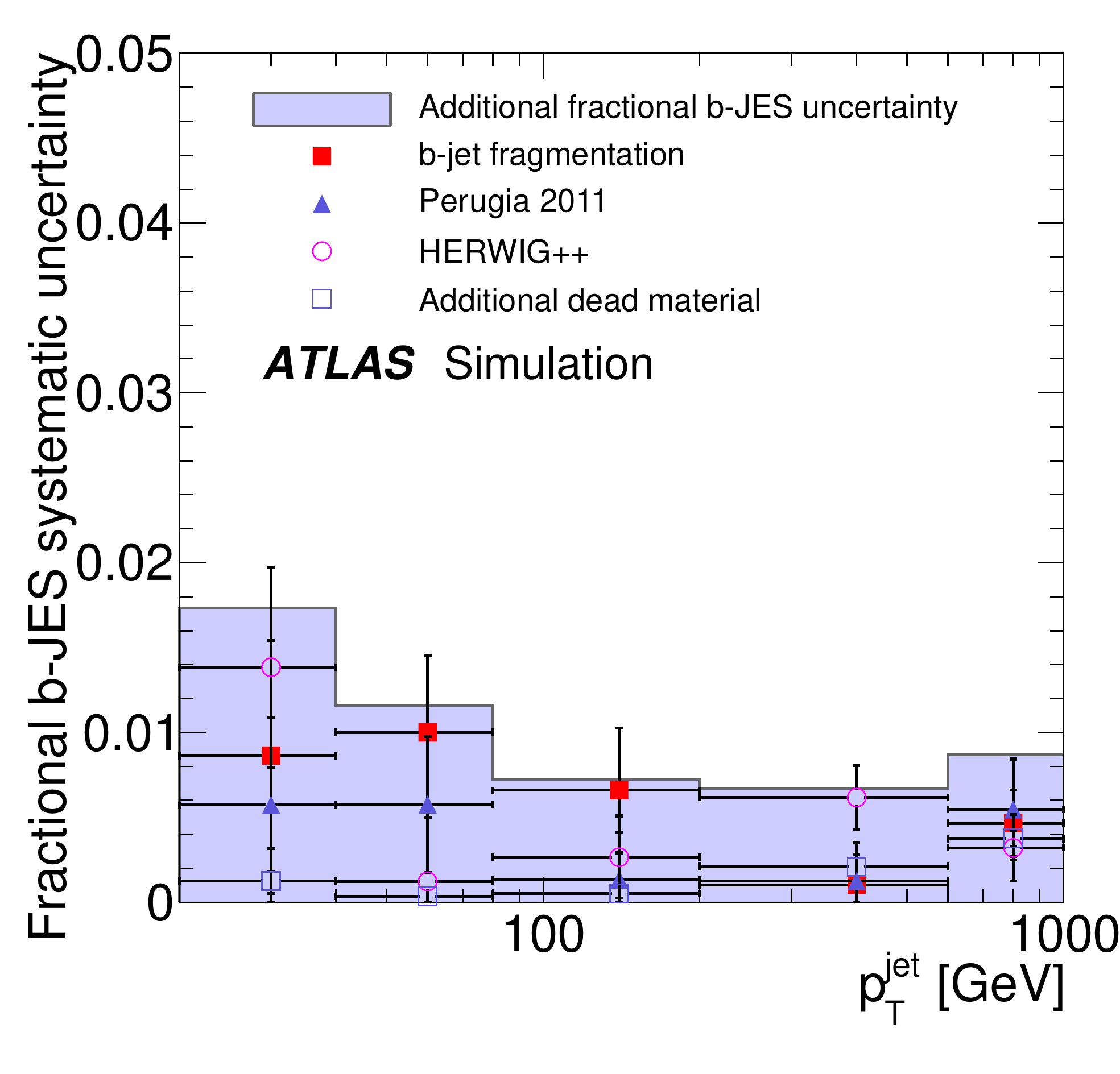} 
\label{fig:systMCSyst}}
\caption[]{Average response to
\bjets{} in some of the different samples used to calculate the 
\bjet{} energy scale systematic uncertainties is depicted in \subref{fig:systMCEscale}. The resulting uncertainties in the ratio of the 
\bjet{} response to the response of jets in an inclusive sample are shown in \subref{fig:systMCSyst}. These results are obtained for 
\bjets{} built with the \antikt{} algorithm with resolution parameter $R = 0.4$. 
}
\label{fig:systMC}
\end{center}
\end{figure*}

The corresponding systematic uncertainties associated with the \bjet{}
energy measurement are shown in \figRef{fig:systMC}\subref{fig:systMCSyst}. These
uncertainties need to be considered in addition to those established
for an inclusive jet sample, since \bjet{}
specific effects  are not taken into account in that analysis. These
uncertainties can be applied to any sample of \bjets, whether a
specific analysis uses tagging or not, and are of a size comparable to
the uncertainties in the \insitu{} measurements presented later in this
paper.

Two key changes are made in this analysis with respect to what is reported in
Ref. \cite{jespaper2010}. The dead material uncertainty, which is large in \figRef{fig:systMC}\subref{fig:systMCEscale},
but does not contribute significantly to the systematic uncertainty reported in 
\figRef{fig:systMC}\subref{fig:systMCSyst}, is calculated as an additional change in the response 
expected from dead material effects for a \bjet{} sample with
respect to an inclusive sample (or a pure light-quark sample for comparable results). 
This is possible in $2011$ because \insitu{} jet energy scale corrections and uncertainties
exist which are already accounting for a potential mis-modelling of the dead material in the \MC{}
simulation. The uncertainty component derived from the propagation of single-particle 
uncertainties to jets is also removed, while it contributes $0.5\%$ in $2010$ data. 
This result relies again on \insitu{} studies, since differences in the calorimeter response between
data and \MC{} simulations are already taken into account in those studies. Residual
effects that could give rise to an additional systematic uncertainty component for \bjets{}  
are constrained using a single-particle evaluation and are shown in \secRef{sec:Summary}.

%
\begin{figure*}[htbp]
\begin{center}
\subfloat[]{
\includegraphics[width=0.48\textwidth]{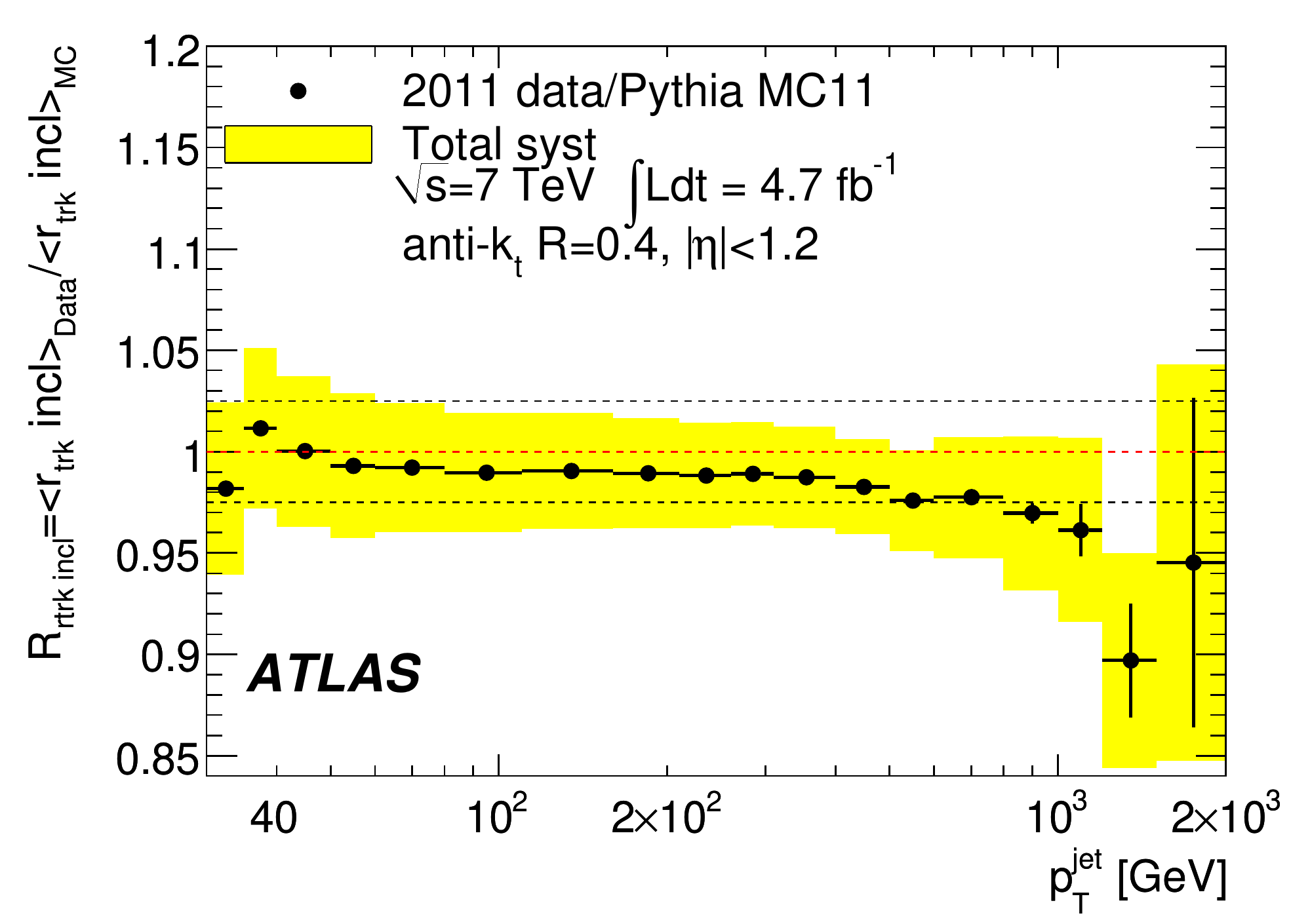} 
\label{fig:tracksJESdijetIncl}}
\subfloat[]{
\includegraphics[width=0.48\textwidth]{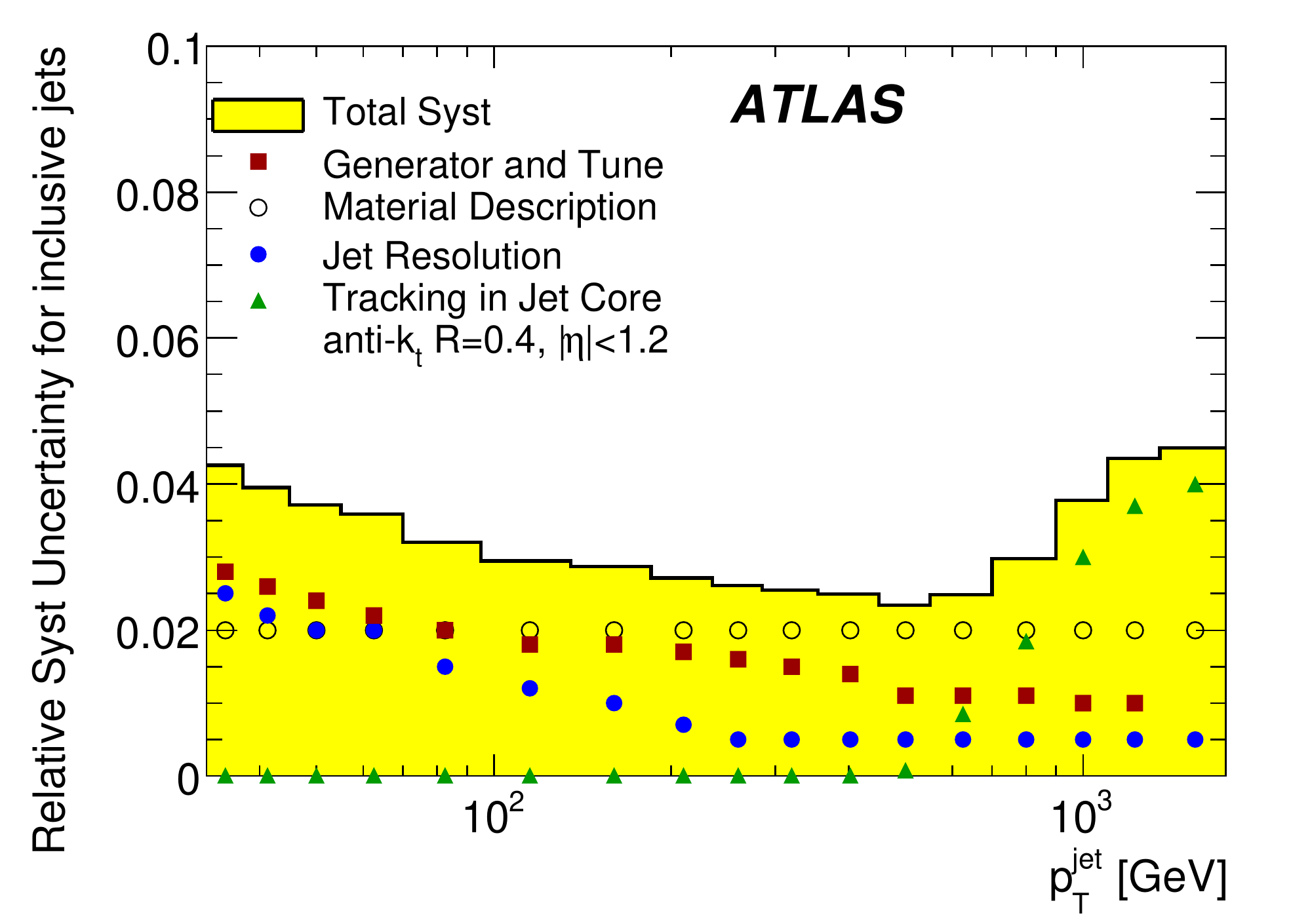} 
\label{fig:tracksJESdijetInclSyst}} \\
\subfloat[]{
\includegraphics[width=0.48\textwidth]{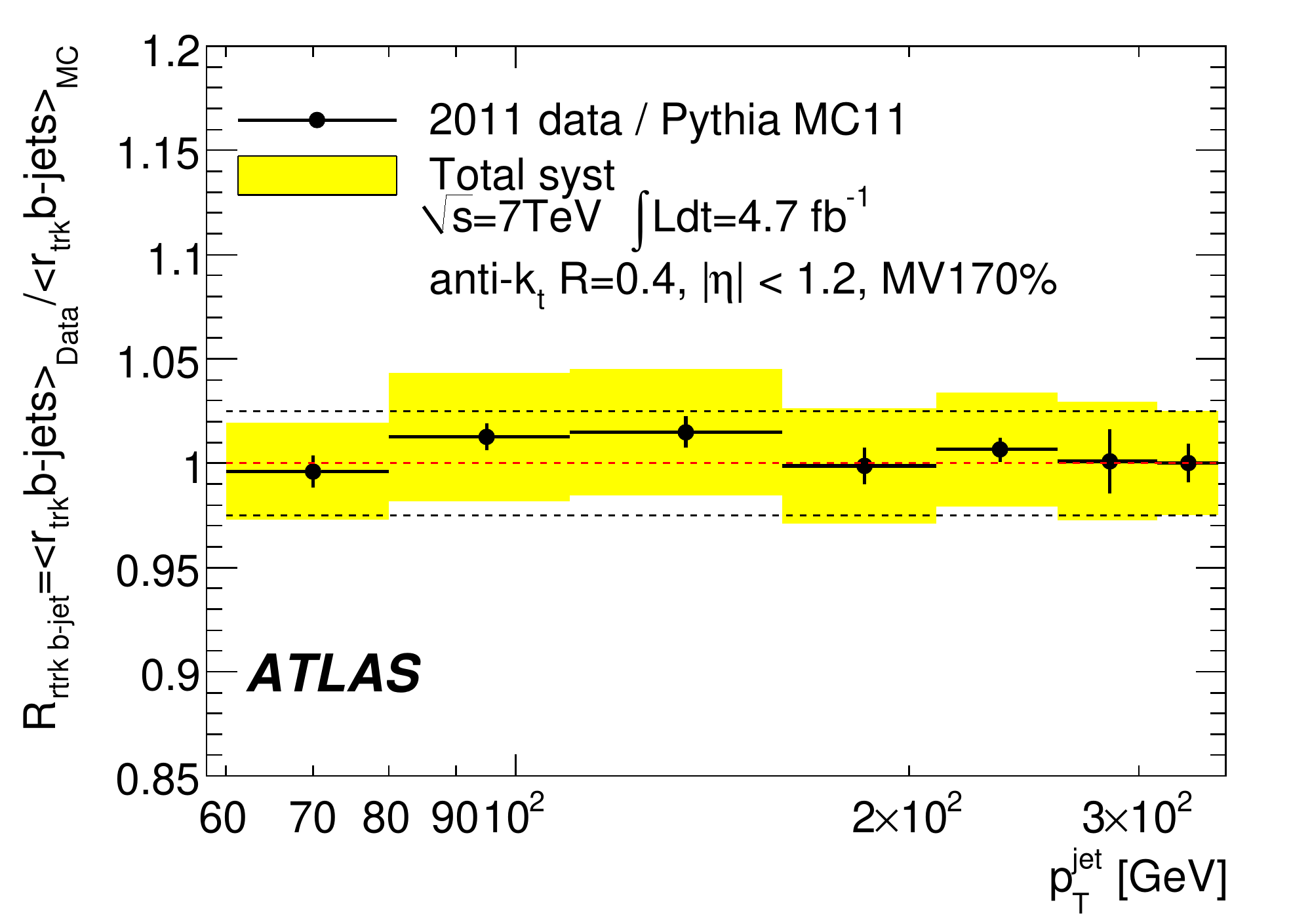} 
\label{fig:tracksJESdijetBtagged}}
\subfloat[]{
\includegraphics[width=0.48\textwidth]{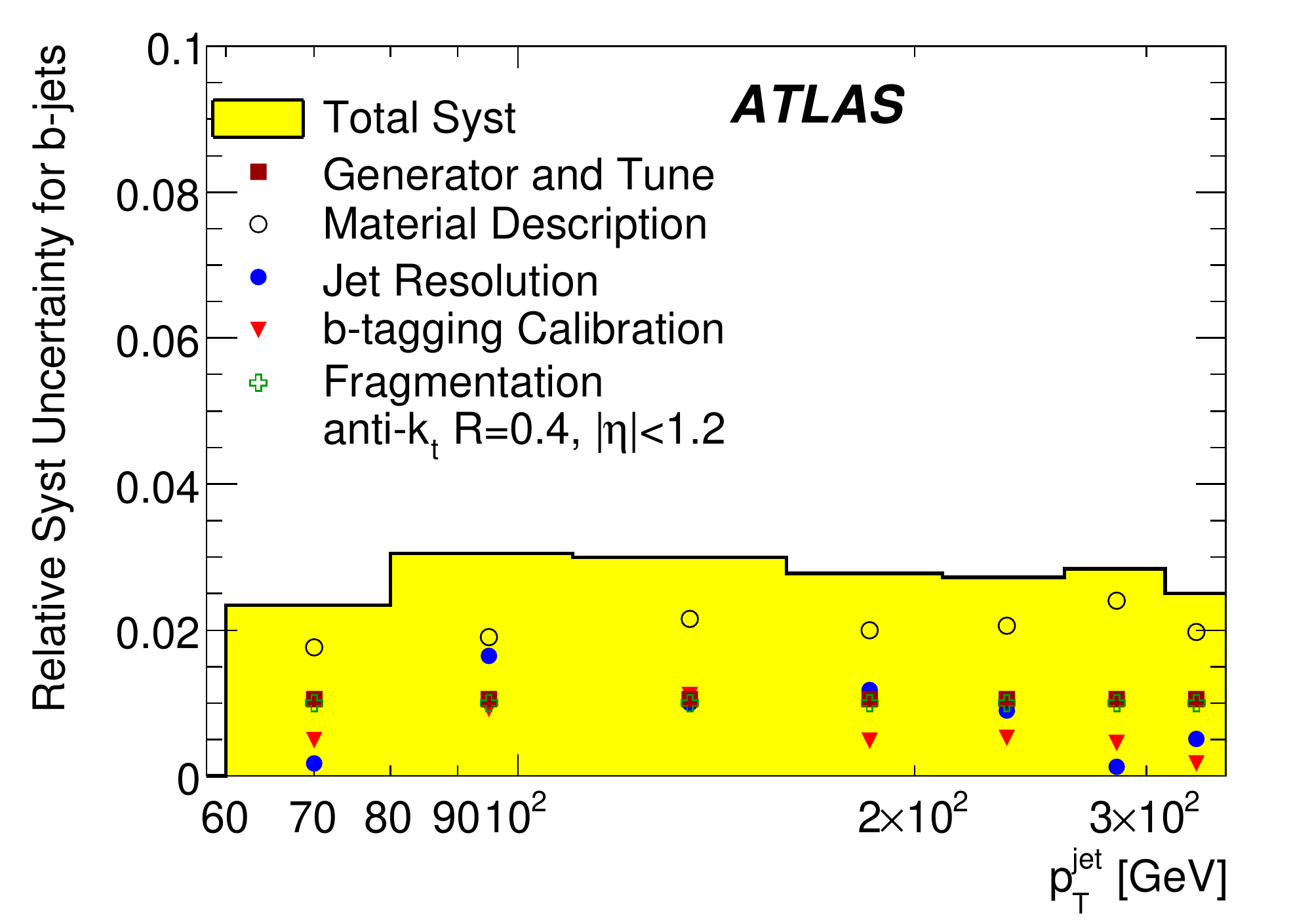} 
\label{fig:tracksJESdijetBtaggedSyst}} \\
\subfloat[]{
\includegraphics[width=0.48\textwidth]{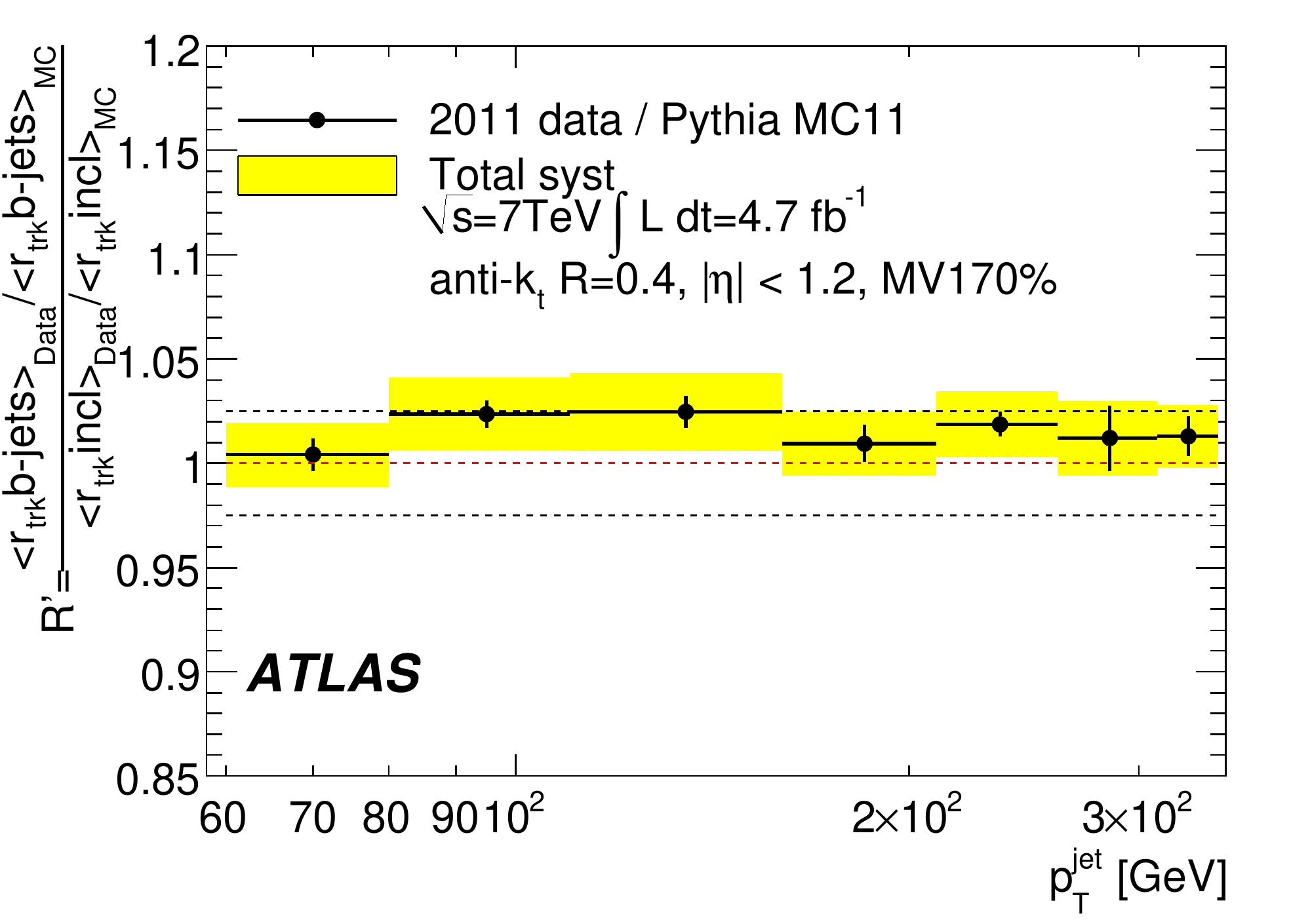} 
\label{fig:tracksJESdijetRatio}}
\subfloat[]{
\includegraphics[width=0.48\textwidth]{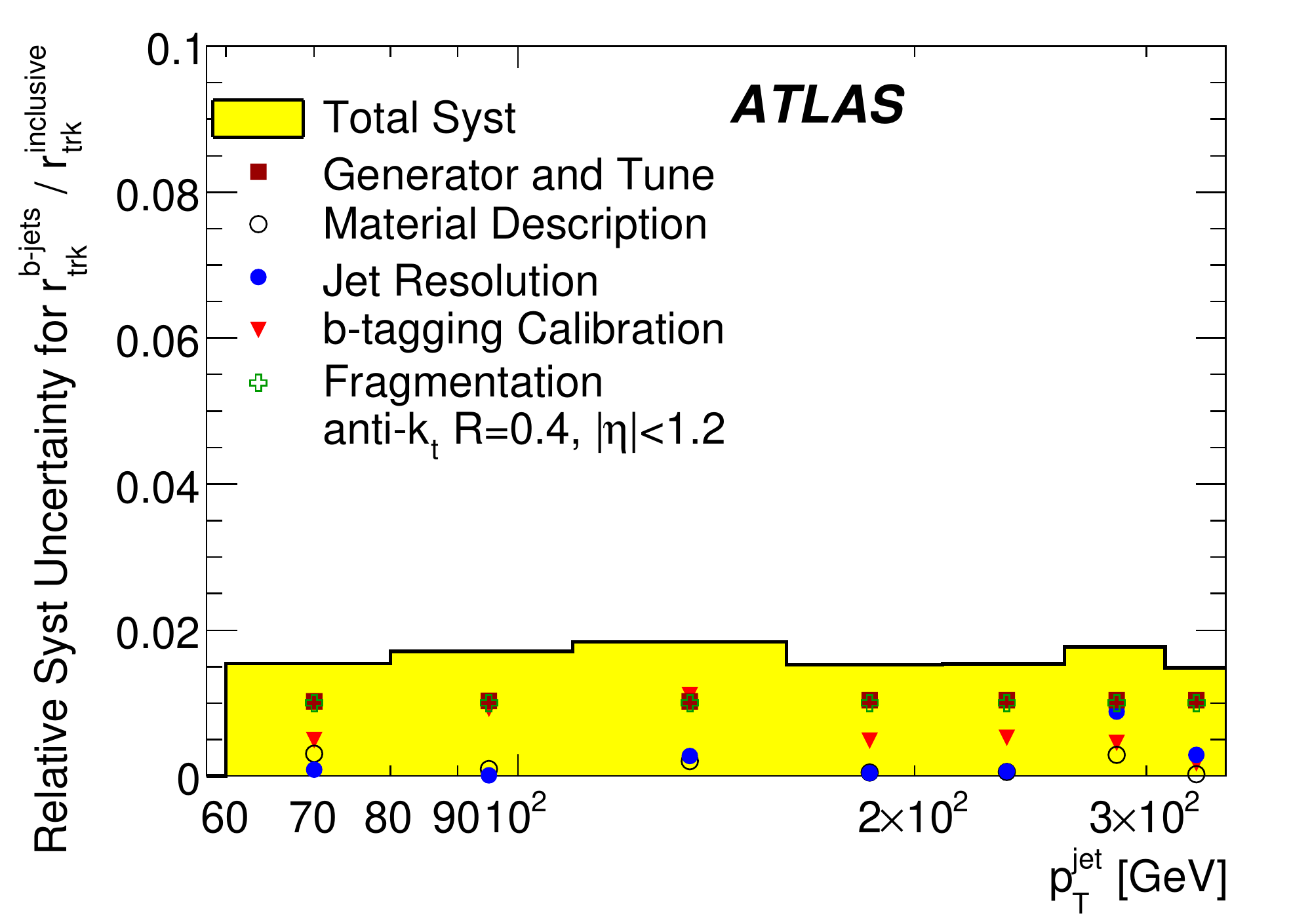} 
\label{fig:tracksJESdijetRatioSyst}}
\caption[]{Ratio of the average $\rtrk$ given in \eqRef{eq:def-rtrk} in data and \MC{}
  simulations for \subref{fig:tracksJESdijetIncl} inclusive jets and \subref{fig:tracksJESdijetBtagged} tagged \bjets{}. In \subref{fig:tracksJESdijetRatio}, the $b$-tagged to inclusive sample ratio variable $R'$ from \eqRef{eq:bjet-rprime} is shown. 
  The contributions of the systematic 
  uncertainties to the total uncertainty in the different measurements are shown in \subref{fig:tracksJESdijetInclSyst}, \subref{fig:tracksJESdijetBtaggedSyst}, and \subref{fig:tracksJESdijetRatioSyst}, respectively. 
  Jets within $|\etajet|<1.2$ are used. }
\label{fig:tracksJESdijet}
\end{center}
\end{figure*}
\begin{figure*}[htbp]
\begin{center}
\subfloat[]{
\includegraphics[width=0.45\textwidth]{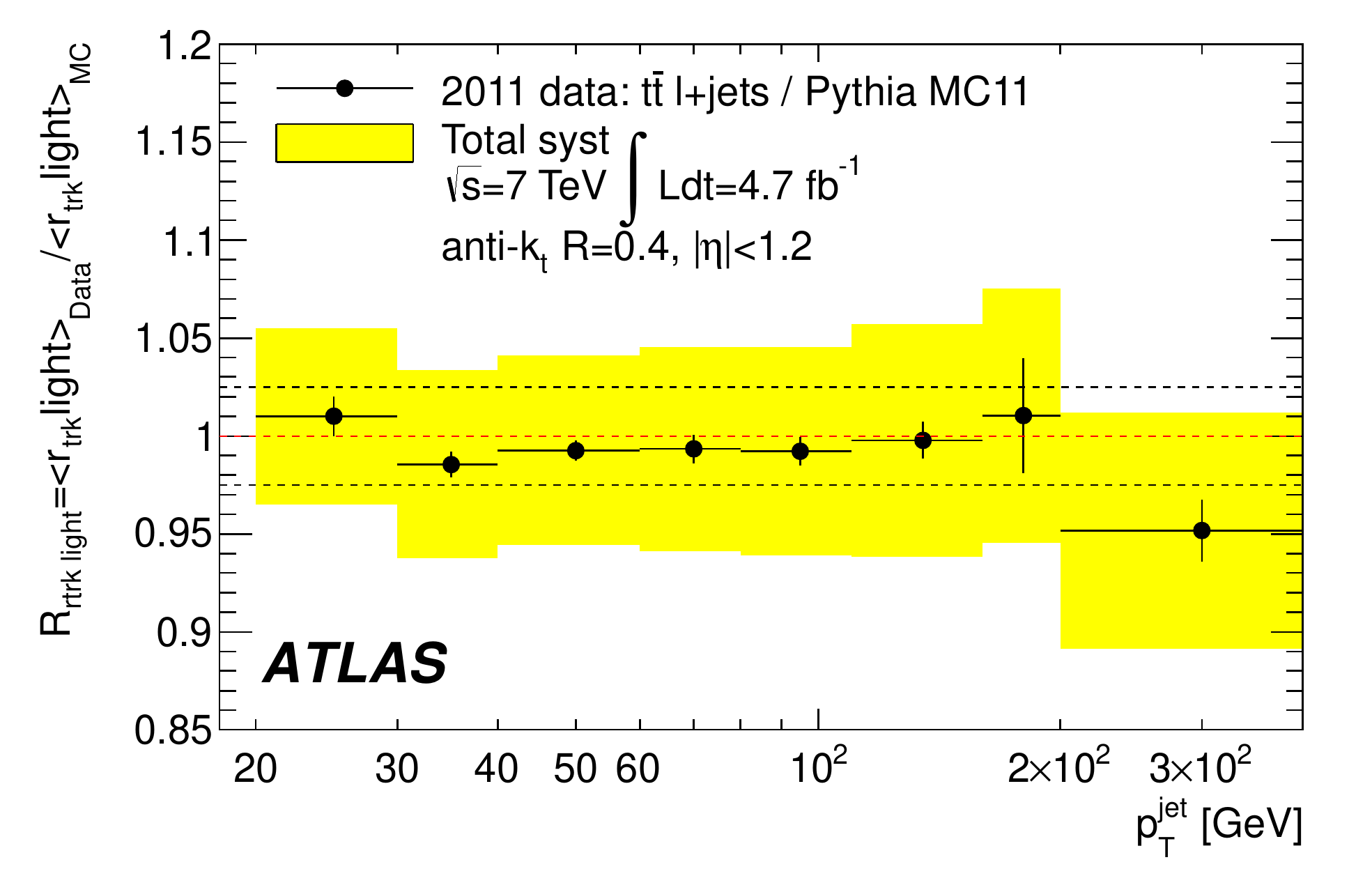} 
\label{fig:tracksJESttbarLight}}
\subfloat[]{
\includegraphics[width=0.45\textwidth]{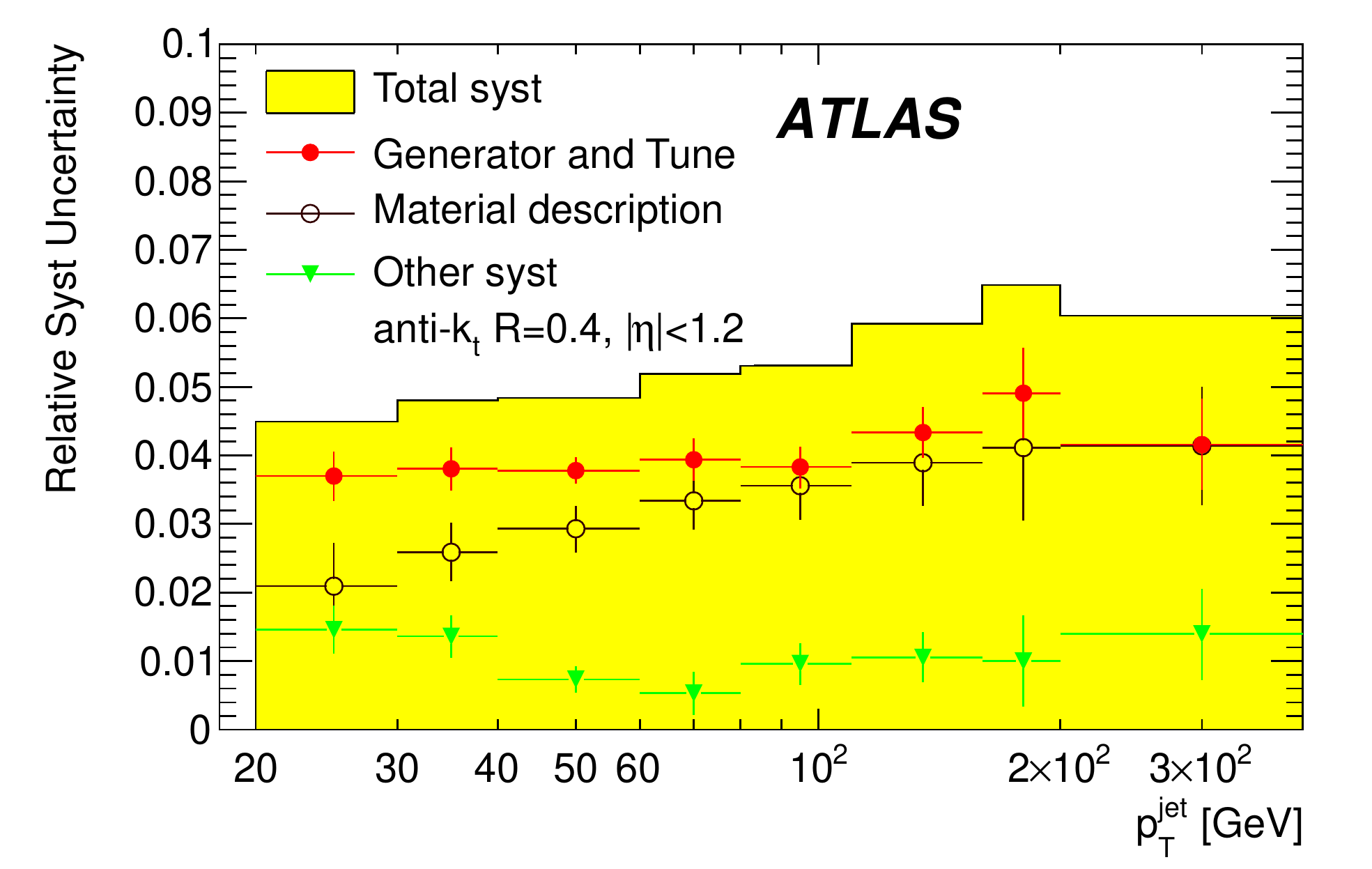} 
\label{fig:tracksJESttbarLightSyst}}\\
\subfloat[]{
\includegraphics[width=0.45\textwidth]{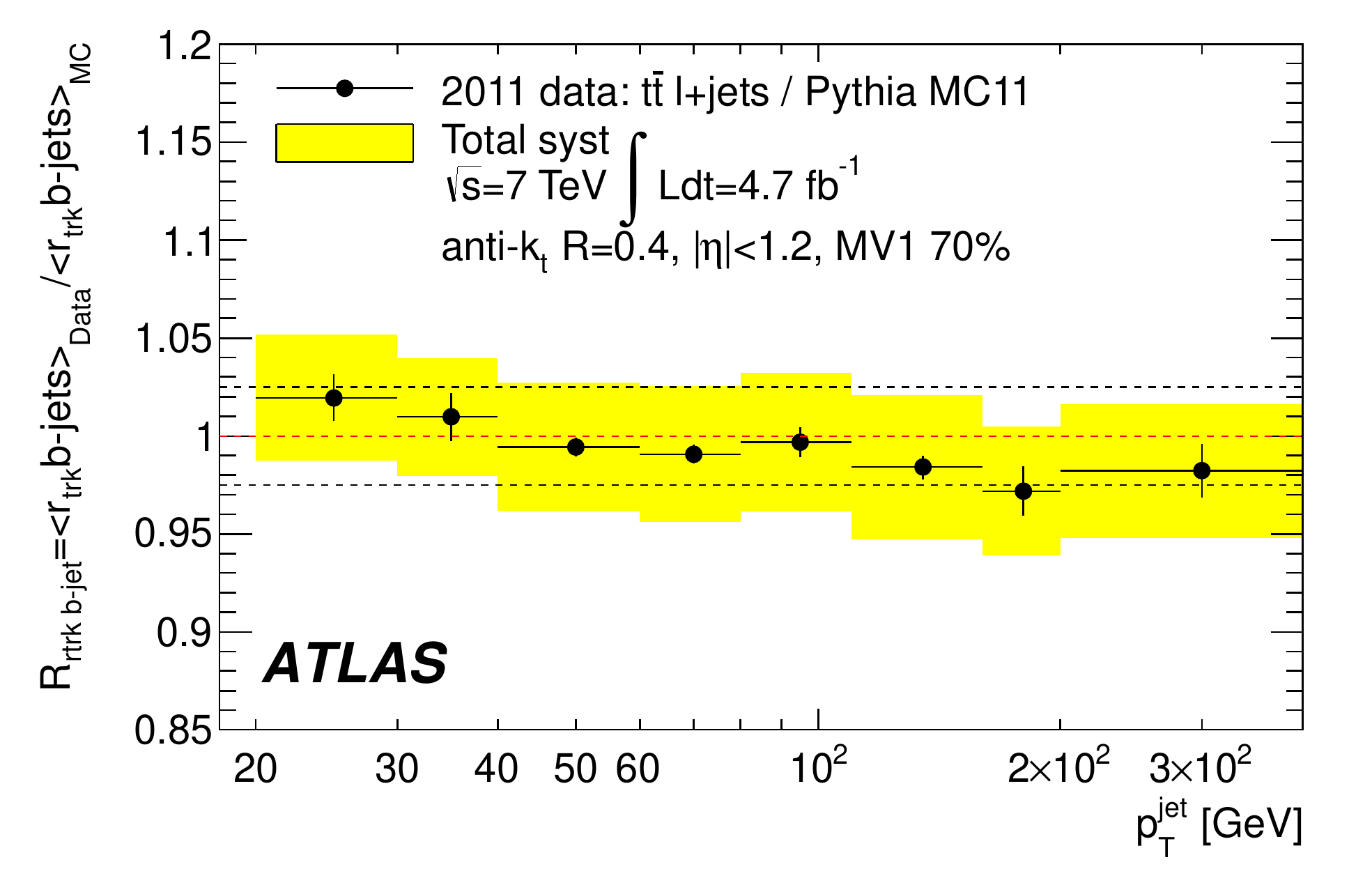} 
\label{fig:tracksJESttbarBtagged}}
\subfloat[]{
\includegraphics[width=0.45\textwidth]{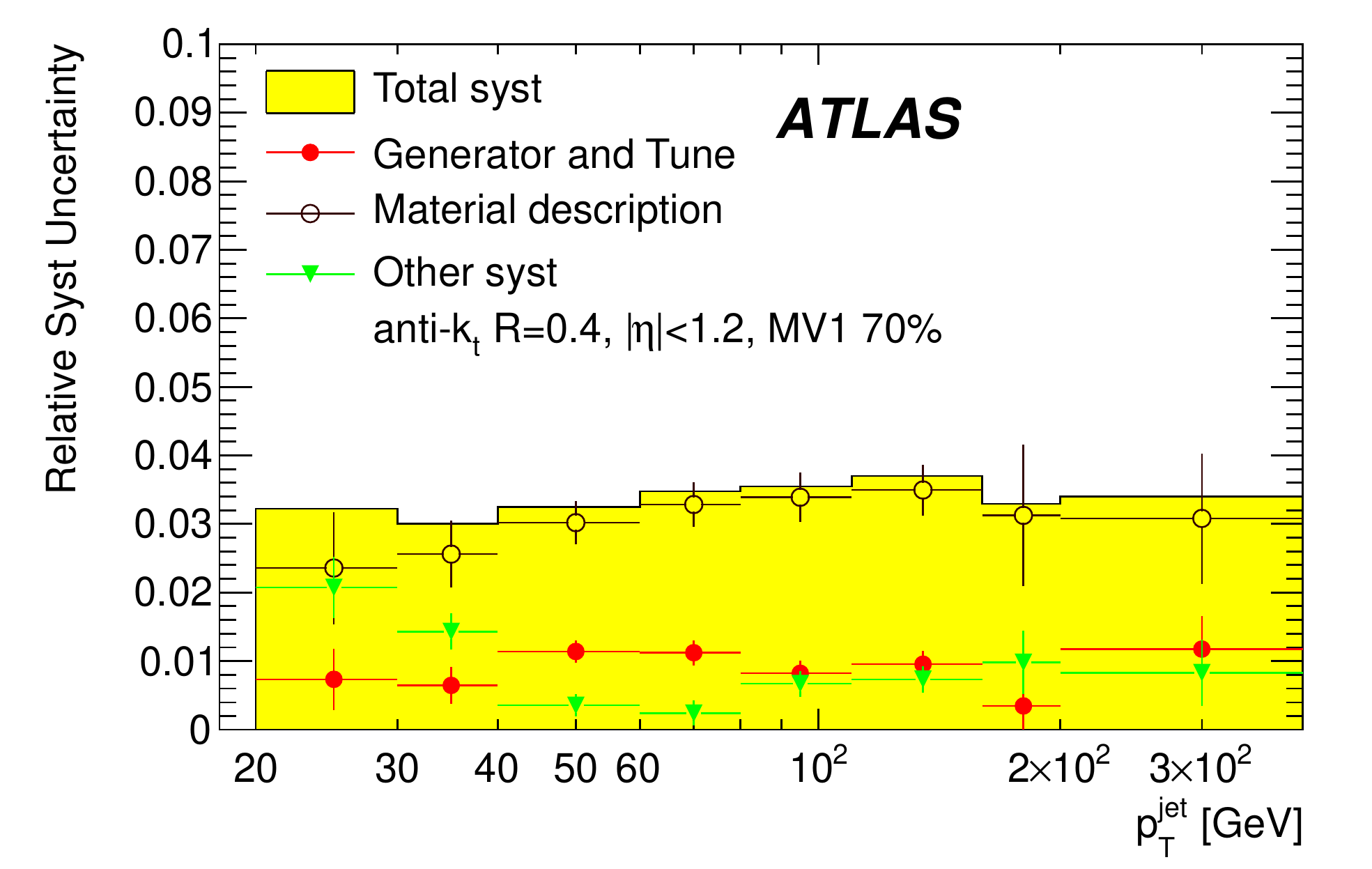} 
\label{fig:tracksJESttbarBtaggedSyst}}\\
\subfloat[]{
\includegraphics[width=0.45\textwidth]{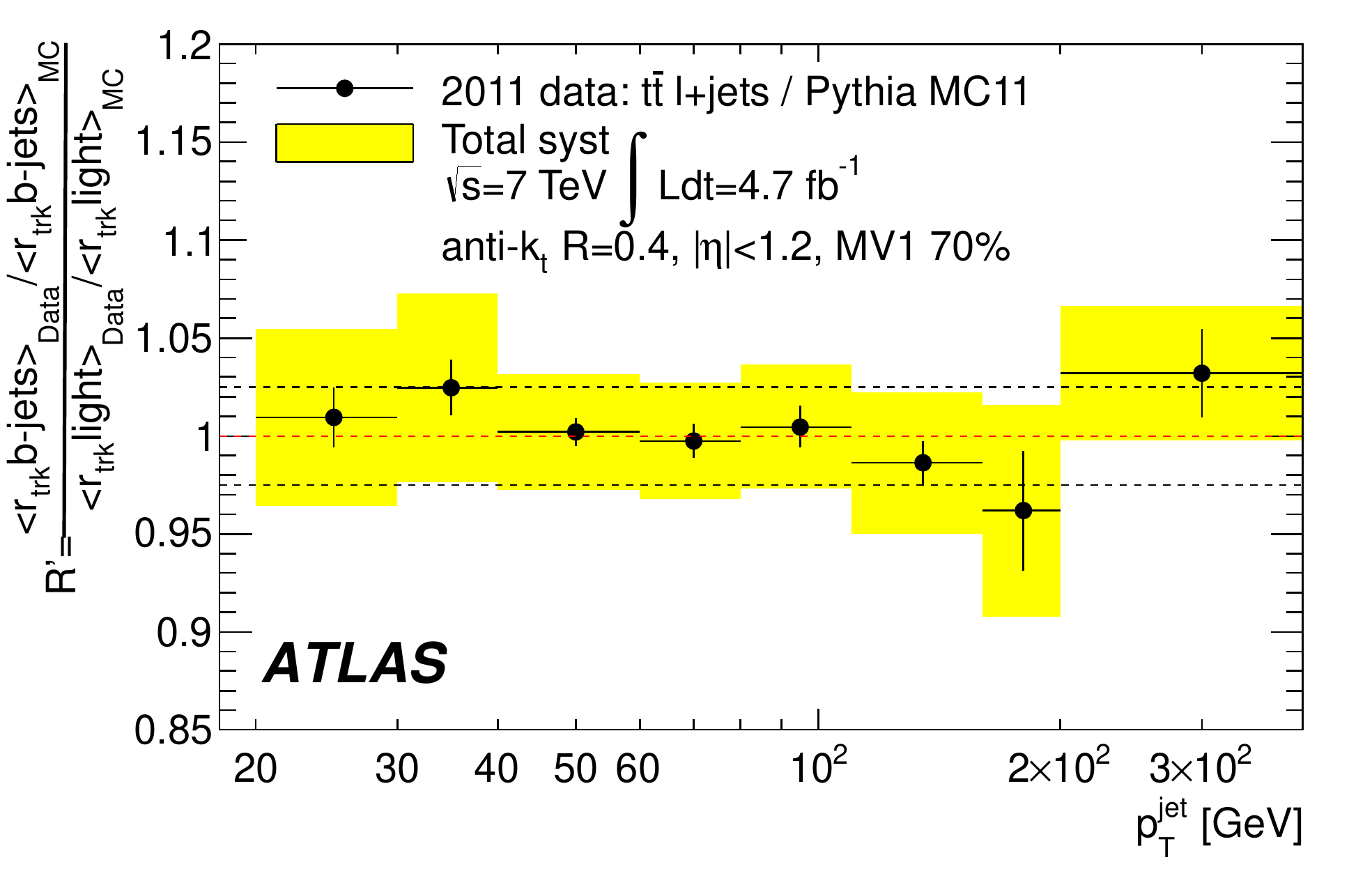} 
\label{fig:tracksJESttbarRatio}}
\subfloat[]{
\includegraphics[width=0.45\textwidth]{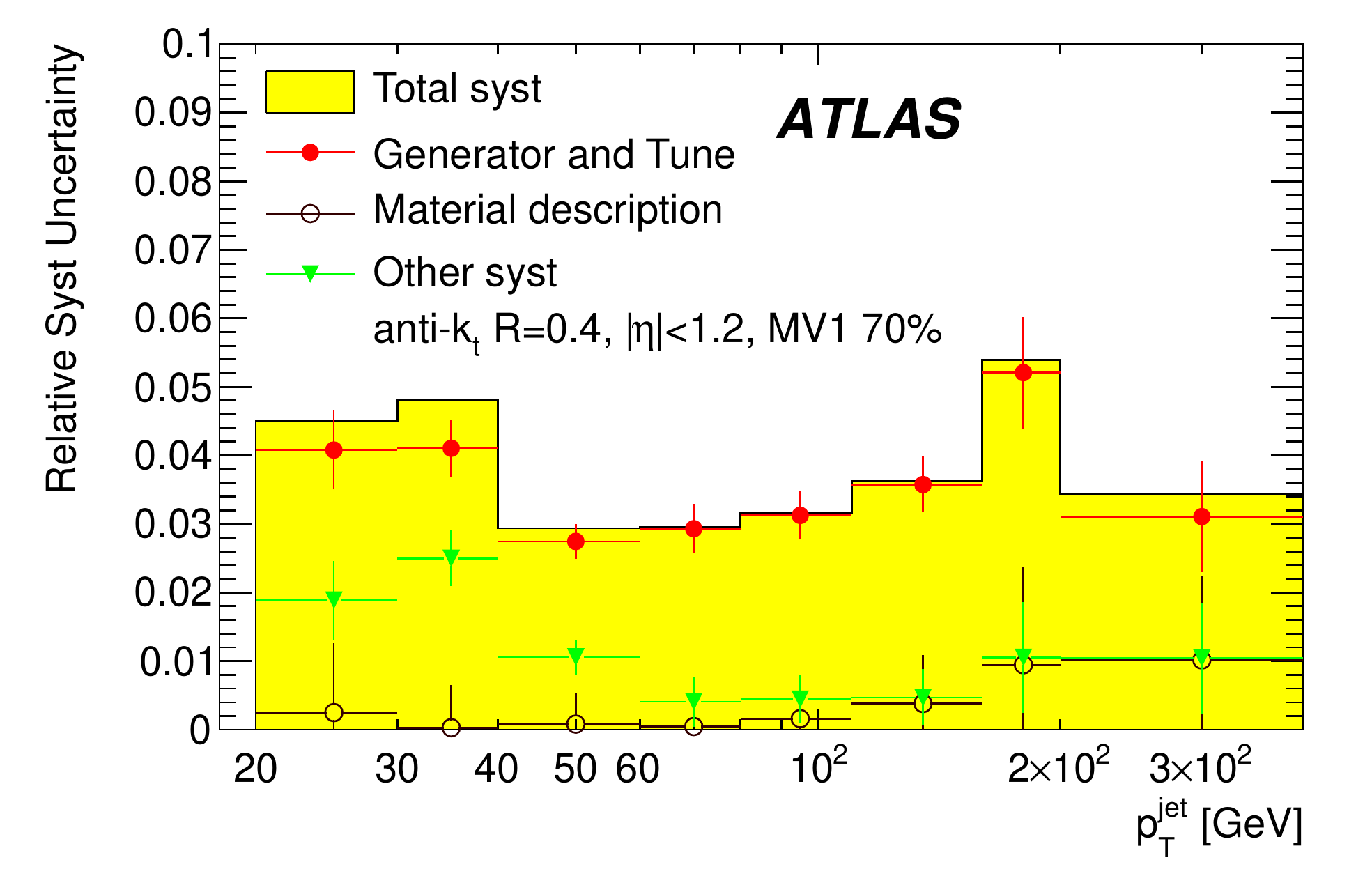} 
\label{fig:tracksJESttbarRatioSyst}}

\caption[]{Ratio of the average $\rtrk$ given in \eqRef{eq:def-rtrk} in $\ttbar$ events in data and \MC{}
  simulations for \subref{fig:tracksJESttbarLight} light-jets and tagged \subref{fig:tracksJESttbarBtagged} \bjets{}. In \subref{fig:tracksJESttbarRatio}, the ratio of
  $R_{\rtrk}$ from \eqRef{eq:def-doublerat} between the \bjet{} and the light-jet sample is shown. The
  total systematic uncertainty is shown as a band, and the dotted lines
  correspond to unity and the $2.5\%$ deviation from unity. The contributions of the systematic 
  uncertainties to the total uncertainty in the different measurements are shown in \subref{fig:tracksJESttbarLightSyst}, \subref{fig:tracksJESttbarBtaggedSyst}, and \subref{fig:tracksJESttbarRatioSyst}, respectively. 
  The
  contributions to the total systematic uncertainty due to the jet
  resolution, $b$-tagging calibration, background contamination and the
  modelling of the initial- and final-state radiation are grouped
  under ``Other systematics". Jets with $|\etajet|<1.2$ are used. }
\label{fig:tracksJESttbar}
\end{center}
\end{figure*}

\subsection{Calorimeter jet energy measurement validation using tracks}
\label{sec:dataValid}
\begin{figure*}[htbp]
\begin{center}
\subfloat[]{
\includegraphics[width=0.45\textwidth]{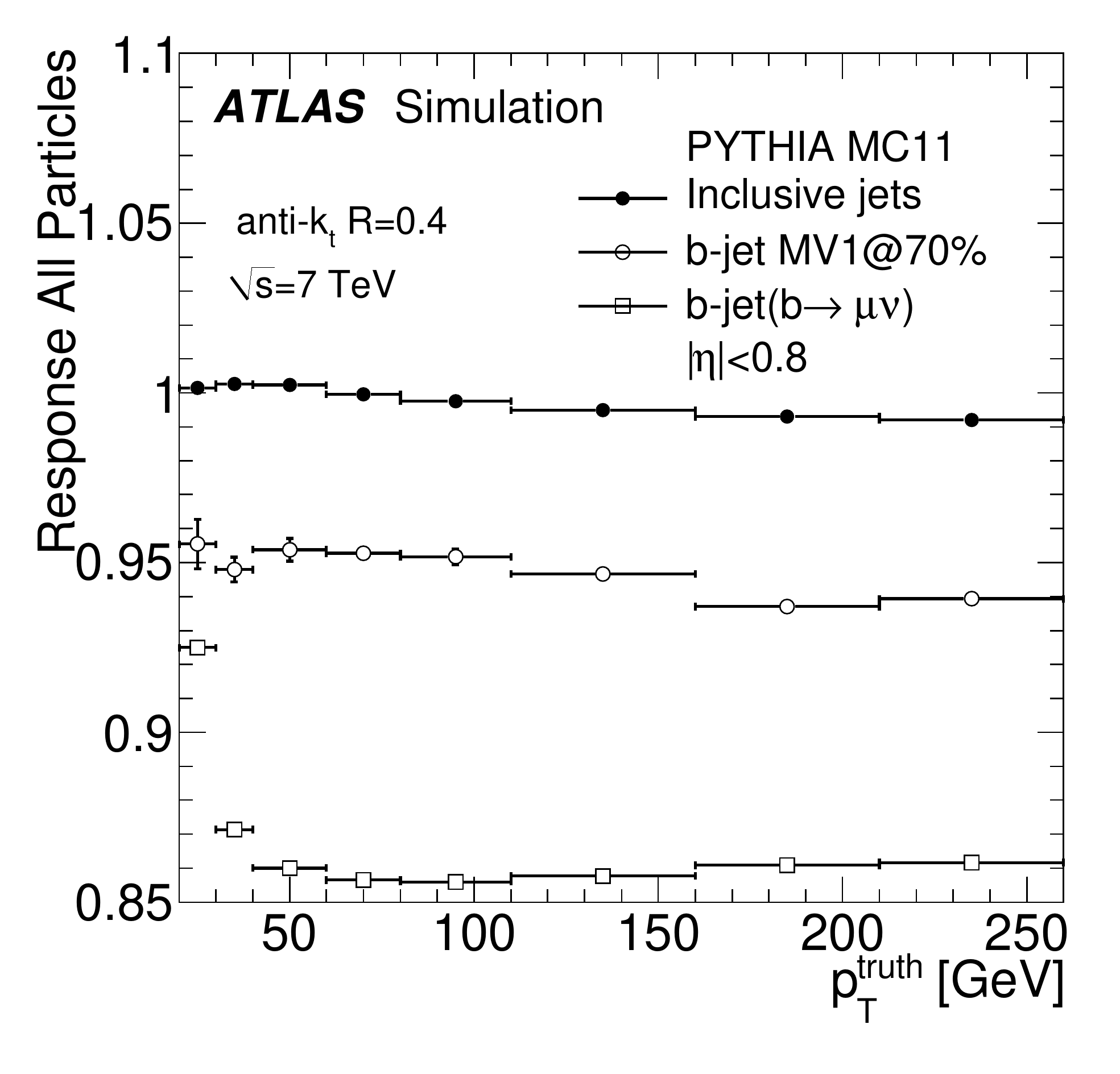} 
\label{fig:semileptCorrResp}}
\subfloat[]{
\includegraphics[width=0.45\textwidth]{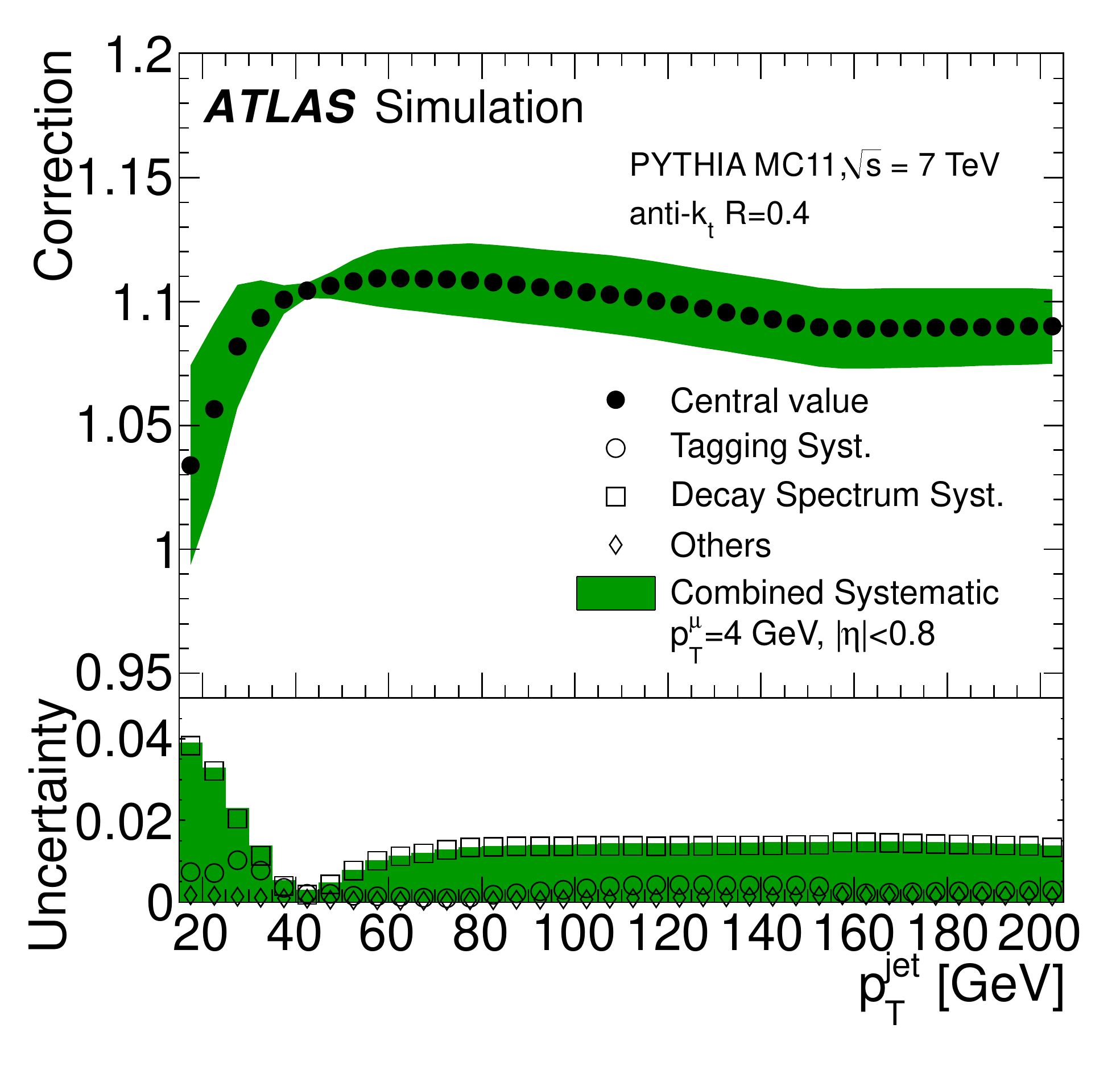} 
\label{fig:semileptCorrSyst}}
\caption[]{Average jet response as a function of true transverse momentum of jets
built using all stable particles, for a sample of inclusive jets (solid circles), a sample of \bjets{}
tagged with the \mvone{} tagging algorithm (open circles) and a sample of semileptonically 
decaying \bjets{} with a reconstructed muon inside (open squares), is shown in \subref{fig:semileptCorrResp}. The resulting
semileptonic correction, as a function of calorimeter jet $\pt$, used to transform the $\pt$ of a jet in 
the semileptonic sample to the $\pt$ of a jet in an inclusive sample of \bjets, is displayed in \subref{fig:semileptCorrSyst}. Associated 
systematic uncertainties are shown around the central value, and the combined uncertainty 
is shown as a coloured band.}
\label{fig:semileptCorr}
\end{center}
\end{figure*}

The calorimeter jet energy scale can be probed by comparing the
measured jet energy to that of a well-calibrated reference object with independent systematic uncertainties.
Charged-par\-ti\-cle tracks are well measured with uncertainties independent
of the calorimeter, and can be associated with jets, are used here. 
The mean value of \rtrk, defined in \eqRef{eq:def-rtrk} is primarily sensitive to the
particle composition of the jet and thus should be well described by any
well-tuned event generator. In computing $\langle\rtrk\rangle$ it is
important to truncate the \rtrk{} distribution (here with $\rtrk<3$) to avoid contributions from fake tracks with unphysically large \pt.

To verify the description of the
calorimeter energy measurement in \MC{} simulations, the double ratio of the
charged-to-total momentum obtained in data to that obtained in Monte
Carlo simulation is studied:
\begin{equation}
  R_{\rm \rtrk} \equiv \frac{\langle{\rtrk}\rangle_{\rm Data} }{\langle {\rtrk}\rangle_{\rm MC}}.
  \label{eq:def-doublerat}
\end{equation}
\index{$R_{\rtrk}$}
The ratio is evaluated for inclusive jets ($R_{\rm \rtrk,
  inclusive}$), $b$-tagged jets ($R_{{\rm \rtrk,}\bjet}$) and
$b$-tagged jets with a reconstructed muon inside ($R^{\mu\nu}_{\rtrk,\bjet}$, in the dijet sample
only).
The calorimeter response ratio $R'$ of $b$-tagged jets relative
to inclusive jets is then defined using \eqRef{eq:def-doublerat} from each respective sample, 
\begin{equation}
  R' \equiv \frac{R_{\rtrk,\bjet}}{R_{\rm \rtrk, inclusive}} . 
  \label{eq:bjet-rprime}
\end{equation}
This ratio is used to test the relative systematic uncertainty between
$b$-tagged and inclusive jets. In the \ttbar{} sample, where the fraction of
\bjets{} is large ($\approx 50\%$), the light jets (non \btagged)
component is used in the denominator instead of the inclusive one.  It is mainly comprised of 
jets from the \Wboson{} boson decay but also to a lesser 
extent of gluon jets from initial- and final-state radiation. 
As a consequence, when comparing the results obtained in the \ttbar{} and the dijet analyses, the difference in terms of jet flavour
components entering the calculation of $R_{\rm \rtrk, inclusive}$ needs to be taken into consideration.

\subsection{Systematic uncertainties}
\label{sec:bjetSystUncert}

Systematic uncertainties in the \rtrk\ measurement arise from the modelling of the jet (and \bjet) fragmentation, $b$-tagging
calibration, jet resolution and track reconstruction efficiency. In addition, for
high-\pt{} jets ($\pt>500$ \GeV) an efficiency loss in the tracking
in the jet core is observed in \MC{} simulations, and a
systematic uncertainty is added to account for potential mis-modelling
of this effect.  These uncertainties are assumed to be uncorrelated.
The resulting fractional systematic uncertainties on $\rtrk$ and $R'$
are shown in Figs. \ref{fig:tracksJESdijet}\subref{fig:tracksJESdijetInclSyst}, \ref{fig:tracksJESdijet}\subref{fig:tracksJESdijetBtaggedSyst}, and \ref{fig:tracksJESdijet}\subref{fig:tracksJESdijetRatioSyst} for the inclusive  jet sample, and in Figs. \ref{fig:tracksJESttbar}\subref{fig:tracksJESttbarLightSyst}, \ref{fig:tracksJESttbar}\subref{fig:tracksJESttbarBtaggedSyst}, and \ref{fig:tracksJESttbar}\subref{fig:tracksJESttbarRatioSyst} for the  \ttbar{} sample.  
They are determined as
follows.
\begin{mylist}
\myitem {\MC{} generator and tunes} These systematic uncertainties
  capture the effects of differences in $\pt^{\rm track}$ caused by
  different fragmentation models. Differences in the calorimeter response, caused by
  the different particle spectra, can also impact the $\rtrk$ measurement in
  certain \MC{} simulations and should not be part of the uncertainty, 
  since such shifts are measurable in the data.  
  The \rtrk{} distribution is, thus, 
  calculated from the various samples described in \secRef{sec:MC} 
  using $\pt^{\rm truth}$ in the
  denominator, even though only small differences are observed in most samples
  when including calorimeter effects, i.e. using the jet \pT{} reconstructed with the calorimeters ($\pt^{\rm calo}$).

  In the top pair analysis, differences between \mcatnlo{} and \powheg+\herwig{} are
  considered as process or generator systematic uncertainties. Fragmentation and 
  decay systematic uncertainties are evaluated 
  taking the difference between \pythia{} and \herwig{}. In the dijet analysis,
  differences between \pythia{} and \herwigpp{} set the systematic uncertainties
  from uncertainties in the decay models. The updated fragmentation tune in \herwigpp{}
  prevents this comparison from being a conservative measure of the \bjet{} fragmentation
  systematic uncertainties. These are evaluated using comparisons 
  to the Bowler--Lund \cite{Bowler:1981sb} and Professor tunes \cite{Buckley:2009bj}.  
%
\myitem {\BTAGGING{} calibration} The scale factors that correct the
$b$-tagging efficiencies in \MC{} simulations to match the measured values are varied 
within their total uncertainty.  
\myitem {Material description} The knowledge of the tracking
  efficiency modelling in \MC{} simulations is evaluated in
  detail in Ref.~\cite{MinBias2}.  The systematic uncertainty on the
  tracking efficiency for isolated tracks increases from $2 \%$
  ($|\etatrk| < 1.3$) to $7 \%$ ($2.3 \le |\etatrk| <2.5$) for tracks
  with $\pt > 500$ \MeV.  The resulting effect on \rtrk{} is about $3 \%$ for $0 \leq |\etajet\,| < 2.1$ 
  and about $4\%$ for $2.1 \leq|\etajet\,| < 2.5$.
\myitem{Tracking in jet core} High track densities in the jet core
influence the tracking efficiency due to shared hits between tracks,
fake tracks and lost tracks.  The number of shared hits is
well described in the \MC{} simulation. The \pt{} carried by fake
tracks is negligible. A relative systematic uncertainty of $50 \%$ on the loss of
efficiency obtained in the simulation is assigned to account for potential
mis-modelling of this effect. 
\myitem {Jet energy resolution} The jet energy resolution in \MC{} 
  simulations is degraded by about $10 \%$. 

\myitem{Background contamination} For the $\ttbar$ sample the analysis
  is repeated including the expected background contamination (except
  the multijet contribution) and the full difference is taken as
  an estimate of the systematic uncertainty.
\end{mylist}

The dominant contributions to the systematic uncertainty in the
$\ttbar$ analysis are due to variations in the detector material and
fragmentation/decay models.  In the dijet sample, the material,
fragmentation and decay uncertainties also dominate the systematic
uncertainties, except at $\pt\gtrsim 500$ \GeV{} where the uncertainty
caused by the loss of efficiency in the jet core dominates.  In
\figRef{fig:tracksJESttbar}, the contributions to the total
systematic uncertainty due to the jet resolution, $b$-tagging
calibration, background contamination and due to the modelling of the
initial- and final-state radiation are labelled as ``other'' systematic
uncertainties.

For $R'$, the tracking components (the material description, impacting
the tracking efficiency) of the systematic uncertainty entering both
the numerator and denominator are correlated and thus approximately
cancel.  A similar consideration holds for the jet energy resolution.
The most significant systematic uncertainties on $R'$ are due to the
choice of the \MC{} generator and the fragmentation and decay models.

\subsection{Results}\label{sec:rtrk}
Figures \ref{fig:tracksJESdijet}\subref{fig:tracksJESdijetIncl} and  \ref{fig:tracksJESdijet}\subref{fig:tracksJESdijetBtagged} show the ratio of the average of the $\rtrk$ distribution in 
data and \MC{} simulations for jets in the inclusive jet sample sample with $|\etajet|< 1.2$.  Figures \ref{fig:tracksJESdijet}\subref{fig:tracksJESdijetInclSyst} and \ref{fig:tracksJESdijet}\subref{fig:tracksJESdijetBtaggedSyst} show the different components of the associated systematic uncertainty, as discussed in \secRef{sec:bjetSystUncert}.

The study in the sample without $b$-tagging 
covers up to approximately $2$~\TeV, and provides a cross
check over
almost the full range of calibrated \pt{} studied \insitu{} through the analyses used
to establish the systematic uncertainty on the jet energy scale in 
\ATLAS. No \pt{} dependence is 
observed and agreement is found between data and \MC{} simulations 
within systematic uncertainties. Similar results are found in higher $|\etajet|$ regions. 

Agreement of the \MC{} simulations with the data 
for the \rtrk\ measurements is found within systematic uncertainties across all \pt{} for inclusive jets 
and for $\ptjet < 400$ \GeV for \btagged{} jets.
The relative response $R'$ between $b$-tagged  and inclusive
jets is shown in \figRef{fig:tracksJESdijet}\subref{fig:tracksJESdijetRatio}  
and the uncertainty band corresponds to the relative \bjet{} energy scale uncertainty
with respect to the inclusive jet sample. \FigRef{fig:tracksJESdijet}\subref{fig:tracksJESdijetRatioSyst} 
shows the different components of the associated systematic uncertainty. 
A difference between data and \MC{} simulations is found but almost covered by
the systematic uncertainties. This difference is partially caused by the overall $1\%$ shift
found in the inclusive sample. 
Similar results are found in the sample of \bjets{} decaying to muons selected
in the dijet sample, with a larger difference between data and \MC{} simulations of up to
$4\%$ in the lowest \pt{} bin probed. 
However, the uncertainties in the modelling are also somewhat larger, limiting 
the constraints on the jet energy scale of these jets to approximately $3\%$. 

The corresponding results from the same study performed in the $\ttbar$ sample
are shown in \figRef{fig:tracksJESttbar}. 

The results in this sample are consistent with those obtained in the dijet sample, except for the
better agreement between data and \MC{} simulations in the light-jet sample, which also
leads to better agreement in the \bjet{} to light-jet sample results. 
The systematic uncertainties are also comparable, despite the different methods used in their
evaluation. 
The uncertainty in the \insitu{} technique used to assess the  \bjet{} energy scale 
is estimated to be approximately $2.5\%$ and $3\%$ in the ranges 
$|\etajet\,| < 1.2$ and $1.2 \leq |\etajet\,| < 2.5$, respectively, 
for jets with $\ptjet < 400$ \GeV{} from these studies.

\subsection{Semileptonic correction and associated uncertainties}
\label{sec:semilCorrMC}
The study of the all-particle response $\Response^{\rm all}$ of \bjets{}, 
i.e. the energy scale calculated with respect to jets built using all stable particles, is also necessary
for many analyses, given that about $40 \%$ of \bjets{} decay
semileptonically, thus having a non-negligible amount of their energy
carried by neutrinos. 
In particular, the study of the $b$-tagging efficiency in a sample of
\bjets{} decaying semileptonically to muons \cite{mv1} requires a
correction that maps the all-particle jet energy scale of that sample
to that of an inclusive sample of \bjets{}.
This correction and its systematic uncertainties are estimated in this section. 
The correction also has applications beyond the $b$-tagging calibration
since it can also be used to improve the reconstruction of \bjets{} identified
as semileptonic. The study of the all-particle energy scale in this
section is performed independently of the study of the calorimeter
energy scale, even though the two are not straightforward to decouple
in \insitu{} studies.

\FigRef{fig:semileptCorr}\subref{fig:semileptCorrResp} shows the all-particle response for an
inclusive jet sample, a sample of \bjets{} tagged with the \mvone{}
algorithm and a sample of \bjets{} containing a muon from a semileptonic $b$ decay. 
The semileptonic \bjets{} sample is selected using hadron-level
information, and no $b$-tagging is imposed. However, the muon is
required to pass kinematic and quality cuts detailed in
Ref. \cite{mv1}. The effect of neutrinos is clearly visible in
both the tagged \bjets{} sample and more significantly in the
semileptonic \bjets{} sample. 
The increase at low $\pt$ in the semileptonic sample
arises from biases created by the muon kinematic cuts.

The response of semileptonically decaying \bjets{} is corrected to
that of an inclusive $b$-tagged jet sample. The correction is
constructed using techniques similar to those used in the \EMJES{}
calibration introduced in
\secRef{sec:jetrecocalib}. This correction is shown in
\figRef{fig:semileptCorr}\subref{fig:semileptCorrSyst}, as a function of calibrated jet $\pt$
for fixed muon $\pt$ and jets with $|\etajet|<0.8$. The correction is
not explicitly dependent on $\pt^{\mu}$ even though it enters in the
calculation of the reconstructed jet $\pt$ used to compute the
correction.

Systematic uncertainties in this correction need to account for our
knowledge of \bjet{} fragmentation and decay, as well as the effect of
the muon spectrum and muon reconstruction.  These uncertainties are presented
in Ref.~\cite{mv1}. Since only one
correction is calculated and used for all tagging algorithms and
operating points commissioned up to date, an additional systematic
uncertainty that covers the spread of the corrections for all these
different operating points is added. All uncertainties are combined in
quadrature. Only the most significant uncertainties are included in \figRef{fig:semileptCorr}\subref{fig:semileptCorrSyst}, namely the uncertainty that arises from the different
correction for different operating points, and the uncertainty that
arises from the limitations in the knowledge of the muon momentum
spectrum in the \cms{} of the decaying hadron. These
uncertainties are estimated by reweighting that spectrum to match a
measurement obtained in $e^+e^-$ scattering \cite{delphiMu}.  Due to
the significant differences between that spectrum and the one found in
\pythia, these variations are considered sufficient.  All other
uncertainties are combined and shown in the figure under the same
curve.

The uncertainty is about $1.5\%$ for most $\pt$ values in the central
region, except at low $\pt$ where it increases to about $4\%$. The
behaviour is similar at larger $\etajet$, except in the most forward bin
($2.1<|\eta|<2.5$), where variations across tagging operating points
cause the uncertainty to increase to about $2\%$.

\begin{figure}[htbp]
\begin{center}
\includegraphics[width=0.49\textwidth]{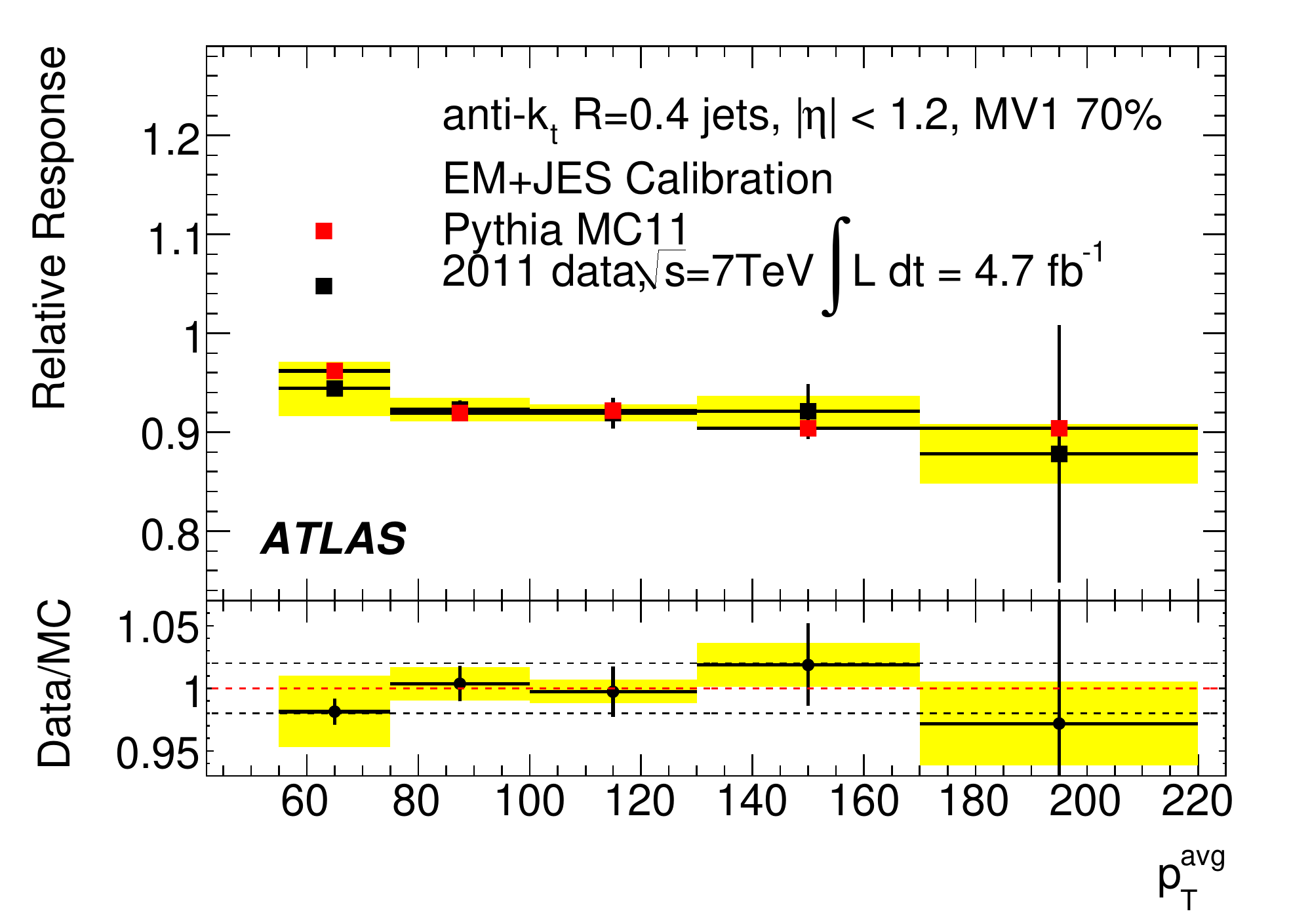} 
\caption{ Relative response of the semileptonic sample with respect to the inclusive \bjet{} 
sample as calculated from the dijet \pt{} asymmetry. The uncertainty band around the data  
denotes systematic uncertainties in the asymmetry measurement. }
\label{fig:neutrinoDijetBalance}
\end{center}
\end{figure}

\subsection{Semileptonic neutrino energy validation using dijet balance}
\label{sec:semilCorr}
The modelling of the energy carried by the neutrino in the inclusive \bjet{} sample and 
in the semileptonic \bjet{} sample can be validated
using the $\pt$ balance of a dijet system. The same technique is used in 
Ref. \cite{jespaper2010} to validate the variation of the calorimeter response as a function  
of different jet properties. The response in data is calculated using the asymmetry in the jet
$\pt$ of the two jets in the dijet system. The two jets are required to be \btagged, and the probe
jet is required to have a selected reconstructed muon within $\DeltaR < 0.4$. 
The relative response, calculated from the asymmetry, is sensitive to
the energy carried by the neutrino, but also to the response differences between the \btagged\  
and semileptonic \bjet{} samples. These differences, however, are 
well modelled in the \MC{} simulation, as shown in \secRef{sec:rtrk}. 

\FigRef{fig:neutrinoDijetBalance} shows the relative response of semileptonic \bjets{} 
with respect to inclusive \bjets{} obtained in data and \MC{} simulations using dijet balance.  

The presence of neutrinos in the \bjet{} decay causes the estimated relative response to
be below $1$. 
The uncertainty band around the 
data represents systematic uncertainties in the imbalance. These are calculated through 
variations in the soft-radiation cut in the selection (i.e. the \pt{} used for the 
veto on the third leading jet) as Section \ref{sec:etaintercaliUncertainty}. 
An additional contribution to the uncertainty is added to the first \pt{} bin to account 
for differences between data and \MC{} simulations in the turn-on of the efficiency 
curve for the muon--jet trigger used in this analysis. 
Agreement is found between data and \MC{} simulations, validating the description of
this process that is exploited to develop the semileptonic correction presented in the previous 
\secRef{sec:semilCorrMC}.

\subsection{Conclusions on heavy-flavour jets}
\label{sec:conclusions}

The uncertainty on the jet energy measurement is studied for light jets as well as 
inclusive and semileptonic \bjets. In the inclusive jet sample
the jet energy scale is probed using tracks associated with jets 
over a wide range of jet $\pt$. Comparisons between data and \MC{}
simulations show agreement within systematic uncertainties of approximately $3\%$
with weak dependence on the transverse momentum of the jets.  
The \bjet{} energy scale is also probed using tracks
associated with \btagged{} jets in the data. The results in the $\ttbar \to
\ljets$ and inclusive jet samples suggest that the jet energy scale of
\bjets{} is well described by the \MC{} simulation, within 
systematic uncertainties of about $2\%$ to $3 \%$. 

In the \MC{} simulation a correction for semileptonic \bjets{} 
decaying to muons is derived, 
which adjusts the transverse momentum measurement  
to that in an inclusive sample of \bjets. The systematic uncertainties on this 
correction are also derived using \MC{} simulations. They are found to be about $2\%$. 
The uncertainty in the jet energy measurement due to effects specific to \bjets{} 
is also determined using Monte
Carlo simulations. This uncertainty ranges from $1\%$ to $3\%$. 

The energy scale
of semileptonic \bjets{} decaying to muons is probed in the
dijet sample in parallel with a study of the energy carried by the accompanying neutrino. 
The latter confirms the results found in \MC{} simulations 
within systematic uncertainties of about $3\%$.

\section{Jet response in problematic calorimeter regions}
\label{sec:badjets}
%
\begin{figure}[h!!]
  \centering
  \includegraphics[width=0.9\linewidth]{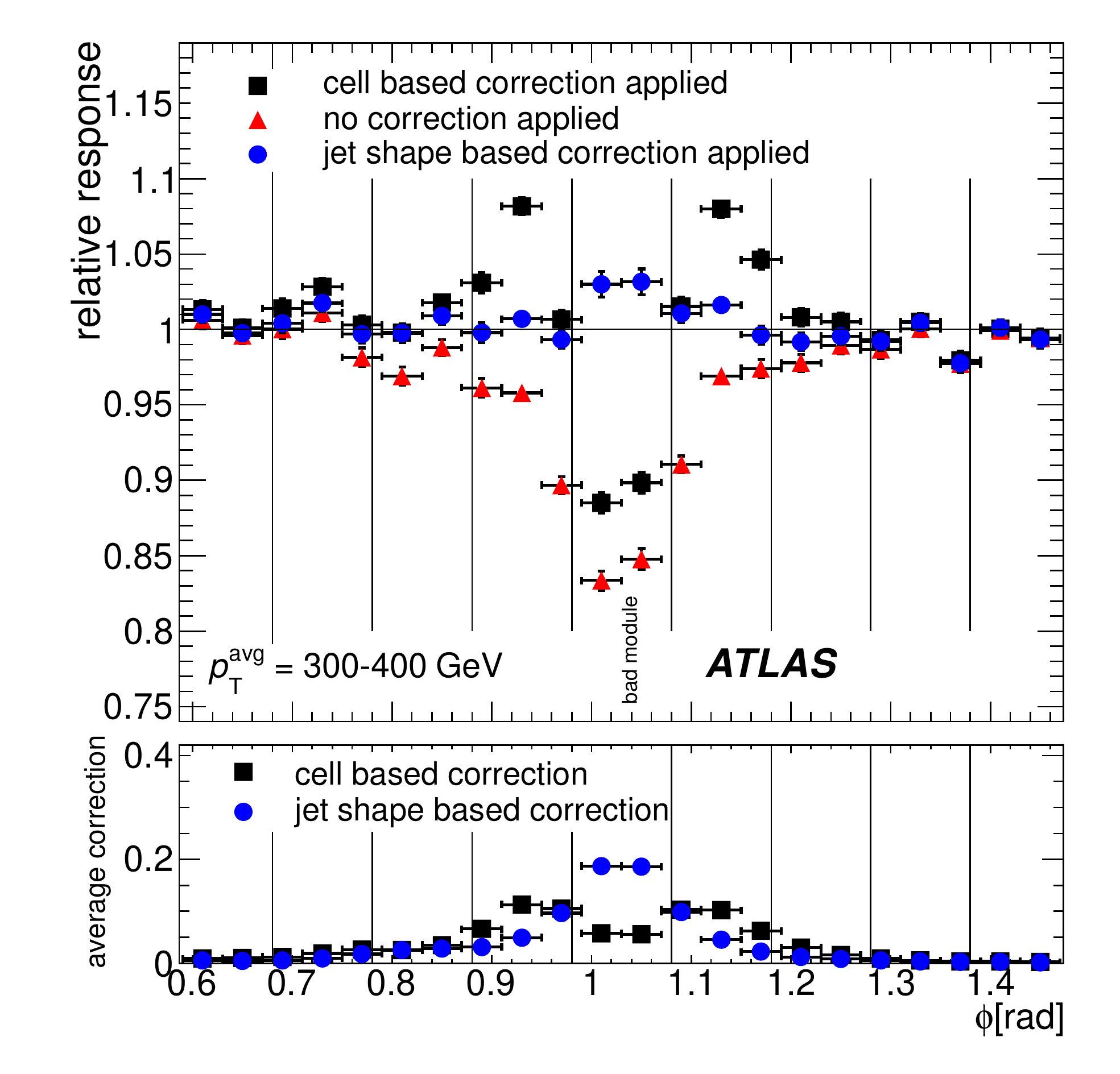}
  \caption{Average relative response of the probe jets with respect to the tag jets as a function of various 
           impact points in the azimuthal direction ($\phi$) of the probe jet.  The average \pt{} of the
           two leading jets is $300 \le \ptavg < 400$~\GeV.
           The vertical solid lines indicate the location of the \Tile{} calorimeter modules. 
           The non-operating \Tile{} calorimeter module is at $\phi= 1.03$. 
           The markers indicate the results for no correction (triangles),
           the cell-based corrections (squares) and the corrections based on the jet shape (circles). 
           The lower part of the figure shows the respective average values of the two 
           corrections as a function of the azimuthal angle of the probe jet.}
  \label{fig:badcalo_data_300_400}
\end{figure}

At the end of the 2011 data taking period $11$ out of $256$ modules of the \ATLAS{} central hadronic \Tile{} calorimeter were not operational.
Moreover, during the data taking, some \Tile{} calorimeter modules occasionally became non-operational for short
periods of time, e.g. due to trips of the high voltage. 
In this section the impact of non-operating \Tile{} modules on the jet energy measured is studied 
using a
tag-and-probe technique based in the \pt{} balance of the two leading jets in the event
following Sect. \ref{sec:etaintercalibrationstandardmethod}.  
The response of the tag jet, required to be in a fully operational part of the calorimeter,
is used to test the response of a probe jet that impinges 
close to and in the region of the non-operating \Tile{} module.  

The performance of two reconstruction algorithms that correct for non-operating parts of the calorimeters
based on the energy deposits in nearby cells or the average transverse jet shape is assessed.
\subsection{Correction algorithms for non-operating calorimeter modules}
\label{sec:dotc}
\subsubsection{Correction based on calorimeter cell energies}
\label{sec:bccc}
This correction is implemented in the standard \ATLAS{} calorimeter energy reconstruction.
It estimates the energy density of a non-operating \Tile{} calorimeter cell on the basis of energy measured 
by the two neighbouring cells that
belong to the same \Tile{} calorimeter layer sub-detector as the non-operating cell.
The energy density of the non-operating cell is estimated as the average (arithmetic mean) of the energy density 
of the neighbouring cells. 
This correction is called \BCHCORRCELL{} correction in the following.

\subsubsection{Corrections based on jet shapes}
\label{sec:bccj}
This correction is applied after jet reconstruction.
The expected average jet shape is used to estimate the energy deposited in the non-operating
\Tile{} calorimeter cells. The correction is derived from \MC{} simulations
where all calorimeter modules are operational.
It is calculated as a function of the transverse momentum and the pseudorapidity of the jet, 
the calorimeter type, the calorimeter layer and the angular distance between the jet axis 
and the cell centre in the \etaphispace{} (\DeltaR{} in \eqRef{eq:deltaRdet}  in Sect. \ref{sec:jetdirections}). 
It is applied for both \LAr{} and \Tile{} calorimeter cells and is called
\BCHCORRJET{} in the following.

In predefined bins of \ptjet{} and \etaDet{} and for all the calorimeter cells that belong to the jets 
the average relative energy (defined as $E_\mathrm{cell} / E_\mathrm{jet}$) in each calorimeter type, 
layer and $\mathrm{d}R$ bin is calculated. 
For all non-operational calorimeter cell in a jet the following correction is calculated:
\begin{displaymath}
\BCHCORRJET{} = \sum_{\mathrm{bad\ cells}} \frac{E_{\mathrm{cell}}}{E_\mathrm{jet}}
\end{displaymath}
and the energy of the jet is corrected with:
\begin{displaymath}
  E_{\mathrm{jet}}^{\mathrm{corrected}} = \frac{E_{\mathrm{jet}}^{\mathrm{uncorrected}}}{1 - \BCHCORRJET}.
\end{displaymath}

\subsection{Performance of the bad calorimeter region corrections}
\label{sec:r}
The performance of the correction methods can be assessed using a tag-and-probe technique in
events with two jets with high transverse momentum. The dependence of the relative
jet response between the tag and the probe jets is studied as a function of the
azimuthal angle of the probe jet.

The tag jet is selected such that it hits a fully operating part of the ATLAS calorimeter
and is inside a central $\eta$ region  
($|\etajet| < 1.6$). 
Jets in the gap between \Tile{} Long Barrel 
and \Tile{} Extended Barrel (i.e. jets with axes pointing to the region
$0.8 \le \etajet < 1.2$)
are excluded.
The probe jet is chosen such that its axis points to the vicinity of the non-operating Tile module. 
Only probe jets with $0.1 \le |\etajet| <0.8$ are used. 

\FigRef{fig:badcalo_data_300_400} shows the jet response of the probe jet in the region
of a missing \Tile{} module and in the neighbouring regions for events where
the average jet \pt{} of the two leading jets is between $300$ and $400$~\GeV.
A decrease of the probe jet response by about $15\%$ is observed in the region with the non-operating 
\Tile{} calorimeter module when no correction is applied. 
This reduces to only about $10\%$ for the cell-based correction. However, an overcorrection by about $10\%$
is observed in the vicinity of the region with the missing \Tile{} module.
The correction based on the jet shape performs much better. 
There is no overcorrection in the vicinity of the problematic module and the probe jet energy 
is compensated much better if the jet axis falls into the module. There is only a small
overcorrection by a few percent in the vicinity of the non-operating module.

\subsubsection{Conclusion on bad calorimeter regions}
\label{sec:badcalorimeterregionconclusions}

The corrections for missing \Tile{} calorimeter modules show a good performance.
The average jet response variations close to the missing calorimeter
are evaluated with a tag-and-probe technique in data.
The jet response variation is about $5$-$10$\%. The correction using jet shape
information shows a better performance than the correction simply averaging
the energy deposition in the neighbouring calorimeter cells.

The Monte Carlo simulation includes the missing \Tile{} calorimeter
modules and describes the jet response variations in data. 
The remaining differences are included in the \JES{} uncertainty
derived from the \insitu{} techniques.

\begin{figure*}[ht!!!!]
  \centering
  \subfloat[\etajet = $0.5$]    {\includegraphics[width=0.49\textwidth]{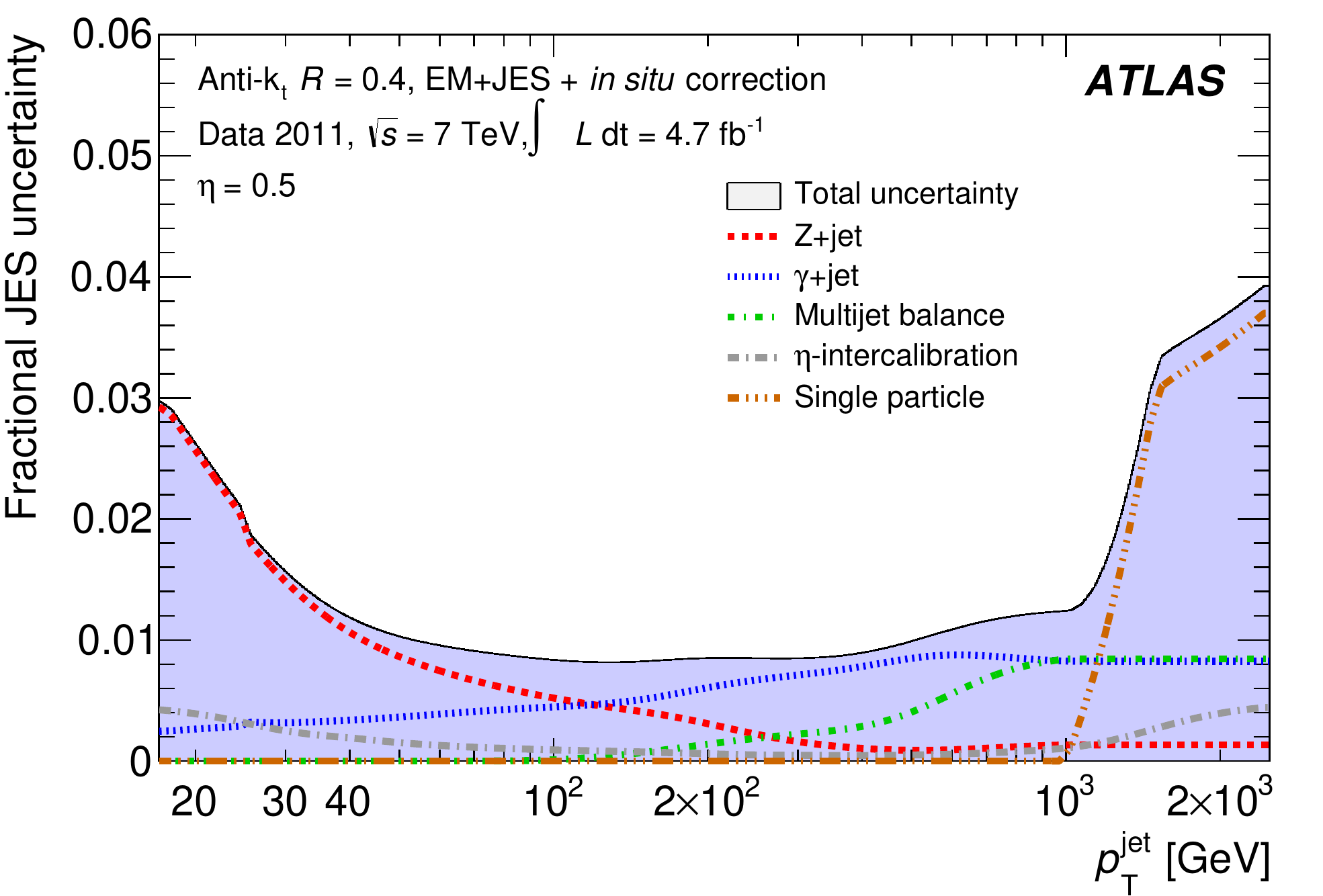}\label{fig:InsituJES_EMJES_0}}
  \subfloat[\etajet = $2.0$]    {\includegraphics[width=0.49\textwidth]{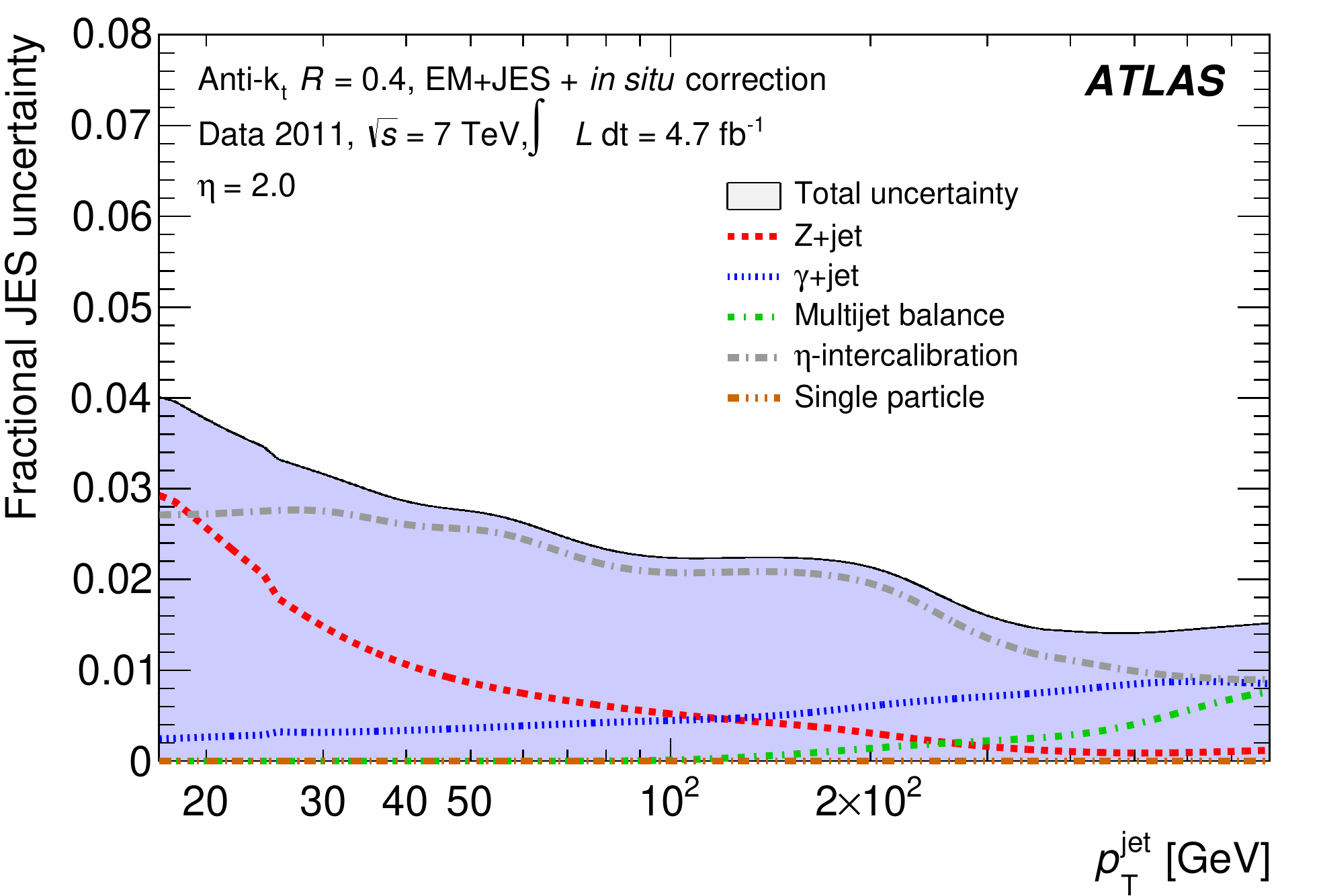}\label{fig:InsituJES_EMJES_1}}\\
  \subfloat[\ptjet = $25$ \GeV] {\includegraphics[width=0.49\textwidth]{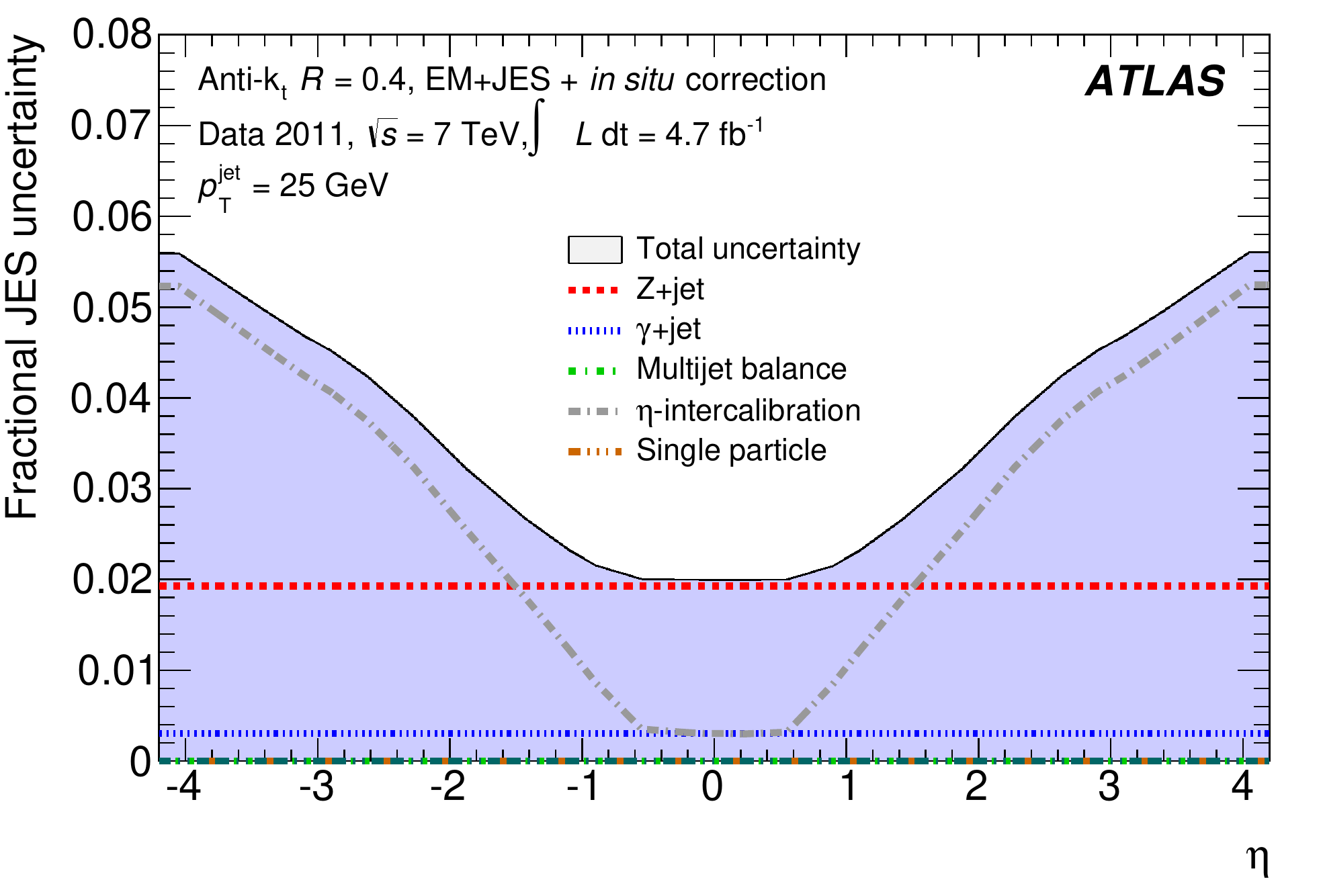}\label{fig:InsituJES_EMJES_2}}
  \subfloat[\ptjet = $300$ \GeV]{\includegraphics[width=0.49\textwidth]{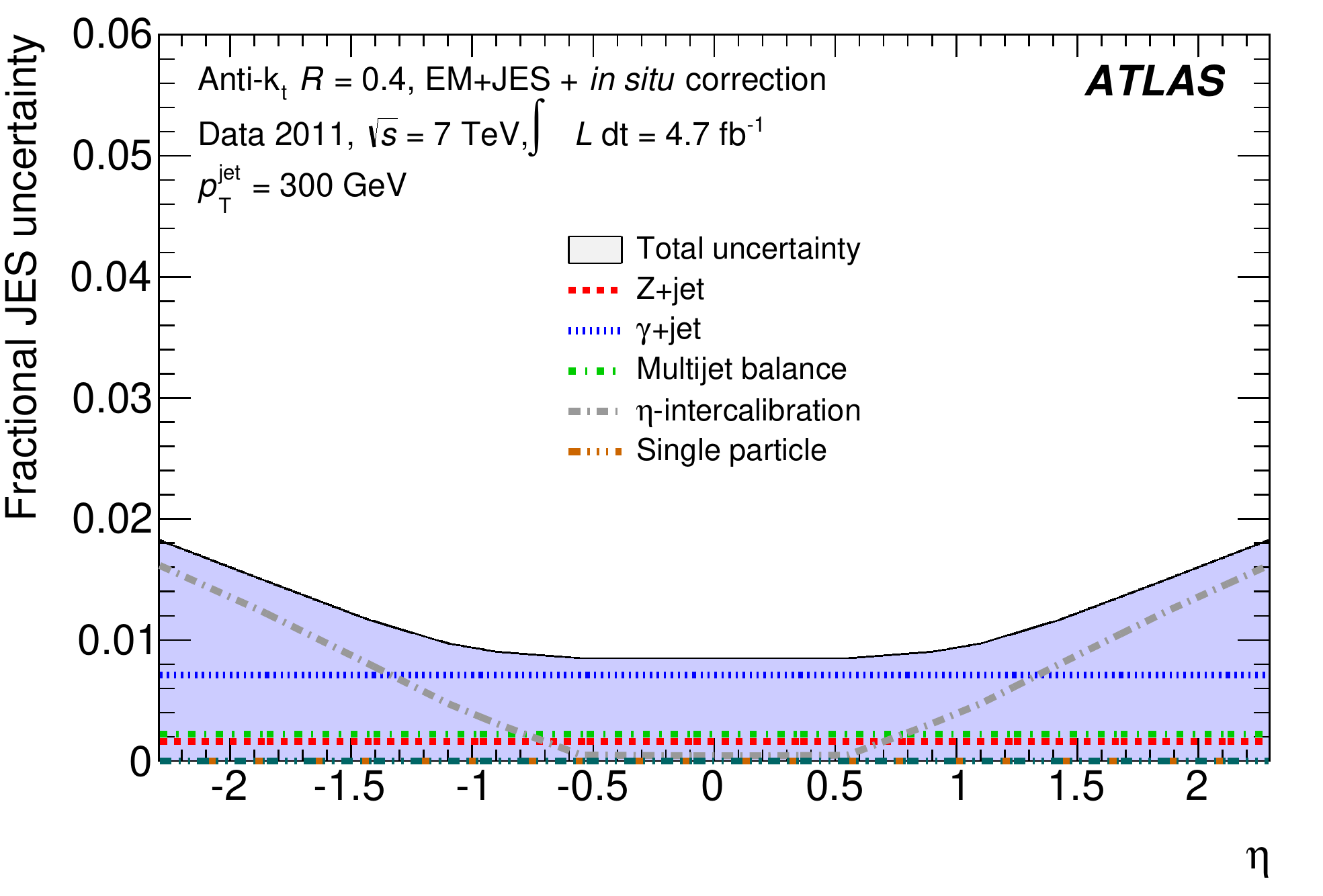}\label{fig:InsituJES_EMJES_3}}
  \caption[]{
    Fractional \insitu{} jet energy scale systematic uncertainty as a function of  (\subref{fig:InsituJES_EMJES_0}, \subref{fig:InsituJES_EMJES_1}) \ptjet{} and     
    (\subref{fig:InsituJES_EMJES_2}, \subref{fig:InsituJES_EMJES_3}) jet pseudorapidity    
    for \antikt{} jets with distance parameter of $R=0.4$ calibrated using the \EMJES{} calibration scheme. 
    The contributions from each \insitu{} method are shown separately.
    Uncertainties from pile-up, flavour, and topology are not included.
    \vspace{7.cm}
    \label{fig:InsituJES_EMJES}
  }
\end{figure*}

\begin{figure*}[ht!]
  \centering
  \subfloat[\etajet = $0.5$]   {\includegraphics[width=0.49\textwidth]{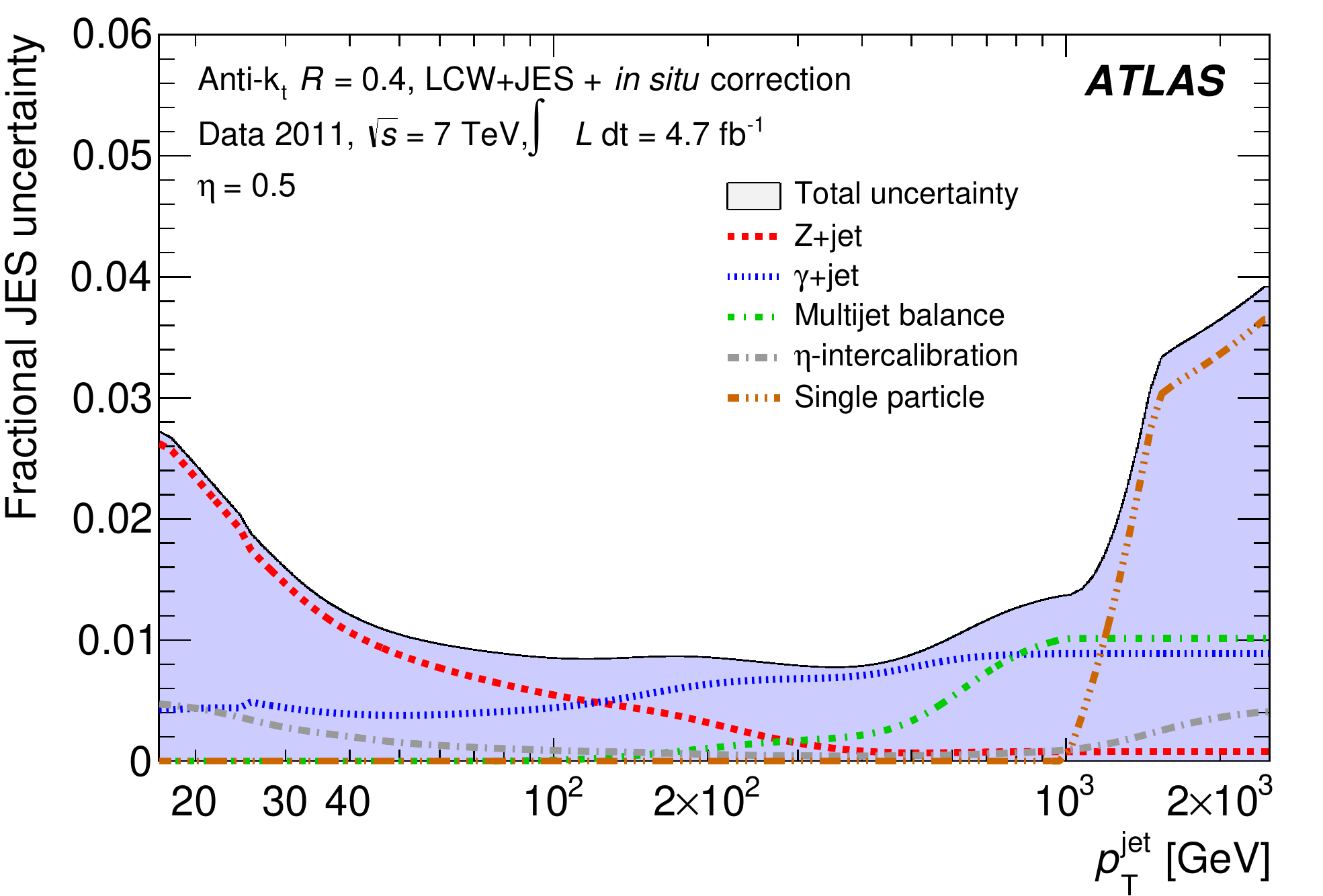}\label{fig:InsituJES_LCJES_0}}
  \subfloat[\etajet = $2.0$]   {\includegraphics[width=0.49\textwidth]{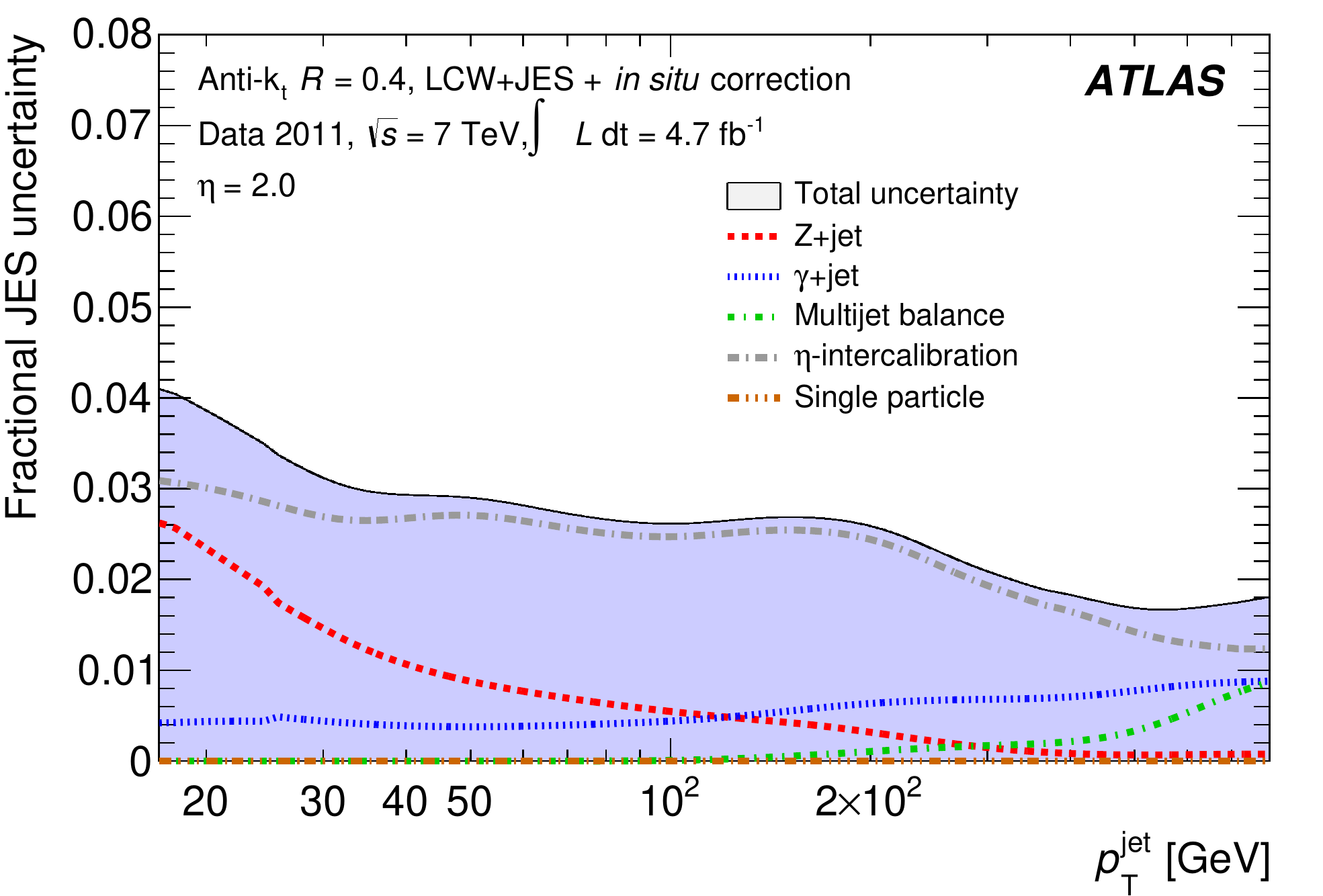}\label{fig:InsituJES_LCJES_1}}\\
  \subfloat[\ptjet = $25$ \GeV]{\includegraphics[width=0.49\textwidth]{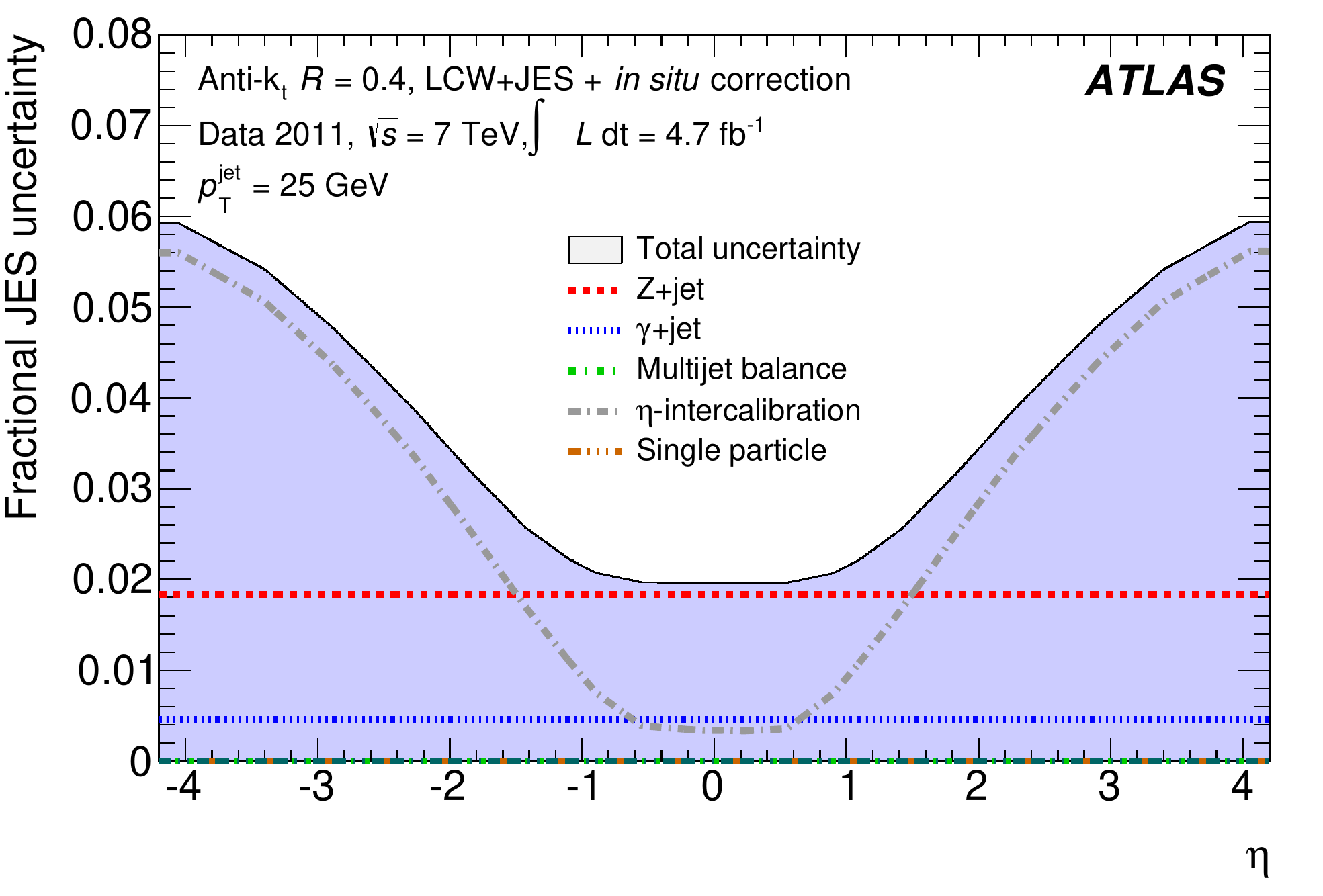}\label{fig:InsituJES_LCJES_2}}
  \subfloat[\ptjet = $300$ \GeV]{\includegraphics[width=0.49\textwidth]{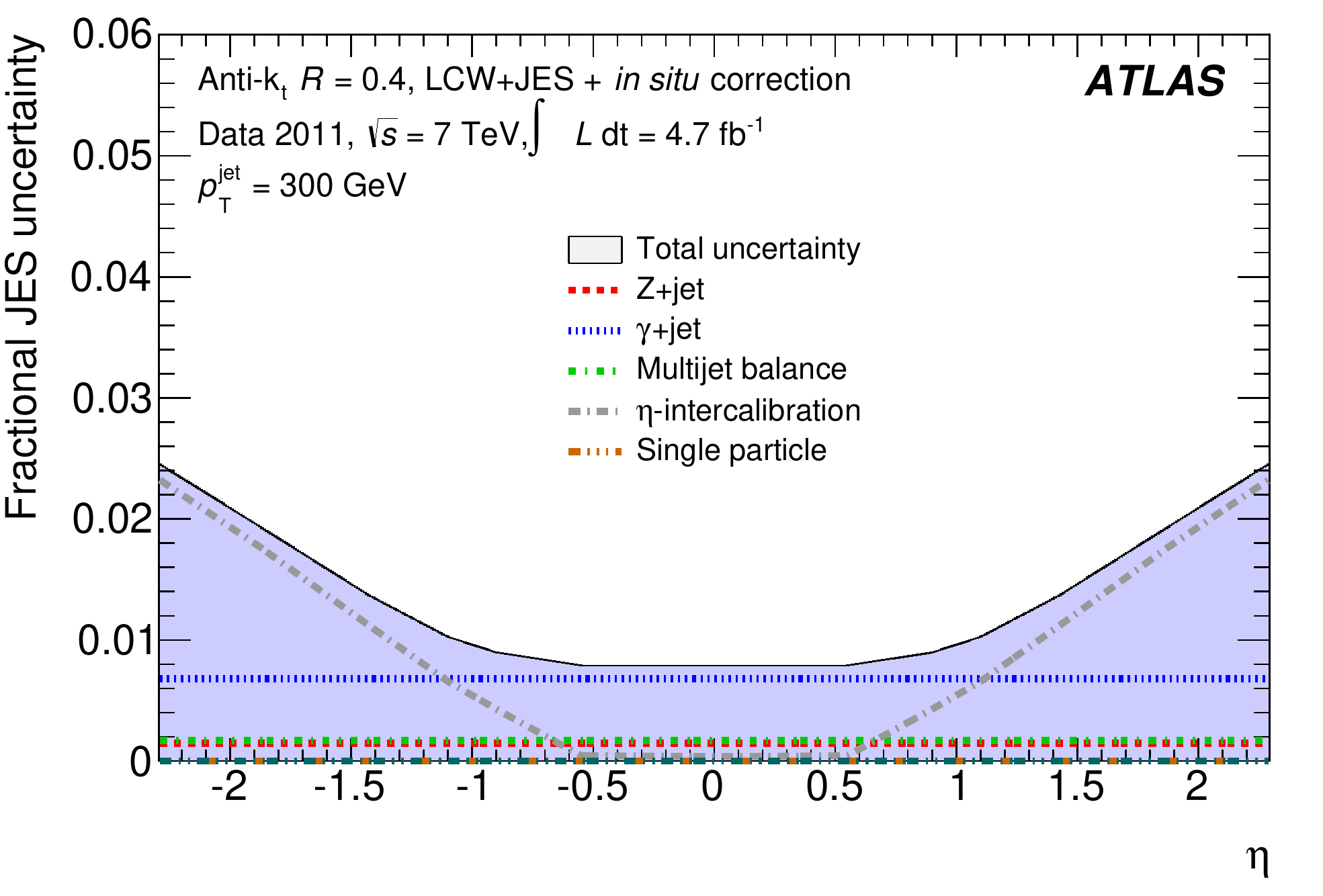}\label{fig:InsituJES_LCJES_3}}
  \caption[]{
    Fractional \insitu{} jet energy scale systematic uncertainty as a function of (\subref{fig:InsituJES_LCJES_0}, \subref{fig:InsituJES_LCJES_1}) \ptjet{} and (\subref{fig:InsituJES_LCJES_2}, \subref{fig:InsituJES_LCJES_3}) jet pseudorapidity 
    for \antikt{} jets with distance parameter of $R=0.4$ calibrated using the \LCWJES{} calibration scheme. 
    The contributions from each \insitu{} method are shown separately.
    Uncertainties from pile-up, flavour, and topology are not included.
    \vspace{7.cm}
    \label{fig:InsituJES_LCJES}
  }
\end{figure*}

\begin{figure*}[ht!]
  \centering
  \subfloat[\etajet = $0.5$]   {\includegraphics[width=0.49\textwidth]{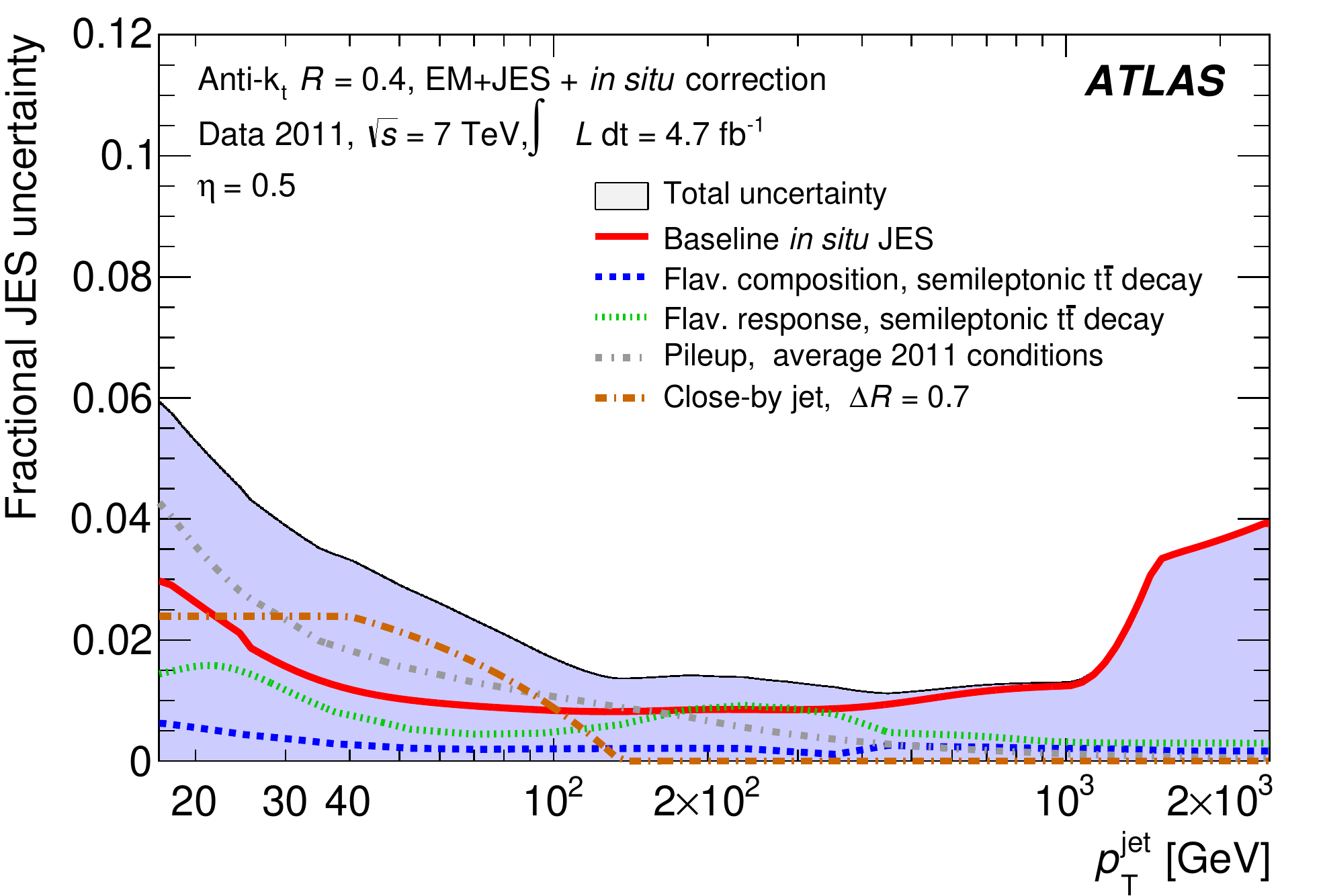}\label{fig:MultiJES_EMJES_TTBar_0}}
  \subfloat[\etajet = $2.0$]   {\includegraphics[width=0.49\textwidth]{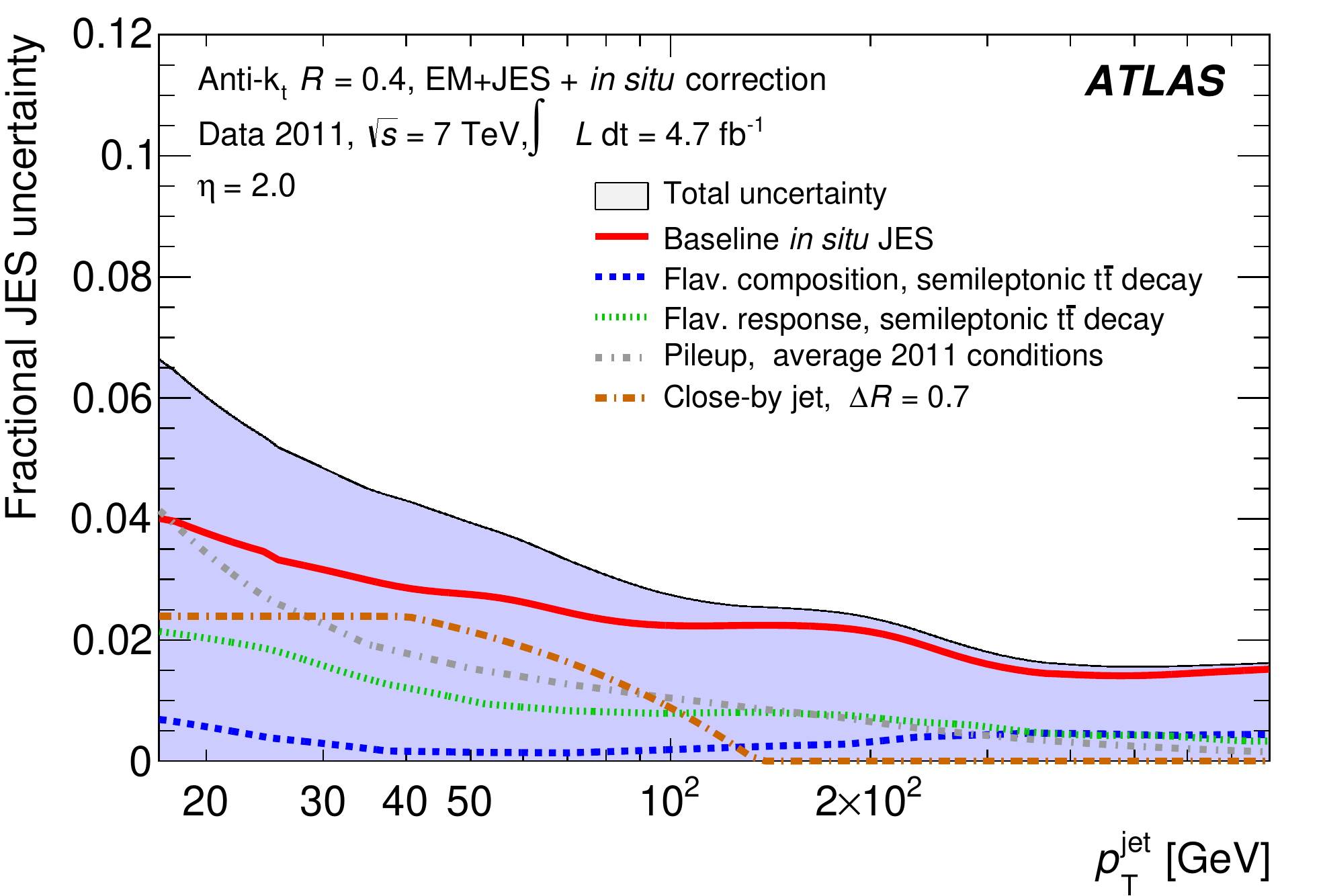}\label{fig:MultiJES_EMJES_TTBar_1}}\\
  \subfloat[\ptjet = $25$ \GeV]{\includegraphics[width=0.49\textwidth]{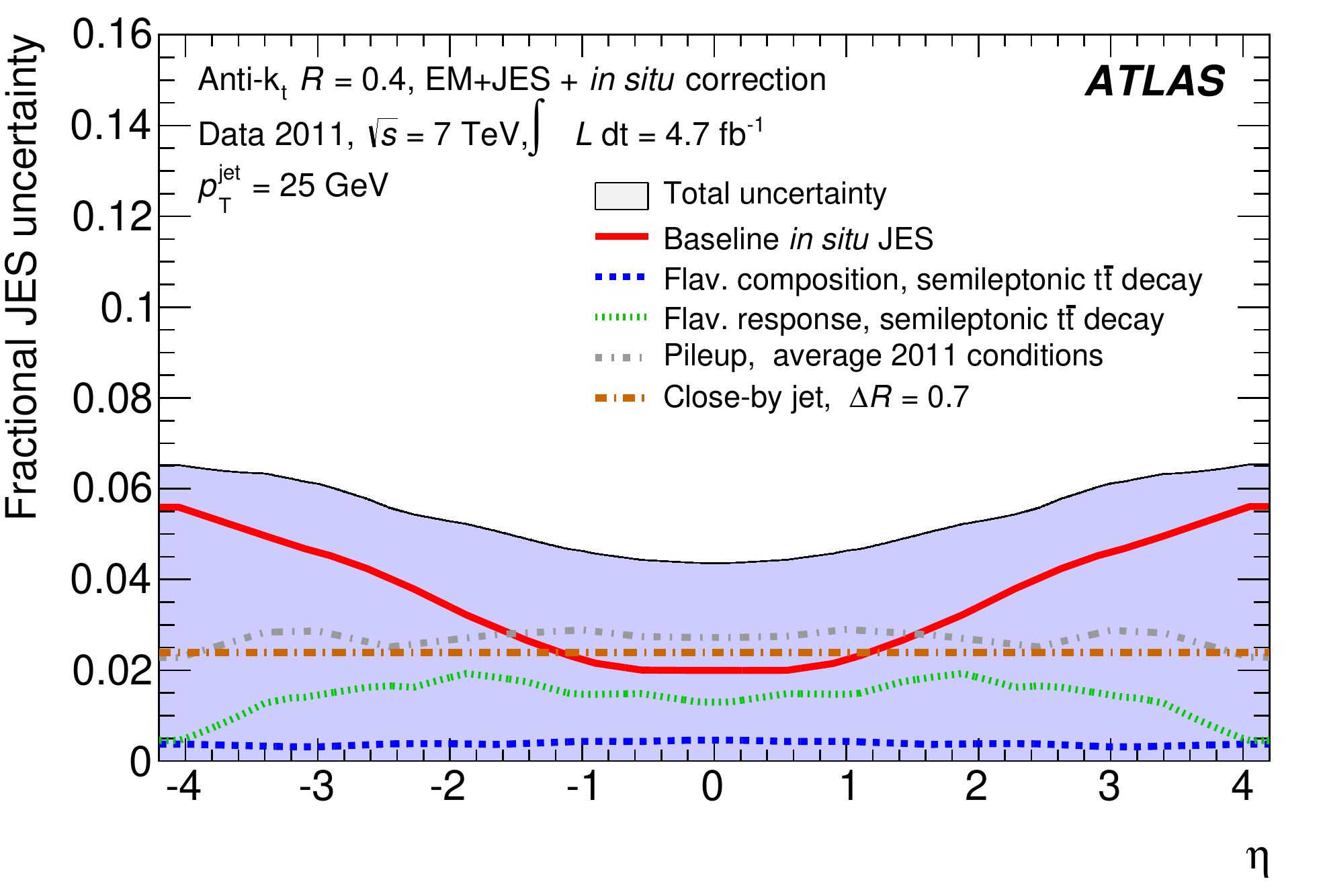}\label{fig:MultiJES_EMJES_TTBar_2}}
  \subfloat[\ptjet = $300$ \GeV]{\includegraphics[width=0.49\textwidth]{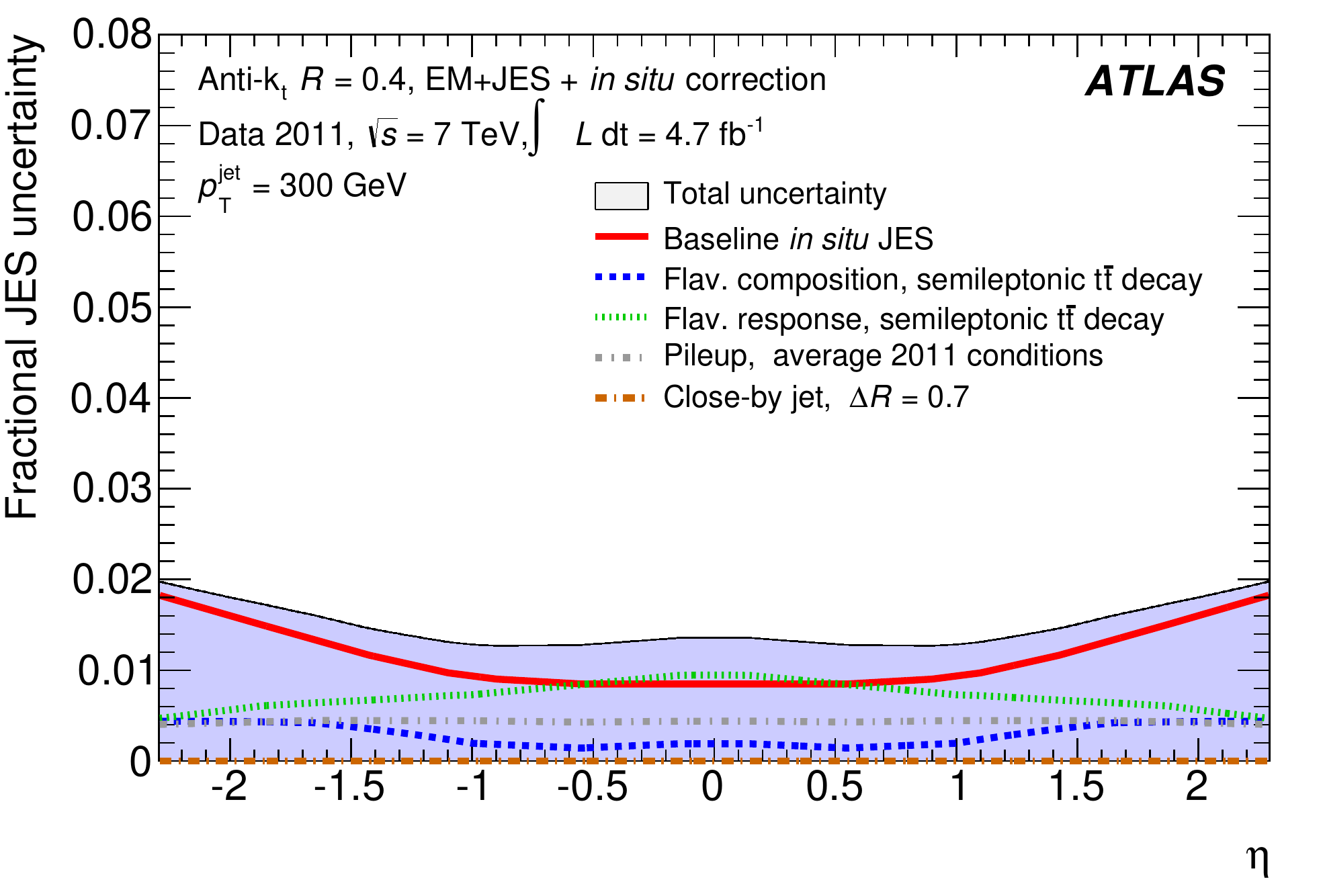}\label{fig:MultiJES_EMJES_TTBar_3}}
  \caption[]{
    Sample-dependent fractional jet energy scale systematic uncertainty as a function of (\subref{fig:MultiJES_EMJES_TTBar_0}, \subref{fig:MultiJES_EMJES_TTBar_1}) \ptjet{} and (\subref{fig:MultiJES_EMJES_TTBar_2}, \subref{fig:MultiJES_EMJES_TTBar_3}) jet pseudorapidity
    for \antikt{} jets with distance parameter of $R=0.4$ calibrated using the \EMJES{} calibration scheme. The uncertainty shown applies
    to semileptonic top-decays with average 2011 pile-up conditions, and does not include the uncertainty on the jet energy scale of \bjets.  
    \vspace{7.cm}
    \label{fig:MultiJES_EMJES_TTBar}
  }
\end{figure*}

\begin{figure*}[h!]
  \centering
  \subfloat[\etajet = $0.5$]   {\includegraphics[width=0.49\textwidth]{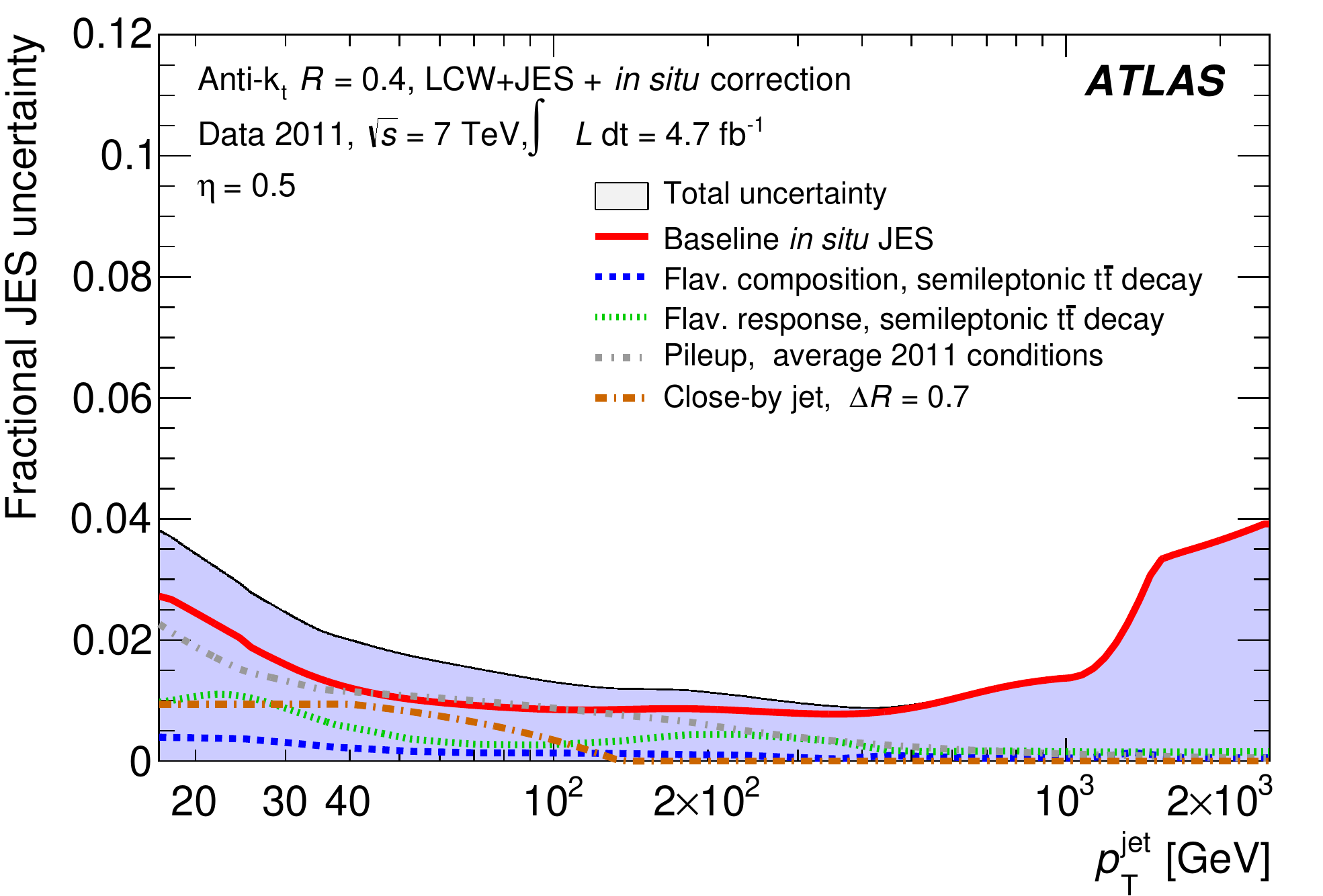}\label{fig:MultiJES_LCJES_TTBar_0}}
  \subfloat[\etajet = $2.0$]   {\includegraphics[width=0.49\textwidth]{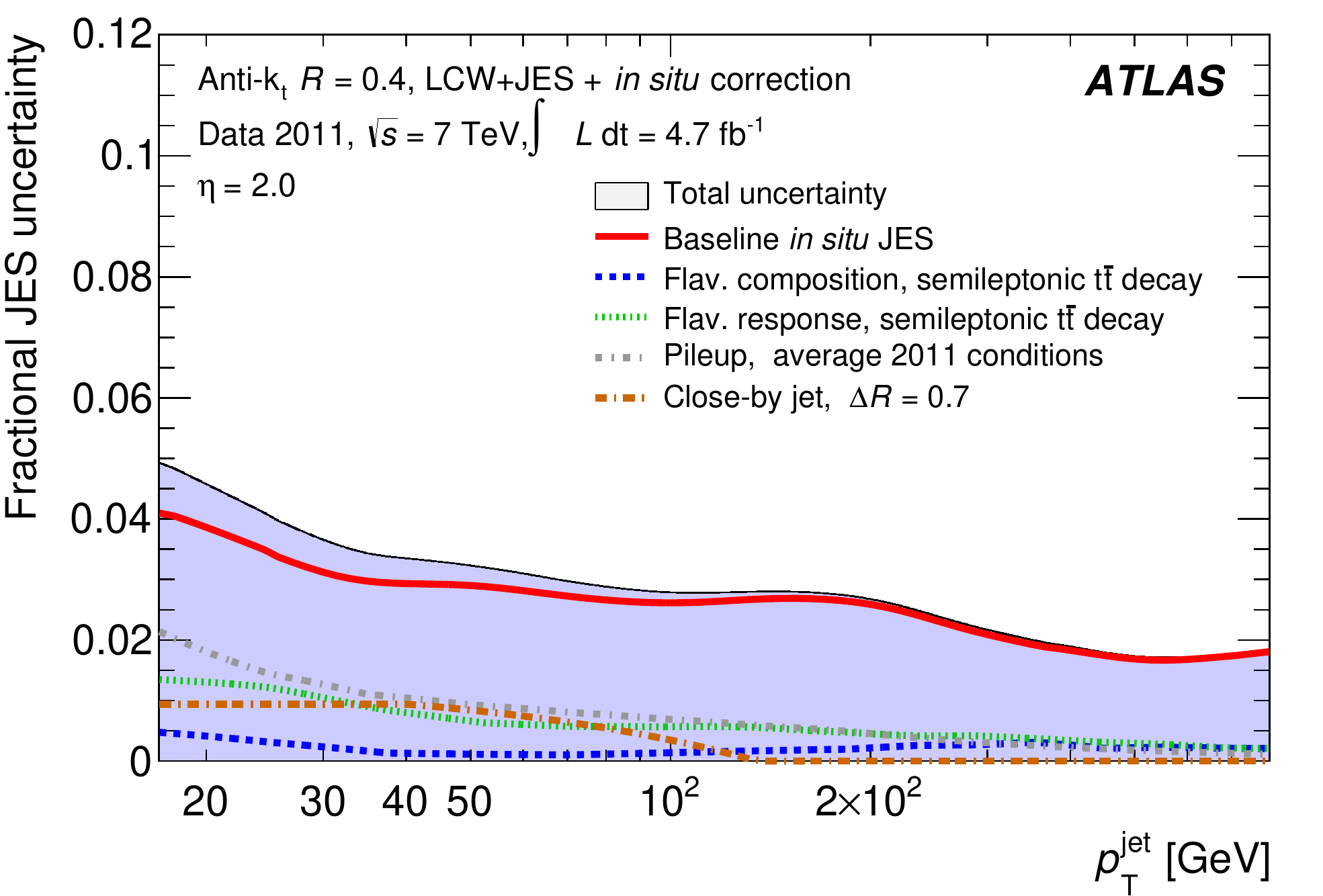}\label{fig:MultiJES_LCJES_TTBar_1}}\\
  \subfloat[\ptjet = $25$ \GeV]{\includegraphics[width=0.49\textwidth]{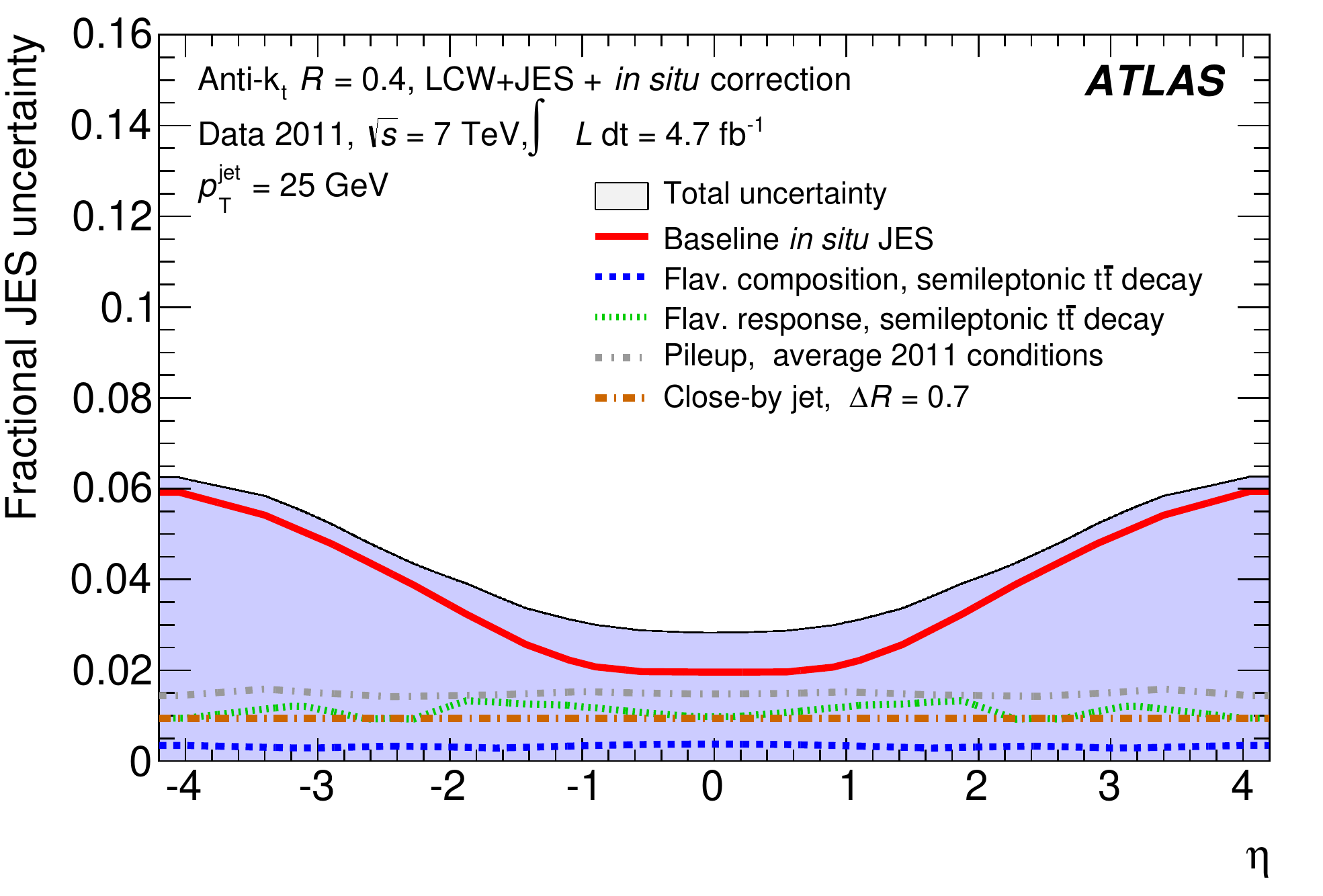}\label{fig:MultiJES_LCJES_TTBar_2}}
  \subfloat[\ptjet = $300$ \GeV]{\includegraphics[width=0.49\textwidth]{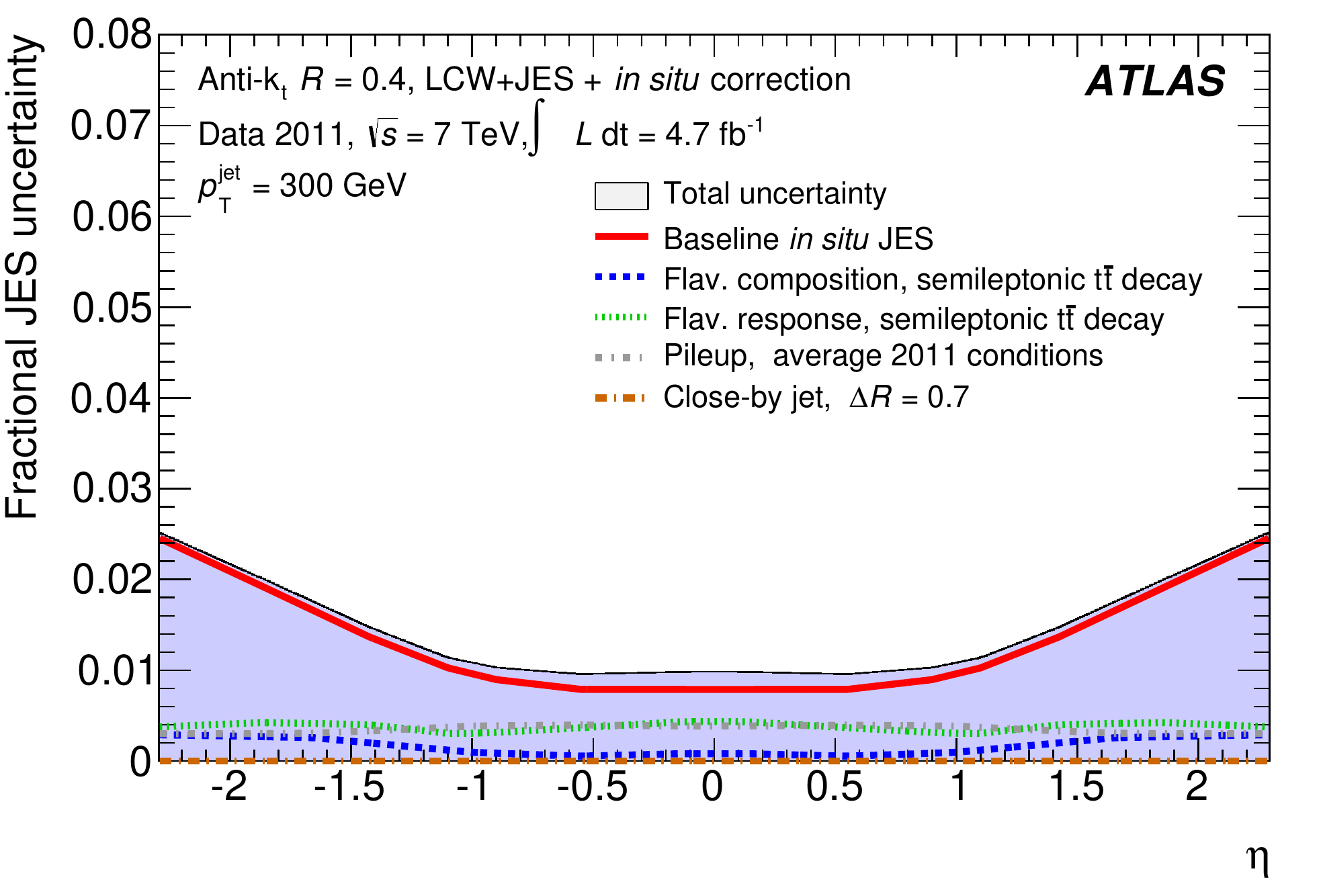}\label{fig:MultiJES_LCJES_TTBar_3}}
  \caption[]{
    Sample-dependent fractional jet energy scale systematic uncertainty as a function of (\subref{fig:MultiJES_LCJES_TTBar_0}, \subref{fig:MultiJES_LCJES_TTBar_1}) \ptjet{} and (\subref{fig:MultiJES_LCJES_TTBar_2}, \subref{fig:MultiJES_LCJES_TTBar_3}) jet pseudorapidity
    for \antikt{} jets with distance parameter of $R=0.4$ calibrated using the \LCWJES{} calibration scheme. The uncertainty shown applies
    to semileptonic top-decays with average 2011 pile-up conditions, and does not include the uncertainty on the jet energy scale of \bjets.  
    \vspace{7.cm}
    \label{fig:MultiJES_LCJES_TTBar}
  }
\end{figure*}

\begin{figure*}[h!]
  \centering
  \subfloat[\etajet = $0.5$]   {\includegraphics[width=0.49\textwidth]{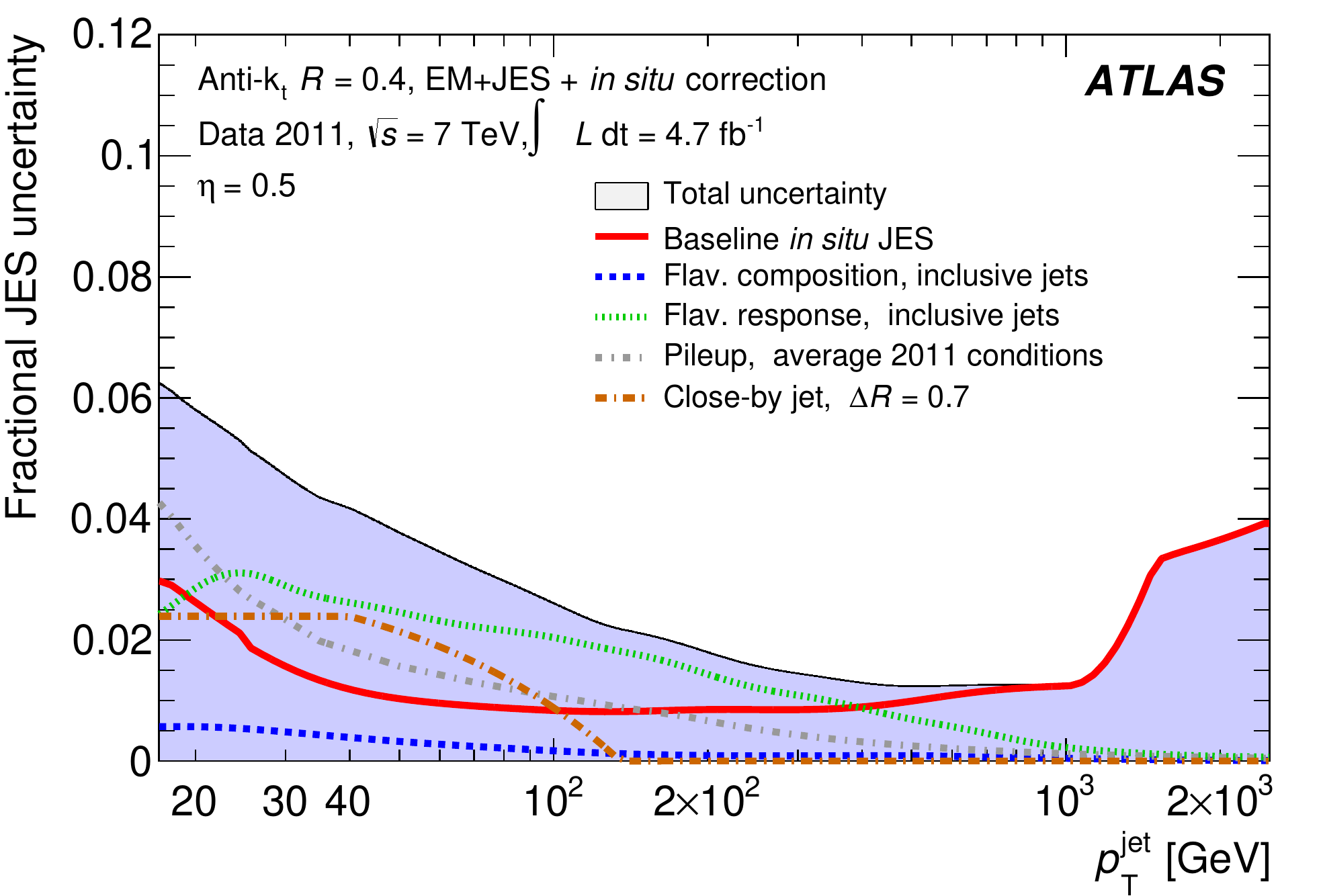}\label{fig:MultiJES_EMJES_IncJets_0}}
  \subfloat[\etajet = $2.0$]   {\includegraphics[width=0.49\textwidth]{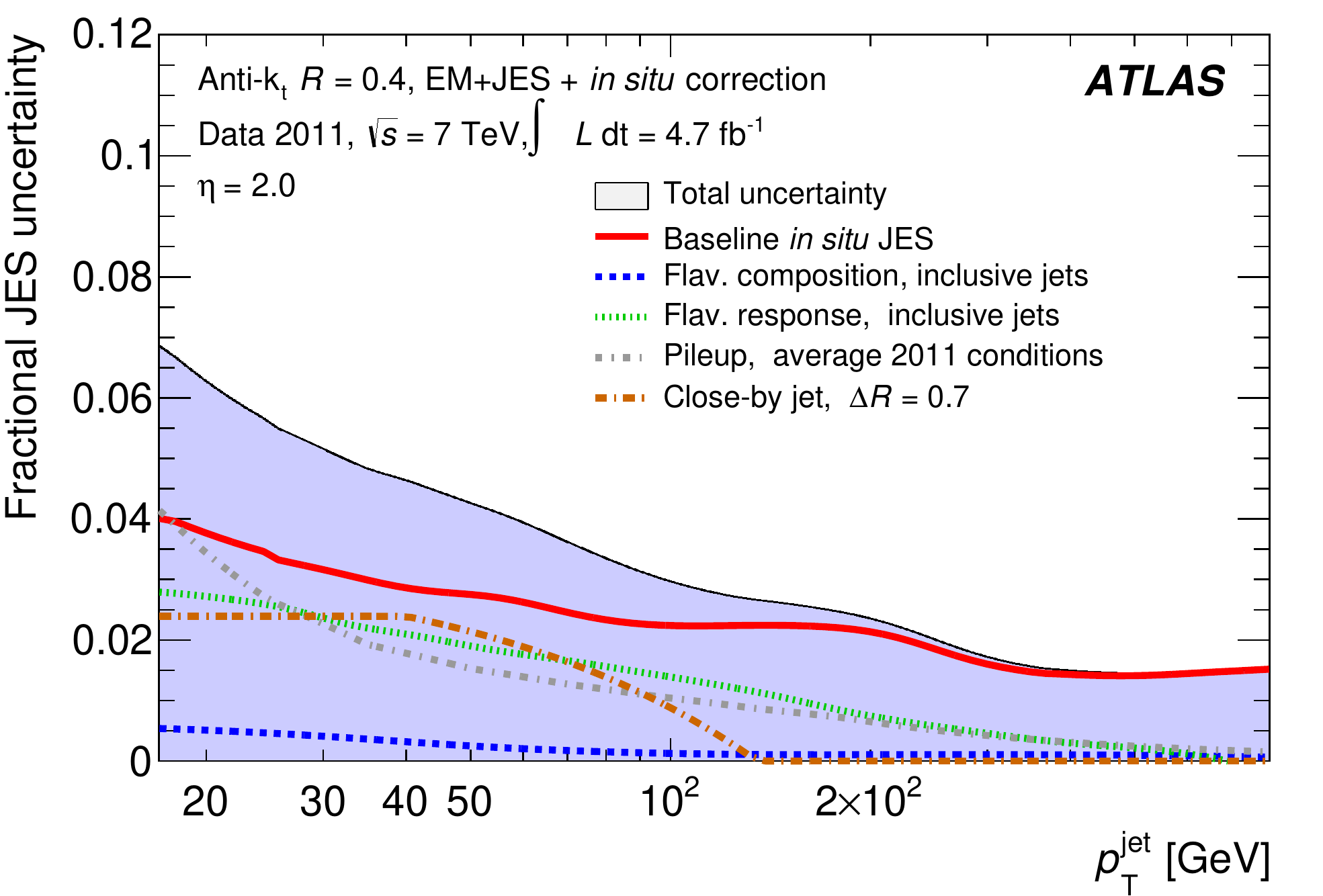}\label{fig:MultiJES_EMJES_IncJets_1}}\\
  \subfloat[\ptjet = $25$ \GeV]{\includegraphics[width=0.49\textwidth]{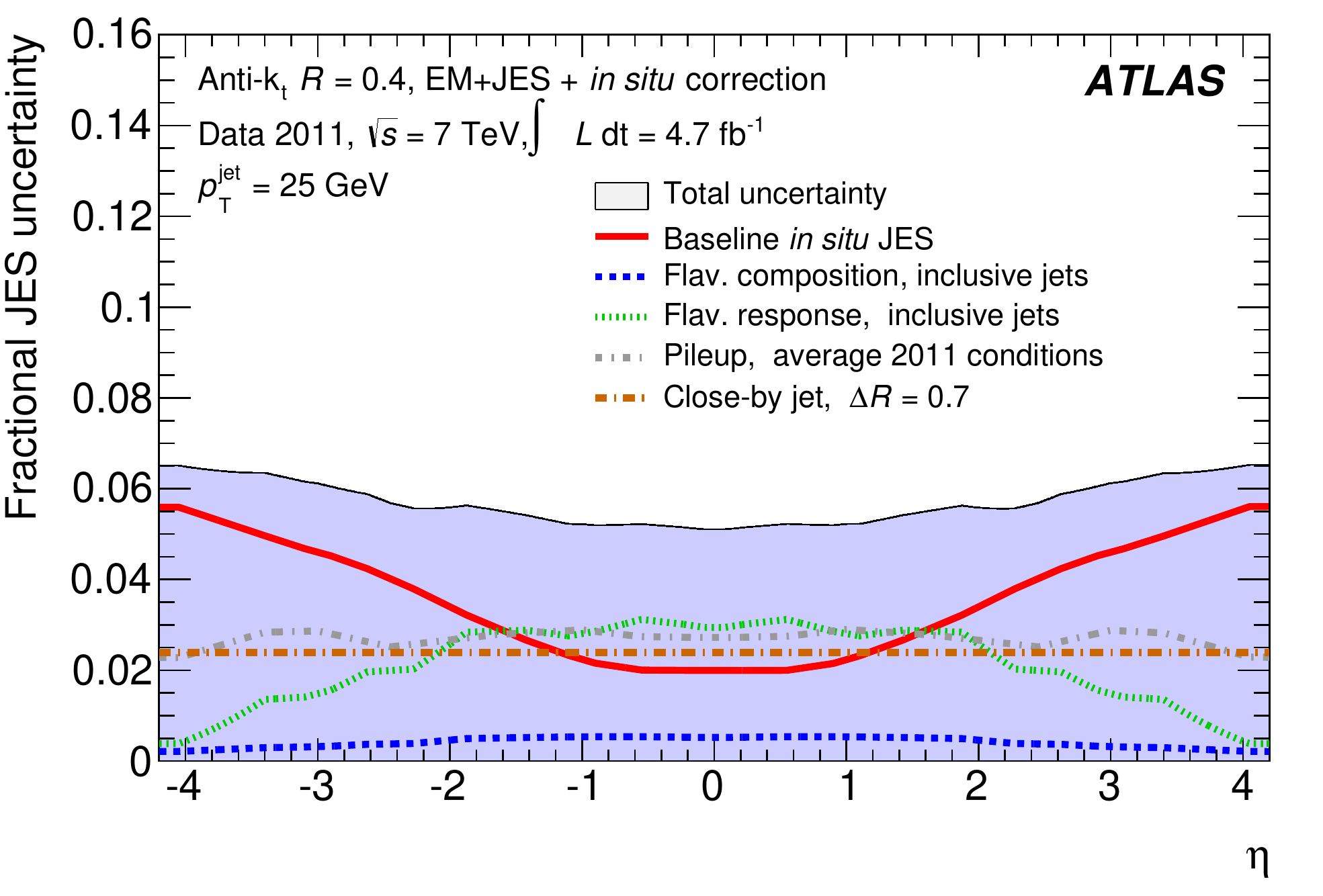}\label{fig:MultiJES_EMJES_IncJets_2}}
  \subfloat[\ptjet = $300$ \GeV]{\includegraphics[width=0.49\textwidth]{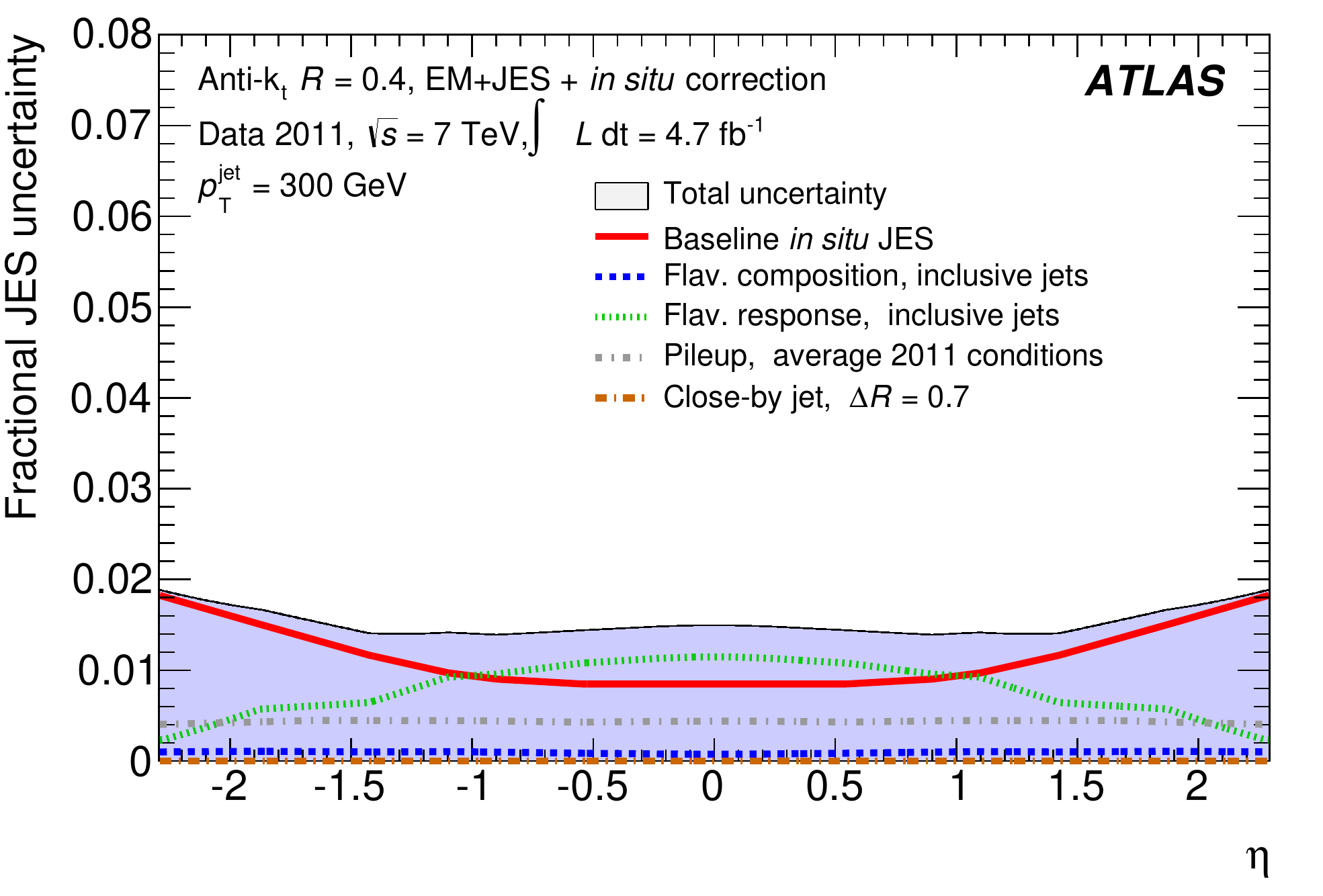}\label{fig:MultiJES_EMJES_IncJets_3}}
  \caption[]{
    Sample-dependent fractional jet energy scale systematic uncertainty as a function of (\subref{fig:MultiJES_EMJES_IncJets_0}, \subref{fig:MultiJES_EMJES_IncJets_1}) \ptjet{} and (\subref{fig:MultiJES_EMJES_IncJets_2}, \subref{fig:MultiJES_EMJES_IncJets_3}) jet pseudorapidity
    for \antikt{} jets with distance parameter of $R=0.4$ calibrated using the \EMJES{} calibration scheme. The uncertainty shown applies
    to inclusive QCD jets with average 2011 pile-up conditions, and does not include the uncertainty on the jet energy scale of \bjets.  
    \vspace{7.cm}
    \label{fig:MultiJES_EMJES_IncJets}
  }
\end{figure*}

\begin{figure*}[h!]
  \centering
  \subfloat[\etajet = $0.5$]   {\includegraphics[width=0.49\textwidth]{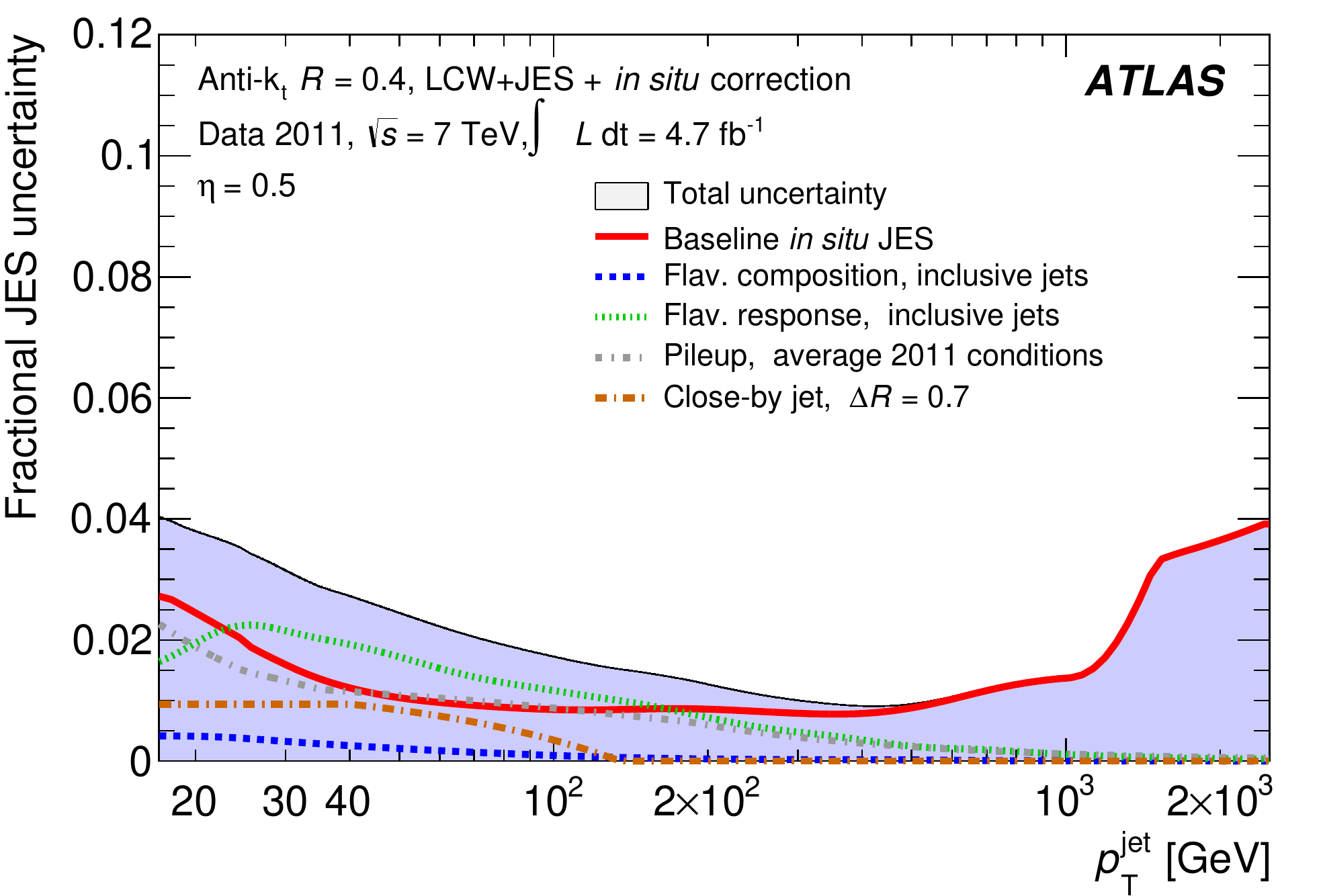}\label{fig:MultiJES_LCJES_IncJets_0}}
  \subfloat[\etajet = $2.0$]   {\includegraphics[width=0.49\textwidth]{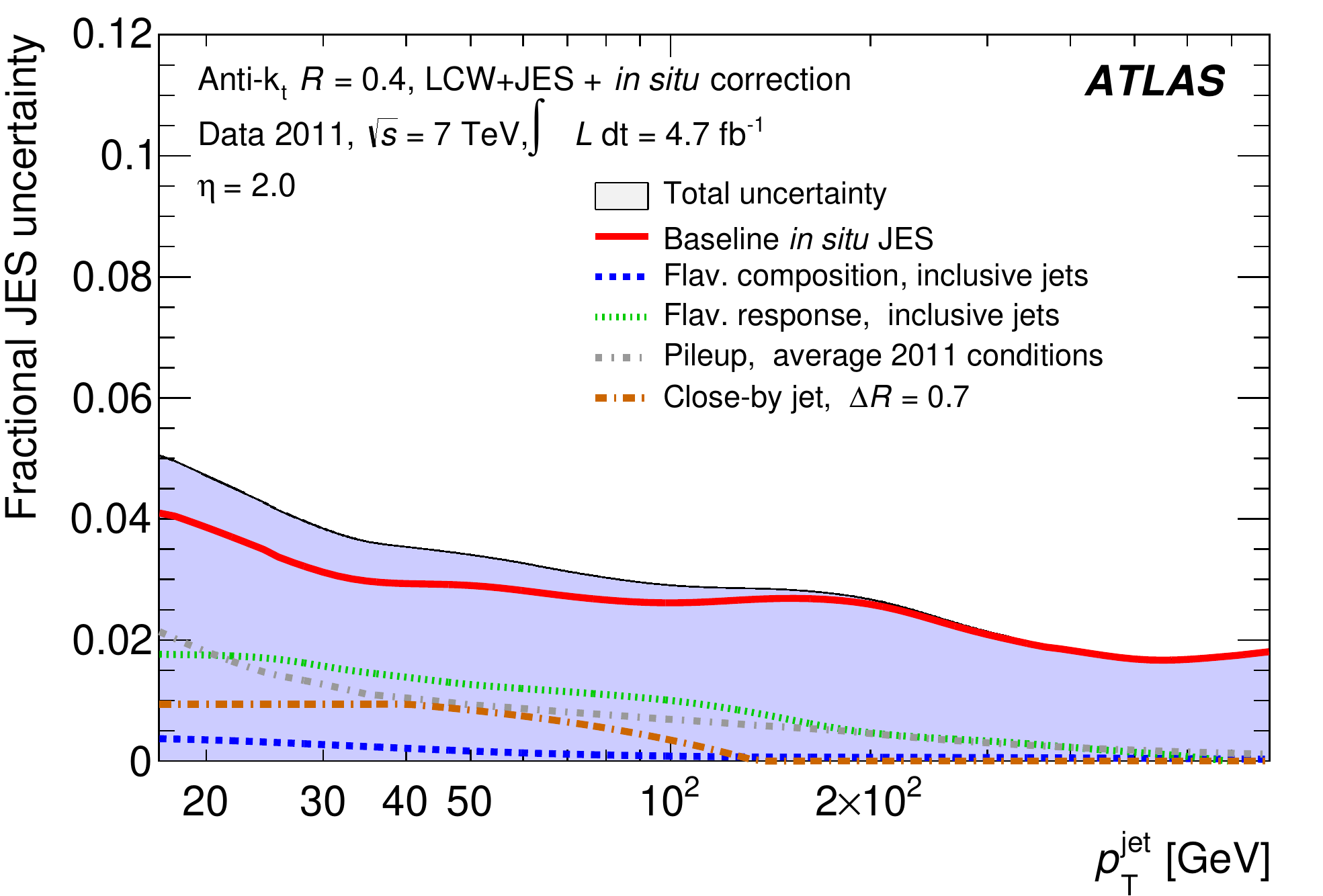}\label{fig:MultiJES_LCJES_IncJets_1}}\\
  \subfloat[\ptjet = $25$ \GeV]{\includegraphics[width=0.49\textwidth]{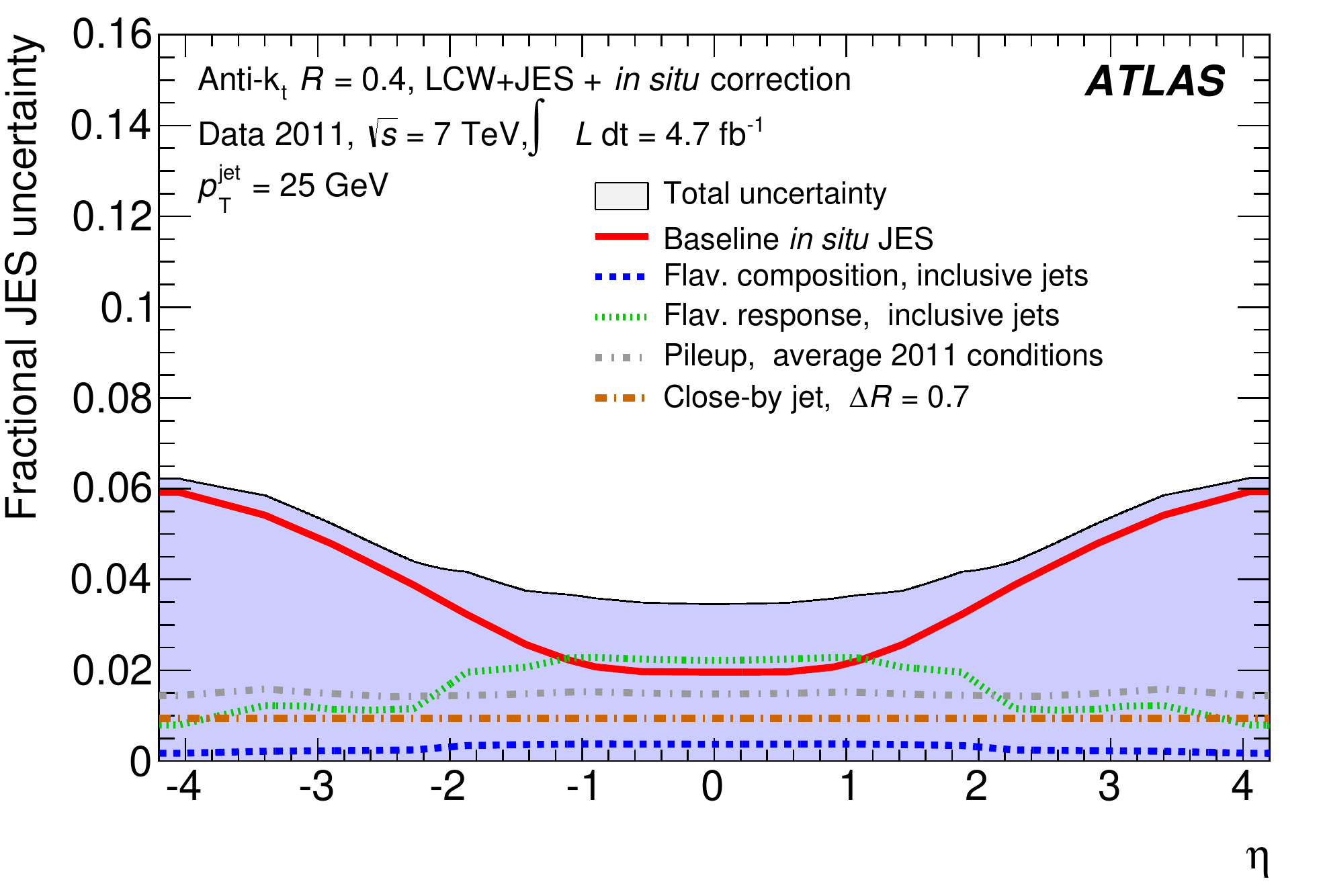}\label{fig:MultiJES_LCJES_IncJets_2}}
  \subfloat[\ptjet = $300$ \GeV]{\includegraphics[width=0.49\textwidth]{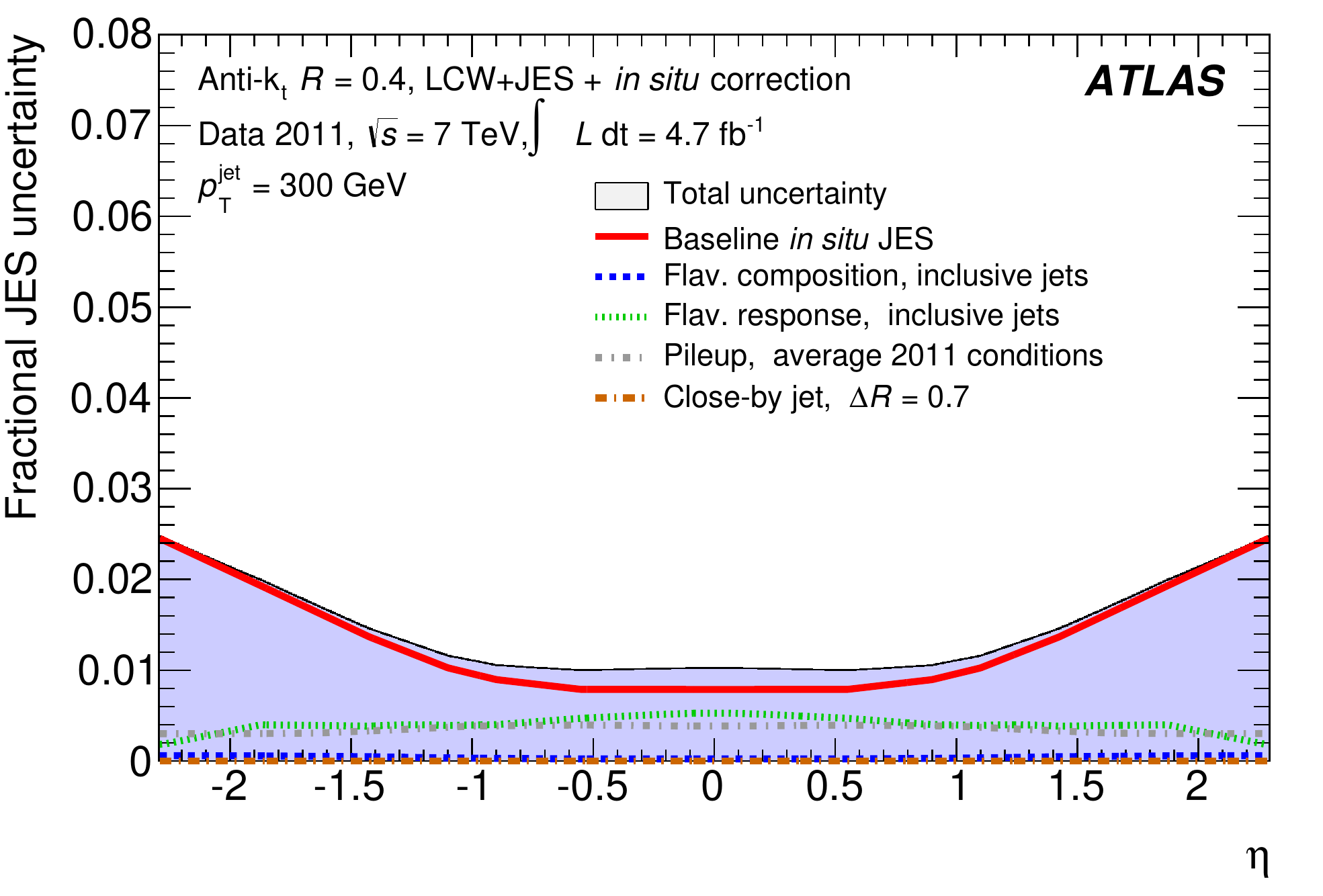}\label{fig:MultiJES_LCJES_IncJets_3}}
  \caption[]{
    Sample-dependent fractional jet energy scale systematic uncertainty as a function of (\subref{fig:MultiJES_LCJES_IncJets_0}, \subref{fig:MultiJES_LCJES_IncJets_1}) \ptjet{} and (\subref{fig:MultiJES_LCJES_IncJets_2}, \subref{fig:MultiJES_LCJES_IncJets_3}) jet pseudorapidity
    for \antikt{} jets with distance parameter of $R=0.4$ calibrated using the \LCWJES{} calibration scheme. The uncertainty shown applies
    to inclusive QCD jets with average 2011 pile-up conditions, and does not include the uncertainty on the jet energy scale of \bjets.  
    \vspace{7.cm}
    \label{fig:MultiJES_LCJES_IncJets}
  }
\end{figure*}

\begin{figure}[ht!]
  \centering
  \subfloat[\EMJES{}]   {\includegraphics[width=0.49\textwidth]{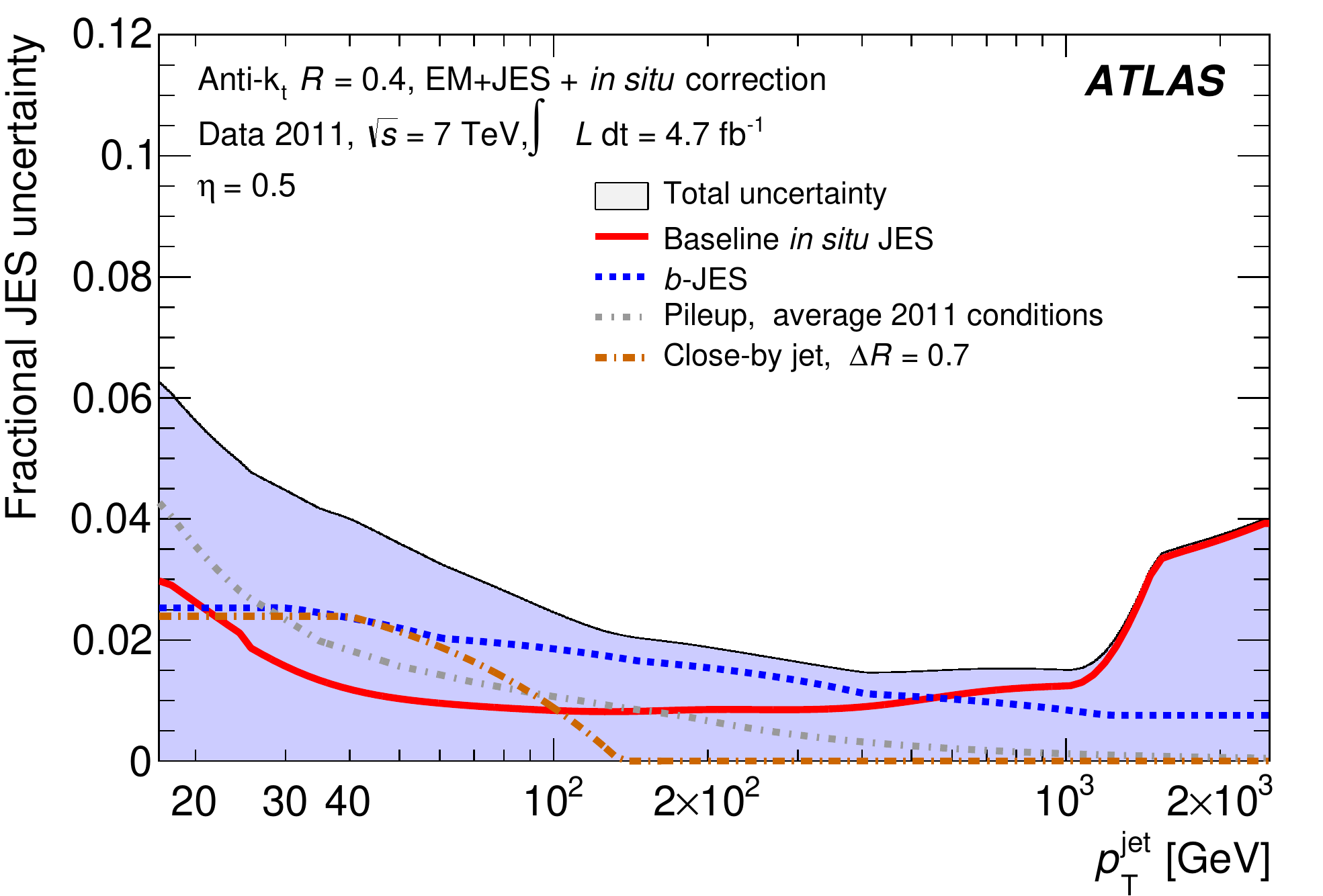}\label{fig:MultijetJES_Summary_eta5_AntiKt4TopoEM_bJES_0}}\\
  \subfloat[\LCWJES{}]   {\includegraphics[width=0.49\textwidth]{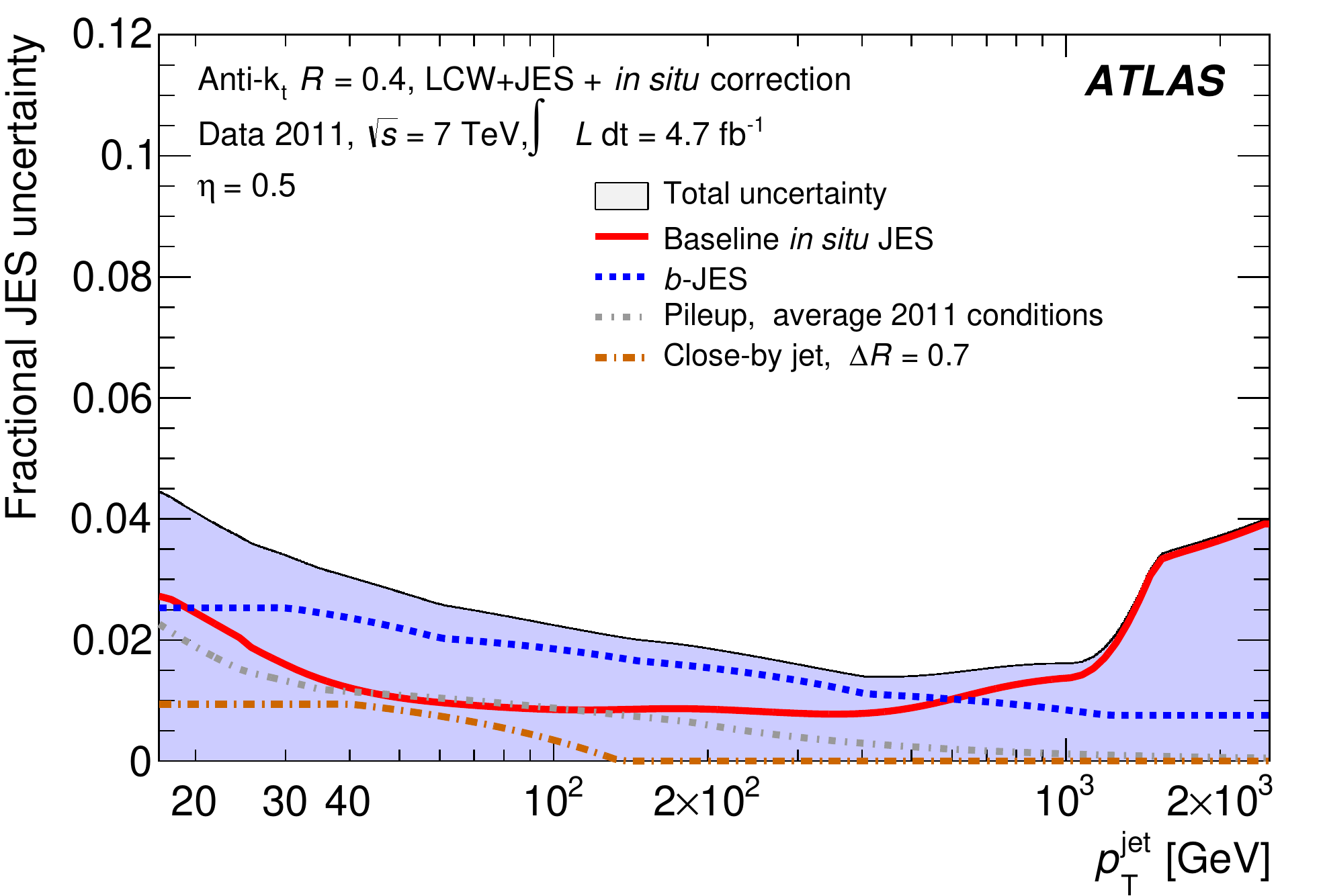}\label{fig:MultijetJES_Summary_eta5_AntiKt4TopoEM_bJES_1}}
  \caption[]{
    Fractional jet energy scale systematic uncertainty as a function of \ptjet{} for \antikt{} jets with distance parameter of $R=0.4$ calibrated using the 
   \subref{fig:MultijetJES_Summary_eta5_AntiKt4TopoEM_bJES_0}  \EMJES{} 
    and \subref{fig:MultijetJES_Summary_eta5_AntiKt4TopoEM_bJES_1} \LCWJES{} calibration schemes. The uncertainty shown applies to \bjets{} with average 2011 pile-up conditions.
  \label{fig:MultijetJES_Summary_eta5_AntiKt4TopoEM_bJES}
  }
\end{figure}

\begin{table*}[hp!!]
\begin{center}
\caption{
  Summary of the \insitu{} \EMJES{} and \LCWJES{} jet energy scale systematic uncertainties for different \ptjet{} and $|\eta|$ values for \antikt{} jets with $R=0.4$ and $R=0.6$.
  These values do not include pile-up, flavour or topology uncertainties.
}
\label{table:JESSystematicstables}
\vskip10pt
\begin{tabular}{|c|c|c|c|c|c|}
\hline

\bf{$|\eta|$ region} & \multicolumn{5}{|c|}{\bf{Fractional \EMJES{} JES uncertainty for $R=0.4$}} \\[1mm]
\hline
\ & \bf{$\ptjet=20$~GeV} & \bf{$\ptjet=40$~GeV} & \bf{$\ptjet =200$ GeV} & \bf{$\ptjet=800$ GeV} & \bf{$\ptjet =1.5$ TeV} \\[1mm]
\hline
$|\eta|=0.1$ & 2.6\% & 1.2\% & 0.8\% & 1.3\% & 3.2\%\\ 
$|\eta|=0.5$ & 2.6\% & 1.2\% & 0.8\% & 1.3\% & 3.2\%\\ 
$|\eta|=1.0$ & 2.8\% & 1.4\% & 1.0\% & 1.3\% & 3.2\%\\ 
$|\eta|=1.5$ & 3.2\% & 2.0\% & 1.5\% & 1.4\% & 3.3\%\\ 
$|\eta|=2.0$ & 3.8\% & 2.9\% & 2.1\% & 1.6\% & \\ 
$|\eta|=2.5$ & 4.3\% & 3.8\% & 2.8\% &  & \\ 
$|\eta|=3.0$ & 4.7\% & 4.5\% & 3.4\% &  & \\ 
$|\eta|=3.5$ & 5.1\% & 4.9\% & 4.6\% &  & \\ 
$|\eta|=4.0$ & 5.7\% & 5.1\% & 4.9\% &  & \\ 
\hline
\end{tabular}

\vskip10pt
\begin{tabular}{|c|c|c|c|c|c|}
\hline

\bf{$|\eta|$ region} & \multicolumn{5}{|c|}{\bf{Fractional \LCWJES{} JES uncertainty for $R=0.4$}} \\[1mm]
\hline
\ & \bf{$\ptjet=20$~GeV} & \bf{$\ptjet=40$~GeV} & \bf{$\ptjet =200$ GeV} & \bf{$\ptjet=800$ GeV} & \bf{$\ptjet =1.5$ TeV} \\[1mm]
\hline
    
$|\eta|=0.1$ & 2.4\% & 1.2\% & 0.8\% & 1.3\% & 3.2\%\\ 
$|\eta|=0.5$ & 2.5\% & 1.2\% & 0.8\% & 1.3\% & 3.2\%\\ 
$|\eta|=1.0$ & 2.6\% & 1.4\% & 1.1\% & 1.3\% & 3.2\%\\ 
$|\eta|=1.5$ & 3.1\% & 2.1\% & 1.7\% & 1.4\% & 3.3\%\\ 
$|\eta|=2.0$ & 3.9\% & 2.9\% & 2.6\% & 1.8\% & \\ 
$|\eta|=2.5$ & 4.6\% & 3.9\% & 3.4\% &  & \\ 
$|\eta|=3.0$ & 5.2\% & 4.6\% & 3.9\% &  & \\ 
$|\eta|=3.5$ & 5.8\% & 5.2\% & 4.5\% &  & \\ 
$|\eta|=4.0$ & 6.2\% & 5.5\% & 5.1\% &  & \\ 
\hline

\end{tabular}

\vskip10pt
\begin{tabular}{|c|c|c|c|c|c|}
\hline

\bf{$|\eta|$ region} & \multicolumn{5}{|c|}{\bf{Fractional \EMJES{} JES uncertainty for $R=0.6$ }} \\[1mm]
\hline
\ & \bf{$\ptjet=20$~GeV} & \bf{$\ptjet=40$~GeV} & \bf{$\ptjet =200$ GeV} & \bf{$\ptjet=800$ GeV} & \bf{$\ptjet =1.5$ TeV} \\[1mm]
\hline

$|\eta|=0.1$ & 2.7\% & 1.4\% & 0.8\% & 1.8\% & 3.3\%\\ 
$|\eta|=0.5$ & 2.7\% & 1.5\% & 0.8\% & 1.8\% & 3.3\%\\ 
$|\eta|=1.0$ & 2.8\% & 1.6\% & 0.9\% & 1.8\% & 3.3\%\\ 
$|\eta|=1.5$ & 3.0\% & 1.9\% & 1.3\% & 1.9\% & 3.3\%\\ 
$|\eta|=2.0$ & 3.6\% & 2.6\% & 1.9\% & 2.0\% & \\ 
$|\eta|=2.5$ & 4.3\% & 3.4\% & 2.4\% &  & \\ 
$|\eta|=3.0$ & 5.2\% & 4.1\% & 3.0\% &  & \\ 
$|\eta|=3.5$ & 5.7\% & 4.7\% & 3.8\% &  & \\ 
$|\eta|=4.0$ & 5.9\% & 4.8\% & 4.6\% &  & \\ 

\hline
\end{tabular}

\vskip10pt
\begin{tabular}{|c|c|c|c|c|c|}
\hline

\bf{$|\eta|$ region} & \multicolumn{5}{|c|}{\bf{Fractional \LCWJES{} JES uncertainty for $R=0.6$}} \\[1mm]
\hline
\ & \bf{$\ptjet=20$~GeV} & \bf{$\ptjet=40$~GeV} & \bf{$\ptjet =200$ GeV} & \bf{$\ptjet=800$ GeV} & \bf{$\ptjet =1.5$ TeV} \\[1mm]
\hline

$|\eta|=0.1$ & 2.3\% & 1.3\% & 0.8\% & 1.6\% & 3.2\%\\ 
$|\eta|=0.5$ & 2.3\% & 1.3\% & 0.8\% & 1.6\% & 3.2\%\\ 
$|\eta|=1.0$ & 2.4\% & 1.4\% & 1.0\% & 1.6\% & 3.2\%\\ 
$|\eta|=1.5$ & 2.7\% & 1.8\% & 1.6\% & 1.7\% & 3.2\%\\ 
$|\eta|=2.0$ & 3.3\% & 2.4\% & 2.2\% & 1.9\% & \\ 
$|\eta|=2.5$ & 4.4\% & 3.3\% & 2.8\% &  & \\ 
$|\eta|=3.0$ & 6.0\% & 4.6\% & 3.3\% &  & \\ 
$|\eta|=3.5$ & 7.0\% & 5.6\% & 3.8\% &  & \\ 
$|\eta|=4.0$ & 7.2\% & 6.0\% & 4.7\% &  & \\ 
\hline

\end{tabular}
\end{center}
\end{table*}

\section{Summary of the total jet energy scale systematic uncertainty}
\label{sec:Summary}
Figures \ref{fig:InsituJES_EMJES} and \ref{fig:InsituJES_LCJES} show the  
fractional jet energy scale uncertainty from the \insitu{} measurements as a function of \ptjet{} 
for four representative values of \etajet{},
and as a function of \etajet{} for two  representative values of \ptjet{}.
The total uncertainty is given by the absolute (\JES) and the relative \insitu{} calibration uncertainties added in quadrature.
For jets in the central region it amounts to $3\%$ at $\ptjet \approx 17$ \GeV, 
falling to $2\%$ at $\ptjet \approx 25$~\GeV, and is below $1\%$ for $55 \leq \ptjet<500$ \GeV.
The uncertainty increases for forward jets ($|\etajet|>1.2$) due to the 
uncertainty on the modelling of the parton radiation altering the dijet \pt{} balance
in the \etaic{} technique.
For very forward low-\pt{} jets ($\pt \approx 25$~\GeV, $|\etajet|\approx 4$), 
the uncertainty can be as large as $6\%$.
The \insitu{} \JES{} uncertainty is similar for the \EMJES{} and \LCWJES{} calibration schemes.

For jets with $\ptjet > 1$~\TeV{} the \JES{} uncertainty is derived from single-hadron 
response measurements~\cite{eppaper2010}, given the large statistical error 
of the multijet balance technique beyond $\ptjet > 1$~\TeV{}.
The uncertainties from the \insitu{} techniques are kept fixed at $\ptjet = 1$~\TeV{}
and subtracted in quadrature from the uncertainty of the single-hadron response measurements,
which is the dominant contribution at high $\ptjet$ in $2010$ and $2011$.

Table~\ref{table:JESSystematicstables} presents a summary of the total \insitu{} \JES{} uncertainties 
in representative $\eta$ and \ptjet{} regions for \antikt{} jets with $R=0.4$ and $R=0.6$ calibrated 
with the \EMJES{} and \LCWJES{} schemes.
%

The total \insitu{} calibration uncertainty 
(labelled ``baseline \insitu{} \JES'') together with the additional uncertainties that depend on the event sample 
used in the physics analysis %
is shown in Figs.~\ref{fig:MultiJES_EMJES_TTBar} to \ref{fig:MultiJES_LCJES_IncJets} for two illustrative samples. 
The procedure to estimate those uncertainties\footnote{If no information on the fraction of gluons or its uncertainty is available
for a given analysis sample, a gluon fraction of $50\%$ with $50\%$ uncertainty is used, representing an unknown flavour composition for the sample.}
is detailed in \secRef{sec:FlavorTopology}. 
 
Figures \ref{fig:MultiJES_EMJES_TTBar} and \ref{fig:MultiJES_LCJES_TTBar} show the 
flavour response uncertainty and the flavour composition uncertainties for light jets in an event sample with 
top-quark pairs decaying semileptonically. 
Semileptonic decays are selected in 
the \MC{} simulation samples based on truth information, and electrons are not considered as jets 
when estimating the jet response. The \MC{} generator used to evaluate 
the sample response and the gluon fraction is \MCAtNLO{}, while the gluon fraction uncertainty
is derived using the difference in gluon fractions between the \ACERMC{} and \PowHeg{} generators. 
The average gluon fraction uncertainty ranges from $2\%$ to $10\%$ depending 
on the jet transverse momentum and pseudorapidity. For differential measurements, 
the gluon fraction and its uncertainty can also be determined as a 
function of the property measured (e.g. number of jets). 
Figure~\ref{fig:MultijetJES_Summary_eta5_AntiKt4TopoEM_bJES}
shows the total uncertainty for \bjets{} in the case of jets with $R =0.4$ calibrated using the \EMJES{} and \LCWJES{} schemes.

Figures~\ref{fig:MultiJES_EMJES_IncJets} and~\ref{fig:MultiJES_LCJES_IncJets} show the flavour uncertainties
for an event sample of inclusive jets. The sample response and gluon fraction are evaluated 
using the \pythia{} nominal sample, while the gluon fraction uncertainty
is derived considering the average difference in the fraction of gluons between the
\pythia{} nominal sample and samples producing using the \PowHeg{} (interfaced with Pythia for parton showering 
and hadronisation) and the \herwigpp{} generators. The gluon fraction uncertainty in the inclusive jet case is up to $7\%$ 
but decreases rapidly with jet \pt{} to less than $2\%$.
 
A conservative topology uncertainty due to close-by jets 
is shown assuming the presence of a close-by jet with $\Rmin{} =0.7$.
The pile-up uncertainties are given for the average conditions of $\Npv = 10$ and $\axing = 8.5$ in the $2011$ dataset, 
with an RMS of $3$ for both \Npv{} and \axing{}. 

The total uncertainty is calculated by adding all uncertainty sources in quadrature. 
The uncertainty for jets calibrated with the \LCWJES{} scheme 
is significantly smaller than the one for the \EMJES{} scheme, mainly because this scheme reduces
the sensitivity to the jet flavour.

\section{Conclusions}
\label{sec:CONCLUSIONS}
The \ATLAS{} jet energy scale (\JES) and its systematic uncertainty are determined for jets 
produced in \pp{} collisions with a \cms{} 
of $\sqrt{s}=7$~\TeV{} using the full $2011$ dataset that corresponds to an 
integrated luminosity of \mylumi. 
Jets are reconstructed from clusters of calorimeter cells with the \antikt{} algorithm with distance parameters 
$R=0.4$ or $R=0.6$.
The uncertainty of the jet energy measurement is evaluated for jets with
calibrated transverse momenta $\ptjet > 20$~\GeV{} and pseudorapidities \AetaRange{4.5}
using a combination of \insitu{} techniques exploiting the
transverse momentum balance between a jet and a reference object.

For central jets (\AetaRange{1.2}) with \ptRange{20}{800},
photons or \Zboson{} bosons are used as reference objects.
A system of low-\pt{} jets is used to extend the \JES{} validation up to  the \TeV{} regime. 
The smallest \JES{} uncertainty of less than $1 \%$ is found for jets with \ptRange{55}{500}. 
For jets with $\pt{} = 20$~\GeV{} the uncertainty is about $3\%$. 
For $\ptjet > 1$~\TeV{} the \JES{} uncertainty is estimated from single-hadron response
measurements \insitu{} and in beam tests and is about $3$\%.
The \JES{} uncertainty for forward jets is derived from dijet \pt{} balance measurements.
The resulting uncertainty is largest for low-\pt{} jets at $|\etajet|=4.5$ and amounts to $6\%$.

From the uncertainties of the \insitu{} techniques used to assess the \JES{} uncertainty,
the correlation of the uncertainties in \ptjet{} and \etajet{} are derived
and made available for physics analysis as a set of systematic uncertainty sources.

The effect of multiple \pp{} interactions is cor\-rec\-ted for as a function of the measured
and the expected numbers of pile-up events, and an uncertainty is evaluated
using \insitu{} techniques.
Additional \JES{} uncertainties due to specific event topologies, such as close-by jets
or selections of event samples with an enhanced content of jets originating from light quarks or gluons, are also discussed. 
These uncertainties depend on the event sample used in a given physics analysis
and are evaluated for representative examples.
For an event sample of semileptonically decaying top-pairs, assuming average $2011$ pile-up conditions, 
the total \JES{} uncertainty accounting for all effects is below $3\%$ for \ptRange{60}{1000} when using the \EMJES{} calibration scheme, 
and it is further reduced to below $2.5\%$ if using the more refined \LCWJES{} calibration scheme. In the case of a
sample of inclusive QCD jets under the same conditions, the total \JES{} uncertainties for the \EMJES{} and \LCWJES{} calibration schemes 
are below $3.5\%$ and $2\%$, respectively.

\section*{Acknowledgement}



\section{Acknowledgements}

We thank CERN for the very successful operation of the LHC, as well as the
support staff from our institutions without whom ATLAS could not be
operated efficiently.

We acknowledge the support of ANPCyT, Argentina; YerPhI, Armenia; ARC,
Australia; BMWFW and FWF, Austria; ANAS, Azerbaijan; SSTC, Belarus; CNPq and FAPESP,
Brazil; NSERC, NRC and CFI, Canada; CERN; CONICYT, Chile; CAS, MOST and NSFC,
China; COLCIENCIAS, Colombia; MSMT CR, MPO CR and VSC CR, Czech Republic;
DNRF, DNSRC and Lundbeck Foundation, Denmark; EPLANET, ERC and NSRF, European Union;
IN2P3-CNRS, CEA-DSM/IRFU, France; GNSF, Georgia; BMBF, DFG, HGF, MPG and AvH
Foundation, Germany; GSRT and NSRF, Greece; ISF, MINERVA, GIF, I-CORE and Benoziyo Center,
Israel; INFN, Italy; MEXT and JSPS, Japan; CNRST, Morocco; FOM and NWO,
Netherlands; BRF and RCN, Norway; MNiSW and NCN, Poland; GRICES and FCT, Portugal; MNE/IFA, Romania; MES of Russia and ROSATOM, Russian Federation; JINR; MSTD,
Serbia; MSSR, Slovakia; ARRS and MIZ\v{S}, Slovenia; DST/NRF, South Africa;
MINECO, Spain; SRC and Wallenberg Foundation, Sweden; SER, SNSF and Cantons of
Bern and Geneva, Switzerland; NSC, Taiwan; TAEK, Turkey; STFC, the Royal
Society and Leverhulme Trust, United Kingdom; DOE and NSF, United States of
America.

The crucial computing support from all WLCG partners is acknowledged
gratefully, in particular from CERN and the ATLAS Tier-1 facilities at
TRIUMF (Canada), NDGF (Denmark, Norway, Sweden), CC-IN2P3 (France),
KIT/GridKA (Germany), INFN-CNAF (Italy), NL-T1 (Netherlands), PIC (Spain),
ASGC (Taiwan), RAL (UK) and BNL (USA) and in the Tier-2 facilities
worldwide.

%
\bibliographystyle{atlasnote}
\bibliography{jetetmiss,bjet-studies,ttbarMC_Samples}

\providecommand{\href}[2]{#2}\begingroup\raggedright\begin{thebibliography}{10%
0}

\bibitem{Cacciari:2008gp}
M.~Cacciari, G.~P. Salam, and G.~Soyez, {\em The anti-k$_t$ jet clustering
  algorithm\/},  \href{http://dx.doi.org/10.1088/1126-6708/2008/04/063}{JHEP
  {\bf 0804} (2008)  063},
\href{http://arxiv.org/abs/0802.1189}{{\tt arXiv:0802.1189 [hep-ph]}}.

\bibitem{atlasjet2010}
{ATLAS} Collaboration, {\em {Measurement of inclusive jet and dijet cross
  sections in proton-proton collisions at 7 TeV centre-of-mass energy with the
  {ATLAS} detector}\/},
  \href{http://dx.doi.org/10.1140/epjc/s10052-010-1512-2}{Eur. Phys. J. {\bf C
  71} (2011)  1512}, \href{http://arxiv.org/abs/1009.5908}{{\tt arXiv:1009.5908
  [hep-ex]}}.

\bibitem{jespaper2010}
{ATLAS} Collaboration, {\em {Jet energy measurement with the ATLAS detector in
  proton-proton collisions at $\sqrt{s}=7$ TeV}\/},
  \href{http://dx.doi.org/10.1140/epjc/s10052-013-2304-2}{Eur. Phys. J. {\bf C
  73} (2013)  2304},
\href{http://arxiv.org/abs/1112.6426}{{\tt arXiv:1112.6426 [hep-ex]}}.

\bibitem{eppaper2010}
{ATLAS} Collaboration, {\em {Single hadron response measurement and calorimeter
  jet energy scale uncertainty with the ATLAS detector at the LHC}\/},
  \href{http://dx.doi.org/10.1140/epjc/s10052-013-2305-1}{Eur. Phys. J. {\bf C
  73} (2013)  2305},
\href{http://arxiv.org/abs/1203.1302}{{\tt arXiv:1203.1302 [hep-ex]}}.

\bibitem{DetectorPaper}
{ATLAS} Collaboration, {\em {T}he {ATLAS} experiment at the {CERN} {L}arge
  {H}adron {C}ollider\/},
  \href{http://dx.doi.org/10.1088/1748-0221/3/08/S08003}{JINST {\bf 3} (2008)
  S08003}.

\bibitem{cdf06}
{CDF} Collaboration, A.~Bhatti et al., {\em {Determination of the jet energy
  scale at the collider detector at Fermilab}\/},
  \href{http://dx.doi.org/10.1016/j.nima.2006.05.269}{Nucl. Instrum. Meth. {\bf
  A 566} (2006)  375--412},
\href{http://arxiv.org/abs/0510047}{{\tt arXiv:0510047 [hep-ex]}}.

\bibitem{ref:D0_MPF}
{D0 Collaboration} Collaboration, B.~Abbott et al., {\em {Determination of the
  absolute jet energy scale in the D0 calorimeters}\/},
  \href{http://dx.doi.org/10.1016/S0168-9002(98)01368-0}{Nucl. Instrum. Meth.
  {\bf A 424} (1999)  352--394},
\href{http://arxiv.org/abs/hep-ex/9805009}{{\tt arXiv:hep-ex/9805009
  [hep-ex]}}.

\bibitem{Abazov:2013hda}
{D0} Collaboration, V.~M. Abazov et al., {\em {Jet energy scale determination
  in the D0 experiment}\/},
\href{http://arxiv.org/abs/1312.6873}{{\tt arXiv:1312.6873 [hep-ex]}}.

\bibitem{cmsJES}
{CMS} Collaboration, {\em {Determination of Jet Energy Calibration and
  Transverse Momentum Resolution in CMS}\/},
  \href{http://dx.doi.org/10.1088/1748-0221/6/11/P11002}{JINST {\bf 6} (2011)
  P11002},
\href{http://arxiv.org/abs/1107.4277}{{\tt arXiv:1107.4277 [physics.ins-det]}}.

\bibitem{D0_jetcross}
{D0} Collaboration, V.~M. Abazov et al., {\em {Measurement of the inclusive jet
  cross section in $p \bar {p}$ collisions at $\sqrt{s}=1.96$ {T}e{V}}\/},
  \href{http://dx.doi.org/10.1103/PhysRevD.85.052006}{Phys. Rev. {\bf D 85}
  (2012)  052006},
\href{http://arxiv.org/abs/1110.3771}{{\tt arXiv:1110.3771 [hep-ex]}}.

\bibitem{cdf_jetcross}
{CDF} Collaboration, T.~Aaltonen et al., {\em {Measurement of the Inclusive Jet
  Cross Section at the Fermilab Tevatron p anti-p Collider Using a Cone-Based
  Jet Algorithm}\/},  \href{http://dx.doi.org/10.1103/PhysRevD.78.052006,
  10.1103/PhysRevD.79.119902}{Phys. Rev. {\bf D 78} (2008)  052006},
\href{http://arxiv.org/abs/0807.2204}{{\tt arXiv:0807.2204 [hep-ex]}}.

\bibitem{ua183}
{UA2} Collaboration, P.~Bagnaia et al., {\em {Measurement of production and
  properties of jets at the CERN anti-p p collider}\/},
  \href{http://dx.doi.org/10.1007/BF01573214}{Z. Phys. {\bf C 20} (1983)
  117--134}.

\bibitem{h1_jes1}
{H1} Collaboration, C.~Adloff et al., {\em {Measurement of neutral and charged
  current cross-sections in positron proton collisions at large momentum
  transfer}\/},  \href{http://dx.doi.org/10.1007/s100520050721}{Eur. Phys. J.
  {\bf C 13} (2000)  609--639},
\href{http://arxiv.org/abs/hep-ex/9908059}{{\tt arXiv:hep-ex/9908059
  [hep-ex]}}.

\bibitem{h1_jes2}
{H1} Collaboration, F.~Aaron et al., {\em {A Precision Measurement of the
  Inclusive ep Scattering Cross Section at HERA}\/},
  \href{http://dx.doi.org/10.1140/epjc/s10052-009-1169-x}{Eur. Phys. J. {\bf C
  64} (2009)  561--587},
\href{http://arxiv.org/abs/0904.3513}{{\tt arXiv:0904.3513 [hep-ex]}}.

\bibitem{zeus_jes1}
{ZEUS} Collaboration, S.~Chekanov et al., {\em {High mass dijet cross-sections
  in photoproduction at HERA}\/},
  \href{http://dx.doi.org/10.1016/S0370-2693(02)01327-8}{Phys. Lett. {\bf B
  531} (2002)  9--27},
\href{http://arxiv.org/abs/hep-ex/0112030}{{\tt arXiv:hep-ex/0112030
  [hep-ex]}}.

\bibitem{zeus_jes2}
{ZEUS} Collaboration, S.~Chekanov et al., {\em {Dijet photoproduction at HERA
  and the structure of the photon}\/},
  \href{http://dx.doi.org/10.1007/s100520200936}{Eur. Phys. J. {\bf C 23}
  (2002)  615--631},
\href{http://arxiv.org/abs/hep-ex/0112029}{{\tt arXiv:hep-ex/0112029
  [hep-ex]}}.

\bibitem{Wing:2002fc}
M.~Wing, {\em {Setting the jet energy scale for the ZEUS calorimeter}\/},
\href{http://arxiv.org/abs/hep-ex/0206036}{{\tt arXiv:hep-ex/0206036
  [hep-ex]}}.

\bibitem{triggerperformance}
{ATLAS} Collaboration, {\em {Performance of the ATLAS Trigger System in
  2010}\/},  \href{http://dx.doi.org/10.1140/epjc/s10052-011-1849-1}{Eur. Phys.
  J. {\bf C 72} (2012)  1849},
\href{http://arxiv.org/abs/1110.1530}{{\tt arXiv:1110.1530 [hep-ex]}}.

\bibitem{LArReadiness_mod}
{ATLAS} Collaboration, {\em {Readiness of the ATLAS Liquid Argon Calorimeter
  for LHC Collisions}\/},
  \href{http://dx.doi.org/10.1140/epjc/s10052-010-1354-y}{Eur. Phys. J. {\bf C
  70} (2010)  723--753},
\href{http://arxiv.org/abs/0912.2642}{{\tt arXiv:0912.2642 [physics.ins-det]}}.

\bibitem{TileReadiness}
{ATLAS} Collaboration, {\em {Readiness of the {ATLAS} Tile calorimeter for
  {LHC} collisions}\/},
  \href{http://dx.doi.org/10.1140/epjc/s10052-010-1508-y}{Eur. Phys. J. {\bf C
  70} (2010)  1193--1236}, \href{http://arxiv.org/abs/1007.5423}{{\tt
  arXiv:1007.5423 [physics.ins-det]}}.

\bibitem{mcforlhc}
A.~Buckley et al., {\em {General-purpose event generators for LHC physics}\/},
  \href{http://dx.doi.org/10.1016/j.physrep.2011.03.005}{Phys. Rept. {\bf 504}
  (2011)  145--233}, \href{http://arxiv.org/abs/1101.2599}{{\tt arXiv:1101.2599
  [hep-ph]}}.

\bibitem{pythia}
T.~Sjostrand, S.~Mrenna, and P.~Z. Skands, {\em {PYTHIA 6.4 physics and
  manual}\/},  \href{http://dx.doi.org/10.1088/1126-6708/2006/05/026}{JHEP {\bf
  0605} (2006)  026},
\href{http://arxiv.org/abs/0603.175}{{\tt arXiv:0603.175 [hep-ph]}}.

\bibitem{pythiapartonshower}
R.~Corke and T.~Sjostrand, {\em {Improved Parton Showers at Large Transverse
  Momenta}\/},  \href{http://dx.doi.org/10.1140/epjc/s10052-010-1409-0}{Eur.
  Phys. J. {\bf C 69} (2010)  1--18},
  \href{http://arxiv.org/abs/1003.2384}{{\tt arXiv:1003.2384 [hep-ph]}}.

\bibitem{Sjostrand:2004ef}
T.~Sjostrand and P.~Z. Skands, {\em {Transverse-momentum-ordered showers and
  interleaved multiple interactions}\/},
  \href{http://dx.doi.org/10.1140/epjc/s2004-02084-y}{Eur. Phys. J. {\bf C 39}
  (2005)  129--154}, \href{http://arxiv.org/abs/0408302}{{\tt arXiv:0408302
  [hep-ph]}}.

\bibitem{lundstring}
B.~Andersson et al., {\em {Parton fragmentation and string dynamics}\/},
  \href{http://dx.doi.org/10.1016/0370-1573(83)90080-7}{Phys. Rep. {\bf 97}
  (1983)  31--145}.

\bibitem{MC11c}
{ATLAS} Collaboration, {\em ATLAS tunes for Pythia6 and Pythia8 for MC11\/},
  \href{https://cdsweb.cern.ch/record/1363300/files/ATL-PHYS-PUB-2011-009.pdf}%
{ATLAS-PHYS-PUB-2011-009}, July, 2011.
\newblock https://cdsweb.cern.ch/record/1363300.

\bibitem{mrstlostar}
A.~Sherstnev and R.~S. Thorne, {\em {Parton Distributions for LO
  Generators}\/},  \href{http://dx.doi.org/10.1140/epjc/s10052-008-0610-x}{Eur.
  Phys. J. {\bf C55} (2008)  553--575},
\href{http://arxiv.org/abs/arXiv:0711.2473[hep-ph]}{{\tt
  arXiv:0711.2473[hep-ph]}}.

\bibitem{Herwigpp}
M.~Bahr et al., {\em Herwig++ physics and manual\/},
  \href{http://dx.doi.org/10.1140/epjc/s10052-008-0798-9}{Eur. Phys. J. {\bf C
  58} (2008)  639--707}, \href{http://arxiv.org/abs/0803.0883}{{\tt
  arXiv:0803.0883 [hep-ph]}}.

\bibitem{herwig3}
G.~Marchesini et al., {\em {Monte Carlo simulation of general hard processes
  with coherent QCD radiation}\/},
  \href{http://dx.doi.org/10.1016/0550-3213(88)90089-2}{Nucl. Phys. {\bf B 310}
  (1988)  461}.

\bibitem{herwig2}
G.~Marchesini et al., {\em {A Monte Carlo event generator for simulating hadron
  emission reactions with interfering gluons}\/},
  \href{http://dx.doi.org/10.1016/0010-4655(92)90055-4}{Comput. Phys. Commun.
  {\bf 67} (1991)  465--508}.

\bibitem{herwig}
G.~Corcella et al., {\em {HERWIG 6.5 release note}\/},
\href{http://arxiv.org/abs/0210213}{{\tt arXiv:0210213 [hep-ph]}}.

\bibitem{herwigclustermodel}
B.~R. Webber, {\em {A QCD model for jet fragmentation including soft gluon
  interference}\/},
  \href{http://dx.doi.org/10.1016/0550-3213(84)90333-X}{Nucl. Phys. {\bf B 238}
  (1984)  492}.

\bibitem{HerwigppUI}
M.~Bahr, S.~Gieseke, and M.~H. Seymour, {\em Simulation of multiple partonic
  interactions in Herwig++\/},
  \href{http://dx.doi.org/10.1088/1126-6708/2008/07/076}{JHEP {\bf 0807} (2008)
   076},
\href{http://arxiv.org/abs/0803.3633}{{\tt arXiv:0803.3633 [hep-ph]}}.

\bibitem{MadGraph}
J.~Alwall, M.~Herquet, F.~Maltoni, O.~Mattelaer, and T.~Stelzer, {\em MadGraph
  5 : Going Beyond\/},  \href{http://dx.doi.org/10.1007/JHEP06(2011)128}{JHEP
  {\bf 1106} (2011)  128}, \href{http://arxiv.org/abs/1106.0522}{{\tt
  arXiv:1106.0522 [hep-ph]}}.

\bibitem{PDF-CTEQ}
J.~Pumplin, D.~Stump, J.~Huston, H.~Lai, P.~M. Nadolsky, et al., {\em {New
  generation of parton distributions with uncertainties from global QCD
  analysis}\/},  \href{http://dx.doi.org/10.1088/1126-6708/2002/07/012}{JHEP
  {\bf 0207} (2002)  012},
\href{http://arxiv.org/abs/hep-ph/0201195}{{\tt arXiv:hep-ph/0201195
  [hep-ph]}}.

\bibitem{MLM}
M.~L. Mangano, M.~Moretti, and R.~Pittau, {\em {Multijet matrix elements and
  shower evolution in hadronic collisions: $W b \bar{b}$ + $n$ jets as a case
  study}\/},  \href{http://dx.doi.org/10.1016/S0550-3213(02)00249-3}{Nucl.
  Phys. {\bf B 632} (2002)  343--362},
\href{http://arxiv.org/abs/hep-ph/0108069}{{\tt arXiv:hep-ph/0108069
  [hep-ph]}}.

\bibitem{alpgen}
M.~L. Mangano, M.~Moretti, F.~Piccinini, R.~Pittau, and A.~D. Polosa, {\em
  {ALPGEN, a generator for hard multiparton processes in hadronic
  collisions}\/},  \href{http://dx.doi.org/10.1088/1126-6708/2003/07/001}{JHEP
  {\bf 0307} (2003)  001},
\href{http://arxiv.org/abs/hep-ph/0206293}{{\tt arXiv:hep-ph/0206293
  [hep-ph]}}.

\bibitem{jimmy}
J.~Butterworth, J.~R. Forshaw, and M.~Seymour, {\em {Multiparton interactions
  in photoproduction at HERA}\/},
  \href{http://dx.doi.org/10.1007/s002880050286}{Z. Phys. {\bf C 72} (1996)
  637--646},
\href{http://arxiv.org/abs/hep-ph/9601371}{{\tt arXiv:hep-ph/9601371
  [hep-ph]}}.

\bibitem{MC11}
{ATLAS} Collaboration, {\em New ATLAS event generator tunes to 2010 data\/},
  \href{http://cdsweb.cern.ch/record/1345343/files/ATL-PHYS-PUB-2011-008.pdf}{%
ATL-PHYS-PUB-2011-008}, May, 2011.
\newblock http://cdsweb.cern.ch/record/1345343.

\bibitem{FRI-0201}
{S. Frixione and B.R. Webber}, {\em Matching NLO QCD computations and parton
  shower simulations\/},
  \href{http://dx.doi.org/10.1088/1126-6708/2002/06/029}{JHEP {\bf 0206} (2002)
   029}, \href{http://arxiv.org/abs/hep-ph/0204244}{{\tt hep-ph/0204244}}.

\bibitem{Lai:2010vv}
H.-L. Lai et al., {\em {New parton distributions for collider physics}\/},
  \href{http://dx.doi.org/10.1103/PhysRevD.82.074024}{Phys.Rev. {\bf D82}
  (2010)  074024},
\href{http://arxiv.org/abs/1007.2241}{{\tt arXiv:1007.2241 [hep-ph]}}.

\bibitem{COR-0001}
{G. Corcella} et al., {\em HERWIG 6: An Event generator for hadron emission
  reactions with interfering gluons (including supersymmetric processes)\/},
  \href{http://dx.doi.org/10.1088/1126-6708/2001/01/010}{JHEP {\bf 0101} (2001)
   010}, \href{http://arxiv.org/abs/hep-ph/0011363}{{\tt hep-ph/0011363}}.

\bibitem{FRI-0701}
S.~Frixione, P.~Nason, and C.~Oleari, {\em {Matching NLO QCD computations with
  Parton Shower simulations: the POWHEG method}\/},
  \href{http://dx.doi.org/10.1088/1126-6708/2007/11/070}{JHEP {\bf 0711} (2007)
   070},
\href{http://arxiv.org/abs/0709.2092}{{\tt arXiv:0709.2092 [hep-ph]}}.

\bibitem{KER-0401}
{B.P.~Kersevan and E.~Richter-W\c{a}s}, {\em The Monte Carlo event generator
  AcerMC version 2.0 with interfaces to PYTHIA 6.2 and HERWIG 6.5\/},   (2004)
  , \href{http://arxiv.org/abs/hep-ph/0405247}{{\tt hep-ph/0405247}}.

\bibitem{mtop2011paper}
{ATLAS} Collaboration, {\em {Measurement of the top quark mass with the
  template method in the $t\bar{t}$ $\to$ lepton + jets channel using ATLAS
  data}\/},  \href{http://dx.doi.org/10.1140/epjc/s10052-012-2046-6}{Eur. Phys.
  J. {\bf C 72} (2012)  2046}, \href{http://arxiv.org/abs/1203.5755}{{\tt
  arXiv:1203.5755 [hep-ex]}}.

\bibitem{topvetopaper2011}
{ATLAS} Collaboration, {\em {Measurement of ttbar production with a veto on
  additional central jet activity in pp collisions at $\sqrt{s} = 7$~{T}e{V}
  using the ATLAS detector}\/},
  \href{http://dx.doi.org/10.1140/epjc/s10052-012-2043-9}{Eur. Phys. J. {\bf C
  72} (2012)  2043}, \href{http://arxiv.org/abs/1203.5015}{{\tt arXiv:1203.5015
  [hep-ex]}}.

\bibitem{pythia8}
T.~Sjostrand, S.~Mrenna, and P.~Z. Skands, {\em {A Brief Introduction to PYTHIA
  8.1}\/},  \href{http://dx.doi.org/10.1016/j.cpc.2008.01.036}{Comput. Phys.
  Commun. {\bf 178} (2008)  852--867},
\href{http://arxiv.org/abs/0710.3820}{{\tt arXiv:0710.3820 [hep-ph]}}.

\bibitem{Corke:2010yf}
R.~Corke and T.~Sjostrand, {\em {Interleaved Parton Showers and Tuning
  Prospects}\/},  \href{http://dx.doi.org/10.1007/JHExP03(2011)032}{JHEP {\bf
  1103} (2011)  032},
\href{http://arxiv.org/abs/1011.1759}{{\tt arXiv:1011.1759 [hep-ph]}}.

\bibitem{Geant4}
{GEANT4} Collaboration, S.~Agostinelli et al., {\em {GEANT4: A simulation
  toolkit}\/},
\href{http://dx.doi.org/10.1016/S0168-9002(03)01368-8}{Nucl. Instrum. Meth.
  {\bf A 506} (2003)  250--303}.

\bibitem{simulation}
{ATLAS} Collaboration, {\em {The {ATLAS} simulation infrastructure}\/},
  \href{http://dx.doi.org/10.1140/epjc/s10052-010-1429-9}{Eur. Phys. J. {\bf C
  70} (2010)  823--874}, \href{http://arxiv.org/abs/1005.4568}{{\tt
  arXiv:1005.4568 [physics.ins-det]}}.

\bibitem{Bertini}
H.~W. Bertini, {\em {Intranuclear-cascade calculation of the secondary nucleon
  spectra from nucleon-nucleus interactions in the energy range 340 to 2900
  {M}e{V} and comparisons with experiment}\/},
\href{http://dx.doi.org/10.1103/PhysRev.188.1711}{Phys. Rev. {\bf A 188} (1969)
   1711--1730}.

\bibitem{Bertini1}
M.~P. Guthrie, R.~G. Alsmiller, and H.~W. Bertini, {\em {Calculation of the
  capture of negative pions in light elements and comparison with experiments
  pertaining to cancer radiotherapy}\/},
\href{http://dx.doi.org/10.1016/0029-554X(68)90054-2}{Nucl. Instrum. Meth. {\bf
  66} (1968)  29--36}.

\bibitem{Bertini2}
M.~P. Guthrie and H.~W. Bertini, {\em {News item results from medium-energy
  intranuclear-cascade calculation}\/},
  \href{http://dx.doi.org/10.1016/0375-9474(71)90710-X}{Nucl. Phys. {\bf A 169}
  (1971)  670--672}.

\bibitem{Bertini3}
N.~V. Stepanov, {\em {Statistical modeling of fission of excited atomic nuclei.
  2. calculation and comparison with experiment}\/},   In Russian, ITEP,
  Moscow, 1988.

\bibitem{QGS}
G.~Folger and J.~Wellisch, {\em {String parton models in GEANT4}\/},  eConf
  {\bf C0303241} (2003)  MOMT007,
\href{http://arxiv.org/abs/nucl-th/0306007}{{\tt arXiv:nucl-th/0306007
  [nucl-th]}}.

\bibitem{QGSP2}
N.~S. Amelin et al., {\em {Transverse flow and collectivity in
  ultrarelativistic heavy ion collisions}\/},
\href{http://dx.doi.org/10.1103/PhysRevLett.67.1523}{Phys. Rev. Lett. {\bf 67}
  (1991)  1523--1526}.

\bibitem{QGSP4}
L.~V. Bravina, L.~P. Csernai, P.~Levai, N.~S. Amelin, and D.~Strottman, {\em
  {Fluid dynamics and quark gluon string model: What we can expect for Au + Au
  collisions at 11.6-A/GeV/c}\/},
\href{http://dx.doi.org/10.1016/0375-9474(94)90669-6}{Nucl. Phys. {\bf A 566}
  (1994)  461--464}.

\bibitem{QGSP3}
N.~S. Amelin, L.~P. Csernai, E.~F. Staubo, and D.~Strottman, {\em {Collectivity
  in ultrarelativistic heavy ion collisions}\/},
\href{http://dx.doi.org/10.1016/0375-9474(92)90598-E}{Nucl. Phys. {\bf A 544}
  (1992)  463--466}.

\bibitem{QGSP5}
L.~V. Bravina, {\em {Scaling violation of transverse flow in heavy ion
  collisions at AGS energies}\/},
\href{http://dx.doi.org/10.1016/0370-2693(94)01560-Y}{Phys. Lett. {\bf B 344}
  (1995)  49--54}.

\bibitem{ATLAS:2010bfa}
{ATLAS} Collaboration, {\em {The simulation principle and performance of the
  ATLAS fast calorimeter simulation FastCaloSim}\/},
  \href{https://cds.cern.ch/record/1300517/files/ATL-PHYS-PUB-2010-013.pdf}{AT%
L-PHYS-PUB-2010-013}, October, 2010.
\newblock https://cds.cern.ch/record/1300517.

\bibitem{Aad:2010ah}
{ATLAS} Collaboration, {\em {The ATLAS Simulation Infrastructure}\/},
  \href{http://dx.doi.org/10.1140/epjc/s10052-010-1429-9}{Eur. Phys. J. {\bf C
  70} (2010)  823--874},
\href{http://arxiv.org/abs/1005.4568}{{\tt arXiv:1005.4568 [physics.ins-det]}}.

\bibitem{EndcapTBelectronPion2002}
C.~Cojocaru et al., {\em Hadronic calibration of the {ATLAS} liquid argon
  end-cap calorimeter in the pseudorapidity region $1.6<|\eta|<1.8$ in beam
  tests\/},  \href{http://dx.doi.org/10.1016/j.nima.2004.05.133}{Nucl. Instrum.
  Meth. {\bf A 531} (2004)  481--514}.

\bibitem{TopoClusters}
W.~Lampl et al., {\em Calorimeter clustering algorithms: description and
  performance\/},
  \href{http://cdsweb.cern.ch/record/1099735}{ATL-LARG-PUB-2008-002}, April,
  2008.
\newblock http://cdsweb.cern.ch/record/1099735.

\bibitem{Cacciari200657}
M.~Cacciari and G.~P. Salam, {\em Dispelling the $N^{3}$ myth for the $k_t$
  jet-finder\/},  \href{http://dx.doi.org/10.1016/j.physletb.2006.08.037}{Phys.
  Lett. {\bf B 641} (2006)  57--61}.

\bibitem{Fastjet}
M.~Cacciari, G.~P. Salam, and G.~Soyez
\newblock \href{http://fastjet.fr/}{http://fastjet.fr/}.

\bibitem{ctb2004electronseoverp}
E.~Abat et al., {\em {Combined performance studies for electrons at the 2004
  {ATLAS} combined test-beam}\/},
\href{http://dx.doi.org/10.1088/1748-0221/5/11/P11006}{JINST {\bf 5} (2010)
  P11006}.

\bibitem{ctb2004electrons}
M.~Aharrouche et al., {\em {Measurement of the response of the {ATLAS} liquid
  argon barrel calorimeter to electrons at the 2004 combined test-beam}\/},
\href{http://dx.doi.org/10.1016/j.nima.2009.12.055}{Nucl. Instrum. Meth. {\bf A
  614} (2010)  400--432}.

\bibitem{LArTB02uniformity}
J.~Colas et al., {\em {Response uniformity of the {ATLAS} liquid argon
  electromagnetic calorimeter}\/},
  \href{http://dx.doi.org/10.1016/j.nima.2007.08.157}{Nucl. Instrum. Meth. {\bf
  A 582} (2007)  429--455},
\href{http://arxiv.org/abs/0709.1094}{{\tt arXiv:0709.1094 [physics.ins-det]}}.

\bibitem{LArTB02linearity}
M.~Aharrouche et al., {\em {Energy linearity and resolution of the {ATLAS}
  electromagnetic barrel calorimeter in an electron test-beam}\/},
  \href{http://dx.doi.org/10.1016/j.nima.2006.07.053}{Nucl. Instrum. Meth. {\bf
  A 568} (2006)  601--623}.

\bibitem{Tile2002}
P.~Adragna et al., {\em {Testbeam studies of production modules of the {ATLAS}
  {T}ile calorimeter}\/},
\href{http://dx.doi.org/10.1016/j.nima.2009.04.009}{Nucl. Instrum. Meth. {\bf A
  606} (2009)  362--394}.

\bibitem{Pinfold:2008zzb}
J.~Pinfold et al., {\em {Performance of the {ATLAS} liquid argon endcap
  calorimeter in the pseudorapidity region $2.5 < |\eta| < 4.0$ in beam
  tests}\/},
\href{http://dx.doi.org/10.1016/j.nima.2008.05.033}{Nucl. Instrum. Meth. {\bf A
  593} (2008)  324--342}.

\bibitem{LArTB02muons}
M.~Aharrouche et al., {\em {Study of the response of {ATLAS} electromagnetic
  liquid argon calorimeters to muons}\/},
\href{http://dx.doi.org/10.1016/j.nima.2009.05.021}{Nucl. Instrum. Meth. {\bf A
  606} (2009)  419--431}.

\bibitem{Atlaselectronpaper}
{ATLAS} Collaboration, {\em {Electron performance measurements with the ATLAS
  detector using the 2010 LHC proton-proton collision data}\/},
  \href{http://dx.doi.org/10.1140/epjc/s10052-012-1909-1}{Eur. Phys. J. {\bf C
  72} (2012)  1909}, \href{http://arxiv.org/abs/1110.3174}{{\tt arXiv:1110.3174
  [hep-ex]}}.

\bibitem{beambackground2011}
{ATLAS} Collaboration, {\em {Characterisation and mitigation of beam-induced
  backgrounds observed in the ATLAS detector during the 2011 proton-proton
  run}\/},  \href{http://dx.doi.org/10.1088/1748-0221/8/07/P07004}{JINST {\bf
  1307} (2013)  P07004},
\href{http://arxiv.org/abs/1303.0223}{{\tt arXiv:1303.0223 [hep-ex]}}.

\bibitem{Atlasetmiss}
{ATLAS} Collaboration, {\em {Performance of Missing Transverse Momentum
  Reconstruction in Proton-Proton Collisions at 7 TeV with ATLAS}\/},
  \href{http://dx.doi.org/10.1140/epjc/s10052-011-1844-6}{Eur. Phys. J. {\bf C
  72} (2012)  1844},
\href{http://arxiv.org/abs/1108.5602}{{\tt arXiv:1108.5602 [hep-ex]}}.

\bibitem{Beringer:1900zz}
{Particle Data Group} Collaboration, J.~Beringer et al., {\em {Review of
  Particle Physics (RPP)}\/},
\href{http://dx.doi.org/10.1103/PhysRevD.86.010001}{Phys.Rev. {\bf D 86} (2012)
   010001}.

\bibitem{Aad:2011eu}
{ATLAS} Collaboration, {\em {Measurement of the Inelastic Proton-Proton
  Cross-Section at $\sqrt{s}=7$ TeV with the ATLAS Detector}\/},
  \href{http://dx.doi.org/10.1038/ncomms1472}{Nature Commun. {\bf 2} (2011)
  463},
\href{http://arxiv.org/abs/1104.0326}{{\tt arXiv:1104.0326 [hep-ex]}}.

\bibitem{Aad:2011dr}
{ATLAS} Collaboration, {\em {Luminosity determination in p-p collisions at
  $\sqrt{s}=7$ {T}eV using the {ATLAS} detector at the LHC}\/},
  \href{http://dx.doi.org/10.1140/epjc/s10052-011-1630-5}{Eur. Phys. J. {\bf C
  71} (2011)  1630}, \href{http://arxiv.org/abs/1101.2185}{{\tt arXiv:1101.2185
  [hep-ex]}}.

\bibitem{Cacciari:2008jetarea}
M.~Cacciari, G.~P. Salam, and G.~Soyez, {\em The Catchment Area of Jets\/},
  \href{http://dx.doi.org/10.1088/1126-6708/2008/04/005}{JHEP {\bf 0804} (2008)
   005},
\href{http://arxiv.org/abs/0802.1188}{{\tt arXiv:0802.1188 [hep-ph]}}.

\bibitem{cscbook}
{ATLAS} Collaboration, {\em {Expected performance of the {ATLAS} experiment -
  detector, trigger and physics}\/},
\href{http://arxiv.org/abs/0901.0512}{{\tt arXiv:0901.0512 [hep-ex]}}.

\bibitem{Lendermann:2009ah}
V.~Lendermann et al., {\em {Combining Triggers in HEP Data Analysis}\/},
  \href{http://dx.doi.org/10.1016/j.nima.2009.03.173}{Nucl. Instrum. Meth. {\bf
  A 604} (2009)  707--718}, \href{http://arxiv.org/abs/0901.4118}{{\tt
  arXiv:0901.4118 [hep-ex]}}.

\bibitem{jerpaper2010}
{ATLAS} Collaboration, {\em {Jet energy resolution in proton-proton collisions
  at $\sqrt{s}=7$ TeV recorded in 2010 with the ATLAS detector}\/},
  \href{http://dx.doi.org/10.1140/epjc/s10052-013-2306-0}{Eur. Phys. J. {\bf C
  73} (2013)  2306},
\href{http://arxiv.org/abs/1210.6210}{{\tt arXiv:1210.6210 [hep-ex]}}.

\bibitem{egammaTrigger2}
{ATLAS} Collaboration, {\em Performance of the Electron and Photon Trigger in
  p-p Collisions at $\sqrt{s} = 7$ \TeV\/},
  \href{http://cdsweb.cern.ch/record/1375551}{ATLAS-CONF-2011-114}, August,
  2011.
\newblock http://cdsweb.cern.ch/record/1375551.

\bibitem{Resolution2010}
{ATLAS} Collaboration, {\em Jet energy resolution from in-situ techniques with
  the ATLAS detector using proton-proton collisions at a centre of mass energy
  $\sqrt{s}$ = 7 {T}eV\/},
  \href{http://cdsweb.cern.ch/record/1281311/files/ATLAS-CONF-2010-054.pdf}{AT%
LAS-CONF-2010-054}, July, 2010.
\newblock http://cdsweb.cern.ch/record/1281311.

\bibitem{ZJetPaper}
{ATLAS} Collaboration, {\em {Measurement of the production cross section for
  Z/gamma* in association with jets in pp collisions at $\sqrt{s}=7 \TeV$ with
  the ATLAS Detector}\/},
  \href{http://dx.doi.org/10.1103/PhysRevD.85.032009}{Phys. Rev. {\bf D 85}
  (2012)  032009}, \href{http://arxiv.org/abs/1111.2690}{{\tt arXiv:1111.2690
  [hep-ex]}}.

\bibitem{asymErrors}
R.~Barlow, {\em Asymmetric Errors\/},  in {\em Proceedings of the PHYSTAT2003
  conference}.
\newblock SLAC, September, 2003.
\newblock \href{http://arxiv.org/abs/physics/0401042}{{\tt
  arXiv:physics/0401042}}.

\bibitem{photon-isolation}
{ATLAS} Collaboration, {\em {Measurement of the inclusive isolated prompt
  photon cross section in pp collisions at $\sqrt{s}$ = 7 TeV with the ATLAS
  detector}\/},  \href{http://dx.doi.org/10.1103/PhysRevD.83.052005}{Phys. Rev.
  {\bf D 83} (2011)  052005}, \href{http://arxiv.org/abs/1012.4389}{{\tt
  arXiv:1012.4389 [hep-ex]}}.

\bibitem{Gabriel:1993ai}
T.~Gabriel, D.~E. Groom, P.~Job, N.~Mokhov, and G.~Stevenson, {\em {Energy
  dependence of hadronic activity}\/},
\href{http://dx.doi.org/10.1016/0168-9002(94)91317-X}{Nucl. Instrum. Meth. {\bf
  A 338} (1994)  336--347}.

\bibitem{Perugia2010}
P.~Z. Skands, {\em {Tuning Monte Carlo generators: The Perugia tunes}\/},
  \href{http://dx.doi.org/10.1103/PhysRevD.82.074018}{Phys. Rev. {\bf D 82}
  (2010)  074018},
\href{http://arxiv.org/abs/1005.3457}{{\tt arXiv:1005.3457 [hep-ph]}}.

\bibitem{CTB04pion}
E.~Abat et al., {\em {Study of energy response and resolution of the {ATLAS}
  barrel calorimeter to hadrons of energies from 20 {G}eV to 350 {G}eV}\/},
\href{http://dx.doi.org/10.1016/j.nima.2010.04.054}{Nucl. Instrum. Meth. {\bf A
  621} (2010)  134--150}.

\bibitem{TauEnergyUncertainty2012}
{ATLAS} Collaboration, {\em Determination of the tau energy scale and the
  associated systematic uncertainty in proton-proton collisions at $\sqrt{s}$ =
  7 TeV with the ATLAS detector at the LHC in 2011\/},
  \href{http://cdsweb.cern.ch/record/1453781/files/ATLAS-CONF-2012-054.pdf}{AT%
LAS-CONF-2012-054}, June, 2012.
\newblock http://cdsweb.cern.ch/record/1453781.

\bibitem{mv1}
{ATLAS collaboration}, {\em Measurement of the b-tag Efficiency in a Sample of
  Jets Containing Muons with 5 fb−1 of Data from the ATLAS Detector\/},
  ~{\href{http://cdsweb.cern.ch/record/1435197}{ATLAS-CONF-2012-043}}, March,
  2012.
\newblock {http://cdsweb.cern.ch/record/1435197}.

\bibitem{muon}
{ATLAS} Collaboration, {\em {Muon reconstruction efficiency and momentum
  resolution of the ATLAS experiment in proton-proton collisions at
  $\sqrt{s}$=7 TeV in 2010}\/},
\href{http://arxiv.org/abs/1404.4562}{{\tt arXiv:1404.4562 [hep-ex]}}.

\bibitem{MET2011}
{ATLAS} Collaboration, {\em {Performance of Missing Transverse Momentum
  Reconsctruction in pp collisions at $\sqrt{s}=7$~TeV with ATLAS}\/},
  \href{http://dx.doi.org/10.1140/epjc/s10052-011-1844-6}{Eur. Phys. J {\bf C
  72} (2011)  1844}, \href{http://arxiv.org/abs/1108.5602}{{\tt arXiv:1108.5602
  [hep-ex]}}.

\bibitem{Schwartz2}
J.~Gallicchio and M.~D. Schwartz, {\em {Quark and Gluon Tagging at the LHC}\/},
   \href{http://dx.doi.org/10.1103/PhysRevLett.107.172001}{Phys. Rev. Lett.
  {\bf 107} (2011)  172001},
\href{http://arxiv.org/abs/1106.3076}{{\tt arXiv:1106.3076 [hep-ph]}}.

\bibitem{ATLASquarkgluon}
{ATLAS} Collaboration, {\em {Light-quark and gluon jet discrimination in pp
  collisions at $\sqrt{s}$ = 7 TeV with the ATLAS detector}\/},
\href{http://arxiv.org/abs/1405.6583}{{\tt arXiv:1405.6583 [hep-ex]}}.

\bibitem{dijetFlav}
{ATLAS} Collaboration, {\em {Measurement of the flavour composition of dijet
  events in $pp$ collisions at $\sqrt{s}=7$ TeV with the ATLAS detector}\/},
  \href{http://dx.doi.org/10.1140/epjc/s10052-013-2301-5}{Eur. Phys. J. {\bf C
  73} (2013)  2301},
\href{http://arxiv.org/abs/1210.0441}{{\tt arXiv:1210.0441 [hep-ex]}}.

\bibitem{Bowler:1981sb}
M.~Bowler, {\em {$e^+$ $e^-$ Production of Heavy Quarks in the String
  Model}\/},  \href{http://dx.doi.org/10.1007/BF01574001}{Z. Phys. {\bf C 11}
  (1981)  169}.

\bibitem{Buckley:2009bj}
A.~Buckley, H.~Hoeth, H.~Lacker, H.~Schulz, and J.~E. von Seggern, {\em
  {Systematic event generator tuning for the LHC}\/},
  \href{http://dx.doi.org/10.1140/epjc/s10052-009-1196-7}{Eur. Phys. J. {\bf C
  65} (2010)  331--357},
\href{http://arxiv.org/abs/0907.2973}{{\tt arXiv:0907.2973 [hep-ph]}}.

\bibitem{MinBias2}
{ATLAS} Collaboration, {\em {Charged-particle multiplicities in p-p
  interactions measured with the {ATLAS} detector at the {LHC}}\/},
  \href{http://dx.doi.org/10.1088/1367-2630/13/5/053033}{New J. Phys. {\bf 13}
  (2011)  053033}, \href{http://arxiv.org/abs/1012.5104}{{\tt arXiv:1012.5104
  [hep-ex]}}.

\bibitem{delphiMu}
{DELPHI} Collaboration, J.~Abdallah et al., {\em {Determination of heavy quark
  non-perturbative parameters from spectral moments in semileptonic B
  decays}\/},  \href{http://dx.doi.org/10.1140/epjc/s2005-02406-7}{Eur. Phys.
  J. {\bf C 45} (2006)  35--59},
\href{http://arxiv.org/abs/hep-ex/0510024}{{\tt arXiv:hep-ex/0510024
  [hep-ex]}}.

\bibitem{JESUncertaintyICHEP}
{ATLAS} Collaboration, {\em Jet energy scale and its systematic uncertainty for
  jets produced in proton-proton collisions at $\sqrt{s}$=7 {T}eV and measured
  with the {ATLAS} detector\/},
  \href{http://cdsweb.cern.ch/record/1281329}{ATLAS-CONF-2010-056}, July, 2010.
\newblock http://cdsweb.cern.ch/record/1281329.

\end{thebibliography}\endgroup
%
\clearpage
\onecolumn
\appendix
\section*{Appendix A: Comparison of the ATLAS JES uncertainty with previous calibrations}
The progress of the JES uncertainty is demonstrated in \figRef{fig:JES_compare}. 
The label ``2011 \insitu'' refers to the 
uncertainty documented in this 
paper, the 
uncertainty estimate on the $2010$ data-set is detailed in Ref.~\cite{jespaper2010} 
while the uncertainty determined before LHC collisions is described in Ref.~\cite{JESUncertaintyICHEP}.
The label ``2010 \insitu'' refers to the 
uncertainty derived from \insitu{} techniques in the $2010$ data-set
that is discussed as cross-check to the uncertainty derived from the
single-hadron response in Ref.~\cite{jespaper2010}.
%
\begin{figure*}[!h]
  \centering
  \subfloat[$R=0.4$ \EMJES]{\includegraphics[width=0.49\textwidth]{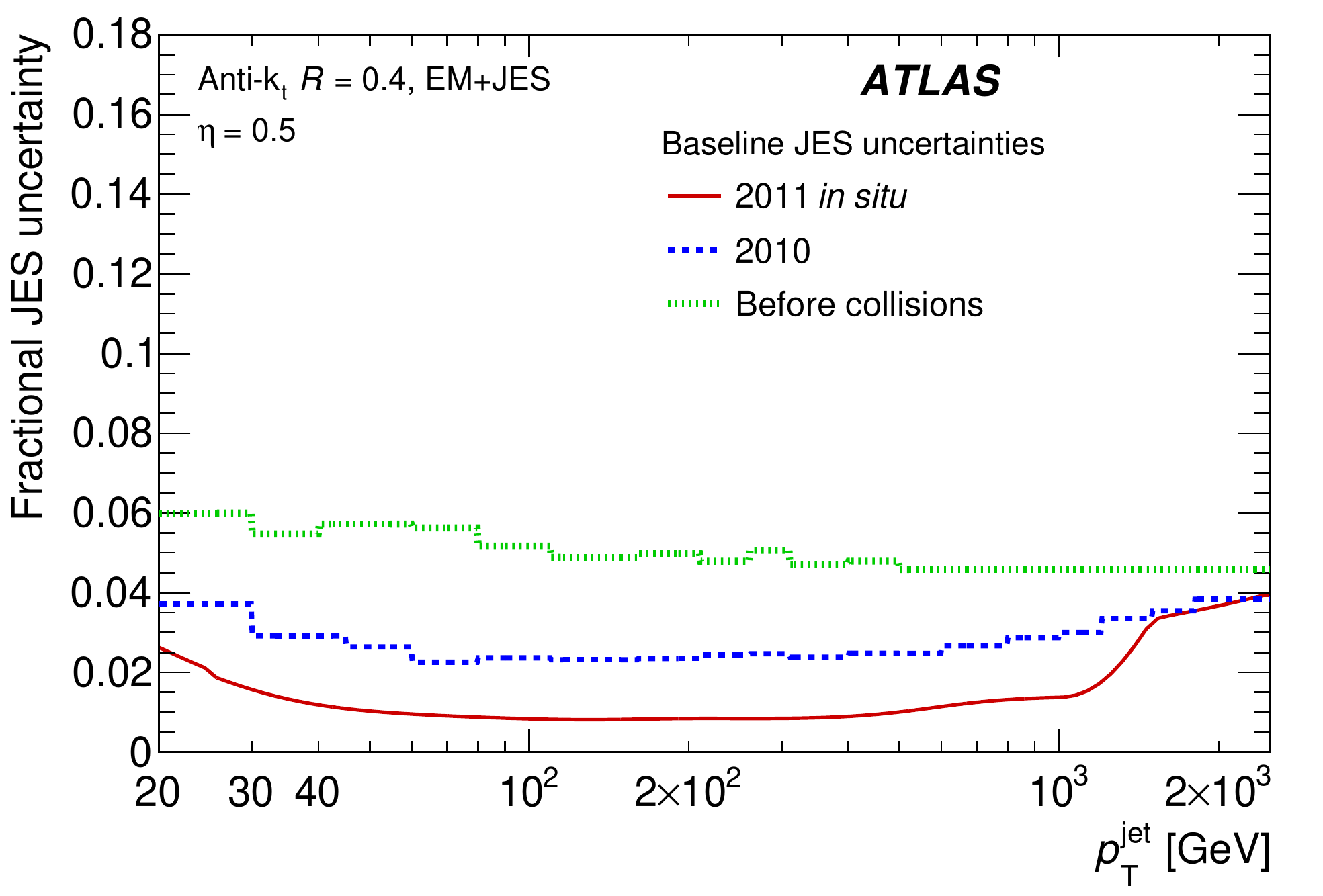}\label{fig:JES_compare_0}}
  \subfloat[$R=0.6$ \EMJES]{\includegraphics[width=0.49\textwidth]{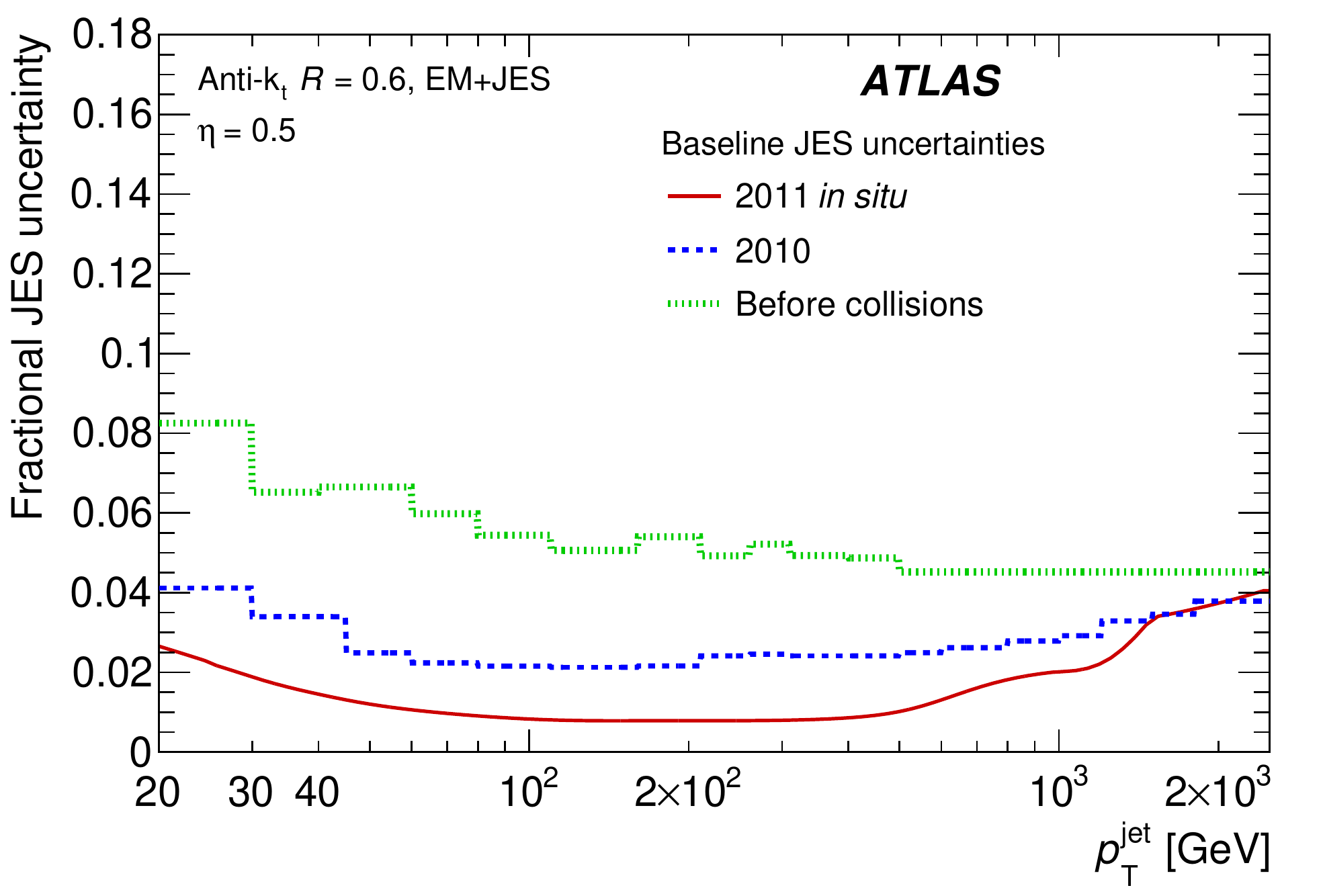}\label{fig:JES_compare_1}}\\
  \subfloat[$R=0.4$ \LCWJES]{\includegraphics[width=0.49\textwidth]{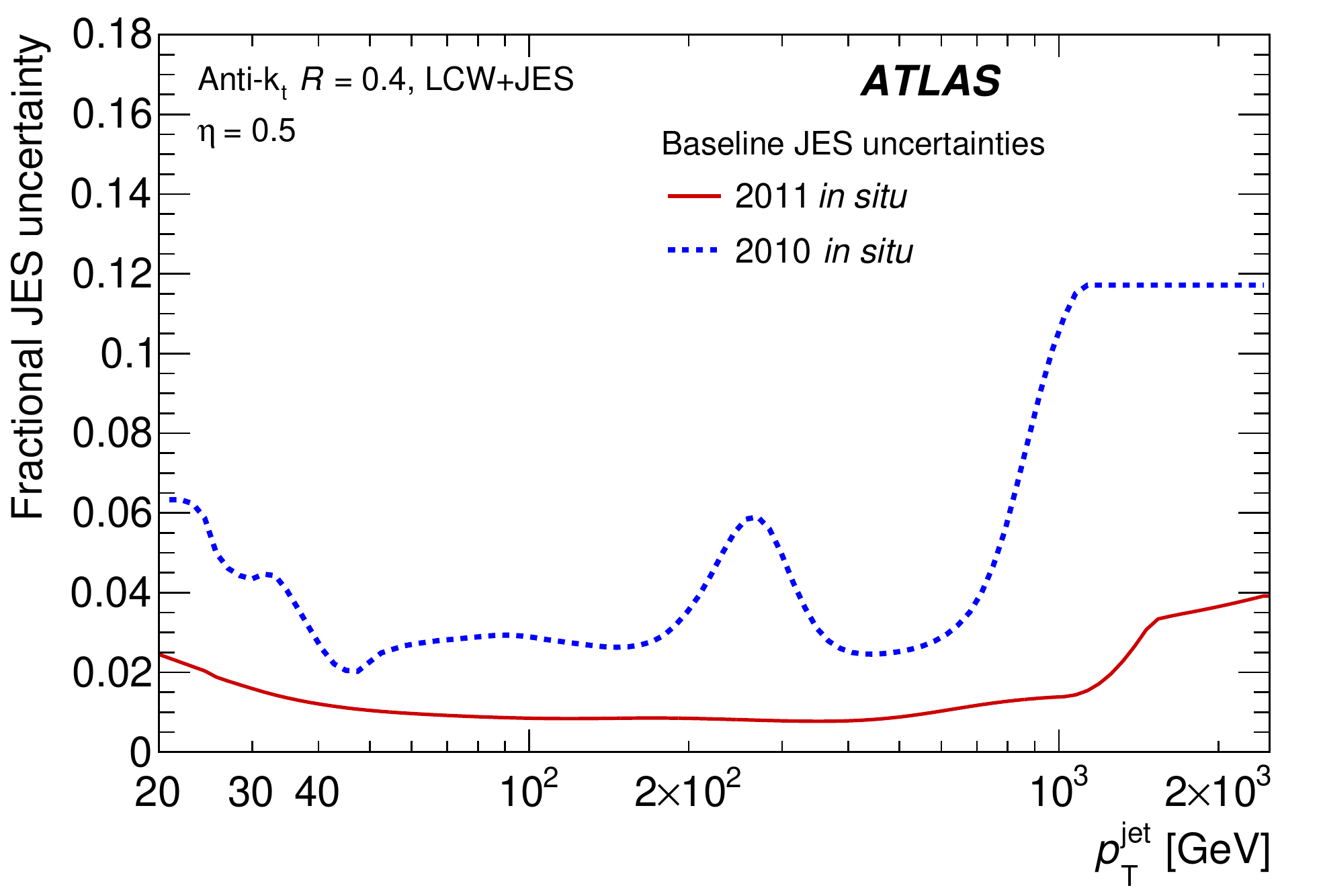}\label{fig:JES_compare_2}}
  \subfloat[$R=0.6$ \LCWJES]{\includegraphics[width=0.49\textwidth]{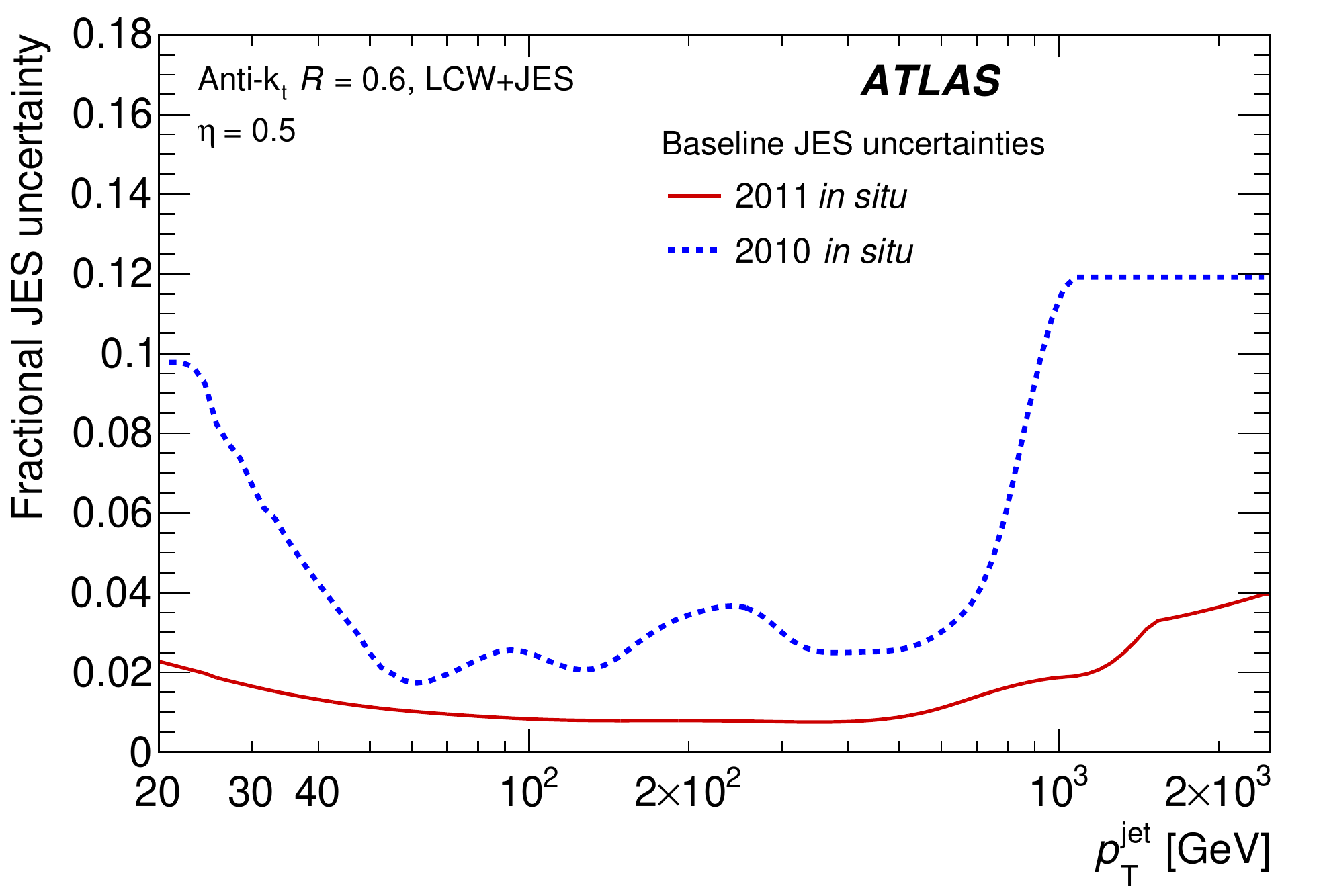}\label{fig:JES_compare_3}}
  \caption[]{
    Fractional jet energy scale systematic uncertainty for inclusive jets as a function of \ptjet{} for jets with (\subref{fig:JES_compare_0}, \subref{fig:JES_compare_2}) $R=0.4$ and (\subref{fig:JES_compare_1}, \subref{fig:JES_compare_3}) $R=0.6$  
    calibrated with the (\subref{fig:JES_compare_0}, \subref{fig:JES_compare_1}) \EMJES{} and (\subref{fig:JES_compare_2}, \subref{fig:JES_compare_3})  \LCWJES{} schemes and with $\eta = 0.5$.
    \label{fig:JES_compare}
  }
\end{figure*}

%
%

\newpage
\begin{flushleft}
{\Large The ATLAS Collaboration}

\bigskip

G.~Aad$^{\rm 48}$,
T.~Abajyan$^{\rm 21}$,
B.~Abbott$^{\rm 112}$,
J.~Abdallah$^{\rm 12}$,
S.~Abdel~Khalek$^{\rm 116}$,
O.~Abdinov$^{\rm 11}$,
R.~Aben$^{\rm 106}$,
B.~Abi$^{\rm 113}$,
M.~Abolins$^{\rm 89}$,
O.S.~AbouZeid$^{\rm 159}$
% ...
\bigskip
\\
% ...
$^{11}$ Institute of Physics, Azerbaijan Academy of Sciences, Baku, Azerbaijan\\
$^{12}$ Institut de F{\'\i}sica d'Altes Energies and Departament de F{\'\i}sica de la Universitat Aut{\`o}noma de Barcelona, Barcelona, Spain\\
% ...
$^{21}$ Physikalisches Institut, University of Bonn, Bonn, Germany\\
% ...
$^{48}$ Fakult{\"a}t f{\"u}r Mathematik und Physik, Albert-Ludwigs-Universit{\"a}t, Freiburg, Germany\\
% ...
$^{89}$ Department of Physics and Astronomy, Michigan State University, East Lansing MI, United States of America\\
% ...
$^{106}$ Nikhef National Institute for Subatomic Physics and University of Amsterdam, Amsterdam, Netherlands\\
% ...
$^{112}$ Homer L. Dodge Department of Physics and Astronomy, University of Oklahoma, Norman OK, United States of America\\
$^{113}$ Department of Physics, Oklahoma State University, Stillwater OK, United States of America\\
% ...
$^{116}$ LAL, Universit{\'e} Paris-Sud and CNRS/IN2P3, Orsay, France\\
% ...
$^{159}$ Department of Physics, University of Toronto, Toronto ON, Canada\\
% ...
$^{*}$ Deceased
\end{flushleft}

%
\end{document}